\documentclass[]{pasj01}
\jyear{2023}

\usepackage{url,lscape}

\definecolor{xlinkcolor}{rgb}{0,0,1}
\definecolor{xurlcolor}{cmyk}{0.78,0.17,0.09,0} 
\makeatletter
\let\old@ssect\@ssect 
\makeatother

\usepackage{natbib,amsmath}
\usepackage[pdfpagelabels=false]{hyperref}
\hypersetup{colorlinks=True,linkcolor=red,citecolor=xlinkcolor,urlcolor=xurlcolor}

\makeatletter
\def\@ssect#1#2#3#4#5#6{%
  \NR@gettitle{#6}
  \old@ssect{#1}{#2}{#3}{#4}{#5}{#6}
}
\makeatother

\begin{document}

\title{ 
Census for the Rest-frame Optical and UV Morphologies of Galaxies at $z=4-10$:
First Phase of Inside-Out Galaxy Formation
}
\author{Yoshiaki Ono,\altaffilmark{1}
Yuichi Harikane,\altaffilmark{1} 
Masami Ouchi,\altaffilmark{2,1,3} 
Kimihiko Nakajima,\altaffilmark{2} 
Yuki Isobe,\altaffilmark{1,4}  
Takatoshi Shibuya,\altaffilmark{5}  
Minami Nakane,\altaffilmark{1,4}   
Hiroya Umeda,\altaffilmark{1,4}    
Yi Xu,\altaffilmark{1,6}     
and 
Yechi Zhang\altaffilmark{1,6}    
}

\altaffiltext{1}{Institute for Cosmic Ray Research, The University of Tokyo, 5-1-5 Kashiwanoha, Kashiwa, Chiba 277-8582, Japan}
\altaffiltext{2}{National Astronomical Observatory of Japan, 2-21-1 Osawa, Mitaka, Tokyo 181-8588, Japan}
\altaffiltext{3}{Kavli Institute for the Physics and Mathematics of the Universe (WPI), The University of Tokyo, 5-1-5 Kashiwanoha, Kashiwa, Chiba, 277-8583, Japan}
\altaffiltext{4}{Department of Physics, Graduate School of Science, The University of Tokyo, 7-3-1 Hongo, Bunkyo, Tokyo 113-0033, Japan}
\altaffiltext{5}{Kitami Institute of Technology, 165 Koen-cho, Kitami, Hokkaido 090-8507, Japan}
\altaffiltext{6}{Department of Astronomy, Graduate School of Science, The University of Tokyo, 7-3-1 Hongo, Bunkyo, Tokyo 113-0033, Japan}

\email{ono@icrr.u-tokyo.ac.jp}


\maketitle

\begin{abstract}
We present the rest-frame optical and UV surface brightness (SB) profiles for 
$149$ galaxies with $M_{\rm opt}< -19.4$ mag at $z=4$--$10$
($29$ of which are spectroscopically confirmed with JWST NIRSpec), 
securing high signal-to-noise ratios of $10$--$135$
with deep JWST NIRCam $1$--$5\mu$m images
obtained by the CEERS survey.
We derive morphologies of our high-$z$ galaxies, 
carefully evaluating the systematics of SB profile measurements
with Monte Carlo simulations as well as the impacts of 
a) AGNs,
b) multiple clumps including galaxy mergers, 
c) spatial resolution differences with previous HST studies, 
and 
d) strong emission lines, e.g., H$\alpha$ and [{\sc Oiii}],  on optical morphologies with medium-band F410M images.
Conducting S\'ersic profile fitting to our high-$z$ galaxy SBs with GALFIT,
we obtain the effective radii of optical $r_{\rm e, opt}$ and UV $r_{\rm e, UV}$ wavelengths
ranging $r_{\rm e, opt}=0.05$--$1.6$ kpc and $r_{\rm e, UV}=0.03$--$1.7$ kpc that are
consistent with previous results within large scatters in the size luminosity relations. 
However, we find the effective radius ratio, $r_{\rm e, opt}/r_{\rm e, UV}$, 
is almost unity, $1.01^{+0.35}_{-0.22}$, over $z=4$--$10$ 
with no signatures of past inside-out star formation such found at $z\sim 0$--$2$. 
There are no spatial offsets exceeding $3\sigma$ between the optical and UV morphology centers in case of no mergers, 
indicative of major star-forming activity only found near a mass center of galaxies at $z\gtrsim 4$
probably experiencing the first phase of inside-out galaxy formation.
\end{abstract}


\section{Introduction} \label{sec:introduction}

\hspace{1em}
Investigating the evolution of galaxy morphologies over cosmic time  
offers valuable insights into understanding galaxy evolution (\citealt{2014ARA&A..52..291C}). 
The launch of the James Webb Space Telescope (JWST; \citealt{2023PASP..135f8001G}) 
has catalyzed significant advancements 
in characterizing the morphological properties of high-$z$ star-forming galaxies 
(SFGs) predominantly selected based on the rest-frame UV spectral shapes 
(\citealt{2022ApJ...938L..17Y}; \citealt{2023ApJ...942L..28T}; \citealt{2023ApJ...951...72O}; 
see also, \citealt{2023arXiv230706336L}). 
Several morphological studies have also been reported for other galaxy types 
including dusty SFGs and quiescent galaxies 
as well as galaxies selected more generally using photometric redshifts 
(\citealt{2022ApJ...937L..33S}; \citealt{2023A&A...676A..26G}; \citealt{2023arXiv230703264V}; 
\citealt{2023arXiv230706994I}; \citealt{2023arXiv230707599L}; \citealt{2023ApJ...954..113Y}). 
Furthermore, comprehensive morphological classifications have been conducted, 
encompassing visual classifications 
(\citealt{2022ApJ...938L...2F}; \citealt{2023ApJ...946L..15K}; \citealt{2023ApJ...955...94F}) 
and classifications employing machine learning techniques 
(\citealt{2023ApJ...942L..42R}; \citealt{2023arXiv230502478H}; \citealt{2023arXiv230617225T}).

To understand the evolution of galaxy morphologies, 
it is important to elucidate how galaxy sizes change across various wavelengths.
In general, the rest-frame UV continuum predominantly traces young massive stars, 
whereas the rest-frame optical continuum captures even less massive stars. 
Notably, the rest-frame UV morphologies of nearby galaxies 
tend to be patchier compared to their rest-frame optical ones  
(e.g., \citealt{2000ApJS..131..441K}; \citealt{2001AJ....122..729K}; \citealt{2002ApJS..143..113W}). 
This trend extends to galaxies up to at least $z\sim1$ 
based on Hubble Space Telescope (HST) data (\citealt{2005ApJ...631..101P}).
\cite{2014ApJ...788...28V} have investigated 
the wavelength dependence of SFG sizes up to $z\sim2$, 
finding that galaxy sizes decrease with longer wavelengths.  
Their results could be interpreted as a sign of inside-out growth, 
considering that longer wavelengths trace older less massive stars. 
However, some previous studies have reported little difference 
in morphologies in the rest-frame UV and optical for high-$z$ galaxies 
(e.g., \citealt{2000RSPTA.358.2001D}; \citealt{2005ApJ...631..101P}; \citealt{2015ApJS..219...15S}). 
Nevertheless, it is worth noting that 
HST data probe the rest-frame optical morphologies of galaxies only up to $z\sim3$.

\begin{figure}
\begin{center}
   \includegraphics[height=0.3\textheight]{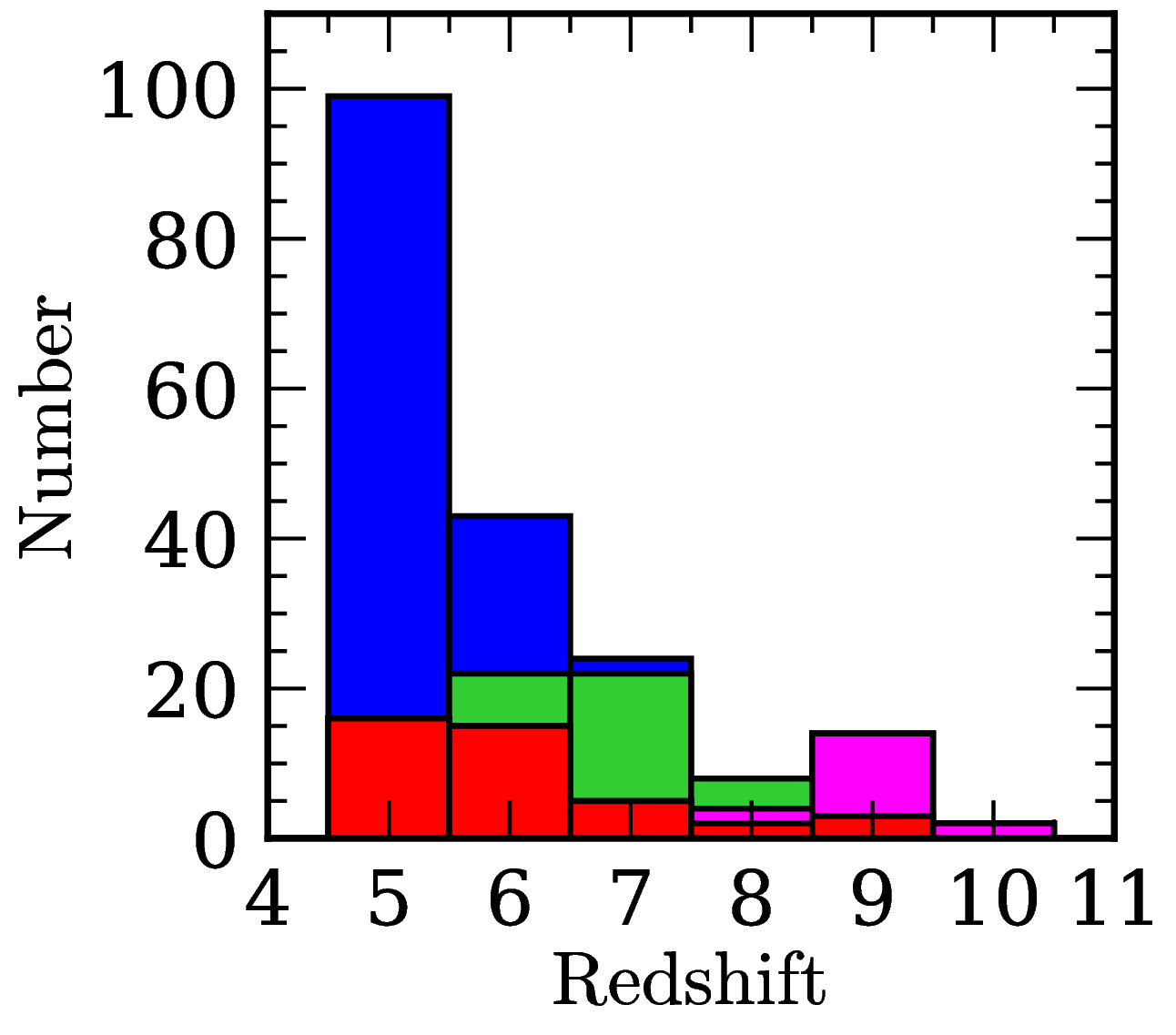}
\caption{
Histogram of redshifts of galaxies in the samples used in this study. 
The red, magenta, blue, and green histograms 
represent the samples of 
\cite{2023ApJS..269...33N}, \cite{2023ApJ...946L..13F}, 
\cite{2015ApJS..219...15S}, and \cite{2015ApJ...803...34B}, respectively. 
}
\label{fig:redshift_histogram}
\end{center}
\end{figure}

From a theoretical standpoint, 
the rest-frame UV and optical morphologies of higher-$z$ galaxies 
have been investigated using cosmological simulations 
that properly take into account the baryonic physics 
including star formation and feedback processes. 
For instance, 
\cite{2018MNRAS.477..219M} have used the results of the FIRE-2 cosmological simulations, 
to study the rest-frame UV and optical sizes of $z=5$--$10$ galaxies 
with halo masses of $10^{8-12} M_\odot$ at $z=5$. 
They have found that the rest-frame optical sizes are significantly larger compared to the rest-frame UV,   
because the rest-frame UV images are dominated by a limited number of bright clumps with young massive stars 
that are often not associated with a substantial stellar mass.
Conversely, 
based on the results of the IllustrisTNG cosmological simulations, 
\cite{2023ApJ...946...71C} have studied the morphologies of galaxies at $z=5$--$6$ 
with stellar masses exceeding $10^9 M_\odot$, 
and reported that their sizes in the rest-frame UV and optical are similar,  
probably due to their young ages. 
Similar results have also been obtained by \cite{2020MNRAS.494.5636W}  
based on the SIMBA cosmological hydrodynamic simulations.

\begin{figure}
\begin{center}
   \includegraphics[width=0.5\textwidth]{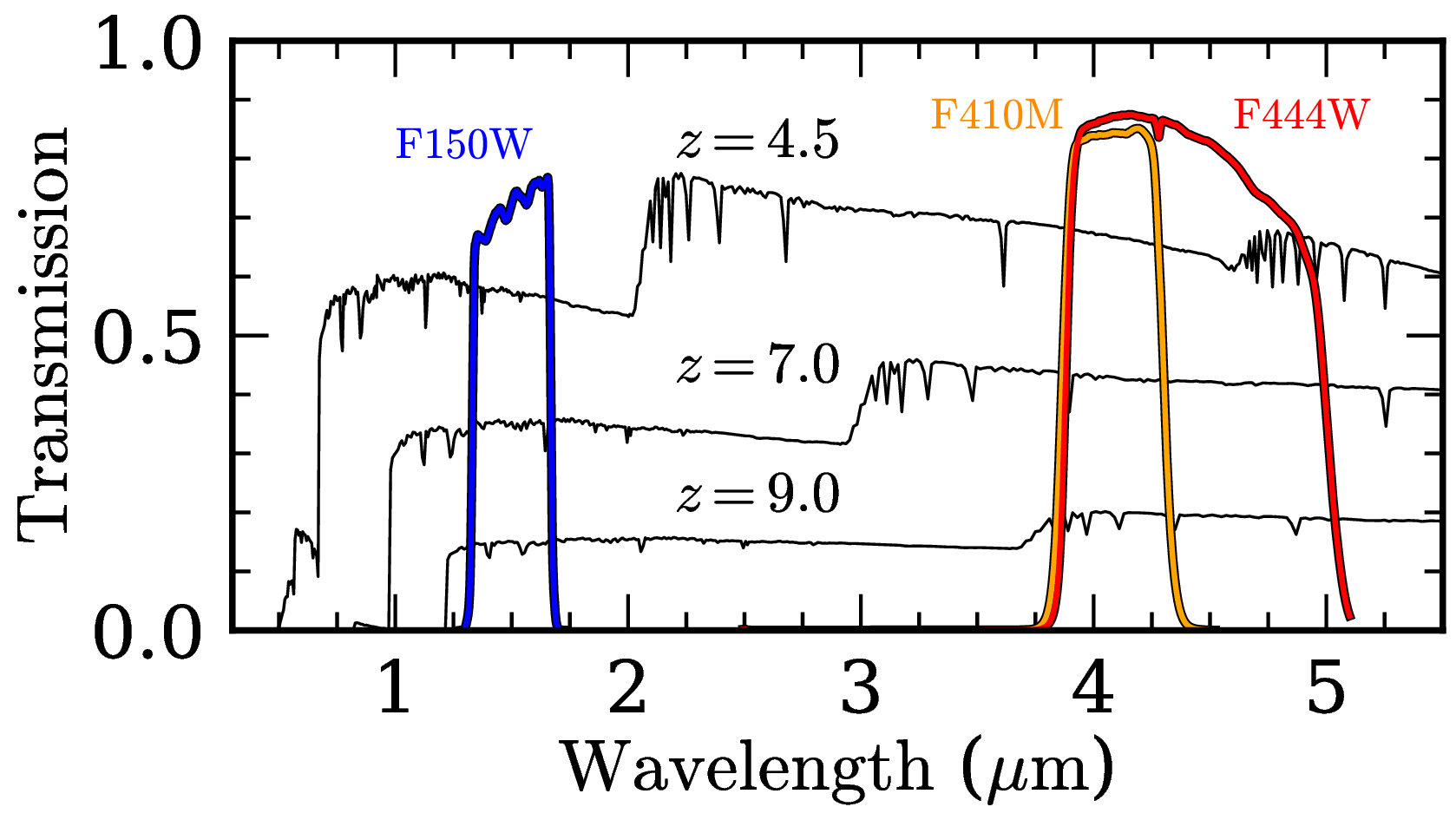}
   \includegraphics[width=0.5\textwidth]{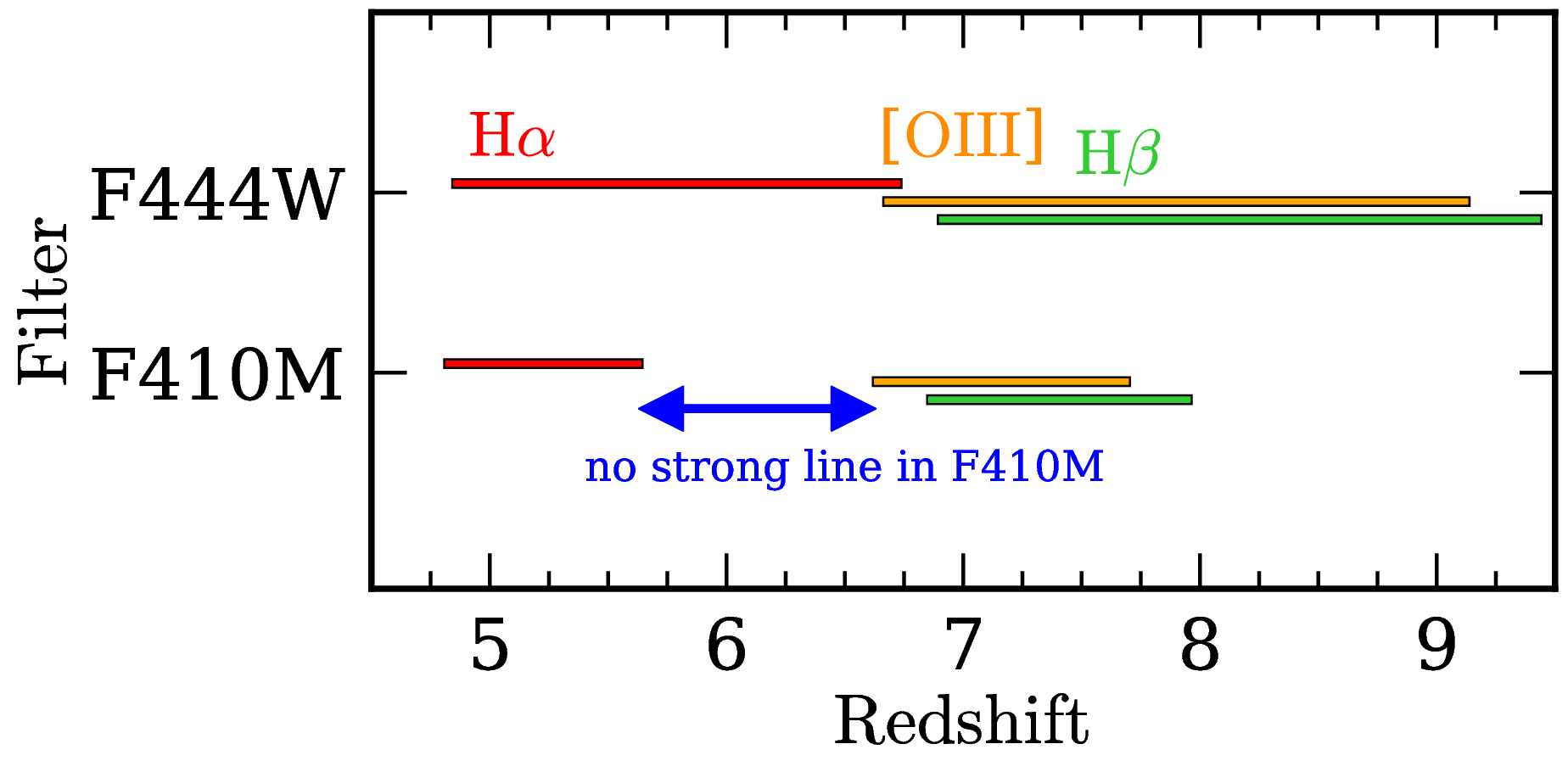}
\caption{
\textbf{Top}: 
Transmissions of two NIRCam broadband filters 
(blue: F150W; red: F444W) 
and one NIRCam medium-band filter 
(orange: F410M) 
together with spectra of SFGs 
from the \cite{2003MNRAS.344.1000B} library 
at three example redshifts of $z=4.5$, $7.0$, and $9.0$ (black lines). 
\textbf{Bottom}: 
Redshift ranges where the rest-frame strong emission lines of 
H$\alpha$ (red), [{\sc Oiii}] (orange), and H$\beta$ (green) 
enter the JWST NIRCam filters of F410M and F444W. 
For F444W, H$\alpha$ and [{\sc Oiii}] are allowed to enter seamlessly 
because of the wide wavelength range, 
while for F410M, there is a redshift gap between the redshift ranges 
where H$\alpha$ and [{\sc Oiii}] enter, i.e., $z\simeq 5.63$--$6.63$. 
}
\label{fig:SED}
\end{center}
\end{figure}

\begin{table*}
{\footnotesize
\caption{Numbers of Sources Included in the Samples and the Numbers of Duplicated Sources between the Samples
}
\begin{center}
\begin{tabular}{ccccc} \hline
	& \cite{2023ApJS..269...33N}	& \cite{2023ApJ...946L..13F}	& \cite{2015ApJS..219...15S}	& \cite{2015ApJ...803...34B} \\ \hline
Number of galaxies 										& $41$ 	& $15$ 	& $106$ 	& $28$ \\
Number of galaxies with S/N$>10$ in F150W 	& $31$ 	& $4$ 	& $105$ 	& $26$ \\
Number of galaxies with S/N$>10$ in F444W 	& $24$ 	& $10$ 	& $82$ 	& $23$ \\
Duplicate with \cite{2023ApJ...946L..13F} 						& $2$ 	& ---		& ---	 	& --- \\
Duplicate with \cite{2015ApJS..219...15S}						& $6$ 	& $0$ 	& ---	 	& --- \\
Duplicate with \cite{2015ApJ...803...34B} 						& $0$ 	& $0$ 	& $5$ 	& --- \\
\hline
\end{tabular}
\end{center}
\label{tab:sample_numbers}
}
\end{table*}

\begin{table*}
{\small
\caption{Limiting Magnitudes and PSF FWHMs of the JWST NIRCam Images Used for the Present Analysis of Source Sizes
}
\begin{center}
\begin{tabular}{cccc} \hline
\multicolumn{4}{c}{$10 \sigma$ Depth / PSF FWHM} \\ 
Field	& F150W	& F410M	& F444W \\ \hline
CEERS1 	& $28.3$ / $0\farcs061$ & $28.1$ / $0\farcs154$ & $28.3$ / $0\farcs161$  \\
CEERS2 	& $28.2$ / $0\farcs062$ & $28.1$ / $0\farcs170$ & $28.6$ / $0\farcs172$  \\
CEERS3 	& $28.4$ / $0\farcs065$ & $28.2$ / $0\farcs155$ & $28.4$ / $0\farcs159$  \\
CEERS4 	& $28.3$ / $0\farcs065$ & $28.2$ / --- & $28.2$ / $0\farcs164$  \\
CEERS5 	& $28.3$ / $0\farcs066$ & $28.2$ / --- & $28.2$ / $0\farcs169$  \\
CEERS6 	& $28.3$ / $0\farcs059$ & $28.2$ / $0\farcs157$ & $28.2$ / $0\farcs166$  \\
CEERS7 	& $28.3$ / $0\farcs062$ & $28.2$ / --- & $28.2$ / ---  \\
CEERS8 	& $28.3$ / $0\farcs061$ & $28.2$ / --- & $28.2$ / $0\farcs165$  \\
CEERS9 	& $28.3$ / $0\farcs065$ & $28.2$ / --- & $28.2$ / $0\farcs157$  \\
CEERS10	& $28.3$ / $0\farcs064$ & $28.2$ / --- & $28.2$ / $0\farcs162$  \\
\hline
\end{tabular}
\end{center}
\label{tab:limitmag}
}
\end{table*}

Thanks to the near-infrared camera onboard JWST, NIRCam  (\citealt{2005SPIE.5904....1R}), 
deep high-resolution images can now be obtained up to around $5\mu$m, 
allowing for the examination of galaxy morphologies in the rest-frame UV and optical
up to $z \simeq 9.5$.
\cite{2022ApJ...938L..17Y} have investigated the sizes of galaxies at $z\sim 7$--$10$ 
in the rest-frame UV and optical based on the JWST GLASS data 
(\citealt{2022ApJ...935..110T}), 
and found that the average ratio of sizes in the rest-frame optical to UV is consistent with unity. 
\cite{2023ApJ...951...72O} have expanded individual high-$z$ galaxy size measurements 
to faint galaxies by stacking their NIRCam images, 
thus further revealing that the size ratio is consistent with unity 
even down to fainter magnitudes. 
However, these studies have not covered a bright magnitude range exceeding $-21$ mag, 
which corresponds to the characteristic luminosity 
of $z\sim3$ galaxies, $L^\ast_{z=3}$ (\citealt{1999ApJ...519....1S}).

In this paper, we utilize JWST NIRCam images 
taken by the Cosmic Evolution Early Release Science (CEERS; \citealt{2022ApJ...940L..55F}) survey 
to determine the sizes of galaxies at $z\simeq4$--$10$ in the rest-frame UV and optical 
across a broad luminosity range. 
We investigate sizes of spectroscopically identified galaxies at $z_{\rm spec} > 4.5$ 
in the CEERS fields compiled by \cite{2023ApJS..269...33N} 
as well as galaxy candidates at similar redshifts 
selected based on photometric redshifts or the dropout technique 
(\citealt{2023ApJ...946L..13F}; \citealt{2015ApJS..219...15S}; \citealt{2015ApJ...803...34B}).

This paper is structured as follows. 
In Section \ref{sec:data}, we introduce the CEERS NIRCam data used in this study, 
and compile samples of galaxies at $z \simeq 4$--$10$ 
found in the CEERS fields in previous studies. 
In Section \ref{sec:analyses}, 
we describe our methodology for two-dimensional (2D) surface brightness (SB) profile fitting 
and conduct Monte Carlo (MC) simulations to correct systematic uncertainties in size and total magnitude measurements  
and to estimate statistical uncertainties in these measurements. 
In addition, we investigate the impact of strong emission lines 
on the rest-frame optical size measurements 
and the effect of the spatial resolution difference between the rest-frame UV and optical 
on the size measurements. 
Section \ref{sec:results} presents the results of our SB profile fittings for the galaxies at $z \simeq 4$--$10$ 
in the rest-frame UV and optical, and compares our results with those in previous work 
such as the size-luminosity relation and size ratio between the rest-frame UV and optical. 
Finally, we summarize this study in Section \ref{sec:summary}. 
Throughout this paper, 
we employ magnitudes in the AB system (\citealt{1983ApJ...266..713O}). 
A flat universe with $\Omega_{\rm m} = 0.3$, $\Omega_\Lambda = 0.7$, 
and $H_0 = 70$ km s$^{-1}$ Mpc$^{-1}$ is adopted. 
In this cosmological model, 
for instance, an angular dimension of $1.0$ arcsec 
corresponds to a physical dimension of 
$6.603$ kpc at $z=4.5$, 
$5.226$ kpc at $z=7.0$, 
and $4.463$ kpc at $z=9.0$ 
(Equation 18 of \citealt{1999astro.ph..5116H}).

\section{Data and Samples} \label{sec:data}

\hspace{1em}
We investigate sizes of galaxies at $z \simeq 4$--$10$ found in the CEERS fields. 
Firstly, we use the spectroscopic sample of \cite{2023ApJS..269...33N}, 
who have compiled spectroscopically identified galaxies in the CEERS fields 
(See also, \citealt{2023ApJ...951L..22A}; \citealt{2023ApJ...949L..25F}; \citealt{2023MNRAS.526.1657T}). 
To increase the number of sources that we examine, 
we incorporate samples of galaxy candidates at similar redshifts 
selected based on photometric redshifts or the dropout technique. 
One is the \cite{2023ApJ...946L..13F} sample at $z \sim 9$, 
which is selected from the JWST and HST data with photometric redshift values of 
$z_{\rm photo} \simeq 8$--$10$. 
In addition, we include the \cite{2015ApJS..219...15S} sample of dropout galaxies at $z \sim 5$--$7$ 
selected in \cite{2016ApJ...821..123H}  
from the HST data obtained by the 
The Cosmic Assembly Near-IR Deep Extragalactic Legacy Survey
(CANDELS; \citealt{2011ApJS..197...35G}; \citealt{2011ApJS..197...36K}). 
To compensate for the small number for $z\sim7$--$8$, 
we further add the $z\sim7$--$8$ samples from \cite{2015ApJ...803...34B}, 
which are dropout galaxies selected based on the HST data. 
Note that there are overlaps between these samples; 
in this paper, we assign a unique ID to each source as listed in Table \ref{tab:sample}.
The number of sources in each sample and the numbers of duplicated sources are summarized in Table \ref{tab:sample_numbers}.
Their redshift distributions are presented in Figure \ref{fig:redshift_histogram}.

To trace their rest-frame UV and optical continuum from galaxies at $z\simeq 4$--$10$, 
we make use of the JWST NIRCam F150W and F444W images, respectively 
(top panel of Figure \ref{fig:SED}). 
We refer the reader to \cite{2023ApJS..265....5H} and 
\textcolor{blue}{Y. Harikane et al. in preparation} 
for details about the imaging data sets. 
At this moment, our reductions of the F444W images for CEERS7 is suboptimal for reasons we do not yet fully understand; 
we therefore do not use it for our analyses. 
Although the wide wavelength coverage of F444W 
allows the capture of strong emission lines such as H$\alpha$ and [{\sc Oiii}]$5008$ 
at $z \gtrsim 5$ as demonstrated in the bottom panel of Figure \ref{fig:SED}, 
this does not significantly affect measurements of galaxy sizes 
as shown later in a comparison of measurements with F444W and F410M (Section \ref{subsec:comparison_f410m_f444w}). 
The pixel scale of the NIRCam images is $0\farcs015$ pix$^{-1}$. 
Their $10 \sigma$ limiting magnitudes are summarized in Table \ref{tab:limitmag}. 
Following the previous work (\citealt{2012ApJ...756L..12M}; \citealt{2023ApJ...951...72O}), 
we select sources with a signal-to-noise ratio (S/N) larger than $10$  
for individual analyses in each band 
based on their apparent magnitudes measured in $0\farcs2$ diameter circular apertures.
For instance, if a galaxy shows an S/N larger than $10$ in the F150W (F444W) image, 
its F150W (F444W) image is used for analysis. 
In cases where a galaxy has an S/N larger than $10$ in F150W 
but has an S/N below $10$ in F444W, 
only the F150W image is utilized for analysis.
The number of sources in each catalog that have an S/N larger than $10$ 
in each band is presented in Table \ref{tab:sample_numbers}.

When measuring galaxy sizes, it is imperative to consider image smearing by 
point spread functions (PSFs). 
To address this, we utilize empirical PSFs generated by stacking bright point sources in the actual NIRCam images. 
Within each field, we select $4$--$9$ unsaturated bright point sources with $\simeq 22$--$24$ mag. 
The PSF FWHM values are listed in Table \ref{tab:limitmag}.

\begin{figure}
\begin{center}
   \includegraphics[width=0.4\textwidth]{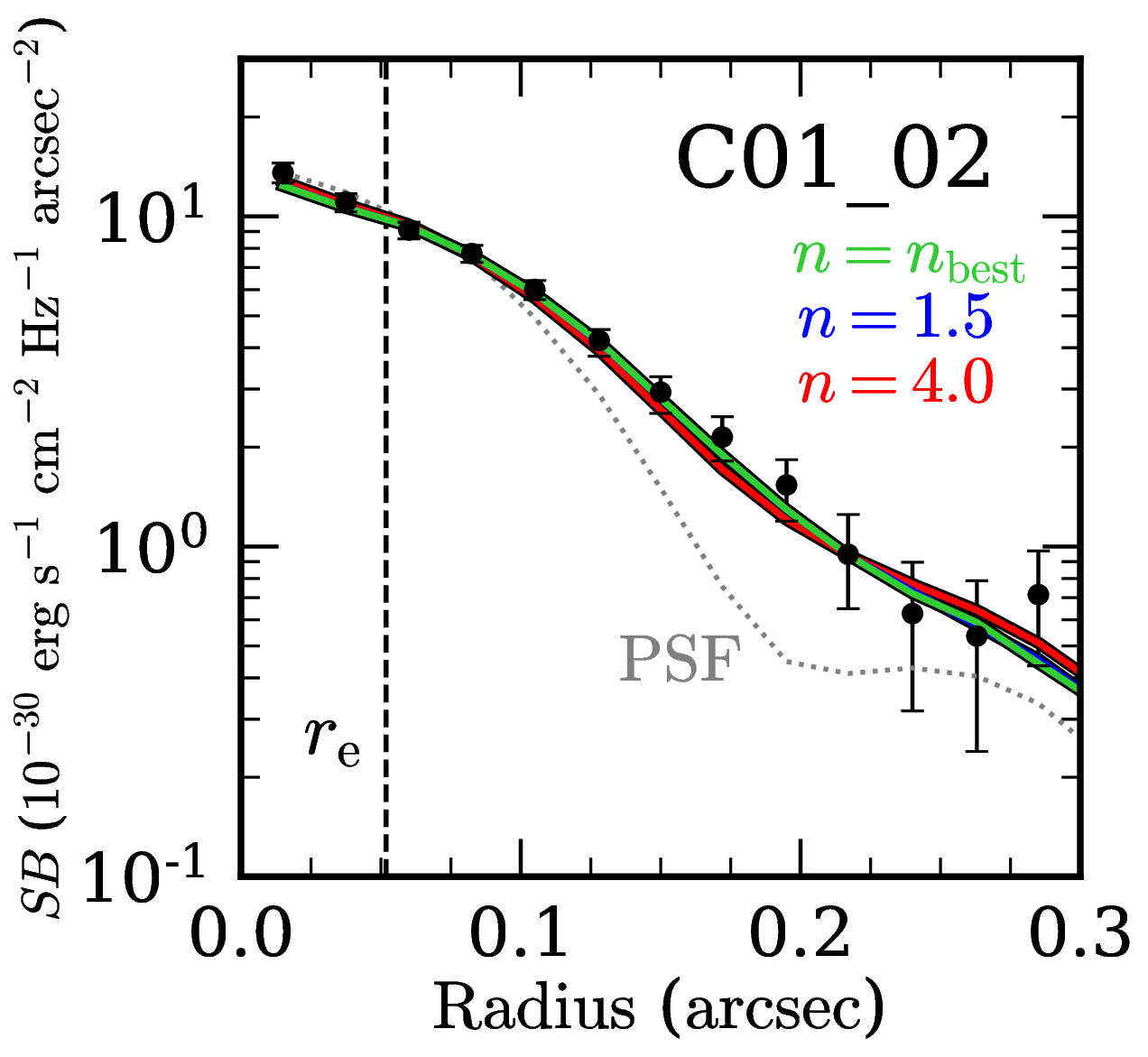}
   \includegraphics[width=0.4\textwidth]{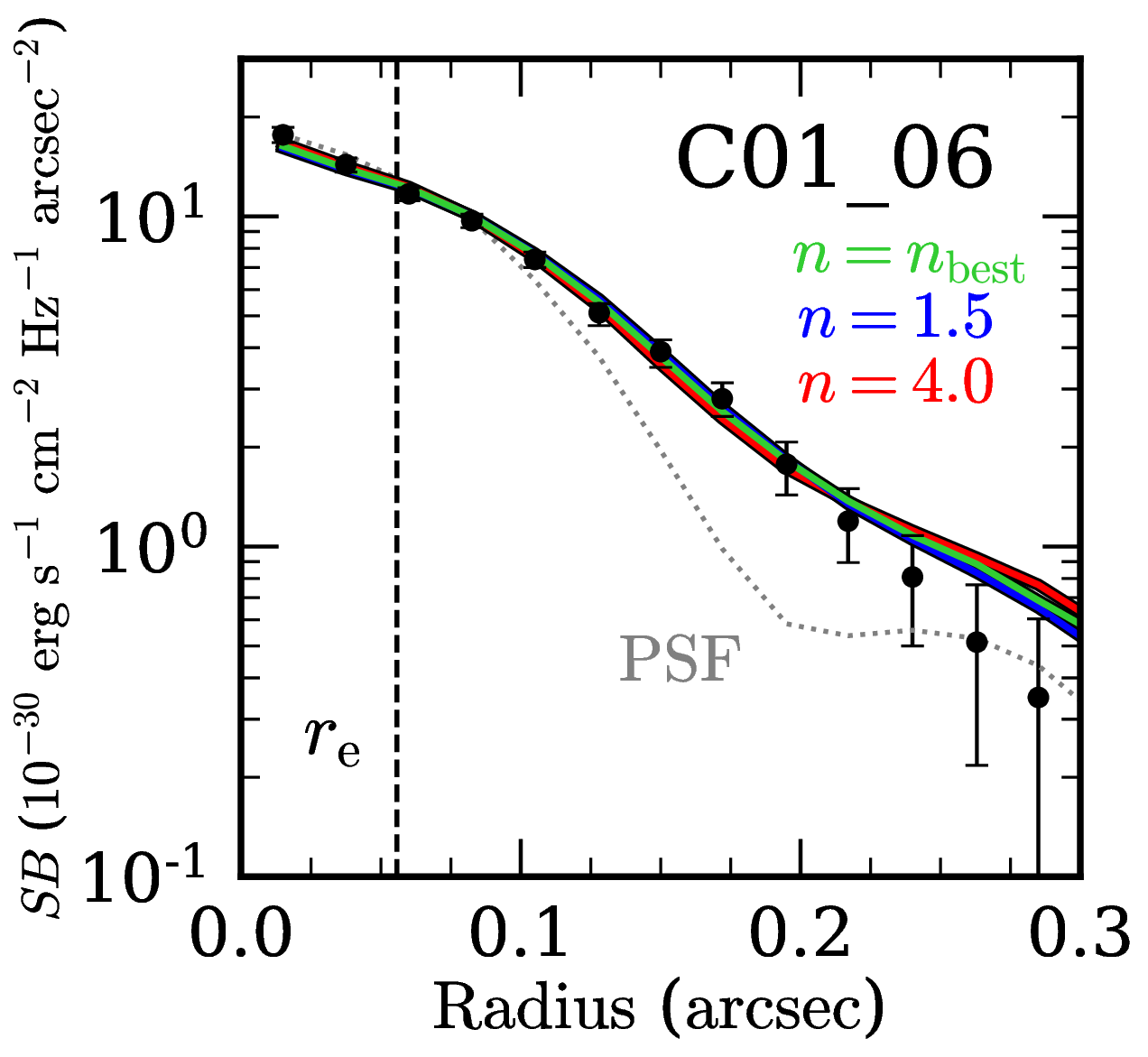}
   \includegraphics[width=0.4\textwidth]{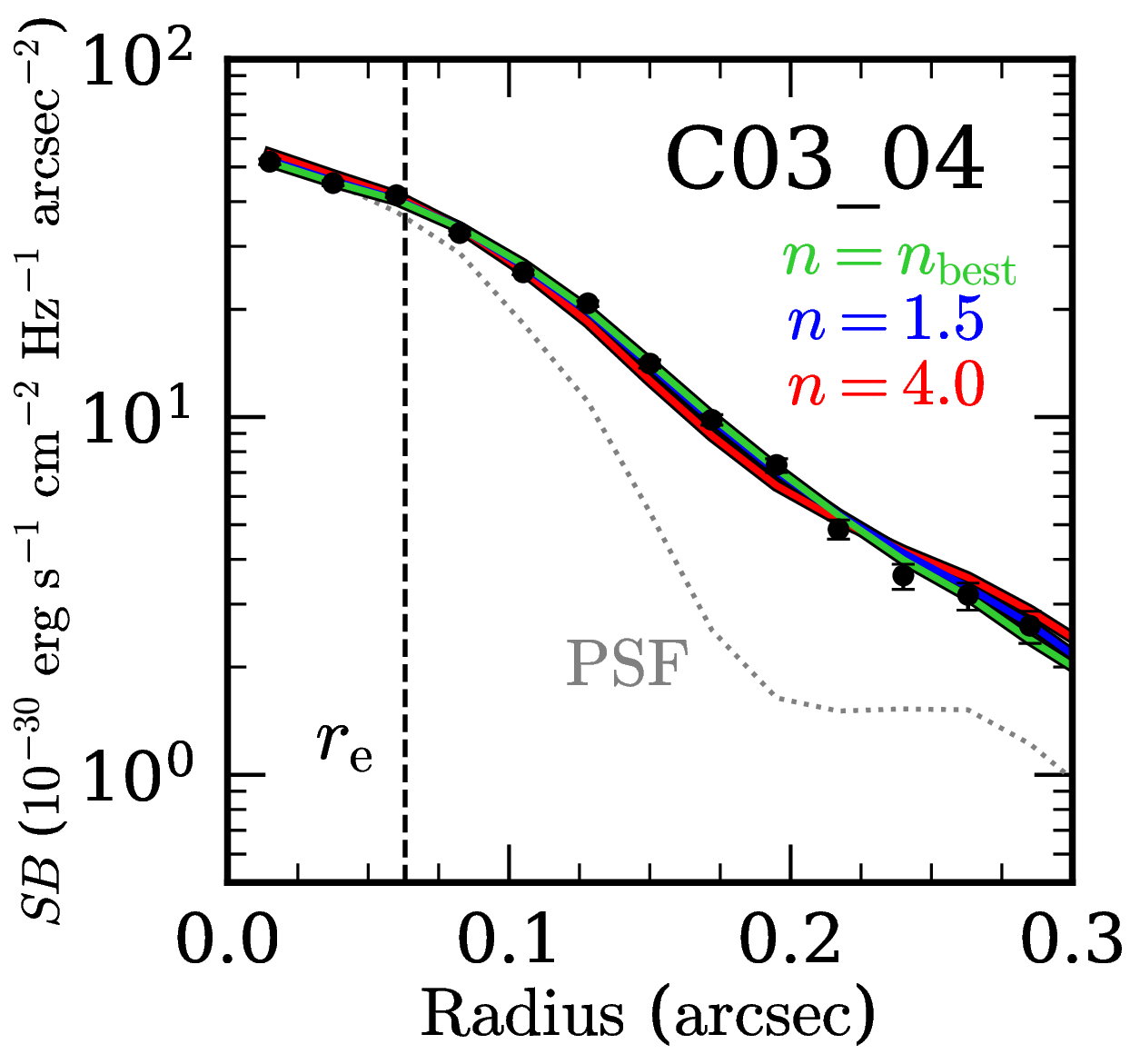}
\caption{
Examples of the rest-frame optical SB radial profiles:  
C01{\_}02 (top), C01{\_}06 (middle), and C03{\_}04 (bottom). 
In each panel, the black circles represent the observed SB profile in F444W. 
The $1\sigma$ uncertainties are calculated from the $68$th percentiles of 
radial profiles obtained at randomly selected positions in the F444W image. 
The green curve corresponds to the best-fit S{\'e}rsic profile 
where the S{\'e}rsic index is allowed to vary as a free parameter 
($n_{\rm best} = 1.4$ for C01{\_}02, 
$n_{\rm best} = 2.1$ for C01{\_}06, 
and 
$n_{\rm best} = 1.1$ for C03{\_}04). 
The blue and red curves denote the best-fit S{\'e}rsic profiles 
with a fixed S{\'e}rsic index of $n=1.5$ and $n=4.0$, respectively. 
The gray dotted curve is the PSF profile whose peak is normalized by the peak of the observed profile. 
The black vertical dashed line represents the best-fit half-light radius, $r_{\rm e}$. 
}
\label{fig:fit_radial_SBs}
\end{center}
\end{figure}

\section{Analyses} \label{sec:analyses}

\subsection{Surface Brightness Profile Fitting} \label{subsec:SB_profile_fitting}

\hspace{1em}
We measure the half-light radii of the high-$z$ galaxies in our compiled sample 
through fitting the S{\'e}rsic profile (\citealt{1968adga.book.....S}) 
to the observed 2D SB profiles. 
The S{\'e}rsic profile is defined as follows: 
\begin{equation}
\Sigma (r)
	= \Sigma_e \exp \left( - b_n \left[ \left( \frac{r}{r_e} \right)^{1/n} -1 \right] \right), 
\end{equation}
where 
$\Sigma_e$ is the surface brightness at the half-light radius $r_e$, 
and $n$ signifies the S{\'e}rsic index.  
The value of $b_n$ is specified such that $r_e$ encapsulates half of the total flux.

We deploy GALFIT version 3 
(\citealt{2002AJ....124..266P,2010AJ....139.2097P}) 
for the profile fitting. 
This software convolves a galaxy surface brightness profile with a PSF profile 
and optimizes the fits using the Levenberg-Marquardt algorithm to minimize $\chi^2$. 
The resultant parameters from GALFIT 
comprise  
the centroid coordinates, 
total magnitude, 
radius along the semi-major axis ($a$),
S{\'e}rsic index ($n$), 
axis ratio ($b/a$), 
and position angle of the fitted object. 
The circularized half-light radius, $r_e = a \sqrt{b/a}$, is calculated 
by utilizing the radius along the semi-major axis and the axis ratio, 
which is commonly used in galaxy size measurements 
in previous studies (e.g., \citealt{2012ApJ...746..162N}; \citealt{2012ApJ...756L..12M}; 
\citealt{2013ApJ...777..155O}; \citealt{2015ApJS..219...15S}; \citealt{2018ApJ...855....4K}). 
SExtractor is deployed to obtain initial parameters for the GALFIT profile fitting. 
With the exception of the S{\'e}rsic index, 
all parameters are set to vary during the profile fitting. 
Noise images are derived from the inverse square root of the weight maps 
and employed to assign weights to individual pixels during the profile fitting.  
Segmentation images generated by SExtractor are used for 
masking extraneous objects around the sources of interest.

We fix the S\'ersic index at $n=1.5$, 
which corresponds to the median value of SFGs with similar UV luminosities to those of our sources 
reported in the previous work (\citealt{2015ApJS..219...15S}).\footnote{Based on 
SB profile fittings for high-$z$ SFGs with GALFIT, 
previous studies have reported that the results remain almost the same 
when the fixed S\'ersic index value is set to $n=1.0$  
(e.g., \citealt{2013ApJ...777..155O}; \citealt{2023ApJ...951...72O}).}
In fact, when we run GALFIT with the S{\'e}rsic index left as a free parameter 
for $29$ bright spectroscopically confirmed galaxies in our sample, 
the median values of the best-fit S{\'e}rsic indices are 
$1.4$ in F150W and $1.0$ in F444W 
(see also, \citealt{2023arXiv230809076S}; 
for lower redshifts, see \citealt{2023ApJ...957...46M}). 
However, the uncertainties of the S{\'e}rsic indices for individual sources would not be small. 
For illustration,
Figure \ref{fig:fit_radial_SBs} displays the observed radial profiles for 
C01{\_}02, C01{\_}06, and C03{\_}04, 
together with the best-fit results 
where the S{\'e}rsic index $n$ is treated as a free parameter 
and where $n$ is fixed at $1.5$ or $4.0$. 
All of these cases fit the observed data points well, 
suggesting that it is challenging to determine the S{\'e}rsic index 
simultaneously with the other parameters using the currently available data. 
Therefore, we fix the value of the S{\'e}rsic index in the SB profile fittings, 
consistent with the approach taken in the previous studies.

\begin{figure*}
\begin{center}
   \includegraphics[width=0.3\textwidth]{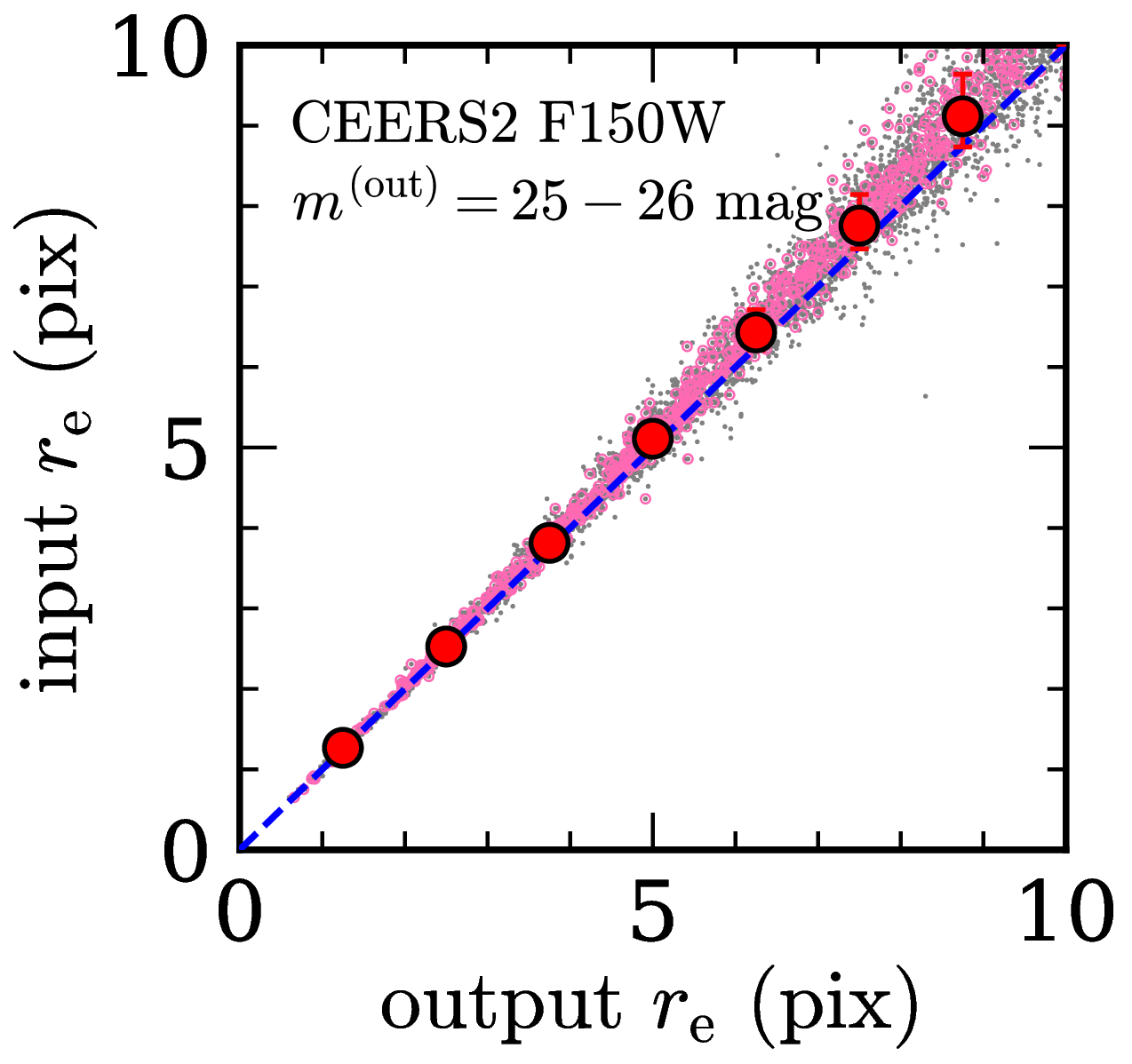}
   \includegraphics[width=0.3\textwidth]{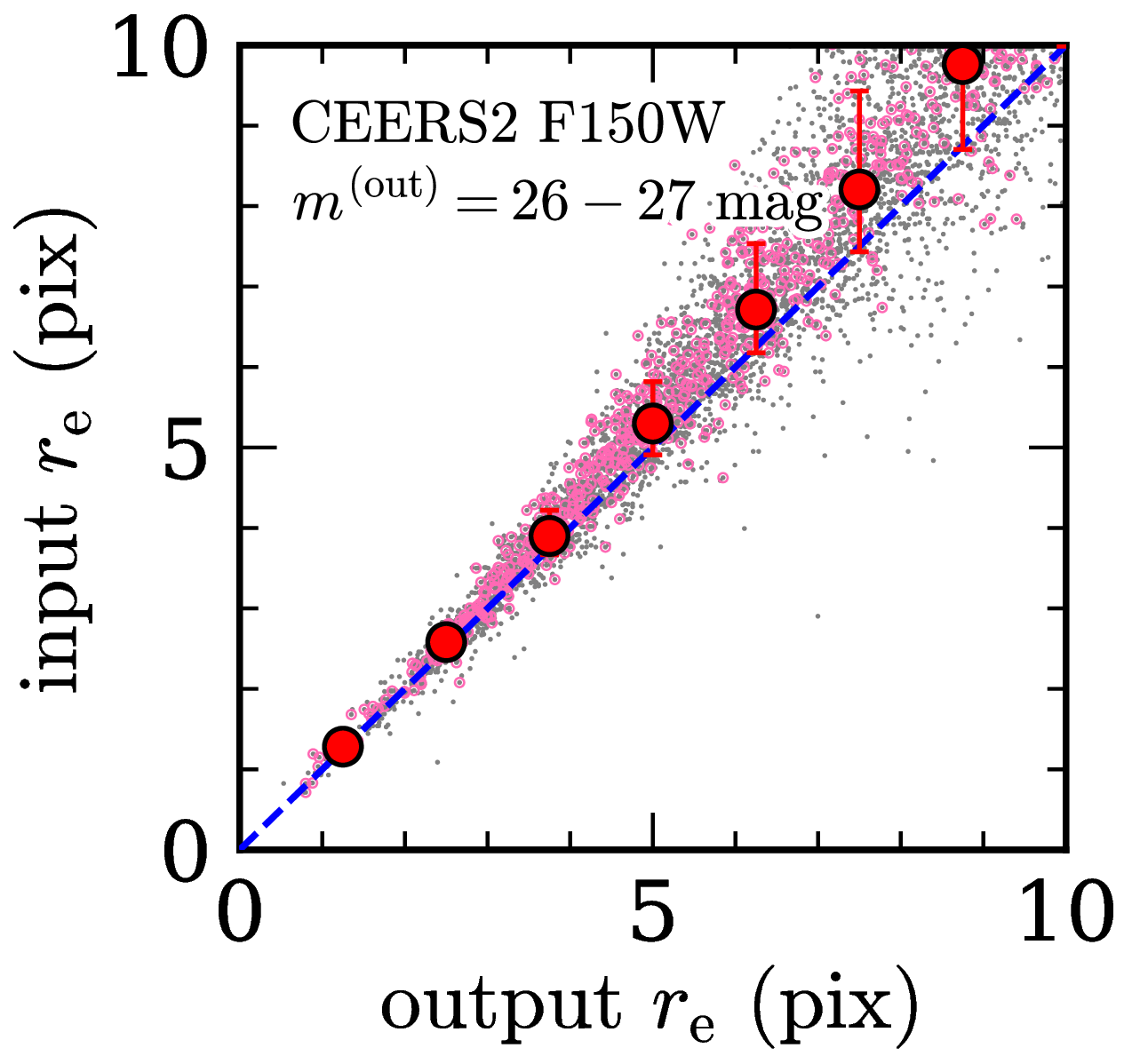}
   \includegraphics[width=0.3\textwidth]{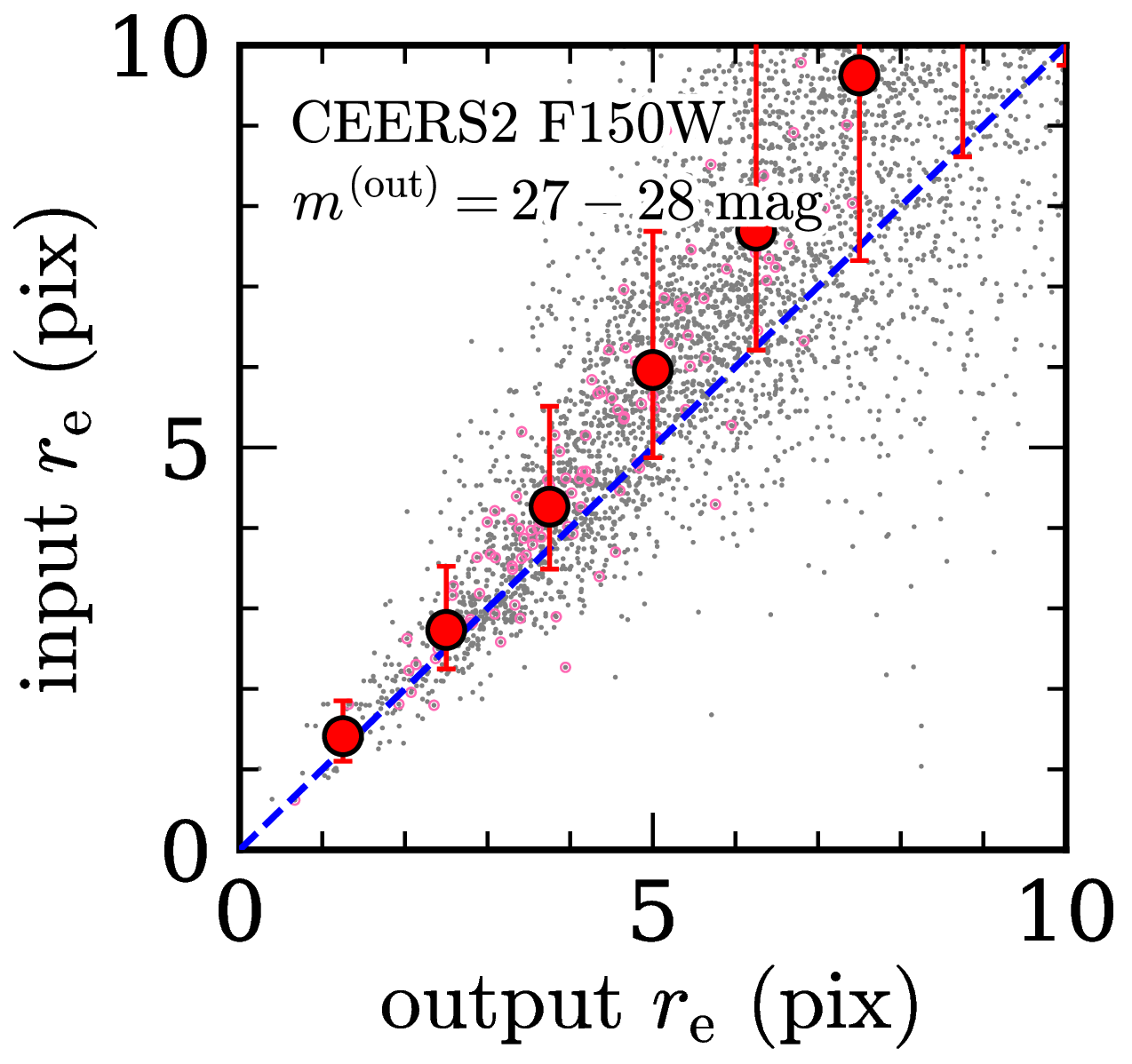}
   \includegraphics[width=0.3\textwidth]{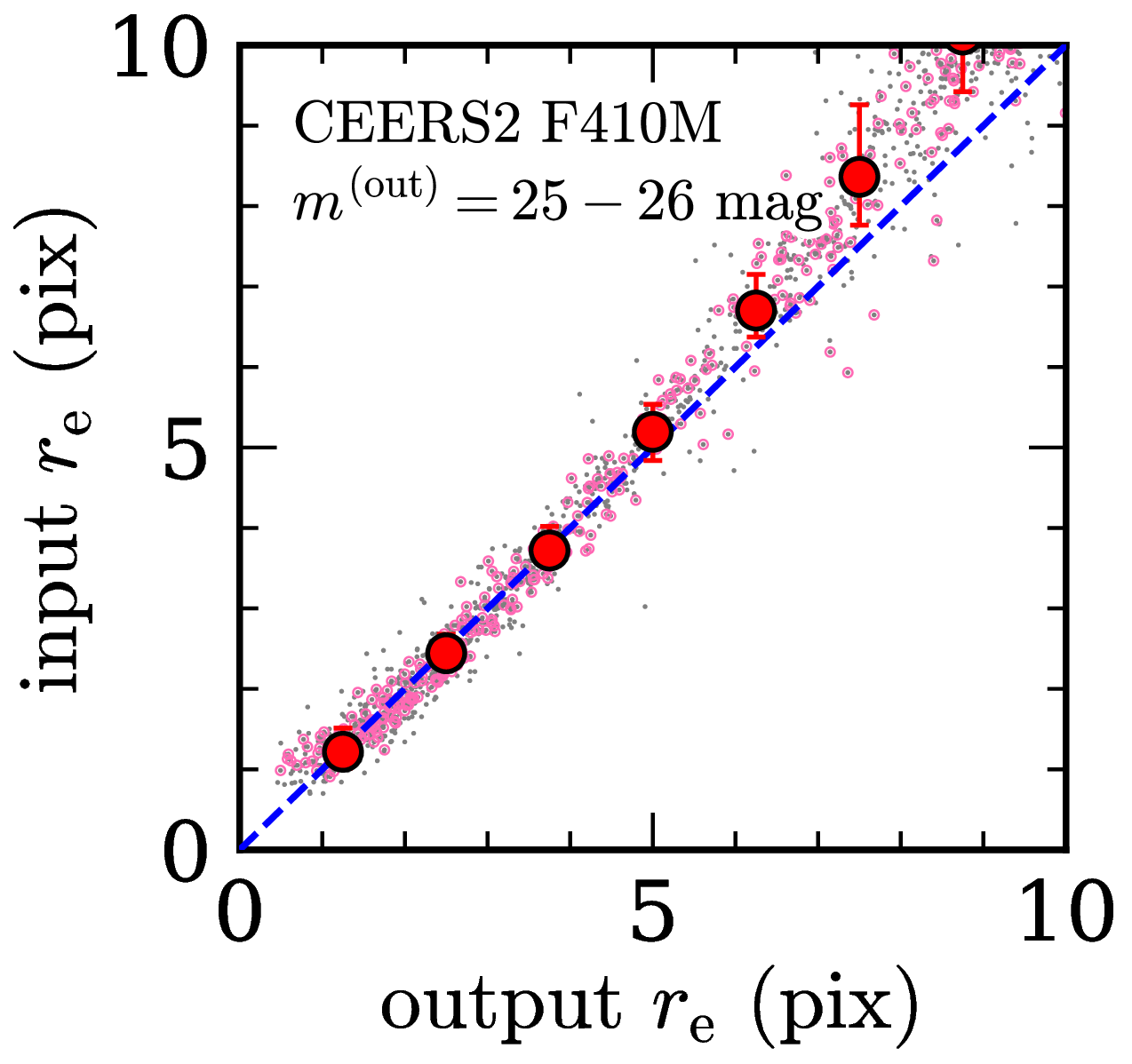}
   \includegraphics[width=0.3\textwidth]{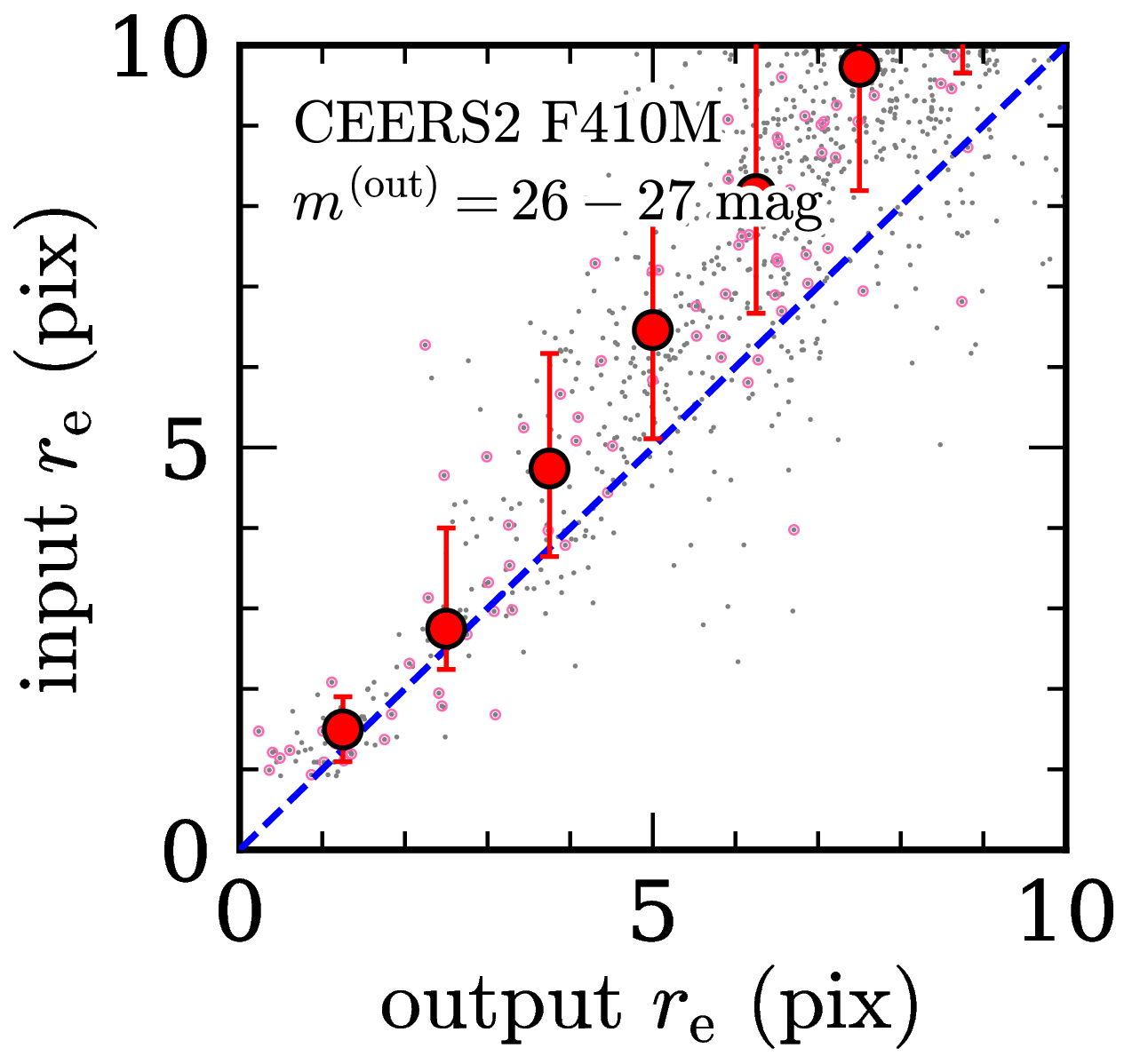}
   \includegraphics[width=0.3\textwidth]{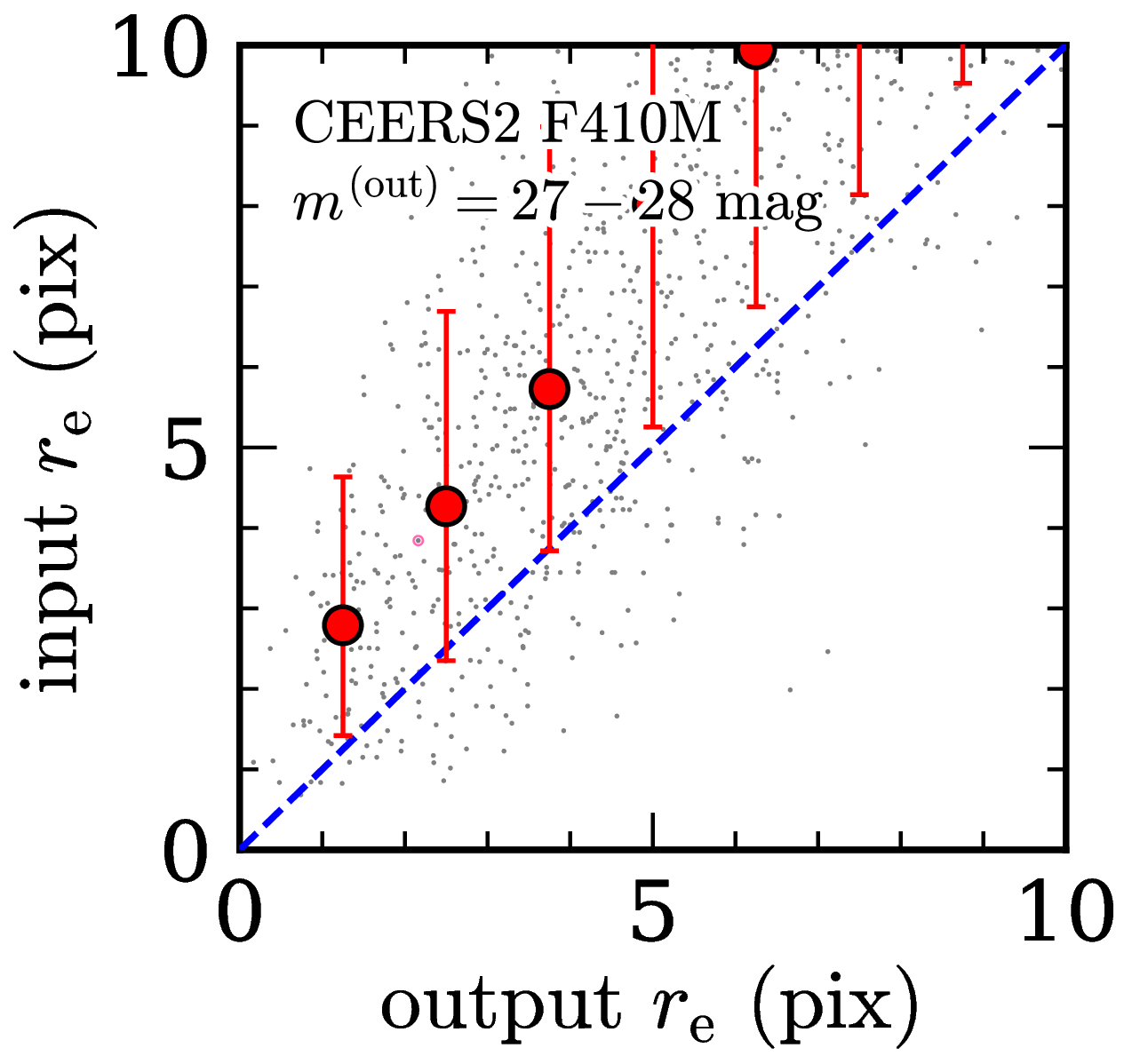}
   \includegraphics[width=0.3\textwidth]{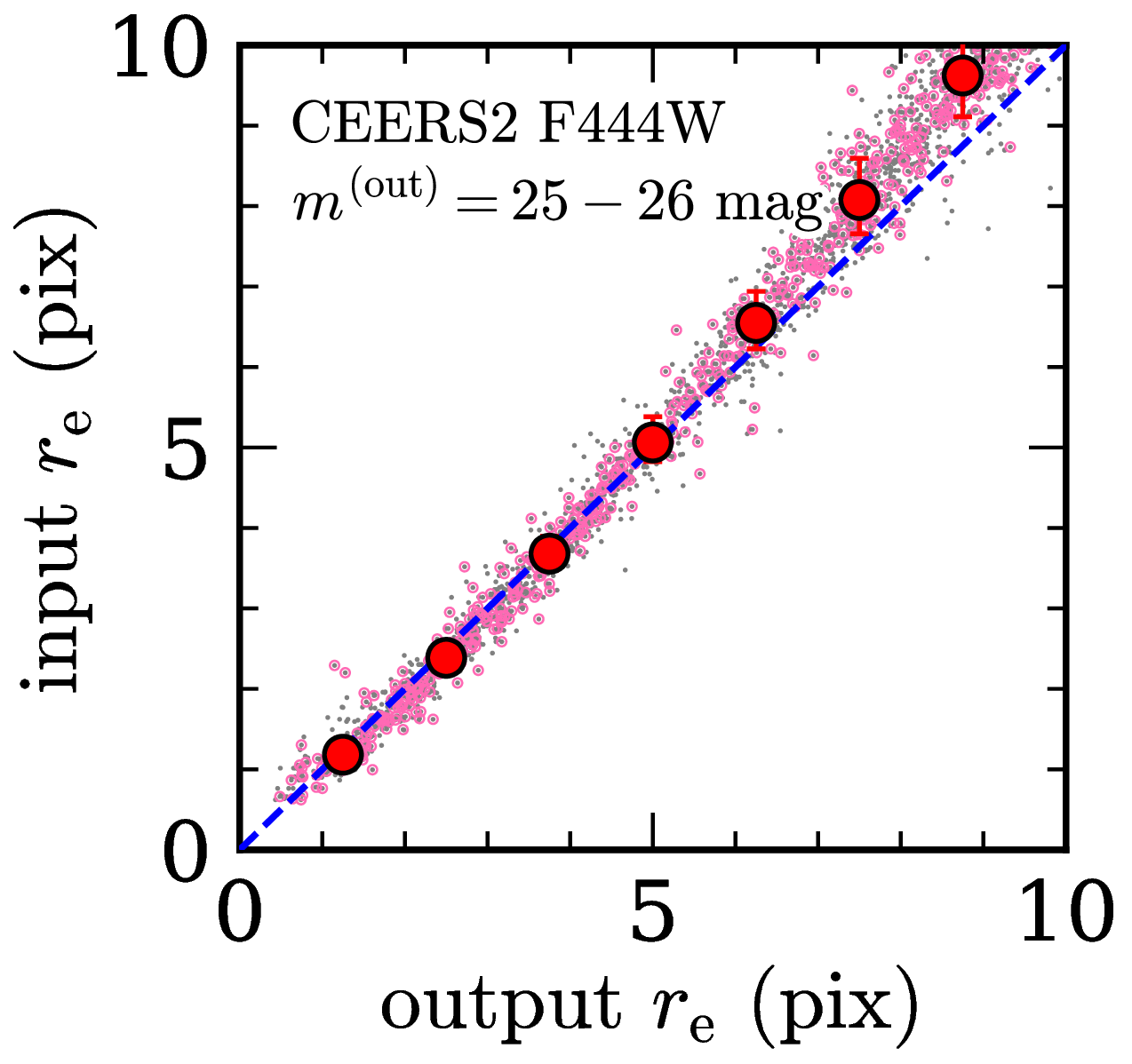}
   \includegraphics[width=0.3\textwidth]{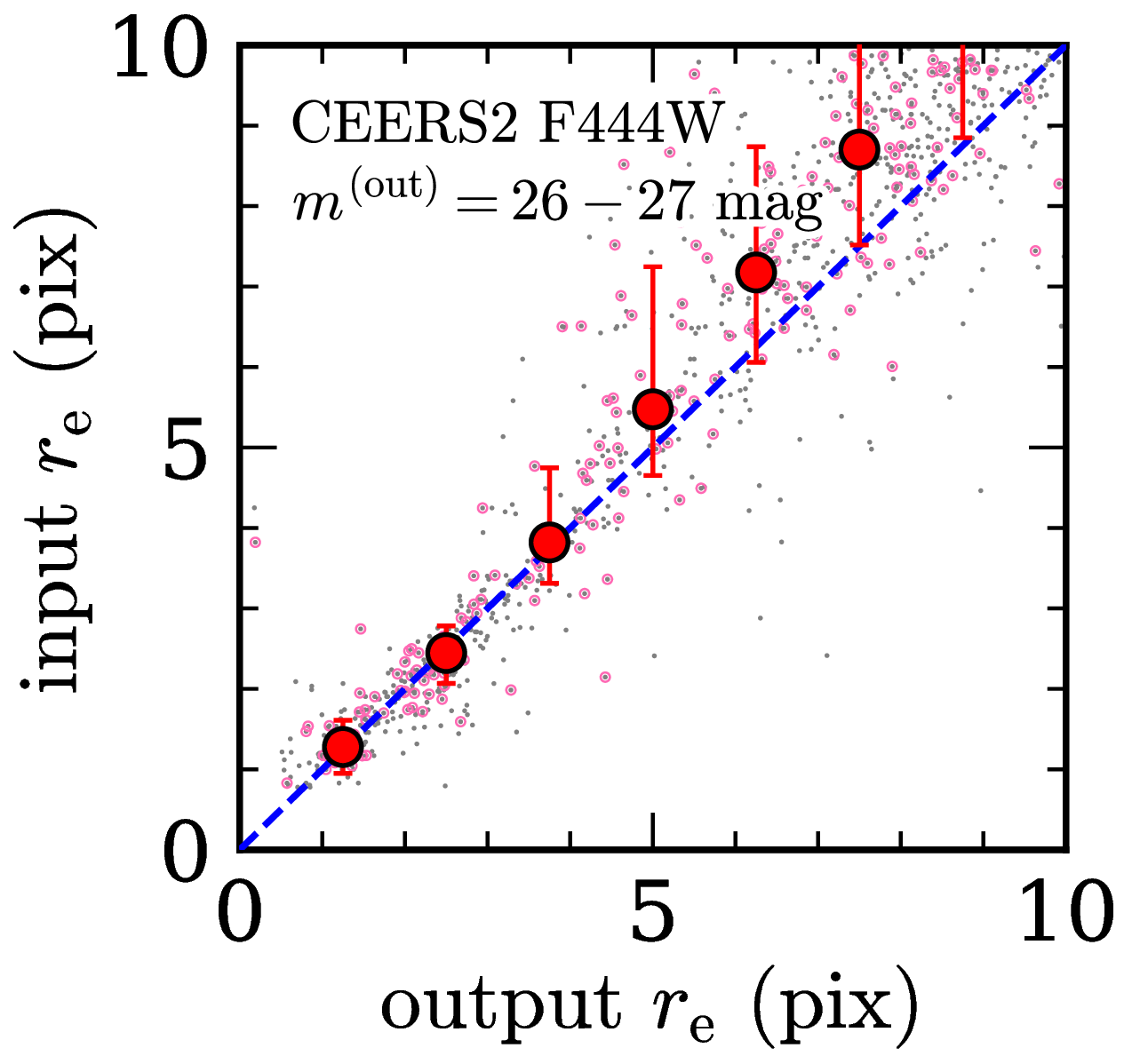}
   \includegraphics[width=0.3\textwidth]{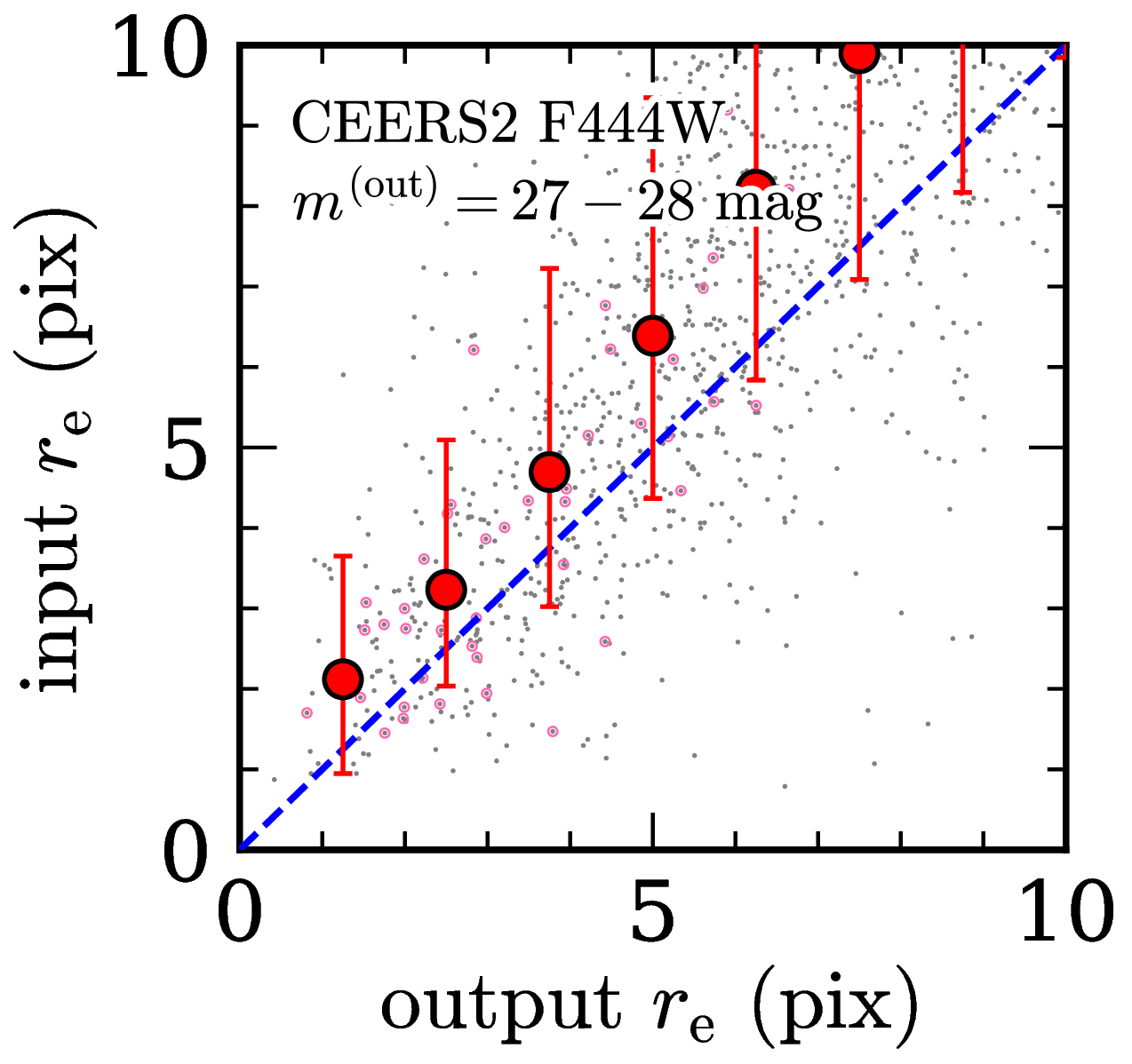}
   \includegraphics[width=0.3\textwidth]{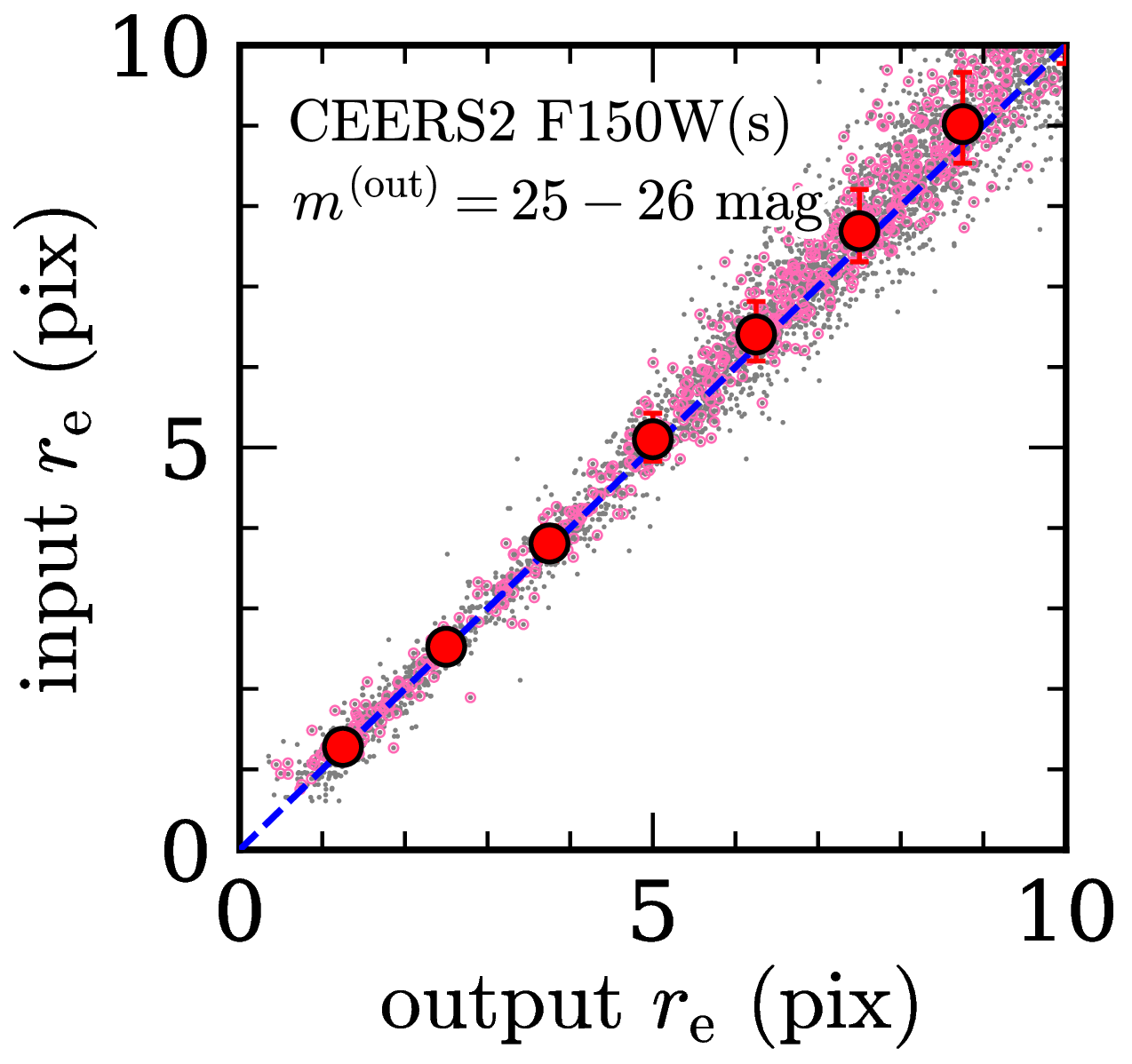}
   \includegraphics[width=0.3\textwidth]{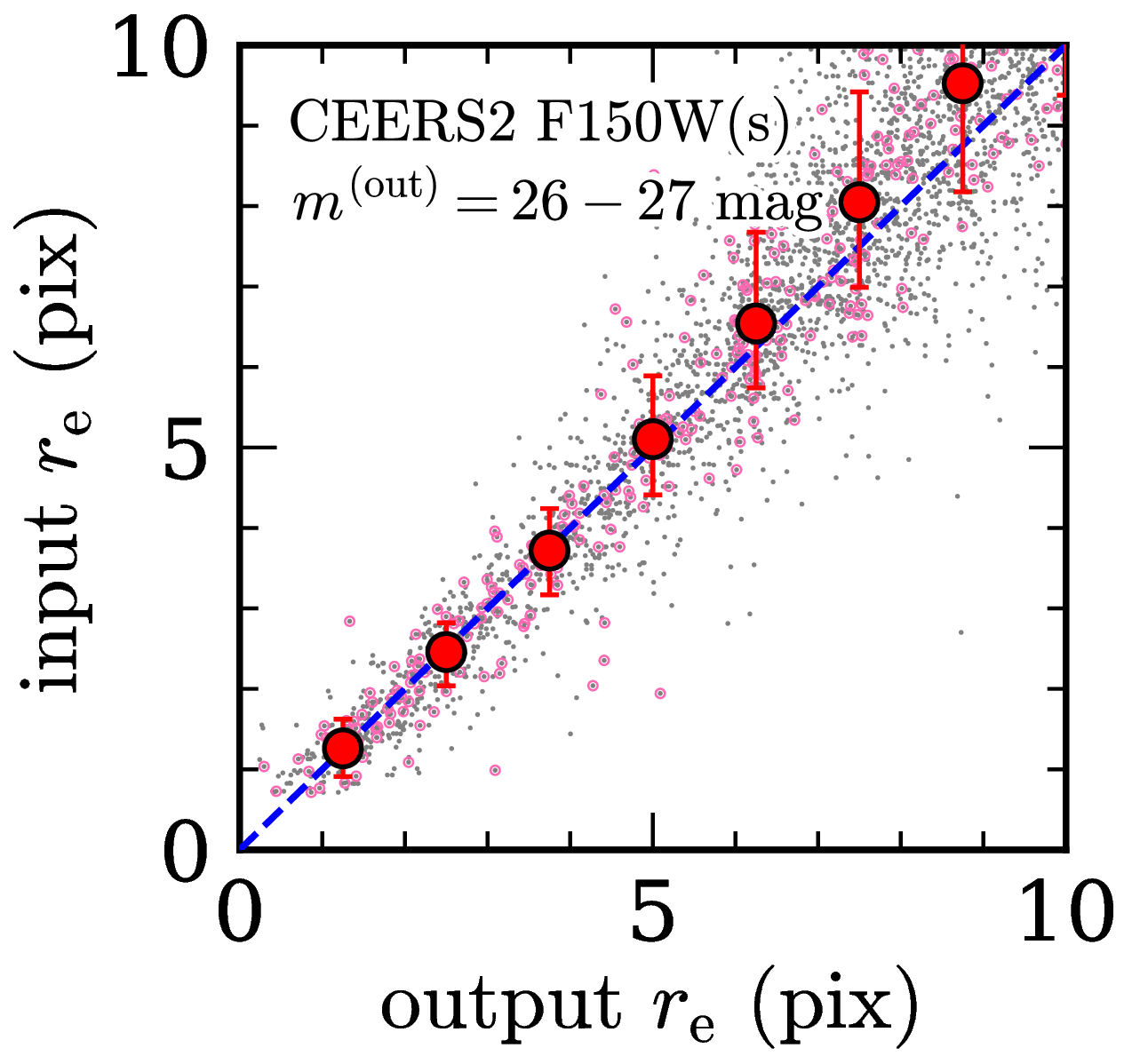}
   \includegraphics[width=0.3\textwidth]{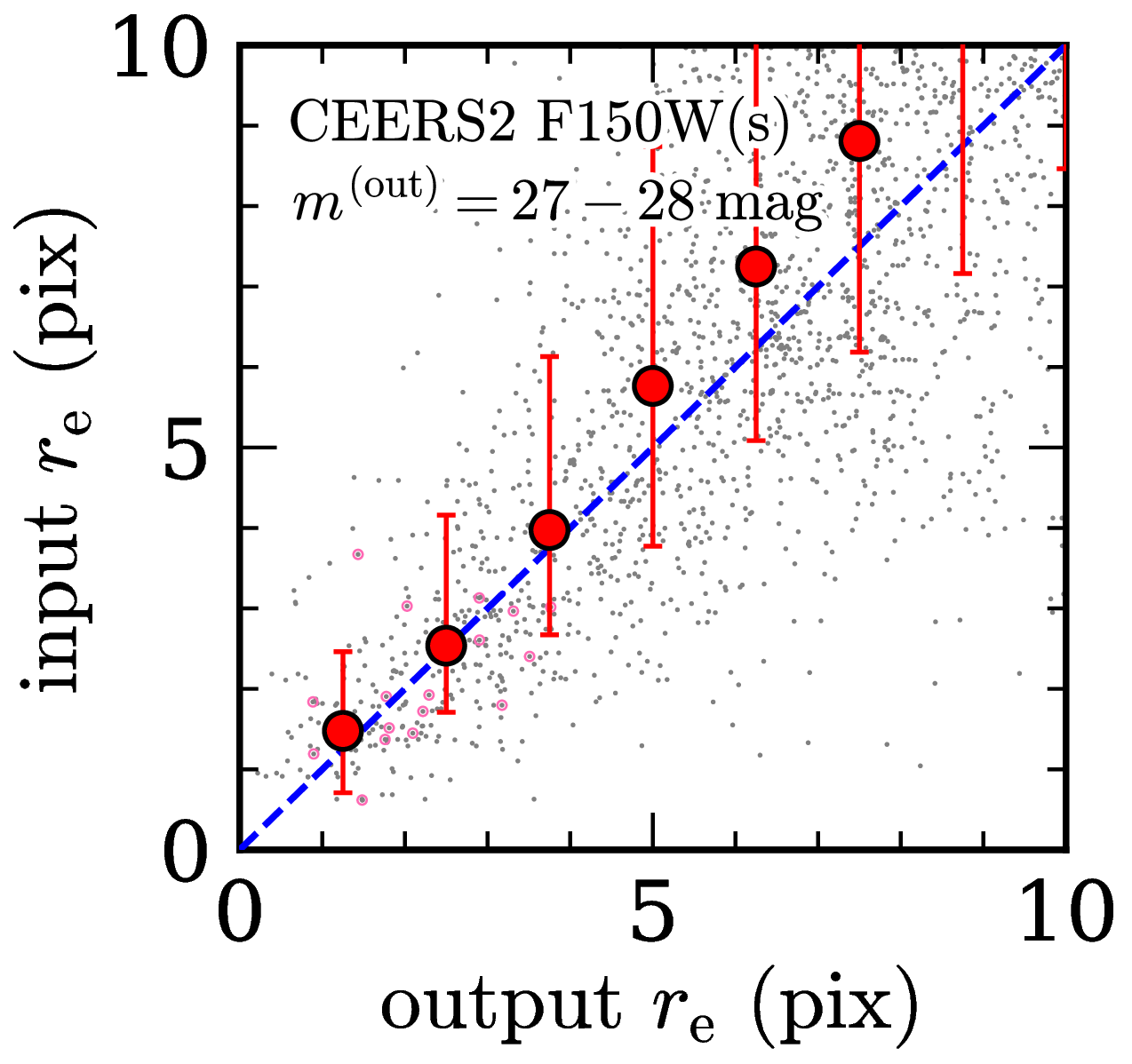}
\caption{
Input circularized radius  vs. output circularized radius 
for output total magnitudes of 
$m^{\rm (out)} = 25$--$26$ mag, $26$--$27$ mag, and $27$--$28$ mag from left to right, 
based on our GALFIT MC simulations  
for the CEERS2 field in F150W, F410M, F444W, and PSF-matched F150W [F150W(s)]. 
The red filled circles and error bars denote 
the median values of the differences between the input and output circularized radii  
and the corresponding 68 percentile ranges, respectively. 
Individual simulated objects are represented with gray dots, 
and those with an aperture magnitude S/N $> 10$ 
are marked with open pink circles.
The blue dashed line illustrates the relationship 
where the input and output circularized radius are equivalent.
}
\label{fig:input_output_re}
\end{center}
\end{figure*}

\begin{figure}
\begin{center}
   \includegraphics[width=0.23\textwidth]{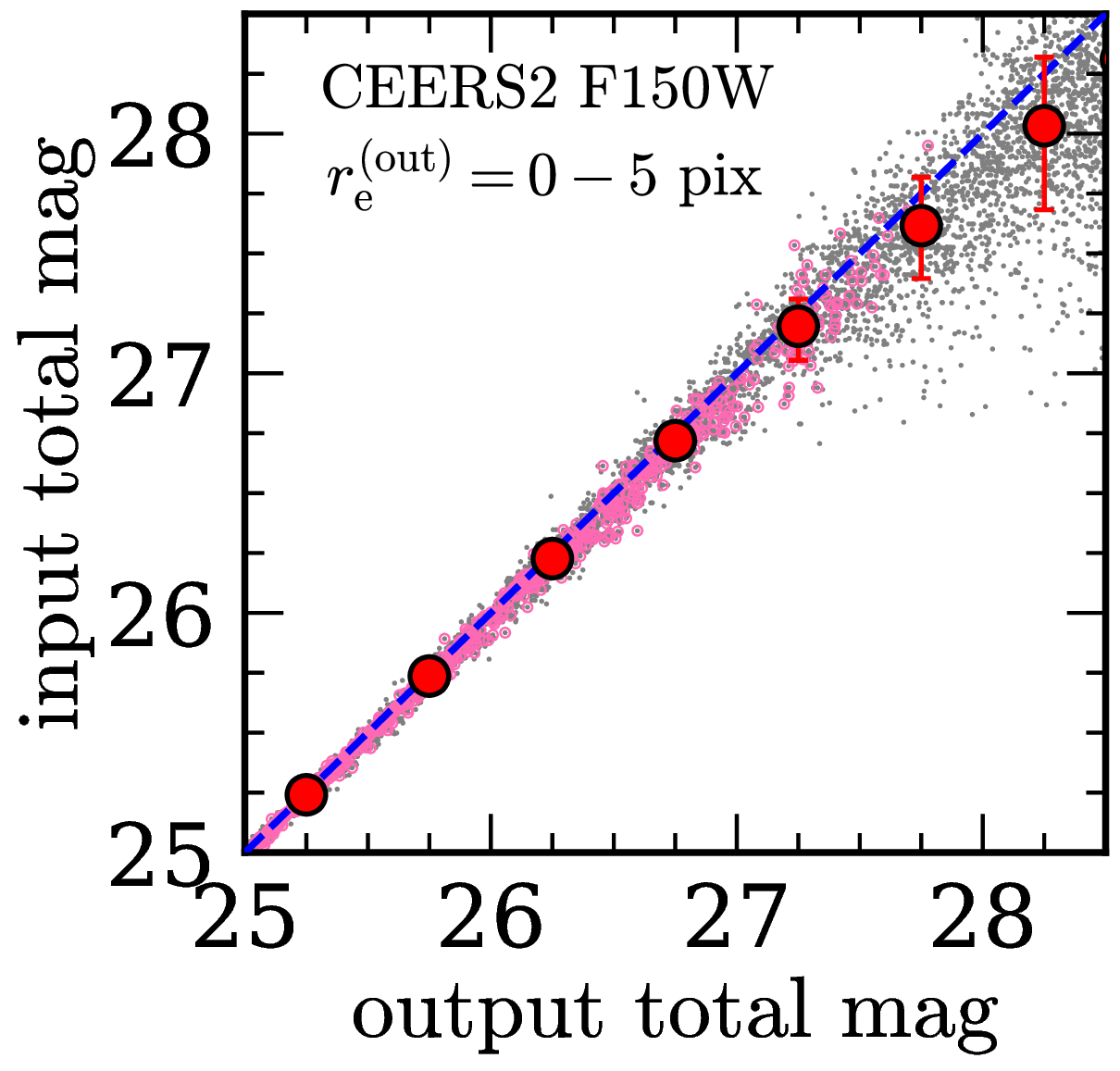}
   \includegraphics[width=0.23\textwidth]{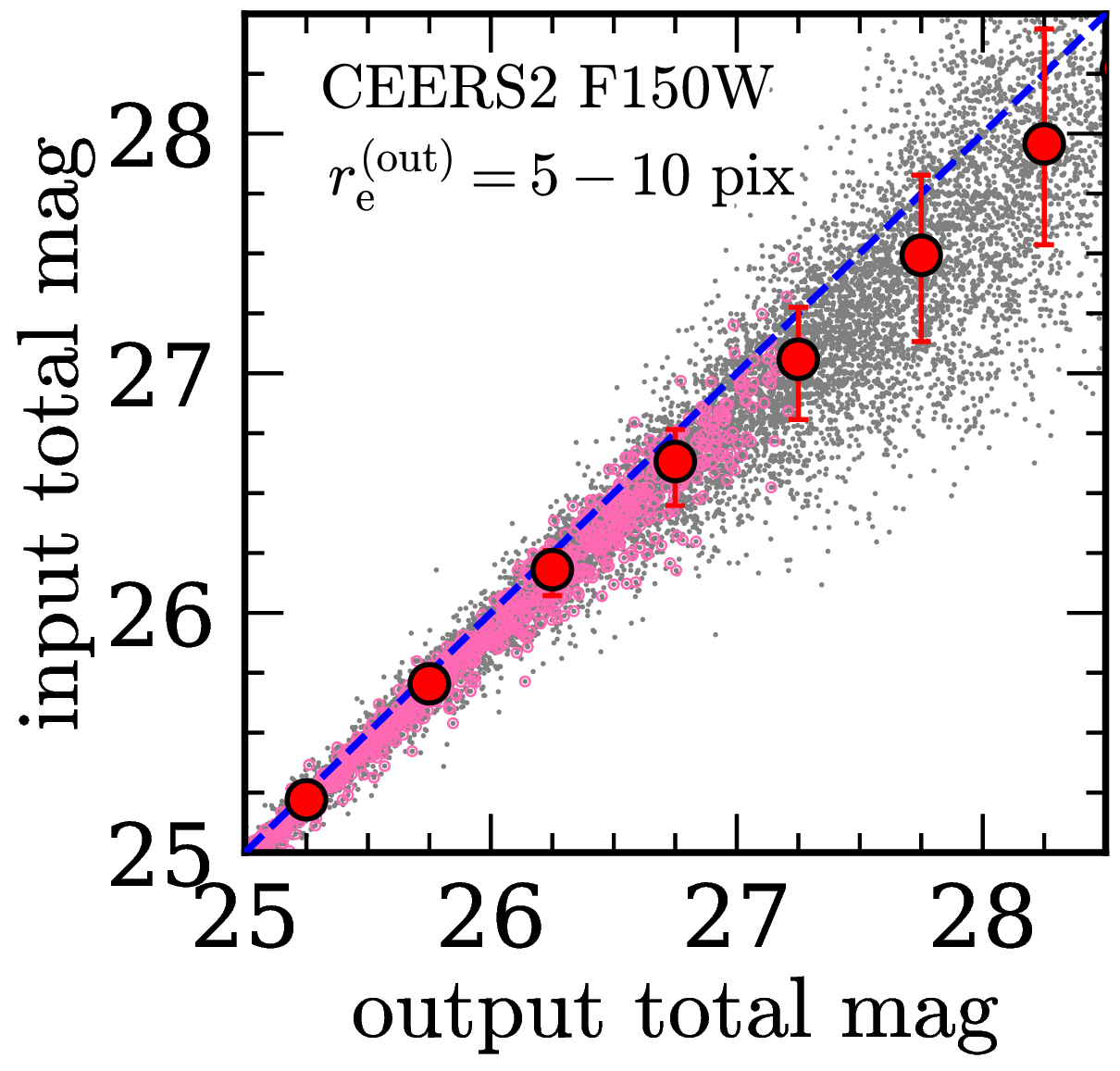}
   \includegraphics[width=0.23\textwidth]{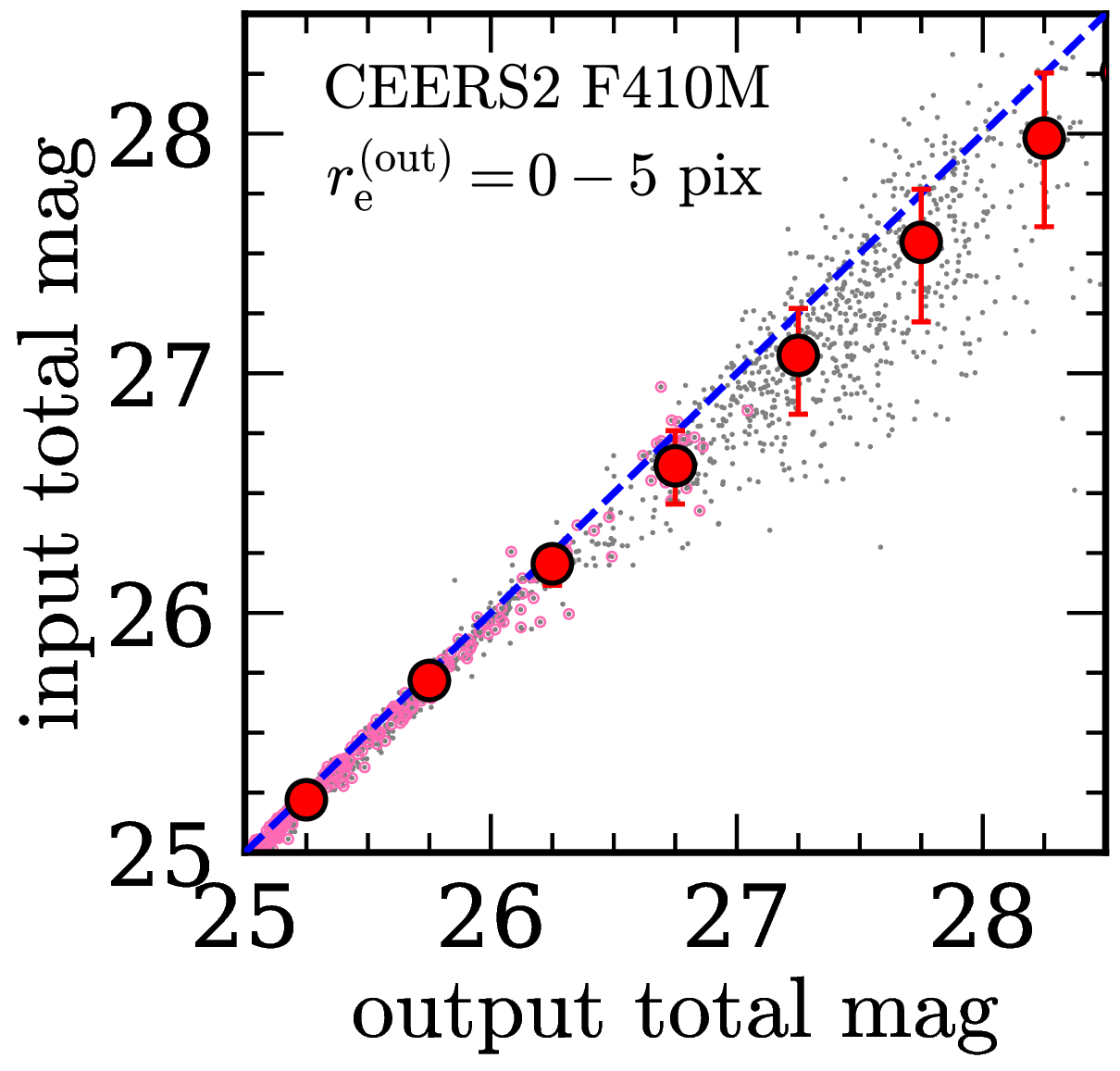}
   \includegraphics[width=0.23\textwidth]{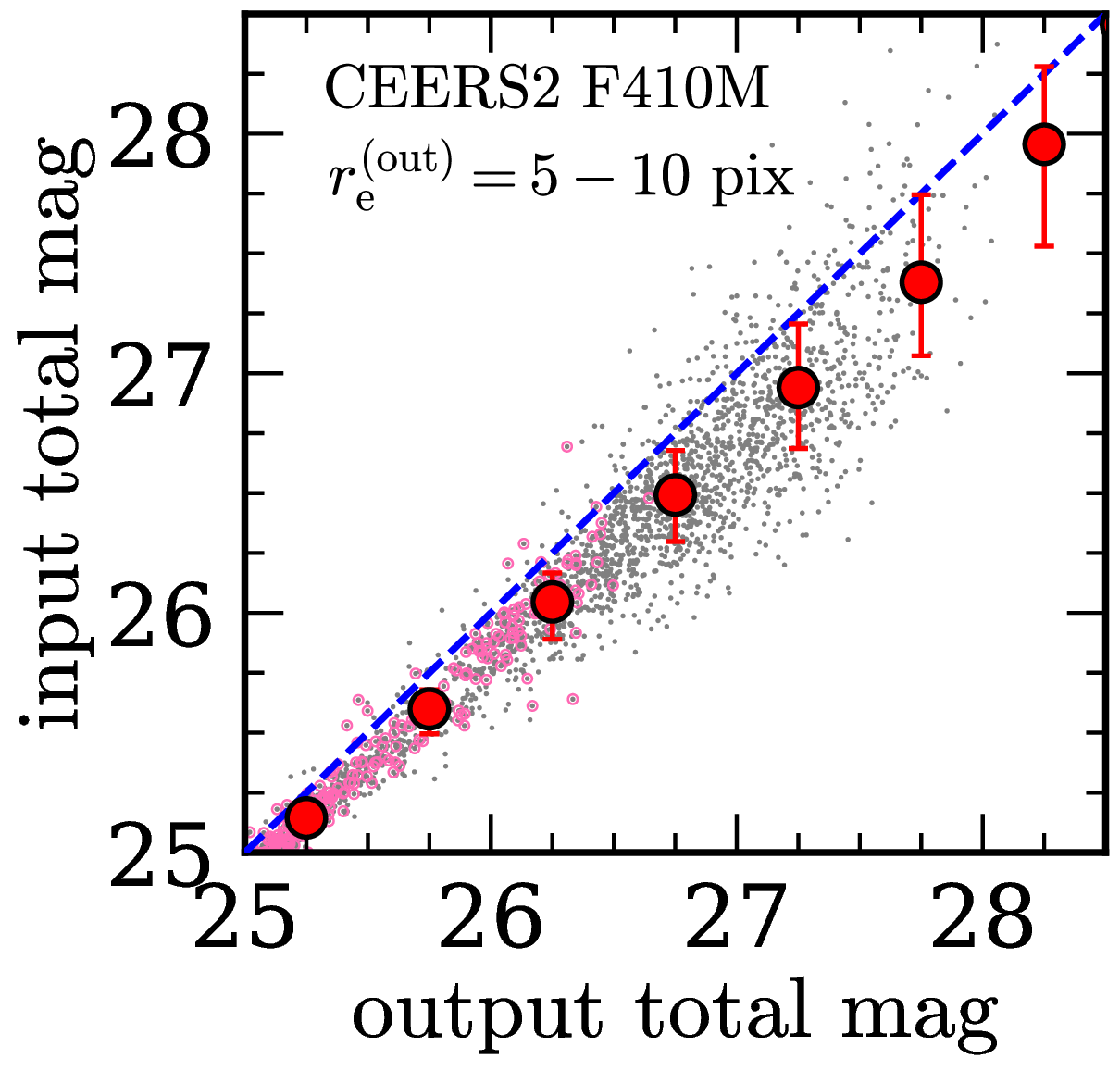}
   \includegraphics[width=0.23\textwidth]{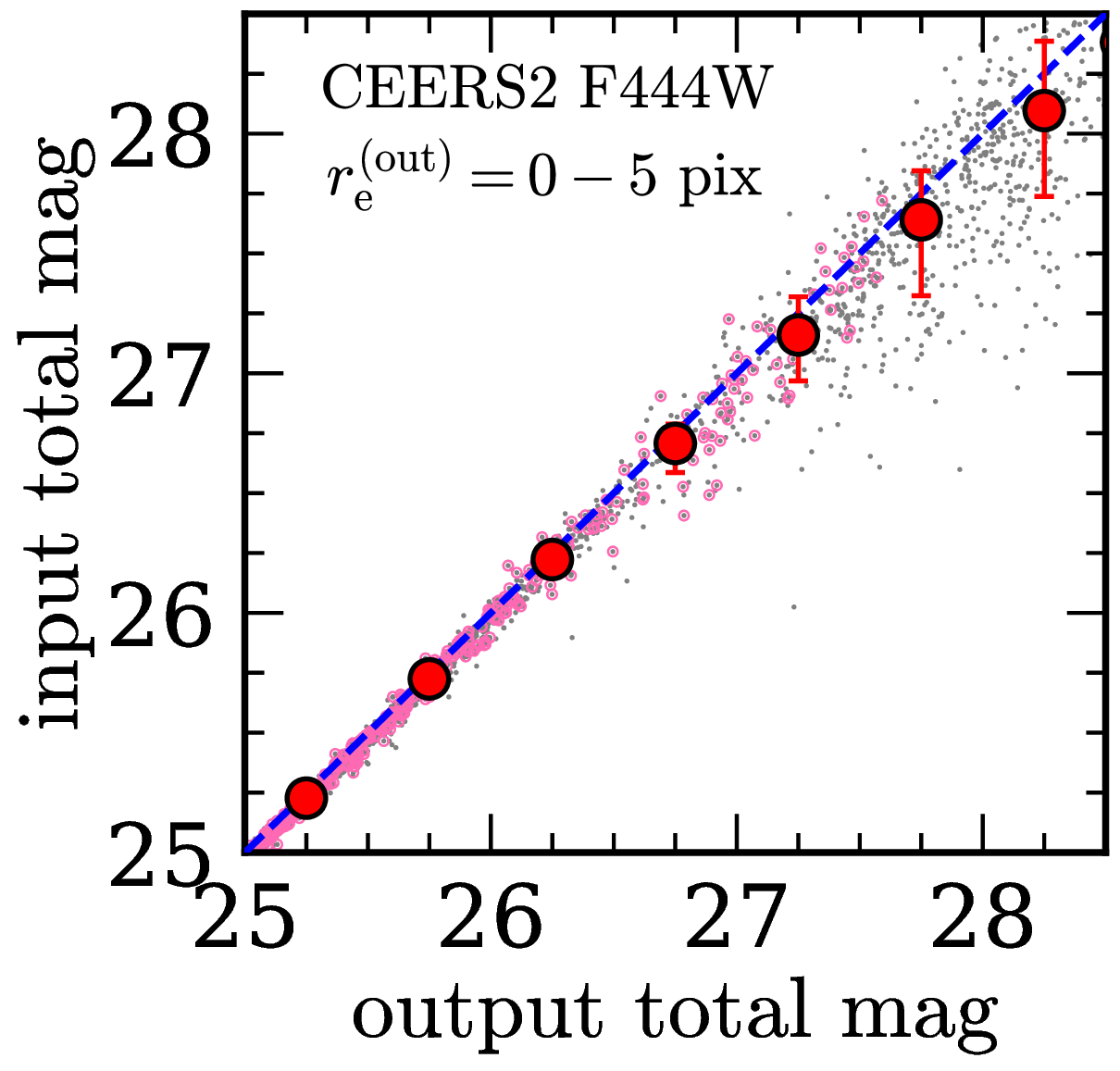}
   \includegraphics[width=0.23\textwidth]{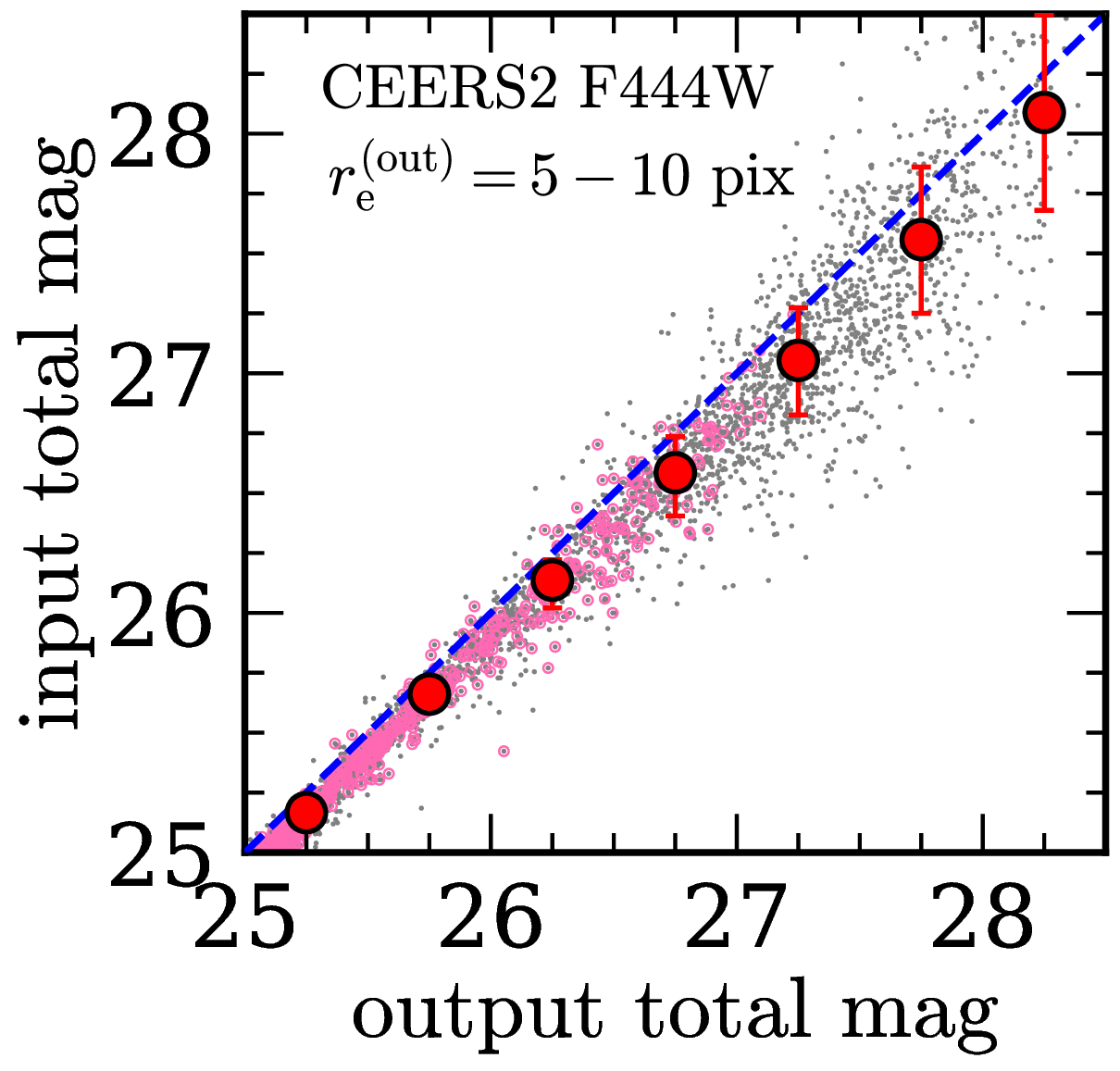}
   \includegraphics[width=0.23\textwidth]{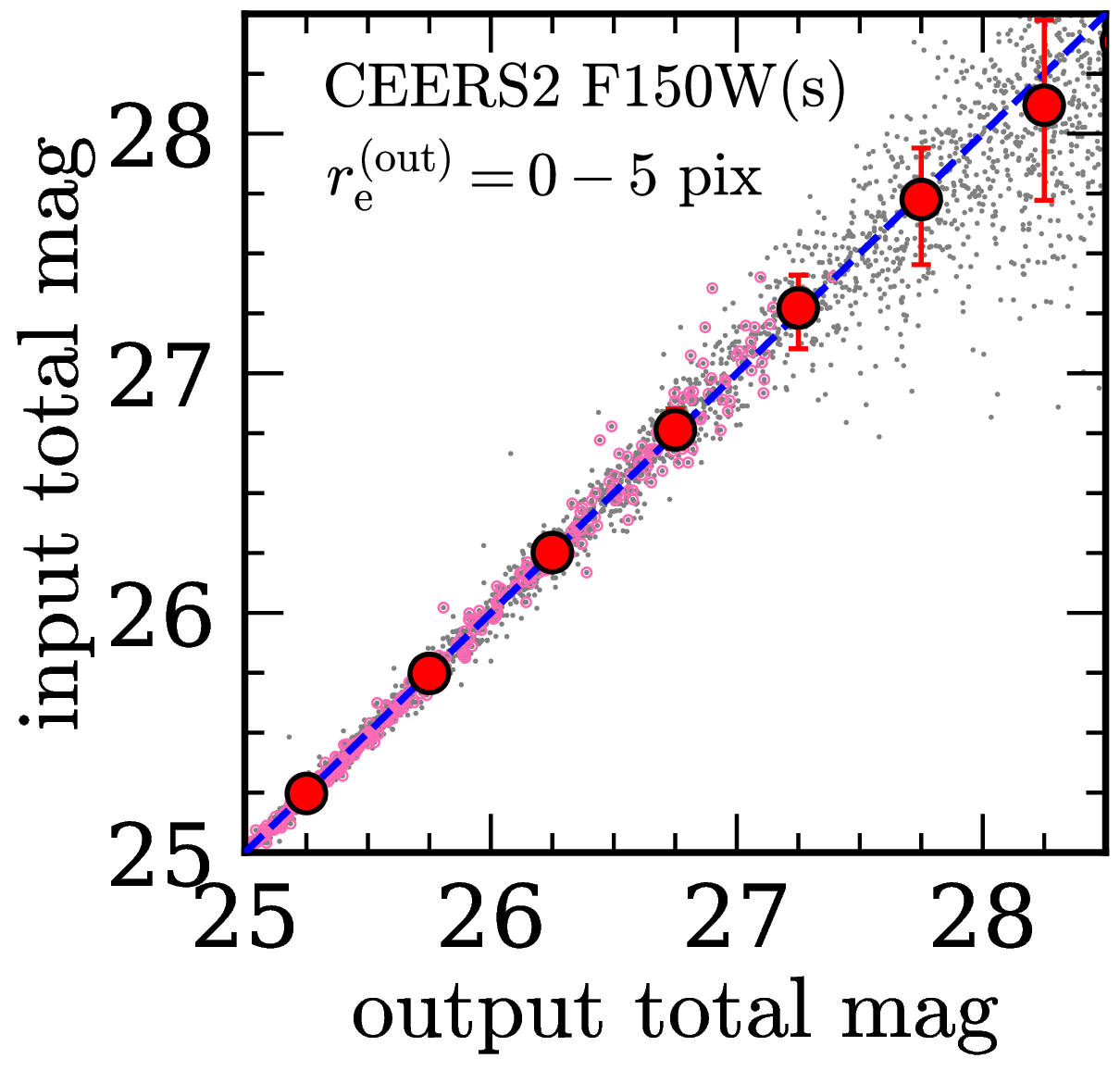}
   \includegraphics[width=0.23\textwidth]{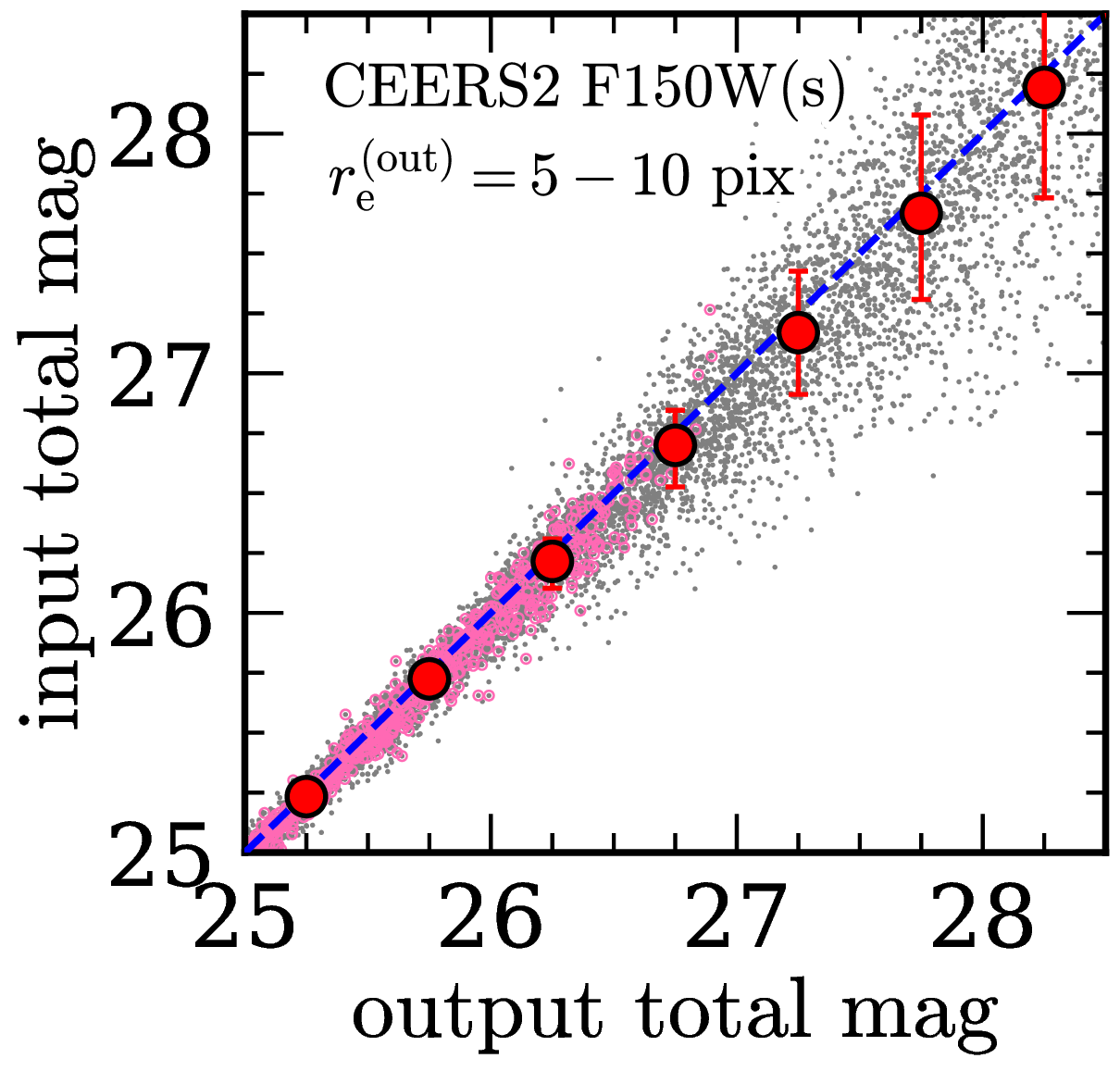}
\caption{
Input total magnitude vs. output total magnitude 
for output half-light radii of 
$r_{\rm e}^{\rm (out)} = 0$--$5$ pixels (left) and $5$--$10$ pixels (right) 
based on our  GALFIT MC simulations  
for the CEERS2 field in F150W, F410M, F444W, and PSF-matched F150W [F150W(s)]. 
The red filled circles and error bars denote 
the median values of the differences between the input and output magnitudes 
and the corresponding 68 percentile ranges, respectively. 
Individual simulated objects are represented by gray dots, 
with an additional open pink circle marking those having aperture magnitude S/N above $10$.
The blue dashed line corresponds to the relationship 
where the input and output magnitudes are equal.
}
\label{fig:input_output_mag}
\end{center}
\end{figure}

\begin{figure*}
\begin{center}
   \includegraphics[width=0.28\textwidth]{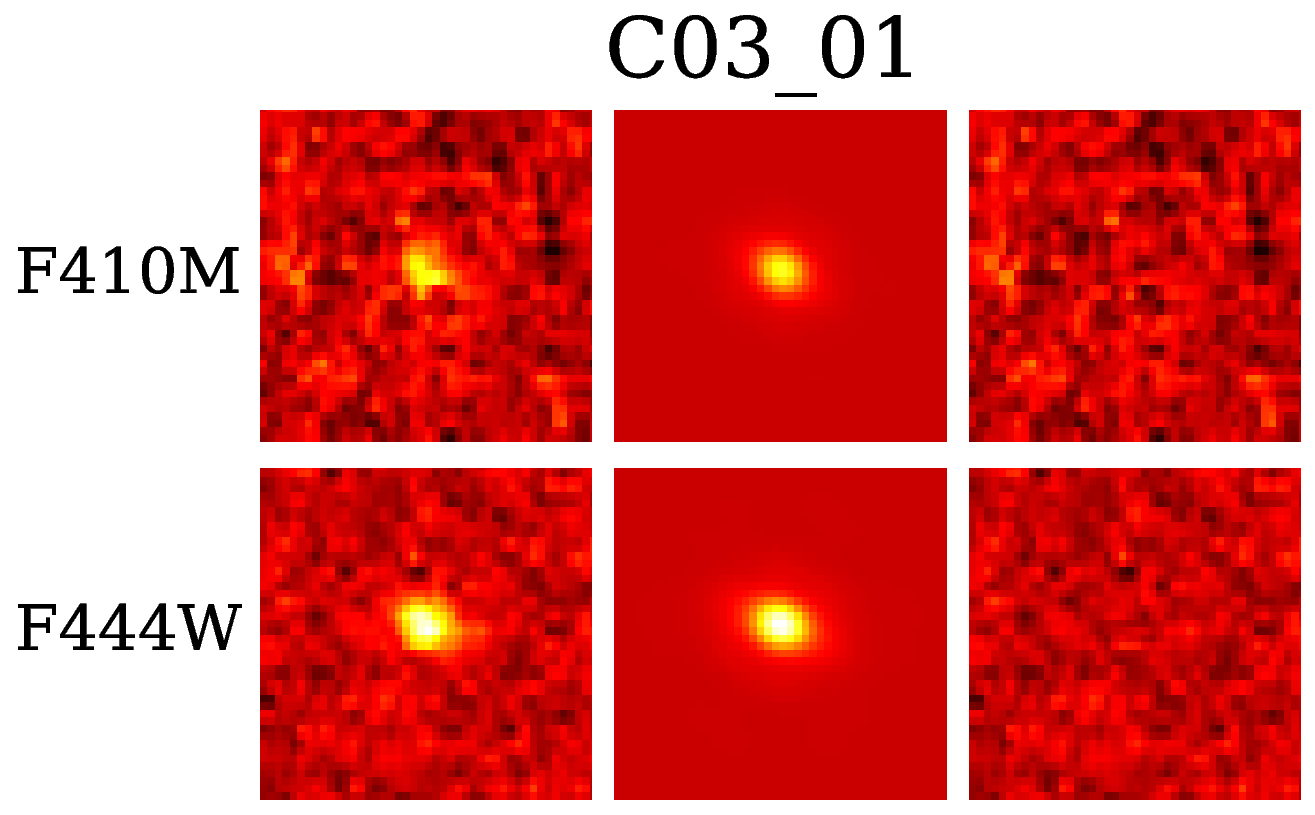}
   \includegraphics[width=0.28\textwidth]{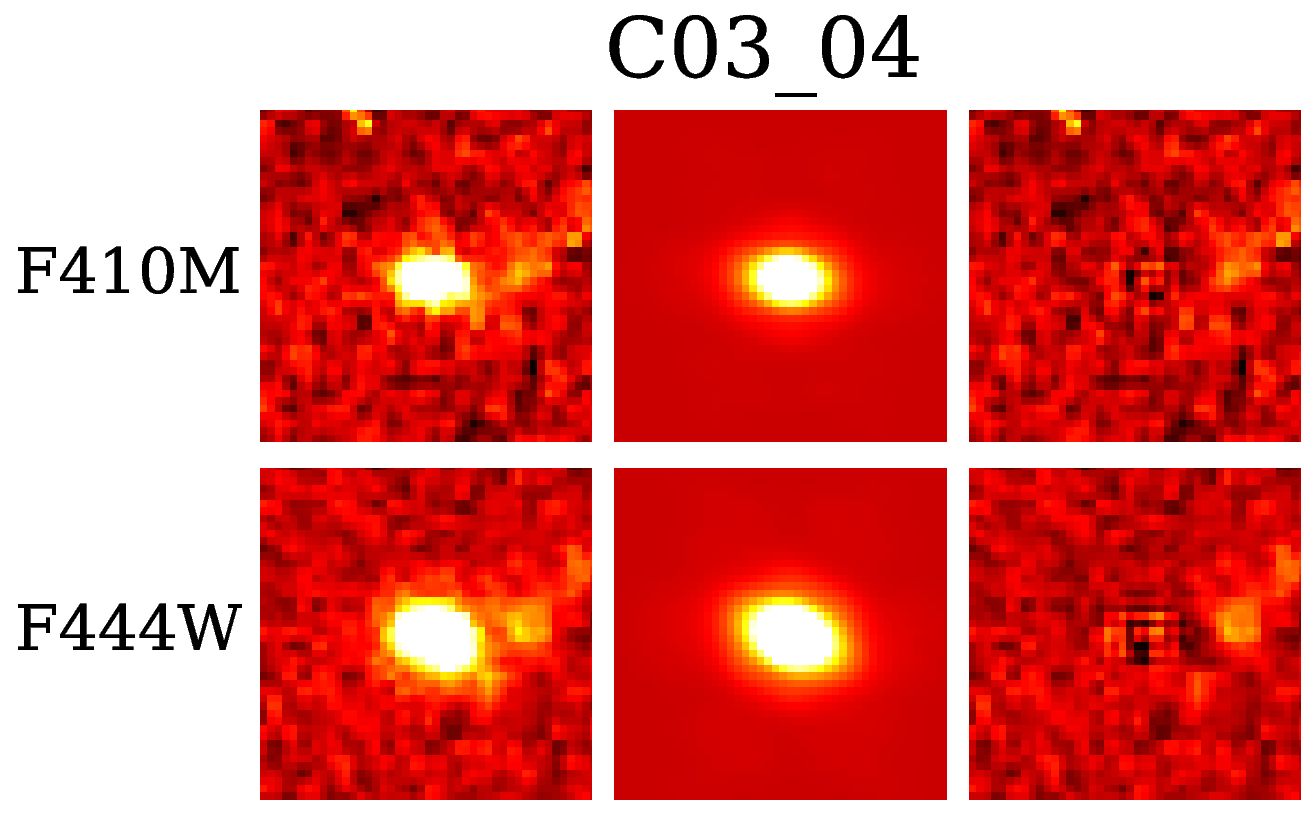}
   \includegraphics[width=0.28\textwidth]{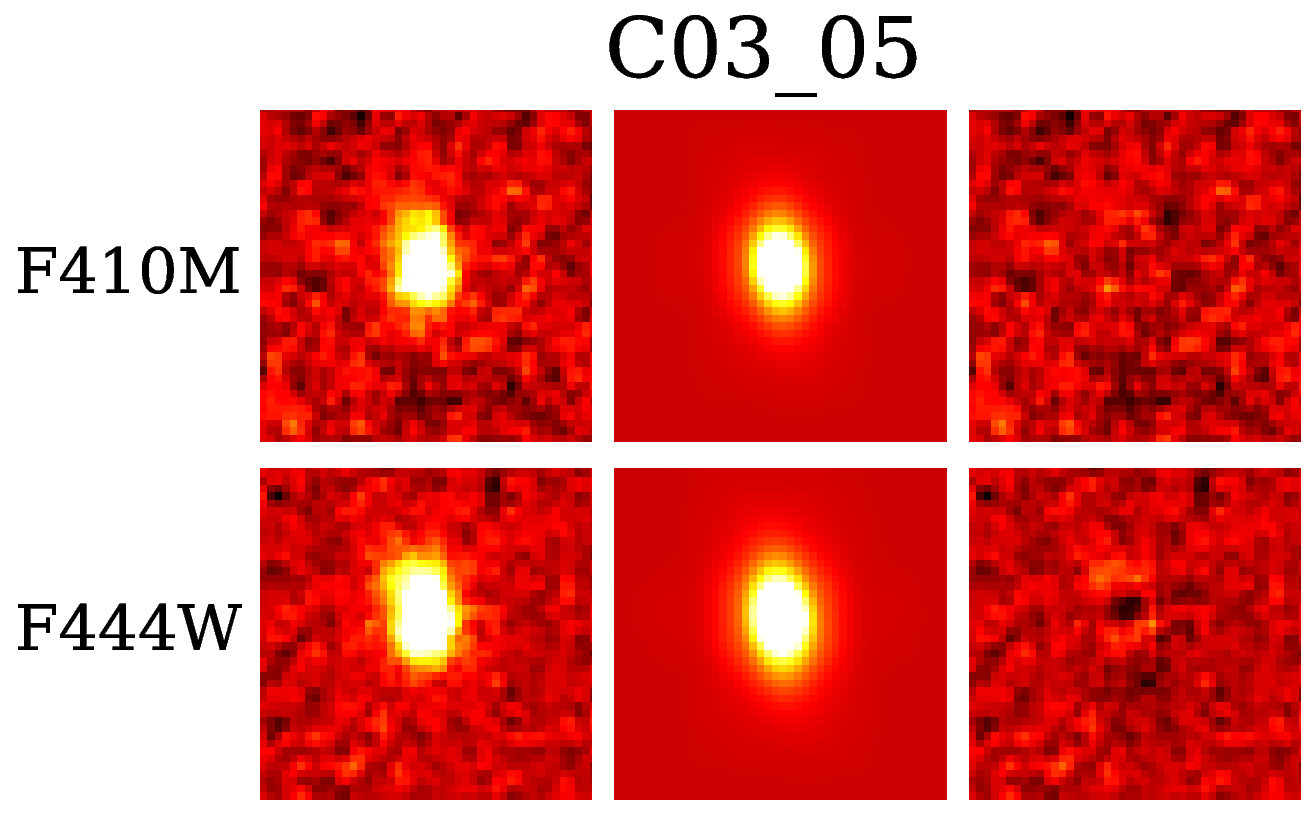}
   \includegraphics[width=0.28\textwidth]{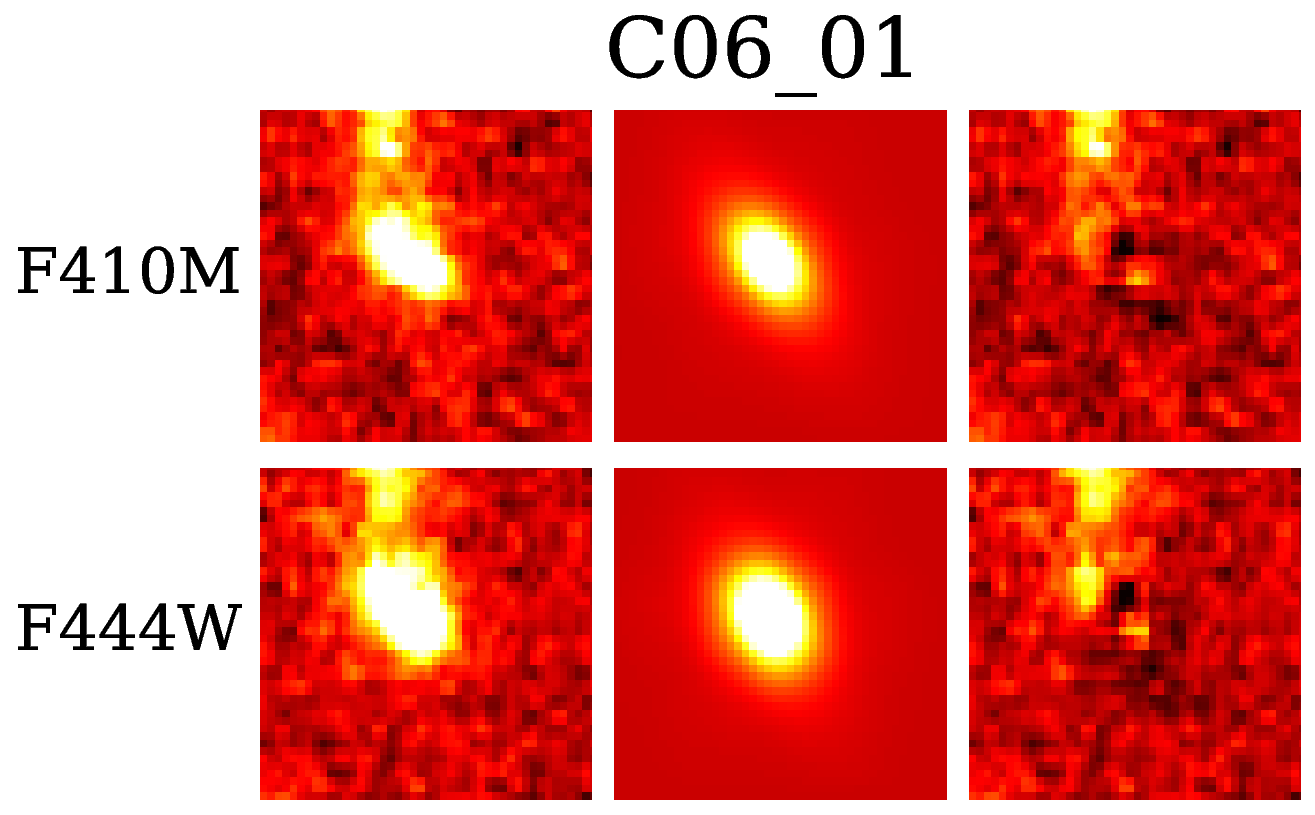}
   \includegraphics[width=0.28\textwidth]{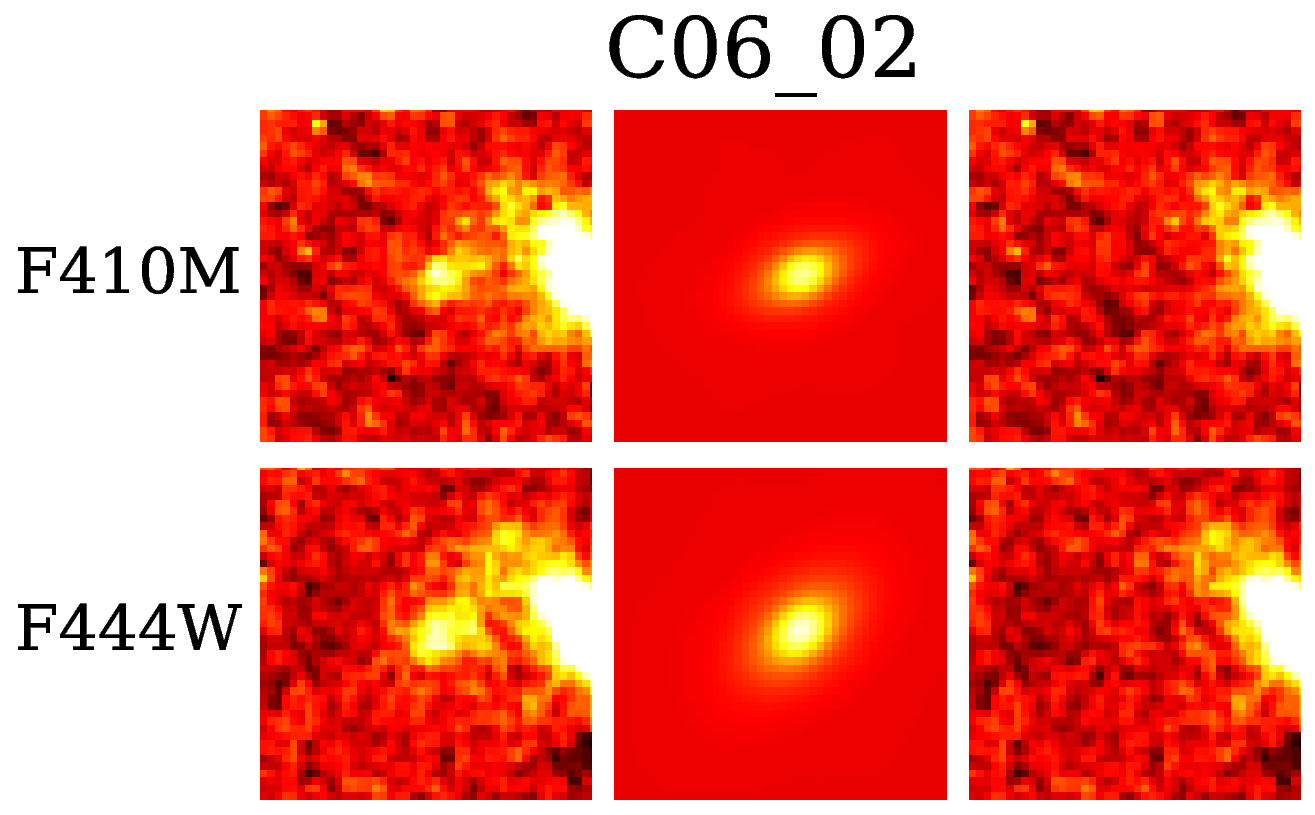}
   \includegraphics[width=0.28\textwidth]{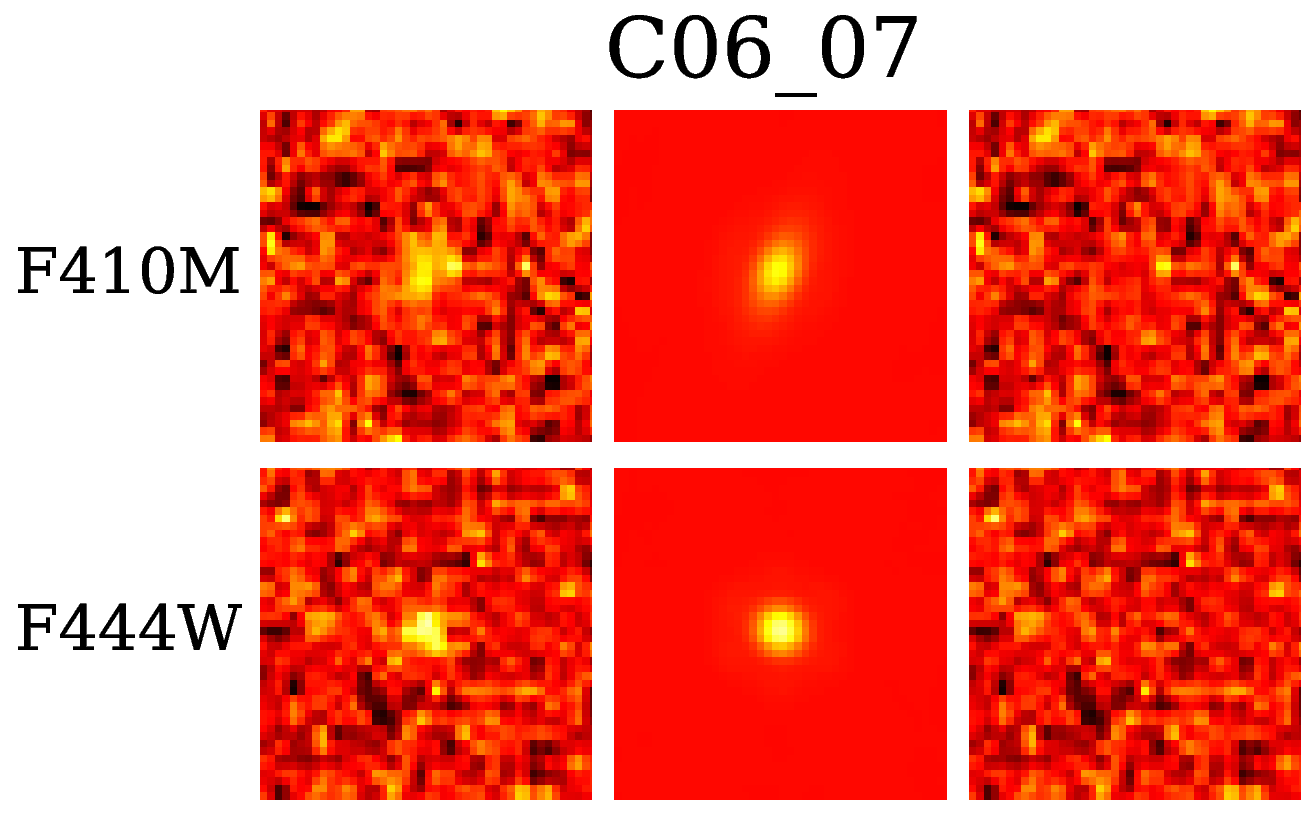}
\caption{
S{\'e}rsic profile fitting results for 
the F410M and F444W images of spectroscopically confirmed galaxies 
within the redshift range $z_{\rm spec} = 5.63$--$6.63$, 
where strong emission lines do not enter the F410M band.  
From left to right, the $1\farcs5 \times 1\farcs5$ cutouts of the original images, 
the best-fit S{\'e}rsic model profile images, 
and the residual images 
that are created by subtracting the best-fit images from the original ones 
are presented.  
}
\label{fig:SB_fitting_results_F410M_F444W}
\end{center}
\end{figure*}

\subsection{Monte Carlo Simulations} \label{subsec:MC_simulation}

\hspace{1em}
Following previous work 
(e.g., \citealt{2013ApJ...777..155O}; \citealt{2015ApJS..219...15S}; \citealt{2023ApJ...951...72O}), 
we execute a series of MC simulations 
to quantify systematic and statistical uncertainties in the profile fitting with GALFIT. 
We use the F150W and F444W images, 
which probe the rest-frame UV and optical emission from our galaxies at $z \simeq 4$--$10$. 
In addition, we utilize the F410M images 
as detailed in Section \ref{subsec:comparison_f410m_f444w}, 
as well as smoothed F150W images 
whose PSF sizes are comparable to those of the F444W images 
as described in Section \ref{subsec:PSF_match}. 
As demonstrated in Table \ref{tab:limitmag},
the depths and PSF FWHMs are almost the same across different pointings of CEERS. 
For simplicity, we conduct MC simulations for each band with respect to CEERS2 
and apply the results to correct systematic uncertainties 
and estimate statistical uncertainties for all the pointings from CEERS1 to CEERS10. 
We use GALFIT to generate galaxy images with a fixed S{\'e}rsic index $n$ of 1.5.  
The half-light radius $r_e$ and total magnitude are randomly selected 
within ranges of $0.5$ to $27.0$ pixels and $24.5$ to $30.0$ mag, respectively. 
These galaxy images are then convolved with a PSF image, 
which is a composite of bright, unsaturated stars selected from each band.
The PSF-convolved galaxy images are subsequently inserted into blank regions 
of the actual NIRCam images and analyzed in the identical manner to our high-$z$ galaxies.

We present the results of size measurements 
for our MC simulated galaxies in Figure \ref{fig:input_output_re}, 
and those of total magnitude measurements in Figure \ref{fig:input_output_mag}. 
The general trends seen in these figures are the same 
as those in our previous results of \cite{2023ApJ...951...72O}. 
Our MC simulations confirm a systematic underestimation of half-light radii 
and overestimation of total magnitudes in GALFIT measurements for fainter objects. 
It is also found that 
the MC simulated galaxies with faint magnitudes and large sizes 
do not meet the S/N $> 10$ criterion for aperture magnitude.
Since our sample is limited to those with an aperture magnitude S/N larger than $10$, 
the systematic effects are not large.\footnote{We confirm that 
the size and magnitude distribution of our high-$z$ galaxies with S/Ns larger than $10$ 
nearly entirely falls within the range covered by the MC simulated galaxies with S/N $> 10$.}
We rectify these systematic effects 
and also use statistical uncertainties in size and total magnitude measurements 
in the same manner as our previous work; 
briefly, we utilize the MC simulation outcomes 
within the same output magnitude (size) bins as our high-$z$ galaxies 
to correct for the output sizes (magnitudes) 
by the median differences between the input and output values 
and to assign the $68$th percentiles in these bins as statistical uncertainties 
(See Section 3 of \citealt{2023ApJ...951...72O} for details).

\begin{figure}
\begin{center}
   \includegraphics[width=0.4\textwidth]{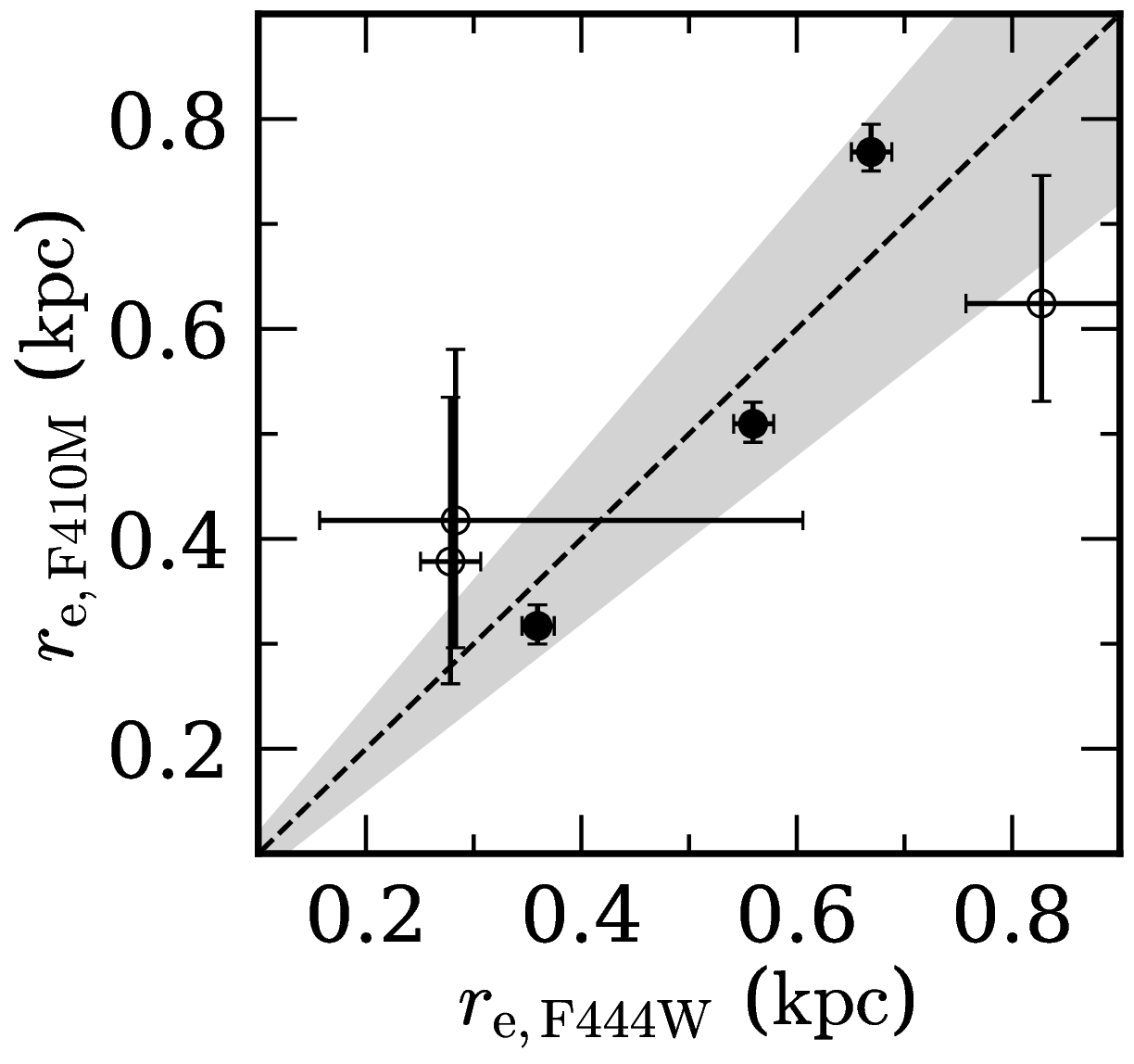}
   \includegraphics[width=0.4\textwidth]{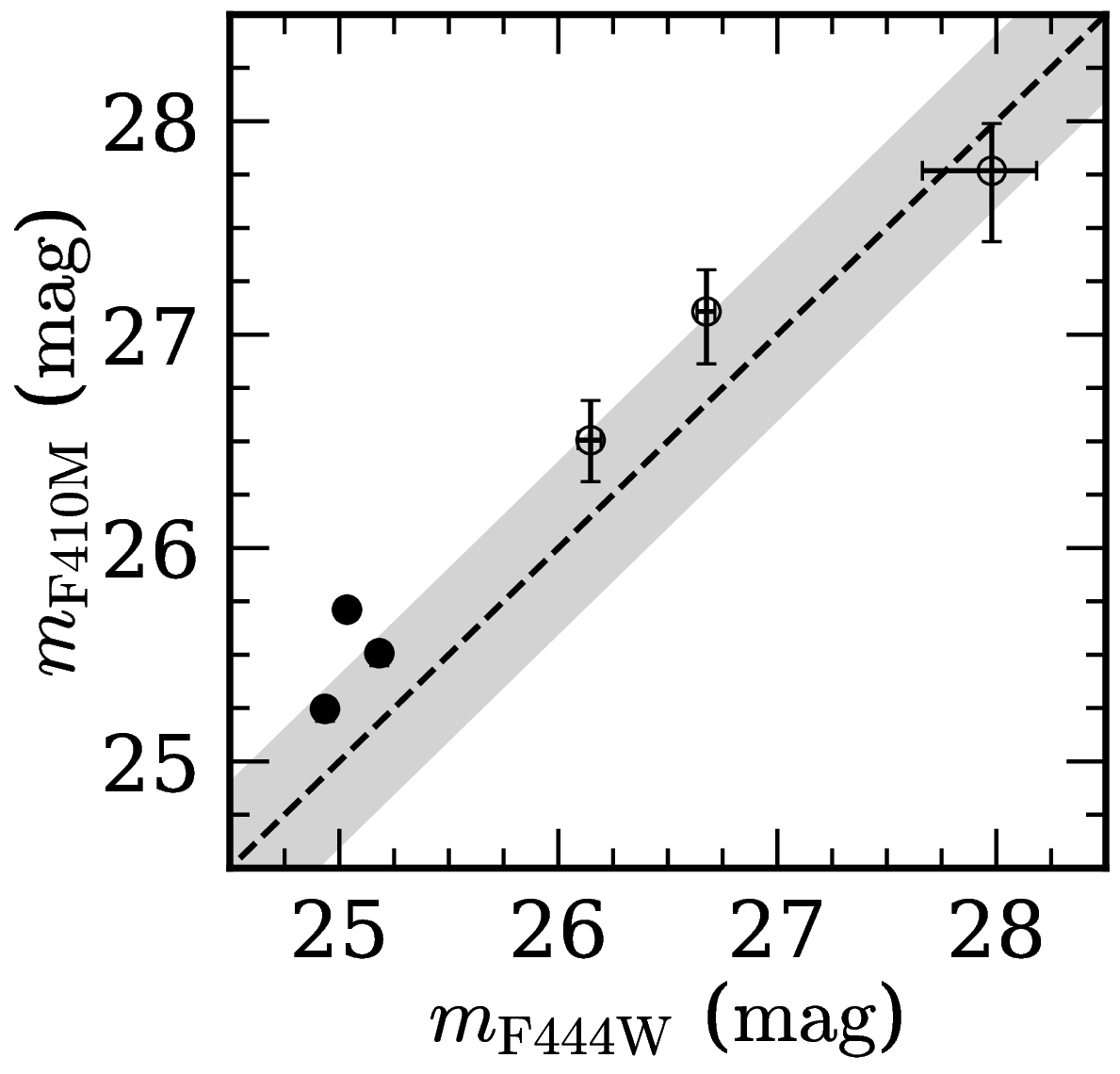}
\caption{
Comparison of SB profile fitting results for F410M and F444W images 
of galaxies with spectroscopic redshifts $z_{\rm spec}=5.63$--$6.63$, 
where strong emission lines do not enter the F410M band. 
The top panel shows the size comparison, while the bottom panel illustrates the total magnitude comparison. 
The filled circles represent bright sources, while the open circles indicate faint sources. 
The dashed line denotes the case where the results from F410M and F444W are equal.
The gray shade in the top (bottom) panel represents the range where the difference between these sizes 
(total magnitudes) is within $\pm20${\%} ($\pm 0.4$ mag).
}
\label{fig:f410m_f444w_re}
\end{center}
\end{figure}

\subsection{Impact of Strong Emission Lines} 
\label{subsec:comparison_f410m_f444w}

\hspace{1em}
As mentioned in Section \ref{sec:data}, 
strong emission lines such as H$\alpha$ and [{\sc Oiii}] from galaxies at $z\gtrsim 5$ 
are included in F444W due to its wide wavelength coverage. 
To assess the impact of the strong emission lines 
on our galaxy size measurements, 
we use images with the medium-band filter F410M. 
As shown in the bottom panel of Figure \ref{fig:SED}, 
for the redshift range of $z \simeq 5.63$--$6.63$, 
F410M does not encompass the strong emission lines, 
allowing us to primarily probe the rest-frame optical continuum.

We perform SB profile fittings 
for galaxies with spectroscopic redshifts of $z_{\rm spec} = 5.63$--$6.63$ in our sample 
using the F410M and F444W images. 
Although the number of sources with spectroscopic redshifts within this range 
is 13 regardless of their S/Ns,
successful convergence of the SB profile fitting is achieved 
for 3 sources each in the CEERS3 and CEERS6 fields. 
For the remaining sources, the SB profile fitting encounters numerical convergence issues 
probably due to their low S/Ns 
(for details, see Section 10 of the GALFIT user's manual).  
Figure \ref{fig:SB_fitting_results_F410M_F444W} displays the results of the SB profile fittings. 
Among them, C03{\_}01, C06{\_}02, and C06{\_}07 are relatively faint, 
but still appear to fit well with the S{\'e}rsic profiles.
Given the limited number of sources explored here, 
we include these faint ones in the plots as well.

The top and bottom panels of Figure \ref{fig:f410m_f444w_re} 
compare the size and total magnitude measurement results, respectively. 
We confirm a broad agreement in sizes derived from F410M and F444W.
While individual size measurements differ by about $1$--$4\sigma$ 
between F410M and F444W, suggesting that sizes may vary to this extent for individual sources, 
the overall average does not seem to be significantly impacted. 
Moreover, particularly for relatively brighter sources indicated with filled circles, 
the difference in size is within $20${\%}. 
On the other hand, 
the total magnitudes measured with F444W are systematically brighter 
than those measured with F410M mostly by 
$\simeq 0.2$--$0.4$ mag,
which is likely due to the strong H$\alpha$ emission line probed with F444W. 
Indeed, since the FWHM of the F444W filter 
transmission\footnote{\url{https://jwst-docs.stsci.edu/jwst-near-infrared-camera/nircam-instrumentation/nircam-filters}} 
is about $1.1\mu$m, 
these magnitude differences roughly correspond to 
rest-frame H$\alpha$ emission line equivalent widths (EWs) of $320$--$700${\AA} 
for galaxies at $z=6$. 
These values roughly align with the average EW values 
for high-$z$ SFGs reported in previous studies: 
$555^{+332}_{-311}$ {\AA} for $z=4.9$ galaxies with a median $M_{\rm UV} = -21.4$ mag 
(Table 4 of \citealt{2018ApJ...859...84H}) 
and $453 \pm 84$ {\AA} for $z=3.8$--$5.3$ galaxies with $-20.5 < M_{\rm UV} < -19.5$ mag 
(Table 3 of \citealt{2019A&A...627A.164L}; 
see also, \citealt{2020MNRAS.493.5120M}; \citealt{2023A&A...672A.186P}).\footnote{C03{\_}04 shows 
a slightly larger magnitude difference of $0.68$ mag. 
This might indicate that this source has a larger H$\alpha$ EW 
experiencing a relatively young starburst compared to the others.} 
Because the H$\beta+$[{\sc Oiii}] EWs are not very different from that of H$\alpha$ 
(the median H$\beta+$[{\sc Oiii}] EW value for galaxies at $z\sim6.5$--$8$ 
with slightly fainter magnitudes is 780{\AA}; \citealt{2023arXiv230605295E}),  
the impact of strong line emission from H$\alpha$ and H$\beta+$[{\sc Oiii}] should be 
comparable.\footnote{Note that the impact of H$\beta+$[{\sc Oiii}] line emission 
on the recovered size and magnitudes for sources will need to be investigated in future work.} 
Since the size and magnitude differences 
due to the strong emission lines are not very large 
owing to the wide wavelength coverage of F444W,  
from the next subsection onwards, 
we utilize the F444W images 
to measure the the rest-frame optical sizes and total magnitudes of our galaxies 
over a more extensive redshift range.

Note that \cite{2023arXiv230607940Z} have selected sources with spatially extended H$\beta+$[{\sc Oiii}] emission 
by combining two NIRCam broadband filters, 
one of which incorporates the H$\beta+$[{\sc Oiii}] emission and the other does not. 
The focus of their study is primarily on the spatially extended SB distributions substantially beyond the effective radii. 
In contrast, our size measurements mainly rely on 
SB profiles within the effective radii with relatively high S/Ns. 
Therefore, there is no conflict between our confirmation and the approach of \cite{2023arXiv230607940Z}. 
Moreover, it is worth noting that 
sources exhibiting spatially extended H$\beta+$[{\sc Oiii}] emission are rare, 
and hence unlikely to significantly impact our general measurements of high-$z$ galaxy sizes.

\begin{figure}
\begin{center}
   \includegraphics[width=0.4\textwidth]{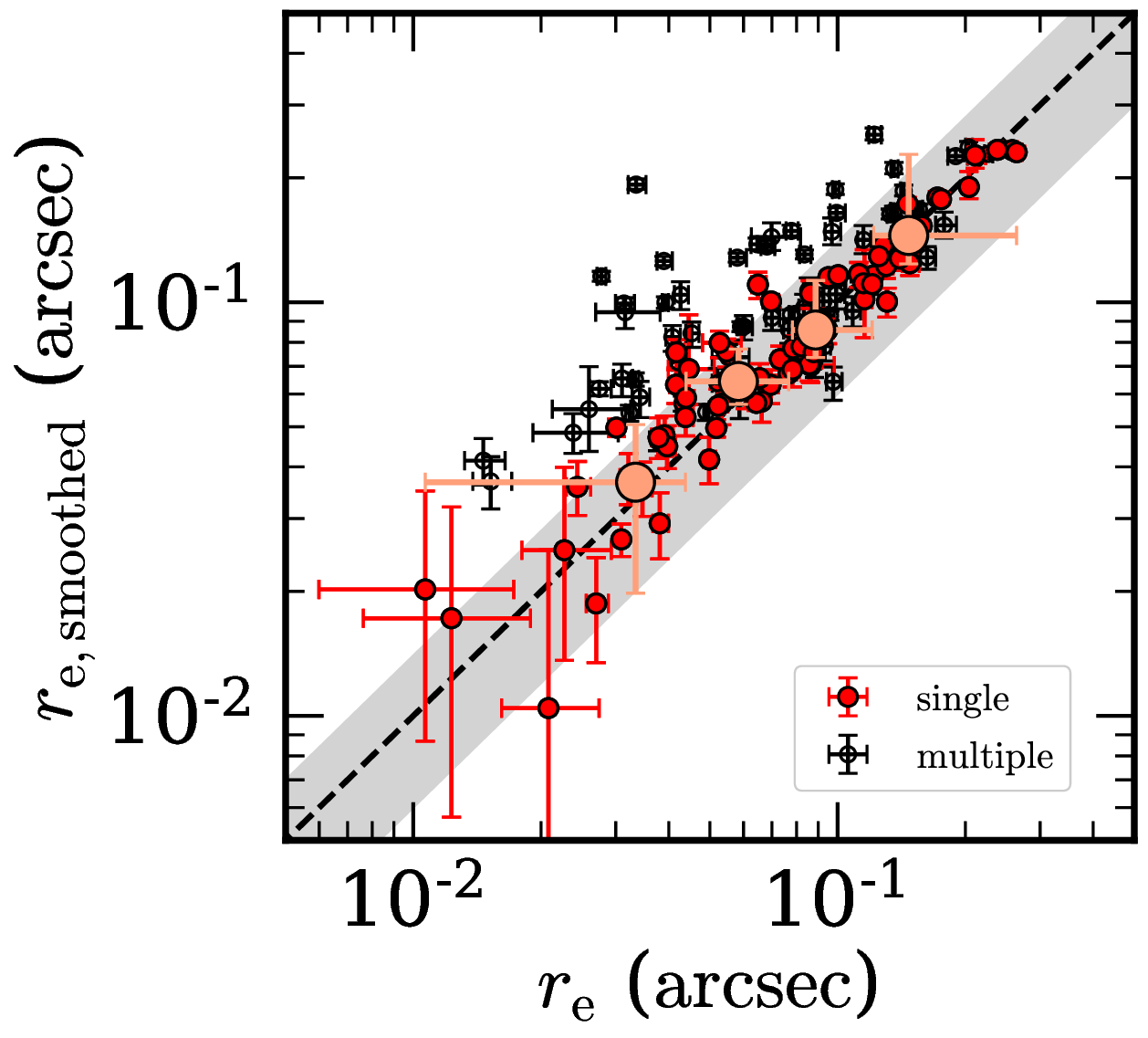}
   \includegraphics[width=0.4\textwidth]{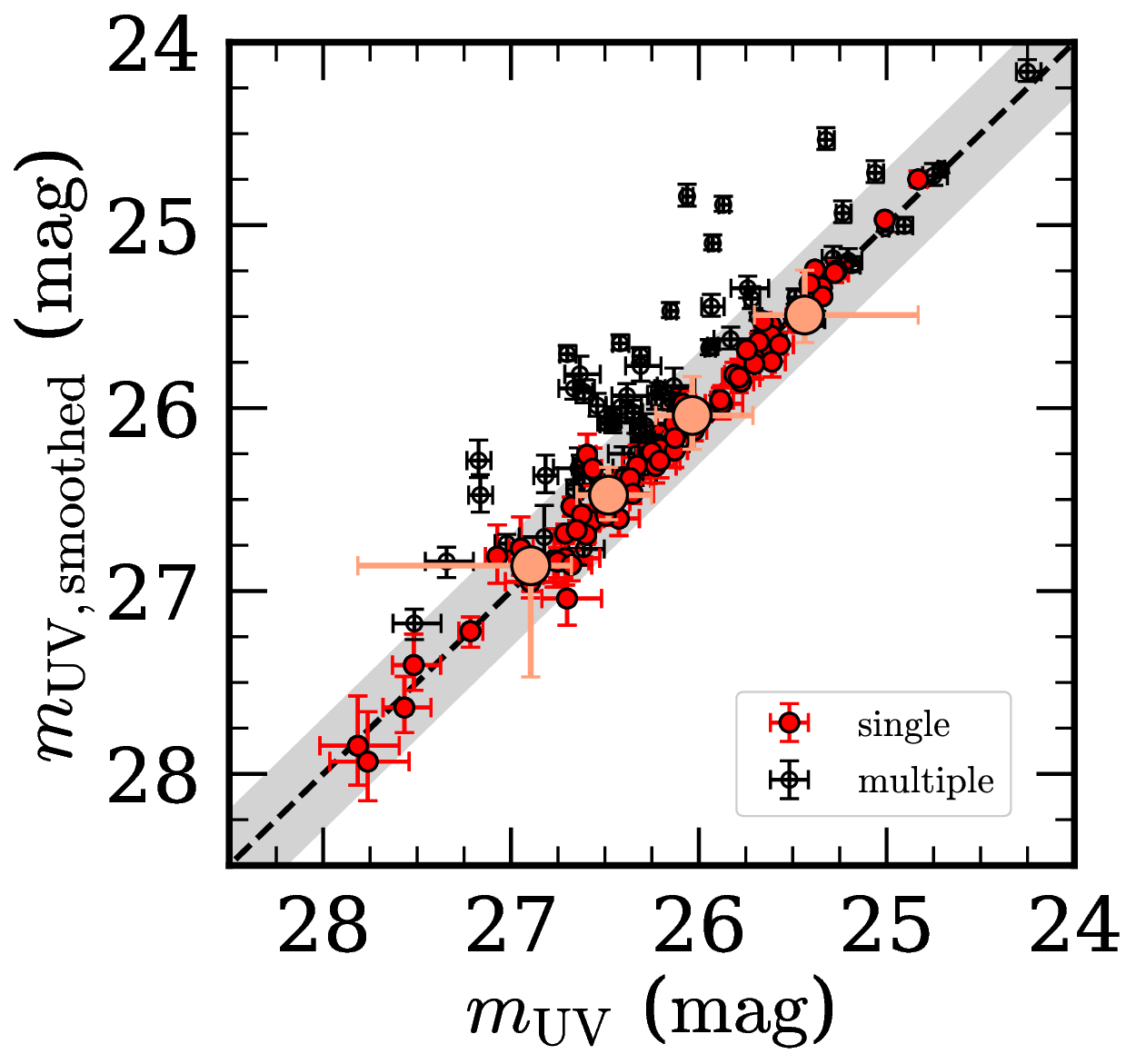}
\caption{
Comparison of SB profile fitting results 
for the galaxies at $z \simeq 4$--$10$ 
with the F150W images before and after PSF matching. 
The top panel presents the results for size, while the bottom panel shows the results for total magnitude. 
The red filled circles represent sources 
where the component fitted with the smoothed F150W image and/or the F444W image 
is the same as the one fitted with the original F150W image (single). 
The black open circles indicate sources 
where the component fitted with the smoothed F150W image and/or the F444W image 
includes surrounding components not fitted in the original F150W image (multiple). 
The large orange filled circles and error bars along the $y$-axis 
represent the median values and 68th percentiles of the results from the PSF-matched images 
for the single sources, 
with the sample divided into quartiles based on the results from the images before PSF matching. 
The error bars along the $x$-axis represent the range of 
size or total magnitude values in each divided sample.
The gray shade in the top (bottom) panel represents the range where the difference between the sizes 
(total magnitudes) is within $\pm40${\%} ($\pm 0.3$ mag).
}
\label{fig:comp_PSFmatched}
\end{center}
\end{figure}

\subsection{Impact of Different Spatial Resolutions in the Rest-frame UV and Optical} 
\label{subsec:PSF_match}

\hspace{1em}
One of the primary topics of interest in this study is the comparison between the rest-frame UV and optical sizes. 
For this purpose, the difference of the spatial resolutions of the F150W and F444W images needs to be taken into account. 
For instance, as presented in Figure \ref{fig:SB_fitting_results}, 
C01{\_}05 has two components near the center of the F150W cutout image, 
and the SB profile fitting is performed for one of the two components. 
However, in the F444W image, these two components are considered as one blended source, 
and the SB profile fitting is performed for the blended source accordingly. 
As such, for fair comparisons of the SB profile fitting results with the F150W and F444W images, 
we need to consider the difference of the spatial resolutions.

We create smoothed F150W images by applying Gaussian kernels to the original F150W images  
so that the PSF sizes of the smoothed F150W images match those of the F444W images. 
We then perform SB profile fittings using GALFIT on the PSF-matched F150W images. 
We also carry out MC simulations for the PSF-matched F150W images 
following the same methodology outlined in Section \ref{subsec:MC_simulation}. 
The results of size and total magnitude measurements for our MC simulated galaxies 
are shown in Figure \ref{fig:input_output_re} and Figure \ref{fig:input_output_mag}, respectively. 
Based on these MC simulation results, 
we correct systematic effects associated with the SB profile fittings 
and evaluate statistical uncertainties in the measurements.

By visually inspecting the output images of the SB profile fittings, 
we assign a flag about blendedness to each object as listed in Table \ref{tab:fitting_results}. 
When the component fitted in the smoothed F150W image and/or the F444W image 
is a single component in the original F150W image, 
we set the flag to 1 (single). 
When the component fitted in the smoothed F150W image and/or the F444W image 
is multiple components in the original F150W image, 
we set the flag to 2 (multiple). 
In other words, we assign a multiple flag to a source 
if visual inspection of the residual image from a single component fit for it 
reveals the presence of additional components. 
These cases often show multiple segments in segmentation maps created by SExtractor. 
Typically, the centers of these multiple components are spatially offset from each other.
For comparisons of the SB profile fitting results in the rest-frame UV and optical, 
we use the results from the original (PSF-matched) F150W images for the single (multiple) flagged sources. 
This approach ensures a fair comparison even with the data having different spatial resolutions.

The question of whether the sizes and total magnitudes obtained from the original F150W images 
can be reproduced when the spatial resolution is as low as that of the F444W images is intriguing.
Figure \ref{fig:comp_PSFmatched} compares the size and total magnitude results 
derived for the original and PSF-matched F150W images. 
The top panel illustrates the comparison of the size measurements. 
For single sources, 
the results are in good agreement across a wide range of sizes. 
This indicates that, even when the spatial resolution is as large as in the F444W image, 
small sizes derived for the original F150W image can still be obtained in the PSF-matched F150W image. 
On the other hand, 
for multiple sources, 
the obtained sizes tend to be larger after PSF matching as expected, 
because more components are fitted in the SB profile fitting after PSF matching.

The bottom panel of Figure \ref{fig:comp_PSFmatched} demonstrates the magnitude measurements 
with the F150W images before and after PSF matching. 
As expected, the results for the single sources align well, 
while the multiple sources tend to be systematically brighter after PSF matching.

\section{Results and Discussion} \label{sec:results}

\subsection{SB Profile Fitting Results for the Rest-frame UV and Optical Continuum} 
\label{subsec:SB_profile_fitting_results}

\begin{figure}
\begin{center}
   \includegraphics[width=0.4\textwidth]{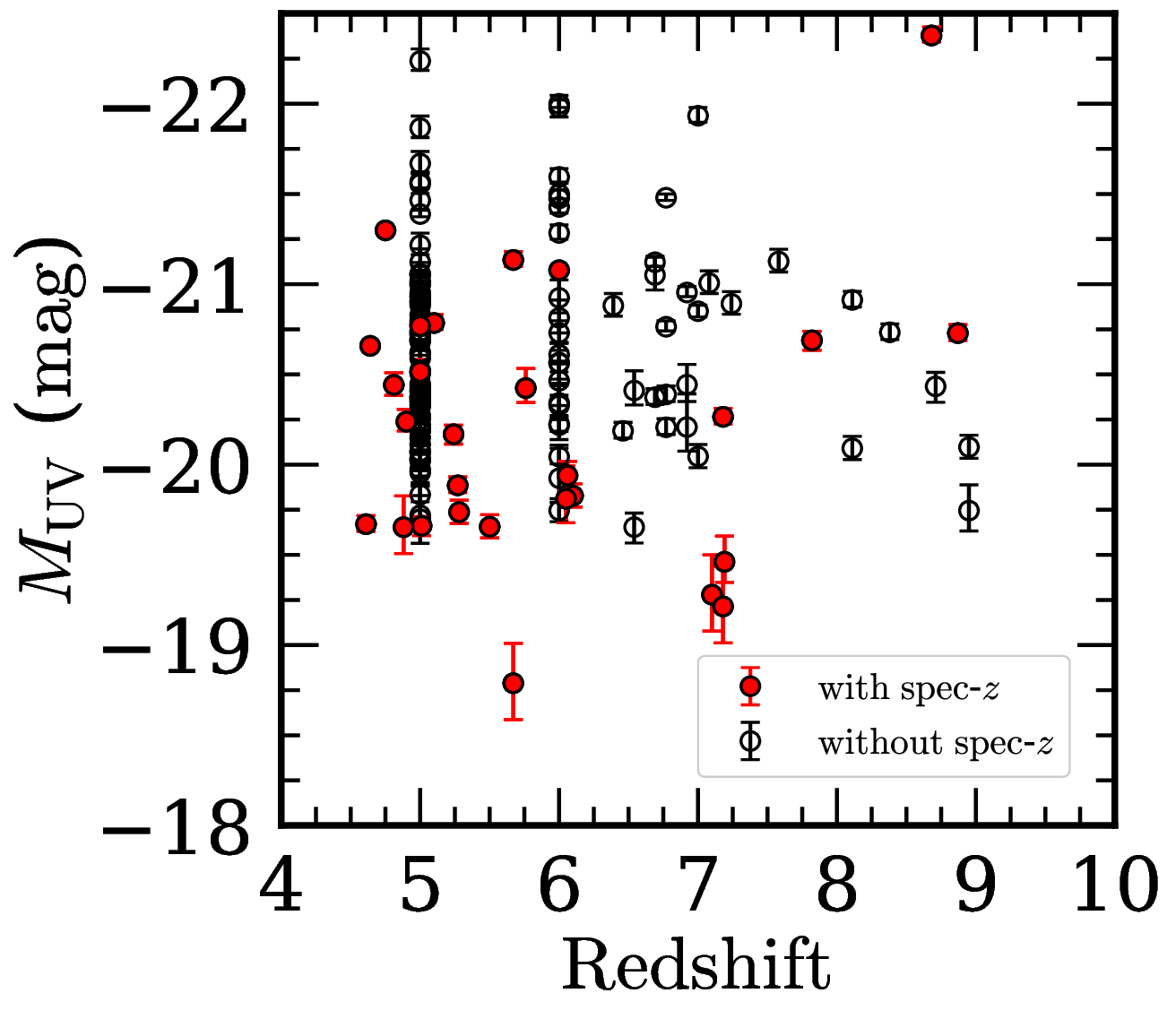}
   \includegraphics[width=0.4\textwidth]{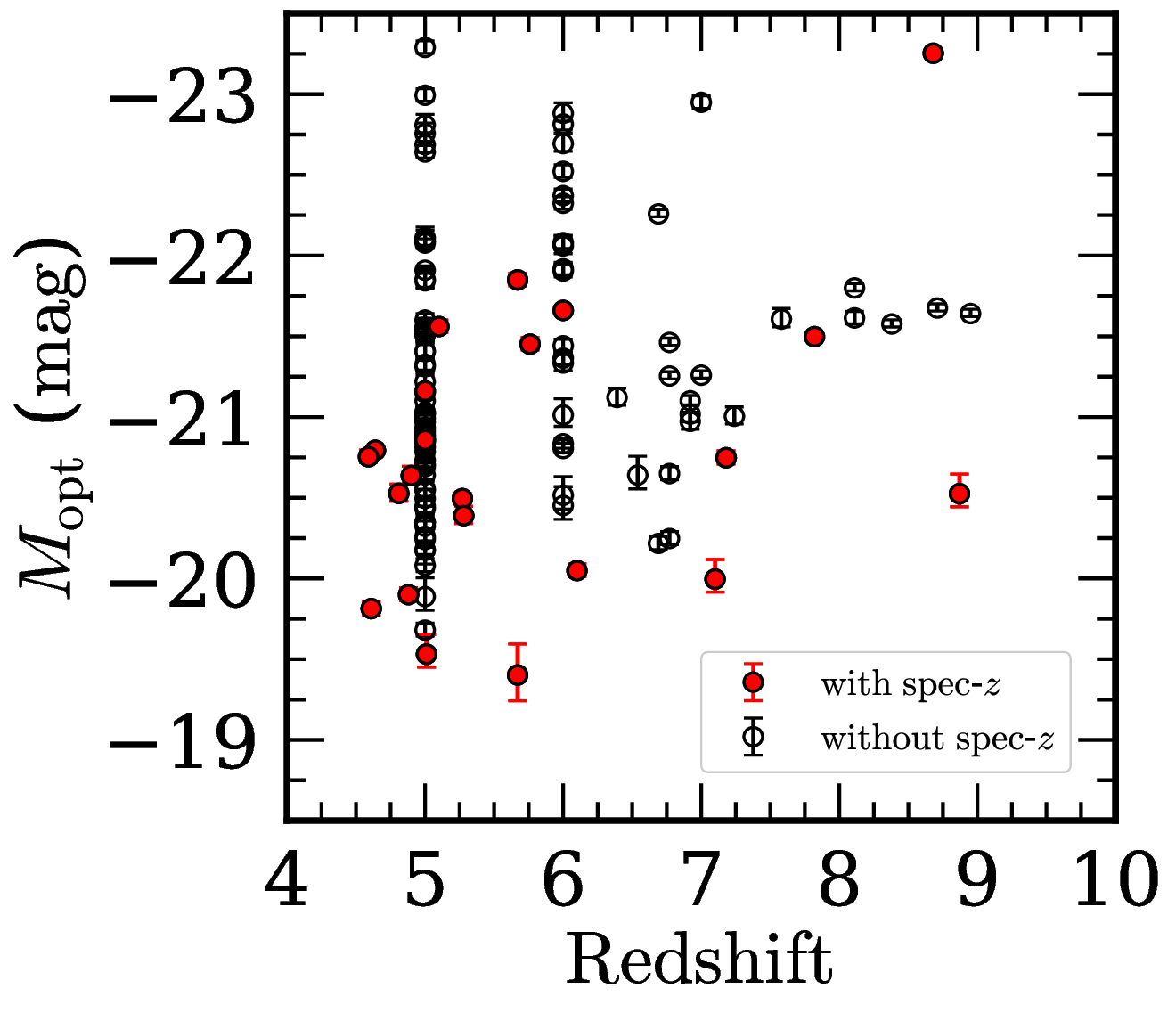}
\caption{
\textbf{Top}: 
Distribution of $M_{\rm UV}$ and redshift for the galaxies at $z\simeq 4$--$10$ 
where the S/Ns are larger than $10$ in the F150W images 
and the SB profile fittings are successfully converged. 
The red filled circles represent spectroscopically confirmed sources, 
while the black open circles indicate sources without spectroscopic confirmation. 
\textbf{Bottom}: 
Same as the top panel but for $M_{\rm opt}$ obtained from the F444W images. 
}
\label{fig:Muv_redshift}
\end{center}
\end{figure}

\hspace{1em}
We perform SB profile fittings for the galaxies at $z \simeq 4$--$10$ in the CEERS fields, 
which are compiled in Section \ref{sec:data}, 
to determine their sizes and total magnitudes in the rest-frame UV and optical. 
The F150W images are used to probe the rest-frame UV continuum, 
whereas the F444W images are used for the rest-frame optical. 
The SB profile fittings are conducted only for sources with S/Ns exceeding $10$ (Section \ref{sec:data}). 
Among them, the SB profile fittings with the F150W (F444W) images successfully converge 
for $29$ and $120$ ($22$ and $96$) sources with and without spectroscopic confirmations, respectively.  
Their distributions on the rest-frame UV absolute magnitude ($M_{\rm UV}$) vs. redshift 
and the the rest-frame optical absolute magnitude ($M_{\rm opt}$) vs. redshift planes 
are presented in Figure \ref{fig:Muv_redshift}. 
The best-fit parameters obtained from the SB profile fittings 
for the individual sources are summarized in Table \ref{tab:fitting_results} 
and the individual fitting result images are shown in Figure \ref{fig:SB_fitting_results}. 
The pseudo-color cutout images of those galaxies 
whose spectroscopic redshifts have been compiled by \cite{2023ApJS..269...33N} 
are shown in Figure \ref{fig:PseudoColors}.\footnote{As can be seen from this figure, 
sources at moderate redshifts often appear green, 
while those at relatively high redshifts often appear red. 
This would be primarily due to the effect of strong emission lines in the rest-frame optical 
(Section \ref{subsec:comparison_f410m_f444w}).}

\begin{figure}
\begin{center}
   \includegraphics[width=0.4\textwidth]{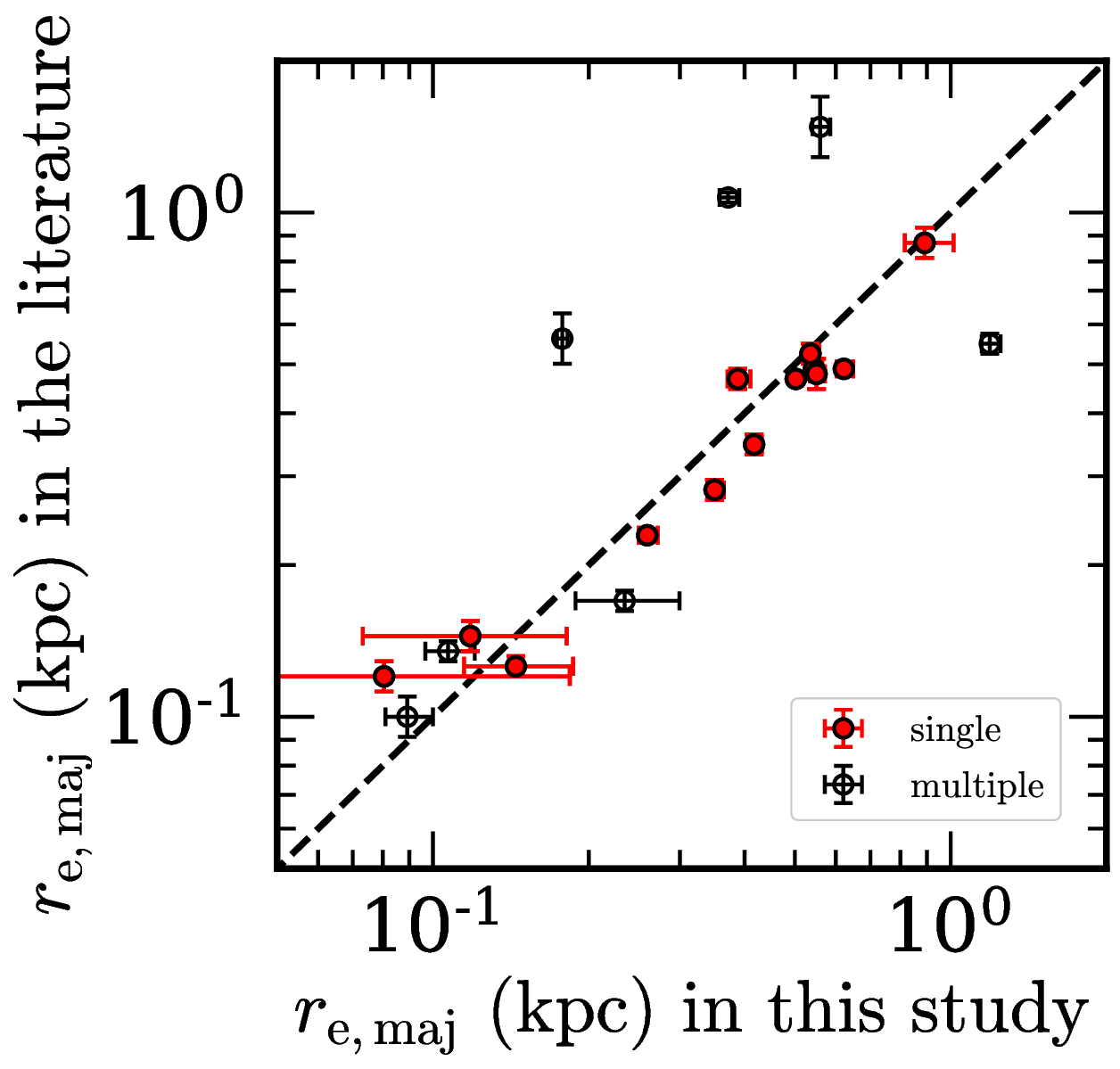}
\caption{
Comparison of size measurement results for individual galaxies in the rest-frame UV. 
The horizontal axis represents our size measurements 
and the vertical axis denotes those in \cite{2023arXiv230805018M}. 
The red filled circles represent sources where the component fitted with JWST NIRCam F444W 
is the same as the one fitted with F150W (single). 
The black open circles indicate sources where the component fitted with F444W 
includes surrounding components not fitted in the F150W image (multiple). 
Given that \cite{2023arXiv230805018M} provide 
only the effective radius along the semi-major axis ($r_{\rm e,maj}$), 
these comparisons are also based on the semi-major-axis effective radius.
The black dashed line is the case where our estimates and those in the literature are equivalent. 
}
\label{fig:comparison_size}
\end{center}
\end{figure}

The morphologies of some of these sources have been investigated in previous work. 
C06{\_}06 is the same as CEERS-AGN-z5-1 in \cite{2023ApJ...942L..17O} 
and CEERS 1670 in \cite{2023ApJ...954L...4K}, 
which is selected as an AGN candidate and confirmed by follow-up spectroscopy.
\cite{2023ApJ...942L..17O} have reported that 
the central SB profile of C06{\_}06 is dominated by the PSF in the F115W image, 
suggesting its size of less than $0.13$ kpc. 
Our obtained size for this source in F150W is $0.09 \pm 0.01$ kpc, 
which is consistent with these previous results. 
Note that the other source reported in \cite{2023ApJ...954L...4K}, 
CEERS 3210, is too faint to be investigated in this study.

In addition to C06{\_}06, 
C03{\_}04 is selected as broad line AGNs by \cite{2023arXiv230311946H} 
based on the broad component detections only in the permitted H$\alpha$ line.\footnote{C03{\_}04 is CEERS 397 
and C06{\_}06 is CEERS 2782 in \cite{2023arXiv230311946H}.} 
In addition, C02{\_}02 (CEERS 1465) and C08{\_}01 (CEERS 1019) are selected as 
possible broad-line AGN candidates in \cite{2023arXiv230311946H} 
(See also, \citealt{2023ApJ...953L..29L}). 
\cite{2023arXiv230311946H} have reported that 
C06{\_}06 shows a compact morphology, 
while the other three sources have extended morphologies. 
In our SB profile fittings, 
although all of these sources are fitted with S{\'e}rsic profiles, 
C06{\_}06 shows the smallest size.\footnote{We exclude 
the F444W fitting results for C06{\_}06 due to issues with SB profile fitting convergence.} 
In this sense, our results are consistent with their findings. 
In what follows, we mark these AGNs and AGN candidates with a special symbol in the figures as necessary.
As mentioned in Section \ref{subsec:size_luminosity_relation}, 
these broad line AGNs and AGN candidates tend to be more compact than normal SFGs. 
It should also be noted that the remaining sources at $z>4.5$ in their study 
are not investigated in this paper due to their faintness 
or encountering numerical convergence issues in the SB profile fittings.

Very recently, based on the data from public extragalactic fields obtained in JWST Cycle 1, 
\cite{2023arXiv230805018M} have compiled a catalog of SFGs at $z=5$--$14$ 
and measured their sizes in the rest-frame UV. 
Their catalog encompasses $20$ of the galaxies that we investigate. 
A comparison of the size measurement results for these galaxies is shown in Figure \ref{fig:comparison_size}. 
Because they only provide the effective radius along the semi-major axis, 
our comparison here is also based on the effective radius along the semi-major axis. 
Based on the single or multiple flags established in Section \ref{subsec:PSF_match}, 
we differentiate the symbols for the data points.
As can be seen from the Figure \ref{fig:comparison_size}, 
when considering only the galaxies labeled single, 
our results align well with theirs. 
However, there are some multiple flagged galaxies that deviate significantly. 
Specifically, for three objects, their results are larger by about $0.3$--$0.5$ dex compared to ours; 
their IDs are C01{\_}05, C06{\_}01, and C08{\_}01 (Figure \ref{fig:SB_fitting_results}). 
Probably, \cite{2023arXiv230805018M} have treated them as a single blended source in their SB profile fittings, 
while we perform the SB profile fittings for one of these components, 
which could explain the significant size discrepancy. 
Despite these exceptions, the results from both studies generally align well.

In \cite{2023arXiv230607940Z}, 
galaxies with spatially extended H$\beta+$[{\sc Oiii}] emission 
are selected based on the emission line images created by subtracting NIRCam broadband images 
with and without emission line contributions. 
Among the selected sources, CEERS 792 is the same as C06{\_}04. 
However, due to the low S/N, 
we do not conduct the SB profile fitting for this particular source in this study.

\begin{figure}
\begin{center}
   \includegraphics[width=0.4\textwidth]{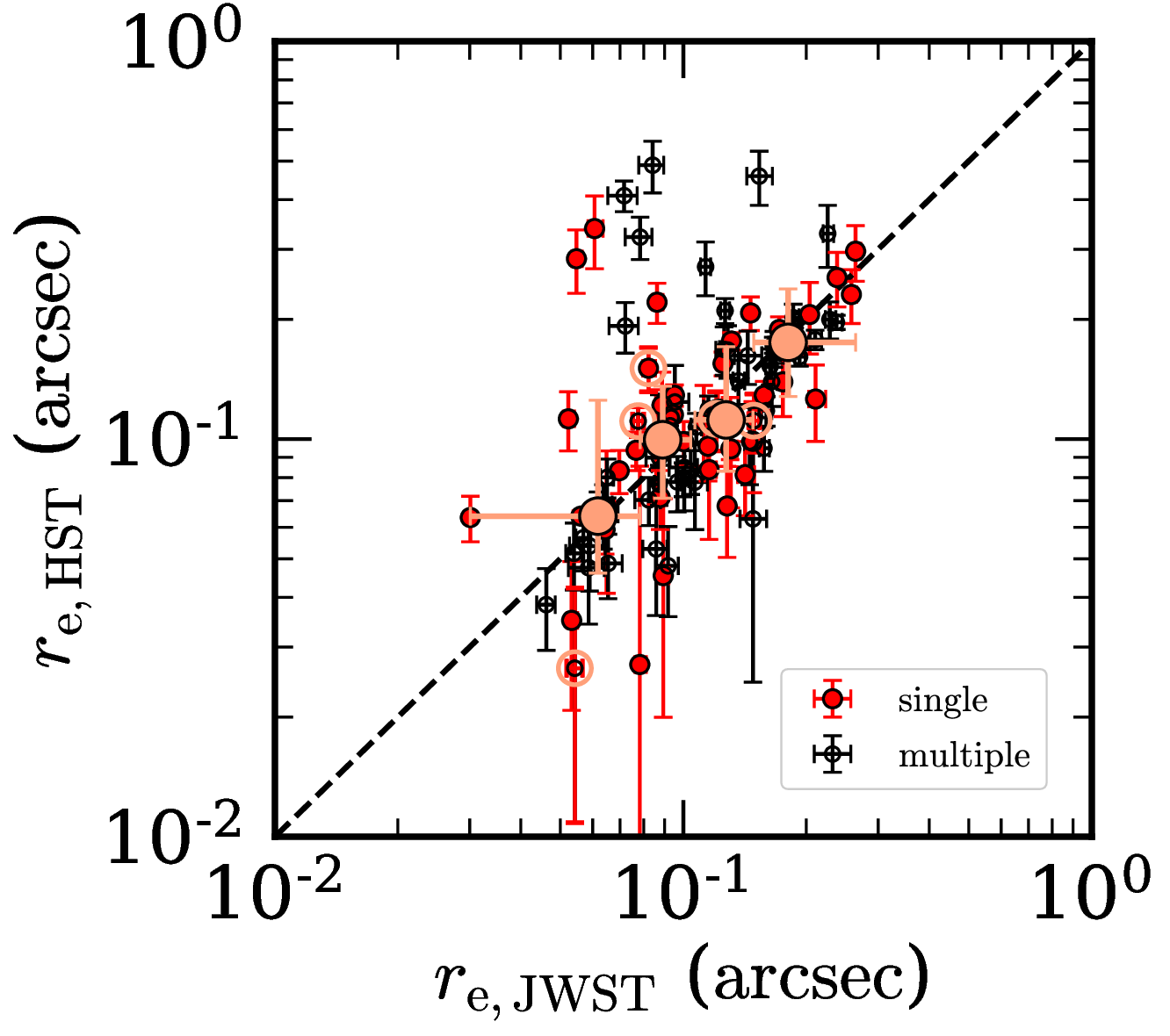}
   \includegraphics[width=0.4\textwidth]{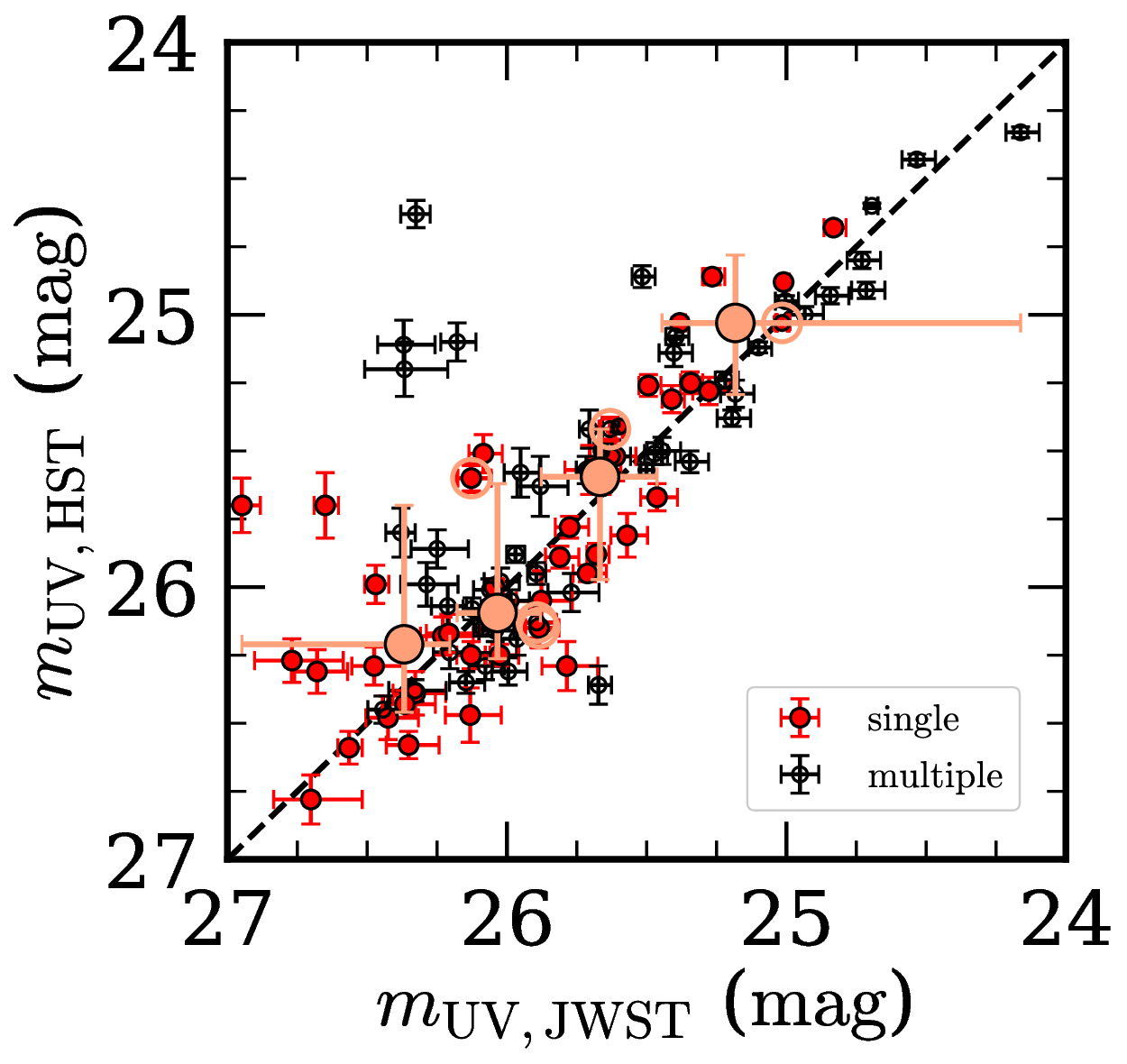}
\caption{
Comparison of SB profile fitting results 
for the galaxies at $z \simeq 4$--$10$ 
based on the rest-frame UV images obtained with HST and JWST. 
The top panel presents the results for size, while the bottom panel shows the results for total magnitude. 
The JWST results are obtained in this study, 
and the HST results are obtained by \cite{2015ApJS..219...15S}. 
The red filled circles represent sources where the component fitted with JWST NIRCam F444W, 
having a spatial resolution similar to HST, 
is the same as the one fitted with F150W (single). 
The black open circles indicate sources where the component fitted with F444W 
includes surrounding components not fitted in the F150W image (multiple). 
For the single (multiple) sources, 
the JWST results are based on the results from the original (PSF-matched) F150W images. 
The large orange open circles denote sources with spectroscopic redshifts. 
The large orange filled circles and error bars along the $y$-axis 
represent the median values and 68th percentiles of the HST results, 
with the sample divided into quartiles based on the results from JWST. 
The error bars along the $x$-axis represent the range of 
size or total magnitude values in each divided sample.
}
\label{fig:comp_HST}
\end{center}
\end{figure}

\subsection{Comparisons between HST and JWST Measurement Results} 

\hspace{1em}
Based on the HST WFC3 images, 
\cite{2015ApJS..219...15S} have performed 
SB profile fittings for the galaxies in their sample 
and estimated their sizes and total magnitudes in the rest-frame UV. 
Because the spatial resolutions of HST and JWST are different, 
it raises a concern whether it is appropriate to compare these SB profile fitting results.

Figure \ref{fig:comp_HST} displays a comparison of the rest-frame UV sizes and total magnitudes 
for the galaxies in the \cite{2015ApJS..219...15S} sample obtained with the HST and JWST images. 
Because of the difference in spatial resolution between HST and JWST, 
some galaxies appearing as single sources in the HST images 
are resolved into multiple components in the JWST images.  
Conveniently, the spatial resolution of HST WFC3 is similar to those of the red filters of JWST NIRCam. 
Therefore, for this comparison with the HST results, 
we use the JWST results from the original or PSF-matched F150W images 
depending on whether the sources are flagged as single or multiple 
in Section \ref{subsec:PSF_match}.

\begin{figure}
\begin{center}
   \includegraphics[width=0.49\textwidth]{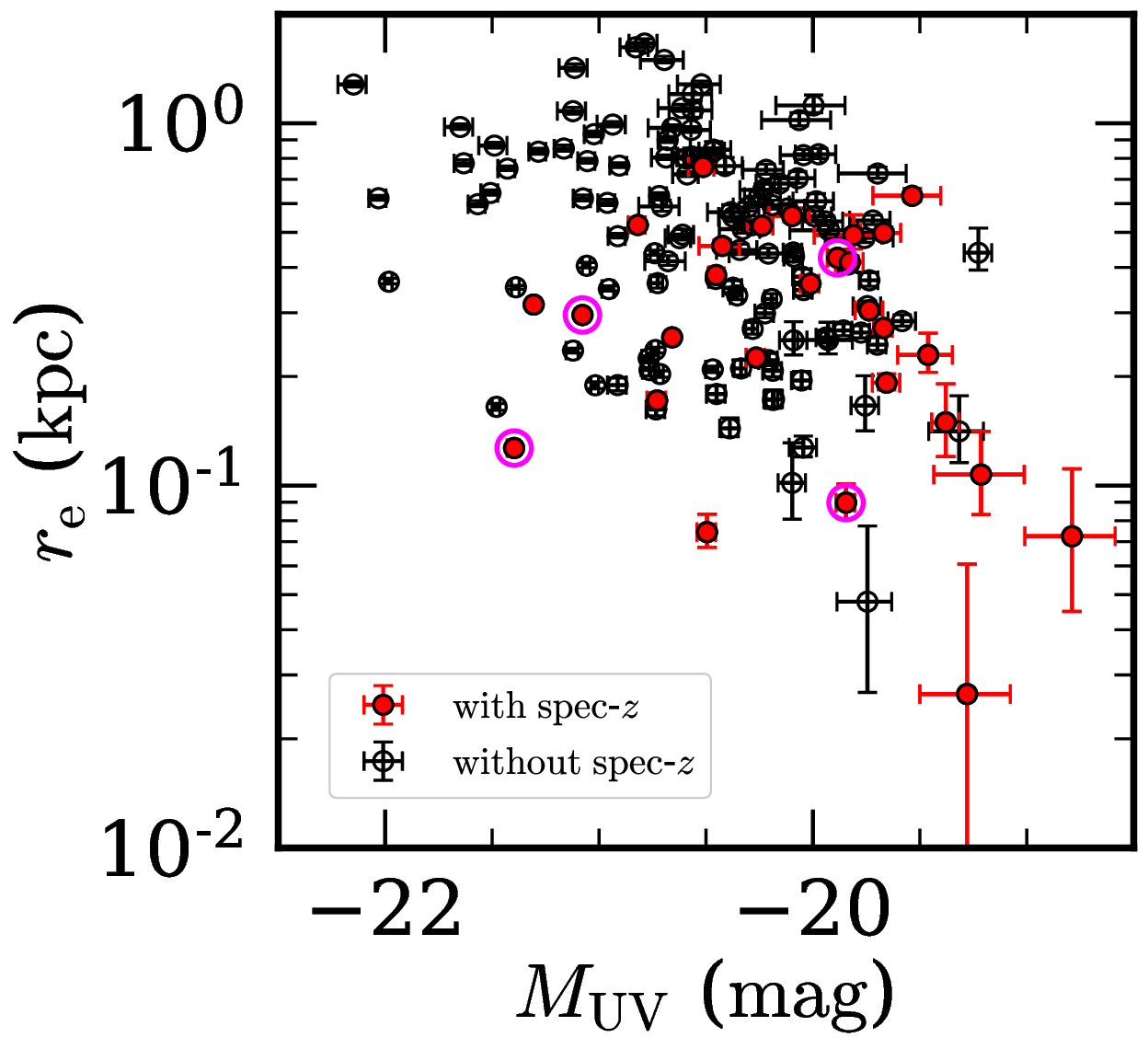}
   \includegraphics[width=0.49\textwidth]{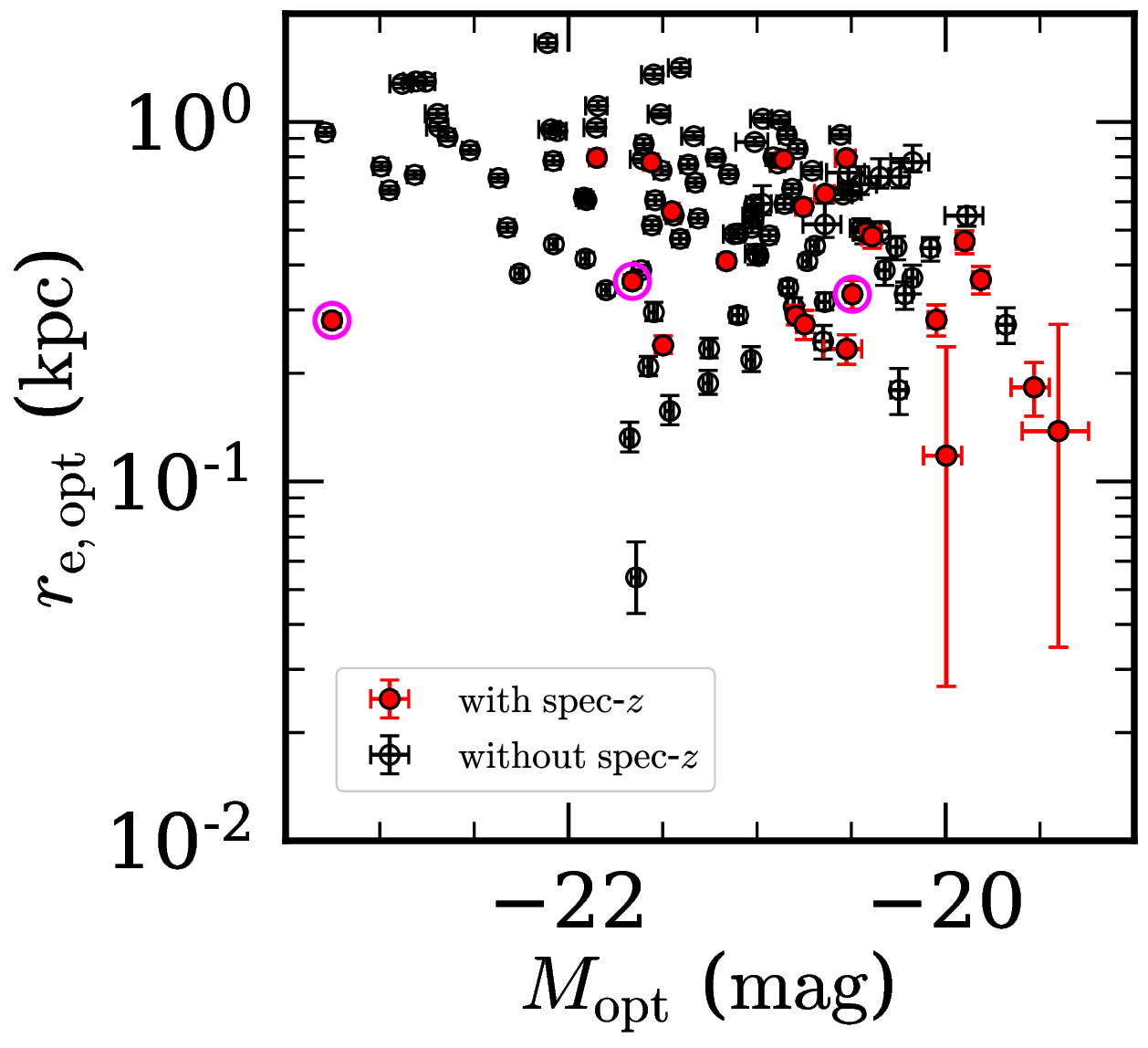}
\caption{
\textbf{Top}: 
Rest-frame UV size $r_{\rm e}$ vs. total magnitude $M_{\rm UV}$ 
for galaxies at $z\simeq 4$--$10$. 
The red filled circles denote sources with spectroscopic redshifts, 
while the black open circles are those without spectroscopic confirmation. 
These sizes and total magnitudes are estimated from the F150W images. 
The large open magenta circles represent broad line AGNs and AGN candidates 
identified in \cite{2023ApJ...954L...4K} and \cite{2023arXiv230311946H} 
(See also, \citealt{2023ApJ...942L..17O}). 
\textbf{Bottom}: 
Same as the top panel but for 
the rest-frame optical size $r_{\rm e,opt}$ vs. total magnitude $M_{\rm opt}$ 
estimated from the F444W images.  
}
\label{fig:re_Muv}
\end{center}
\end{figure}

As depicted in the top panel of Figure \ref{fig:comp_HST}, 
the size results obtained from HST and JWST images roughly coincide. 
Likewise, as presented in the bottom panel of Figure \ref{fig:comp_HST}, 
the total magnitude results also largely agree. 
Although individual outliers exist that deviate significantly, such sources are rare, 
and on average the results show good agreement.  
In the following subsections, we compare our JWST results 
with previously results obtained from the HST images, 
along with previous JWST results.

\subsection{Size--Luminosity Relation in the Rest-frame UV and Optical} 
\label{subsec:size_luminosity_relation}

\hspace{1em}
We investigate the relationship between the size and luminosity of our galaxies. 
The top (bottom) panel of Figure \ref{fig:re_Muv} shows 
the rest-frame UV (optical) sizes and total magnitudes for galaxies at $z\simeq 4.5-9.5$, 
which are estimated from our SB profile fittings with the F150W (F444W) images in this study. 
Here we directly use the optical total magnitudes obtained with the F444W images; 
these measurements would be systematically brighter 
due to the influence of strong emission lines such as H$\alpha$ and [{\sc Oiii}] 
as examined in Section \ref{subsec:comparison_f410m_f444w}.

Both panels confirm a trend that fainter sources tend to be smaller in size 
reported in the literature 
(\citealt{2010ApJ...709L..21O}; \citealt{2012A&A...547A..51G}; \citealt{2013ApJ...777..155O}; \citealt{2013ApJ...765...68H}; 
\citealt{2013ApJ...773..153J}; \citealt{2015ApJ...804..103K}; \citealt{2015ApJ...808....6H}; \citealt{2015ApJS..219...15S}; 
\citealt{2016MNRAS.457..440C}; \citealt{2017MNRAS.466.3612B}; \citealt{2017ApJ...843...41B}; \citealt{2018ApJ...855....4K}; 
\citealt{2019ApJ...882...42B}; \citealt{2020AJ....160..154H}; \citealt{2022ApJ...927...81B}; \citealt{2022ApJ...938L..17Y}; \citealt{2023ApJ...951...72O}). 
Additionally, the spectroscopically confirmed sources appear to be more compact, 
probably because the more compact a source is, 
the easier it is to detect their emission lines by spectroscopy. 
The broad line AGNs and AGN candidates except for C02{\_}02 appear to be relatively compact,  
which would be attributed to the bright emission from the central core 
compared to the host galaxy components.

\begin{figure*}
\begin{center}
   \includegraphics[width=0.33\textwidth]{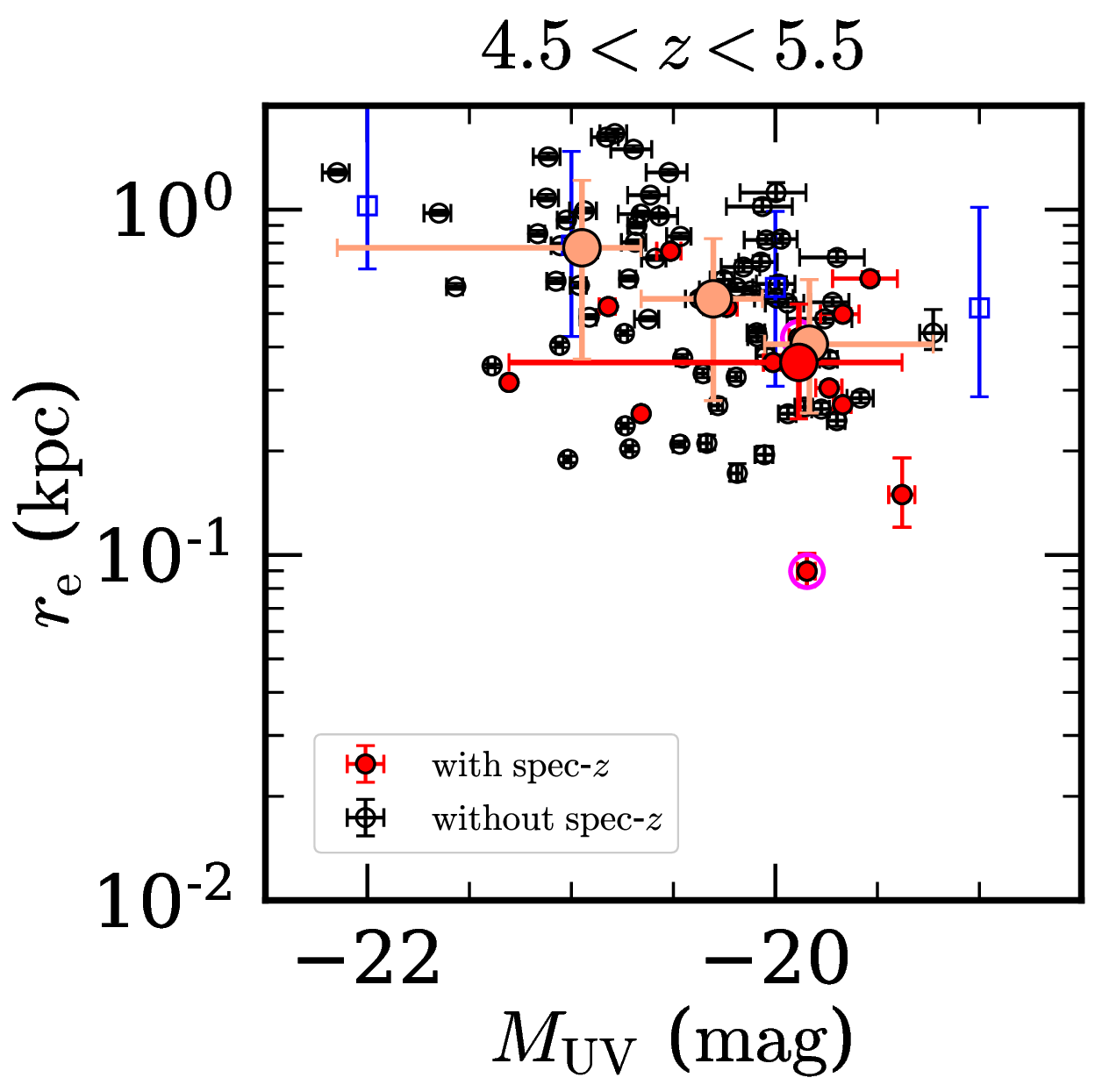}
   \includegraphics[width=0.33\textwidth]{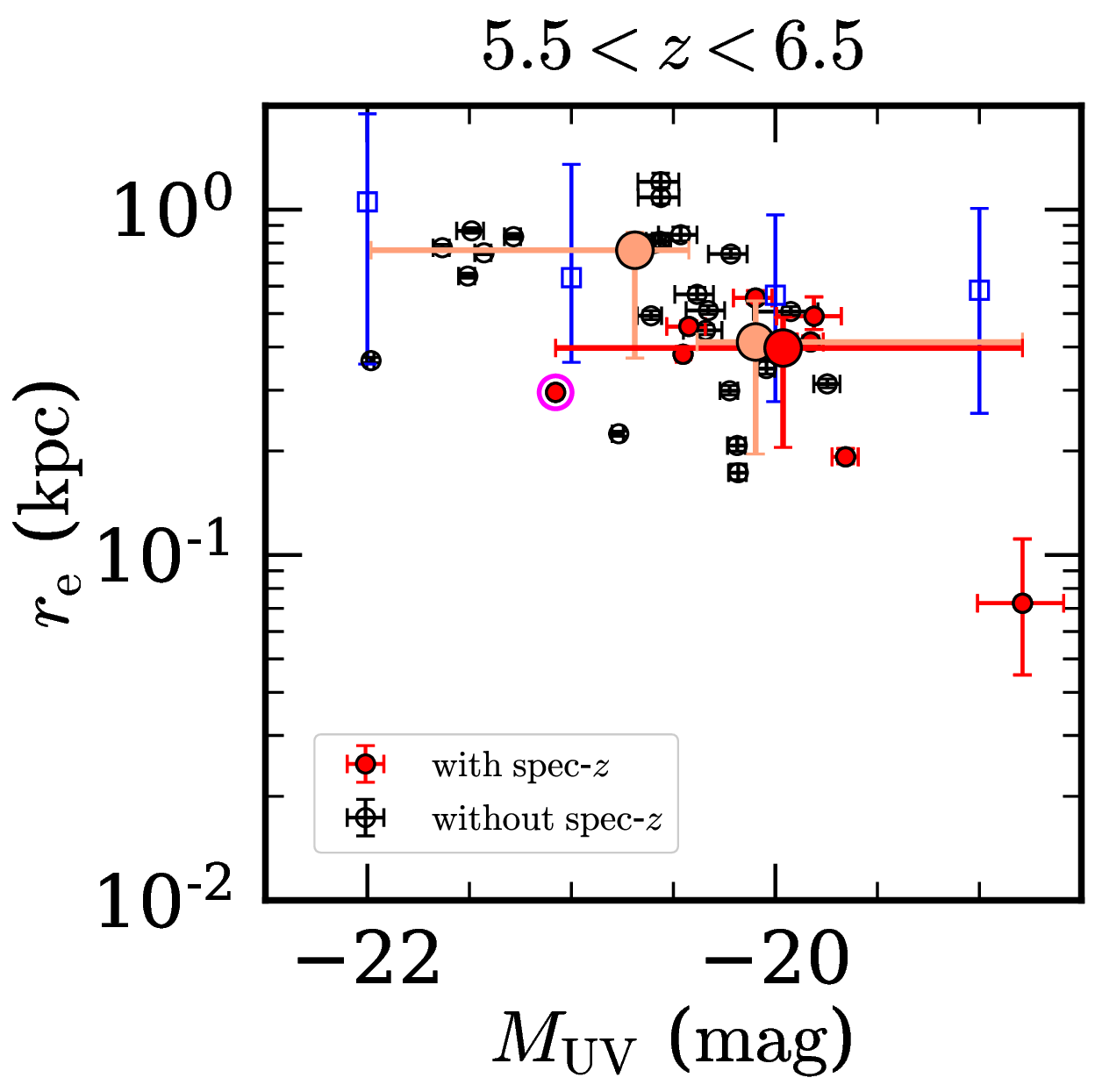}
   \includegraphics[width=0.33\textwidth]{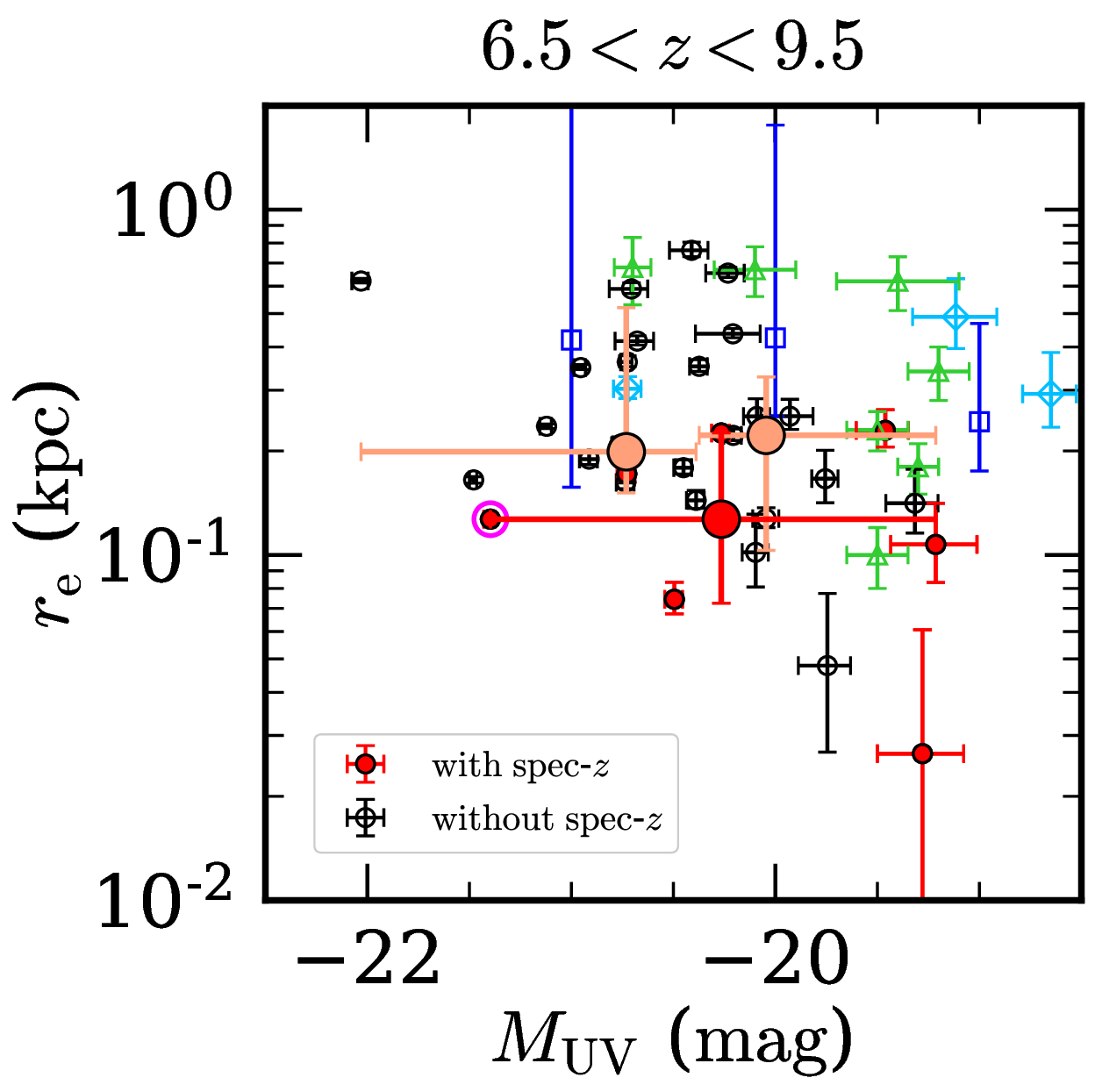}
\caption{
Same as the top panel of Figure \ref{fig:re_Muv}, 
but the data are divided into bins of redshift:  
from left to right, 
$z=4.5$--$5.5$, $z=5.5$--$6.5$, and $z=6.5$--$9.5$. 
The large orange filled circles and error bars along the $y$-axis 
represent the median values and 68th percentiles of the subsamples,  
which are divided into two or three groups based on $M_{\rm UV}$. 
The large red filled circles and error bars along the $y$-axis 
denote the median values and 68th percentiles of the galaxies with spectroscopic redshifts. 
The error bars of the large symbols along the $x$-axis 
denote the range of total magnitude values in each sample.
The blue open squares, cyan open diamonds, and green open triangles are 
previous results for galaxies at similar redshifts 
obtained by  \cite{2015ApJS..219...15S}, \cite{2023ApJ...951...72O}, 
and \cite{2022ApJ...938L..17Y}, respectively. 
}
\label{fig:re_Muv_z}
\end{center}
\end{figure*}

The previous work has shown that the sizes of galaxies decrease with increasing redshifts on average 
(\citealt{2004ApJ...600L.107F}; \citealt{2004ApJ...611L...1B}; \citealt{2006ApJ...653...53B}; 
\citealt{2008ApJ...673..686H}; \citealt{2010ApJ...709L..21O}; \citealt{2012A&A...547A..51G}; 
\citealt{2013ApJ...777..155O}; \citealt{2013ApJ...765...68H}; \citealt{2013ApJ...773..153J}; 
\citealt{2015ApJ...804..103K}; \citealt{2015ApJ...808....6H}; \citealt{2015ApJS..219...15S}; 
\citealt{2016MNRAS.457..440C}; \citealt{2017ApJ...834L..11A}; \citealt{2017MNRAS.466.3612B}; 
\citealt{2017ApJ...843...41B}; \citealt{2018ApJ...855....4K}; \citealt{2020AJ....160..154H}; 
\citealt{2022ApJ...927...81B}; \citealt{2023ApJ...951...72O}). 
Thus, we show the rest-frame UV sizes and total magnitudes divided into three redshift ranges 
in Figure \ref{fig:re_Muv_z}. 
For each redshift bin, we split the samples into three or two groups based on the total magnitude 
and calculate the median sizes and its 68th percentiles. 
These values are plotted together and summarized in Table \ref{tab:re_Muv}. 
For all of these subsamples, we confirm the trend that fainter sources have smaller sizes, 
although this trend is not very apparent in the median values for the $6.5<z<9.5$ subsample.
For comparison, we also show the results of \cite{2015ApJS..219...15S}, 
who have investigated the rest-frame UV sizes of high-$z$ galaxies using HST images, 
as well as the results of \cite{2022ApJ...938L..17Y} and \cite{2023ApJ...951...72O}  
based on JWST images. 
We find that our results are broadly consistent with these previous results 
when comparing at similar total magnitudes. 
Note that the results from \cite{2023arXiv230805018M} are not plotted here, 
because their measurements for the circularized radius are not available.

\begin{table}
{\footnotesize
\caption{Median Sizes and Total Magnitudes in the Rest-frame UV and Optical 
of the Subsamples Divided into Three Redshift Bins 
}
\begin{center}
\begin{tabular}{cccccc} \hline
Redshift			& flag 	& $M_{\rm UV}$	& $r_{\rm e}$	& $M_{\rm opt}$	& $r_{\rm e,opt}$ \\ 
				&		& (mag)			& (kpc)		& (mag)			& (kpc)	\\ 
(1) & (2) & (3) & (4) & (5) & (6) \\ \hline
$z = 4.5$--$5.5$ 	& 1	& $-19.89$ & $0.36^{+0.17}_{-0.11}$ & $-20.58$ & $0.47^{+0.30}_{-0.15}$ \\
				& 2 	& $-19.83$ & $0.41^{+0.22}_{-0.15}$ & $-20.35$ & $0.49^{+0.21}_{-0.15}$ \\ 
				& 2	& $-20.30$ & $0.55^{+0.27}_{-0.27}$ & $-20.89$ & $0.59^{+0.23}_{-0.17}$ \\ 
				& 2 	& $-20.95$ & $0.77^{+0.44}_{-0.41}$ & $-21.72$ & $0.89^{+0.41}_{-0.39}$ \\ 
$z = 5.5$--$6.5$	& 1	& $-19.96$ & $0.40^{+0.09}_{-0.19}$ & $-21.45$ & $0.36^{+0.29}_{-0.13}$ \\ 
				& 2	& $-20.10$ & $0.41^{+0.13}_{-0.22}$ & $-20.92$ & $0.56^{+0.23}_{-0.26}$ \\ 
				& 2 	& $-20.69$ & $0.76^{+0.10}_{-0.39}$ & $-22.20$ & $0.75^{+0.22}_{-0.16}$ \\ 
$z = 6.5$--$9.5$	& 1	& $-20.27$ & $0.13^{+0.10}_{-0.05}$ & $-20.75$ & $0.24^{+0.04}_{-0.05}$ \\ 
				& 2 	& $-20.05$ & $0.22^{+0.11}_{-0.12}$ & $-20.70$ & $0.28^{+0.24}_{-0.06}$ \\ 
				& 2 	& $-20.73$ & $0.20^{+0.32}_{-0.05}$ & $-21.63$ & $0.26^{+0.19}_{-0.11}$ \\ 
\hline
\end{tabular}
\end{center}
Note. (1) Redshift range. 
(2) Flag for spectroscopic redshifts; 1: galaxies with spectroscopic confirmation, 
2:  galaxies with and without spectroscopic confirmation. 
(3) Median rest-frame UV total magnitude. 
(4) Median size in the rest-frame UV and its 68th percentiles. 
(5) Median rest-frame optical total magnitude. 
(6) Median size in the rest-frame optical and its 68th percentiles. 
\label{tab:re_Muv}
}
\end{table}

\begin{table}
{\small
\caption{Median Optical-to-UV Size Ratios and UV Total Magnitudes 
of the Subsamples Divided into Two Redshift Bins 
}
\begin{center}
\begin{tabular}{cccc} \hline
Redshift	& flag 	& $M_{\rm UV}$	& $r_{\rm e,opt} / r_{\rm e,UV}$ \\ 
		& 		& (mag)			& 	\\ \hline
(1) & (2) & (3) & (4) \\ \hline
$z = 4.5$--$5.5$ & 1 & $-20.24$ & $0.97^{+0.29}_{-0.22}$ \\ 
			  & 2	& $-20.17$ & $0.95^{+0.49}_{-0.17}$ \\ 
			  & 2	& $-20.90$ & $1.00^{+0.25}_{-0.15}$ \\ 
$z = 5.5$--$6.5$ & 1 & $-20.42$ & $1.22^{+0.25}_{-0.35}$ \\ 
			  & 2 & $-21.36$ & $0.95^{+0.28}_{-0.10}$ \\ 
			  & 2	& $-20.34$ & $1.08^{+0.48}_{-0.24}$ \\ 
$z = 6.5$--$9.5$ & 1 & $-20.69$ & $1.33^{+1.14}_{-0.36}$ \\ 
			  & 2 & $-20.93$ & $1.13^{+0.28}_{-0.53}$ \\ 
			  & 2	& $-20.32$ & $1.20^{+0.18}_{-0.32}$ \\ 
\hline
\end{tabular}
\end{center}
Note. (1) Redshift range. 
(2) Flag for spectroscopic redshifts; 1: galaxies with spectroscopic confirmation, 
2:  galaxies with and without spectroscopic confirmation. 
(3) Median rest-frame UV total magnitude. 
(4) Median ratio of the rest-frame optical size to the UV size and its 68th percentiles. 
\label{tab:re_ratio}
}
\end{table}

\begin{figure*}
\begin{center}
   \includegraphics[width=0.33\textwidth]{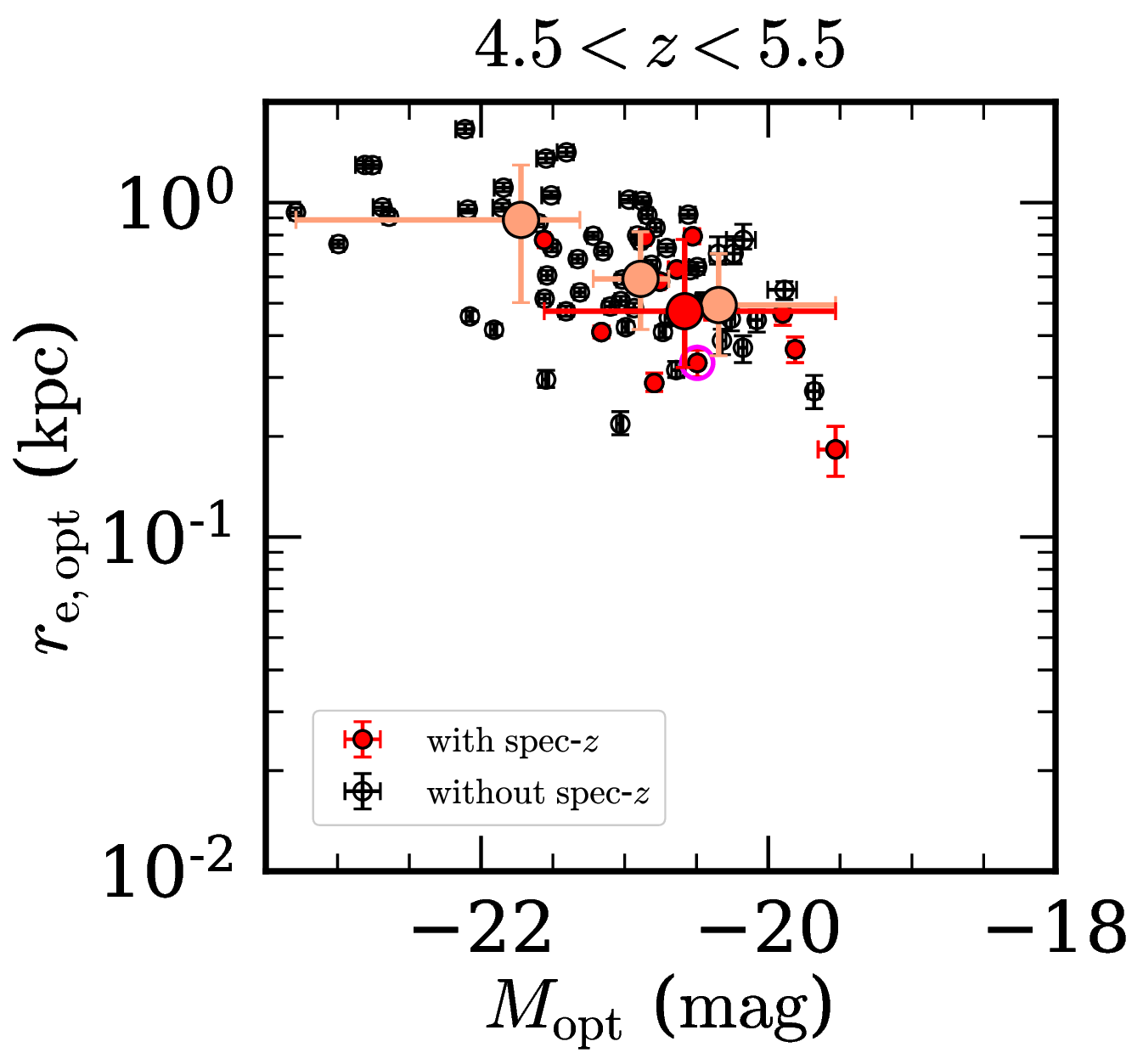}
   \includegraphics[width=0.33\textwidth]{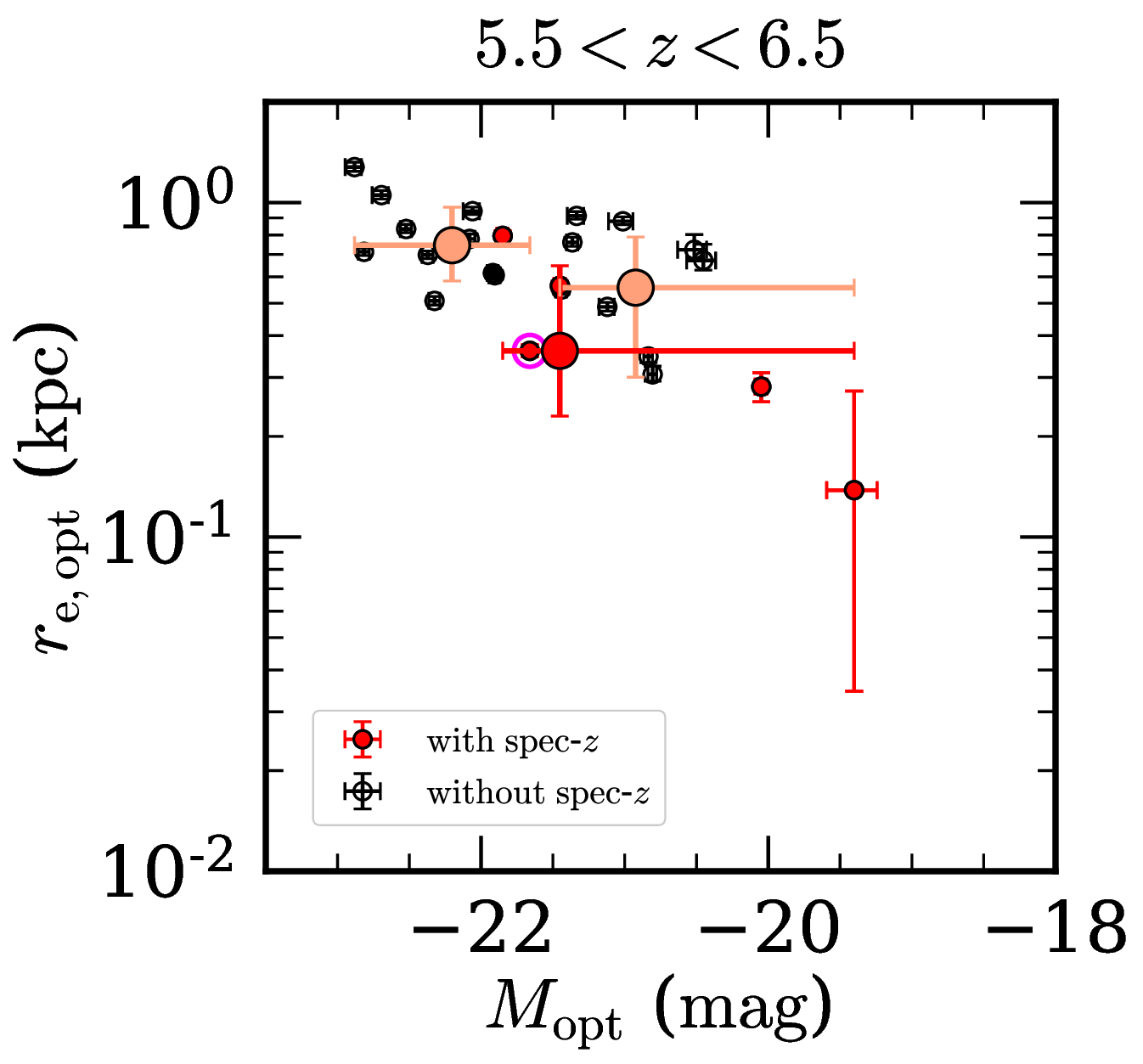}
   \includegraphics[width=0.33\textwidth]{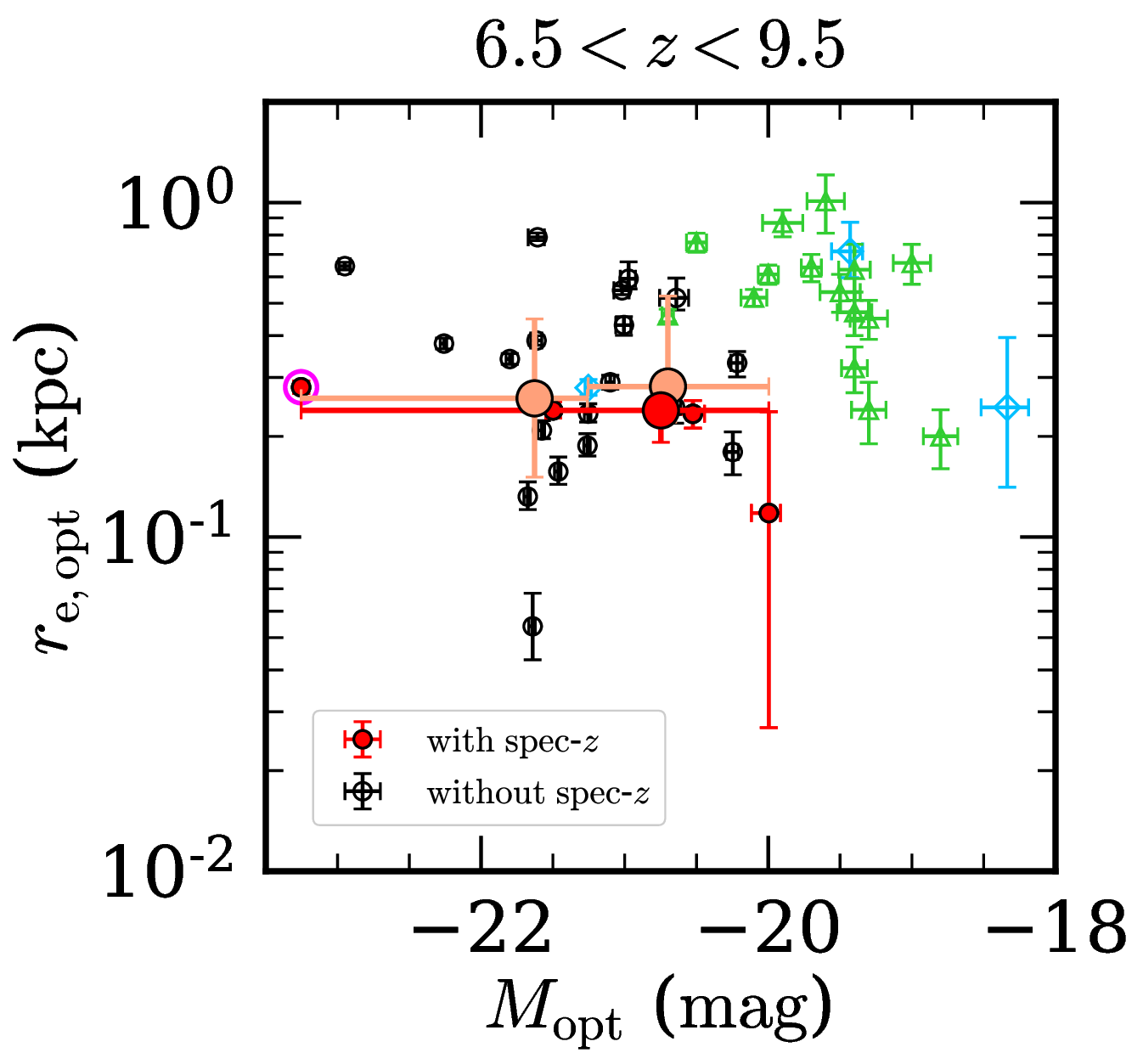}
\caption{
Same as the bottom panel of Figure \ref{fig:re_Muv}, 
but the data are divided into bins of redshift:  
from left to right, 
$z=4.5$--$5.5$, $z=5.5$--$6.5$, and $z=6.5$--$9.5$. 
The large orange filled circles and error bars along the $y$-axis 
represent the median values and 68th percentiles of the subsamples,  
which are divided into two or three groups based on $M_{\rm opt}$. 
The large red filled circles and error bars along the $y$-axis 
denote the median values and 68th percentiles of the galaxies with spectroscopic redshifts. 
The error bars of the large symbols along the $x$-axis 
denote the range of total magnitude values in each divided sample.
The cyan open diamonds and green open triangles are 
previous results for galaxies at similar redshifts 
obtained by  \cite{2023ApJ...951...72O} and \cite{2022ApJ...938L..17Y}, respectively. 
}
\label{fig:re_Mopt_z}
\end{center}
\end{figure*}

In Figure \ref{fig:re_Mopt_z}, 
we show the rest-frame optical sizes and total magnitudes divided into three redshift ranges 
in the same way as Figure \ref{fig:re_Muv_z}. 
For the $4.5 < z < 5.5$ subsample, we confirm the trend that fainter sources are smaller in size. 
However, such a trend is not significantly seen in the remaining subsamples. 
For the $6.5<z<9.5$ subsample, 
we find that our results are broadly consistent with the previous results 
of \cite{2022ApJ...938L..17Y} and \cite{2023ApJ...951...72O} 
at around $M_{\rm opt} \simeq -21$ mag.  
It should be noted that, in these previous studies, the SB profile fittings have been performed 
for galaxies fainter than those investigated in this study 
thanks to the GLASS data, which are deeper than the CEERS data used in this study. 
Additionally, we confirm that our results are on average in broad agreement with 
the very recent results of \cite{2023arXiv230809076S}.

\subsection{Size Ratio and Spatial Offset between the Rest-frame Optical and UV} 

\begin{figure*}
\begin{center}
   \includegraphics[width=0.33\textwidth]{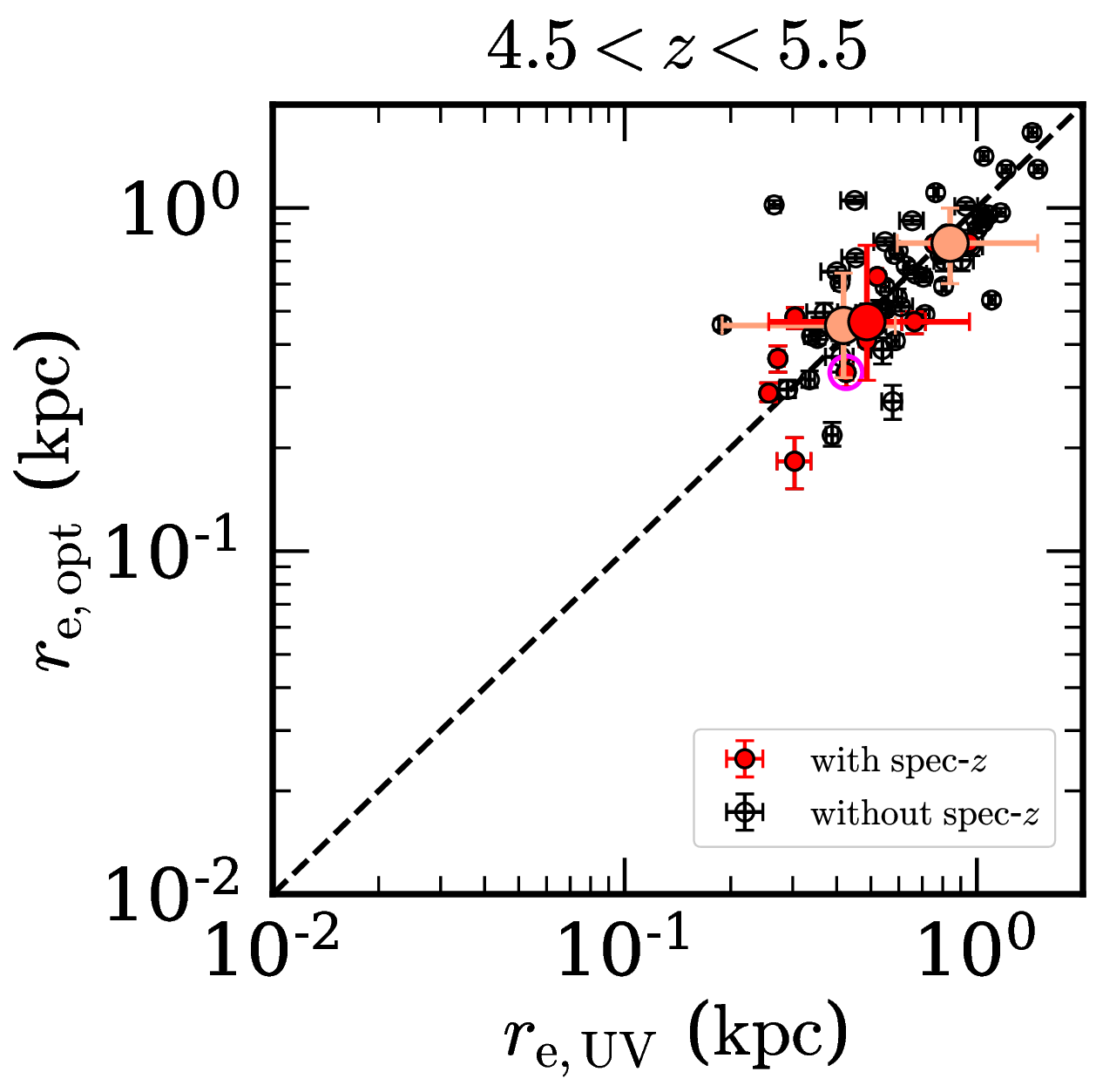}
   \includegraphics[width=0.33\textwidth]{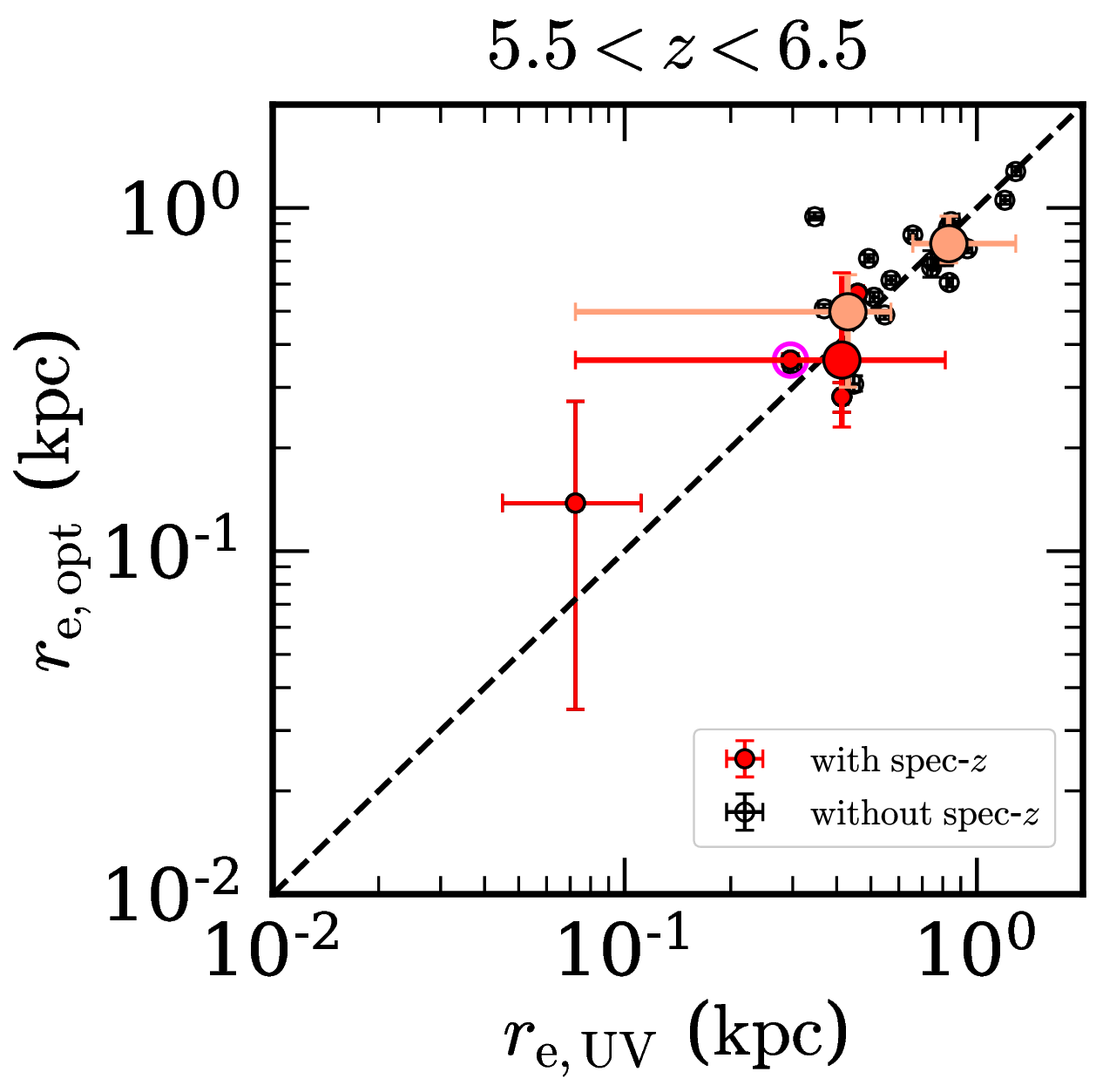}
   \includegraphics[width=0.33\textwidth]{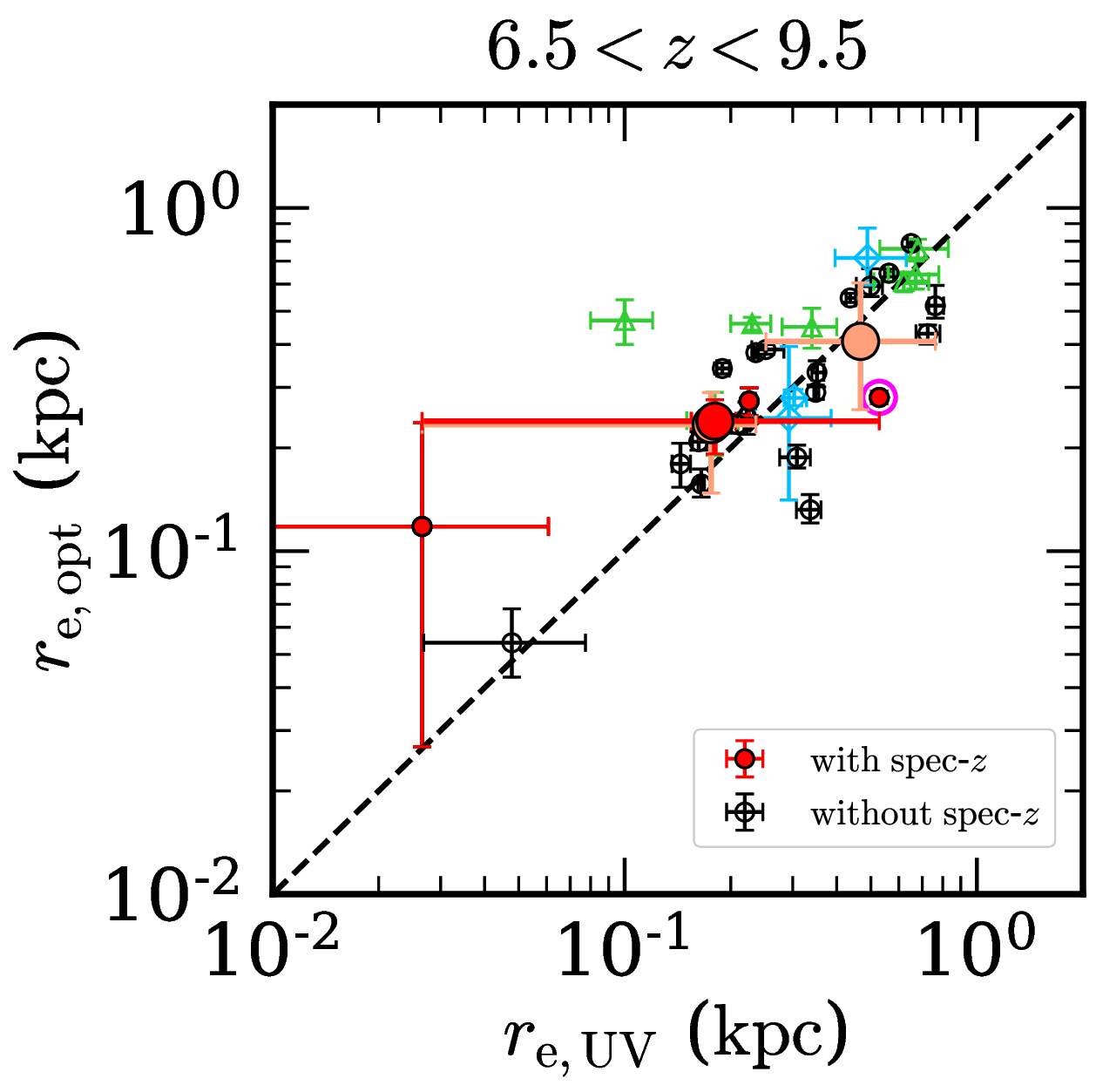}
   \includegraphics[width=0.33\textwidth]{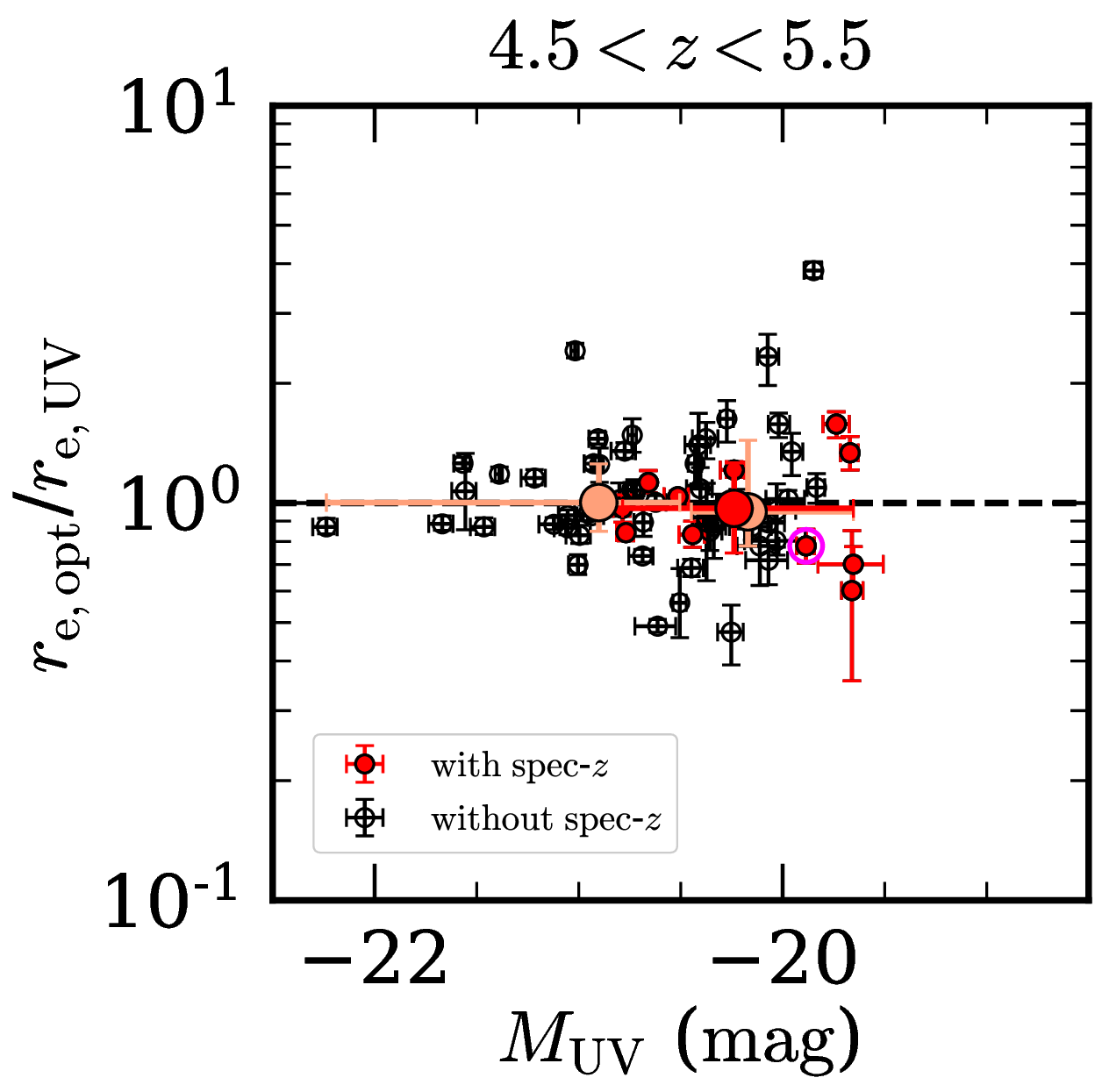}
   \includegraphics[width=0.33\textwidth]{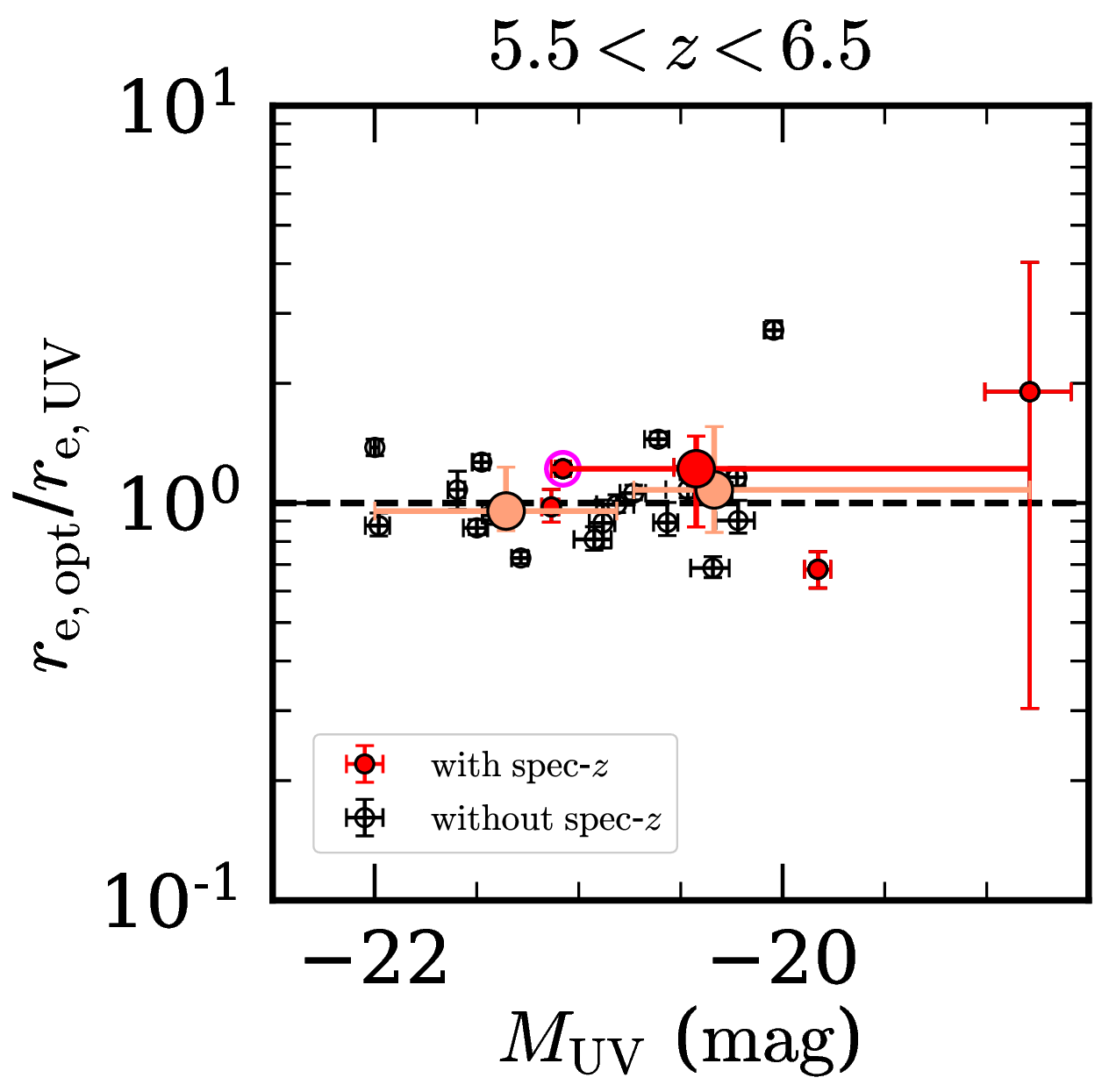}
   \includegraphics[width=0.33\textwidth]{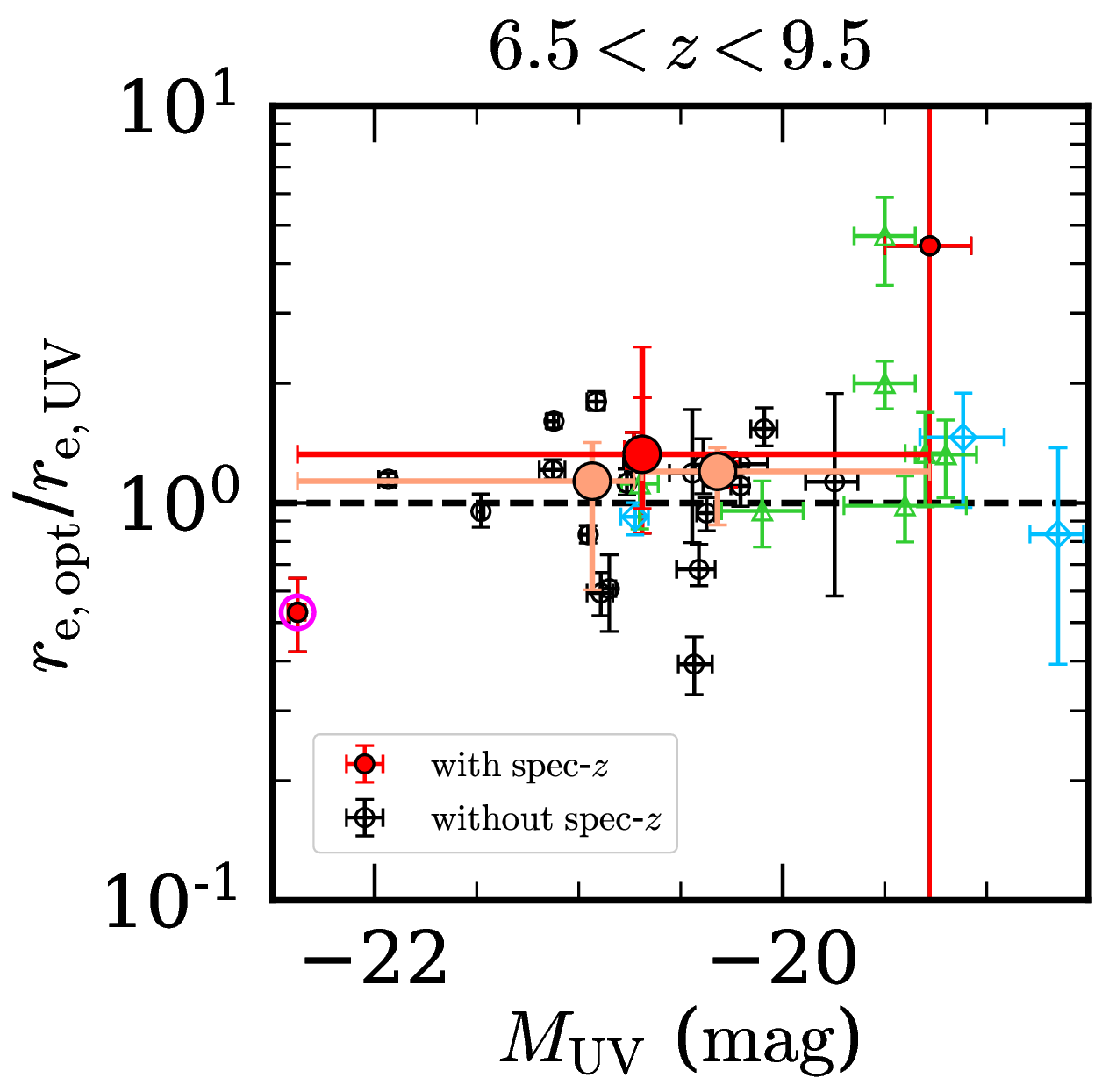}
\caption{\textbf{Top}: 
Comparison of rest-frame UV and optical sizes 
for galaxies at $z=4.5$--$5.5$, $z=5.5$--$6.5$, and $z=6.5$--$9.5$, from left to right. 
The red filled circles are spectroscopically confirmed sources,
while the black open circles are not spectroscopically confirmed.
Due to the different spatial resolutions,
we use the values obtained from the original F150W images for sources flagged as single,
and the values from the PSF-matched F150W images for sources flagged as multiple.
The large orange filled circles and error bars along the $y$-axis 
denote the median values and 68th percentiles of the rest-frame optical sizes 
for the samples divided into two based on the UV size.
The large red filled circles and error bars along the $y$-axis 
denote the median values and 68th percentiles of the galaxies with spectroscopic redshifts. 
The error bars of the large symbols along the $x$-axis 
represent the range of UV sizes in each subsample.
The black dashed line corresponds to the unity size ratio.
The large open magenta circles represent broad line AGNs and AGN candidates 
identified in \cite{2023arXiv230311946H}. 
The cyan open diamonds and green open triangles are 
previous results for galaxies at similar redshifts 
obtained by  \cite{2023ApJ...951...72O} and \cite{2022ApJ...938L..17Y}, respectively. 
\textbf{Bottom}: 
Same as the top panel, 
but the $x$-axis represents the rest-frame UV magnitude,
and the $y$-axis is the ratio of rest-frame optical to UV sizes.
}
\label{fig:re_ratio}
\end{center}
\end{figure*}

\begin{figure}
\begin{center}
   \includegraphics[width=0.4\textwidth]{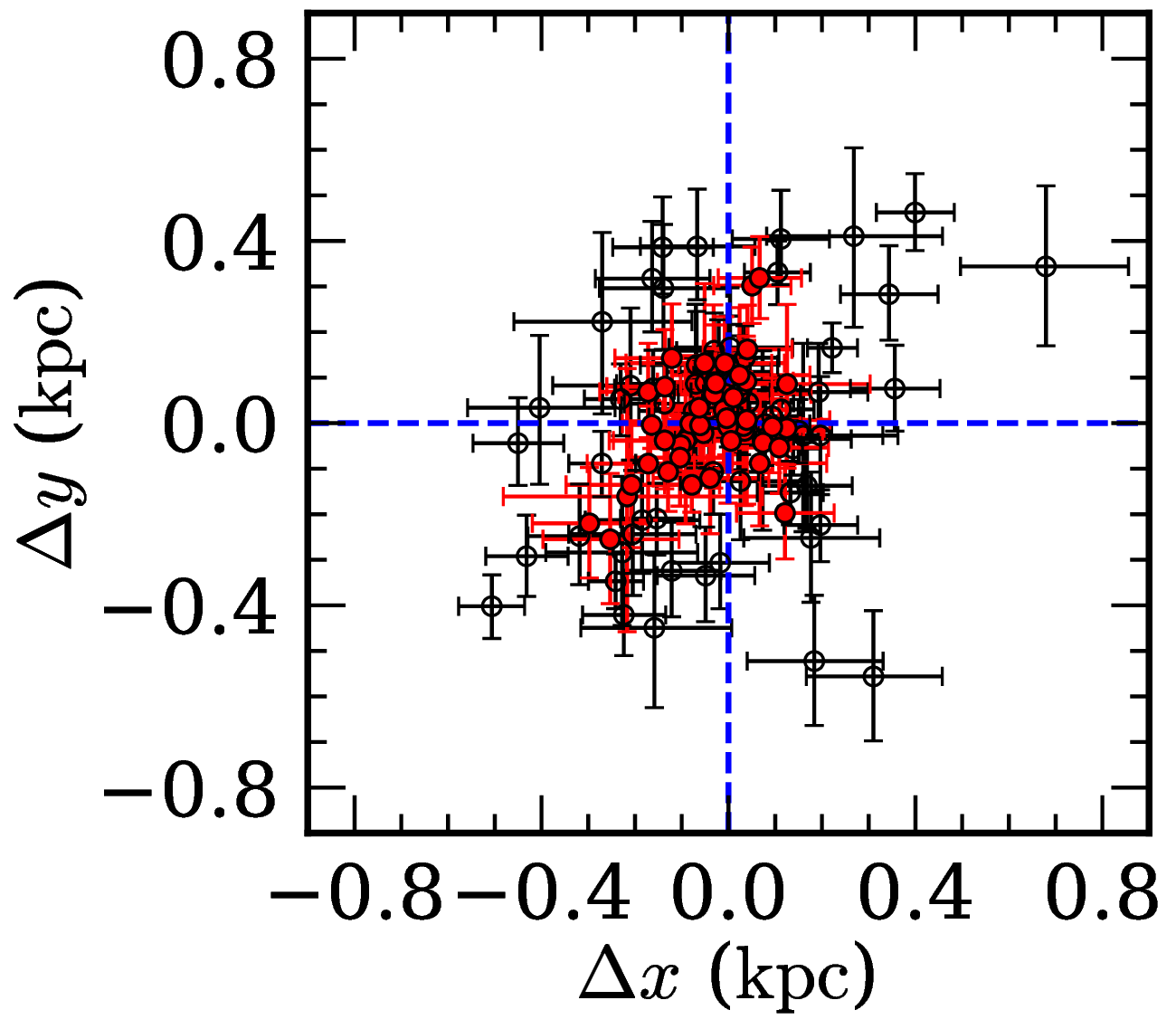}
\caption{Spatial offsets between the rest-frame UV and optical emissions for SFGs at $z=4$--$10$. 
The red filled (black open) circles represent 
the offsets between the F150W and F444W images 
along the $xy$-directions for individual galaxies 
without (with possible) signs of galaxy merger and/or tidal interactions. 
The horizontal and vertical blue dashed lines represent the cases where the offsets are zero.
}
\label{fig:spatial_offsets}
\end{center}
\end{figure}

\hspace{1em}
The main topic of this study is to compare the sizes of high-$z$ galaxies in the rest-frame UV and optical. 
When comparing galaxy sizes in different wavelengths, 
it is necessary to properly consider the wavelength dependence 
of the spatial resolution.\footnote{As mentioned in Section \ref{subsec:PSF_match}, 
some sources are resolved into multiple clumps in the rest-frame UV. 
To determine whether they appears clumpier in the rest-frame UV 
compared to the rest-frame optical, as seen in nearby galaxies
(e.g., \citealt{2000ApJS..131..441K}; \citealt{2001AJ....122..729K}; \citealt{2002ApJS..143..113W}), 
or high-$z$ galaxies have clumpy morphology in the rest-frame optical as well, 
higher spatial resolution imaging in the rest-frame optical is required.} 
As described in Section \ref{subsec:PSF_match}, 
to ensure a fair comparison of the SB profile fitting results in the rest-frame UV and optical, 
we use the results from the original (PSF-matched) F150W images for the single (multiple) flagged sources.

The top panels of Figure \ref{fig:re_ratio} plot the sizes in the rest-frame UV and optical of the $z\simeq4$--$10$ galaxies 
estimated from the SB profile fittings with the F150W and F444W images. 
The bottom panels of Figure \ref{fig:re_ratio} present the size ratio against the total magnitude 
in the rest-frame UV for these galaxies. 
In these plots, we divide the sample into three redshift ranges to examine any difference due to redshift. 
Based on these panels, we find that the size ratio is around unity, 
with some galaxies being larger than 1 and others smaller. 
In a similar manner to Figure \ref{fig:re_Muv_z} and Figure \ref{fig:re_Mopt_z}, 
the sample is divided in half based on size or total magnitude, 
and their median sizes or median size ratios with its 68th percentiles are also shown in Figure \ref{fig:re_ratio}. 
The median size ratios and their corresponding total magnitudes are summarized in Table \ref{tab:re_ratio}. 
In these redshift ranges, we find that the average size ratios align with unity regardless of $M_{\rm UV}$. 
For the highest redshift range, 
our results are in agreement with the previous results obtained for galaxies at $z \simeq 7$--$9$ 
(\citealt{2022ApJ...938L..17Y}; \citealt{2023ApJ...951...72O}). 
Our results demonstrate that the average size ratios of galaxies at lower redshifts of $z \simeq 4.5$--$6.5$ are also unity. 
The median size ratio of all the galaxies at $z\simeq4$--$10$ investigated in this study is 
$r_{\rm e.opt}/r_{\rm e,UV} = 1.01^{+0.35}_{-0.22}$.\footnote{If we do not consider the flag about blendedness, 
we obtain a median size ratio of $r_{\rm e.opt}/r_{\rm e,UV} = 1.18^{+0.76}_{-0.33}$. 
This value is still consistent with unity; 
however, because multiple components are only fitted in the rest-frame optical 
for about half of the sources, 
the size ratio on average becomes larger and its $68$th percentile also increases.} 
Very recently, \cite{2023arXiv230805018M} have reported that 
the sizes in the rest-frame UV and optical for $z=5$--$10$ galaxies are comparable, 
which is consistent with our results.

Utilizing the HST CANDELS data, 
\cite{2014ApJ...788...28V} have demonstrated that 
the sizes of SFGs at $z\sim0$--$2$ decrease with increasing wavelength.  
Interestingly, the difference in size becomes smaller at higher redshifts. 
Their results can be attributed to younger average ages of SFGs at higher redshifts,  
resulting in a convergence of the spatial distributions 
of young massive stars traced in the UV 
and somewhat older less massive stars traced in the optical 
(see also, \citealt{2015ApJS..219...15S}). 
The combination of our results with the previous JWST results 
indicate that there is no significant difference in the sizes of SFGs at $z\simeq4$--$10$ 
in the rest-frame UV and optical across a wide magnitude range, 
which is in line with the idea of younger average ages of SFGs at higher redshifts. 
Indeed, recent spectral energy distribution (SED) analyses for SFGs at comparable redshifts 
have reported remarkably young ages 
(e.g., \citealt{2023ApJ...949L..18P}; \citealt{2023ApJ...949L..25F}).

Furthermore, Figure \ref{fig:spatial_offsets} illustrates the spatial offsets 
between the positions of the rest-frame UV and optical emission of 
$45$ galaxies in our sample 
for which sizes are successfully measured with both the F150W and F444W images 
showing no signs of merger activity (red filled circles). 
The determination of the presence of signs 
of mergers or tidal interactions is based on visual inspection. 
Primarily, sources flagged with the multiple component flag as presented in Table \ref{tab:fitting_results} 
are considered to show signs of mergers or tidal interactions. 
Additionally, sources that are also resolved into multiple components 
in the smoothed F150W images and/or F444W images are classified 
as exhibiting signs of mergers or tidal interactions as well, 
such as C2{\_}20 and C09{\_}01.\footnote{Most of 
these multiple components are not spectroscopically confirmed yet, 
and therefore, further verification is needed to ascertain whether they are physically associated or not.} 
We find that, although $16$ sources exhibit spatial offsets exceeding $1\sigma$, 
none surpass the $3\sigma$ threshold. 
If the positional offset follows a Gaussian distribution centered at zero, 
the expected number of sources within and outside the $1\sigma$ range are 
$45 \times 0.683 = 30.735$ and $45 \times 0.317 = 14.265$, respectively, 
resulting in a chi-squared goodness of fit of $0.309$. 
Since the chi-squared value for a $95${\%} confidence level  
with 1 degree of freedom is $3.84$, 
our results do not contradict the assumption of a Gaussian distribution centered at zero, 
implying on average no significant spatial offset between the rest-frame UV and optical. 
This suggests that, for SFGs at $z=4$--$10$ with no clear signs of galaxy merger, 
the primary star-forming activity occurs near the mass centers of the galaxies.

Our results demonstrate that, for SFGs at $z=4$--$10$, 
the sizes and positions traced in the rest-frame UV and optical continuum 
are not significantly different. 
This contrasts with observations at lower redshifts $z\lesssim 2$, 
where galaxies exhibit smaller sizes at longer wavelengths, 
suggesting inside-out growth (e.g., \citealt{2014ApJ...788...28V}). 
Based on this concept, our results might imply that 
SFGs at $z=4$--$10$ are in the early stages of inside-out galaxy formation, 
where the distribution of young massive and less massive stars is not yet differentiated, 
indicative of the first phase of this process. 
More precisely, future studies might offer clear insights 
through spatially resolved SED analyses 
(e.g., \citealt{2023ApJ...948..126G}; \citealt{2023arXiv230300306B})
for galaxies over a wide range of magnitudes and redshifts.

In Figure \ref{fig:spatial_offsets}, we additionally present the results of the spatial offsets 
in the rest-frame UV and optical 
for sources whose sizes are measured with both the F150W and F444W images 
exhibiting possible signs of merger and/or tidal interactions 
(depicted as black open circles). 
Among the $67$ sources that we analyze, 
$48$ have an offset from the origin greater than $1\sigma$, 
yielding a chi-squared goodness of fit of $49.4$. 
This rejects the hypothesis that the data follows a Gaussian distribution 
centered at the origin at a confidence level of $95${\%}, 
possibly implying that 
SFGs at $z=4$--$10$ accompanied by multiple components 
on average show significant offsets between the positions in the rest-frame UV and optical. 
Although the observed offsets of these sources seem to be less conspicuous than 
those reported in the simulation results of \cite{2018MNRAS.477..219M} (see their Figure 2),
a quantitative comparison in the future might provide valuable insights 
into the physical processes involved in galaxy formation.

\section{Summary} \label{sec:summary}

\hspace{1em}
In this study, we have performed SB profile fittings 
for the $149$ galaxies at $z\simeq4$--$10$ with S/Ns exceeding $10$, 
$29$ of which have been spectroscopically confirmed, 
using the deep JWST NIRCam images obtained by the CEERS survey. 
The F150W and F444W images allow for 
probing their rest-frame UV and optical continuum emission. 
Although the S{\'e}rsic index is typically around $1.0$--$1.5$ in the rest-frame UV and optical, 
we have fixed it in the SB profile fittings in the same manner as previous work 
due to the large uncertainties for individual galaxies. 
Following our previous approach (\citealt{2023ApJ...951...72O}), 
we have performed MC simulations 
to correct for the systematic underestimates of 
galaxy sizes and total luminosities particularly for faint sources. 
Our primary results are as follows: 

\vspace{0.5em}

\begin{enumerate}

\item We have carefully examined 
the impact of strong emission lines on galaxy size measurements 
with galaxies at $z_{\rm spec} = 5.63$--$6.63$, 
utilizing images with the medium-band filter F410M and the broadband filter F444W, 
one of which incorporates strong emission lines and the other does not.  
We have concluded that 
the size differences due to the effect of strong emission lines are minor, 
enabling the use of the F444W images to measure galaxy sizes across a more extensive redshift range. 

\vspace{0.5em}

\item We have also taken into account 
the difference in spatial resolutions between the F150W and F444W images  
for a fair comparison between the rest-frame UV and optical sizes of galaxies, 
by smoothing the F150W images to match the PSF sizes of the F444W images. 
The SB profile fitting results from the original F150W images and the PSF-matched F150W images 
have demonstrated that, even when the spatial resolution differs by a factor of about 3, 
the size and total magnitude measurements are in good agreement. 

\vspace{0.5em}

\item Based on the comparisons between the rest-frame UV sizes and total magnitudes of galaxies 
observed by the HST and JWST, 
considering the difference of their spatial resolutions, 
these two measurements broadly align with each other with few outliers, 
confirming the availability of the previously obtained HST results 
(\citealt{2015ApJS..219...15S}) in comparison with JWST measurements. 

\vspace{0.5em}

\item We have investigated the relationship between the size and luminosity of 
galaxies at $z=4$--$10$ both in the rest-frame UV and optical, 
confirming the trend that fainter sources are smaller in size on average 
as reported in the previous studies 
(e.g., \citealt{2010ApJ...709L..21O}; \citealt{2012A&A...547A..51G}; \citealt{2013ApJ...777..155O}).  

\vspace{0.5em}

\item After careful consideration of the spatial resolution differences 
in the rest-frame UV and optical images, 
we have found that the optical-to-UV size ratios of galaxies at $z=4$--$10$ 
are on average around unity across a wide range of magnitudes,   
likely because of young average ages of SFGs at high redshifts. 

\vspace{0.5em}

\item We have found no significant spatial offsets 
between the rest-frame UV and optical for $z=4$--$10$ SFGs 
with no clear signs of merger, 
implying that the primary star-forming activity in these galaxies is 
likely occurring near their mass centers, 
which can be explained by the possibility that 
these galaxies are experiencing the initial stages of inside-out galaxy formation. 

\end{enumerate}

\section*{Acknowledgements}

\hspace{1em}
This work is based on observations
made with the NASA/ESA/CSA James Webb Space Telescope. 
The data were obtained from the Mikulski Archive for Space Telescopes 
at the Space Telescope Science Institute, 
which is operated by the Association of Universities for Research in Astronomy, Inc., 
under NASA contract NAS 5-03127 for JWST.
This work was partially performed using the computer facilities of
the Institute for Cosmic Ray Research, The University of Tokyo. 
This research made use of 
GALFIT (\citealt{2002AJ....124..266P}; \citealt{2010AJ....139.2097P}), 
SExtractor (\citealt{1996A&AS..117..393B}), 
IRAF (\citealt{1986SPIE..627..733T,1993ASPC...52..173T}),\footnote{IRAF is distributed by the National Optical Astronomy Observatory, 
which is operated by the Association of Universities for Research in Astronomy (AURA) 
under a cooperative agreement with the National Science Foundation.} 
SAOImage DS9 \citep{2003ASPC..295..489J},
Numpy \citep{2020Natur.585..357H}, 
Matplotlib \citep{2007CSE.....9...90H}, 
Scipy \citep{2020NatMe..17..261V}, 
Astropy \citep{2013A&A...558A..33A,2018AJ....156..123A},\footnote{\url{http://www.astropy.org}} 
and 
Ned Wright's Javascript Cosmology Calculator \citep{2006PASP..118.1711W}.\footnote{\url{http://www.astro.ucla.edu/~wright/CosmoCalc.html}}
This work was supported 
by the World Premier International
Research Center Initiative (WPI Initiative), MEXT, Japan, 
as well as 
KAKENHI Grant Numbers 
15K17602, 
15H02064, 
17H01110, 
17H01114, 
19K14752, 
20H00180, 
21H04467, 
21J20785, 
and 22K03670 
through the Japan Society for the Promotion of Science (JSPS). 
This work was partially supported by the joint research program of 
the Institute for Cosmic Ray Research (ICRR), University of Tokyo.

\bibliographystyle{apj}
\bibliography{ref}

\clearpage

\appendix

\setcounter{section}{0}
\renewcommand{\thesection}{A}
\setcounter{table}{0}
\renewcommand{\thetable}{A.\arabic{table}}
\setcounter{figure}{0}
\renewcommand{\thefigure}{A.\arabic{figure}}

\section{IDs of the Galaxies Analyzed in This Study} \label{sec:Sample}

\hspace{1em}
In Section \ref{sec:data}, 
we have compiled galaxies with redshifts ranging from $z\simeq 4$ to $10$ 
in the CEERS fields found in previous studies
(\citealt{2023ApJS..269...33N}; \citealt{2023ApJ...946L..13F}; 
\citealt{2015ApJS..219...15S}; \citealt{2015ApJ...803...34B}). 
Table \ref{tab:sample} summarizes their unique IDs assigned in this study, 
their original IDs in the previous catalogs, 
and their overlaps between the previous catalogs. 
Additionally, 
we include a flag in Table \ref{tab:sample} 
to indicate if these galaxies are also listed in catalogs from the following studies by different groups: 
\cite{2023MNRAS.518.6011D}; \cite{2023ApJ...949L..25F}; \cite{2023MNRAS.523.1009B}; 
\cite{2023MNRAS.526.1657T}; \cite{2023arXiv230406658H}. 
One should be aware that this additional flag is not comprehensive, 
as it does not cover all published literature to date. 
It should also be noted that these catalogs have a preference for 
focusing exclusively on higher-$z$ ($z \gtrsim 7$) sources 
compared to the redshift range covered in our study.

{\tiny
\begin{longtable}{*{12}{c}}
\caption{$z \simeq 4$--$10$ Galaxies in the CEERS Fields Reported in the Literature.
Note. Other references: 1: \cite{2023MNRAS.518.6011D}; 
2: \cite{2023ApJ...949L..25F}; 
3: \cite{2023MNRAS.523.1009B}; 
4: \cite{2023MNRAS.526.1657T}; 
5: \cite{2023arXiv230406658H}. 
}
\label{tab:sample}


\hline
ID	& R.A.	& Decl.	&  \multicolumn{2}{c}{\cite{2023ApJS..269...33N}}	&  \multicolumn{2}{c}{\cite{2023ApJ...946L..13F}}	&  \multicolumn{2}{c}{\cite{2015ApJ...803...34B}}	&  \multicolumn{2}{c}{\cite{2015ApJS..219...15S}} 
	& Other \\
	& (deg)	& (deg)	& ID	& $z_{\rm spec}$			& ID & $z_{\rm photo}$				& ID & $z_{\rm photo}$		& ID & $z_{\rm photo}$ 
	& References \\ 
\hline

\endhead
\endfoot
\endlastfoot
C01{\_}01 & $215.00522$ & $52.99652$ & P5P{\_}00003 & 8.01 & 3908 & 9.04 & --- & --- & --- & --- & 2,4 \\ 
C01{\_}02 & $215.01173$ & $52.98822$ & P5P{\_}00007 & 8.87 & 6059 & 9.01 & --- & --- & --- & --- & 2,5 \\  
C01{\_}03 & $215.00117$ & $53.01119$ & P5P{\_}00044 & 7.10 & --- & --- & --- & --- & --- & --- & 4 \\ 
C01{\_}04 & $215.01562$ & $53.01176$ & P5P{\_}00067 & 6.21 & --- & --- & --- & --- & --- & --- & --- \\ 
C01{\_}05 & $215.01090$ & $53.01324$ & P5P{\_}01912 & 5.10 & --- & --- & --- & --- & --- & --- & --- \\ 
C01{\_}06 & $214.99828$ & $52.99467$ & P5P{\_}01953 & 4.61 & --- & --- & --- & --- & --- & --- & --- \\ 
C01{\_}07 & $214.98878$ & $52.99797$ & P5P{\_}03584 & 4.64 & --- & --- & --- & --- & --- & --- & --- \\ 
C01{\_}08 & $214.96063$ & $52.94051$ & --- & --- & --- & --- & --- & --- & 26647 & 5 & --- \\ 
C01{\_}09 & $215.01605$ & $52.98248$ & --- & --- & --- & --- & --- & --- & 27229 & 5 & --- \\ 
C01{\_}10 & $215.02613$ & $52.99457$ & --- & --- & --- & --- & --- & --- & 28414 & 5 & --- \\ 
C01{\_}11 & $215.01527$ & $52.98672$ & --- & --- & --- & --- & --- & --- & 28427 & 5 & --- \\ 
C01{\_}12 & $215.01114$ & $52.98909$ & --- & --- & --- & --- & --- & --- & 29656 & 5 & --- \\ 
C01{\_}13 & $214.97204$ & $52.96254$ & --- & --- & --- & --- & --- & --- & 29855 & 5 & --- \\  
C01{\_}14 & $214.97481$ & $52.96650$ & --- & --- & --- & --- & --- & --- & 30425 & 5 & --- \\  
C01{\_}15 & $214.99405$ & $52.98085$ & --- & --- & --- & --- & --- & --- & 30550 & 5 & --- \\  
C01{\_}16 & $214.94419$ & $52.96760$ & --- & --- & --- & --- & --- & --- & 35779 & 5 & --- \\  
C01{\_}17 & $214.93204$ & $52.95899$ & --- & --- & --- & --- & --- & --- & 35809 & 5 & --- \\  
C01{\_}18 & $214.92475$ & $52.95584$ & --- & --- & --- & --- & --- & --- & 36262 & 5 & --- \\  
C01{\_}19 & $215.00974$ & $52.98144$ & --- & --- & --- & --- & --- & --- & 28026 & 6 & --- \\  
C01{\_}20 & $215.01415$ & $53.01115$ & --- & --- & --- & --- & --- & --- & 34318 & 6 & --- \\  
C01{\_}21 & $214.95194$ & $52.97174$ & --- & --- & 1875 & 8.92 & --- & --- & --- & --- & --- \\ 
C01{\_}22 & $214.99440$ & $52.98938$ & --- & --- & 3858 & 8.95 & --- & --- & --- & --- & 2,3,5 \\  
C01{\_}23 & $215.00537$ & $52.99670$ & --- & --- & 3910 & 9.55 & --- & --- & --- & --- & 2,3 \\  
C01{\_}24 & $214.95008$ & $52.94927$ & --- & --- & 5534 & 8.62 & --- & --- & --- & --- & --- \\  
C01{\_}25 & $214.96672$ & $52.96829$ & --- & --- & 4143 & 8.98 & --- & --- & --- & --- & --- \\  
C01{\_}26 & $214.99475$ & $53.00781$ & --- & --- & --- & --- & EGSY-9587400281 & 7.58 & --- & --- & --- \\ 
C02{\_}01 & $214.87255$ & $52.87595$ & P4M{\_}00323 & 5.67 & --- & --- & --- & --- & --- & --- & --- \\  
C02{\_}02 & $214.88802$ & $52.88826$ & P4M{\_}01465 & 5.27 & --- & --- & --- & --- & --- & --- & --- \\  
C02{\_}03 & $214.85964$ & $52.88814$ & P4P{\_}02000 & 4.81 & --- & --- & --- & --- & 31247 & 5 & --- \\ 
C02{\_}04 & $214.95381$ & $52.93373$ & --- & --- & --- & --- & --- & --- & 26115 & 5 & --- \\  
C02{\_}05 & $214.94847$ & $52.93859$ & --- & --- & --- & --- & --- & --- & 28205 & 5 & --- \\  
C02{\_}06 & $214.88124$ & $52.89609$ & --- & --- & --- & --- & --- & --- & 29440 & 5 & --- \\  
C02{\_}07 & $214.85954$ & $52.88801$ & --- & --- & --- & --- & --- & --- & 31167 & 5 & --- \\  
C02{\_}08 & $214.91837$ & $52.93187$ & --- & --- & --- & --- & --- & --- & 31703 & 5 & --- \\  
C02{\_}09 & $214.91887$ & $52.93921$ & --- & --- & --- & --- & --- & --- & 33230 & 5 & --- \\  
C02{\_}10 & $214.87215$ & $52.90633$ & --- & --- & --- & --- & --- & --- & 33438 & 5 & --- \\  
C02{\_}11 & $214.91523$ & $52.94503$ & --- & --- & --- & --- & --- & --- & 35205 & 5 & --- \\  
C02{\_}12 & $214.92457$ & $52.91873$ & --- & --- & --- & --- & EGSZ-9419055074 & 6.69 & 27521 & 6 & --- \\ 
C02{\_}13 & $214.93474$ & $52.94714$ & --- & --- & --- & --- & --- & --- & 32583 & 6 & --- \\ 
C02{\_}14 & $214.84477$ & $52.89210$ & --- & --- & 2402 & 8.71 & --- & --- & --- & --- & --- \\ 
C02{\_}15 & $214.87614$ & $52.88083$ & --- & --- & 7534 & 8.95 & --- & --- & --- & --- & --- \\ 
C02{\_}16 & $214.86160$ & $52.90460$ & --- & --- & 2324 & 9.58 & --- & --- & --- & --- & 1,5 \\ 
C02{\_}17 & $214.90224$ & $52.93937$ & --- & --- & 1298 & 8.53 & --- & --- & --- & --- & --- \\ 
C02{\_}18 & $214.84617$ & $52.89400$ & --- & --- & 2274 & 8.62 & --- & --- & --- & --- & --- \\ 
C02{\_}19 & $214.90763$ & $52.94461$ & --- & --- & 1075 & 8.38 & --- & --- & --- & --- & --- \\ 
C02{\_}20 & $214.89087$ & $52.89331$ & --- & --- & --- & --- & EGSZ-9338153359 & 6.92 & --- & --- & --- \\ 
C02{\_}21 & $214.88100$ & $52.89125$ & --- & --- & --- & --- & EGSZ-9314453285 & 6.92 & --- & --- & --- \\ 
C02{\_}22 & $214.89608$ & $52.92519$ & --- & --- & --- & --- & EGSZ-9350655307 & 6.77 & --- & --- & --- \\ 
C02{\_}23 & $214.86304$ & $52.88947$ & --- & --- & --- & --- & EGSZ-9271353221 & 6.77 & --- & --- & --- \\ 
C03{\_}01 & $214.80652$ & $52.87874$ & P4M{\_}00355 & 6.10 & --- & --- & --- & --- & --- & --- & --- \\ 
C03{\_}02 & $214.81273$ & $52.88145$ & P4P{\_}00362 & 6.05 & --- & --- & --- & --- & --- & --- & --- \\ 
C03{\_}03 & $214.81970$ & $52.87967$ & P4M{\_}00381 & 5.51 & --- & --- & --- & --- & --- & --- & --- \\ 
C03{\_}04 & $214.83622$ & $52.88258$ & P4M{\_}00397 & 6.00 & --- & --- & --- & --- & --- & --- & --- \\ 
C03{\_}05 & $214.82900$ & $52.87561$ & P4M{\_}00403 & 5.76 & --- & --- & --- & --- & --- & --- & --- \\ 
C03{\_}06 & $214.83935$ & $52.88247$ & P4M{\_}00407 & 7.03 & --- & --- & --- & --- & --- & --- & 4 \\ 
C03{\_}07 & $214.82538$ & $52.86297$ & P4P{\_}00439 & 7.18 & --- & --- & --- & --- & --- & --- & --- \\ 
C03{\_}08 & $214.81304$ & $52.83415$ & P4P{\_}00498 & 7.18 & --- & --- & --- & --- & --- & --- & 4 \\ 
C03{\_}09 & $214.81168$ & $52.83714$ & P4P{\_}02116 & 5.28 & --- & --- & --- & --- & --- & --- & --- \\ 
C03{\_}10 & $214.86453$ & $52.87087$ & P4P{\_}02362 & 5.32 & --- & --- & --- & --- & --- & --- & --- \\ 
C03{\_}11 & $214.79396$ & $52.82025$ & --- & --- & --- & --- & --- & --- & 26215 & 5 & --- \\ 
C03{\_}12 & $214.81201$ & $52.83682$ & --- & --- & --- & --- & --- & --- & 27005 & 5 & --- \\ 
C03{\_}13 & $214.80701$ & $52.83825$ & --- & --- & --- & --- & --- & --- & 28123 & 5 & --- \\ 
C03{\_}14 & $214.80686$ & $52.83829$ & --- & --- & --- & --- & --- & --- & 28217 & 5 & --- \\ 
C03{\_}15 & $214.82586$ & $52.85515$ & --- & --- & --- & --- & --- & --- & 29028 & 5 & --- \\ 
C03{\_}16 & $214.82518$ & $52.85907$ & --- & --- & --- & --- & --- & --- & 30133 & 5 & --- \\ 
C03{\_}17 & $214.79000$ & $52.83466$ & --- & --- & --- & --- & --- & --- & 30300 & 5 & --- \\ 
C03{\_}18 & $214.77238$ & $52.82342$ & --- & --- & --- & --- & --- & --- & 30572 & 5 & --- \\ 
C03{\_}19 & $214.79449$ & $52.83954$ & --- & --- & --- & --- & --- & --- & 30668 & 5 & --- \\ 
C03{\_}20 & $214.76766$ & $52.83329$ & --- & --- & --- & --- & --- & --- & 33751 & 5 & --- \\ 
C03{\_}21 & $214.76385$ & $52.83544$ & --- & --- & --- & --- & --- & --- & 34851 & 5 & --- \\ 
C03{\_}22 & $214.83347$ & $52.88902$ & --- & --- & --- & --- & --- & --- & 35846 & 5 & --- \\ 
C03{\_}23 & $214.74571$ & $52.83906$ & --- & --- & --- & --- & --- & --- & 37651 & 5 & --- \\ 
C03{\_}24 & $214.83556$ & $52.87748$ & --- & --- & --- & --- & --- & --- & 32502 & 6 & --- \\ 
C03{\_}25 & $214.76480$ & $52.82764$ & --- & --- & --- & --- & --- & --- & 32881 & 6 & --- \\ 
C03{\_}26 & $214.83068$ & $52.88777$ & --- & --- & 1748 & 8.77 & --- & --- & --- & --- & 2 \\ 
C03{\_}27 & $214.79396$ & $52.84158$ & --- & --- & --- & --- & EGSY-9105550297 & 8.11 & --- & --- & --- \\ 
C04{\_}01 & $214.74666$ & $52.74776$ & --- & --- & --- & --- & --- & --- & 17063 & 5 & --- \\ 
C04{\_}02 & $214.79050$ & $52.78157$ & --- & --- & --- & --- & --- & --- & 17722 & 5 & --- \\ 
C04{\_}03 & $214.80567$ & $52.79693$ & --- & --- & --- & --- & --- & --- & 18783 & 5 & --- \\ 
C04{\_}04 & $214.72571$ & $52.76288$ & --- & --- & --- & --- & --- & --- & 24140 & 5 & --- \\ 
C04{\_}05 & $214.78409$ & $52.80859$ & --- & --- & --- & --- & --- & --- & 25149 & 5 & --- \\ 
C04{\_}06 & $214.71657$ & $52.76452$ & --- & --- & --- & --- & --- & --- & 26086 & 5 & --- \\ 
C04{\_}07 & $214.75336$ & $52.74102$ & --- & --- & --- & --- & --- & --- & 14598 & 6 & --- \\ 
C04{\_}08 & $214.79315$ & $52.77049$ & --- & --- & --- & --- & --- & --- & 14850 & 6 & --- \\ 
C04{\_}09 & $214.78767$ & $52.77317$ & --- & --- & --- & --- & --- & --- & 16300 & 6 & --- \\ 
C04{\_}10 & $214.79033$ & $52.77294$ & --- & --- & --- & --- & EGSZ-9096846226 & 6.54 & --- & --- & --- \\ 
C05{\_}01 & $214.88608$ & $52.87689$ & P4P{\_}01534 & 4.59 & --- & --- & --- & --- & 24123 & 5 & --- \\ 
C05{\_}02 & $214.96740$ & $52.93296$ & P9M{\_}01025 & 8.71 & --- & --- & --- & --- & --- & --- & 4,5 \\ 
C05{\_}03 & $214.87854$ & $52.87414$ & P10M{\_}00515 & 5.66 & --- & --- & --- & --- & --- & --- & --- \\ 
C05{\_}04 & $214.92186$ & $52.88563$ & --- & --- & --- & --- & --- & --- & 20105 & 5 & --- \\ 
C05{\_}05 & $214.96585$ & $52.93406$ & --- & --- & --- & --- & --- & --- & 24219 & 7 & --- \\ 
C05{\_}06 & $214.98904$ & $52.91969$ & --- & --- & --- & --- & EGSZ-9573755109 & 6.69 & --- & --- & --- \\ 
C05{\_}07 & $214.94558$ & $52.90025$ & --- & --- & --- & --- & EGSZ-9469454009 & 6.92 & --- & --- & --- \\ 
C05{\_}08 & $214.94675$ & $52.90056$ & --- & --- & --- & --- & EGSZ-9472254020 & 7.24 & --- & --- & --- \\ 
C05{\_}09 & $214.97017$ & $52.91647$ & --- & --- & --- & --- & EGSZ-9528454593 & 6.46 & --- & --- & --- \\ 
C06{\_}01 & $214.86441$ & $52.85366$ & P4P{\_}00545 & 5.67 & --- & --- & --- & --- & --- & --- & --- \\ 
C06{\_}02 & $214.86725$ & $52.83674$ & P4M{\_}00603 & 6.06 & --- & --- & --- & --- & --- & --- & --- \\ 
C06{\_}03 & $214.87646$ & $52.83942$ & P4M{\_}00618 & 6.05 & --- & --- & --- & --- & --- & --- & --- \\ 
C06{\_}04 & $214.87177$ & $52.83317$ & P4M{\_}00792 & 6.26 & --- & --- & --- & --- & --- & --- & --- \\ 
C06{\_}05 & $214.88299$ & $52.84042$ & P4M{\_}01027 & 7.82 & --- & --- & --- & --- & --- & --- & 4 \\ 
C06{\_}06 & $214.82345$ & $52.83028$ & P4M{\_}02782 & 5.24 & --- & --- & --- & --- & --- & --- & --- \\ 
C06{\_}07 & $214.87854$ & $52.87414$ & P10M{\_}00515 & 5.66 & --- & --- & --- & --- & --- & --- & --- \\ 
C06{\_}08 & $214.90369$ & $52.84492$ & P10M{\_}00670 & 5.80 & --- & --- & --- & --- & --- & --- & --- \\ 
C06{\_}09 & $214.85224$ & $52.80917$ & --- & --- & --- & --- & --- & --- & 13963 & 5 & --- \\ 
C06{\_}10 & $214.89508$ & $52.84799$ & --- & --- & --- & --- & --- & --- & 15866 & 5 & --- \\ 
C06{\_}11 & $214.84701$ & $52.81407$ & --- & --- & --- & --- & --- & --- & 15930 & 5 & --- \\ 
C06{\_}12 & $214.82490$ & $52.83610$ & --- & --- & --- & --- & --- & --- & 24754 & 5 & --- \\ 
C06{\_}13 & $214.85514$ & $52.82081$ & --- & --- & --- & --- & --- & --- & 16208 & 6 & --- \\ 
C06{\_}14 & $214.88813$ & $52.85899$ & --- & --- & 4012 & 8.89 & --- & --- & --- & --- & 2 \\ 
C06{\_}15 & $214.80217$ & $52.80589$ & --- & --- & --- & --- & EGSZ-9125248212 & 6.54 & --- & --- & --- \\ 
C06{\_}16 & $214.80625$ & $52.81275$ & --- & --- & --- & --- & EGSZ-9135048459 & 6.77 & --- & --- & --- \\ 
C06{\_}17 & $214.85917$ & $52.85364$ & --- & --- & --- & --- & EGSZ-9262051131 & 6.39 & --- & --- & --- \\ 
C07{\_}01 & $215.09088$ & $52.95152$ & P8M{\_}01358 & 5.50 & --- & --- & --- & --- & --- & --- & --- \\ 
C07{\_}02 & $215.07996$ & $52.95677$ & P8M{\_}01449 & 4.75 & --- & --- & --- & --- & 10735 & 5 & --- \\ 
C07{\_}03 & $215.08602$ & $52.95219$ & P8P{\_}11383 & 5.07 & --- & --- & --- & --- & --- & --- & --- \\ 
C07{\_}04 & $215.13529$ & $52.99323$ & --- & --- & --- & --- & --- & --- & 10203 & 5 & --- \\ 
C07{\_}05 & $215.14644$ & $52.97027$ & --- & --- & --- & --- & --- & --- & 3639 & 5 & --- \\ 
C07{\_}06 & $215.13511$ & $52.96719$ & --- & --- & --- & --- & --- & --- & 4678 & 5 & --- \\ 
C07{\_}07 & $215.14402$ & $52.95935$ & --- & --- & --- & --- & --- & --- & 1744 & 5 & --- \\ 
C07{\_}08 & $215.14620$ & $52.95540$ & --- & --- & --- & --- & --- & --- & 872 & 5 & --- \\ 
C07{\_}09 & $215.09061$ & $52.95167$ & --- & --- & --- & --- & --- & --- & 7923 & 5 & --- \\ 
C07{\_}10 & $215.10084$ & $52.93848$ & --- & --- & --- & --- & --- & --- & 3761 & 5 & --- \\ 
C07{\_}11 & $215.08727$ & $52.94106$ & --- & --- & --- & --- & --- & --- & 6407 & 5 & --- \\ 
C07{\_}12 & $215.10358$ & $52.93191$ & --- & --- & --- & --- & --- & --- & 1984 & 5 & --- \\ 
C07{\_}13 & $215.07971$ & $52.93832$ & --- & --- & --- & --- & --- & --- & 6795 & 5 & --- \\ 
C07{\_}14 & $215.08314$ & $52.91996$ & --- & --- & --- & --- & --- & --- & 2462 & 5 & --- \\ 
C07{\_}15 & $215.12673$ & $52.98391$ & --- & --- & --- & --- & --- & --- & 9529 & 6 & --- \\ 
C07{\_}16 & $215.07467$ & $52.94472$ & --- & --- & --- & --- & EGSZ-0179256410 & 6.69 & --- & --- & --- \\ 
C08{\_}01 & $215.03540$ & $52.89067$ & P9M{\_}01019 & 8.68 & --- & --- & --- & --- & --- & --- & 4,5 \\ 
C08{\_}02 & $215.03972$ & $52.90160$ & P9M{\_}01038 & 7.19 & --- & --- & --- & --- & --- & --- & 4 \\ 
C08{\_}03 & $215.03055$ & $52.90259$ & P9M{\_}01324 & 5.01 & --- & --- & --- & --- & --- & --- & --- \\ 
C08{\_}04 & $215.03318$ & $52.89002$ & --- & --- & --- & --- & --- & --- & 3610 & 5 & --- \\ 
C08{\_}05 & $215.04419$ & $52.89879$ & --- & --- & --- & --- & --- & --- & 3733 & 5 & --- \\ 
C08{\_}06 & $215.00895$ & $52.87673$ & --- & --- & --- & --- & --- & --- & 4326 & 5 & --- \\ 
C08{\_}07 & $215.06373$ & $52.91751$ & --- & --- & --- & --- & --- & --- & 4736 & 5 & --- \\ 
C08{\_}08 & $214.97835$ & $52.85675$ & --- & --- & --- & --- & --- & --- & 4743 & 5 & --- \\ 
C08{\_}09 & $215.03308$ & $52.89651$ & --- & --- & --- & --- & --- & --- & 5021 & 5 & --- \\ 
C08{\_}10 & $214.98738$ & $52.86896$ & --- & --- & --- & --- & --- & --- & 6056 & 5 & --- \\ 
C08{\_}11 & $215.04264$ & $52.91202$ & --- & --- & --- & --- & --- & --- & 6813 & 5 & --- \\ 
C08{\_}12 & $215.01083$ & $52.90101$ & --- & --- & --- & --- & --- & --- & 9263 & 5 & --- \\ 
C08{\_}13 & $214.97813$ & $52.87957$ & --- & --- & --- & --- & --- & --- & 9635 & 5 & --- \\ 
C08{\_}14 & $215.00995$ & $52.91071$ & --- & --- & --- & --- & --- & --- & 11586 & 5 & --- \\ 
C08{\_}15 & $215.03564$ & $52.89226$ & --- & --- & --- & --- & EGSZ-0085553321 & 7.00 & 3320 & 6 & --- \\ 
C08{\_}16 & $215.04600$ & $52.89818$ & --- & --- & --- & --- & --- & --- & 3441 & 6 & --- \\ 
C08{\_}17 & $215.00943$ & $52.87537$ & --- & --- & --- & --- & --- & --- & 3989 & 6 & --- \\ 
C08{\_}18 & $215.00769$ & $52.87418$ & --- & --- & --- & --- & --- & --- & 4076 & 6 & --- \\ 
C08{\_}19 & $214.98723$ & $52.87465$ & --- & --- & --- & --- & --- & --- & 7282 & 6 & --- \\ 
C08{\_}20 & $214.98694$ & $52.87462$ & --- & --- & --- & --- & --- & --- & 7371 & 6 & --- \\ 
C08{\_}21 & $214.98797$ & $52.87950$ & --- & --- & --- & --- & EGSZ-9571152461 & 6.46 & 8177 & 6 & --- \\ 
C08{\_}22 & $215.03394$ & $52.91384$ & --- & --- & --- & --- & --- & --- & 8534 & 6 & --- \\ 
C08{\_}23 & $215.03210$ & $52.91902$ & --- & --- & --- & --- & --- & --- & 9935 & 6 & --- \\ 
C08{\_}24 & $214.95842$ & $52.87517$ & --- & --- & --- & --- & EGSZ-9500252305 & 6.54 & 11721 & 6 & --- \\ 
C08{\_}25 & $215.03721$ & $52.90675$ & --- & --- & --- & --- & EGSZ-0089354243 & 7.00 & --- & --- & --- \\ 
C08{\_}26 & $215.03704$ & $52.89194$ & --- & --- & --- & --- & EGSZ-0088953310 & 7.00 & --- & --- & --- \\ 
C08{\_}27 & $215.00600$ & $52.90536$ & --- & --- & --- & --- & EGSZ-0014454193 & 6.69 & --- & --- & --- \\ 
C08{\_}28 & $214.97996$ & $52.86114$ & --- & --- & --- & --- & EGSZ-9551951401 & 7.08 & --- & --- & --- \\ 
C08{\_}29 & $214.97192$ & $52.86922$ & --- & --- & --- & --- & EGSZ-9532652092 & 6.39 & --- & --- & --- \\ 
C08{\_}30 & $214.99875$ & $52.85542$ & --- & --- & --- & --- & EGSY-9597051195 & 8.38 & --- & --- & --- \\ 
C08{\_}31 & $214.98958$ & $52.86658$ & --- & --- & --- & --- & EGSY-9575051597 & 8.11 & --- & --- & --- \\ 
C09{\_}01 & $214.95990$ & $52.83118$ & P10M{\_}01207 & 4.90 & --- & --- & --- & --- & 2131 & 5 & --- \\ 
C09{\_}02 & $214.94755$ & $52.83709$ & P10M{\_}01289 & 4.88 & --- & --- & --- & --- & --- & --- & --- \\ 
C09{\_}03 & $214.95516$ & $52.84290$ & P10M{\_}01294 & 5.00 & --- & --- & --- & --- & 5342 & 5 & --- \\ 
C09{\_}04 & $214.94379$ & $52.85006$ & P10M{\_}01374 & 5.00 & --- & --- & --- & --- & 8518 & 5 & --- \\ 
C09{\_}05 & $214.92738$ & $52.81349$ & --- & --- & --- & --- & --- & --- & 3324 & 5 & --- \\ 
C09{\_}06 & $214.95796$ & $52.83598$ & --- & --- & --- & --- & --- & --- & 3426 & 5 & --- \\ 
C09{\_}07 & $214.93801$ & $52.83249$ & --- & --- & --- & --- & --- & --- & 5763 & 5 & --- \\ 
C09{\_}08 & $214.94369$ & $52.85014$ & --- & --- & --- & --- & --- & --- & 8598 & 5 & --- \\ 
C10{\_}01 & $214.83230$ & $52.74412$ & --- & --- & --- & --- & --- & --- & 2973 & 5 & --- \\ 
C10{\_}02 & $214.81900$ & $52.75974$ & --- & --- & --- & --- & --- & --- & 8275 & 5 & --- \\ 
C10{\_}03 & $214.80973$ & $52.75440$ & --- & --- & --- & --- & --- & --- & 8514 & 5 & --- \\ 
C10{\_}04 & $214.83919$ & $52.77684$ & --- & --- & --- & --- & --- & --- & 8846 & 5 & --- \\ 
C10{\_}05 & $214.84428$ & $52.78964$ & --- & --- & --- & --- & --- & --- & 10641 & 5 & --- \\ 
C10{\_}06 & $214.83044$ & $52.78370$ & --- & --- & --- & --- & --- & --- & 11720 & 5 & --- \\ 
C10{\_}07 & $214.83851$ & $52.79330$ & --- & --- & --- & --- & --- & --- & 12509 & 5 & --- \\ 
C10{\_}08 & $214.80272$ & $52.76821$ & --- & --- & --- & --- & --- & --- & 12664 & 5 & --- \\ 
C10{\_}09 & $214.85085$ & $52.77673$ & --- & --- & --- & --- & EGSZ-9242146357 & 6.84 & 7028 & 7 & --- \\ 
C10{\_}10 & $214.83100$ & $52.74914$ & --- & --- & --- & --- & EGSZ-9194444569 & 6.77 & --- & --- & --- \\ 
\hline
\end{longtable}
}

\setcounter{section}{0}
\renewcommand{\thesection}{B}
\setcounter{table}{0}
\renewcommand{\thetable}{B.\arabic{table}}
\setcounter{figure}{0}
\renewcommand{\thefigure}{B.\arabic{figure}}

\section{SB Fitting Results for Individual Galaxies} \label{sec:SB_Fitting_Results}

\hspace{1em}
In Section \ref{sec:results}, 
we have performed SB profile fittings for the high-$z$ galaxies in the CEERS fields. 
The sizes and total magnitudes obtained from the SB profile fittings for individual galaxies 
are summarized in Table \ref{tab:fitting_results}. 
In addition, the SB profile fitting result images for each object 
are presented in Figure \ref{fig:SB_fitting_results}.
Moreover, the pseudo-color images of those galaxies whose spectroscopic redshifts are 
compiled by \cite{2023ApJS..269...33N} are shown in Figure \ref{fig:PseudoColors}.

Upon scrutinizing the output images of the SB profile fittings, 
we allocate a blendedness designation for each subject as listed in Table \ref{tab:fitting_results}. 
This process involves identifying 
whether the fitted component in either the smoothed F150W and/or the F444W image 
corresponds to a single or multiple sources in the original F150W image. 
For instances where the fitted component in the F444W image is a single component in the original F150W image, 
we label it as `single' with a flag value of 1. 
In contrast, when the fitted component in the F444W image is multiple components in the original F150W image, 
the classification becomes `multiple', marked by a flag value of 2. 
Essentially, this means that a source is marked as having multiple components 
if the residual image after a single component fit indicates the existence of other components.
This approach ensures a fair comparison even with the data having different spatial resolutions.

\begin{landscape}
\setlength{\topmargin}{0mm}
\setlength{\textheight}{170mm}
\setlength{\tabcolsep}{3pt}
\begin{tiny}
\begin{longtable}{ccccccccccccc}
\caption{GALFIT Results. 
Note. (1) ID. 
(2) Total apparent UV magnitude in F150W. 
(3) Total absolute UV magnitude  in F150W. 
(4) UV size measured in F150W. 
(5) Total apparent UV magnitude measured in the smoothed F150W images.  
(6) Total absolute UV magnitude measured in the smoothed F150W images.
(7) UV size measured in the smoothed F150W images. 
(8) Total apparent optical magnitude in F444W. 
(9) Total absolute optical magnitude in F444W. 
(10) Optical size measured in F444W. 
(11) Optical-to-UV size ratio measured with the F150W and F444W images. 
(12) Optical-to-UV size ratio measured with the smoothed F150W and F444W images. 
(13) Flag about blendedness. 
1: the component fitted in the smoothed F150W image and/or the F444W image is a single component in the original F150W image. 
2: the component fitted in the smoothed F150W image and/or the F444W image is multiple components in the original F150W image. 
0: too faint or the SB profile fitting encounters numerical convergence issues.  
\label{tab:fitting_results}}
\hline
\hline
    ID
    &  \multicolumn{3}{c}{F150W}
    &  \multicolumn{3}{c}{F150W (PSF matched)}
    &  \multicolumn{3}{c}{F444W} 
    & $\dfrac{r_{\rm e,opt}}{r_{\rm e,UV}}$
    & $\dfrac{r_{\rm e,opt}}{r_{\rm e,UV}}$
    & flag \\
    
    &  $m_{\rm UV}$
    &  $M_{\rm UV}$
    &  $r_{\rm e}$ 
    &  $m_{\rm UV}$ 
    &  $M_{\rm UV}$ 
    &  $r_{\rm e}$ 
    &  $m_{\rm opt}$ 
    &  $M_{\rm opt}$ 
    &  $r_{\rm e,opt}$ 
    &  
    &  (PSF matched)
    &  
\\
    
    &  (mag)
    &  (mag)
    &  (kpc)
    &  (mag)
    &  (mag)
    &  (kpc)
    &  (mag)
    &  (mag)
    &  (kpc)
    &  
    &  
    &  
 \\
(1)     
    &  (2)
    &  (3)
    &  (4)
    &  (5)
    &  (6)
    &  (7)
    &  (8)
    &  (9)
    &  (10)
    &  (11)
    &  (12) 
    &  (13)
 \\
\hline
\endhead
C01{\_}02 & $26.57^{+0.04}_{-0.05}$ & $-20.73^{+0.04}_{-0.05}$ & $0.17^{+0.01}_{-0.01}$ & $26.62^{+0.05}_{-0.05}$ & $-20.68^{+0.05}_{-0.05}$ & $0.13^{+0.02}_{-0.02}$ & $26.77^{+0.08}_{-0.12}$ & $-20.53^{+0.08}_{-0.12}$ & $0.23^{+0.02}_{-0.02}$ & $1.36^{+0.15}_{-0.14}$ & $1.77^{+0.28}_{-0.28}$ & 1 \\ 
C01{\_}03 & $27.68^{+0.20}_{-0.22}$ & $-19.28^{+0.20}_{-0.22}$ & $0.03^{+0.03}_{-0.02}$ & --- & --- & --- & $26.96^{+0.08}_{-0.12}$ & $-20.00^{+0.08}_{-0.12}$ & $0.12^{+0.12}_{-0.09}$ & $4.44^{+7.25}_{-5.30}$ & --- & 1 \\ 
C01{\_}05 & $26.42^{+0.04}_{-0.05}$ & $-20.01^{+0.04}_{-0.05}$ & $0.36^{+0.02}_{-0.01}$ & $25.65^{+0.04}_{-0.05}$ & $-20.78^{+0.04}_{-0.05}$ & $0.80^{+0.02}_{-0.02}$ & $24.87^{+0.04}_{-0.04}$ & $-21.56^{+0.04}_{-0.04}$ & $0.77^{+0.02}_{-0.02}$ & $2.14^{+0.12}_{-0.10}$ & $0.97^{+0.08}_{-0.07}$ & 2 \\ 
C01{\_}06 & $26.60^{+0.04}_{-0.05}$ & $-19.67^{+0.04}_{-0.05}$ & $0.27^{+0.01}_{-0.01}$ & $26.25^{+0.09}_{-0.11}$ & $-20.01^{+0.09}_{-0.11}$ & $0.49^{+0.04}_{-0.04}$ & $26.45^{+0.04}_{-0.04}$ & $-19.81^{+0.04}_{-0.04}$ & $0.36^{+0.03}_{-0.03}$ & $1.34^{+0.13}_{-0.13}$ & $0.74^{+0.15}_{-0.16}$ & 1 \\ 
C01{\_}07 & $25.62^{+0.01}_{-0.02}$ & $-20.66^{+0.01}_{-0.02}$ & $0.26^{+0.00}_{-0.00}$ & $25.55^{+0.02}_{-0.02}$ & $-20.73^{+0.02}_{-0.02}$ & $0.30^{+0.02}_{-0.02}$ & $25.48^{+0.02}_{-0.03}$ & $-20.79^{+0.02}_{-0.03}$ & $0.29^{+0.02}_{-0.02}$ & $1.13^{+0.08}_{-0.07}$ & $0.97^{+0.10}_{-0.09}$ & 1 \\ 
C01{\_}08 & $25.78^{+0.05}_{-0.07}$ & $-20.62^{+0.05}_{-0.07}$ & $0.48^{+0.01}_{-0.01}$ & $25.86^{+0.03}_{-0.03}$ & $-20.54^{+0.03}_{-0.03}$ & $0.42^{+0.02}_{-0.01}$ & $25.46^{+0.04}_{-0.04}$ & $-20.93^{+0.04}_{-0.04}$ & $0.48^{+0.02}_{-0.02}$ & $1.00^{+0.04}_{-0.03}$ & $1.14^{+0.05}_{-0.05}$ & 1 \\ 
C01{\_}09 & $24.83^{+0.03}_{-0.04}$ & $-21.57^{+0.03}_{-0.04}$ & $0.60^{+0.01}_{-0.01}$ & $24.75^{+0.04}_{-0.05}$ & $-21.65^{+0.04}_{-0.05}$ & $0.72^{+0.02}_{-0.02}$ & $23.41^{+0.04}_{-0.04}$ & $-22.99^{+0.04}_{-0.04}$ & $0.75^{+0.02}_{-0.02}$ & $1.26^{+0.04}_{-0.04}$ & $1.04^{+0.05}_{-0.05}$ & 1 \\ 
C01{\_}10 & $25.94^{+0.02}_{-0.03}$ & $-20.45^{+0.02}_{-0.03}$ & $0.37^{+0.01}_{-0.01}$ & $25.67^{+0.04}_{-0.05}$ & $-20.73^{+0.04}_{-0.05}$ & $0.55^{+0.02}_{-0.01}$ & $25.38^{+0.04}_{-0.04}$ & $-21.02^{+0.04}_{-0.04}$ & $0.59^{+0.02}_{-0.02}$ & $1.58^{+0.07}_{-0.06}$ & $1.07^{+0.07}_{-0.07}$ & 2 \\ 
C01{\_}11 & $26.36^{+0.08}_{-0.11}$ & $-20.04^{+0.08}_{-0.11}$ & $0.82^{+0.03}_{-0.02}$ & --- & --- & --- & --- & --- & --- & --- & --- & 0 \\ 
C01{\_}12 & $26.66^{+0.04}_{-0.05}$ & $-19.74^{+0.04}_{-0.05}$ & $0.37^{+0.02}_{-0.01}$ & $26.44^{+0.05}_{-0.05}$ & $-19.96^{+0.05}_{-0.05}$ & $0.37^{+0.03}_{-0.04}$ & $26.05^{+0.05}_{-0.06}$ & $-20.35^{+0.05}_{-0.06}$ & $0.50^{+0.03}_{-0.04}$ & $1.35^{+0.11}_{-0.11}$ & $1.35^{+0.15}_{-0.17}$ & 2 \\ 
C01{\_}13 & $26.33^{+0.15}_{-0.18}$ & $-20.06^{+0.15}_{-0.18}$ & $1.02^{+0.05}_{-0.04}$ & $26.25^{+0.09}_{-0.11}$ & $-20.15^{+0.09}_{-0.11}$ & $0.81^{+0.05}_{-0.05}$ & $26.05^{+0.05}_{-0.06}$ & $-20.35^{+0.05}_{-0.06}$ & $0.70^{+0.09}_{-0.05}$ & $0.69^{+0.09}_{-0.05}$ & $0.87^{+0.10}_{-0.06}$ & 2 \\ 
C01{\_}14 & $26.35^{+0.04}_{-0.05}$ & $-20.05^{+0.04}_{-0.05}$ & $0.20^{+0.01}_{-0.01}$ & $26.03^{+0.03}_{-0.03}$ & $-20.37^{+0.03}_{-0.03}$ & $0.41^{+0.03}_{-0.04}$ & $26.22^{+0.03}_{-0.03}$ & $-20.18^{+0.03}_{-0.03}$ & $0.37^{+0.03}_{-0.04}$ & $1.89^{+0.20}_{-0.20}$ & $0.90^{+0.23}_{-0.26}$ & 2 \\ 
C01{\_}15 & $26.42^{+0.08}_{-0.11}$ & $-19.98^{+0.08}_{-0.11}$ & $0.66^{+0.03}_{-0.02}$ & --- & --- & --- & --- & --- & --- & --- & --- & 0 \\ 
C01{\_}16 & $25.49^{+0.03}_{-0.04}$ & $-20.91^{+0.03}_{-0.04}$ & $0.76^{+0.01}_{-0.01}$ & $25.51^{+0.04}_{-0.05}$ & $-20.89^{+0.04}_{-0.05}$ & $0.73^{+0.02}_{-0.02}$ & $24.56^{+0.05}_{-0.07}$ & $-21.84^{+0.05}_{-0.07}$ & $1.11^{+0.04}_{-0.03}$ & $1.45^{+0.05}_{-0.04}$ & $1.52^{+0.06}_{-0.06}$ & 1 \\ 
C01{\_}17 & $26.21^{+0.02}_{-0.03}$ & $-20.19^{+0.02}_{-0.03}$ & $0.17^{+0.01}_{-0.01}$ & $25.89^{+0.03}_{-0.03}$ & $-20.50^{+0.03}_{-0.03}$ & $0.39^{+0.02}_{-0.01}$ & $25.37^{+0.02}_{-0.03}$ & $-21.03^{+0.02}_{-0.03}$ & $0.22^{+0.02}_{-0.02}$ & $1.26^{+0.14}_{-0.11}$ & $0.56^{+0.12}_{-0.10}$ & 2 \\ 
C01{\_}18 & $26.09^{+0.05}_{-0.07}$ & $-20.30^{+0.05}_{-0.07}$ & $0.59^{+0.02}_{-0.01}$ & $26.18^{+0.06}_{-0.07}$ & $-20.22^{+0.06}_{-0.07}$ & $0.49^{+0.03}_{-0.04}$ & $25.97^{+0.05}_{-0.06}$ & $-20.43^{+0.05}_{-0.06}$ & $0.49^{+0.03}_{-0.04}$ & $0.84^{+0.06}_{-0.06}$ & $1.00^{+0.08}_{-0.09}$ & 2 \\ 
C01{\_}19 & $25.06^{+0.03}_{-0.04}$ & $-21.63^{+0.03}_{-0.04}$ & $0.78^{+0.02}_{-0.01}$ & $24.71^{+0.05}_{-0.07}$ & $-21.98^{+0.05}_{-0.07}$ & $1.20^{+0.04}_{-0.03}$ & $24.00^{+0.05}_{-0.07}$ & $-22.69^{+0.05}_{-0.07}$ & $1.05^{+0.03}_{-0.03}$ & $1.36^{+0.05}_{-0.04}$ & $0.88^{+0.07}_{-0.05}$ & 2 \\ 
C01{\_}20 & $25.27^{+0.03}_{-0.04}$ & $-21.43^{+0.03}_{-0.04}$ & $0.75^{+0.01}_{-0.01}$ & $25.25^{+0.04}_{-0.05}$ & $-21.45^{+0.04}_{-0.05}$ & $0.78^{+0.03}_{-0.02}$ & $24.32^{+0.04}_{-0.04}$ & $-22.37^{+0.04}_{-0.04}$ & $0.70^{+0.02}_{-0.02}$ & $0.93^{+0.03}_{-0.03}$ & $0.89^{+0.04}_{-0.04}$ & 1 \\ 
C01{\_}22 & $27.22^{+0.06}_{-0.07}$ & $-20.10^{+0.06}_{-0.07}$ & $0.10^{+0.03}_{-0.02}$ & $27.22^{+0.09}_{-0.08}$ & $-20.09^{+0.09}_{-0.08}$ & $0.11^{+0.07}_{-0.05}$ & --- & --- & --- & --- & --- & 1 \\ 
C01{\_}26 & $26.38^{+0.08}_{-0.11}$ & $-20.68^{+0.08}_{-0.11}$ & $0.42^{+0.01}_{-0.01}$ & $25.93^{+0.06}_{-0.07}$ & $-21.13^{+0.06}_{-0.07}$ & $0.65^{+0.02}_{-0.02}$ & $25.45^{+0.05}_{-0.07}$ & $-21.61^{+0.05}_{-0.07}$ & $0.79^{+0.02}_{-0.02}$ & $1.89^{+0.09}_{-0.07}$ & $1.21^{+0.07}_{-0.06}$ & 2 \\ 
C02{\_}01 & $27.82^{+0.20}_{-0.22}$ & $-18.79^{+0.20}_{-0.22}$ & $0.07^{+0.04}_{-0.03}$ & $27.85^{+0.21}_{-0.27}$ & $-18.76^{+0.21}_{-0.27}$ & $0.10^{+0.09}_{-0.07}$ & $27.20^{+0.16}_{-0.19}$ & $-19.40^{+0.16}_{-0.19}$ & $0.14^{+0.14}_{-0.10}$ & $1.91^{+2.13}_{-1.60}$ & $1.36^{+2.48}_{-1.92}$ & 1 \\ 
C02{\_}02 & $26.60^{+0.04}_{-0.05}$ & $-19.89^{+0.04}_{-0.05}$ & $0.43^{+0.02}_{-0.01}$ & $26.69^{+0.05}_{-0.05}$ & $-19.79^{+0.05}_{-0.05}$ & $0.39^{+0.03}_{-0.04}$ & $25.99^{+0.03}_{-0.03}$ & $-20.49^{+0.03}_{-0.03}$ & $0.33^{+0.03}_{-0.03}$ & $0.78^{+0.08}_{-0.07}$ & $0.86^{+0.10}_{-0.11}$ & 1 \\ 
C02{\_}03 & $26.67^{+0.08}_{-0.11}$ & $-19.67^{+0.08}_{-0.11}$ & $0.50^{+0.02}_{-0.01}$ & $25.89^{+0.06}_{-0.07}$ & $-20.44^{+0.06}_{-0.07}$ & $0.95^{+0.03}_{-0.03}$ & $25.81^{+0.05}_{-0.06}$ & $-20.53^{+0.05}_{-0.06}$ & $0.79^{+0.02}_{-0.02}$ & $1.59^{+0.07}_{-0.06}$ & $0.83^{+0.07}_{-0.06}$ & 2 \\ 
C02{\_}04 & $25.01^{+0.01}_{-0.02}$ & $-21.39^{+0.01}_{-0.02}$ & $0.35^{+0.00}_{-0.00}$ & $24.97^{+0.02}_{-0.02}$ & $-21.43^{+0.02}_{-0.02}$ & $0.41^{+0.02}_{-0.01}$ & $24.49^{+0.02}_{-0.03}$ & $-21.91^{+0.02}_{-0.03}$ & $0.42^{+0.02}_{-0.02}$ & $1.18^{+0.05}_{-0.05}$ & $1.01^{+0.07}_{-0.06}$ & 1 \\ 
C02{\_}05 & $26.43^{+0.08}_{-0.11}$ & $-19.97^{+0.08}_{-0.11}$ & $0.82^{+0.03}_{-0.02}$ & $26.60^{+0.09}_{-0.11}$ & $-19.79^{+0.09}_{-0.11}$ & $0.63^{+0.05}_{-0.05}$ & $25.61^{+0.04}_{-0.04}$ & $-20.78^{+0.04}_{-0.04}$ & $0.84^{+0.03}_{-0.02}$ & $1.02^{+0.05}_{-0.04}$ & $1.33^{+0.09}_{-0.09}$ & 1 \\ 
C02{\_}06 & $26.33^{+0.08}_{-0.11}$ & $-20.07^{+0.08}_{-0.11}$ & $0.70^{+0.03}_{-0.02}$ & $26.31^{+0.09}_{-0.11}$ & $-20.08^{+0.09}_{-0.11}$ & $0.74^{+0.05}_{-0.05}$ & $25.86^{+0.05}_{-0.06}$ & $-20.54^{+0.05}_{-0.06}$ & $0.63^{+0.02}_{-0.02}$ & $0.89^{+0.05}_{-0.04}$ & $0.85^{+0.07}_{-0.07}$ & 1 \\ 
C02{\_}07 & $25.71^{+0.05}_{-0.07}$ & $-20.69^{+0.05}_{-0.07}$ & $0.80^{+0.01}_{-0.01}$ & $25.71^{+0.04}_{-0.05}$ & $-20.69^{+0.04}_{-0.05}$ & $0.86^{+0.03}_{-0.03}$ & $25.54^{+0.04}_{-0.04}$ & $-20.85^{+0.04}_{-0.04}$ & $0.59^{+0.02}_{-0.02}$ & $0.74^{+0.03}_{-0.03}$ & $0.69^{+0.04}_{-0.03}$ & 1 \\ 
C02{\_}08 & $26.04^{+0.02}_{-0.03}$ & $-20.35^{+0.02}_{-0.03}$ & $0.33^{+0.01}_{-0.01}$ & $26.05^{+0.03}_{-0.03}$ & $-20.35^{+0.03}_{-0.03}$ & $0.36^{+0.03}_{-0.04}$ & $25.76^{+0.03}_{-0.03}$ & $-20.64^{+0.03}_{-0.03}$ & $0.32^{+0.02}_{-0.02}$ & $0.94^{+0.07}_{-0.05}$ & $0.88^{+0.11}_{-0.11}$ & 1 \\ 
C02{\_}09 & $25.28^{+0.06}_{-0.07}$ & $-21.12^{+0.06}_{-0.07}$ & $1.08^{+0.03}_{-0.02}$ & $25.26^{+0.05}_{-0.07}$ & $-21.14^{+0.05}_{-0.07}$ & $1.13^{+0.04}_{-0.03}$ & $24.31^{+0.05}_{-0.07}$ & $-22.09^{+0.05}_{-0.07}$ & $0.95^{+0.03}_{-0.02}$ & $0.88^{+0.04}_{-0.03}$ & $0.85^{+0.04}_{-0.03}$ & 1 \\ 
C02{\_}10 & $26.82^{+0.06}_{-0.07}$ & $-19.58^{+0.06}_{-0.07}$ & $0.28^{+0.01}_{-0.01}$ & $26.37^{+0.09}_{-0.11}$ & $-20.03^{+0.09}_{-0.11}$ & $0.53^{+0.03}_{-0.04}$ & $25.95^{+0.05}_{-0.06}$ & $-20.45^{+0.05}_{-0.06}$ & $0.51^{+0.03}_{-0.04}$ & $1.78^{+0.13}_{-0.14}$ & $0.96^{+0.16}_{-0.18}$ & 2 \\ 
C02{\_}12 & $26.37^{+0.08}_{-0.11}$ & $-20.33^{+0.08}_{-0.11}$ & $0.51^{+0.02}_{-0.01}$ & $26.38^{+0.09}_{-0.11}$ & $-20.31^{+0.09}_{-0.11}$ & $0.52^{+0.03}_{-0.04}$ & $25.25^{+0.04}_{-0.04}$ & $-21.44^{+0.04}_{-0.04}$ & $0.55^{+0.02}_{-0.02}$ & $1.08^{+0.05}_{-0.04}$ & $1.05^{+0.07}_{-0.08}$ & 1 \\ 
C02{\_}13 & $26.77^{+0.13}_{-0.18}$ & $-19.92^{+0.13}_{-0.18}$ & $0.51^{+0.02}_{-0.01}$ & $26.83^{+0.15}_{-0.17}$ & $-19.86^{+0.15}_{-0.17}$ & $0.48^{+0.03}_{-0.04}$ & --- & --- & --- & --- & --- & 1 \\ 
C02{\_}14 & $27.34^{+0.11}_{-0.14}$ & $-19.93^{+0.11}_{-0.14}$ & $0.25^{+0.03}_{-0.02}$ & $26.84^{+0.09}_{-0.08}$ & $-20.43^{+0.09}_{-0.08}$ & $0.34^{+0.02}_{-0.03}$ & $25.60^{+0.02}_{-0.03}$ & $-21.68^{+0.02}_{-0.03}$ & $0.13^{+0.01}_{-0.01}$ & $0.52^{+0.08}_{-0.06}$ & $0.39^{+0.07}_{-0.06}$ & 2 \\ 
C02{\_}15 & $27.57^{+0.11}_{-0.14}$ & $-19.75^{+0.11}_{-0.14}$ & $0.05^{+0.03}_{-0.02}$ & $27.64^{+0.14}_{-0.17}$ & $-19.68^{+0.14}_{-0.17}$ & $0.09^{+0.07}_{-0.05}$ & $25.67^{+0.02}_{-0.03}$ & $-21.64^{+0.02}_{-0.03}$ & $0.05^{+0.01}_{-0.01}$ & $1.13^{+0.76}_{-0.55}$ & $0.60^{+0.87}_{-0.69}$ & 1 \\ 
C02{\_}20 & $26.71^{+0.13}_{-0.18}$ & $-20.21^{+0.13}_{-0.18}$ & $0.44^{+0.02}_{-0.01}$ & $26.82^{+0.15}_{-0.17}$ & $-20.10^{+0.15}_{-0.17}$ & $0.40^{+0.03}_{-0.03}$ & $25.90^{+0.05}_{-0.06}$ & $-21.02^{+0.05}_{-0.06}$ & $0.55^{+0.02}_{-0.02}$ & $1.25^{+0.06}_{-0.05}$ & $1.36^{+0.10}_{-0.11}$ & 1 \\ 
C02{\_}21 & $27.16^{+0.06}_{-0.07}$ & $-19.76^{+0.06}_{-0.07}$ & $0.17^{+0.03}_{-0.02}$ & $26.48^{+0.09}_{-0.11}$ & $-20.44^{+0.09}_{-0.11}$ & $0.50^{+0.04}_{-0.04}$ & $25.95^{+0.05}_{-0.06}$ & $-20.97^{+0.05}_{-0.06}$ & $0.59^{+0.07}_{-0.04}$ & $3.56^{+0.86}_{-0.58}$ & $1.19^{+0.53}_{-0.39}$ & 2 \\ 
C02{\_}22 & $26.12^{+0.02}_{-0.03}$ & $-20.76^{+0.02}_{-0.03}$ & $0.21^{+0.01}_{-0.01}$ & $26.10^{+0.03}_{-0.03}$ & $-20.79^{+0.03}_{-0.03}$ & $0.25^{+0.03}_{-0.03}$ & $25.63^{+0.02}_{-0.03}$ & $-21.25^{+0.02}_{-0.03}$ & $0.23^{+0.02}_{-0.01}$ & $1.12^{+0.09}_{-0.08}$ & $0.92^{+0.15}_{-0.14}$ & 1 \\ 
C02{\_}23 & $26.68^{+0.04}_{-0.05}$ & $-20.21^{+0.04}_{-0.05}$ & $0.22^{+0.01}_{-0.01}$ & $26.54^{+0.05}_{-0.05}$ & $-20.35^{+0.05}_{-0.05}$ & $0.34^{+0.03}_{-0.03}$ & $26.24^{+0.04}_{-0.04}$ & $-20.65^{+0.04}_{-0.04}$ & $0.25^{+0.03}_{-0.03}$ & $1.10^{+0.13}_{-0.12}$ & $0.73^{+0.15}_{-0.16}$ & 1 \\ 
C03{\_}01 & $26.89^{+0.06}_{-0.07}$ & $-19.83^{+0.06}_{-0.07}$ & $0.41^{+0.02}_{-0.01}$ & $26.91^{+0.09}_{-0.08}$ & $-19.81^{+0.09}_{-0.08}$ & $0.41^{+0.03}_{-0.04}$ & $26.67^{+0.04}_{-0.04}$ & $-20.05^{+0.04}_{-0.04}$ & $0.28^{+0.03}_{-0.03}$ & $0.68^{+0.07}_{-0.07}$ & $0.68^{+0.08}_{-0.09}$ & 1 \\ 
C03{\_}04 & $25.62^{+0.01}_{-0.02}$ & $-21.08^{+0.01}_{-0.02}$ & $0.30^{+0.00}_{-0.00}$ & $25.60^{+0.02}_{-0.02}$ & $-21.09^{+0.02}_{-0.02}$ & $0.28^{+0.01}_{-0.01}$ & $25.03^{+0.02}_{-0.03}$ & $-21.66^{+0.02}_{-0.03}$ & $0.36^{+0.02}_{-0.01}$ & $1.22^{+0.05}_{-0.05}$ & $1.27^{+0.08}_{-0.08}$ & 1 \\ 
C03{\_}05 & $26.20^{+0.08}_{-0.11}$ & $-20.42^{+0.08}_{-0.11}$ & $0.46^{+0.02}_{-0.01}$ & $26.22^{+0.09}_{-0.11}$ & $-20.41^{+0.09}_{-0.11}$ & $0.45^{+0.03}_{-0.04}$ & $25.18^{+0.04}_{-0.04}$ & $-21.45^{+0.04}_{-0.04}$ & $0.56^{+0.02}_{-0.02}$ & $1.23^{+0.06}_{-0.05}$ & $1.25^{+0.10}_{-0.11}$ & 1 \\ 
C03{\_}07 & $27.76^{+0.20}_{-0.22}$ & $-19.21^{+0.20}_{-0.22}$ & $0.11^{+0.03}_{-0.02}$ & $27.93^{+0.21}_{-0.27}$ & $-19.04^{+0.21}_{-0.27}$ & $0.05^{+0.08}_{-0.06}$ & --- & --- & --- & --- & --- & 1 \\ 
C03{\_}08 & $26.71^{+0.04}_{-0.05}$ & $-20.27^{+0.04}_{-0.05}$ & $0.23^{+0.01}_{-0.01}$ & $26.69^{+0.05}_{-0.05}$ & $-20.29^{+0.05}_{-0.05}$ & $0.27^{+0.03}_{-0.03}$ & $26.23^{+0.04}_{-0.04}$ & $-20.75^{+0.04}_{-0.04}$ & $0.27^{+0.03}_{-0.03}$ & $1.21^{+0.12}_{-0.12}$ & $1.01^{+0.17}_{-0.16}$ & 1 \\ 
C03{\_}09 & $26.75^{+0.06}_{-0.07}$ & $-19.74^{+0.06}_{-0.07}$ & $0.30^{+0.01}_{-0.01}$ & $26.84^{+0.09}_{-0.08}$ & $-19.65^{+0.09}_{-0.08}$ & $0.25^{+0.03}_{-0.03}$ & $26.10^{+0.05}_{-0.06}$ & $-20.39^{+0.05}_{-0.06}$ & $0.48^{+0.03}_{-0.03}$ & $1.58^{+0.12}_{-0.12}$ & $1.89^{+0.23}_{-0.23}$ & 1 \\ 
C03{\_}11 & $25.57^{+0.06}_{-0.07}$ & $-20.83^{+0.06}_{-0.07}$ & $1.62^{+0.05}_{-0.04}$ & $25.65^{+0.05}_{-0.07}$ & $-20.75^{+0.05}_{-0.07}$ & $1.47^{+0.05}_{-0.04}$ & --- & --- & --- & --- & --- & 1 \\ 
C03{\_}12 & $25.23^{+0.03}_{-0.04}$ & $-21.16^{+0.03}_{-0.04}$ & $0.85^{+0.02}_{-0.01}$ & $24.93^{+0.05}_{-0.07}$ & $-21.46^{+0.05}_{-0.07}$ & $1.04^{+0.03}_{-0.03}$ & $23.76^{+0.04}_{-0.04}$ & $-22.64^{+0.04}_{-0.04}$ & $0.91^{+0.03}_{-0.02}$ & $1.07^{+0.04}_{-0.03}$ & $0.87^{+0.05}_{-0.04}$ & 2 \\ 
C03{\_}13 & $25.29^{+0.06}_{-0.07}$ & $-21.11^{+0.06}_{-0.07}$ & $1.42^{+0.04}_{-0.03}$ & $25.18^{+0.05}_{-0.07}$ & $-21.22^{+0.05}_{-0.07}$ & $1.44^{+0.05}_{-0.04}$ & $24.29^{+0.05}_{-0.07}$ & $-22.11^{+0.05}_{-0.07}$ & $1.66^{+0.06}_{-0.05}$ & $1.17^{+0.05}_{-0.04}$ & $1.16^{+0.06}_{-0.04}$ & 2 \\ 
C03{\_}14 & $25.83^{+0.09}_{-0.11}$ & $-20.57^{+0.09}_{-0.11}$ & $0.96^{+0.02}_{-0.01}$ & $25.62^{+0.05}_{-0.07}$ & $-20.77^{+0.05}_{-0.07}$ & $1.05^{+0.03}_{-0.03}$ & $24.99^{+0.05}_{-0.07}$ & $-21.41^{+0.05}_{-0.07}$ & $1.41^{+0.05}_{-0.04}$ & $1.48^{+0.06}_{-0.05}$ & $1.35^{+0.07}_{-0.06}$ & 2 \\ 
C03{\_}15 & $26.46^{+0.04}_{-0.05}$ & $-19.94^{+0.04}_{-0.05}$ & $0.26^{+0.01}_{-0.01}$ & $26.06^{+0.06}_{-0.07}$ & $-20.34^{+0.06}_{-0.07}$ & $0.52^{+0.03}_{-0.04}$ & $25.71^{+0.02}_{-0.03}$ & $-20.69^{+0.02}_{-0.03}$ & $0.45^{+0.02}_{-0.02}$ & $1.76^{+0.10}_{-0.09}$ & $0.87^{+0.13}_{-0.15}$ & 2 \\ 
C03{\_}16 & $26.57^{+0.04}_{-0.05}$ & $-19.83^{+0.04}_{-0.05}$ & $0.41^{+0.02}_{-0.01}$ & $26.33^{+0.09}_{-0.11}$ & $-20.07^{+0.09}_{-0.11}$ & $0.69^{+0.05}_{-0.05}$ & $26.32^{+0.04}_{-0.04}$ & $-20.08^{+0.04}_{-0.04}$ & $0.45^{+0.03}_{-0.04}$ & $1.09^{+0.09}_{-0.09}$ & $0.64^{+0.11}_{-0.12}$ & 1 \\ 
C03{\_}17 & $26.64^{+0.13}_{-0.18}$ & $-19.76^{+0.13}_{-0.18}$ & $0.48^{+0.02}_{-0.01}$ & $26.33^{+0.09}_{-0.11}$ & $-20.07^{+0.09}_{-0.11}$ & $0.54^{+0.03}_{-0.04}$ & $26.08^{+0.03}_{-0.03}$ & $-20.32^{+0.03}_{-0.03}$ & $0.39^{+0.03}_{-0.04}$ & $0.80^{+0.07}_{-0.08}$ & $0.72^{+0.08}_{-0.09}$ & 2 \\ 
C03{\_}18 & $26.12^{+0.02}_{-0.03}$ & $-20.28^{+0.02}_{-0.03}$ & $0.27^{+0.01}_{-0.01}$ & $25.97^{+0.03}_{-0.03}$ & $-20.43^{+0.03}_{-0.03}$ & $0.36^{+0.02}_{-0.01}$ & $26.14^{+0.03}_{-0.03}$ & $-20.26^{+0.03}_{-0.03}$ & $0.45^{+0.03}_{-0.04}$ & $1.66^{+0.14}_{-0.14}$ & $1.26^{+0.14}_{-0.15}$ & 2 \\ 
C03{\_}19 & $26.54^{+0.04}_{-0.05}$ & $-19.86^{+0.04}_{-0.05}$ & $0.27^{+0.01}_{-0.01}$ & $25.99^{+0.06}_{-0.07}$ & $-20.41^{+0.06}_{-0.07}$ & $0.66^{+0.05}_{-0.05}$ & $25.56^{+0.04}_{-0.04}$ & $-20.84^{+0.04}_{-0.04}$ & $0.92^{+0.03}_{-0.02}$ & $3.42^{+0.19}_{-0.14}$ & $1.40^{+0.28}_{-0.28}$ & 2 \\ 
C03{\_}20 & $26.70^{+0.13}_{-0.18}$ & $-19.70^{+0.13}_{-0.18}$ & $0.73^{+0.03}_{-0.02}$ & $27.04^{+0.15}_{-0.17}$ & $-19.36^{+0.15}_{-0.17}$ & $0.64^{+0.20}_{-0.12}$ & --- & --- & --- & --- & --- & 1 \\ 
C03{\_}21 & $26.62^{+0.04}_{-0.05}$ & $-19.78^{+0.04}_{-0.05}$ & $0.26^{+0.01}_{-0.01}$ & $26.33^{+0.05}_{-0.05}$ & $-20.07^{+0.05}_{-0.05}$ & $0.45^{+0.03}_{-0.04}$ & $24.89^{+0.05}_{-0.07}$ & $-21.51^{+0.05}_{-0.07}$ & $1.05^{+0.03}_{-0.02}$ & $3.98^{+0.21}_{-0.16}$ & $2.34^{+0.32}_{-0.36}$ & 2 \\ 
C03{\_}22 & $25.88^{+0.09}_{-0.11}$ & $-20.52^{+0.09}_{-0.11}$ & $1.28^{+0.03}_{-0.02}$ & $25.98^{+0.08}_{-0.10}$ & $-20.42^{+0.08}_{-0.10}$ & $1.19^{+0.11}_{-0.08}$ & --- & --- & --- & --- & --- & 1 \\ 
C03{\_}23 & $26.31^{+0.08}_{-0.11}$ & $-20.08^{+0.08}_{-0.11}$ & $0.79^{+0.03}_{-0.02}$ & --- & --- & --- & --- & --- & --- & --- & --- & 0 \\ 
C03{\_}24 & $24.71^{+0.01}_{-0.02}$ & $-21.98^{+0.01}_{-0.02}$ & $0.37^{+0.01}_{-0.00}$ & $24.69^{+0.02}_{-0.02}$ & $-22.00^{+0.02}_{-0.02}$ & $0.37^{+0.01}_{-0.01}$ & $24.37^{+0.04}_{-0.04}$ & $-22.32^{+0.04}_{-0.04}$ & $0.51^{+0.02}_{-0.01}$ & $1.39^{+0.05}_{-0.04}$ & $1.38^{+0.07}_{-0.06}$ & 2 \\ 
C03{\_}25 & $26.13^{+0.05}_{-0.07}$ & $-20.57^{+0.05}_{-0.07}$ & $0.81^{+0.04}_{-0.03}$ & $26.23^{+0.09}_{-0.11}$ & $-20.46^{+0.09}_{-0.11}$ & $0.76^{+0.07}_{-0.06}$ & $26.18^{+0.09}_{-0.12}$ & $-20.52^{+0.09}_{-0.12}$ & $0.72^{+0.08}_{-0.04}$ & $0.89^{+0.11}_{-0.07}$ & $0.95^{+0.13}_{-0.09}$ & 1 \\ 
C03{\_}27 & $26.25^{+0.04}_{-0.05}$ & $-20.91^{+0.04}_{-0.05}$ & $0.19^{+0.01}_{-0.01}$ & $26.24^{+0.03}_{-0.03}$ & $-20.92^{+0.03}_{-0.03}$ & $0.21^{+0.03}_{-0.03}$ & $25.36^{+0.02}_{-0.03}$ & $-21.80^{+0.02}_{-0.03}$ & $0.34^{+0.01}_{-0.01}$ & $1.80^{+0.11}_{-0.09}$ & $1.59^{+0.23}_{-0.22}$ & 1 \\ 
C04{\_}02 & $25.38^{+0.01}_{-0.02}$ & $-21.02^{+0.01}_{-0.02}$ & $0.19^{+0.00}_{-0.00}$ & $25.24^{+0.02}_{-0.02}$ & $-21.16^{+0.02}_{-0.02}$ & $0.31^{+0.01}_{-0.02}$ & $24.32^{+0.02}_{-0.03}$ & $-22.08^{+0.02}_{-0.03}$ & $0.46^{+0.02}_{-0.02}$ & $2.42^{+0.11}_{-0.10}$ & $1.46^{+0.15}_{-0.14}$ & 1 \\ 
C04{\_}03 & $26.15^{+0.05}_{-0.07}$ & $-20.25^{+0.05}_{-0.07}$ & $0.63^{+0.03}_{-0.02}$ & $25.95^{+0.06}_{-0.07}$ & $-20.45^{+0.06}_{-0.07}$ & $0.71^{+0.02}_{-0.02}$ & $25.30^{+0.04}_{-0.04}$ & $-21.10^{+0.04}_{-0.04}$ & $0.49^{+0.02}_{-0.02}$ & $0.78^{+0.05}_{-0.04}$ & $0.69^{+0.04}_{-0.03}$ & 2 \\ 
C04{\_}04 & $25.66^{+0.01}_{-0.02}$ & $-20.74^{+0.01}_{-0.02}$ & $0.24^{+0.00}_{-0.00}$ & $25.60^{+0.02}_{-0.02}$ & $-20.80^{+0.02}_{-0.02}$ & $0.29^{+0.01}_{-0.02}$ & $24.85^{+0.02}_{-0.03}$ & $-21.55^{+0.02}_{-0.03}$ & $0.30^{+0.02}_{-0.02}$ & $1.25^{+0.08}_{-0.07}$ & $1.02^{+0.10}_{-0.09}$ & 2 \\ 
C04{\_}05 & $25.68^{+0.03}_{-0.04}$ & $-20.72^{+0.03}_{-0.04}$ & $0.63^{+0.01}_{-0.01}$ & $25.64^{+0.04}_{-0.05}$ & $-20.76^{+0.04}_{-0.05}$ & $0.73^{+0.02}_{-0.02}$ & $25.07^{+0.04}_{-0.04}$ & $-21.33^{+0.04}_{-0.04}$ & $0.68^{+0.02}_{-0.02}$ & $1.08^{+0.04}_{-0.03}$ & $0.93^{+0.04}_{-0.04}$ & 1 \\ 
C04{\_}06 & $26.27^{+0.04}_{-0.05}$ & $-20.13^{+0.04}_{-0.05}$ & $0.46^{+0.02}_{-0.01}$ & --- & --- & --- & $24.85^{+0.05}_{-0.07}$ & $-21.55^{+0.05}_{-0.07}$ & $1.35^{+0.05}_{-0.04}$ & $2.95^{+0.16}_{-0.12}$ & --- & 2 \\ 
C04{\_}07 & $26.47^{+0.04}_{-0.05}$ & $-20.22^{+0.04}_{-0.05}$ & $0.30^{+0.01}_{-0.01}$ & $26.47^{+0.05}_{-0.05}$ & $-20.23^{+0.05}_{-0.05}$ & $0.32^{+0.03}_{-0.04}$ & $25.86^{+0.03}_{-0.03}$ & $-20.83^{+0.03}_{-0.03}$ & $0.35^{+0.02}_{-0.01}$ & $1.16^{+0.07}_{-0.06}$ & $1.08^{+0.12}_{-0.14}$ & 1 \\ 
C04{\_}08 & $26.48^{+0.08}_{-0.11}$ & $-20.22^{+0.08}_{-0.11}$ & $0.74^{+0.03}_{-0.02}$ & $26.51^{+0.09}_{-0.11}$ & $-20.18^{+0.09}_{-0.11}$ & $0.70^{+0.04}_{-0.05}$ & $26.24^{+0.09}_{-0.12}$ & $-20.45^{+0.09}_{-0.12}$ & $0.67^{+0.08}_{-0.04}$ & $0.90^{+0.11}_{-0.06}$ & $0.96^{+0.12}_{-0.08}$ & 1 \\ 
C04{\_}09 & $26.23^{+0.08}_{-0.11}$ & $-20.46^{+0.08}_{-0.11}$ & $0.85^{+0.04}_{-0.03}$ & $26.32^{+0.09}_{-0.11}$ & $-20.38^{+0.09}_{-0.11}$ & $0.71^{+0.04}_{-0.05}$ & $25.36^{+0.05}_{-0.07}$ & $-21.33^{+0.05}_{-0.07}$ & $0.91^{+0.03}_{-0.02}$ & $1.08^{+0.07}_{-0.05}$ & $1.29^{+0.08}_{-0.08}$ & 1 \\ 
C04{\_}10 & $26.42^{+0.08}_{-0.11}$ & $-20.41^{+0.08}_{-0.11}$ & $0.76^{+0.04}_{-0.03}$ & $26.48^{+0.09}_{-0.11}$ & $-20.35^{+0.09}_{-0.11}$ & $0.69^{+0.04}_{-0.04}$ & $26.19^{+0.09}_{-0.12}$ & $-20.64^{+0.09}_{-0.12}$ & $0.52^{+0.08}_{-0.04}$ & $0.68^{+0.11}_{-0.06}$ & $0.75^{+0.11}_{-0.07}$ & 1 \\ 
C05{\_}01 & $26.94^{+0.06}_{-0.07}$ & $-19.32^{+0.06}_{-0.07}$ & $0.27^{+0.01}_{-0.01}$ & --- & --- & --- & $25.50^{+0.04}_{-0.04}$ & $-20.75^{+0.04}_{-0.04}$ & $0.58^{+0.02}_{-0.02}$ & $2.17^{+0.12}_{-0.10}$ & --- & 2 \\ 
C05{\_}04 & $25.74^{+0.09}_{-0.11}$ & $-20.66^{+0.09}_{-0.11}$ & $0.97^{+0.02}_{-0.01}$ & $25.35^{+0.05}_{-0.07}$ & $-21.05^{+0.05}_{-0.07}$ & $1.00^{+0.03}_{-0.03}$ & $24.80^{+0.04}_{-0.04}$ & $-21.60^{+0.04}_{-0.04}$ & $0.87^{+0.03}_{-0.02}$ & $0.89^{+0.04}_{-0.03}$ & $0.87^{+0.04}_{-0.03}$ & 2 \\ 
C05{\_}05 & $26.49^{+0.04}_{-0.05}$ & $-20.45^{+0.04}_{-0.05}$ & $0.18^{+0.01}_{-0.01}$ & $26.09^{+0.03}_{-0.03}$ & $-20.85^{+0.03}_{-0.03}$ & $0.31^{+0.03}_{-0.03}$ & $25.68^{+0.02}_{-0.03}$ & $-21.26^{+0.02}_{-0.03}$ & $0.19^{+0.02}_{-0.01}$ & $1.05^{+0.11}_{-0.08}$ & $0.61^{+0.13}_{-0.13}$ & 2 \\ 
C05{\_}06 & $26.49^{+0.04}_{-0.05}$ & $-20.37^{+0.04}_{-0.05}$ & $0.35^{+0.02}_{-0.01}$ & $26.45^{+0.05}_{-0.05}$ & $-20.41^{+0.05}_{-0.05}$ & $0.35^{+0.03}_{-0.03}$ & $26.65^{+0.04}_{-0.04}$ & $-20.22^{+0.04}_{-0.04}$ & $0.33^{+0.03}_{-0.03}$ & $0.94^{+0.09}_{-0.09}$ & $0.94^{+0.11}_{-0.13}$ & 1 \\ 
C05{\_}07 & $25.96^{+0.02}_{-0.03}$ & $-20.95^{+0.02}_{-0.03}$ & $0.35^{+0.01}_{-0.00}$ & $26.03^{+0.03}_{-0.03}$ & $-20.89^{+0.03}_{-0.03}$ & $0.30^{+0.03}_{-0.03}$ & $25.82^{+0.03}_{-0.03}$ & $-21.10^{+0.03}_{-0.03}$ & $0.29^{+0.01}_{-0.01}$ & $0.83^{+0.04}_{-0.04}$ & $0.96^{+0.09}_{-0.10}$ & 1 \\ 
C05{\_}08 & $26.28^{+0.08}_{-0.11}$ & $-20.71^{+0.08}_{-0.11}$ & $0.59^{+0.02}_{-0.02}$ & $26.10^{+0.06}_{-0.07}$ & $-20.89^{+0.06}_{-0.07}$ & $0.73^{+0.06}_{-0.05}$ & $25.99^{+0.05}_{-0.06}$ & $-21.00^{+0.05}_{-0.06}$ & $0.43^{+0.03}_{-0.03}$ & $0.73^{+0.05}_{-0.05}$ & $0.59^{+0.07}_{-0.07}$ & 2 \\ 
C05{\_}09 & $26.62^{+0.04}_{-0.05}$ & $-20.19^{+0.04}_{-0.05}$ & $0.21^{+0.01}_{-0.01}$ & $26.58^{+0.05}_{-0.05}$ & $-20.23^{+0.05}_{-0.05}$ & $0.26^{+0.03}_{-0.03}$ & --- & --- & --- & --- & --- & 1 \\ 
C06{\_}01 & $26.15^{+0.02}_{-0.03}$ & $-20.45^{+0.02}_{-0.03}$ & $0.38^{+0.02}_{-0.01}$ & $25.47^{+0.04}_{-0.05}$ & $-21.13^{+0.04}_{-0.05}$ & $0.81^{+0.03}_{-0.02}$ & $24.75^{+0.04}_{-0.04}$ & $-21.85^{+0.04}_{-0.04}$ & $0.80^{+0.03}_{-0.02}$ & $2.09^{+0.12}_{-0.09}$ & $0.98^{+0.10}_{-0.08}$ & 2 \\ 
C06{\_}02 & $26.61^{+0.08}_{-0.11}$ & $-20.10^{+0.08}_{-0.11}$ & $0.55^{+0.03}_{-0.02}$ & $26.77^{+0.09}_{-0.08}$ & $-19.94^{+0.09}_{-0.08}$ & $0.37^{+0.03}_{-0.04}$ & --- & --- & --- & --- & --- & 2 \\ 
C06{\_}03 & $26.90^{+0.13}_{-0.18}$ & $-19.81^{+0.13}_{-0.18}$ & $0.49^{+0.07}_{-0.04}$ & $26.95^{+0.09}_{-0.08}$ & $-19.76^{+0.09}_{-0.08}$ & $0.40^{+0.03}_{-0.04}$ & --- & --- & --- & --- & --- & 1 \\ 
C06{\_}05 & $26.61^{+0.04}_{-0.05}$ & $-20.50^{+0.04}_{-0.05}$ & $0.07^{+0.01}_{-0.01}$ & $26.42^{+0.05}_{-0.05}$ & $-20.69^{+0.05}_{-0.05}$ & $0.18^{+0.03}_{-0.03}$ & $25.61^{+0.02}_{-0.03}$ & $-21.50^{+0.02}_{-0.03}$ & $0.24^{+0.02}_{-0.01}$ & $3.21^{+0.44}_{-0.34}$ & $1.33^{+0.51}_{-0.49}$ & 2 \\ 
C06{\_}06 & $26.63^{+0.04}_{-0.05}$ & $-19.84^{+0.04}_{-0.05}$ & $0.09^{+0.01}_{-0.01}$ & $26.31^{+0.05}_{-0.05}$ & $-20.17^{+0.05}_{-0.05}$ & $0.25^{+0.03}_{-0.03}$ & --- & --- & --- & --- & --- & 2 \\ 
C06{\_}09 & $26.21^{+0.08}_{-0.11}$ & $-20.19^{+0.08}_{-0.11}$ & $0.59^{+0.02}_{-0.01}$ & $26.29^{+0.09}_{-0.11}$ & $-20.11^{+0.09}_{-0.11}$ & $0.54^{+0.03}_{-0.04}$ & $26.51^{+0.09}_{-0.12}$ & $-19.89^{+0.09}_{-0.12}$ & $0.55^{+0.03}_{-0.04}$ & $0.92^{+0.06}_{-0.06}$ & $1.02^{+0.08}_{-0.09}$ & 1 \\ 
C06{\_}10 & $25.46^{+0.06}_{-0.07}$ & $-20.94^{+0.06}_{-0.07}$ & $0.99^{+0.02}_{-0.01}$ & $25.53^{+0.05}_{-0.07}$ & $-20.87^{+0.05}_{-0.07}$ & $0.96^{+0.03}_{-0.03}$ & $25.84^{+0.05}_{-0.06}$ & $-20.56^{+0.05}_{-0.06}$ & $0.92^{+0.03}_{-0.02}$ & $0.93^{+0.04}_{-0.03}$ & $0.96^{+0.04}_{-0.03}$ & 1 \\ 
C06{\_}11 & $25.34^{+0.03}_{-0.04}$ & $-21.05^{+0.03}_{-0.04}$ & $0.79^{+0.01}_{-0.01}$ & $25.34^{+0.04}_{-0.05}$ & $-21.06^{+0.04}_{-0.05}$ & $0.81^{+0.02}_{-0.02}$ & $25.69^{+0.05}_{-0.06}$ & $-20.71^{+0.05}_{-0.06}$ & $0.73^{+0.02}_{-0.02}$ & $0.93^{+0.03}_{-0.03}$ & $0.90^{+0.04}_{-0.03}$ & 1 \\ 
C06{\_}12 & $26.41^{+0.08}_{-0.11}$ & $-19.98^{+0.08}_{-0.11}$ & $0.61^{+0.03}_{-0.02}$ & $26.00^{+0.06}_{-0.07}$ & $-20.40^{+0.06}_{-0.07}$ & $0.93^{+0.07}_{-0.07}$ & $25.52^{+0.05}_{-0.07}$ & $-20.88^{+0.05}_{-0.07}$ & $1.01^{+0.03}_{-0.02}$ & $1.66^{+0.10}_{-0.07}$ & $1.09^{+0.14}_{-0.12}$ & 2 \\ 
C06{\_}13 & $26.10^{+0.05}_{-0.07}$ & $-20.59^{+0.05}_{-0.07}$ & $0.80^{+0.04}_{-0.03}$ & $25.96^{+0.06}_{-0.07}$ & $-20.73^{+0.06}_{-0.07}$ & $0.83^{+0.03}_{-0.02}$ & $25.68^{+0.07}_{-0.10}$ & $-21.01^{+0.07}_{-0.10}$ & $0.88^{+0.03}_{-0.02}$ & $1.10^{+0.07}_{-0.05}$ & $1.06^{+0.05}_{-0.04}$ & 2 \\ 
C06{\_}15 & $27.51^{+0.11}_{-0.14}$ & $-19.32^{+0.11}_{-0.14}$ & $0.14^{+0.04}_{-0.03}$ & $27.18^{+0.09}_{-0.08}$ & $-19.65^{+0.09}_{-0.08}$ & $0.30^{+0.08}_{-0.06}$ & --- & --- & --- & --- & --- & 2 \\ 
C06{\_}16 & $26.50^{+0.04}_{-0.05}$ & $-20.39^{+0.04}_{-0.05}$ & $0.14^{+0.01}_{-0.01}$ & $26.59^{+0.05}_{-0.05}$ & $-20.30^{+0.05}_{-0.05}$ & $0.10^{+0.03}_{-0.03}$ & $26.64^{+0.04}_{-0.04}$ & $-20.25^{+0.04}_{-0.04}$ & $0.18^{+0.03}_{-0.03}$ & $1.25^{+0.20}_{-0.19}$ & $1.80^{+0.41}_{-0.40}$ & 1 \\ 
C07{\_}01 & $26.90^{+0.06}_{-0.07}$ & $-19.66^{+0.06}_{-0.07}$ & $0.19^{+0.01}_{-0.01}$ & $26.86^{+0.09}_{-0.08}$ & $-19.70^{+0.09}_{-0.08}$ & $0.23^{+0.03}_{-0.03}$ & --- & --- & --- & --- & --- & 1 \\ 
C07{\_}02 & $25.01^{+0.01}_{-0.02}$ & $-21.31^{+0.01}_{-0.02}$ & $0.32^{+0.00}_{-0.00}$ & $25.02^{+0.02}_{-0.02}$ & $-21.30^{+0.02}_{-0.02}$ & $0.35^{+0.02}_{-0.02}$ & --- & --- & --- & --- & --- & 2 \\ 
C07{\_}04 & $25.61^{+0.06}_{-0.07}$ & $-20.79^{+0.06}_{-0.07}$ & $1.66^{+0.05}_{-0.04}$ & $25.75^{+0.08}_{-0.10}$ & $-20.65^{+0.08}_{-0.10}$ & $1.45^{+0.05}_{-0.04}$ & --- & --- & --- & --- & --- & 1 \\ 
C07{\_}06 & $26.70^{+0.04}_{-0.05}$ & $-19.70^{+0.04}_{-0.05}$ & $0.24^{+0.01}_{-0.01}$ & $25.70^{+0.04}_{-0.05}$ & $-20.69^{+0.04}_{-0.05}$ & $0.79^{+0.02}_{-0.02}$ & --- & --- & --- & --- & --- & 2 \\ 
C07{\_}07 & $25.70^{+0.09}_{-0.11}$ & $-20.70^{+0.09}_{-0.11}$ & $1.49^{+0.04}_{-0.03}$ & $25.76^{+0.08}_{-0.10}$ & $-20.64^{+0.08}_{-0.10}$ & $1.47^{+0.05}_{-0.04}$ & --- & --- & --- & --- & --- & 1 \\ 
C07{\_}08 & $26.68^{+0.08}_{-0.11}$ & $-19.72^{+0.08}_{-0.11}$ & $0.54^{+0.02}_{-0.01}$ & $26.86^{+0.09}_{-0.08}$ & $-19.54^{+0.09}_{-0.08}$ & $0.44^{+0.03}_{-0.04}$ & --- & --- & --- & --- & --- & 1 \\ 
C07{\_}09 & $25.34^{+0.01}_{-0.02}$ & $-21.06^{+0.01}_{-0.02}$ & $0.40^{+0.01}_{-0.01}$ & $25.39^{+0.02}_{-0.02}$ & $-21.01^{+0.02}_{-0.02}$ & $0.36^{+0.02}_{-0.01}$ & --- & --- & --- & --- & --- & 1 \\ 
C07{\_}10 & $26.24^{+0.08}_{-0.11}$ & $-20.16^{+0.08}_{-0.11}$ & $0.68^{+0.03}_{-0.02}$ & $26.21^{+0.06}_{-0.07}$ & $-20.19^{+0.06}_{-0.07}$ & $0.60^{+0.05}_{-0.05}$ & --- & --- & --- & --- & --- & 2 \\ 
C07{\_}12 & $26.28^{+0.08}_{-0.11}$ & $-20.12^{+0.08}_{-0.11}$ & $0.58^{+0.02}_{-0.01}$ & $26.20^{+0.06}_{-0.07}$ & $-20.19^{+0.06}_{-0.07}$ & $0.63^{+0.05}_{-0.05}$ & --- & --- & --- & --- & --- & 2 \\ 
C07{\_}13 & $25.32^{+0.03}_{-0.04}$ & $-21.08^{+0.03}_{-0.04}$ & $0.62^{+0.01}_{-0.01}$ & $24.53^{+0.05}_{-0.07}$ & $-21.87^{+0.05}_{-0.07}$ & $1.18^{+0.04}_{-0.03}$ & --- & --- & --- & --- & --- & 2 \\ 
C07{\_}14 & $25.81^{+0.05}_{-0.07}$ & $-20.59^{+0.05}_{-0.07}$ & $0.72^{+0.01}_{-0.01}$ & $25.82^{+0.06}_{-0.07}$ & $-20.58^{+0.06}_{-0.07}$ & $0.70^{+0.02}_{-0.02}$ & --- & --- & --- & --- & --- & 1 \\ 
C07{\_}15 & $26.95^{+0.06}_{-0.07}$ & $-19.75^{+0.06}_{-0.07}$ & $0.31^{+0.01}_{-0.01}$ & $26.77^{+0.15}_{-0.17}$ & $-19.93^{+0.15}_{-0.17}$ & $0.43^{+0.03}_{-0.04}$ & --- & --- & --- & --- & --- & 1 \\ 
C07{\_}16 & $26.63^{+0.08}_{-0.11}$ & $-20.23^{+0.08}_{-0.11}$ & $0.65^{+0.03}_{-0.02}$ & $25.82^{+0.08}_{-0.10}$ & $-21.05^{+0.08}_{-0.10}$ & $1.36^{+0.05}_{-0.04}$ & --- & --- & --- & --- & --- & 2 \\ 
C08{\_}01 & $25.87^{+0.02}_{-0.03}$ & $-21.40^{+0.02}_{-0.03}$ & $0.13^{+0.00}_{-0.00}$ & $24.89^{+0.04}_{-0.05}$ & $-22.38^{+0.04}_{-0.05}$ & $0.53^{+0.01}_{-0.01}$ & $24.01^{+0.02}_{-0.03}$ & $-23.25^{+0.02}_{-0.03}$ & $0.28^{+0.01}_{-0.01}$ & $2.21^{+0.11}_{-0.10}$ & $0.53^{+0.12}_{-0.11}$ & 2 \\ 
C08{\_}02 & $27.52^{+0.11}_{-0.14}$ & $-19.46^{+0.11}_{-0.14}$ & $0.23^{+0.03}_{-0.02}$ & $27.40^{+0.14}_{-0.17}$ & $-19.57^{+0.14}_{-0.17}$ & $0.35^{+0.13}_{-0.06}$ & --- & --- & --- & --- & --- & 1 \\ 
C08{\_}03 & $27.02^{+0.06}_{-0.07}$ & $-19.38^{+0.06}_{-0.07}$ & $0.15^{+0.04}_{-0.03}$ & $26.74^{+0.05}_{-0.05}$ & $-19.66^{+0.05}_{-0.05}$ & $0.30^{+0.03}_{-0.03}$ & $26.87^{+0.08}_{-0.12}$ & $-19.53^{+0.08}_{-0.12}$ & $0.18^{+0.03}_{-0.03}$ & $1.22^{+0.40}_{-0.32}$ & $0.60^{+0.25}_{-0.25}$ & 2 \\ 
C08{\_}04 & $26.55^{+0.04}_{-0.05}$ & $-19.85^{+0.04}_{-0.05}$ & $0.27^{+0.01}_{-0.01}$ & --- & --- & --- & $25.43^{+0.05}_{-0.07}$ & $-20.97^{+0.05}_{-0.07}$ & $1.02^{+0.03}_{-0.02}$ & $3.84^{+0.20}_{-0.15}$ & --- & 1 \\ 
C08{\_}05 & $24.75^{+0.06}_{-0.07}$ & $-21.65^{+0.06}_{-0.07}$ & $0.98^{+0.02}_{-0.01}$ & $24.73^{+0.05}_{-0.07}$ & $-21.67^{+0.05}_{-0.07}$ & $1.05^{+0.03}_{-0.03}$ & $23.11^{+0.04}_{-0.04}$ & $-23.29^{+0.04}_{-0.04}$ & $0.93^{+0.03}_{-0.02}$ & $0.96^{+0.04}_{-0.03}$ & $0.89^{+0.04}_{-0.03}$ & 2 \\ 
C08{\_}06 & $26.06^{+0.02}_{-0.03}$ & $-20.34^{+0.02}_{-0.03}$ & $0.21^{+0.01}_{-0.01}$ & $24.84^{+0.05}_{-0.07}$ & $-21.56^{+0.05}_{-0.07}$ & $1.21^{+0.04}_{-0.03}$ & $23.64^{+0.05}_{-0.07}$ & $-22.76^{+0.05}_{-0.07}$ & $1.29^{+0.04}_{-0.03}$ & $6.15^{+0.38}_{-0.30}$ & $1.07^{+0.26}_{-0.21}$ & 2 \\ 
C08{\_}07 & $25.37^{+0.03}_{-0.04}$ & $-21.02^{+0.03}_{-0.04}$ & $0.93^{+0.02}_{-0.01}$ & $25.52^{+0.04}_{-0.05}$ & $-20.88^{+0.04}_{-0.05}$ & $0.79^{+0.02}_{-0.02}$ & $25.18^{+0.04}_{-0.04}$ & $-21.22^{+0.04}_{-0.04}$ & $0.80^{+0.02}_{-0.02}$ & $0.85^{+0.03}_{-0.02}$ & $1.00^{+0.03}_{-0.03}$ & 2 \\ 
C08{\_}08 & $25.72^{+0.03}_{-0.04}$ & $-20.68^{+0.03}_{-0.04}$ & $0.90^{+0.02}_{-0.01}$ & $25.40^{+0.05}_{-0.07}$ & $-21.00^{+0.05}_{-0.07}$ & $1.17^{+0.04}_{-0.03}$ & $23.71^{+0.05}_{-0.07}$ & $-22.69^{+0.05}_{-0.07}$ & $0.97^{+0.03}_{-0.02}$ & $1.08^{+0.04}_{-0.03}$ & $0.83^{+0.05}_{-0.04}$ & 2 \\ 
C08{\_}09 & $25.79^{+0.09}_{-0.11}$ & $-20.61^{+0.09}_{-0.11}$ & $1.10^{+0.03}_{-0.02}$ & $25.83^{+0.08}_{-0.10}$ & $-20.57^{+0.08}_{-0.10}$ & $1.12^{+0.04}_{-0.03}$ & $25.09^{+0.04}_{-0.04}$ & $-21.31^{+0.04}_{-0.04}$ & $0.54^{+0.02}_{-0.02}$ & $0.49^{+0.02}_{-0.02}$ & $0.48^{+0.02}_{-0.02}$ & 1 \\ 
C08{\_}10 & $26.35^{+0.04}_{-0.05}$ & $-20.05^{+0.04}_{-0.05}$ & $0.38^{+0.02}_{-0.01}$ & $26.02^{+0.06}_{-0.07}$ & $-20.38^{+0.06}_{-0.07}$ & $0.55^{+0.03}_{-0.04}$ & $25.49^{+0.04}_{-0.04}$ & $-20.91^{+0.04}_{-0.04}$ & $0.80^{+0.02}_{-0.02}$ & $2.11^{+0.12}_{-0.09}$ & $1.45^{+0.14}_{-0.16}$ & 2 \\ 
C08{\_}11 & $24.25^{+0.06}_{-0.07}$ & $-22.15^{+0.06}_{-0.07}$ & $1.28^{+0.03}_{-0.02}$ & $24.16^{+0.05}_{-0.07}$ & $-22.24^{+0.05}_{-0.07}$ & $1.49^{+0.05}_{-0.04}$ & $23.59^{+0.05}_{-0.07}$ & $-22.81^{+0.05}_{-0.07}$ & $1.30^{+0.04}_{-0.03}$ & $1.01^{+0.04}_{-0.03}$ & $0.87^{+0.04}_{-0.03}$ & 2 \\ 
C08{\_}12 & $25.93^{+0.05}_{-0.07}$ & $-20.47^{+0.05}_{-0.07}$ & $0.84^{+0.02}_{-0.01}$ & $25.45^{+0.05}_{-0.07}$ & $-20.95^{+0.05}_{-0.07}$ & $1.03^{+0.03}_{-0.03}$ & $24.55^{+0.05}_{-0.07}$ & $-21.85^{+0.05}_{-0.07}$ & $0.96^{+0.03}_{-0.02}$ & $1.15^{+0.04}_{-0.03}$ & $0.94^{+0.05}_{-0.04}$ & 2 \\ 
C08{\_}13 & $25.68^{+0.01}_{-0.02}$ & $-20.71^{+0.01}_{-0.02}$ & $0.20^{+0.00}_{-0.00}$ & $25.50^{+0.02}_{-0.02}$ & $-20.90^{+0.02}_{-0.02}$ & $0.34^{+0.01}_{-0.02}$ & $25.41^{+0.02}_{-0.03}$ & $-20.99^{+0.02}_{-0.03}$ & $0.42^{+0.02}_{-0.02}$ & $2.09^{+0.09}_{-0.09}$ & $1.25^{+0.12}_{-0.12}$ & 2 \\ 
C08{\_}14 & $26.11^{+0.05}_{-0.07}$ & $-20.29^{+0.05}_{-0.07}$ & $0.52^{+0.02}_{-0.01}$ & $26.04^{+0.06}_{-0.07}$ & $-20.36^{+0.06}_{-0.07}$ & $0.61^{+0.05}_{-0.05}$ & $24.84^{+0.04}_{-0.04}$ & $-21.56^{+0.04}_{-0.04}$ & $0.52^{+0.02}_{-0.02}$ & $0.99^{+0.05}_{-0.04}$ & $0.85^{+0.09}_{-0.09}$ & 2 \\ 
C08{\_}15 & $25.20^{+0.06}_{-0.07}$ & $-21.49^{+0.06}_{-0.07}$ & $0.87^{+0.02}_{-0.01}$ & $25.19^{+0.05}_{-0.07}$ & $-21.50^{+0.05}_{-0.07}$ & $0.90^{+0.03}_{-0.02}$ & $24.61^{+0.04}_{-0.04}$ & $-22.08^{+0.04}_{-0.04}$ & $0.78^{+0.03}_{-0.02}$ & $0.90^{+0.04}_{-0.03}$ & $0.87^{+0.04}_{-0.03}$ & 2 \\ 
C08{\_}16 & $26.13^{+0.09}_{-0.11}$ & $-20.56^{+0.09}_{-0.11}$ & $1.20^{+0.07}_{-0.05}$ & $26.09^{+0.08}_{-0.10}$ & $-20.61^{+0.08}_{-0.10}$ & $1.29^{+0.12}_{-0.09}$ & --- & --- & --- & --- & --- & 1 \\ 
C08{\_}17 & $25.19^{+0.03}_{-0.04}$ & $-21.51^{+0.03}_{-0.04}$ & $0.64^{+0.01}_{-0.01}$ & $25.22^{+0.04}_{-0.05}$ & $-21.48^{+0.04}_{-0.05}$ & $0.66^{+0.02}_{-0.02}$ & $24.17^{+0.04}_{-0.04}$ & $-22.52^{+0.04}_{-0.04}$ & $0.83^{+0.03}_{-0.02}$ & $1.30^{+0.05}_{-0.04}$ & $1.27^{+0.06}_{-0.05}$ & 2 \\ 
C08{\_}18 & $26.13^{+0.09}_{-0.11}$ & $-20.56^{+0.09}_{-0.11}$ & $1.08^{+0.07}_{-0.05}$ & $25.88^{+0.08}_{-0.10}$ & $-20.81^{+0.08}_{-0.10}$ & $1.29^{+0.04}_{-0.03}$ & $23.81^{+0.05}_{-0.07}$ & $-22.88^{+0.05}_{-0.07}$ & $1.28^{+0.04}_{-0.04}$ & $1.18^{+0.09}_{-0.06}$ & $0.99^{+0.06}_{-0.04}$ & 2 \\ 
C08{\_}19 & $26.08^{+0.05}_{-0.07}$ & $-20.61^{+0.05}_{-0.07}$ & $0.49^{+0.02}_{-0.01}$ & $25.98^{+0.06}_{-0.07}$ & $-20.72^{+0.06}_{-0.07}$ & $0.60^{+0.04}_{-0.05}$ & $23.88^{+0.04}_{-0.04}$ & $-22.82^{+0.04}_{-0.04}$ & $0.71^{+0.02}_{-0.02}$ & $1.45^{+0.06}_{-0.05}$ & $1.19^{+0.11}_{-0.12}$ & 1 \\ 
C08{\_}20 & $26.65^{+0.04}_{-0.05}$ & $-20.04^{+0.04}_{-0.05}$ & $0.35^{+0.02}_{-0.01}$ & $26.67^{+0.05}_{-0.05}$ & $-20.03^{+0.05}_{-0.05}$ & $0.35^{+0.03}_{-0.04}$ & $24.63^{+0.05}_{-0.07}$ & $-22.06^{+0.05}_{-0.07}$ & $0.94^{+0.03}_{-0.02}$ & $2.72^{+0.15}_{-0.11}$ & $2.68^{+0.25}_{-0.28}$ & 1 \\ 
C08{\_}21 & $25.93^{+0.02}_{-0.03}$ & $-20.77^{+0.02}_{-0.03}$ & $0.22^{+0.00}_{-0.00}$ & $25.10^{+0.04}_{-0.05}$ & $-21.59^{+0.04}_{-0.05}$ & $0.57^{+0.02}_{-0.02}$ & $24.78^{+0.04}_{-0.04}$ & $-21.92^{+0.04}_{-0.04}$ & $0.62^{+0.02}_{-0.02}$ & $2.74^{+0.10}_{-0.09}$ & $1.08^{+0.12}_{-0.12}$ & 2 \\ 
C08{\_}22 & $25.41^{+0.03}_{-0.04}$ & $-21.28^{+0.03}_{-0.04}$ & $0.83^{+0.02}_{-0.01}$ & $25.32^{+0.05}_{-0.07}$ & $-21.37^{+0.05}_{-0.07}$ & $0.99^{+0.04}_{-0.03}$ & $24.79^{+0.04}_{-0.04}$ & $-21.90^{+0.04}_{-0.04}$ & $0.61^{+0.02}_{-0.02}$ & $0.73^{+0.03}_{-0.02}$ & $0.61^{+0.03}_{-0.03}$ & 1 \\ 
C08{\_}23 & $26.35^{+0.08}_{-0.11}$ & $-20.34^{+0.08}_{-0.11}$ & $0.45^{+0.02}_{-0.01}$ & $26.47^{+0.05}_{-0.05}$ & $-20.23^{+0.05}_{-0.05}$ & $0.39^{+0.03}_{-0.04}$ & $25.89^{+0.03}_{-0.03}$ & $-20.81^{+0.03}_{-0.03}$ & $0.31^{+0.02}_{-0.01}$ & $0.69^{+0.05}_{-0.04}$ & $0.78^{+0.07}_{-0.07}$ & 1 \\ 
C08{\_}24 & $26.31^{+0.08}_{-0.11}$ & $-20.38^{+0.08}_{-0.11}$ & $0.57^{+0.03}_{-0.02}$ & $25.77^{+0.08}_{-0.10}$ & $-20.92^{+0.08}_{-0.10}$ & $0.94^{+0.03}_{-0.02}$ & $25.33^{+0.04}_{-0.04}$ & $-21.36^{+0.04}_{-0.04}$ & $0.76^{+0.03}_{-0.02}$ & $1.34^{+0.08}_{-0.06}$ & $0.81^{+0.06}_{-0.05}$ & 2 \\ 
C08{\_}25 & $26.89^{+0.06}_{-0.07}$ & $-20.05^{+0.06}_{-0.07}$ & $0.13^{+0.01}_{-0.01}$ & $26.87^{+0.09}_{-0.08}$ & $-20.07^{+0.09}_{-0.08}$ & $0.19^{+0.03}_{-0.03}$ & --- & --- & --- & --- & --- & 1 \\ 
C08{\_}27 & $25.74^{+0.02}_{-0.03}$ & $-21.12^{+0.02}_{-0.03}$ & $0.24^{+0.00}_{-0.00}$ & $25.68^{+0.02}_{-0.02}$ & $-21.18^{+0.02}_{-0.02}$ & $0.32^{+0.01}_{-0.01}$ & $24.61^{+0.02}_{-0.03}$ & $-22.26^{+0.02}_{-0.03}$ & $0.38^{+0.01}_{-0.01}$ & $1.61^{+0.07}_{-0.06}$ & $1.20^{+0.09}_{-0.09}$ & 1 \\ 
C08{\_}28 & $26.23^{+0.02}_{-0.03}$ & $-20.73^{+0.02}_{-0.03}$ & $0.36^{+0.02}_{-0.01}$ & $25.95^{+0.06}_{-0.07}$ & $-21.01^{+0.06}_{-0.07}$ & $0.52^{+0.02}_{-0.02}$ & --- & --- & --- & --- & --- & 2 \\ 
C08{\_}29 & $26.61^{+0.04}_{-0.05}$ & $-20.18^{+0.04}_{-0.05}$ & $0.17^{+0.01}_{-0.01}$ & $25.91^{+0.06}_{-0.07}$ & $-20.88^{+0.06}_{-0.07}$ & $0.55^{+0.02}_{-0.02}$ & $25.67^{+0.05}_{-0.06}$ & $-21.12^{+0.05}_{-0.06}$ & $0.49^{+0.02}_{-0.01}$ & $2.82^{+0.19}_{-0.15}$ & $0.89^{+0.13}_{-0.12}$ & 2 \\ 
C08{\_}30 & $26.48^{+0.04}_{-0.05}$ & $-20.73^{+0.04}_{-0.05}$ & $0.16^{+0.01}_{-0.01}$ & $26.52^{+0.05}_{-0.05}$ & $-20.69^{+0.05}_{-0.05}$ & $0.17^{+0.03}_{-0.02}$ & $25.64^{+0.02}_{-0.03}$ & $-21.58^{+0.02}_{-0.03}$ & $0.21^{+0.01}_{-0.01}$ & $1.28^{+0.11}_{-0.09}$ & $1.25^{+0.21}_{-0.20}$ & 1 \\ 
C08{\_}31 & $27.07^{+0.06}_{-0.07}$ & $-20.09^{+0.06}_{-0.07}$ & $0.25^{+0.03}_{-0.02}$ & $26.81^{+0.15}_{-0.17}$ & $-20.35^{+0.15}_{-0.17}$ & $0.38^{+0.03}_{-0.03}$ & $25.55^{+0.04}_{-0.04}$ & $-21.61^{+0.04}_{-0.04}$ & $0.39^{+0.01}_{-0.01}$ & $1.54^{+0.20}_{-0.14}$ & $1.01^{+0.12}_{-0.13}$ & 1 \\ 
C09{\_}01 & $26.13^{+0.05}_{-0.07}$ & $-20.24^{+0.05}_{-0.07}$ & $0.52^{+0.02}_{-0.01}$ & $26.16^{+0.06}_{-0.07}$ & $-20.20^{+0.06}_{-0.07}$ & $0.50^{+0.03}_{-0.04}$ & $25.73^{+0.05}_{-0.06}$ & $-20.64^{+0.05}_{-0.06}$ & $0.63^{+0.02}_{-0.02}$ & $1.21^{+0.06}_{-0.05}$ & $1.27^{+0.09}_{-0.10}$ & 1 \\ 
C09{\_}02 & $26.82^{+0.13}_{-0.18}$ & $-19.54^{+0.13}_{-0.18}$ & $0.63^{+0.03}_{-0.02}$ & $26.71^{+0.15}_{-0.17}$ & $-19.65^{+0.15}_{-0.17}$ & $0.66^{+0.05}_{-0.05}$ & $26.46^{+0.04}_{-0.04}$ & $-19.90^{+0.04}_{-0.04}$ & $0.47^{+0.03}_{-0.04}$ & $0.74^{+0.06}_{-0.06}$ & $0.70^{+0.08}_{-0.08}$ & 2 \\ 
C09{\_}03 & $25.88^{+0.05}_{-0.07}$ & $-20.51^{+0.05}_{-0.07}$ & $0.76^{+0.01}_{-0.01}$ & $25.96^{+0.06}_{-0.07}$ & $-20.44^{+0.06}_{-0.07}$ & $0.70^{+0.02}_{-0.02}$ & $25.54^{+0.04}_{-0.04}$ & $-20.86^{+0.04}_{-0.04}$ & $0.79^{+0.02}_{-0.02}$ & $1.04^{+0.03}_{-0.03}$ & $1.13^{+0.04}_{-0.04}$ & 1 \\ 
C09{\_}04 & $25.58^{+0.03}_{-0.04}$ & $-20.82^{+0.03}_{-0.04}$ & $0.52^{+0.01}_{-0.01}$ & $25.63^{+0.04}_{-0.05}$ & $-20.77^{+0.04}_{-0.05}$ & $0.49^{+0.02}_{-0.01}$ & $25.24^{+0.02}_{-0.03}$ & $-21.16^{+0.02}_{-0.03}$ & $0.41^{+0.02}_{-0.02}$ & $0.78^{+0.04}_{-0.03}$ & $0.84^{+0.04}_{-0.04}$ & 2 \\ 
C09{\_}05 & $26.46^{+0.08}_{-0.11}$ & $-19.94^{+0.08}_{-0.11}$ & $0.54^{+0.02}_{-0.01}$ & $26.08^{+0.06}_{-0.07}$ & $-20.32^{+0.06}_{-0.07}$ & $0.67^{+0.05}_{-0.05}$ & $25.90^{+0.05}_{-0.06}$ & $-20.50^{+0.05}_{-0.06}$ & $0.64^{+0.02}_{-0.02}$ & $1.20^{+0.06}_{-0.05}$ & $0.96^{+0.10}_{-0.10}$ & 2 \\ 
C09{\_}06 & $25.93^{+0.02}_{-0.03}$ & $-20.47^{+0.02}_{-0.03}$ & $0.21^{+0.00}_{-0.00}$ & $25.66^{+0.02}_{-0.02}$ & $-20.74^{+0.02}_{-0.02}$ & $0.41^{+0.02}_{-0.01}$ & $24.86^{+0.04}_{-0.04}$ & $-21.54^{+0.04}_{-0.04}$ & $0.61^{+0.02}_{-0.02}$ & $2.89^{+0.11}_{-0.11}$ & $1.48^{+0.15}_{-0.14}$ & 2 \\ 
C09{\_}07 & $27.17^{+0.06}_{-0.07}$ & $-19.23^{+0.06}_{-0.07}$ & $0.44^{+0.07}_{-0.05}$ & $26.29^{+0.09}_{-0.11}$ & $-20.11^{+0.09}_{-0.11}$ & $0.90^{+0.07}_{-0.07}$ & $26.16^{+0.05}_{-0.06}$ & $-20.24^{+0.05}_{-0.06}$ & $0.70^{+0.09}_{-0.05}$ & $1.60^{+0.34}_{-0.20}$ & $0.78^{+0.24}_{-0.16}$ & 2 \\ 
C09{\_}08 & $26.41^{+0.08}_{-0.11}$ & $-19.99^{+0.08}_{-0.11}$ & $0.55^{+0.02}_{-0.01}$ & $26.38^{+0.05}_{-0.05}$ & $-20.02^{+0.05}_{-0.05}$ & $0.45^{+0.03}_{-0.04}$ & $25.25^{+0.04}_{-0.04}$ & $-21.15^{+0.04}_{-0.04}$ & $0.72^{+0.02}_{-0.02}$ & $1.29^{+0.06}_{-0.05}$ & $1.58^{+0.10}_{-0.12}$ & 2 \\ 
C10{\_}01 & $26.21^{+0.02}_{-0.03}$ & $-20.19^{+0.02}_{-0.03}$ & $0.33^{+0.01}_{-0.01}$ & $26.13^{+0.03}_{-0.03}$ & $-20.27^{+0.03}_{-0.03}$ & $0.40^{+0.03}_{-0.04}$ & $25.59^{+0.04}_{-0.04}$ & $-20.81^{+0.04}_{-0.04}$ & $0.65^{+0.02}_{-0.02}$ & $1.99^{+0.09}_{-0.08}$ & $1.63^{+0.18}_{-0.21}$ & 2 \\ 
C10{\_}02 & $25.66^{+0.01}_{-0.02}$ & $-20.74^{+0.01}_{-0.02}$ & $0.44^{+0.01}_{-0.01}$ & $25.53^{+0.04}_{-0.05}$ & $-20.87^{+0.04}_{-0.05}$ & $0.63^{+0.02}_{-0.02}$ & $24.99^{+0.04}_{-0.04}$ & $-21.41^{+0.04}_{-0.04}$ & $0.47^{+0.02}_{-0.02}$ & $1.08^{+0.04}_{-0.04}$ & $0.75^{+0.05}_{-0.05}$ & 1 \\ 
C10{\_}03 & $25.44^{+0.03}_{-0.04}$ & $-20.96^{+0.03}_{-0.04}$ & $0.60^{+0.01}_{-0.01}$ & $25.47^{+0.04}_{-0.05}$ & $-20.93^{+0.04}_{-0.05}$ & $0.58^{+0.02}_{-0.01}$ & $24.89^{+0.04}_{-0.04}$ & $-21.51^{+0.04}_{-0.04}$ & $0.73^{+0.02}_{-0.02}$ & $1.21^{+0.04}_{-0.04}$ & $1.26^{+0.05}_{-0.04}$ & 2 \\ 
C10{\_}04 & $26.31^{+0.04}_{-0.05}$ & $-20.09^{+0.04}_{-0.05}$ & $0.44^{+0.02}_{-0.01}$ & $26.15^{+0.06}_{-0.07}$ & $-20.25^{+0.06}_{-0.07}$ & $0.58^{+0.03}_{-0.04}$ & $26.72^{+0.04}_{-0.04}$ & $-19.68^{+0.04}_{-0.04}$ & $0.27^{+0.03}_{-0.03}$ & $0.62^{+0.08}_{-0.07}$ & $0.47^{+0.08}_{-0.08}$ & 2 \\ 
C10{\_}05 & $25.49^{+0.03}_{-0.04}$ & $-20.91^{+0.03}_{-0.04}$ & $0.49^{+0.01}_{-0.01}$ & $25.40^{+0.04}_{-0.05}$ & $-21.00^{+0.04}_{-0.05}$ & $0.59^{+0.02}_{-0.01}$ & $25.67^{+0.02}_{-0.03}$ & $-20.73^{+0.02}_{-0.03}$ & $0.41^{+0.02}_{-0.02}$ & $0.84^{+0.04}_{-0.03}$ & $0.70^{+0.04}_{-0.04}$ & 2 \\ 
C10{\_}06 & $26.40^{+0.15}_{-0.18}$ & $-20.00^{+0.15}_{-0.18}$ & $1.12^{+0.08}_{-0.05}$ & $26.37^{+0.14}_{-0.15}$ & $-20.03^{+0.14}_{-0.15}$ & $0.96^{+0.07}_{-0.07}$ & $26.22^{+0.09}_{-0.12}$ & $-20.17^{+0.09}_{-0.12}$ & $0.77^{+0.09}_{-0.05}$ & $0.69^{+0.09}_{-0.05}$ & $0.80^{+0.09}_{-0.06}$ & 2 \\ 
C10{\_}07 & $26.31^{+0.04}_{-0.05}$ & $-20.09^{+0.04}_{-0.05}$ & $0.43^{+0.02}_{-0.01}$ & $25.72^{+0.04}_{-0.05}$ & $-20.68^{+0.04}_{-0.05}$ & $0.86^{+0.03}_{-0.03}$ & $25.51^{+0.04}_{-0.04}$ & $-20.89^{+0.04}_{-0.04}$ & $0.77^{+0.02}_{-0.02}$ & $1.79^{+0.09}_{-0.07}$ & $0.89^{+0.08}_{-0.07}$ & 2 \\ 
C10{\_}08 & $26.03^{+0.05}_{-0.07}$ & $-20.37^{+0.05}_{-0.07}$ & $0.55^{+0.02}_{-0.01}$ & $26.12^{+0.06}_{-0.07}$ & $-20.28^{+0.06}_{-0.07}$ & $0.48^{+0.03}_{-0.04}$ & $25.37^{+0.04}_{-0.04}$ & $-21.03^{+0.04}_{-0.04}$ & $0.51^{+0.02}_{-0.02}$ & $0.92^{+0.04}_{-0.04}$ & $1.07^{+0.07}_{-0.08}$ & 1 \\ 
C10{\_}09 & $24.91^{+0.03}_{-0.04}$ & $-22.03^{+0.03}_{-0.04}$ & $0.62^{+0.01}_{-0.01}$ & $25.00^{+0.04}_{-0.05}$ & $-21.93^{+0.04}_{-0.05}$ & $0.56^{+0.02}_{-0.02}$ & $23.99^{+0.04}_{-0.04}$ & $-22.95^{+0.04}_{-0.04}$ & $0.65^{+0.02}_{-0.02}$ & $1.04^{+0.03}_{-0.03}$ & $1.15^{+0.04}_{-0.04}$ & 2 \\ 
C10{\_}10 & $25.41^{+0.01}_{-0.02}$ & $-21.48^{+0.01}_{-0.02}$ & $0.16^{+0.00}_{-0.00}$ & $25.48^{+0.02}_{-0.02}$ & $-21.41^{+0.02}_{-0.02}$ & $0.14^{+0.01}_{-0.01}$ & $25.42^{+0.02}_{-0.03}$ & $-21.46^{+0.02}_{-0.03}$ & $0.16^{+0.02}_{-0.01}$ & $0.95^{+0.10}_{-0.08}$ & $1.10^{+0.13}_{-0.12}$ & 1 \\ 
\hline
\end{longtable}
\end{tiny}
\end{landscape}

\begin{figure*}
\begin{center}
   \includegraphics[height=0.14\textheight]{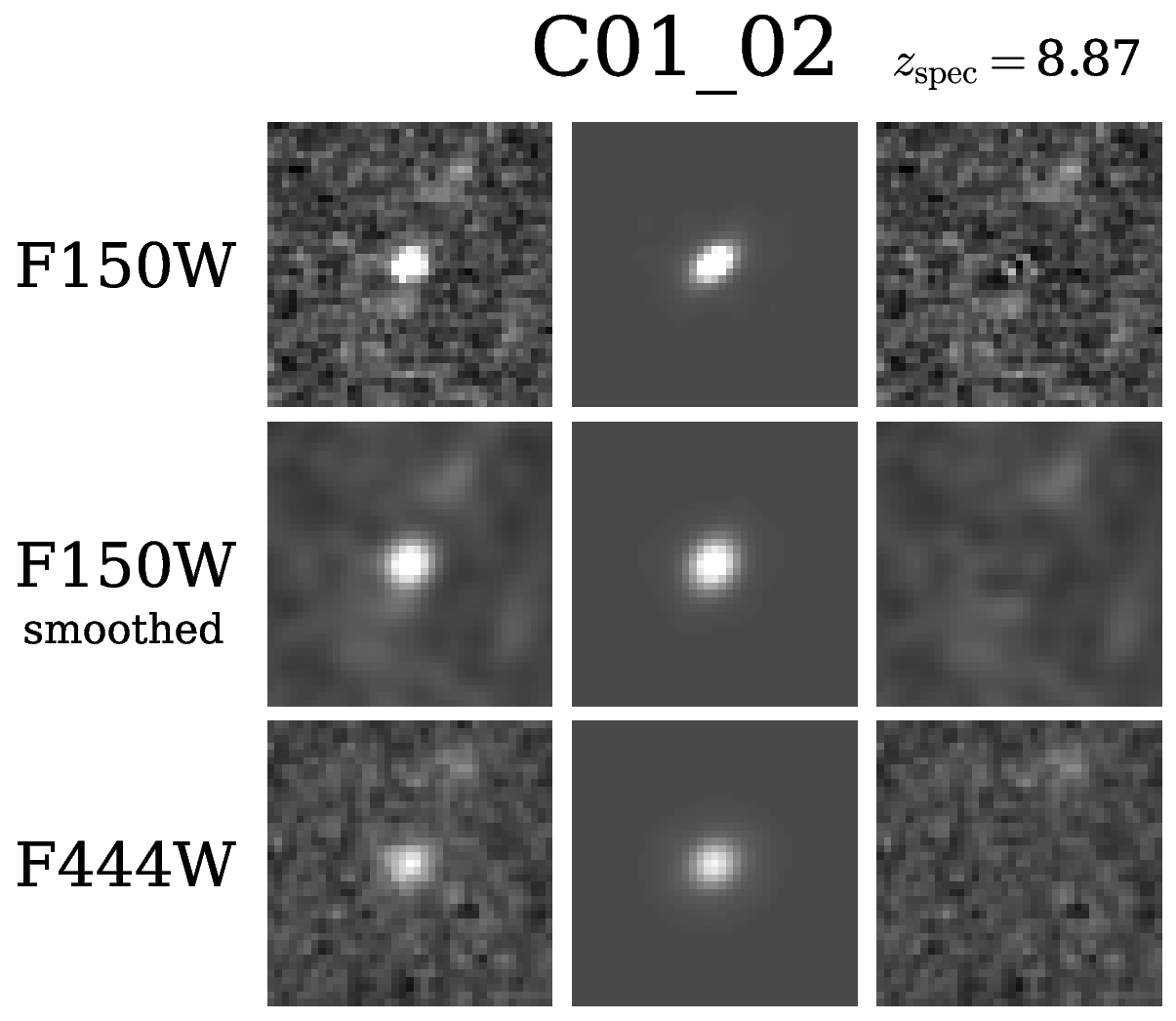}
   \includegraphics[height=0.14\textheight]{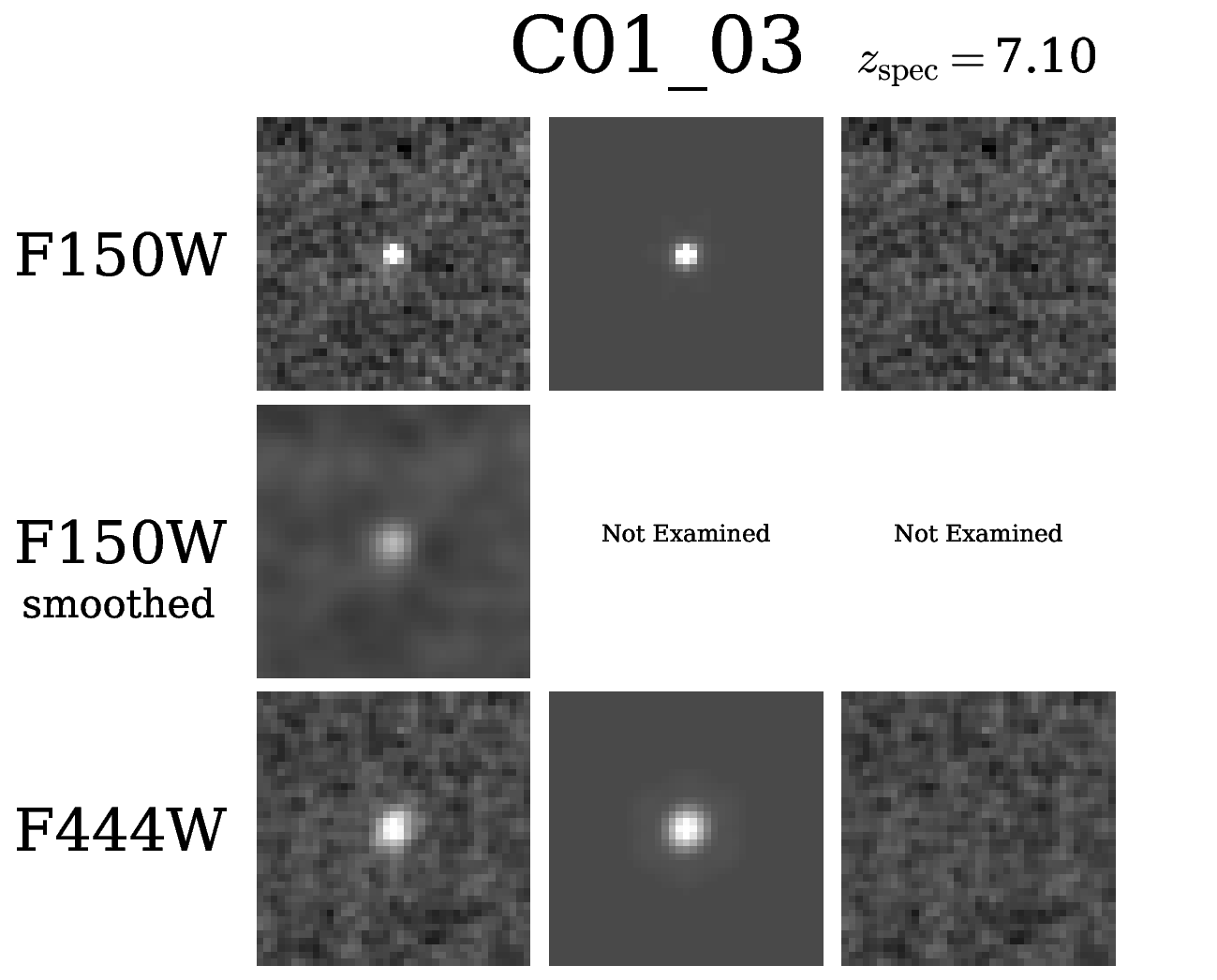}
   \includegraphics[height=0.14\textheight]{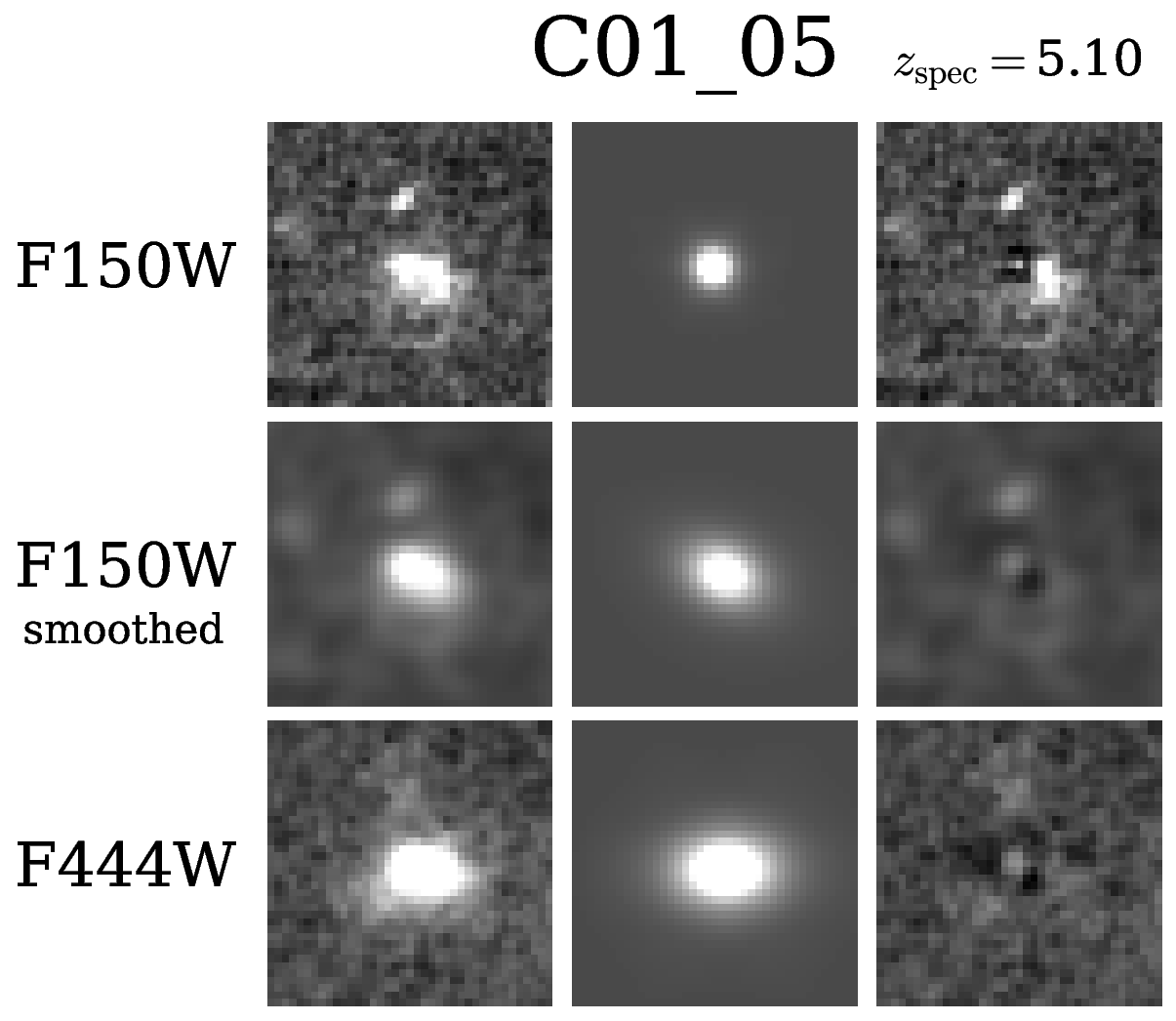}
   \includegraphics[height=0.14\textheight]{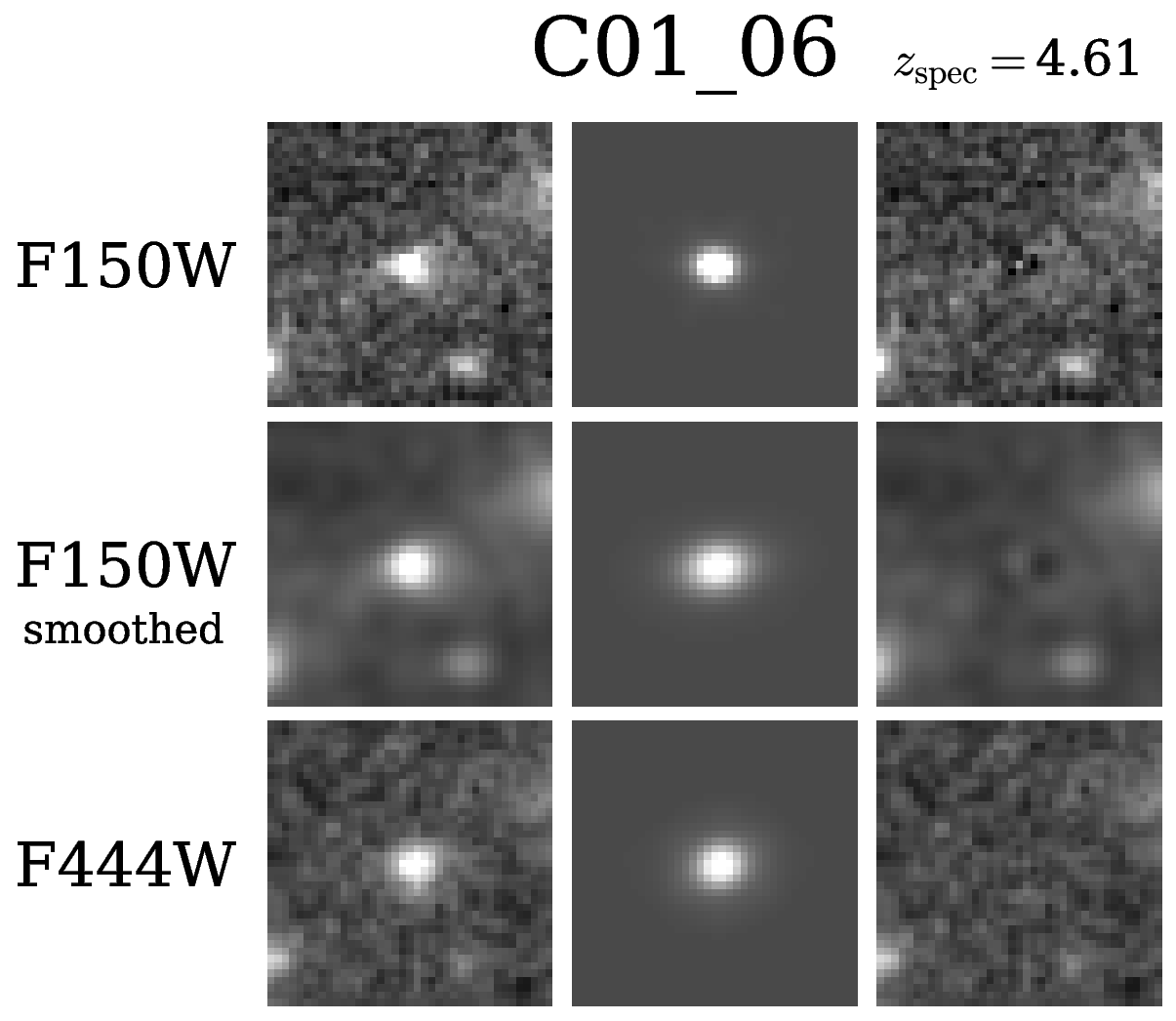}
   \includegraphics[height=0.14\textheight]{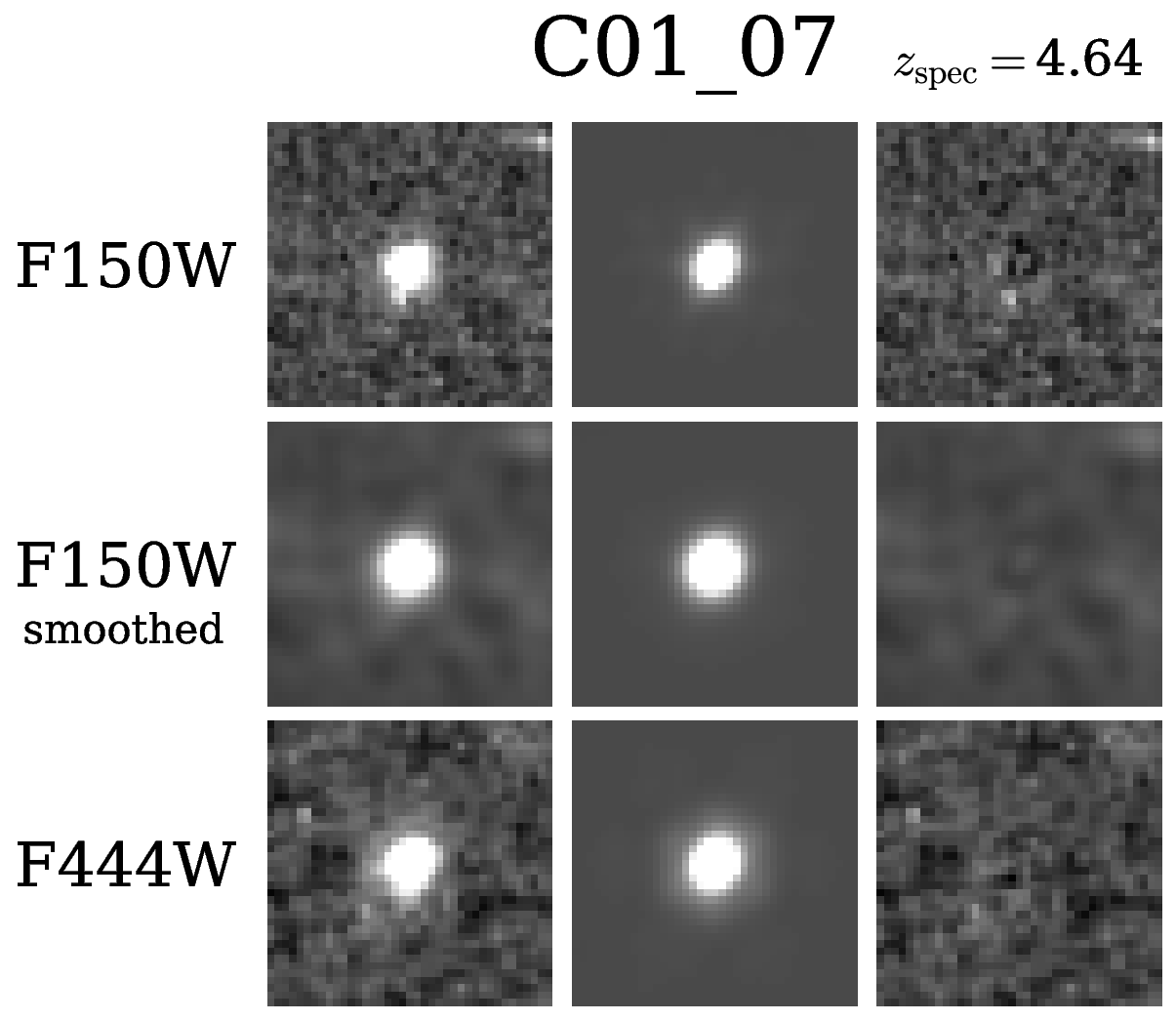}
   \includegraphics[height=0.14\textheight]{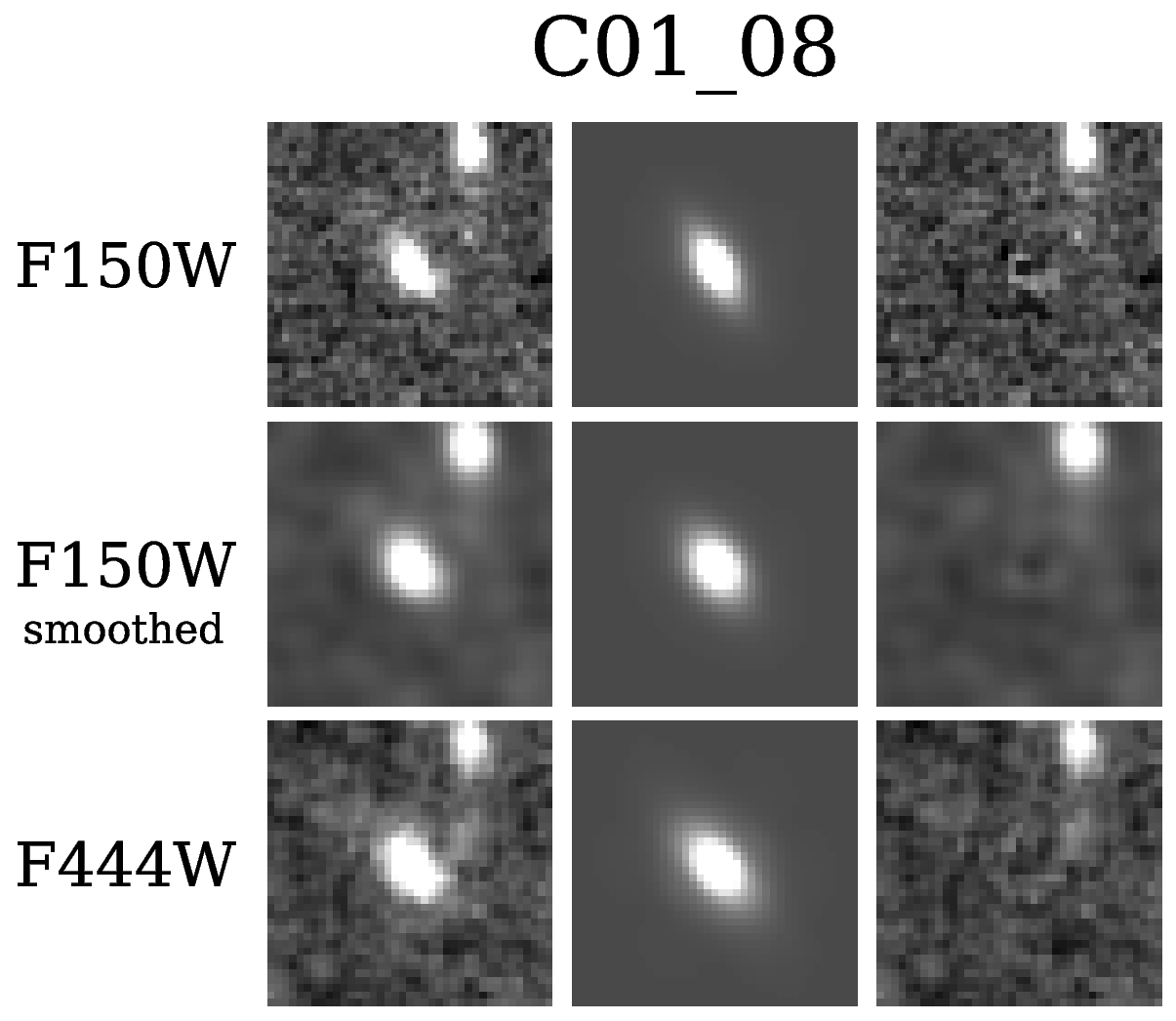}
   \includegraphics[height=0.14\textheight]{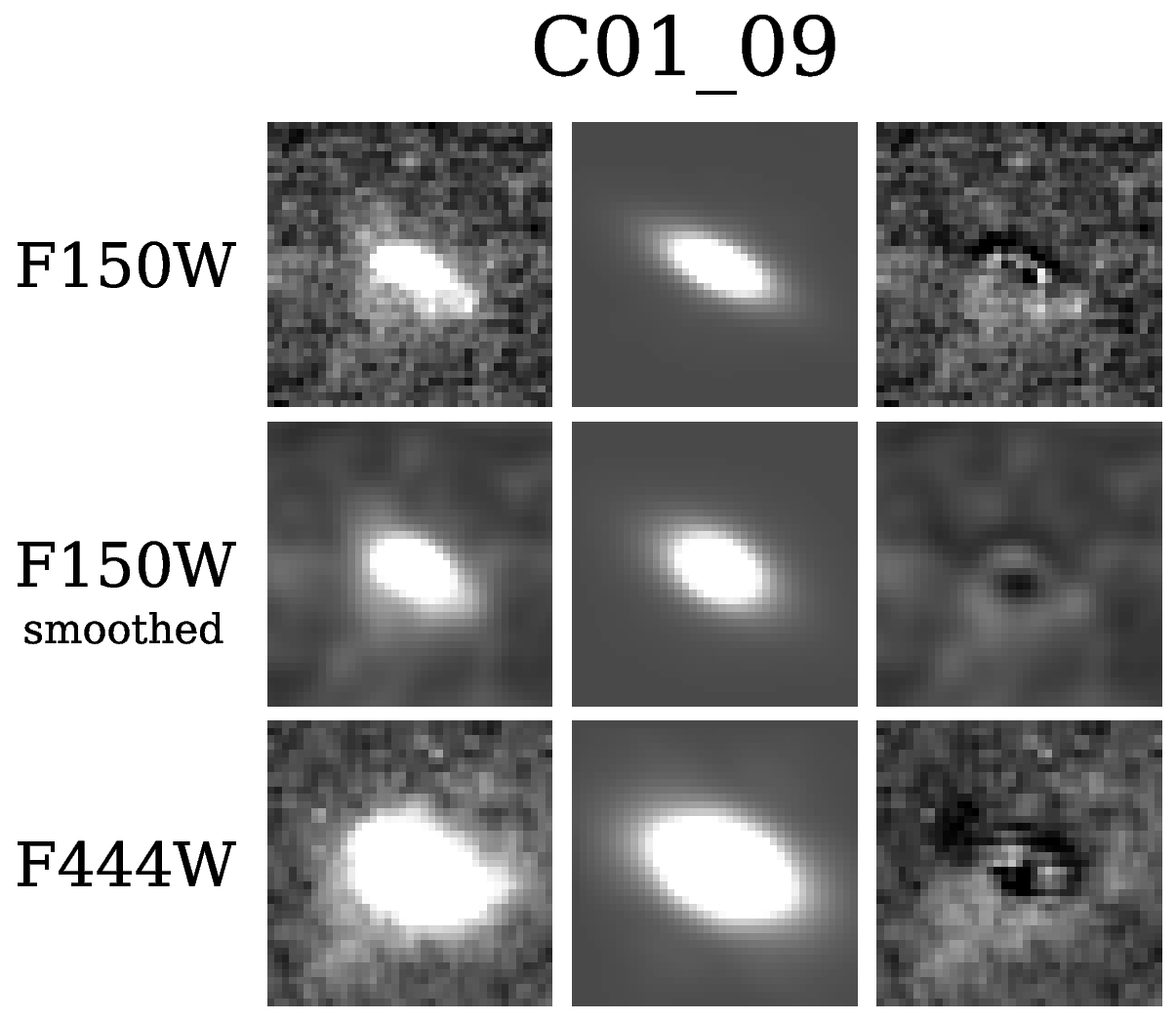}
   \includegraphics[height=0.14\textheight]{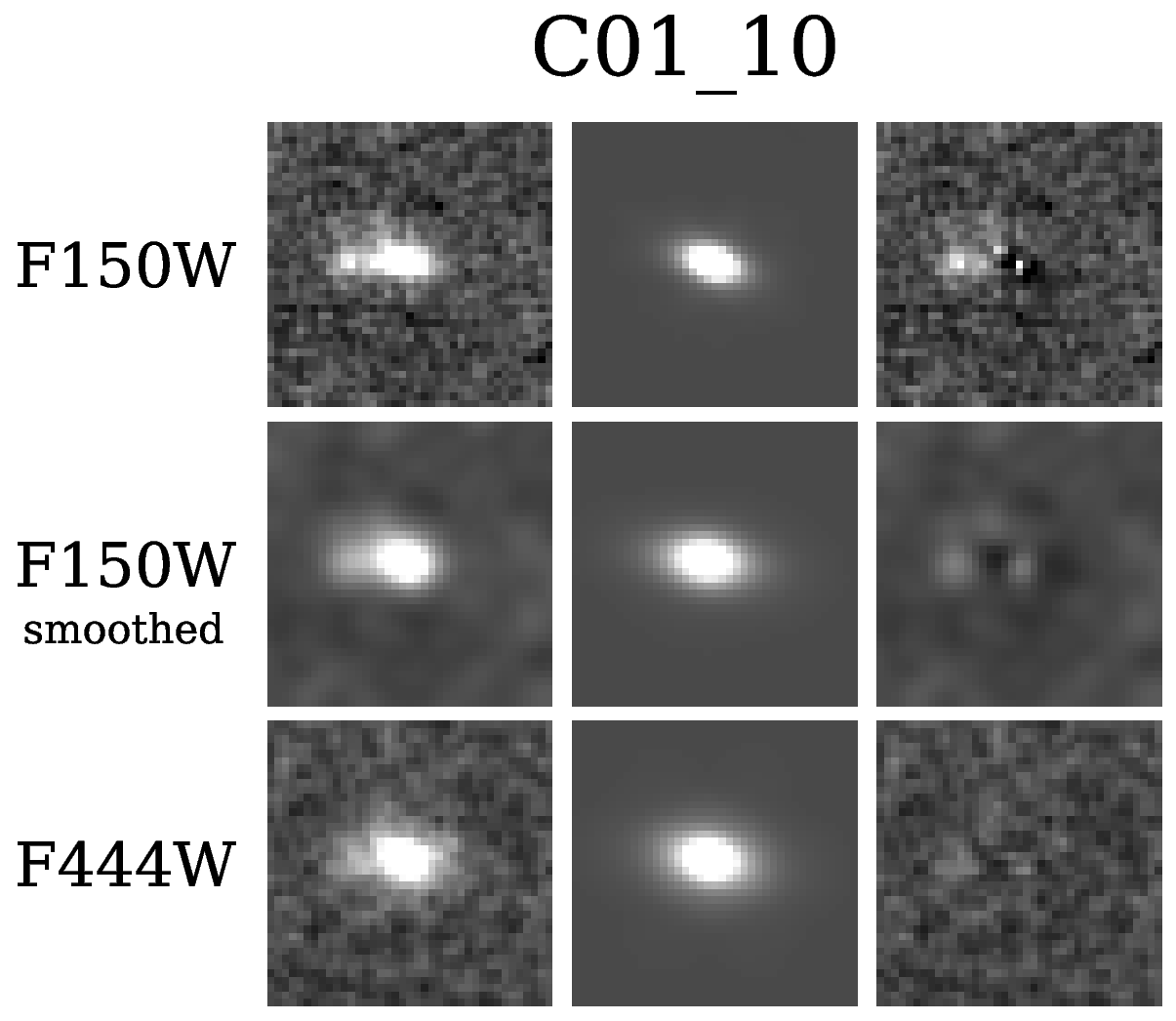}
   \includegraphics[height=0.14\textheight]{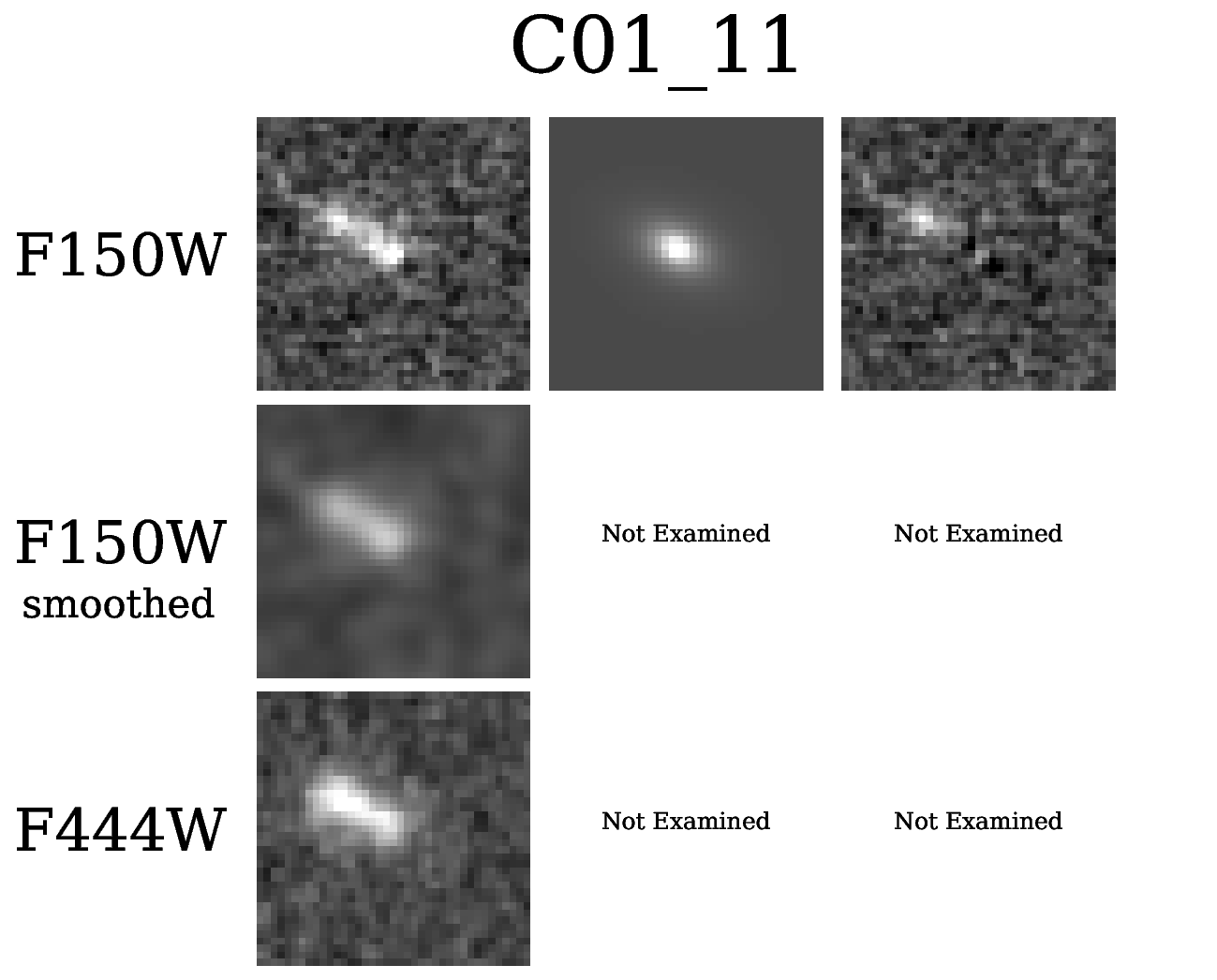}
   \includegraphics[height=0.14\textheight]{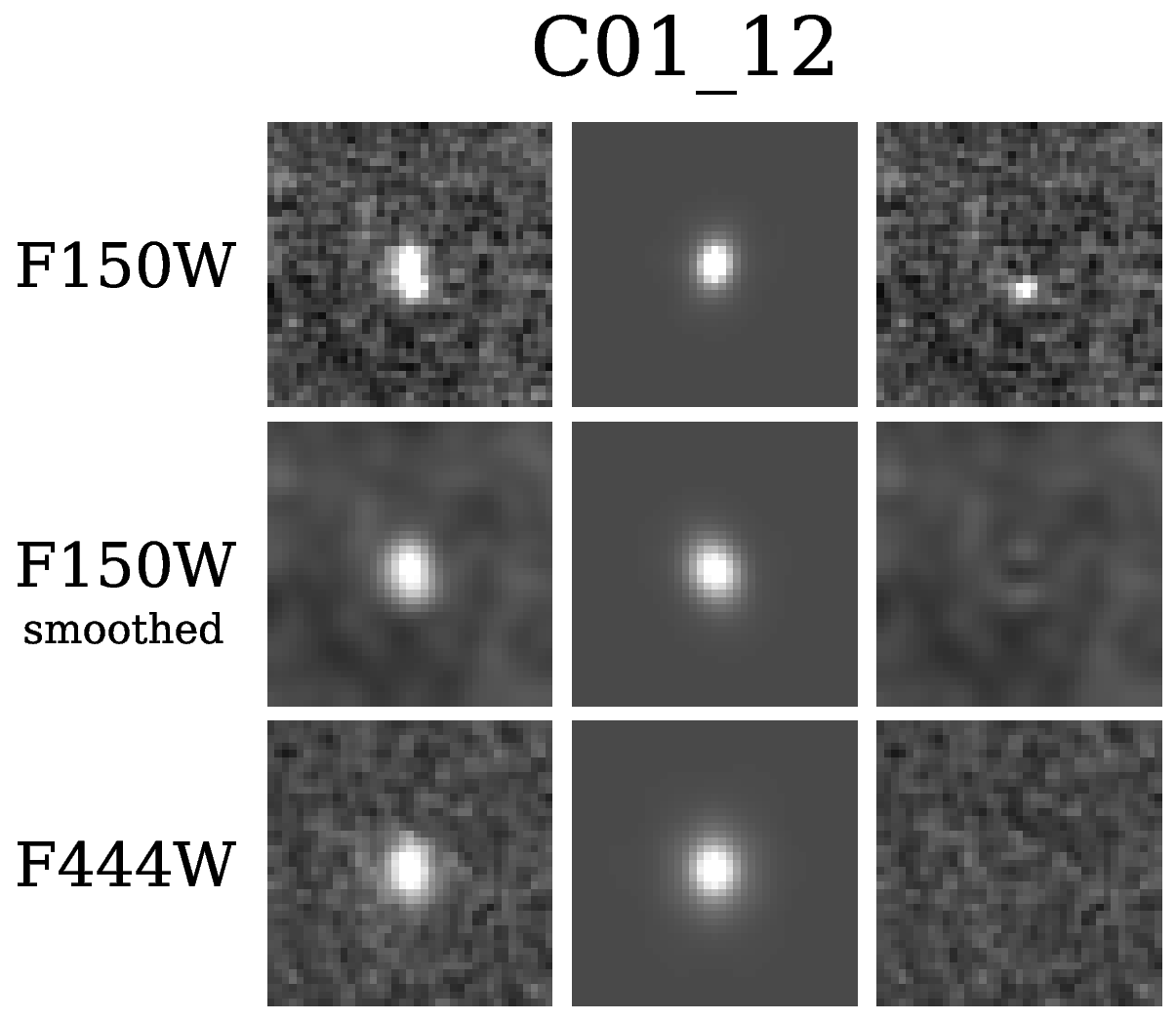}
   \includegraphics[height=0.14\textheight]{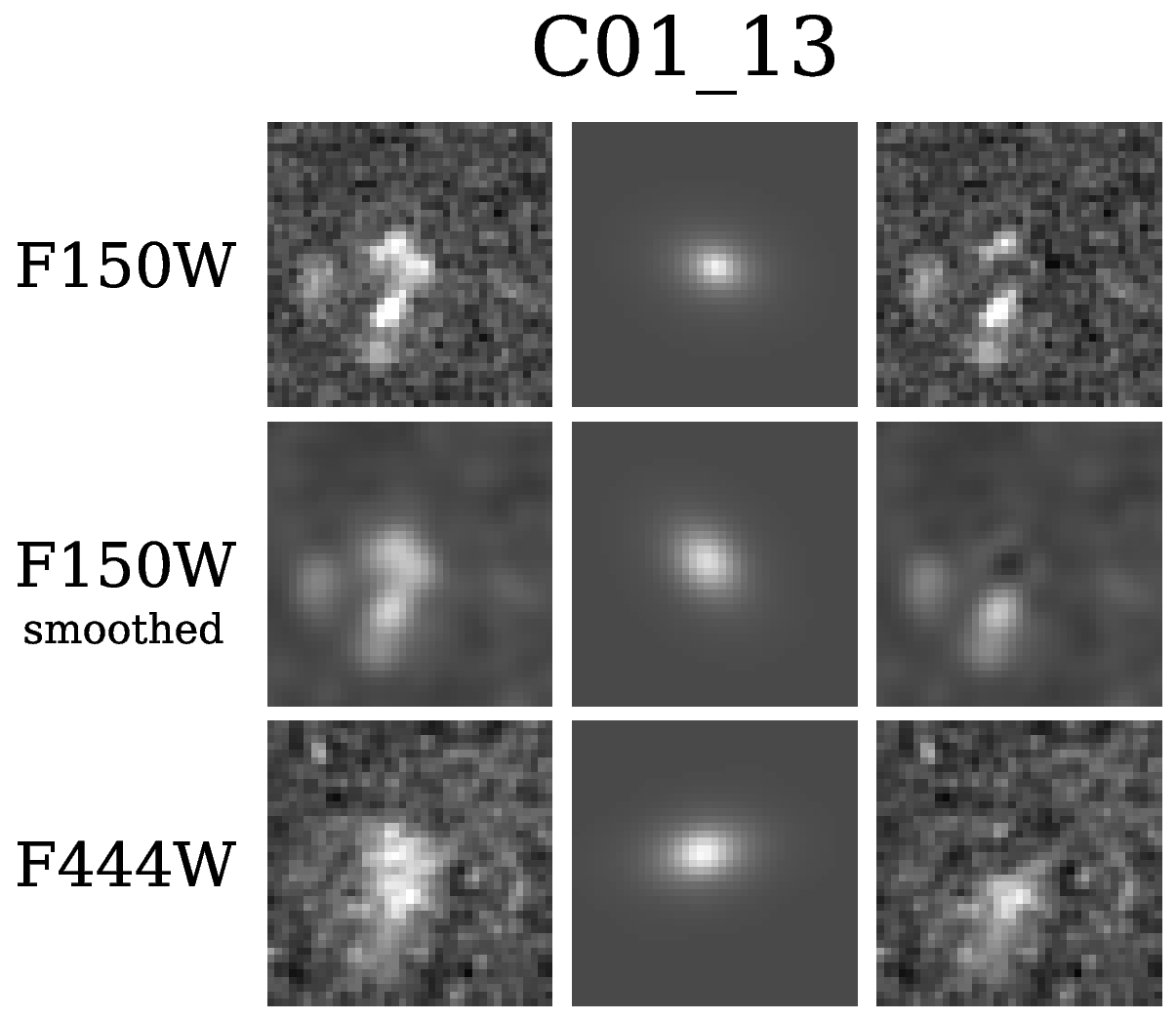}
   \includegraphics[height=0.14\textheight]{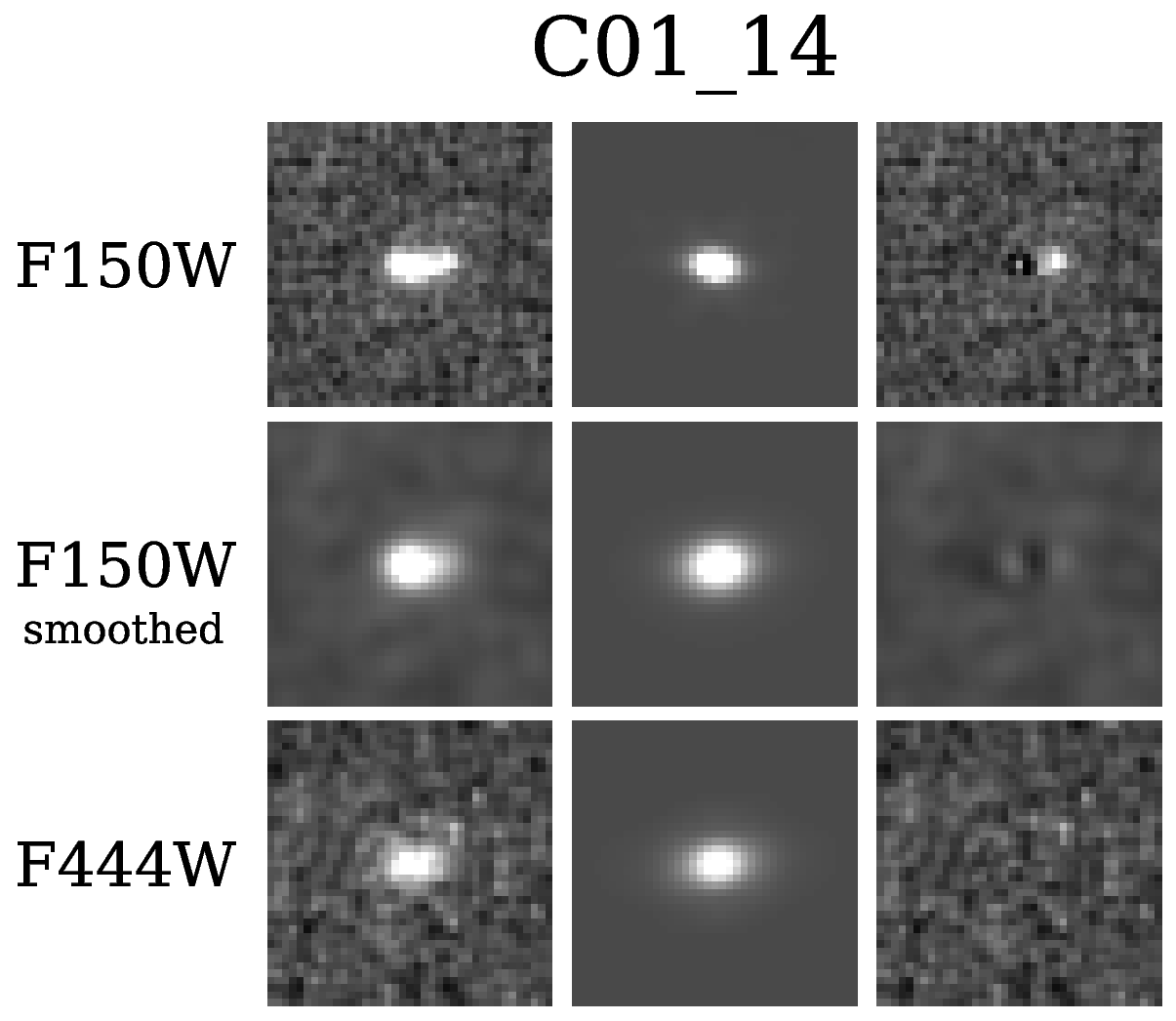}
   \includegraphics[height=0.14\textheight]{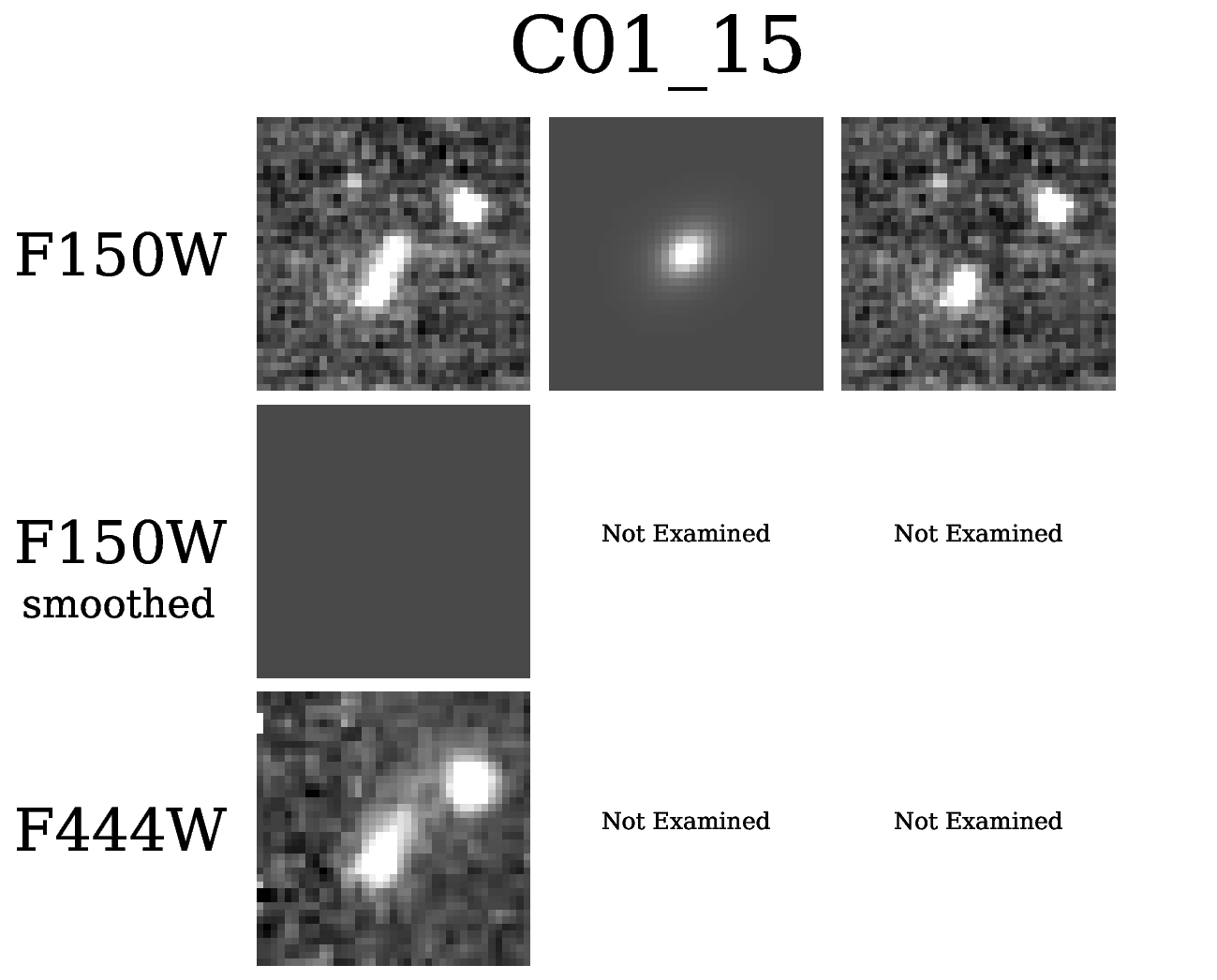}
   \includegraphics[height=0.14\textheight]{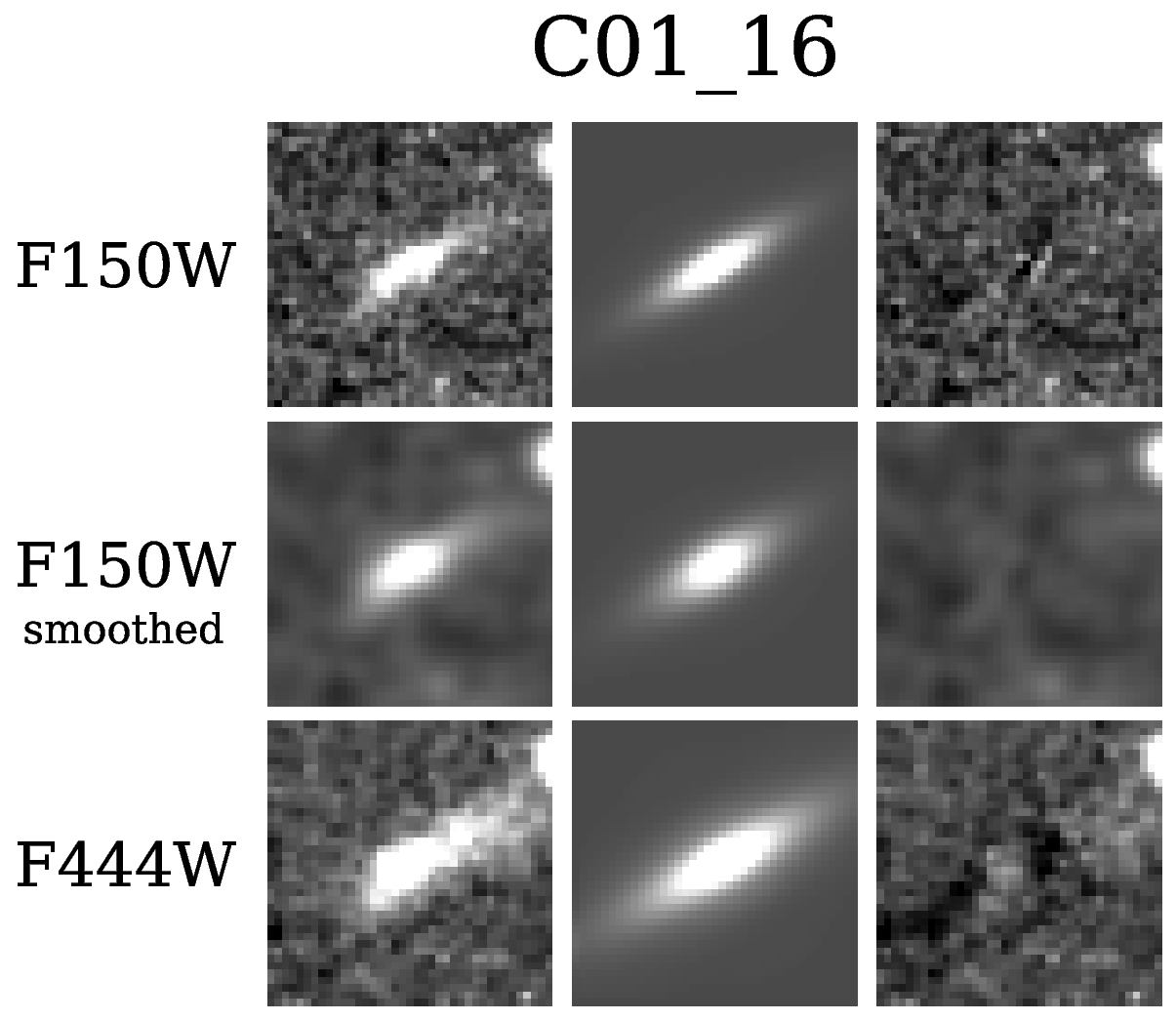}
   \includegraphics[height=0.14\textheight]{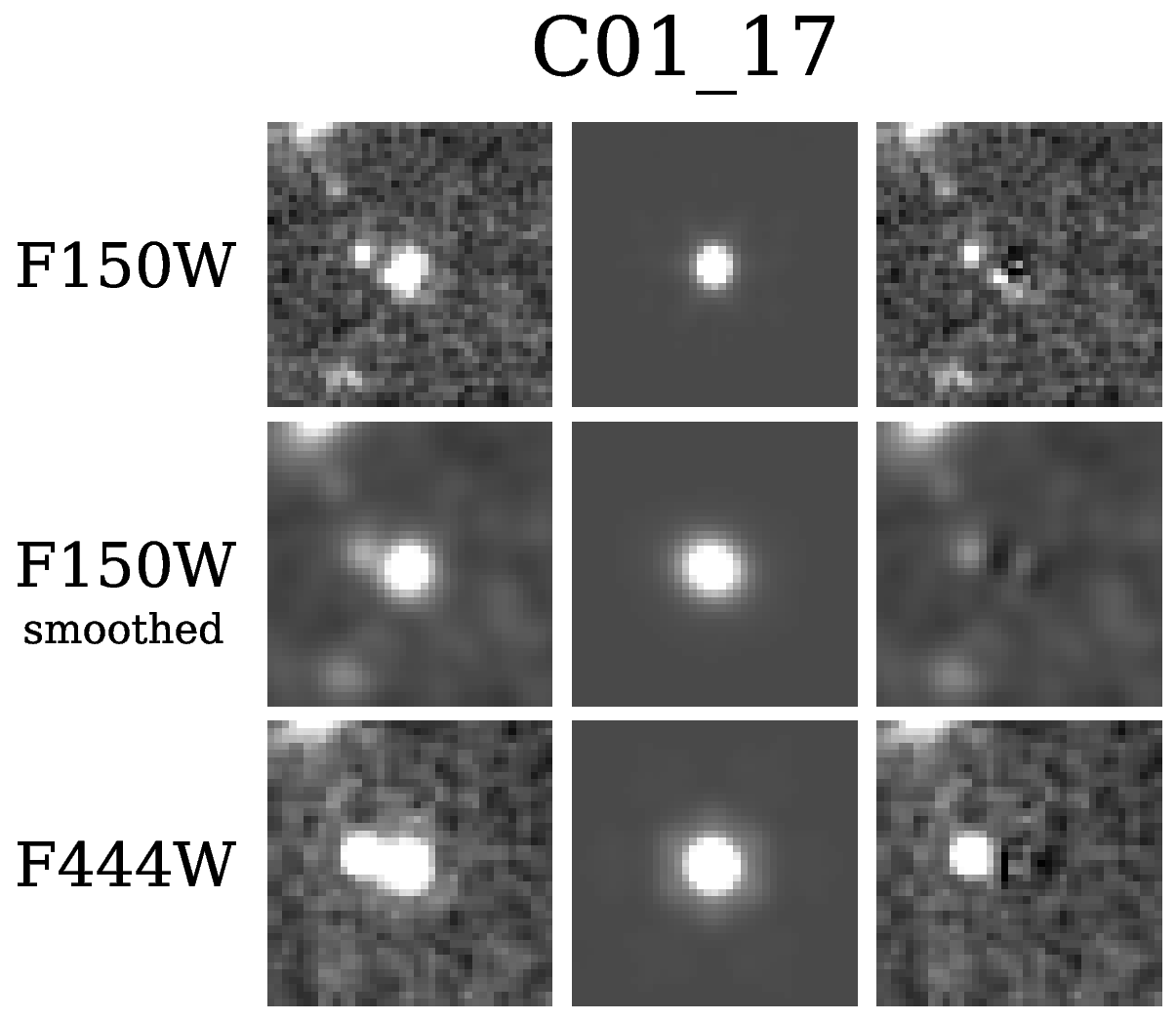}
   \includegraphics[height=0.14\textheight]{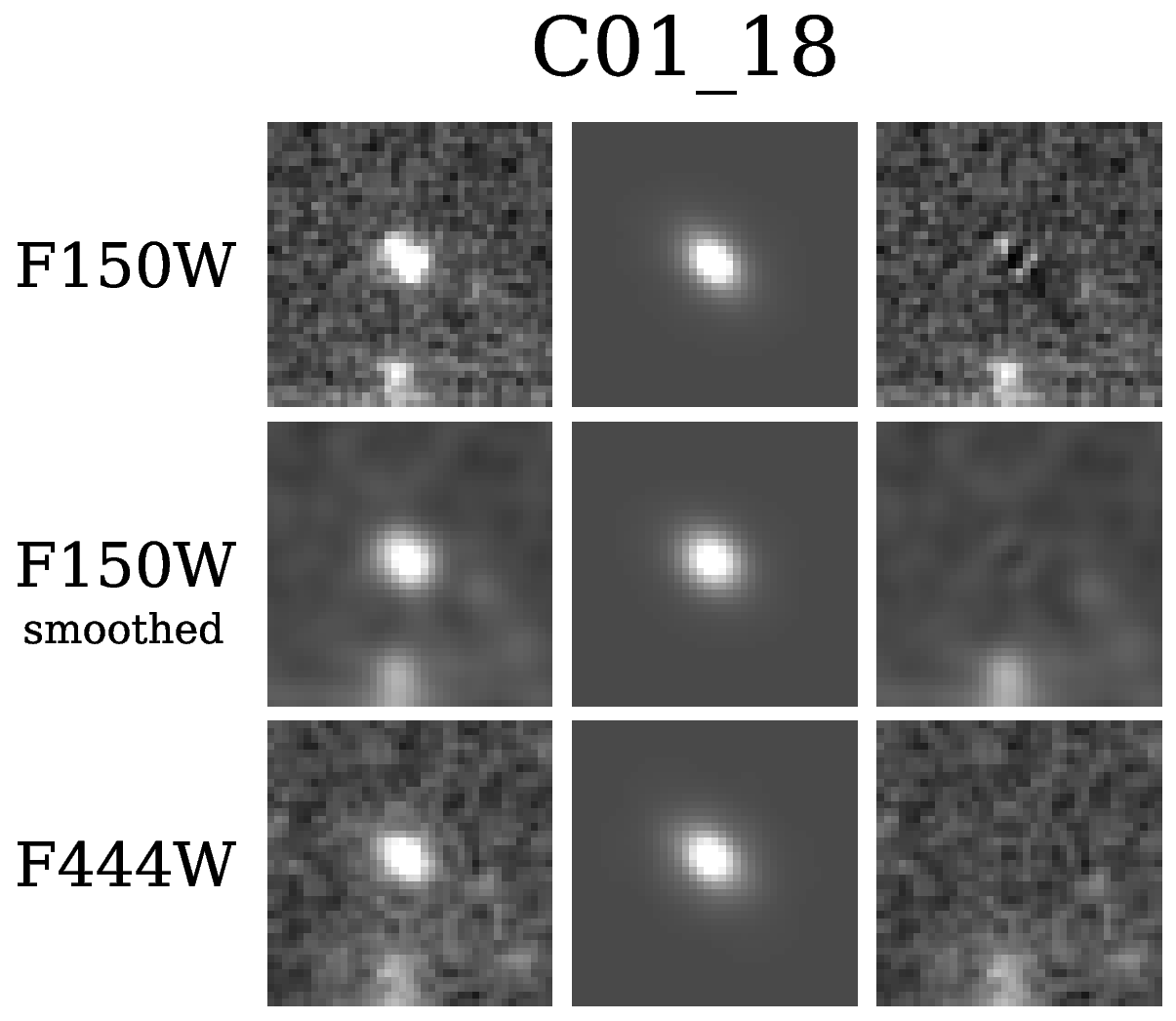}
   \includegraphics[height=0.14\textheight]{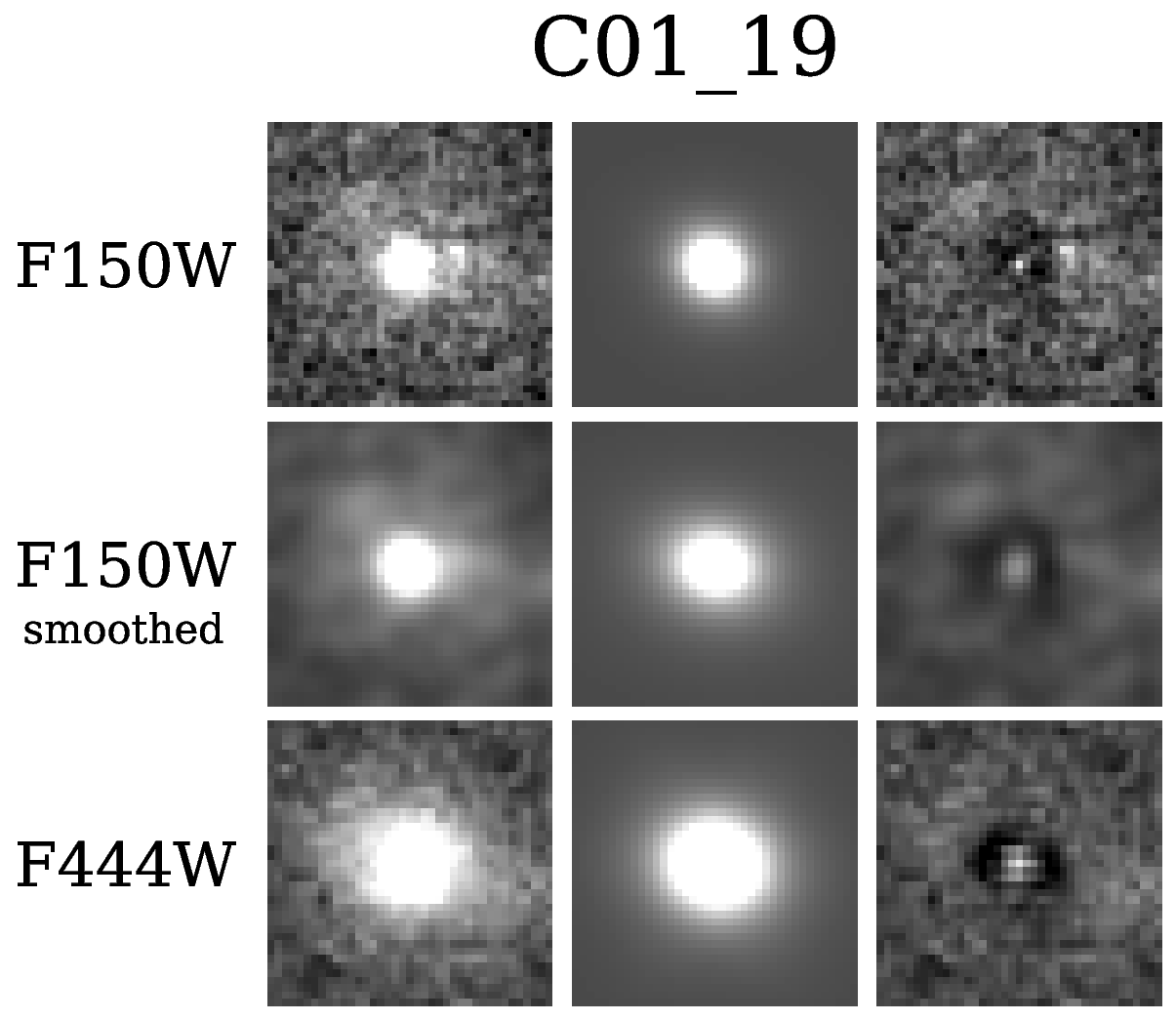}
   \includegraphics[height=0.14\textheight]{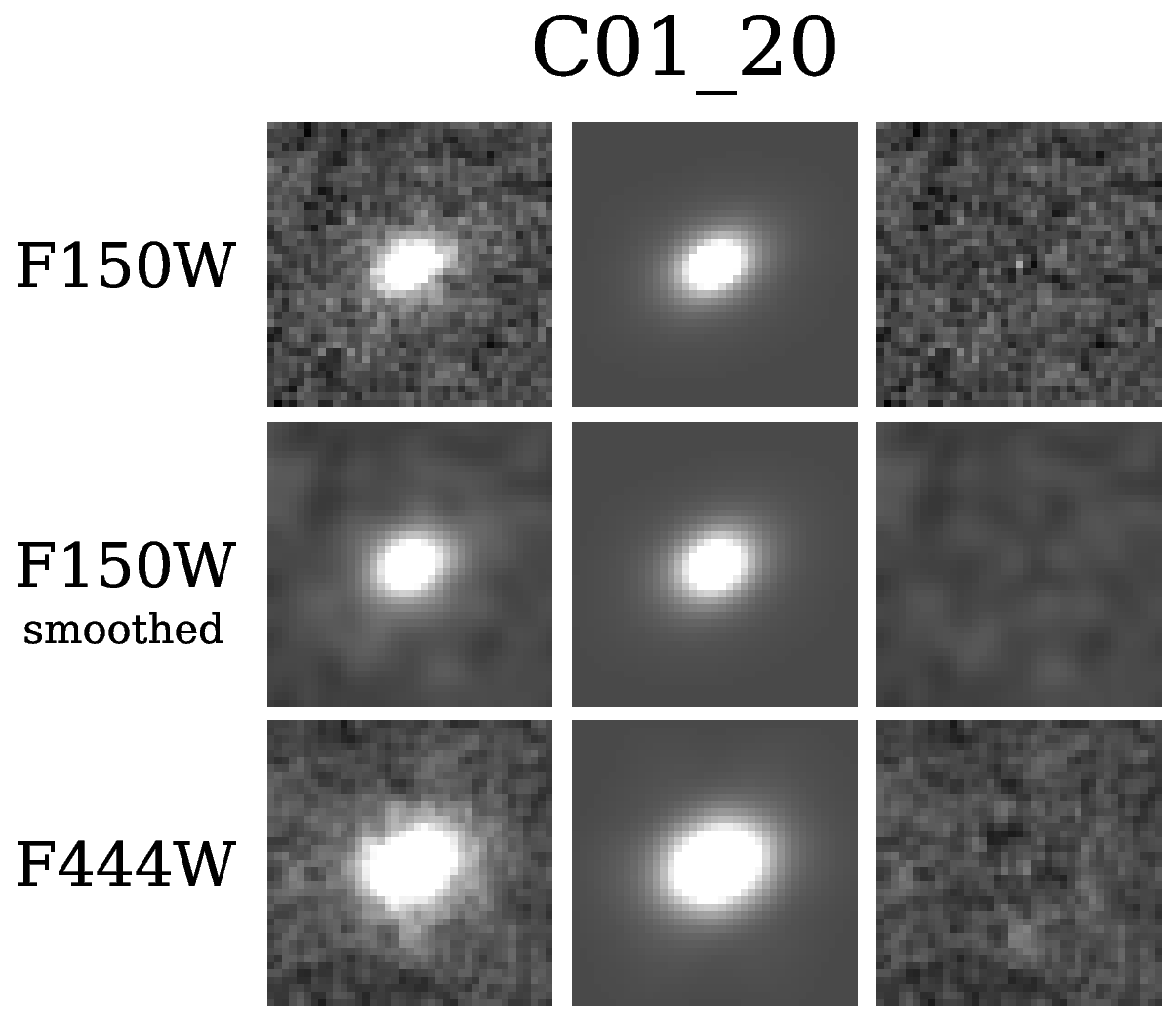}
   \includegraphics[height=0.14\textheight]{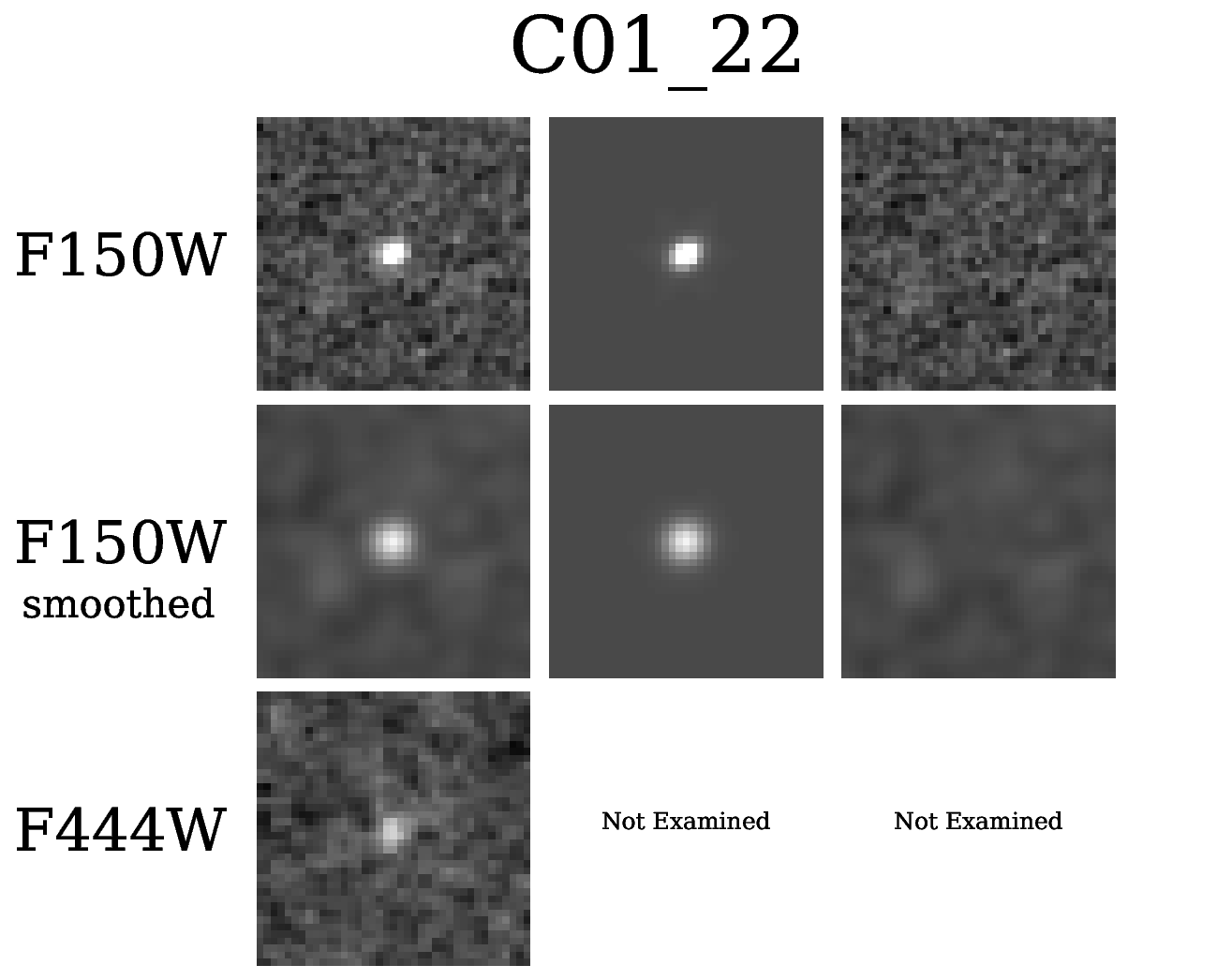}
   \includegraphics[height=0.14\textheight]{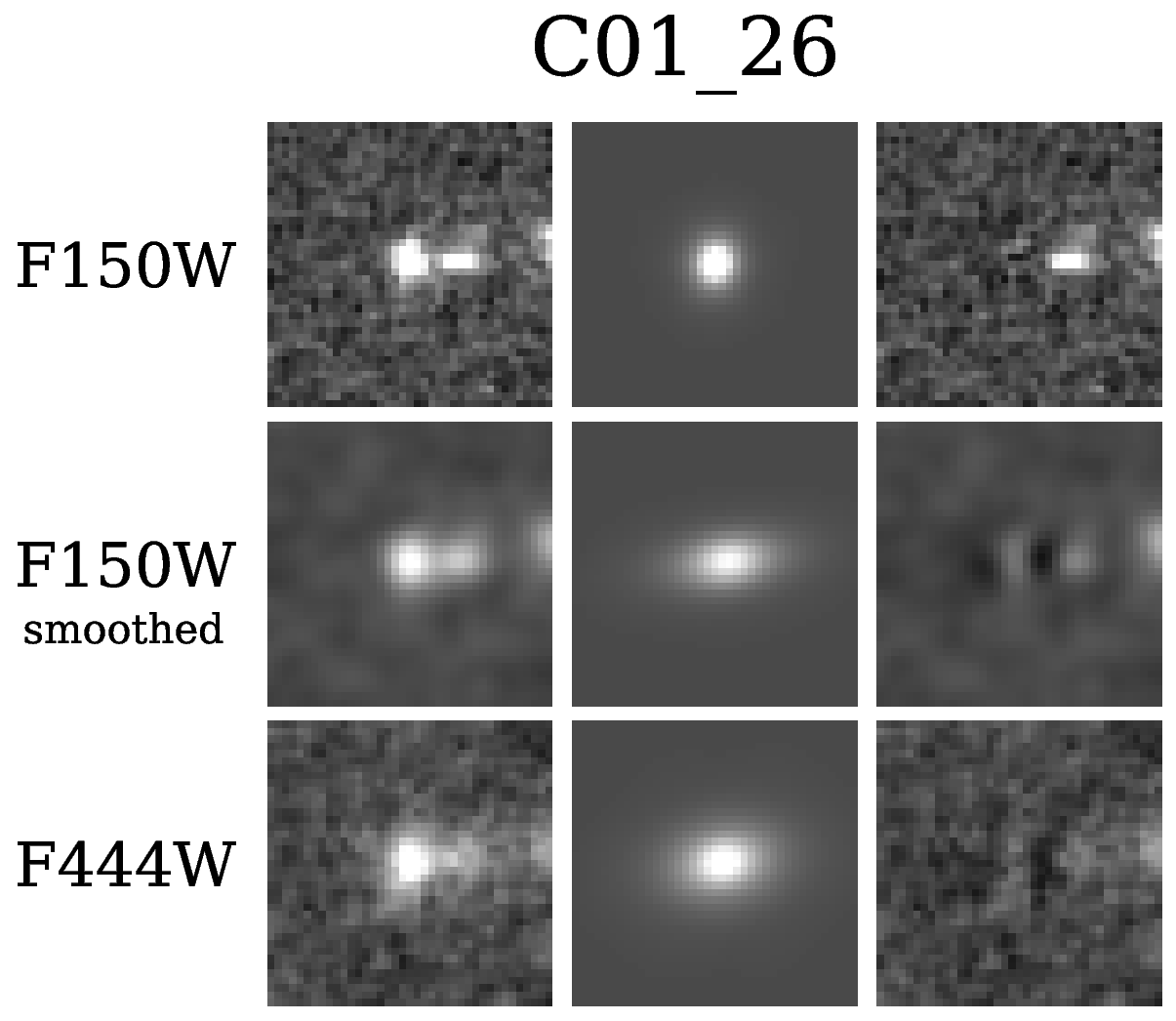}
   \includegraphics[height=0.14\textheight]{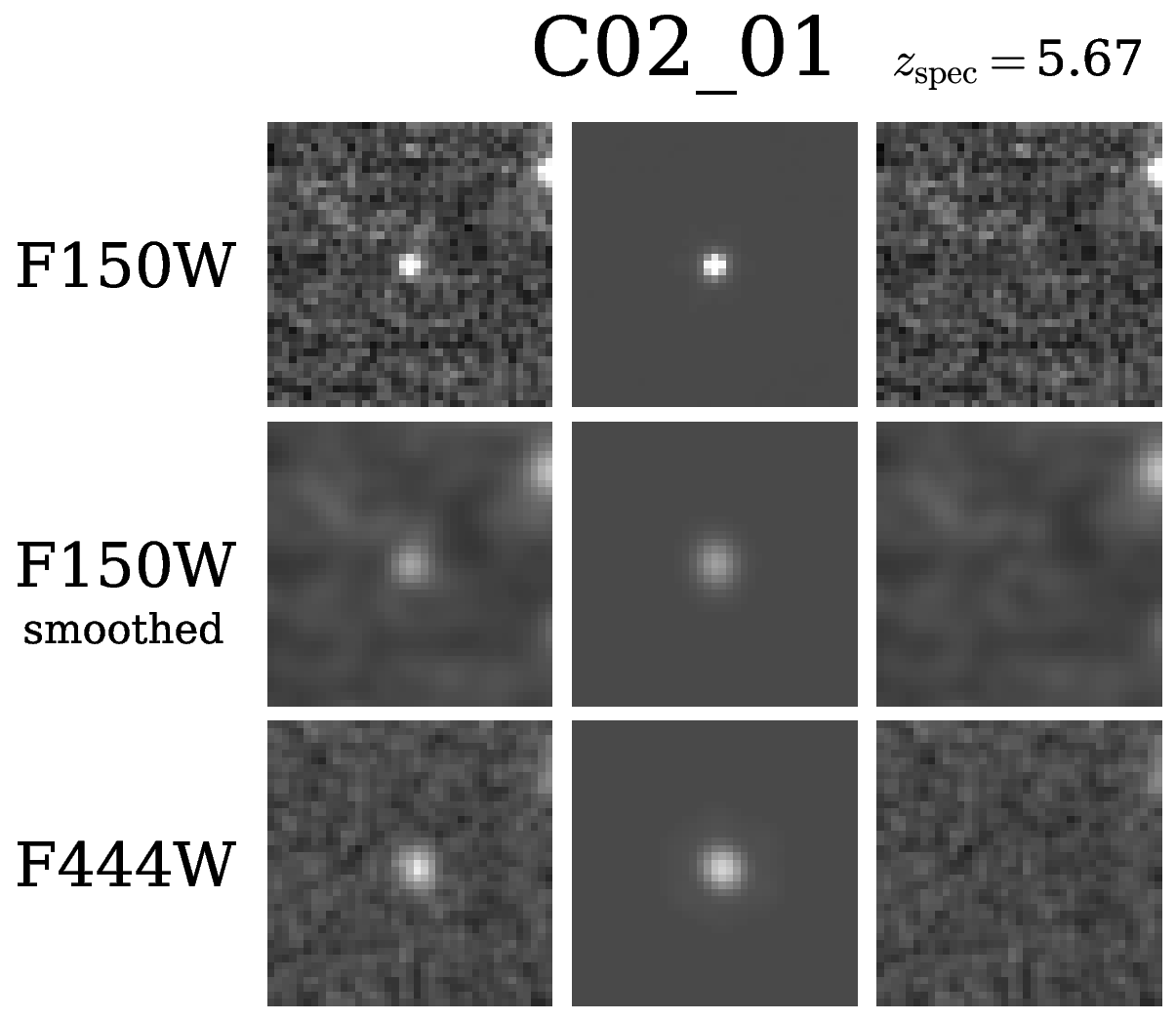}
   \includegraphics[height=0.14\textheight]{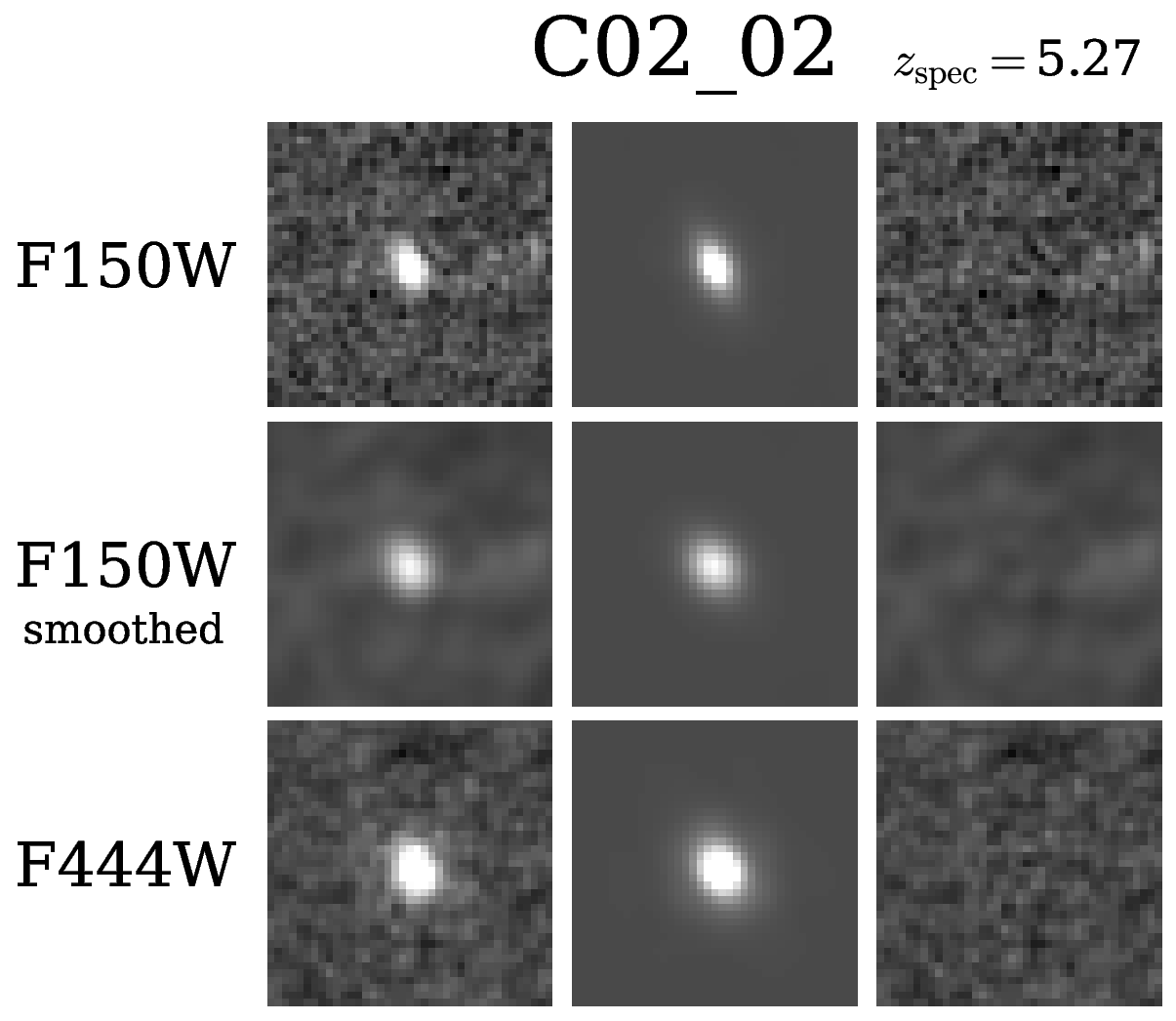}
   \includegraphics[height=0.14\textheight]{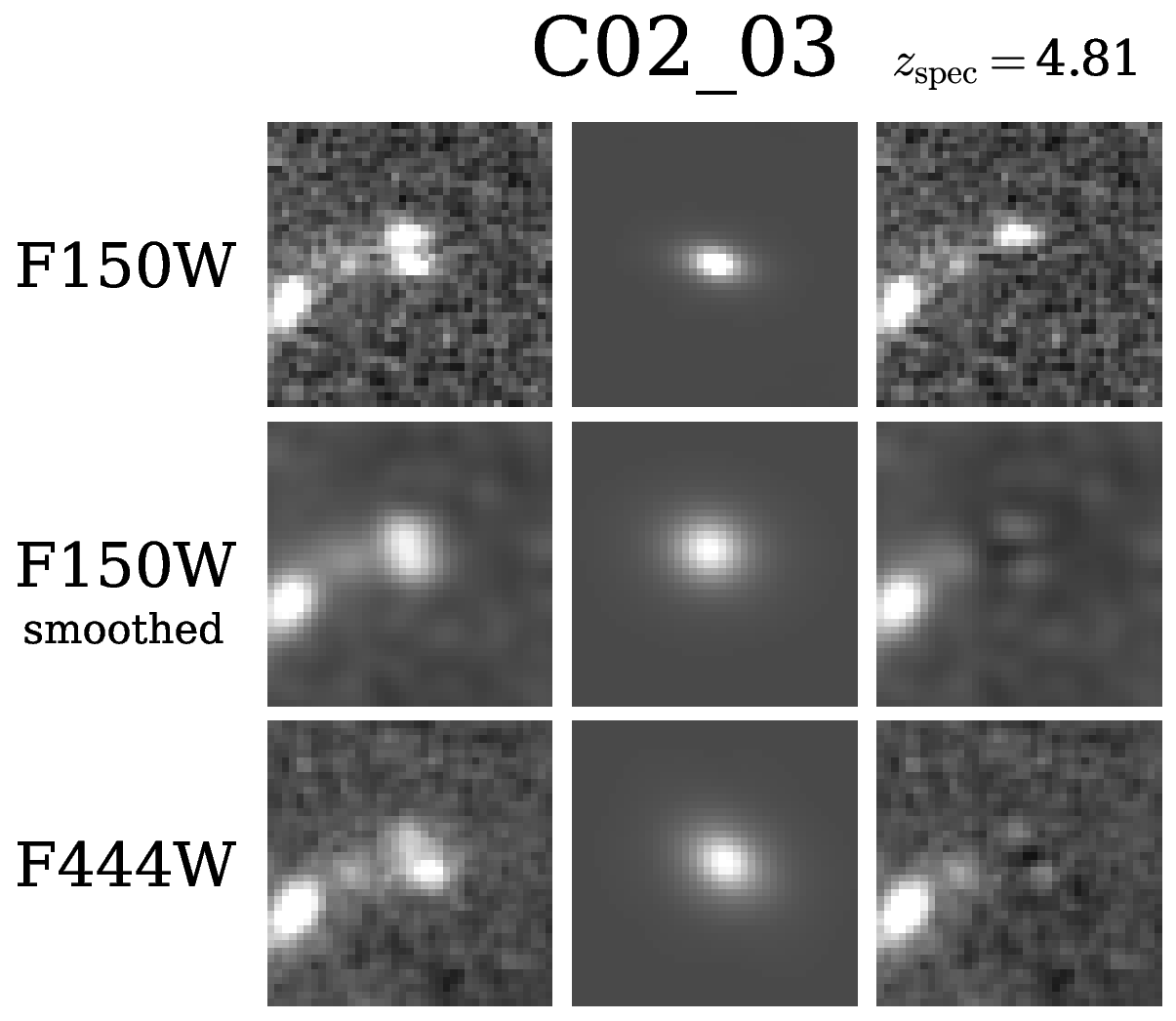}
   \includegraphics[height=0.14\textheight]{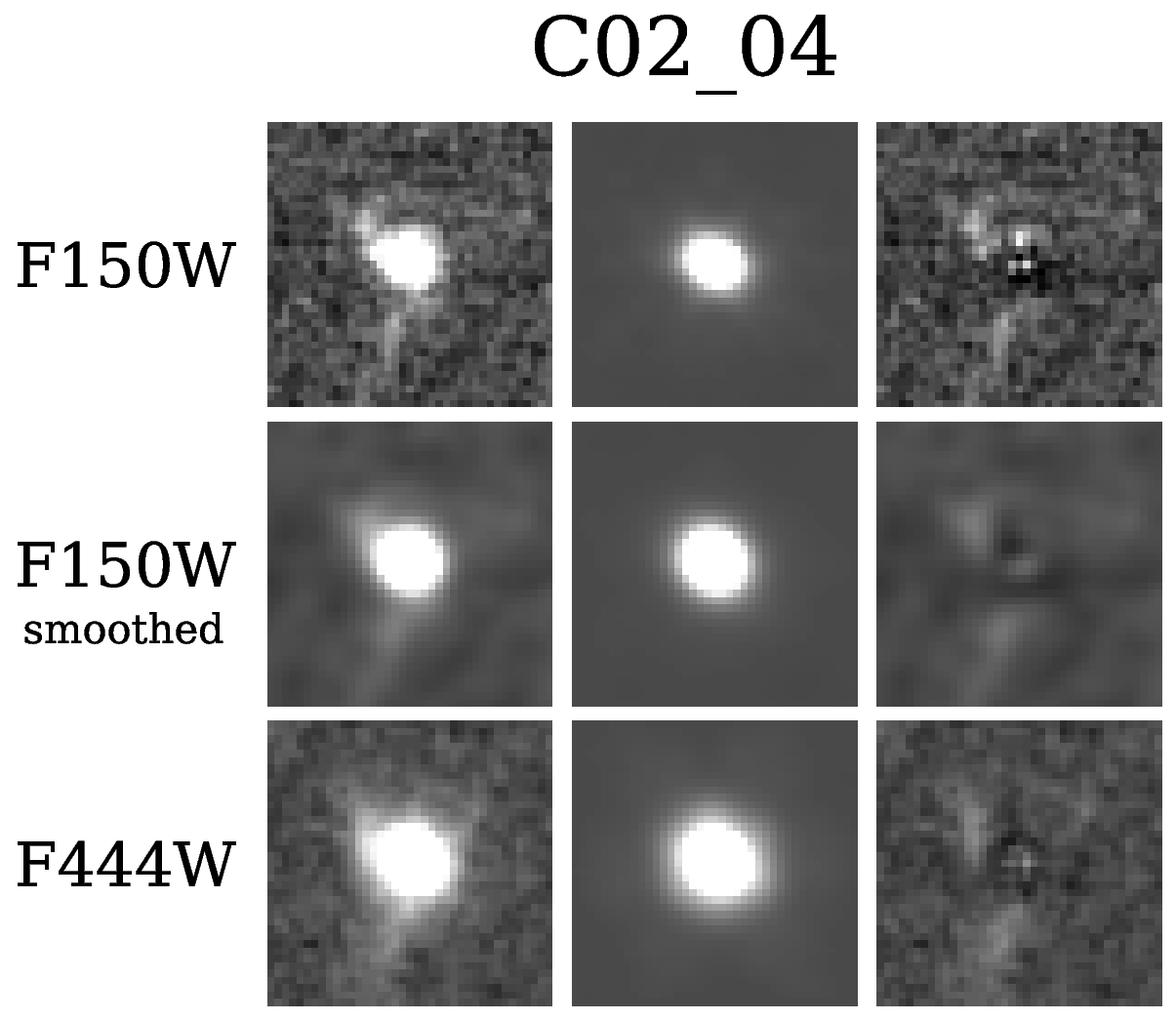}
\caption{
S\'ersic profile fitting results for bright sources in our sample. 
For each source, 
from top to bottom, 
the fitting results for F150W, PSF-matched F150W, and F444W are presented. 
From left to right, 
the $1 \farcs 5 \times 1 \farcs 5$ cutouts of the original image, 
the best-fit S\'ersic model profile images, 
and 
the residual images that are produced by subtracting the best-fit images from the original ones  
are shown. 
}
\label{fig:SB_fitting_results}
\end{center}
\end{figure*}

\addtocounter{figure}{-1}
\begin{figure*}
\begin{center}
   \includegraphics[height=0.14\textheight]{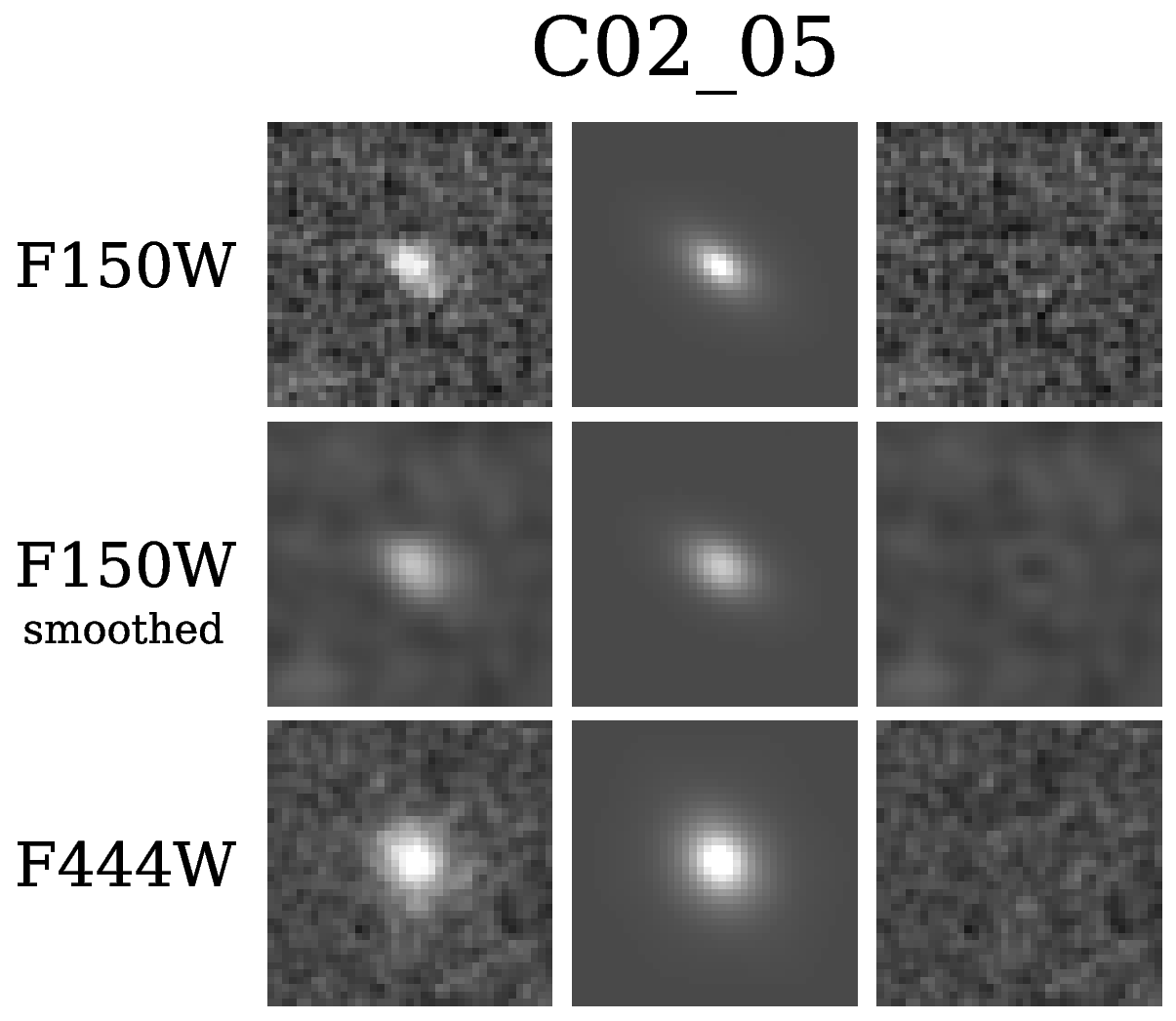}
   \includegraphics[height=0.14\textheight]{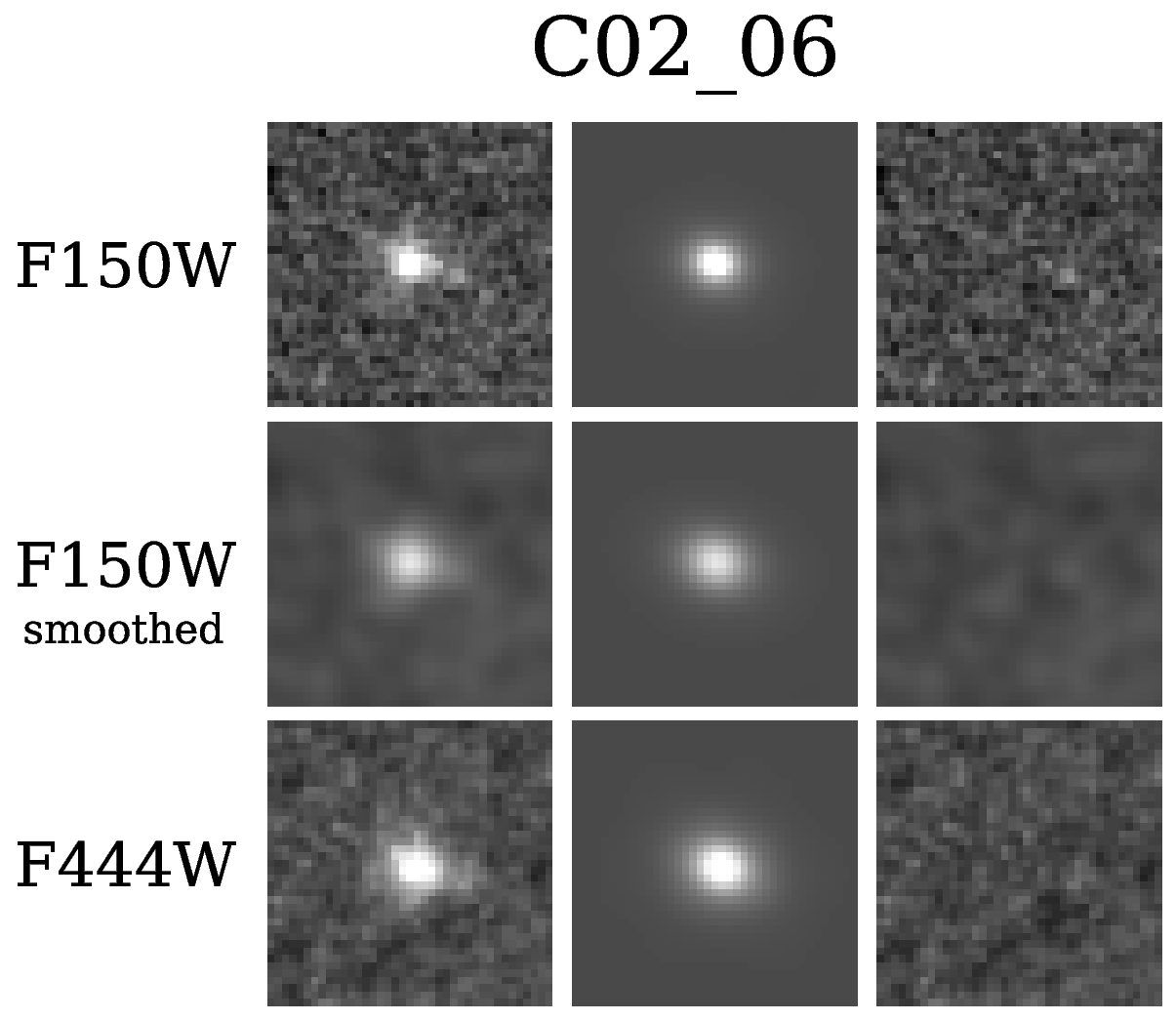}
   \includegraphics[height=0.14\textheight]{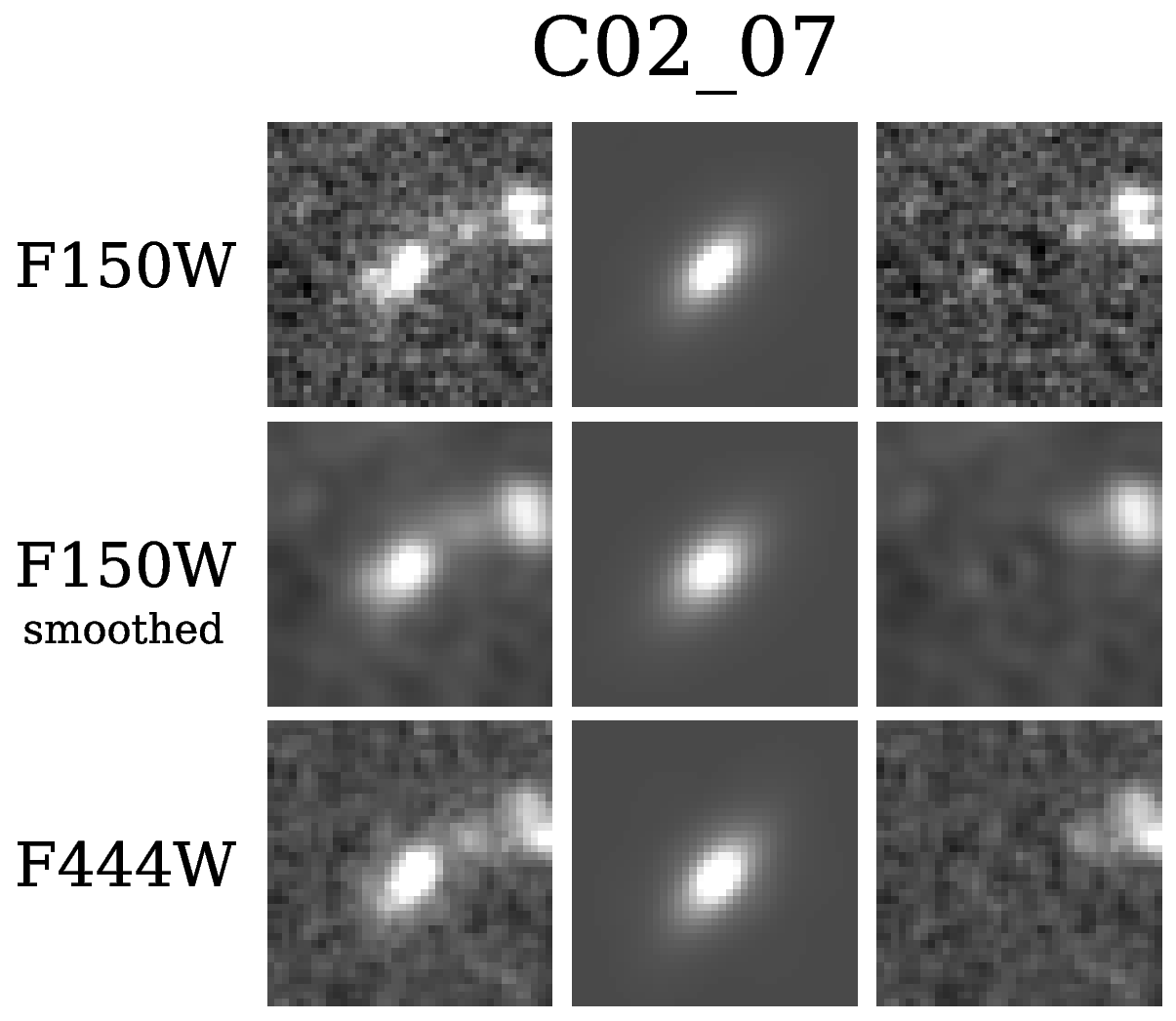}
   \includegraphics[height=0.14\textheight]{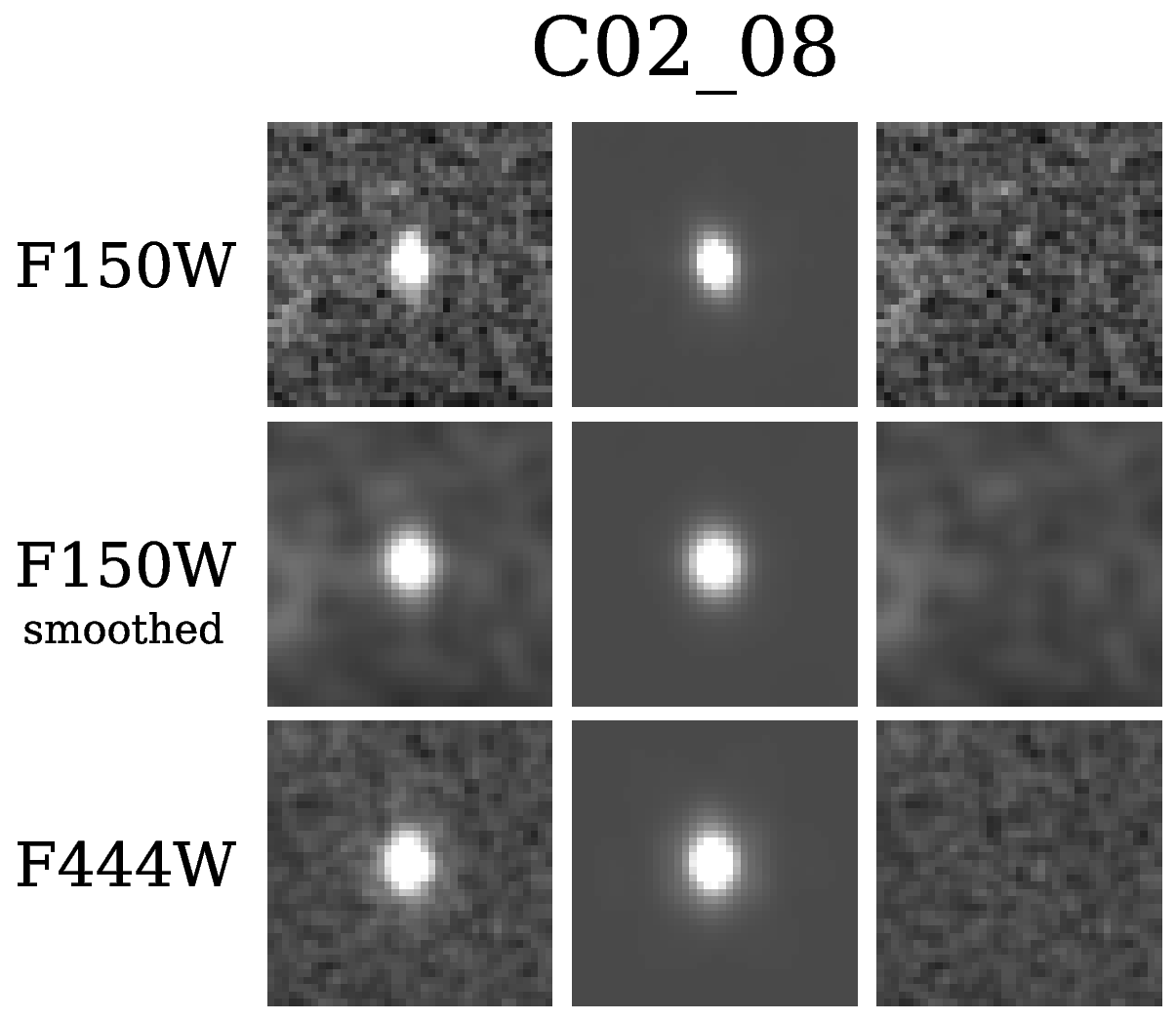}
   \includegraphics[height=0.14\textheight]{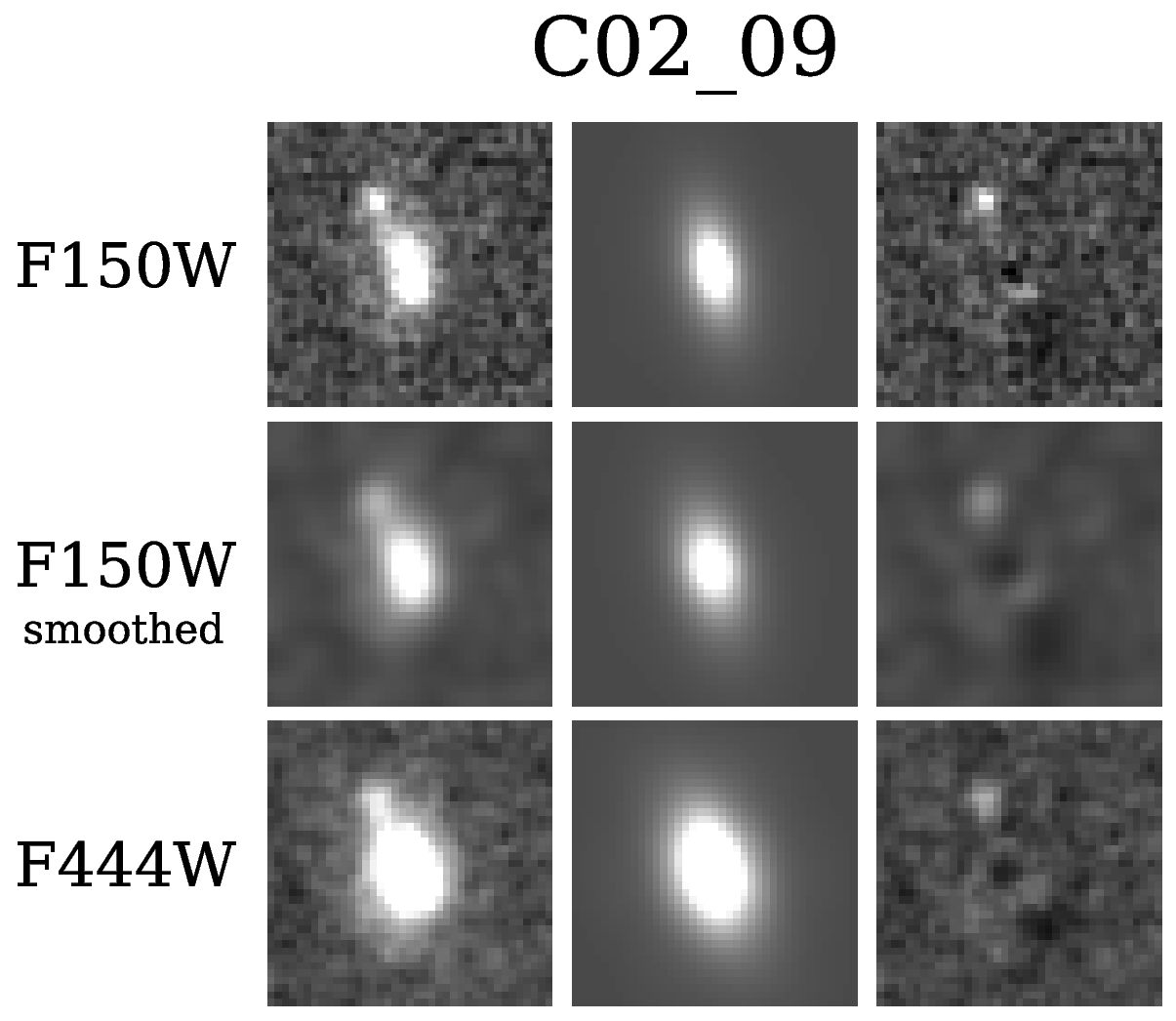}
   \includegraphics[height=0.14\textheight]{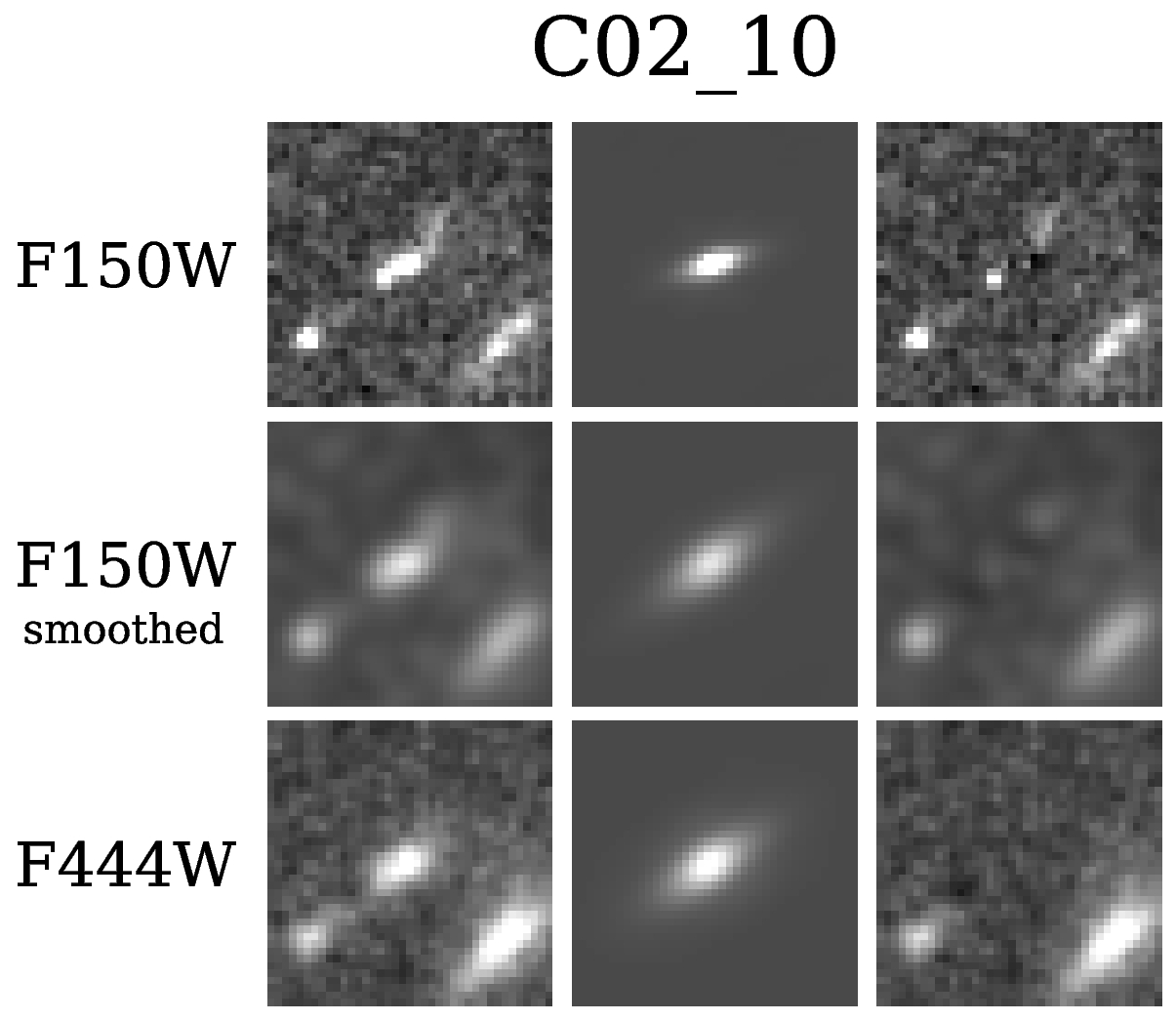}
   \includegraphics[height=0.14\textheight]{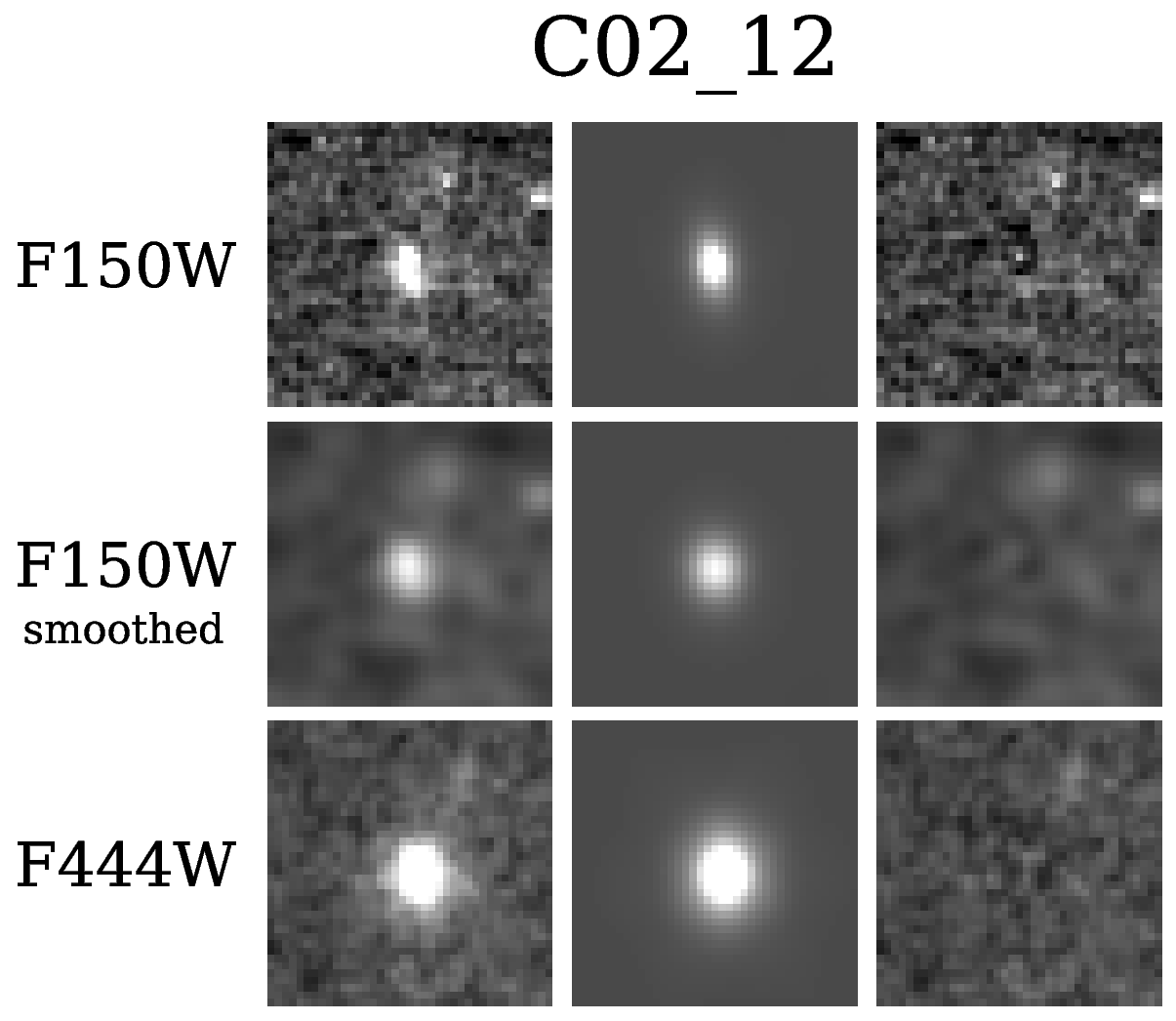}
   \includegraphics[height=0.14\textheight]{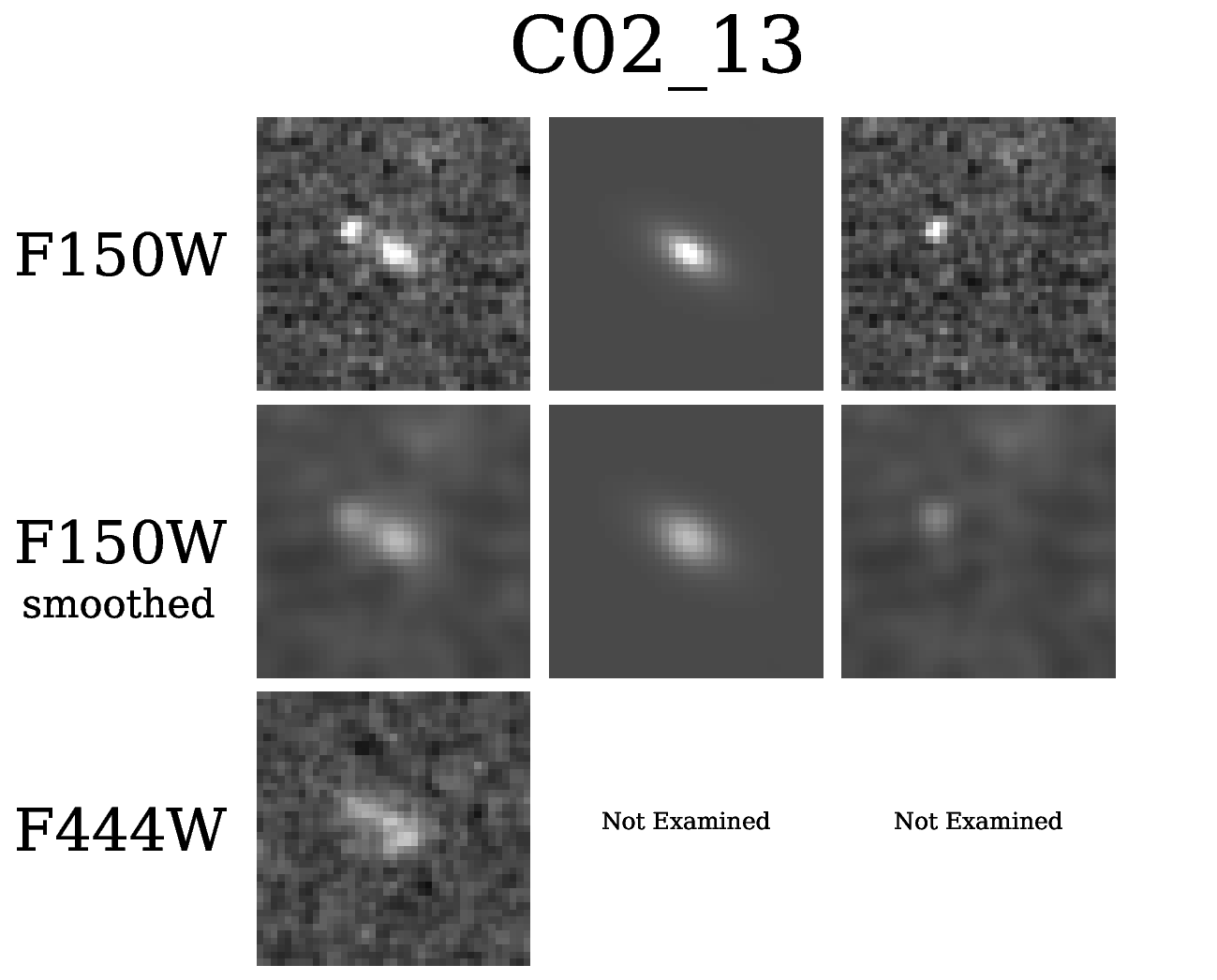}
   \includegraphics[height=0.14\textheight]{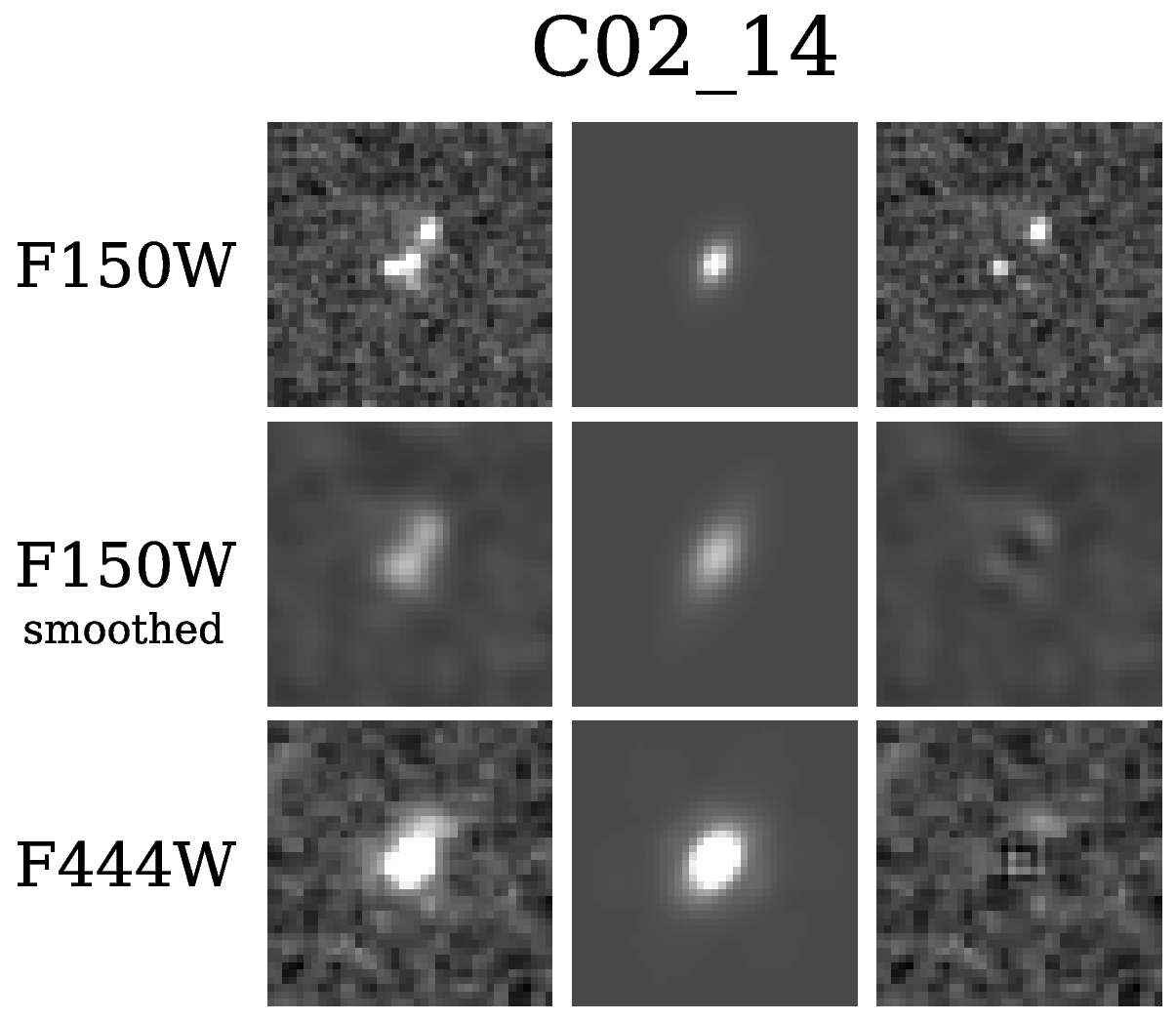}
   \includegraphics[height=0.14\textheight]{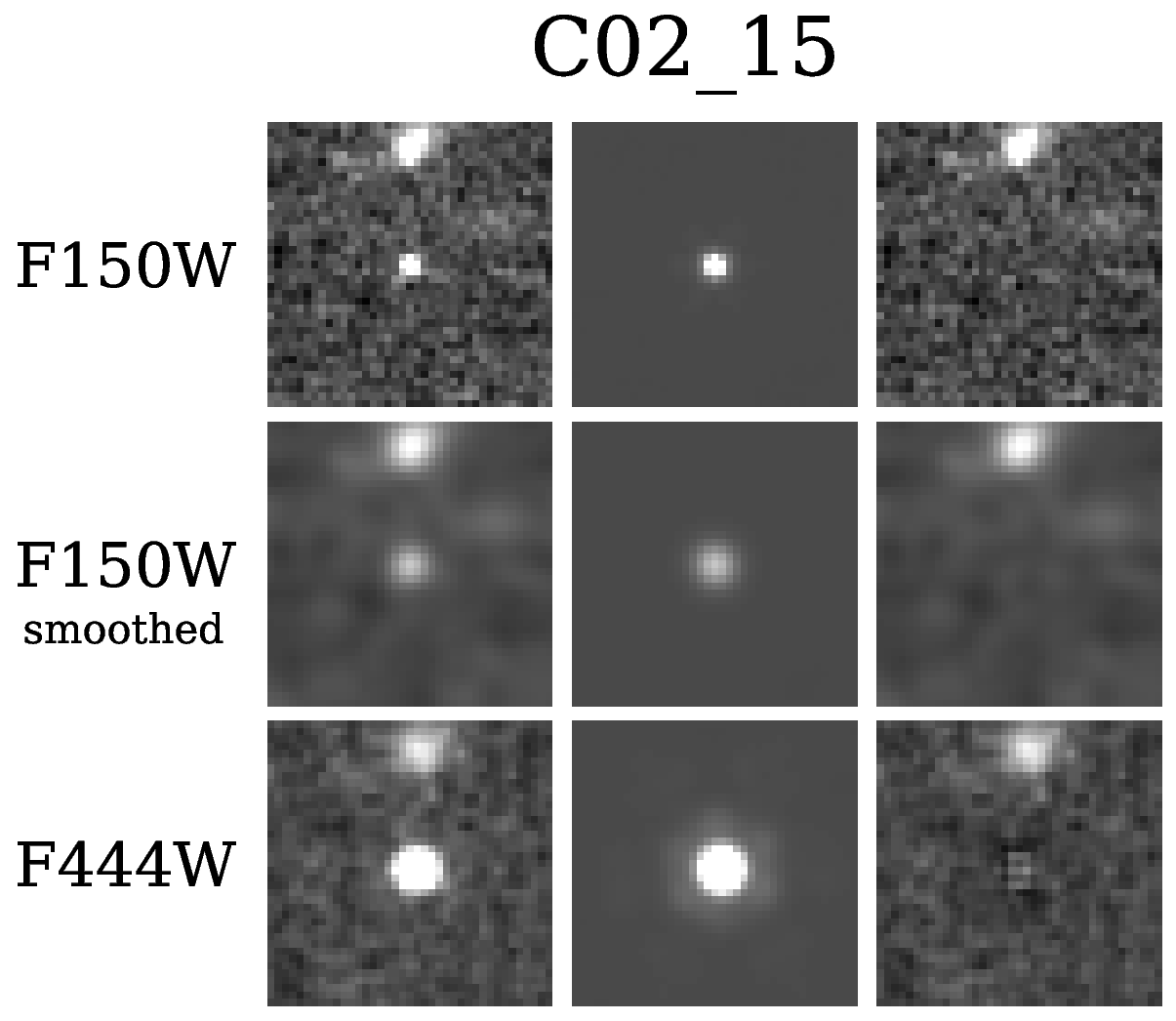}
   \includegraphics[height=0.14\textheight]{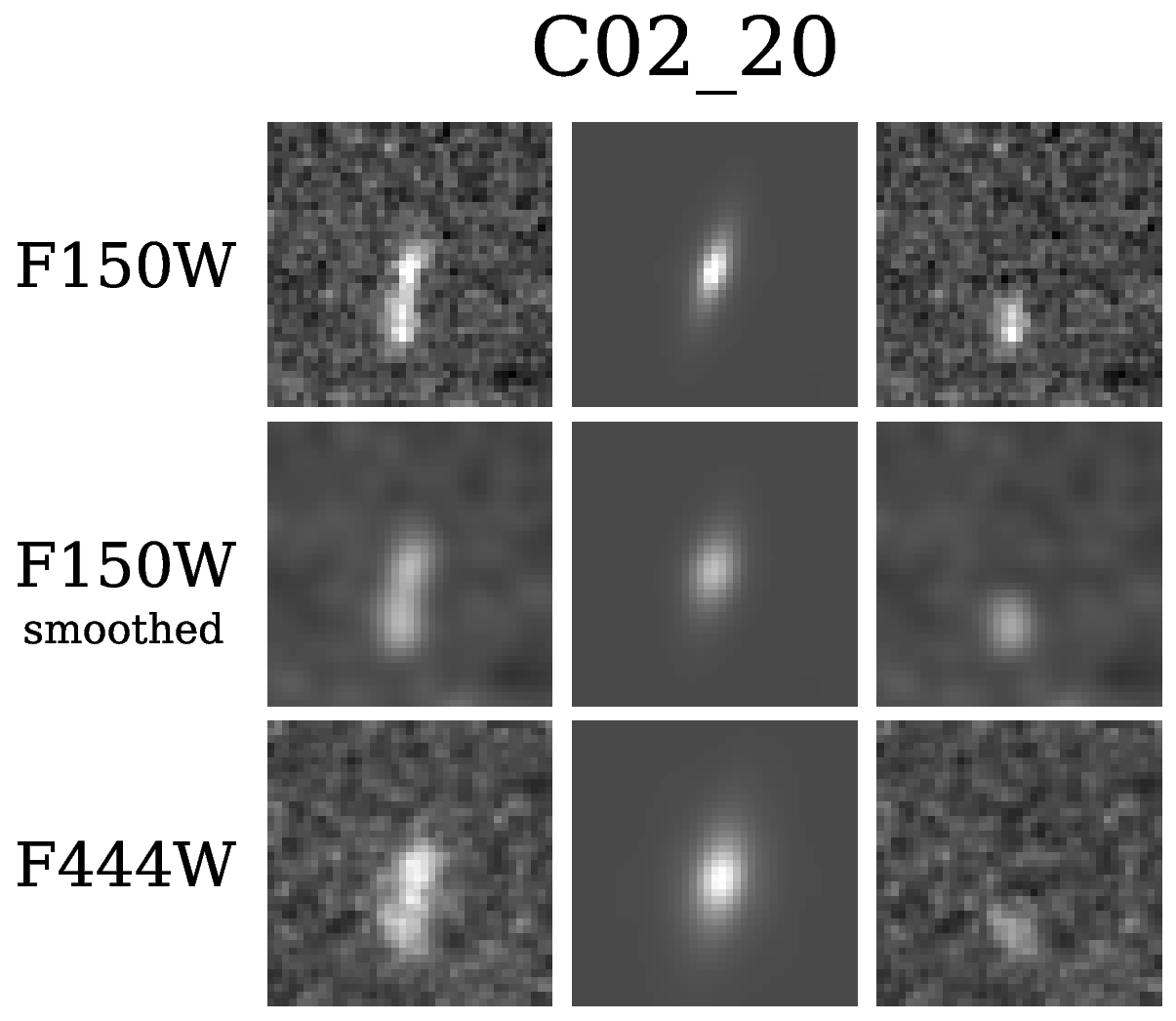}
   \includegraphics[height=0.14\textheight]{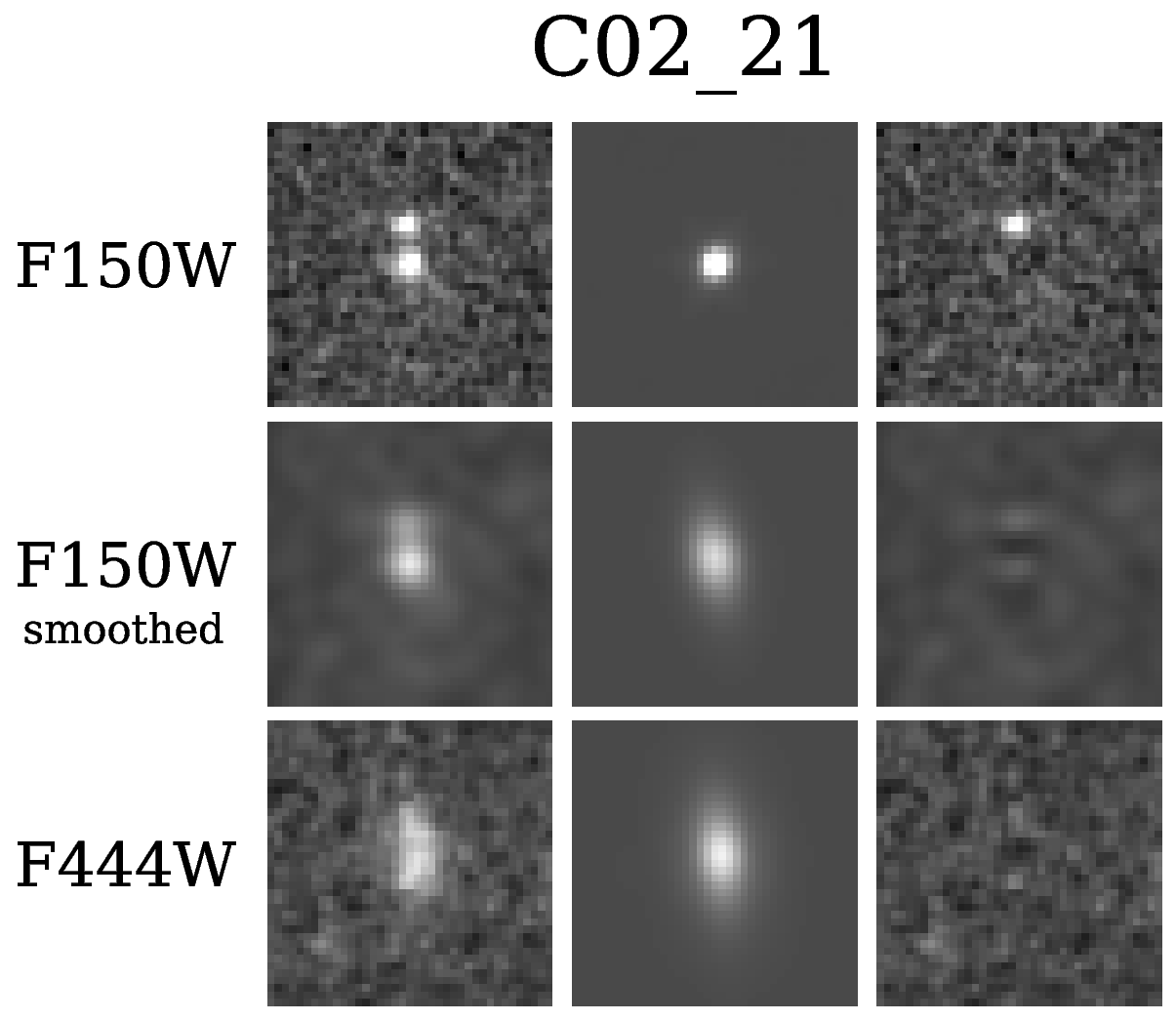}
   \includegraphics[height=0.14\textheight]{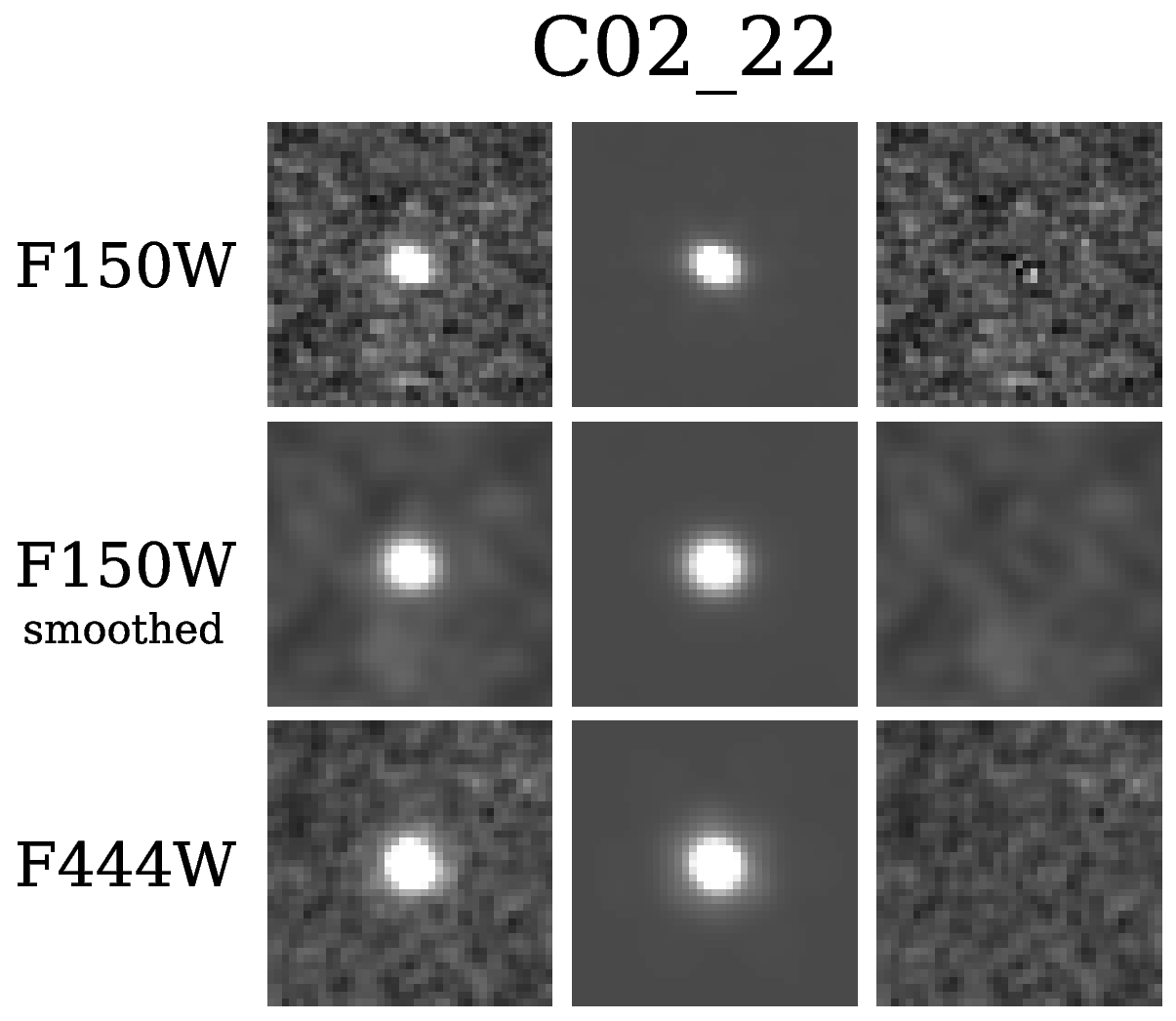}
   \includegraphics[height=0.14\textheight]{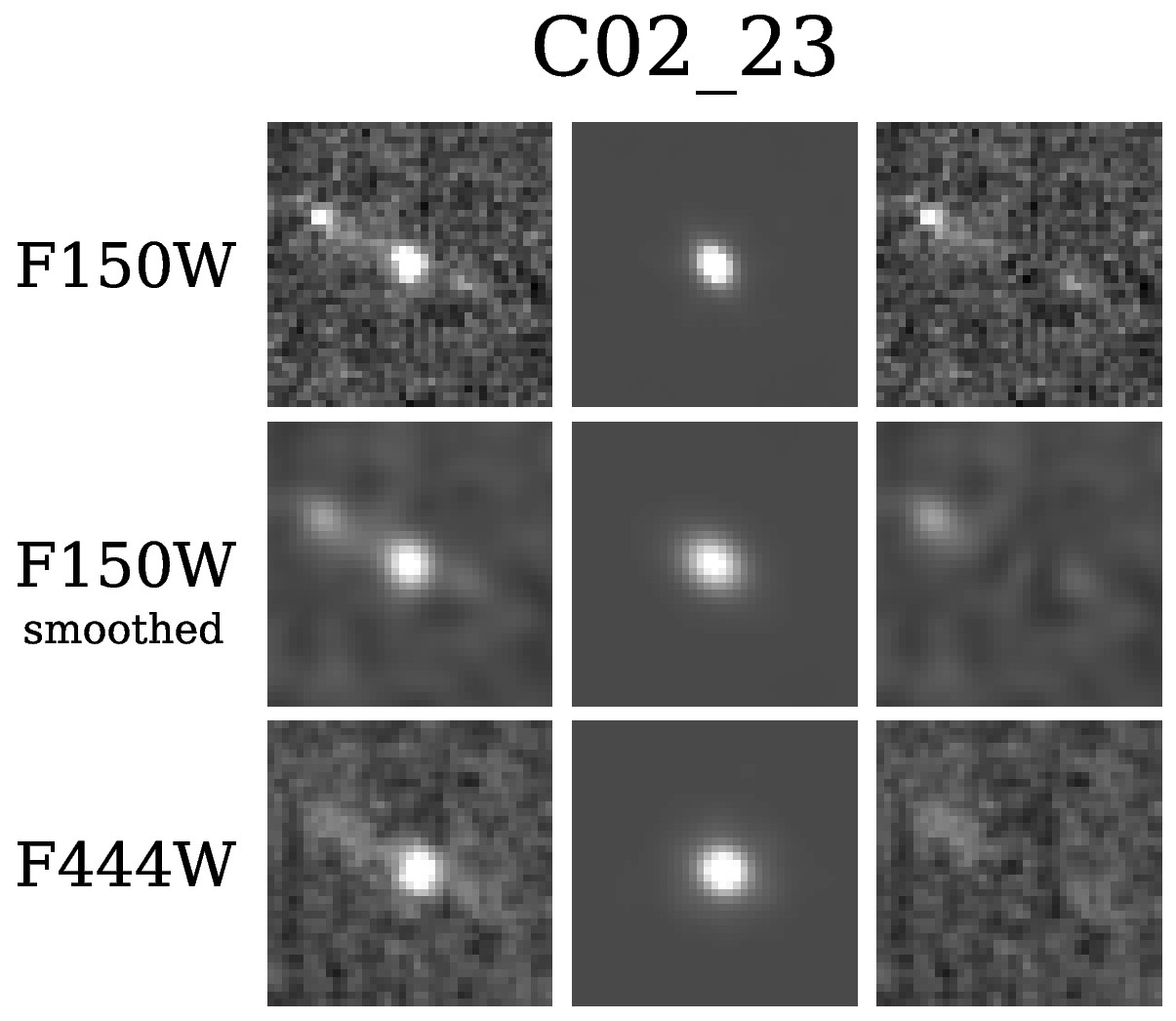}
   \includegraphics[height=0.14\textheight]{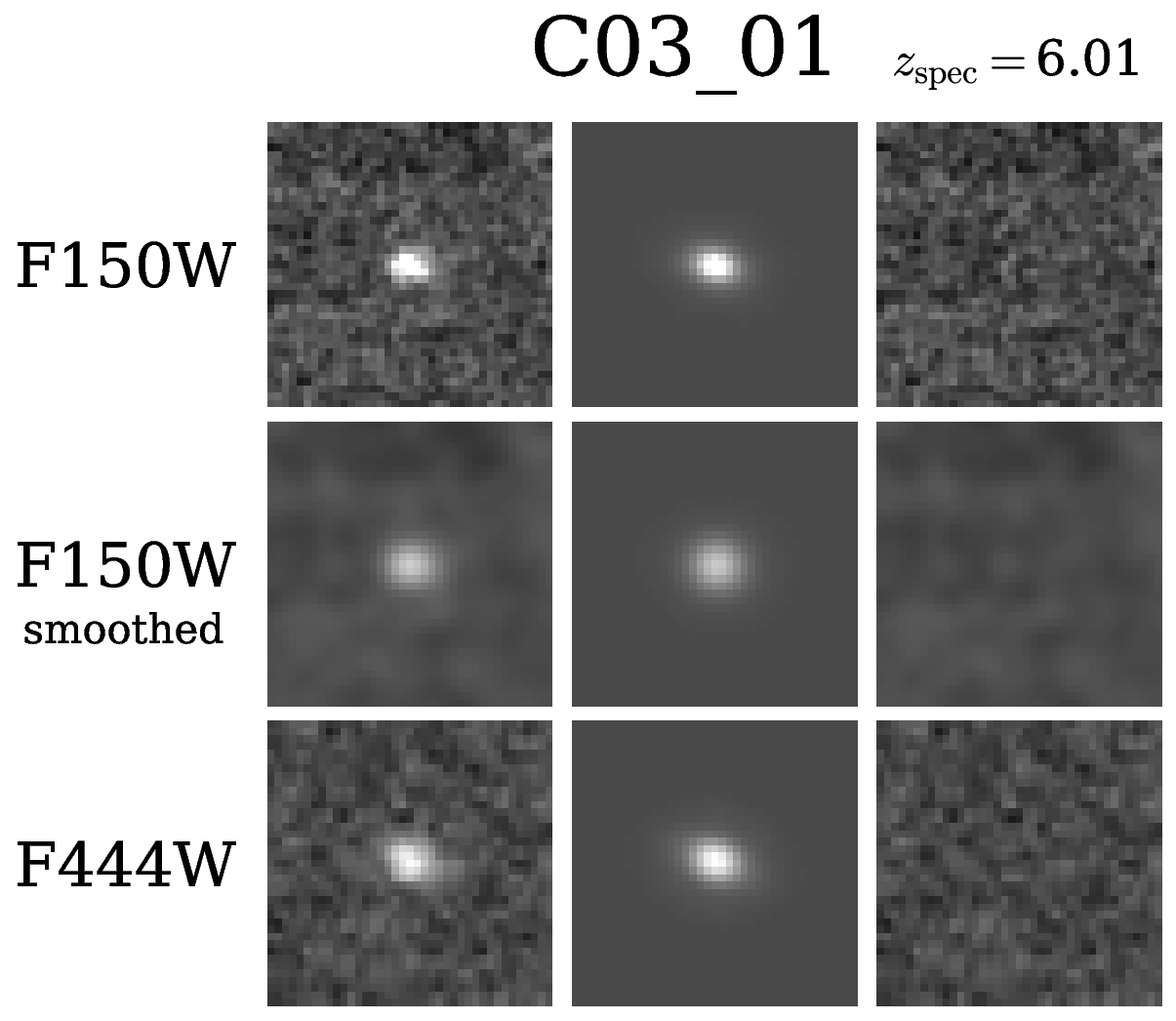}
   \includegraphics[height=0.14\textheight]{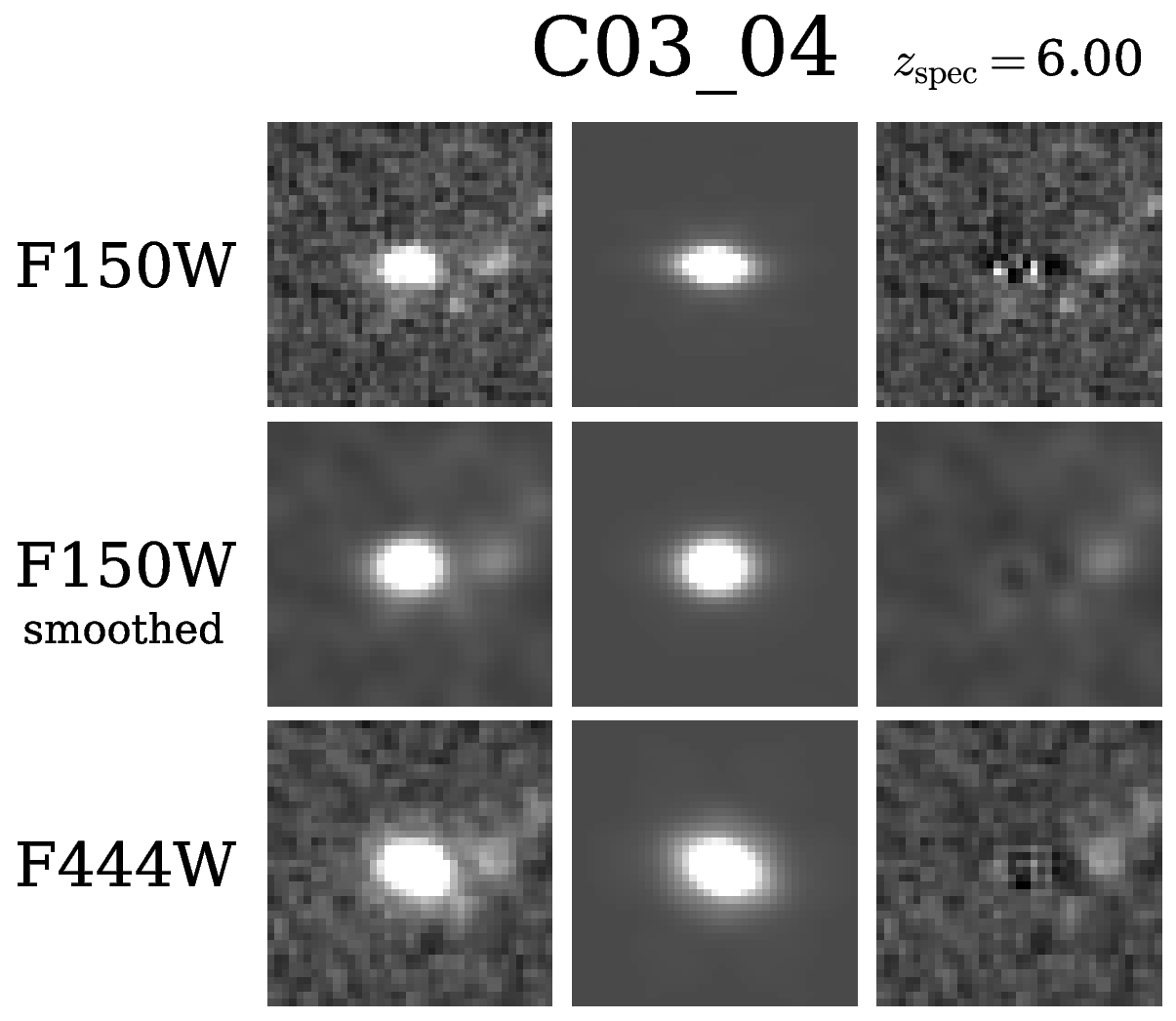}
   \includegraphics[height=0.14\textheight]{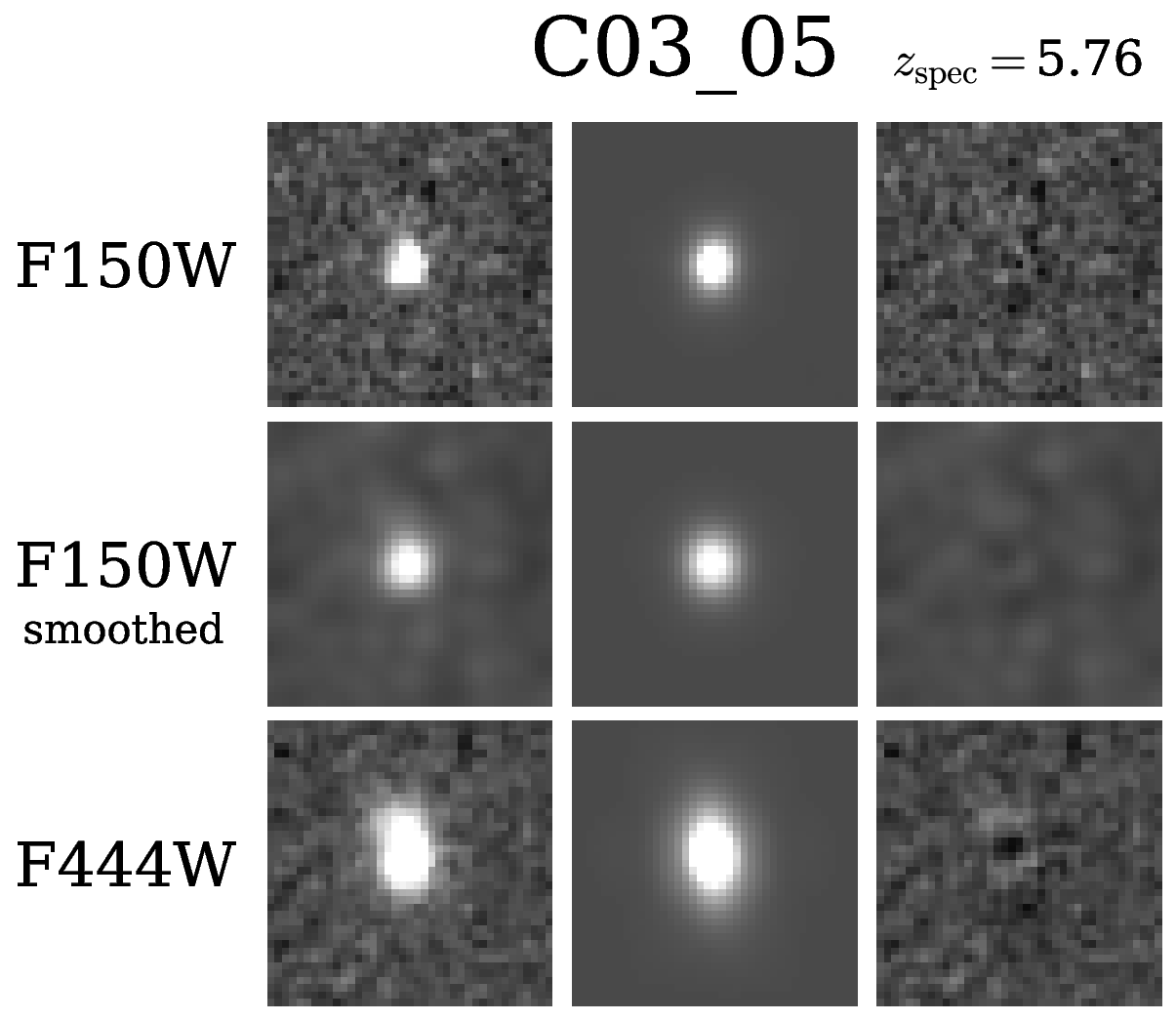}
   \includegraphics[height=0.14\textheight]{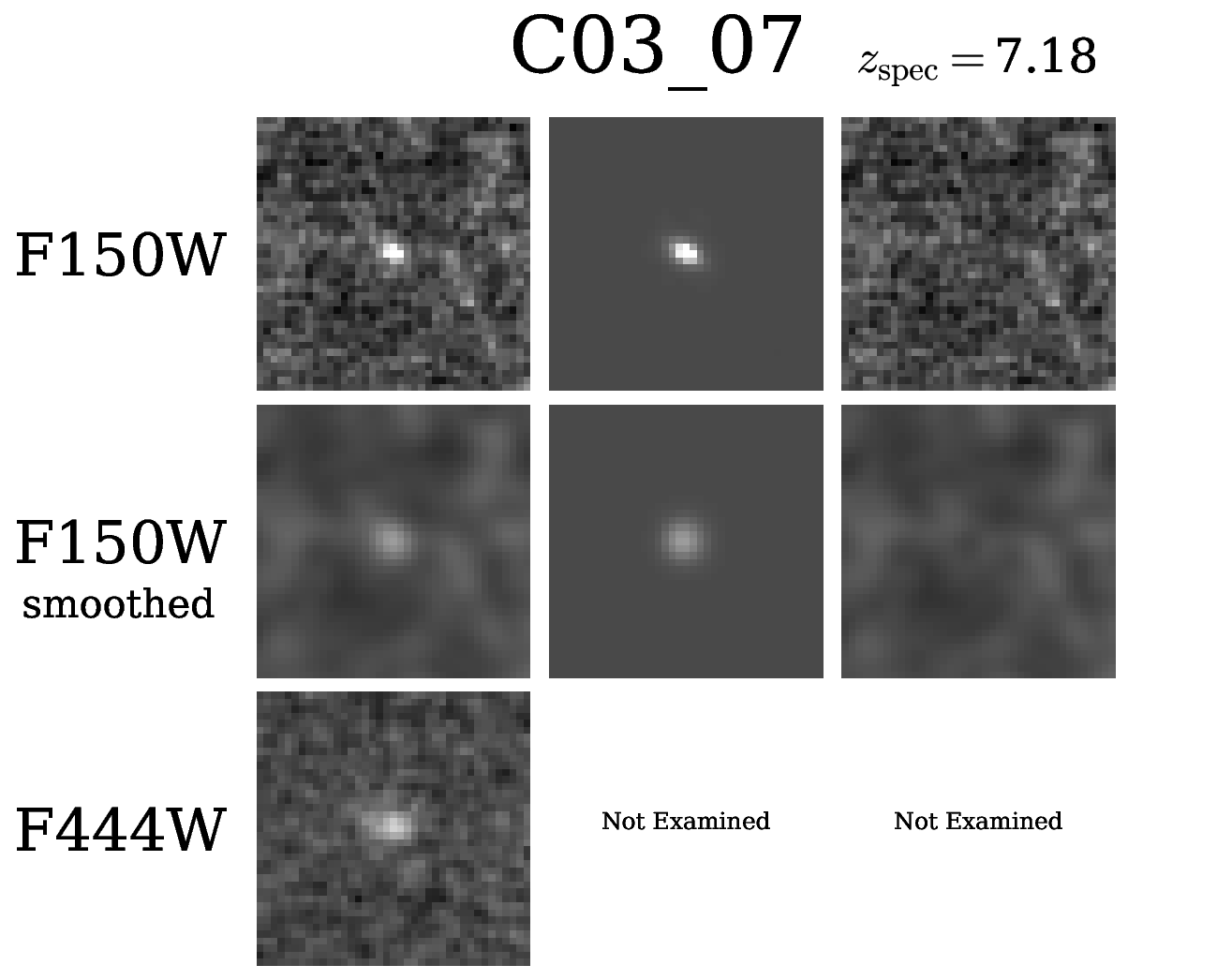}
   \includegraphics[height=0.14\textheight]{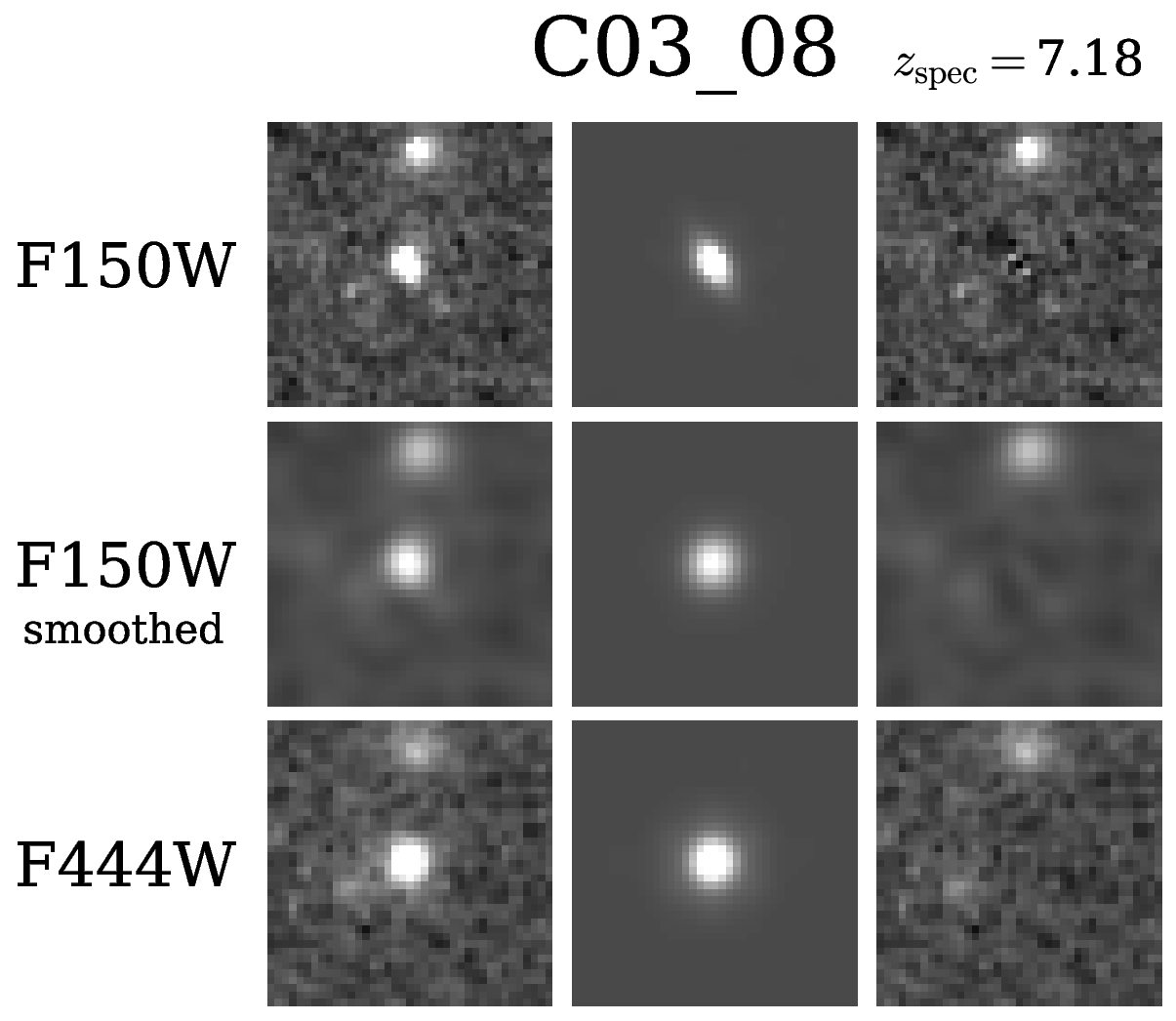}
   \includegraphics[height=0.14\textheight]{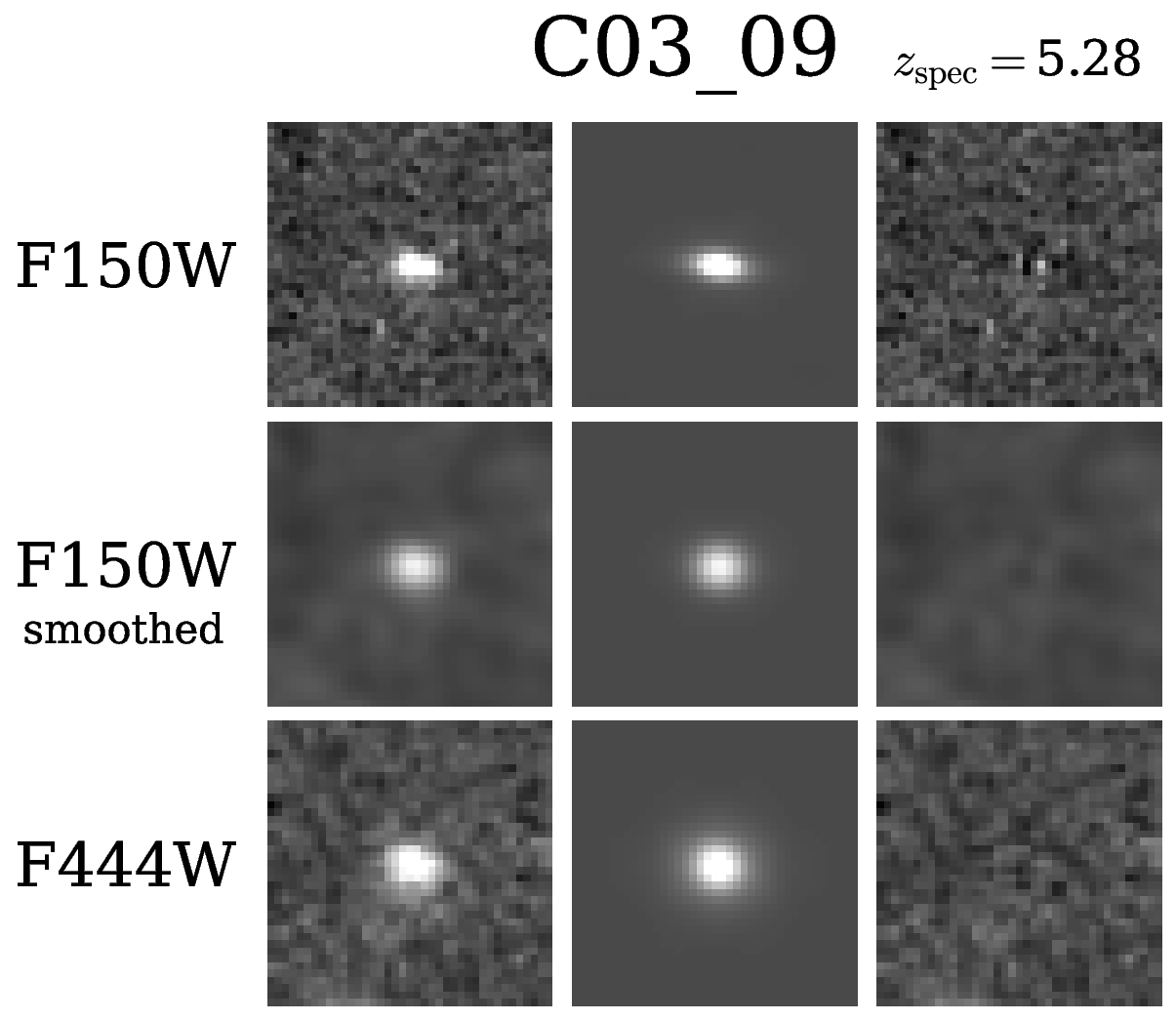}
   \includegraphics[height=0.14\textheight]{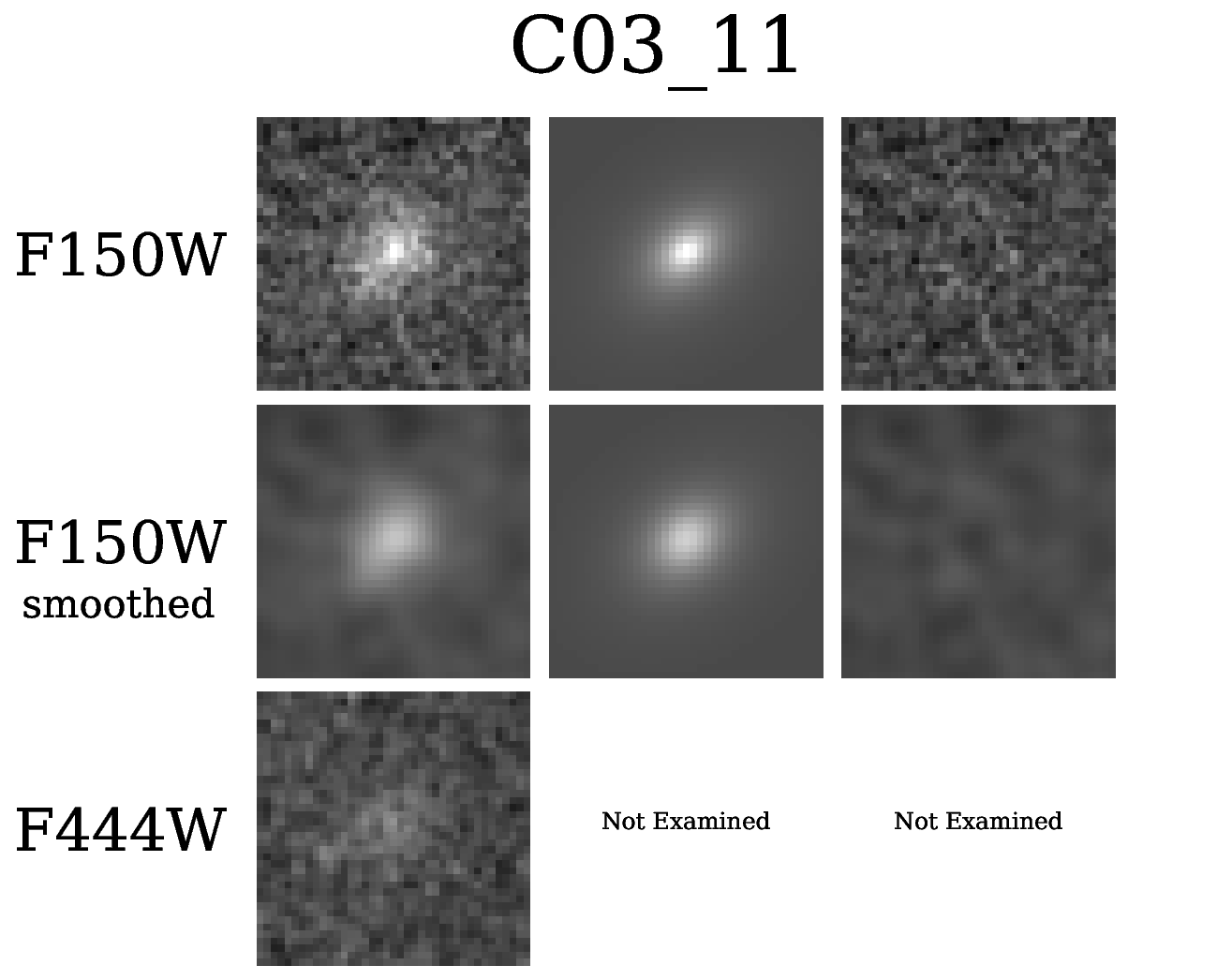}
   \includegraphics[height=0.14\textheight]{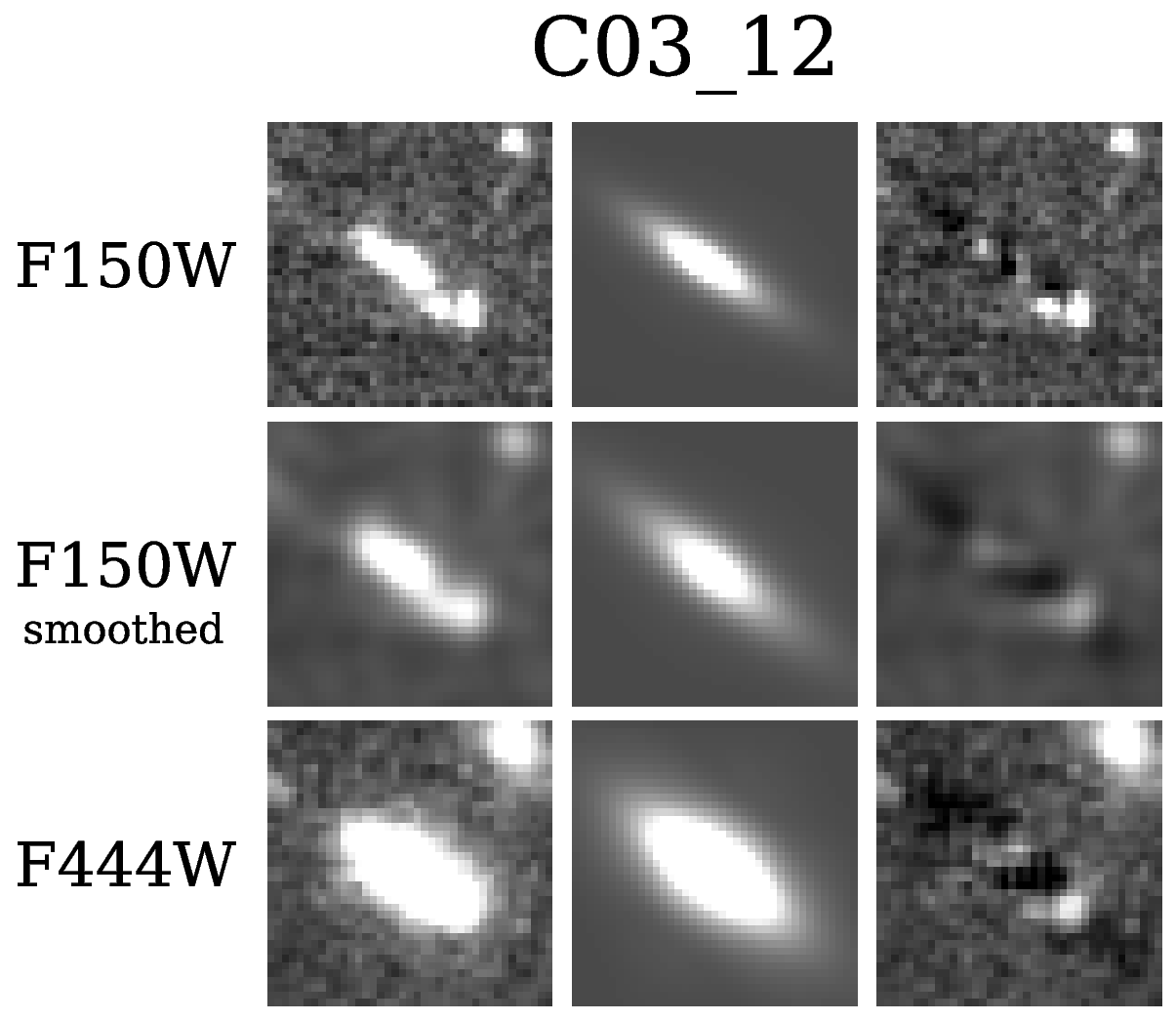}
   \includegraphics[height=0.14\textheight]{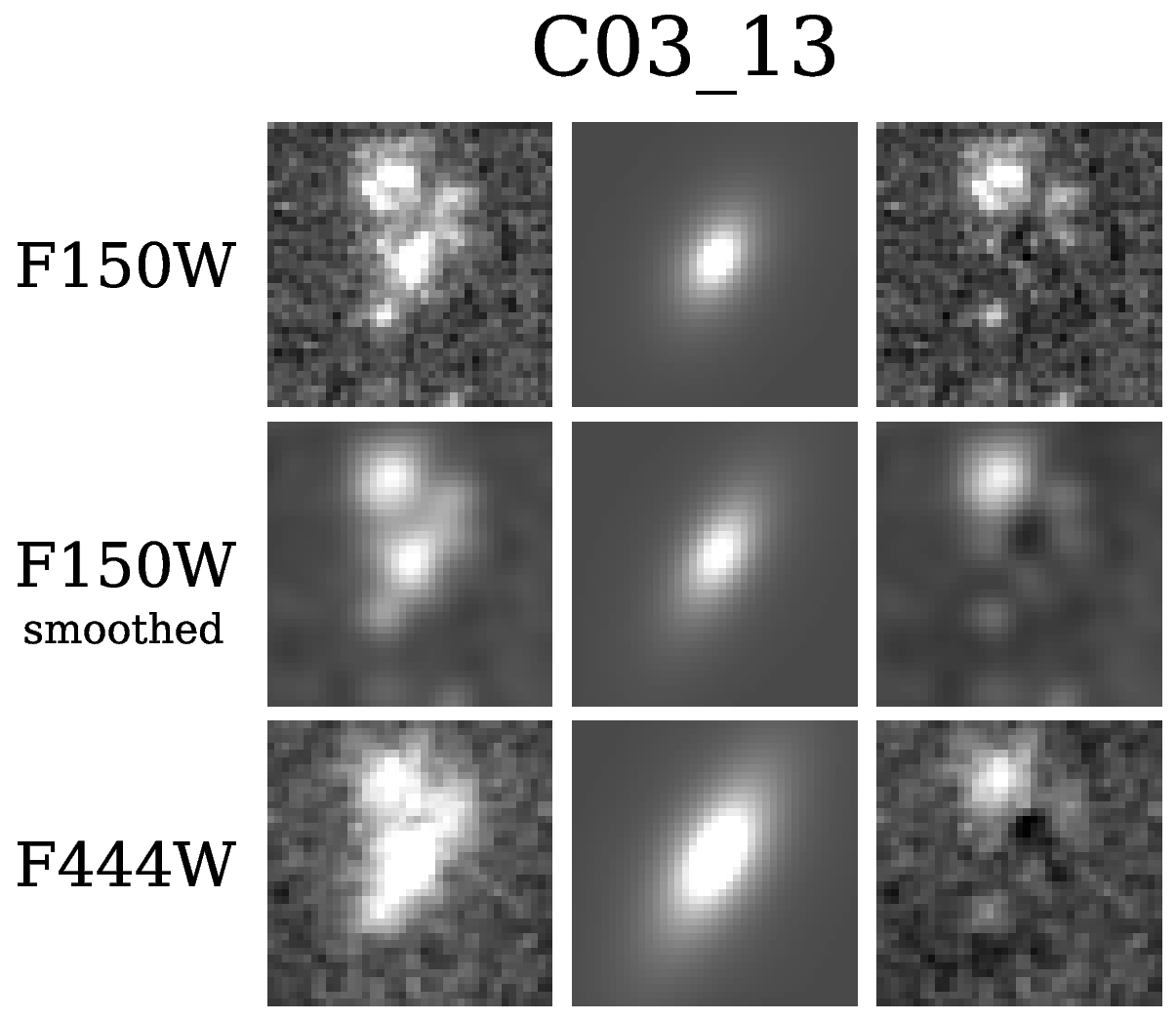}
   \includegraphics[height=0.14\textheight]{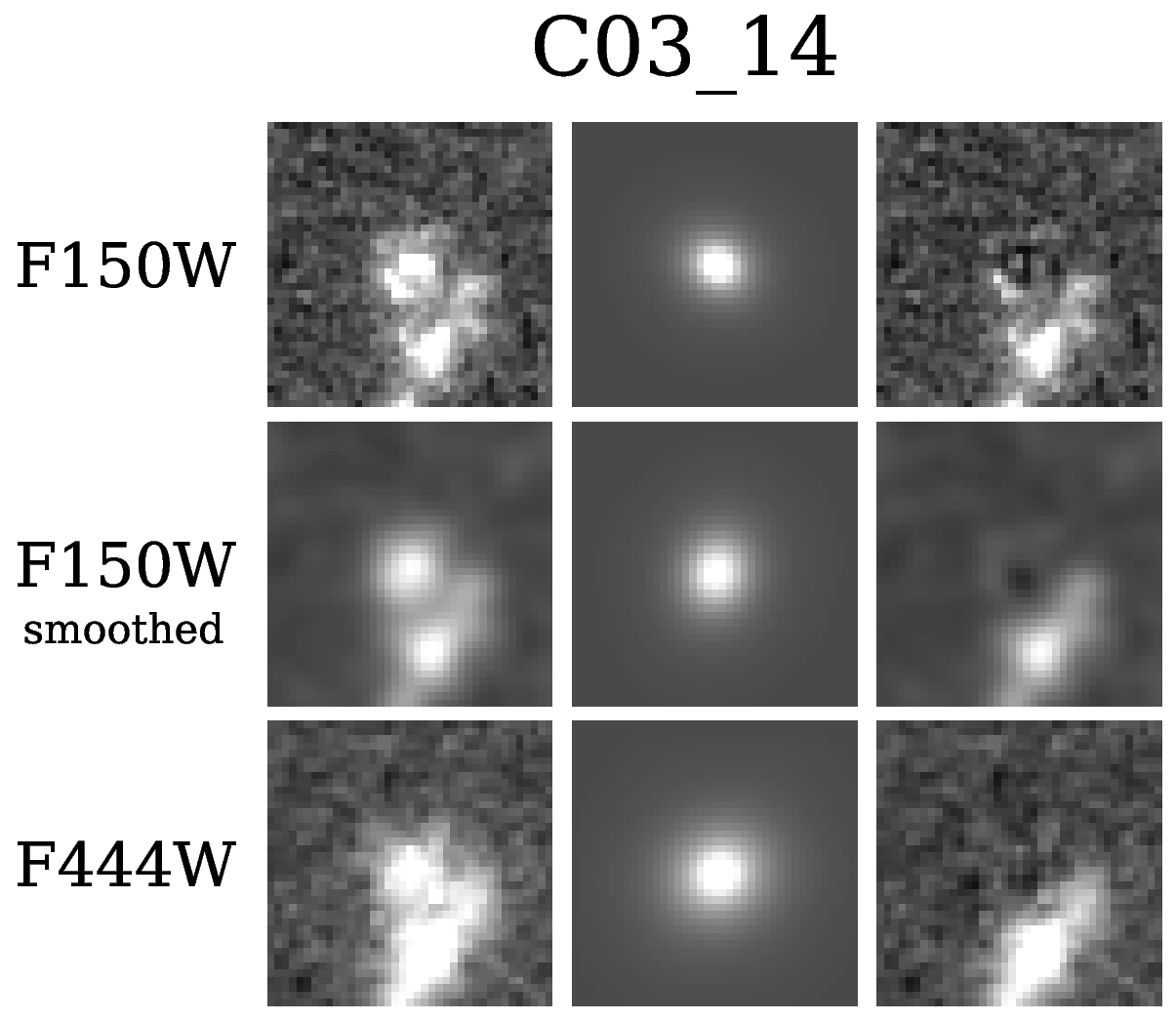}
\caption{
(Continued)
}
\end{center}
\end{figure*}

\addtocounter{figure}{-1}
\begin{figure*}
\begin{center}
   \includegraphics[height=0.14\textheight]{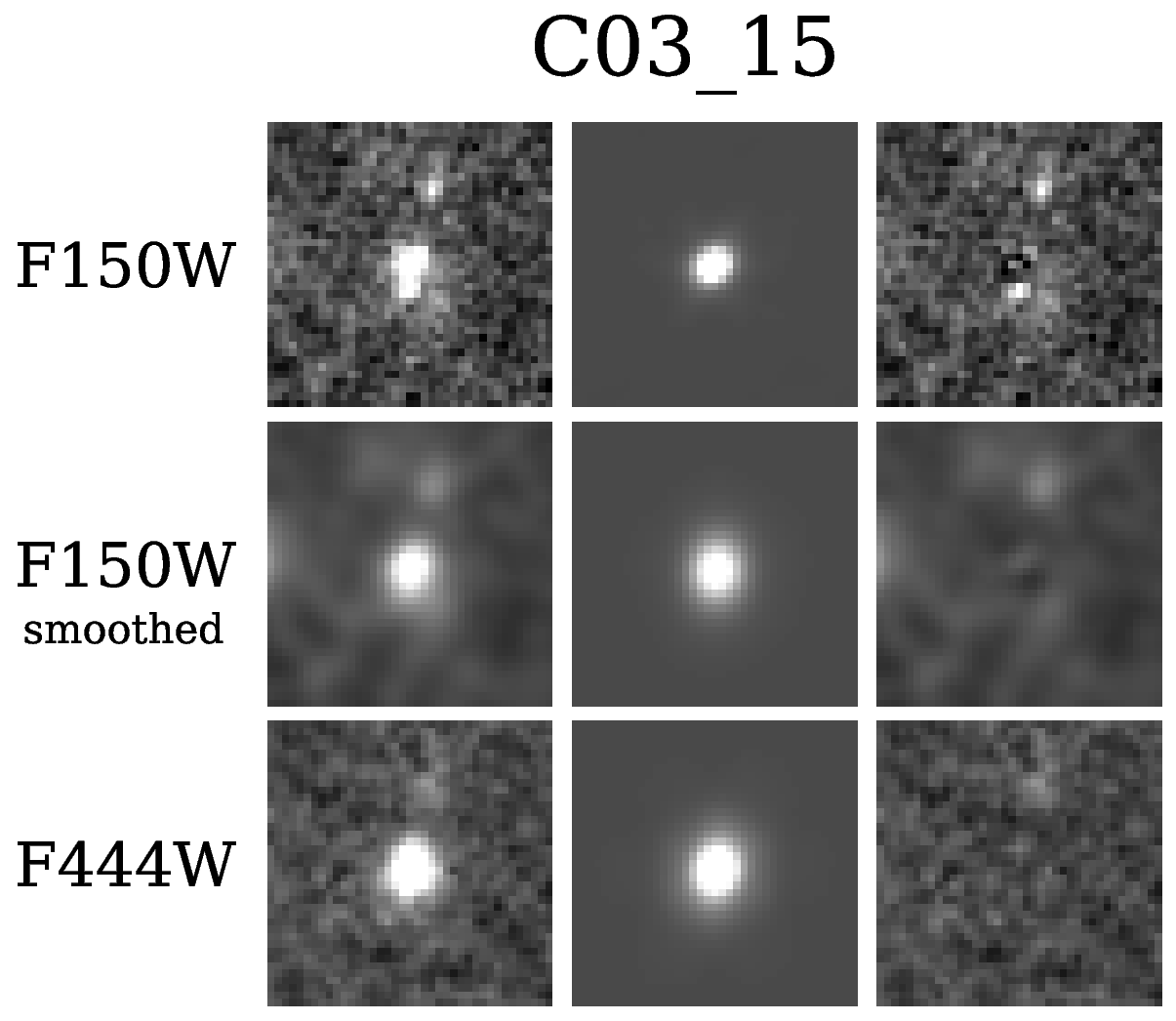}
   \includegraphics[height=0.14\textheight]{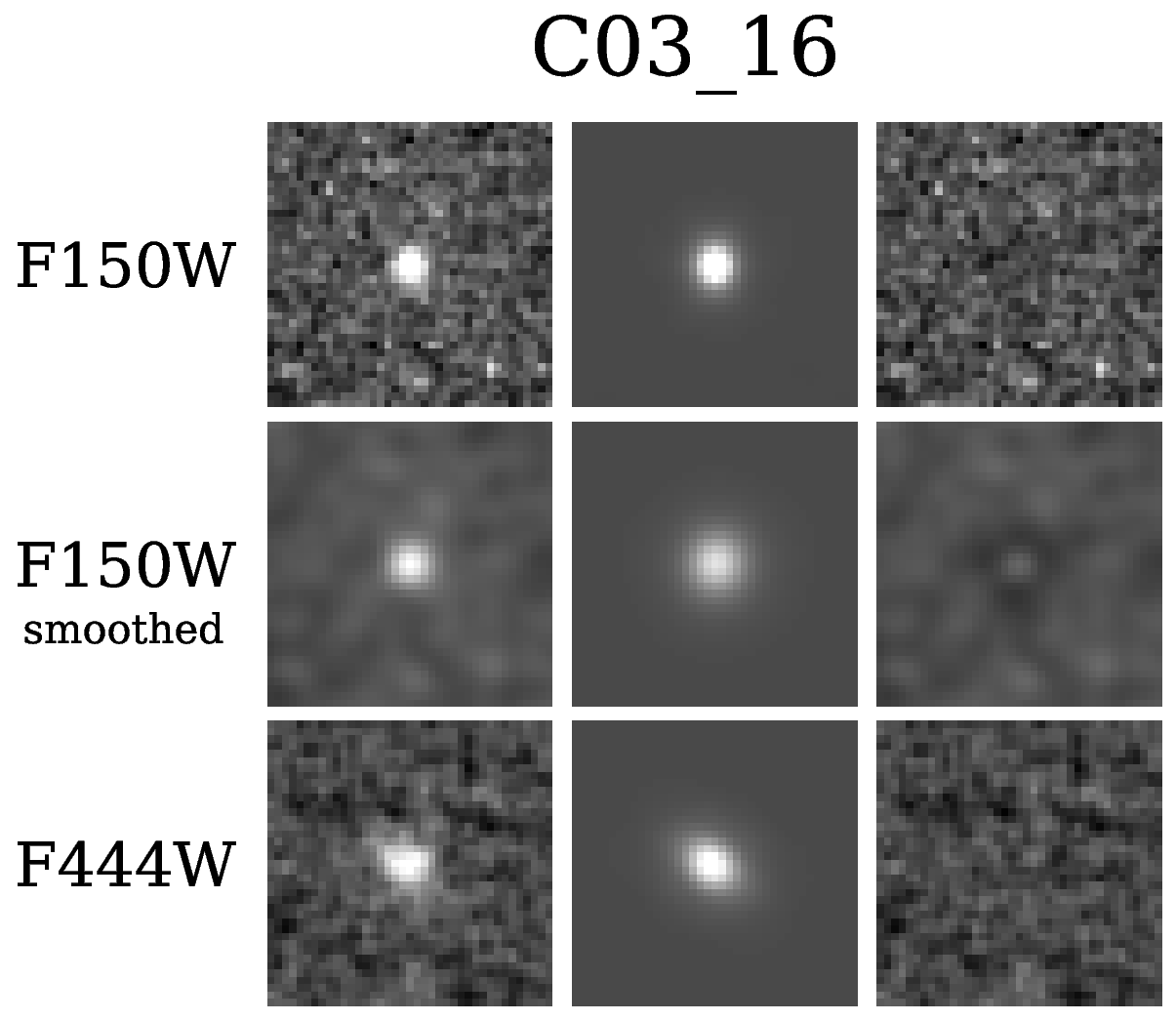}
   \includegraphics[height=0.14\textheight]{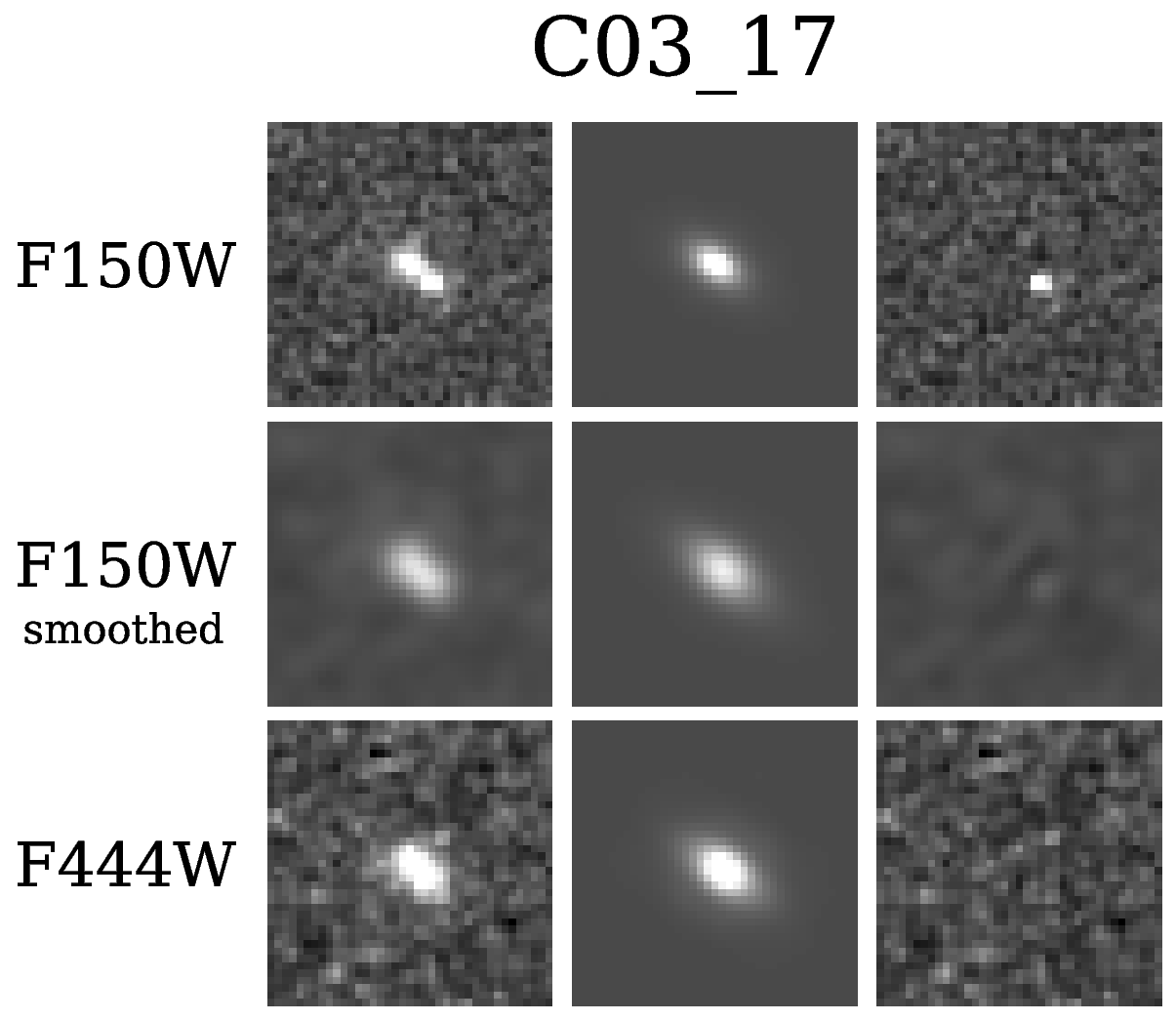}
   \includegraphics[height=0.14\textheight]{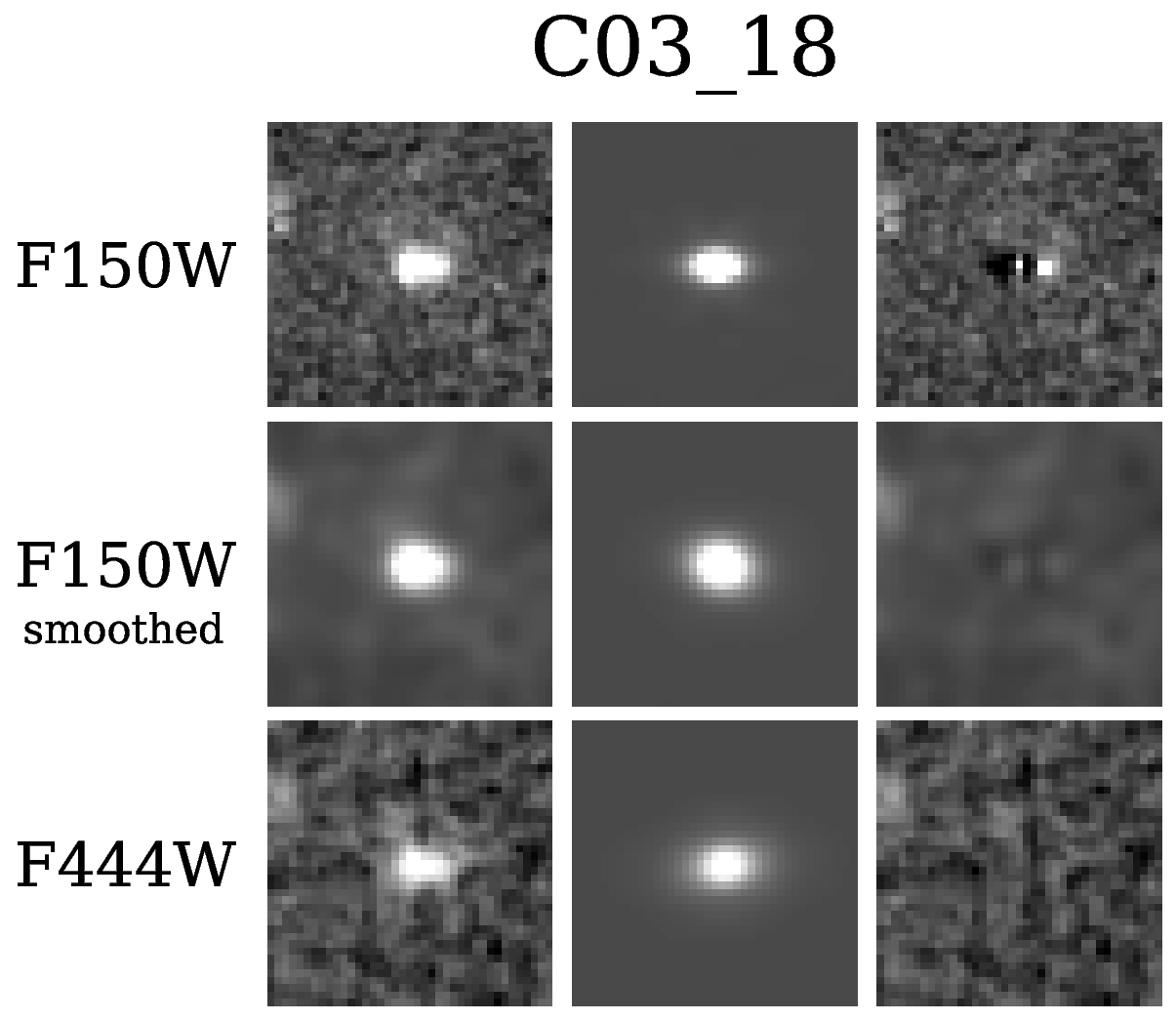}
   \includegraphics[height=0.14\textheight]{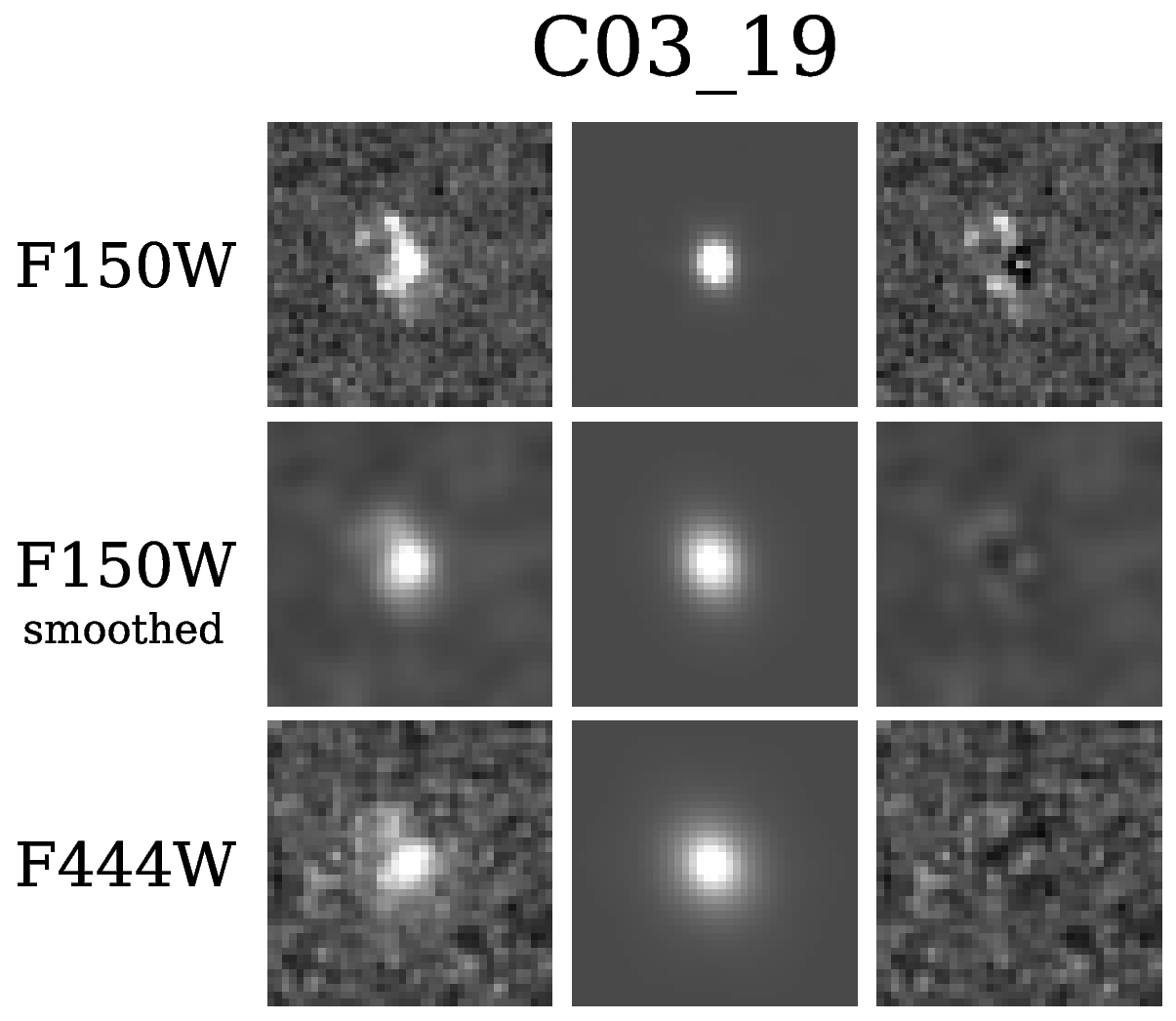}
   \includegraphics[height=0.14\textheight]{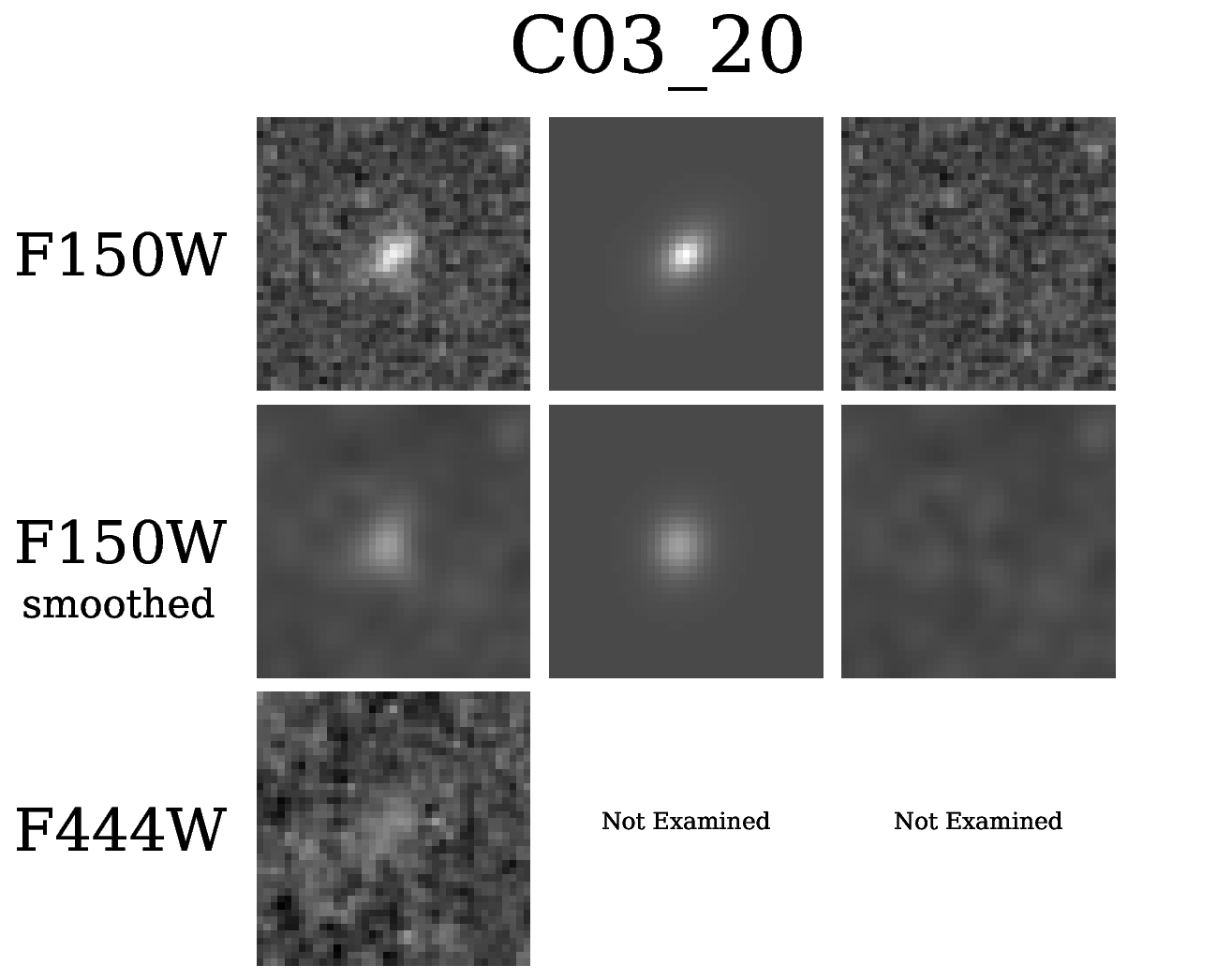}
   \includegraphics[height=0.14\textheight]{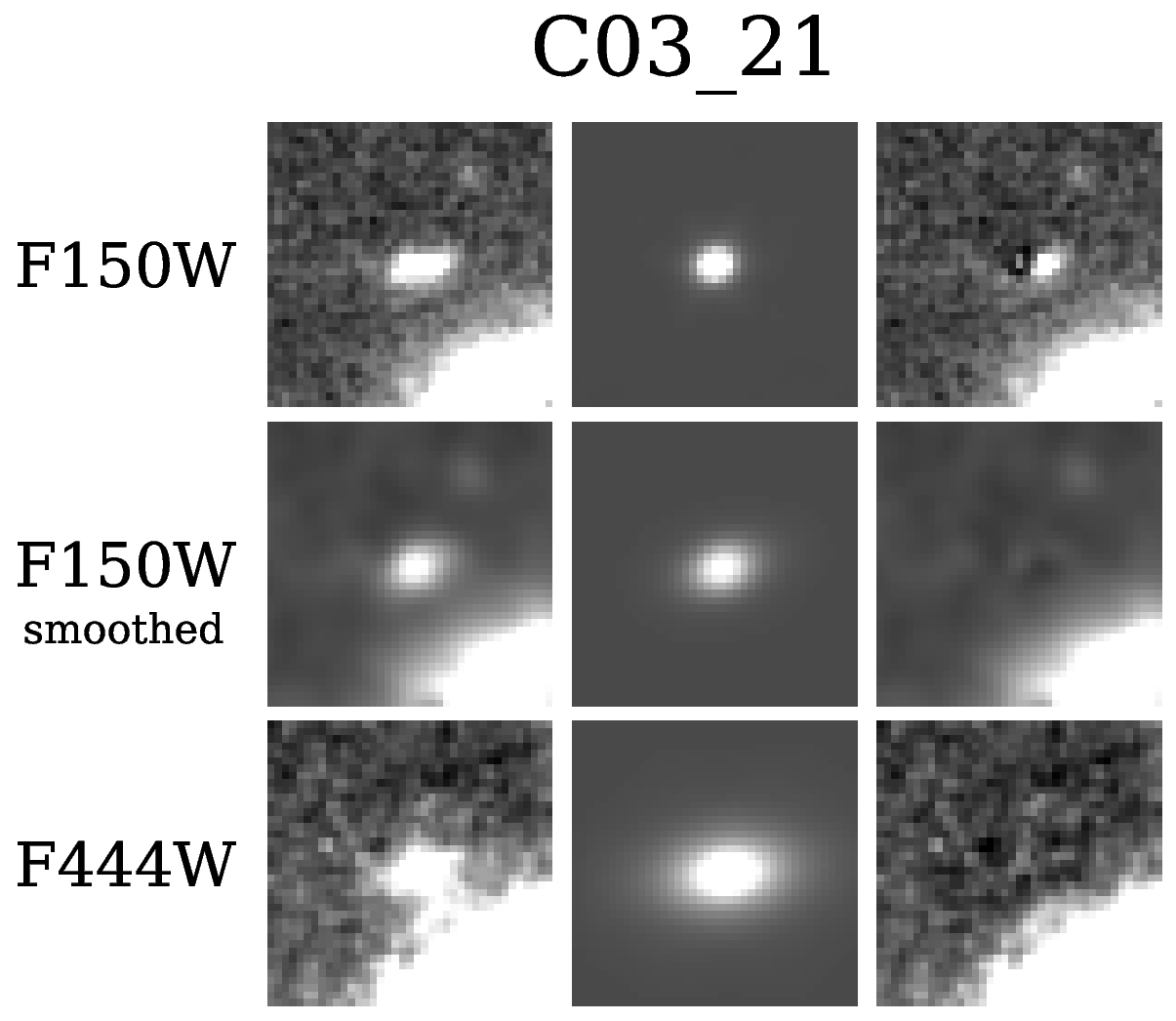}
   \includegraphics[height=0.14\textheight]{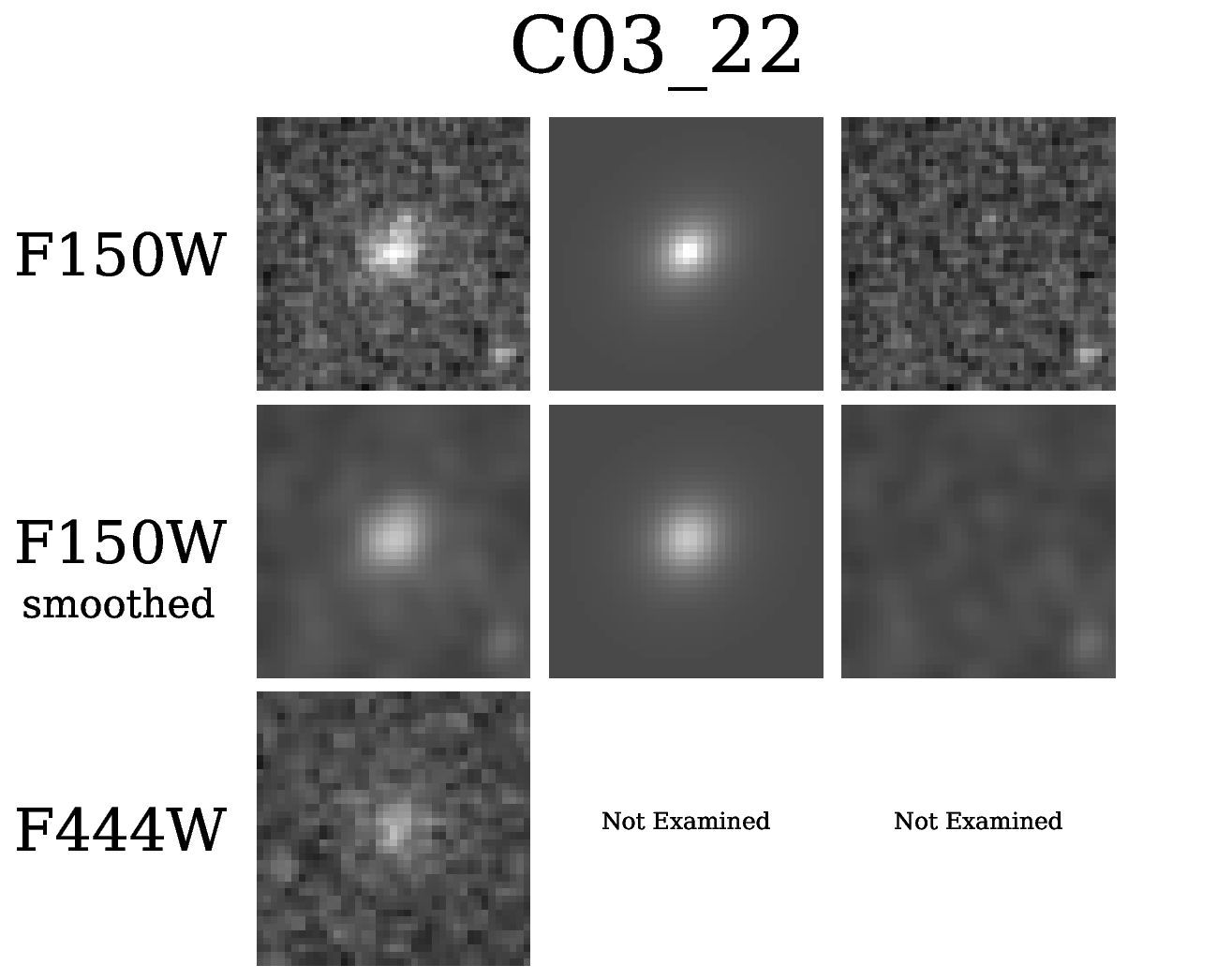}
   \includegraphics[height=0.14\textheight]{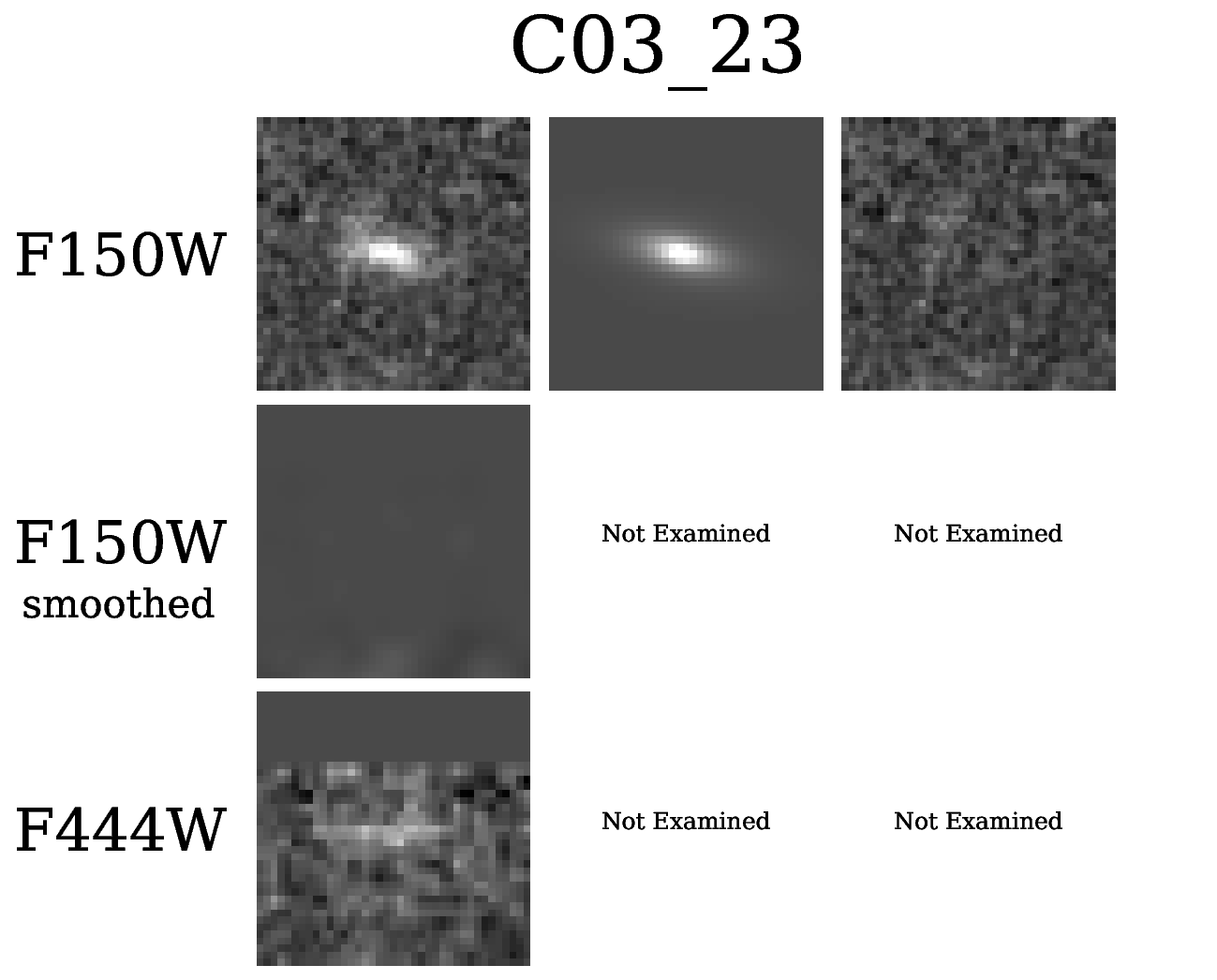}
   \includegraphics[height=0.14\textheight]{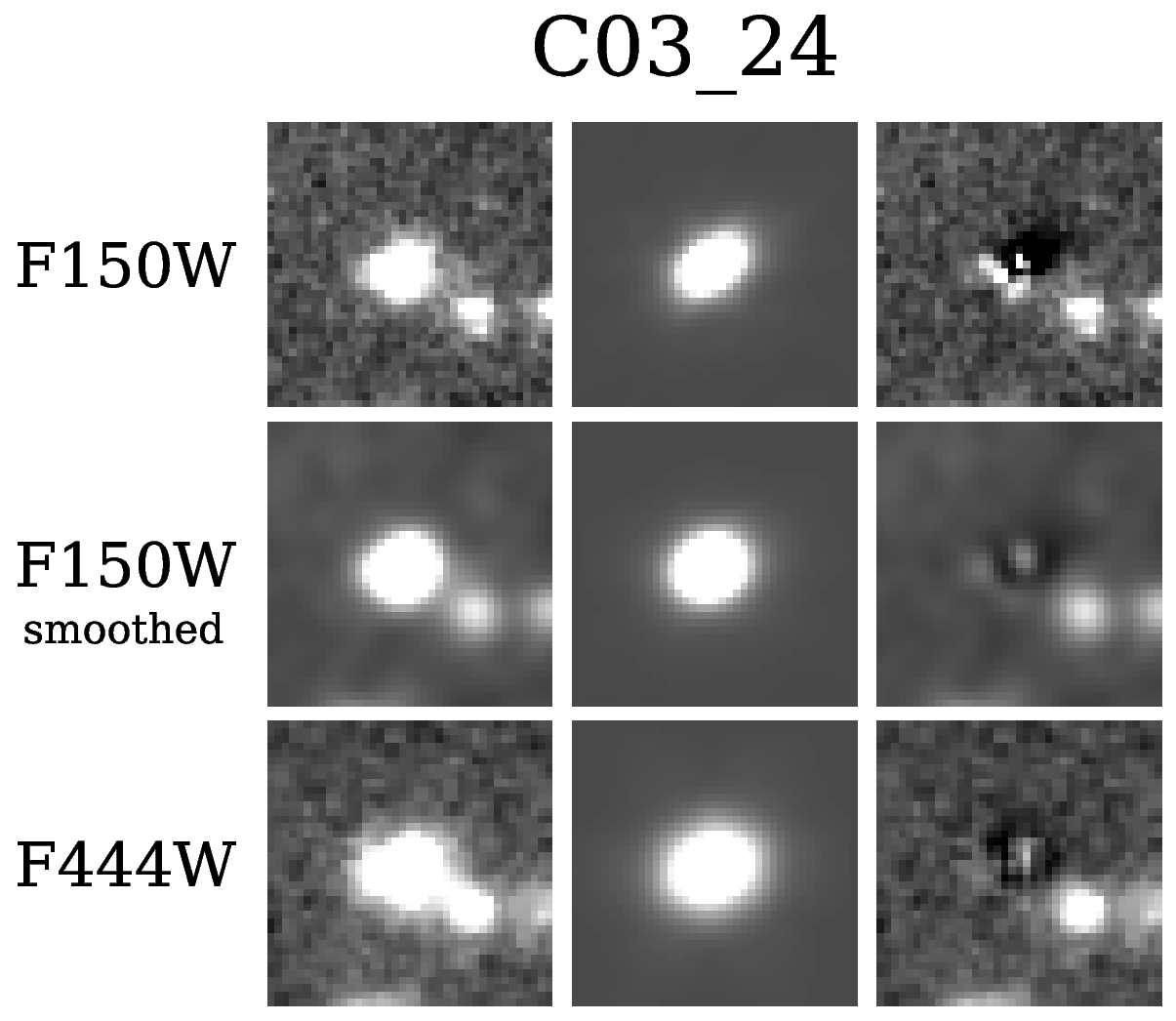}
   \includegraphics[height=0.14\textheight]{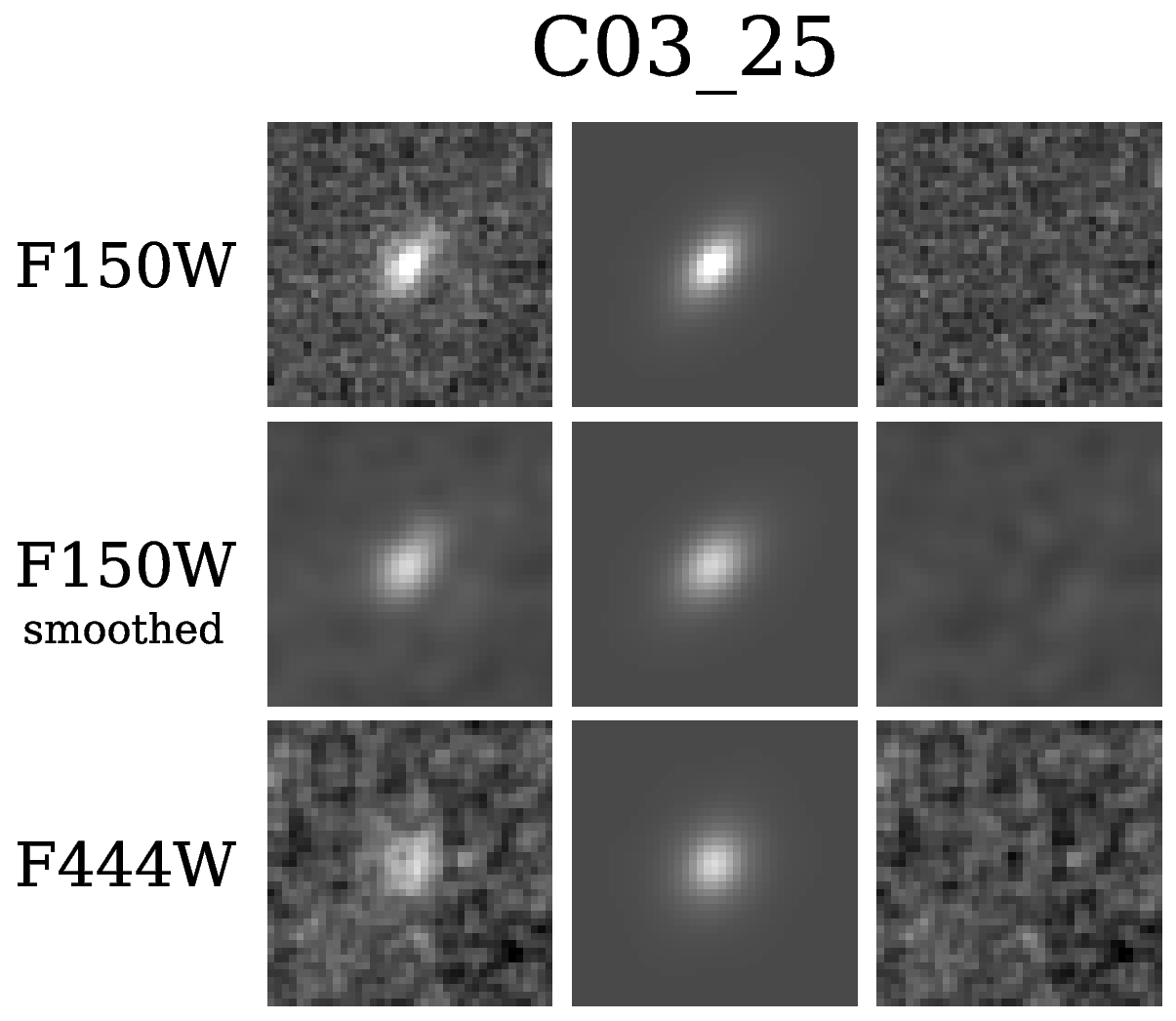}
   \includegraphics[height=0.14\textheight]{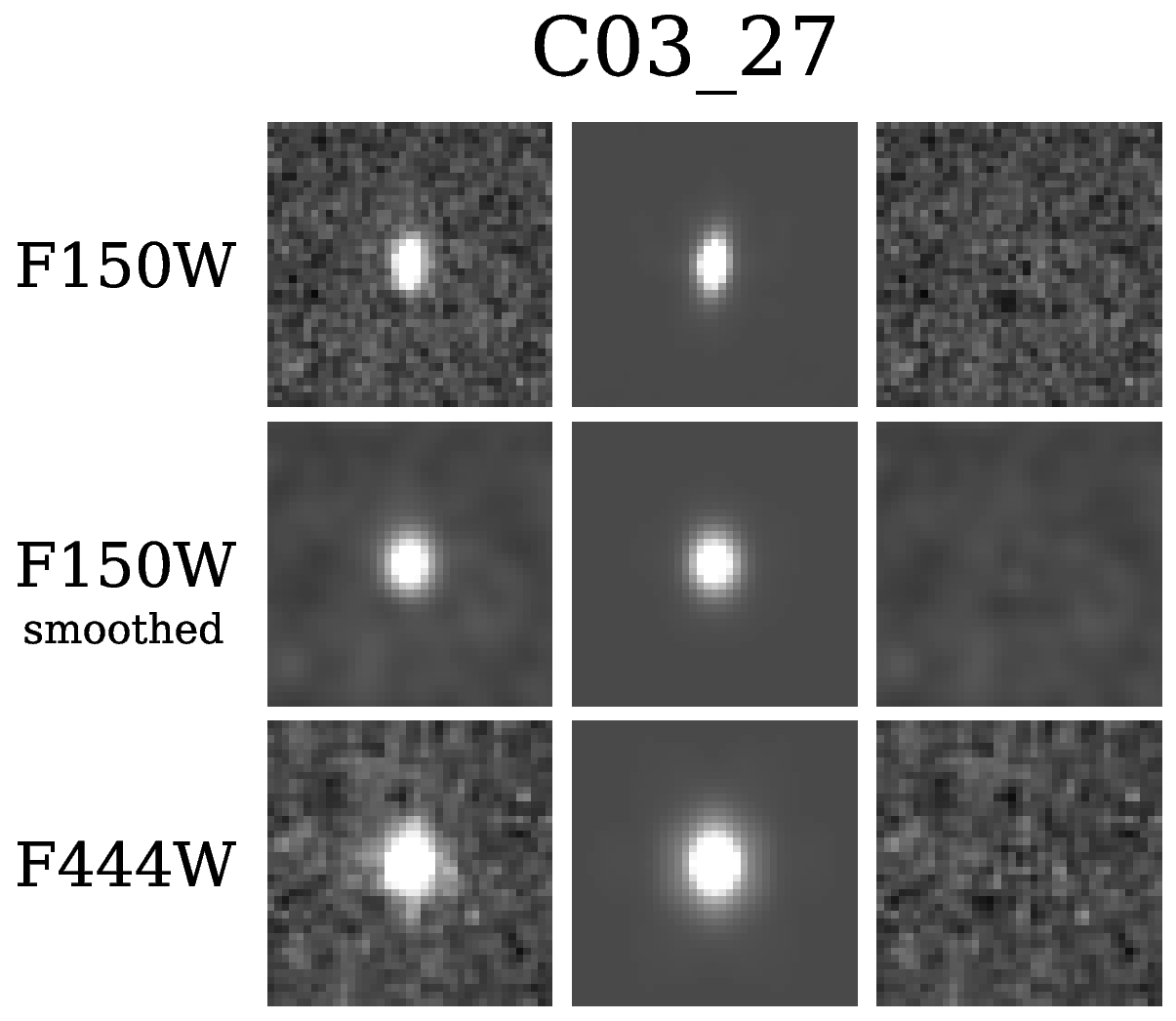}
   \includegraphics[height=0.14\textheight]{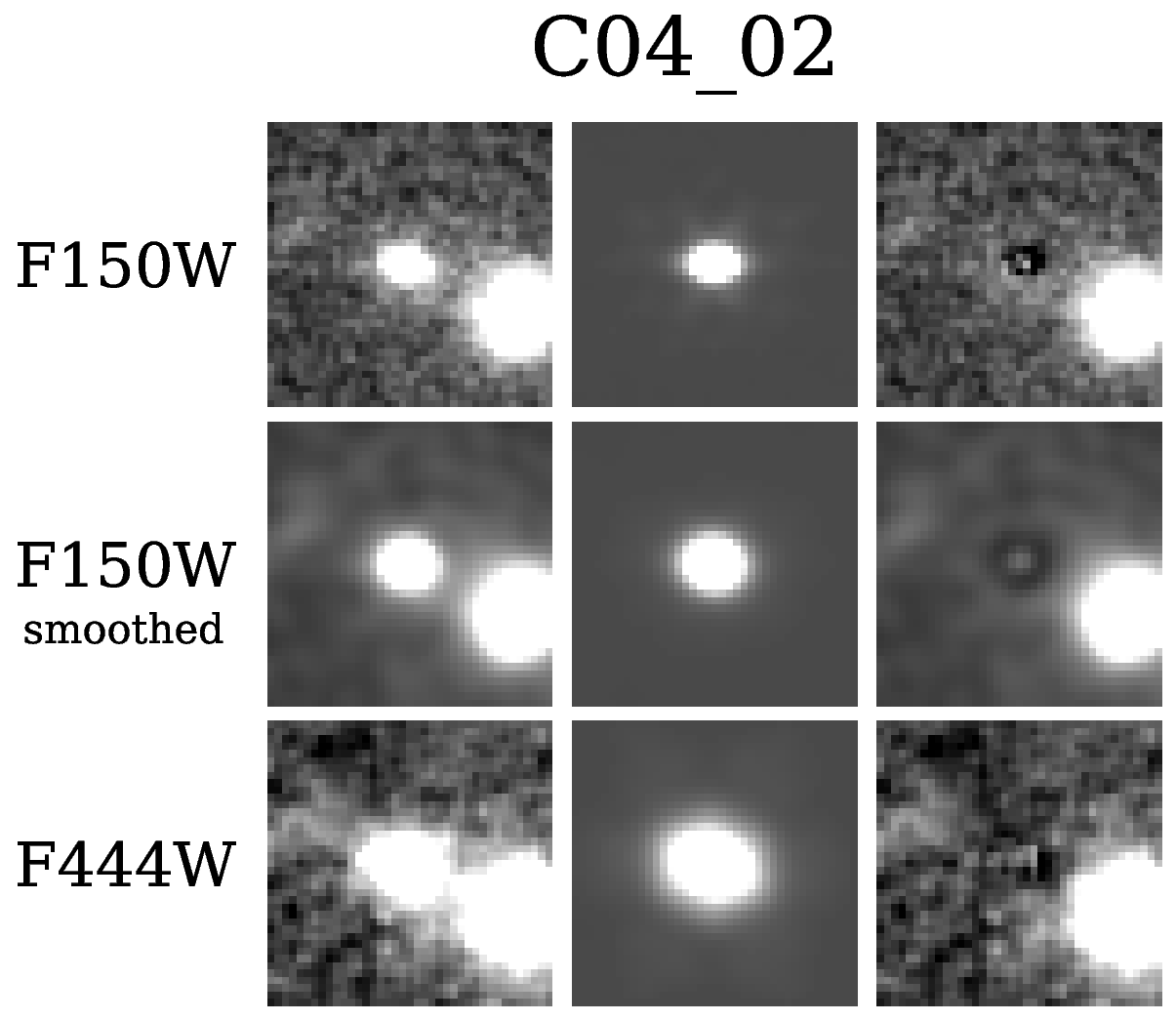}
   \includegraphics[height=0.14\textheight]{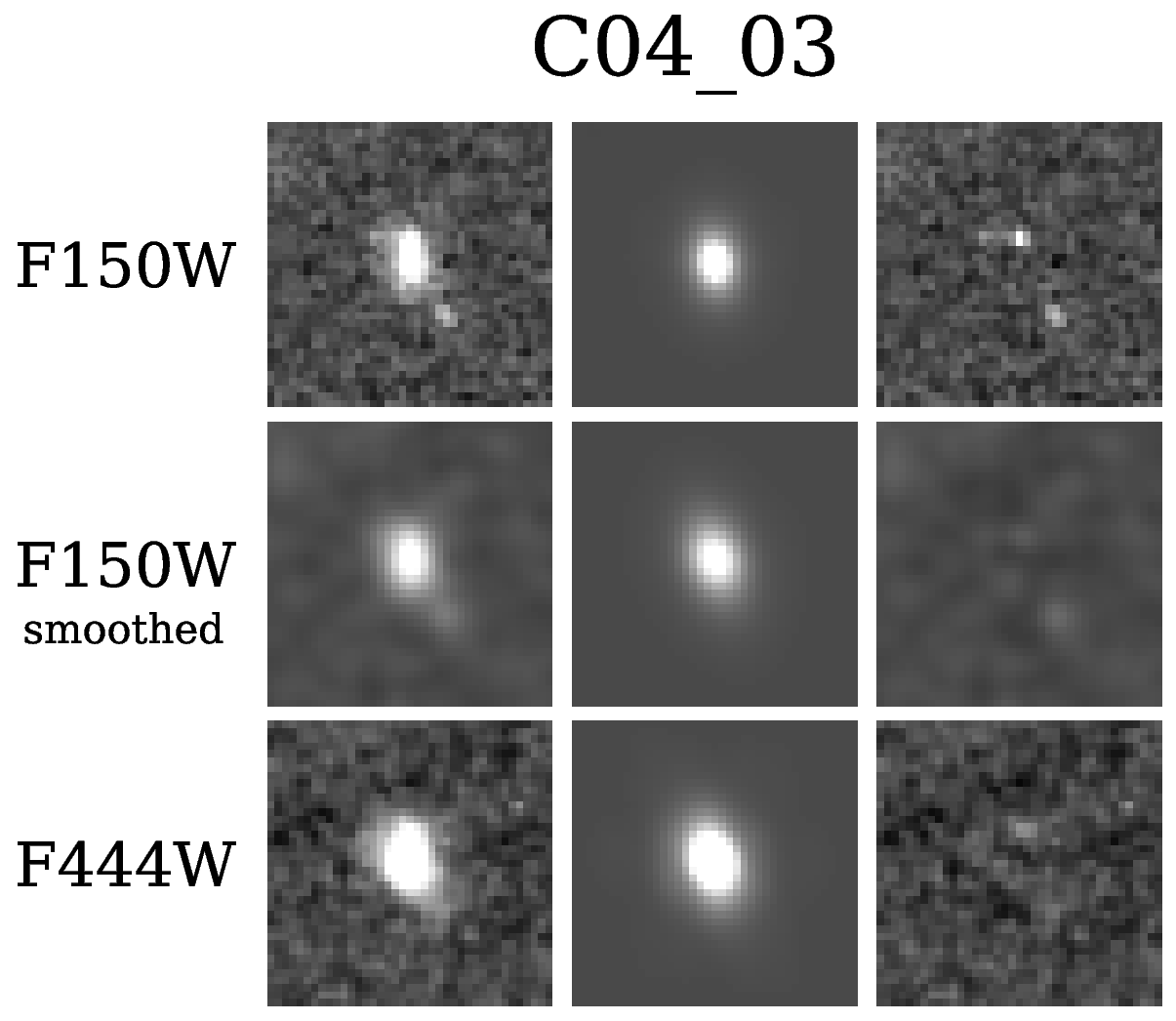}
   \includegraphics[height=0.14\textheight]{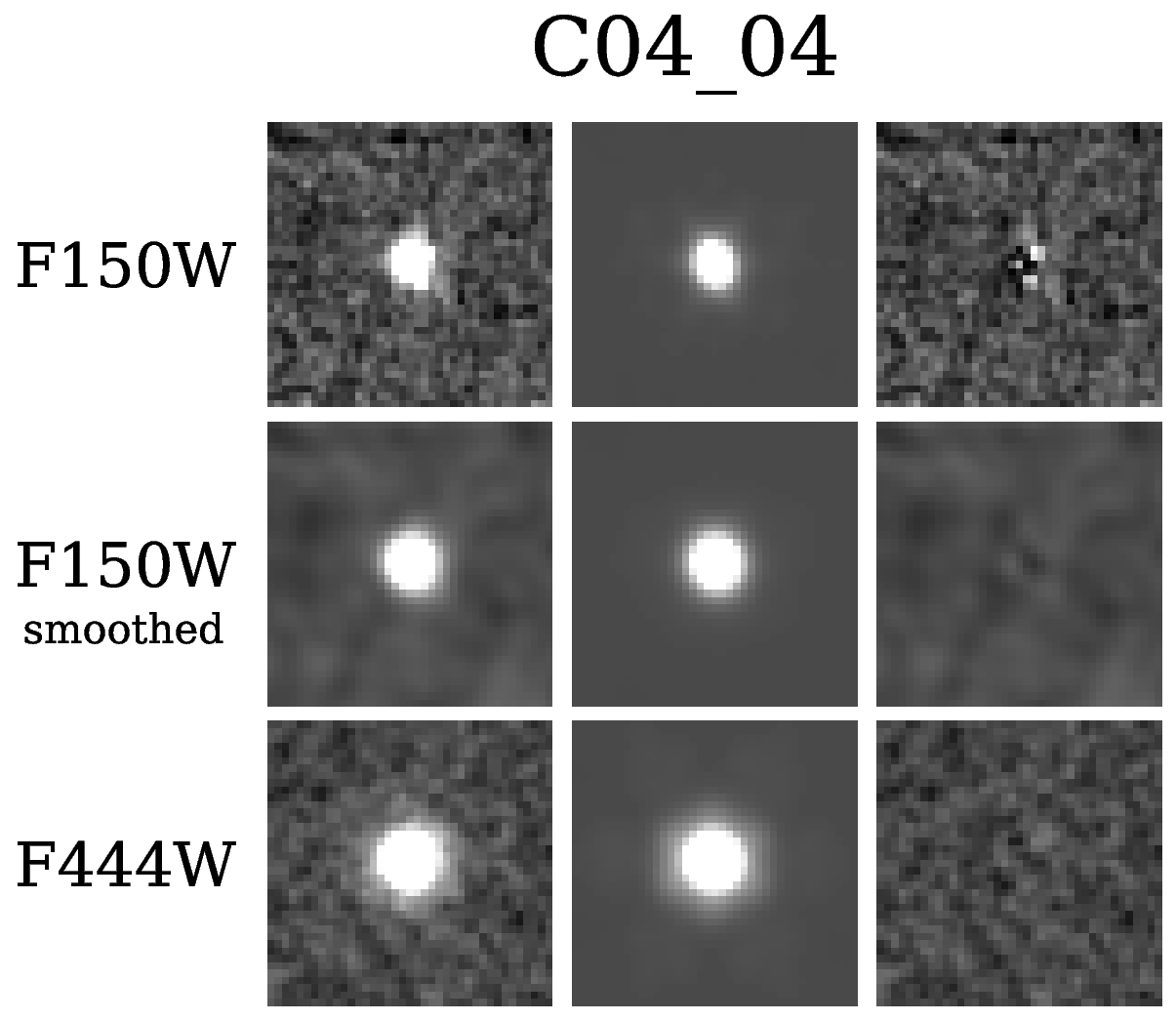}
   \includegraphics[height=0.14\textheight]{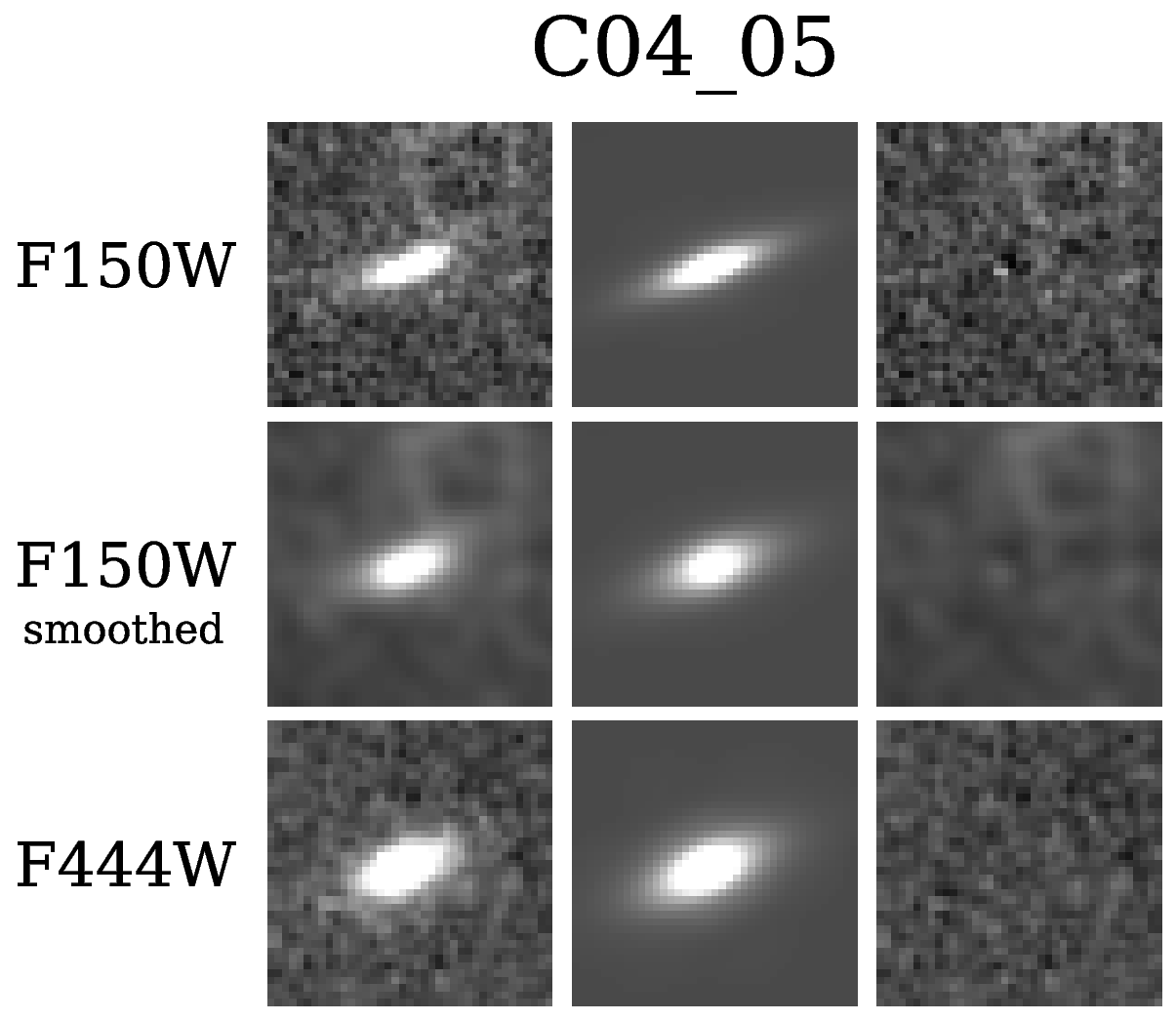}
   \includegraphics[height=0.14\textheight]{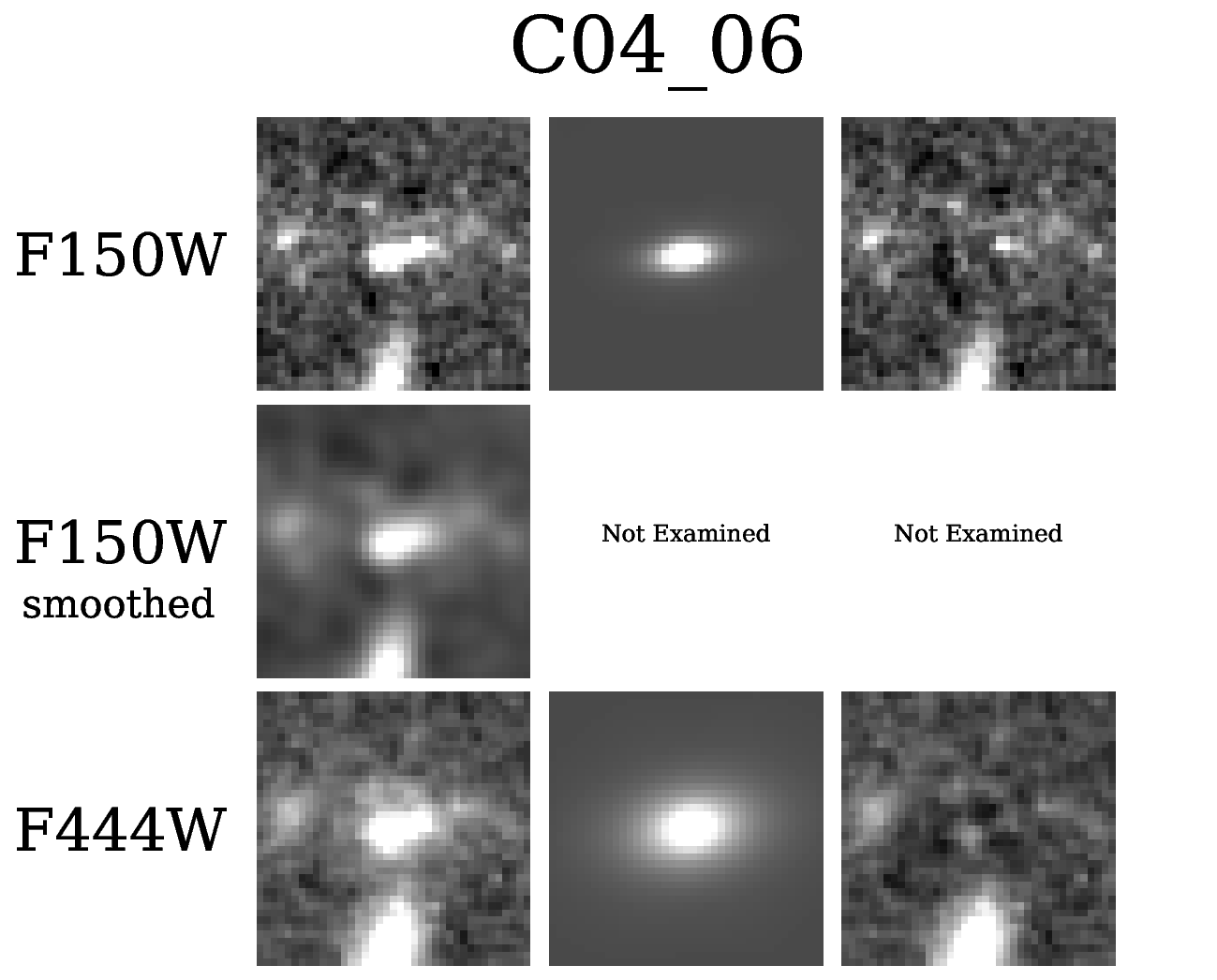}
   \includegraphics[height=0.14\textheight]{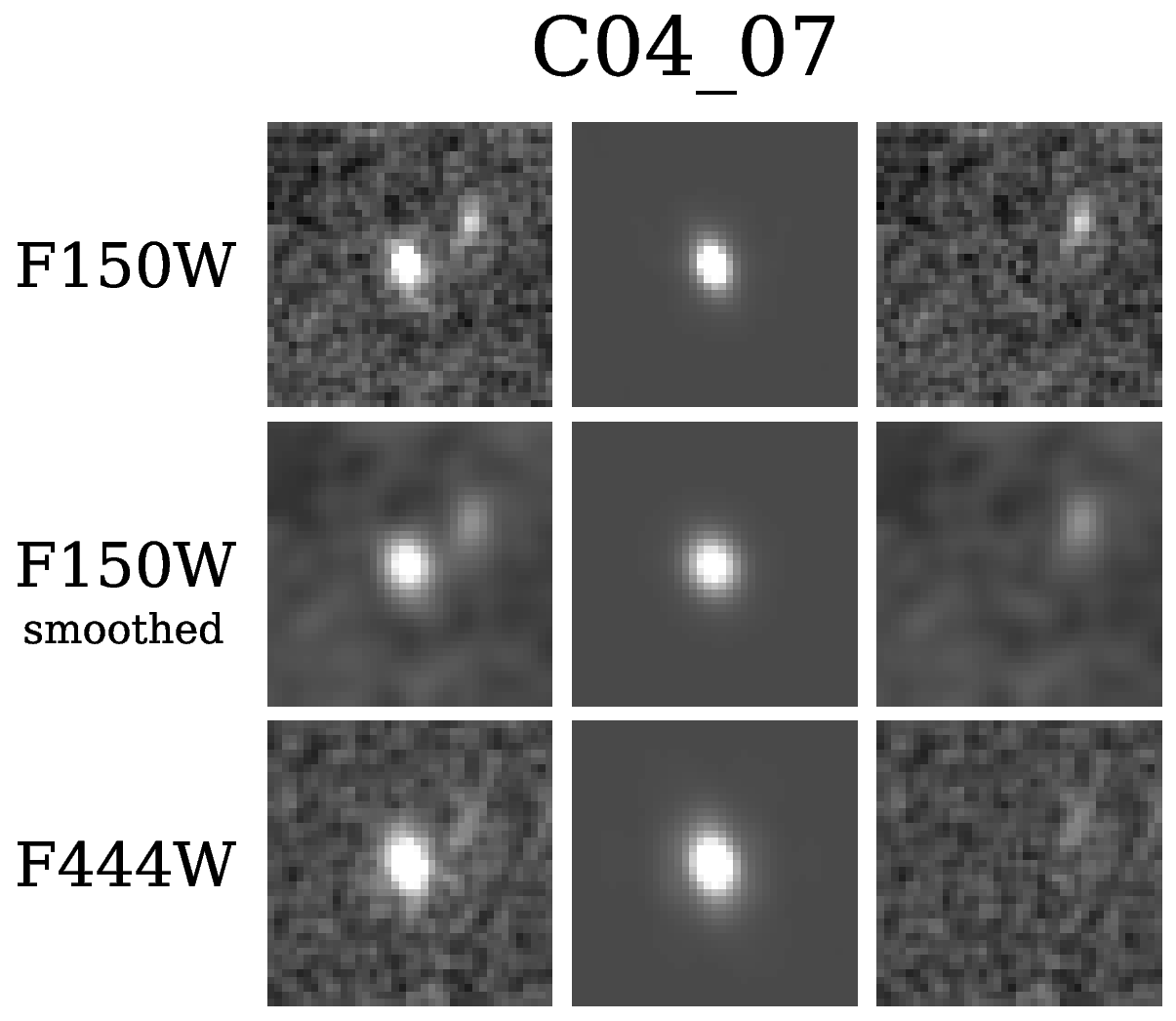}
   \includegraphics[height=0.14\textheight]{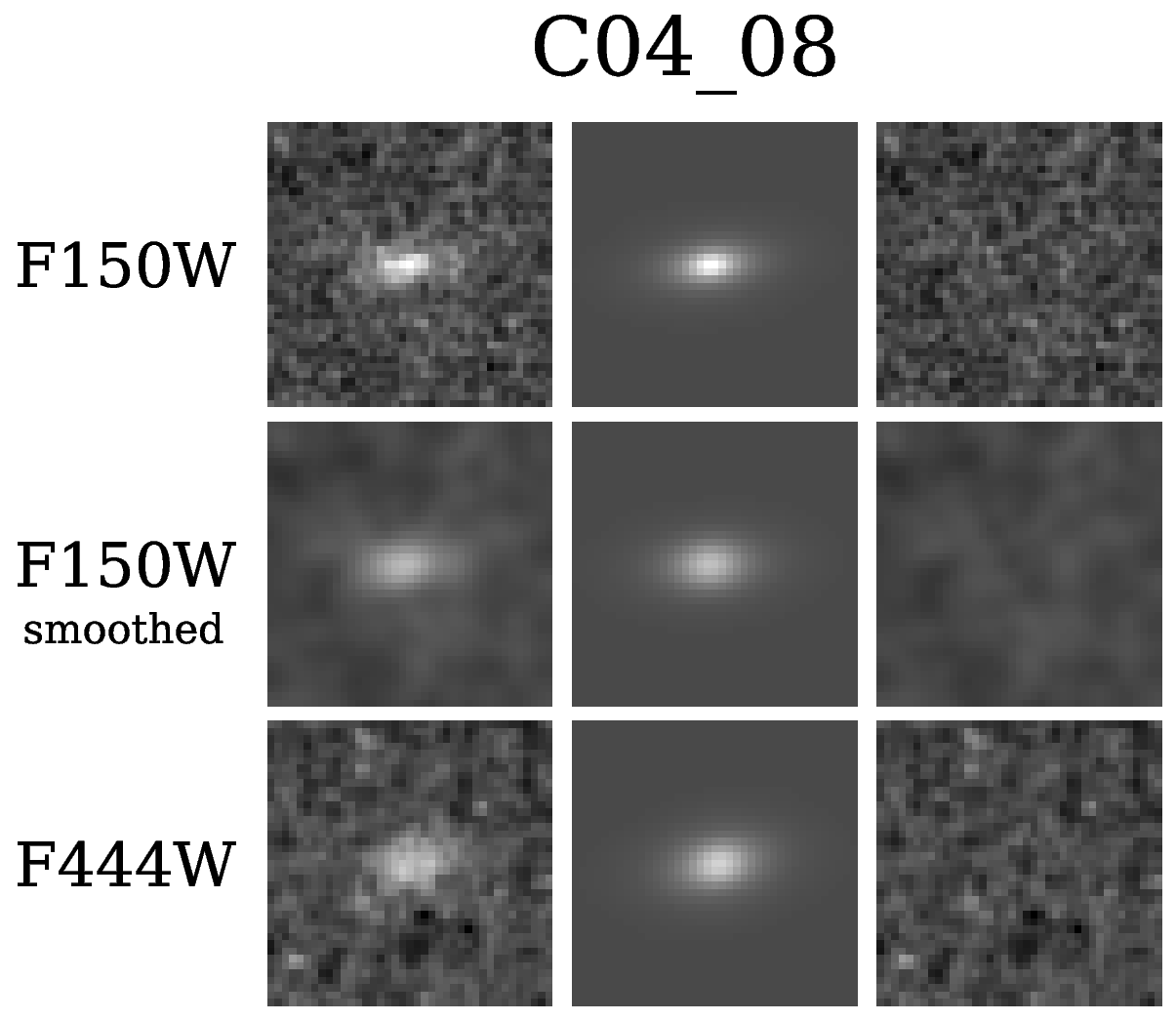}
   \includegraphics[height=0.14\textheight]{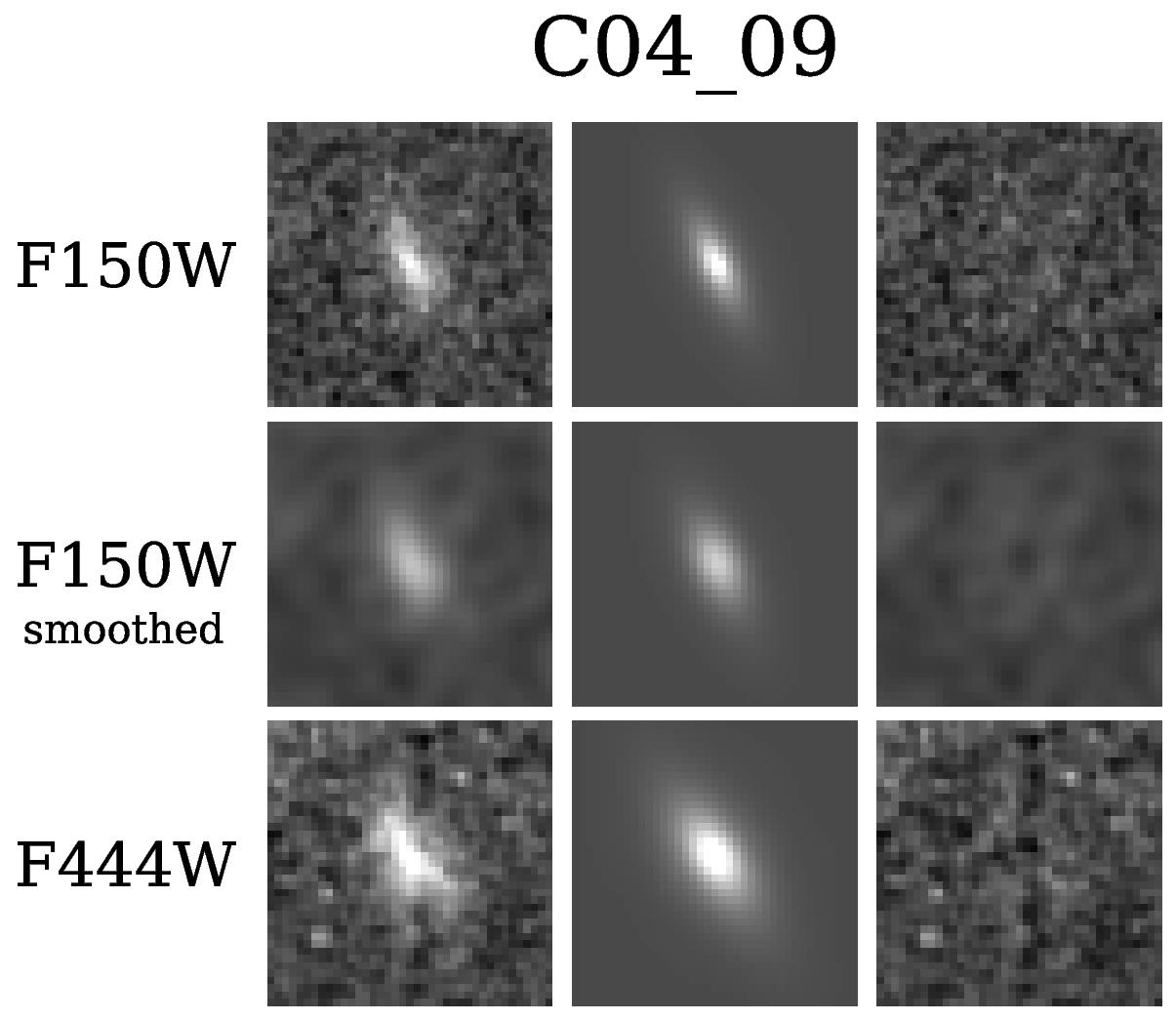}
   \includegraphics[height=0.14\textheight]{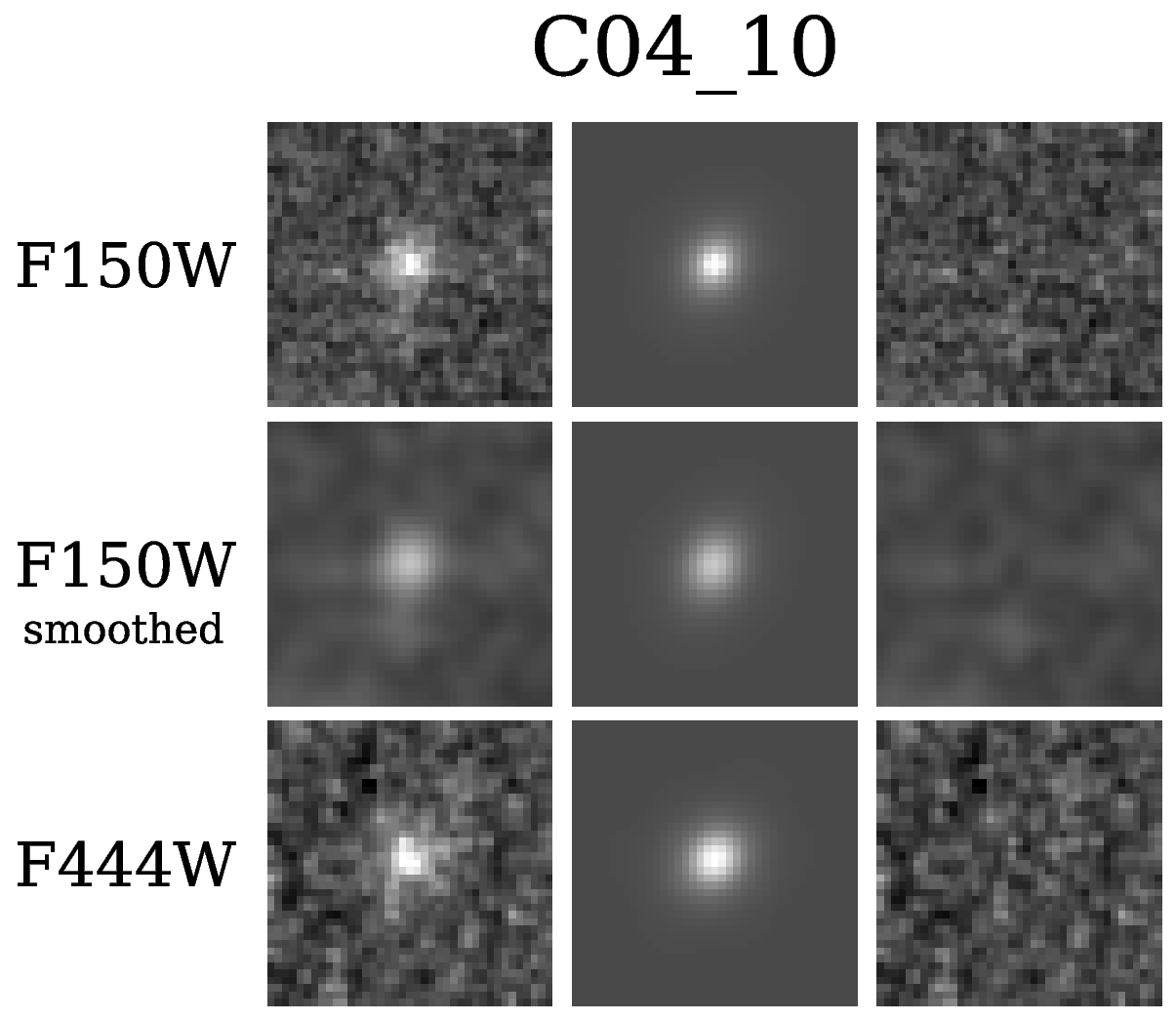}
   \includegraphics[height=0.14\textheight]{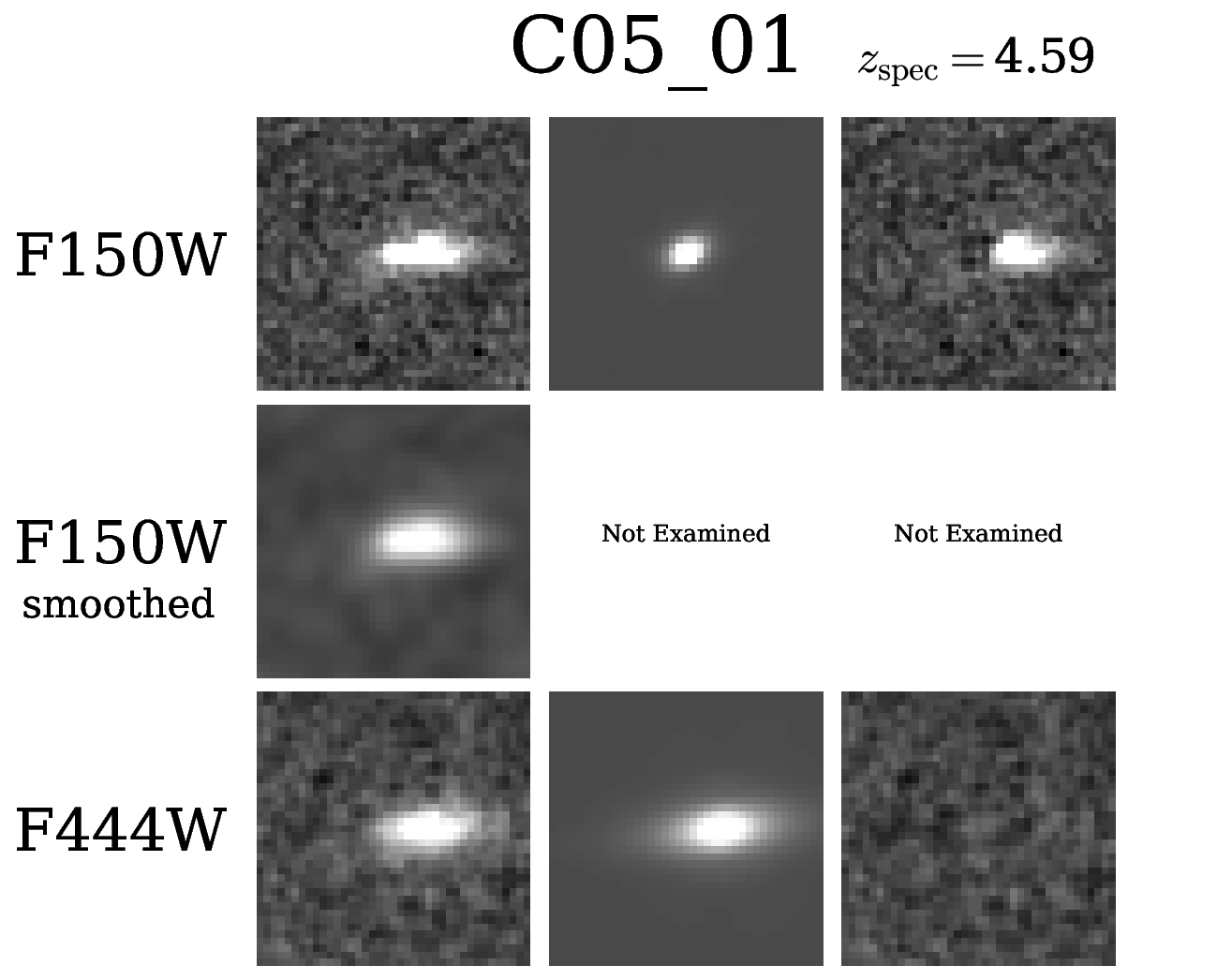}
   \includegraphics[height=0.14\textheight]{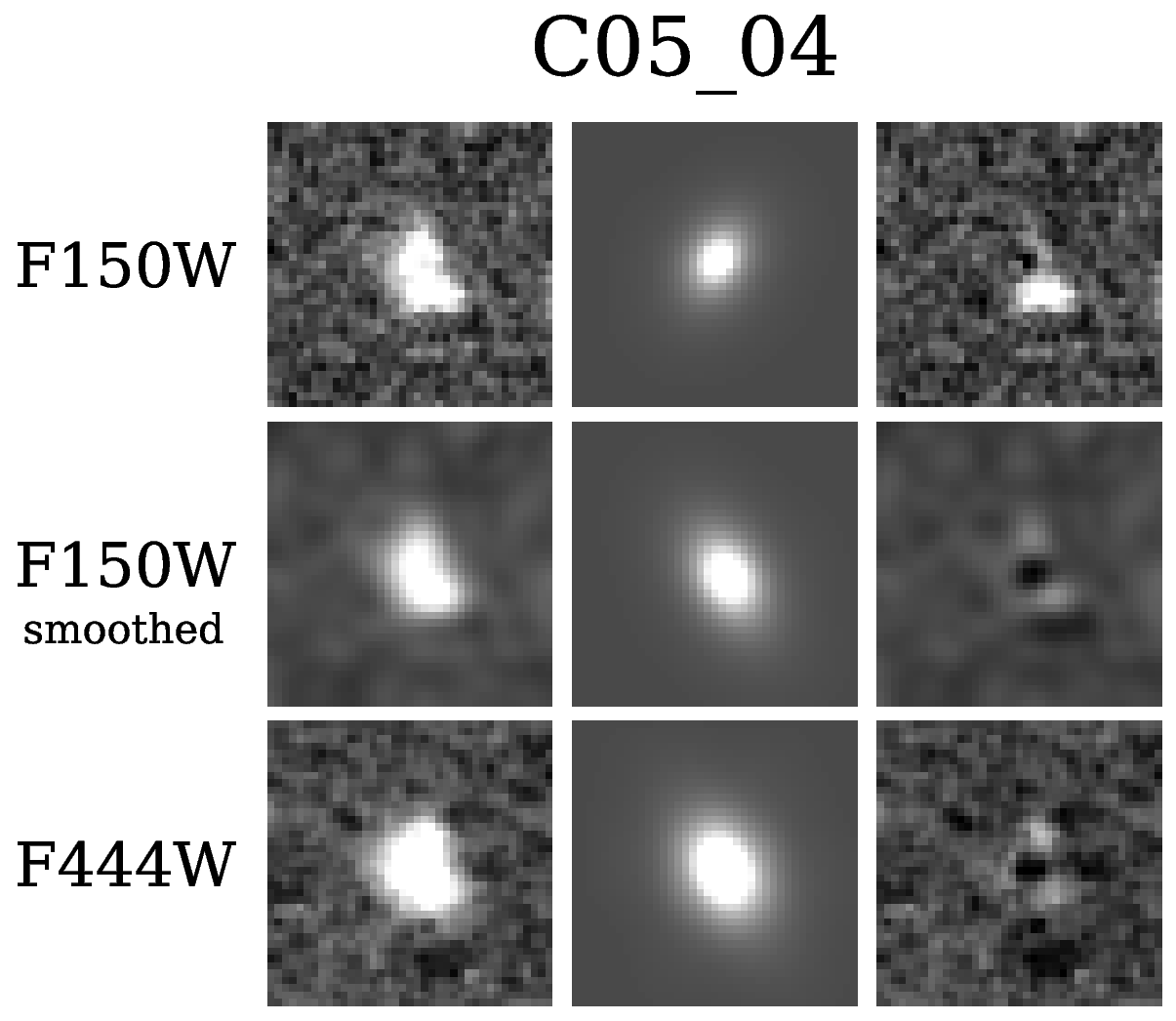}
   \includegraphics[height=0.14\textheight]{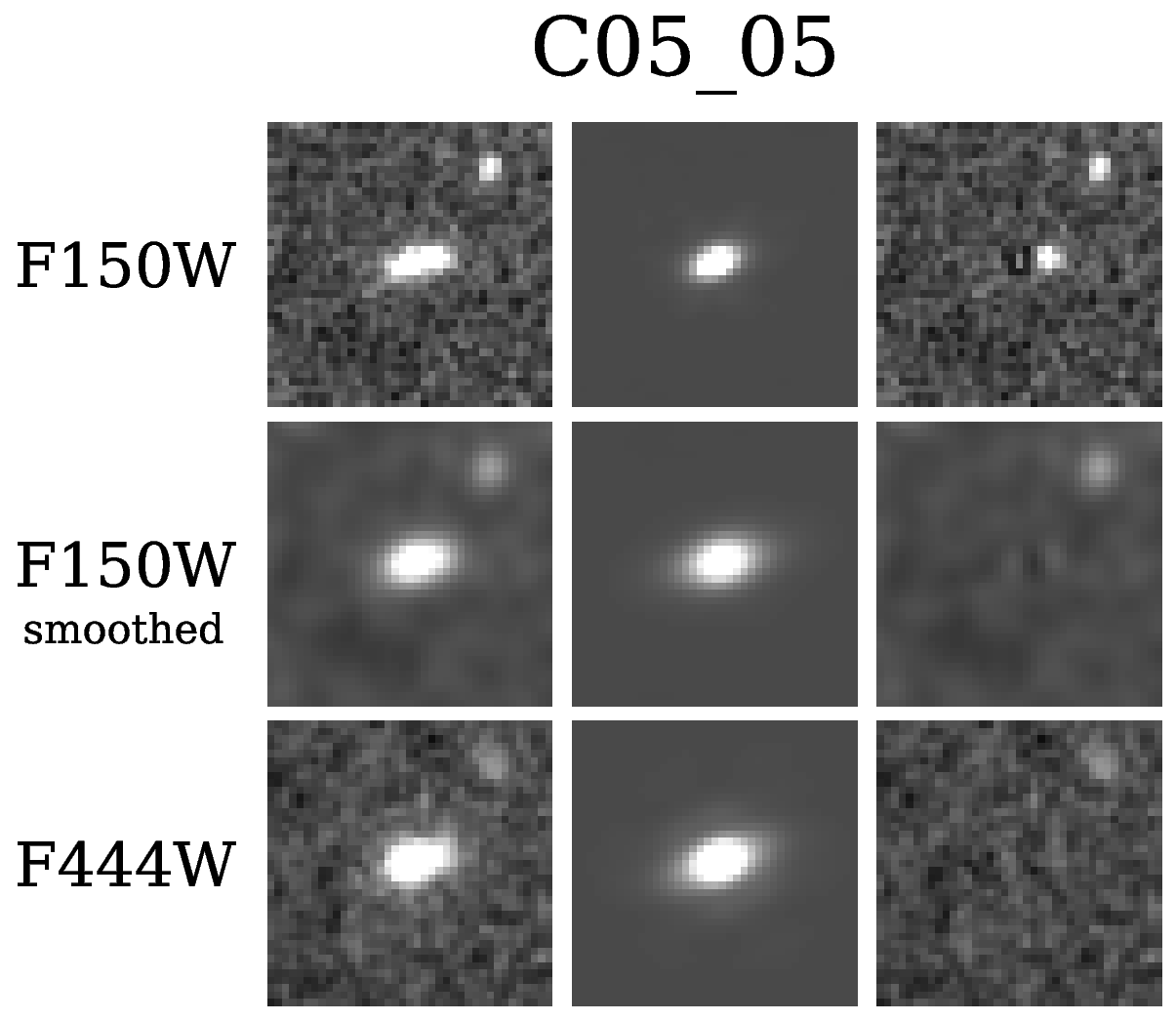}
\caption{
(Continued)
}
\end{center}
\end{figure*}

\addtocounter{figure}{-1}
\begin{figure*}
\begin{center}
   \includegraphics[height=0.14\textheight]{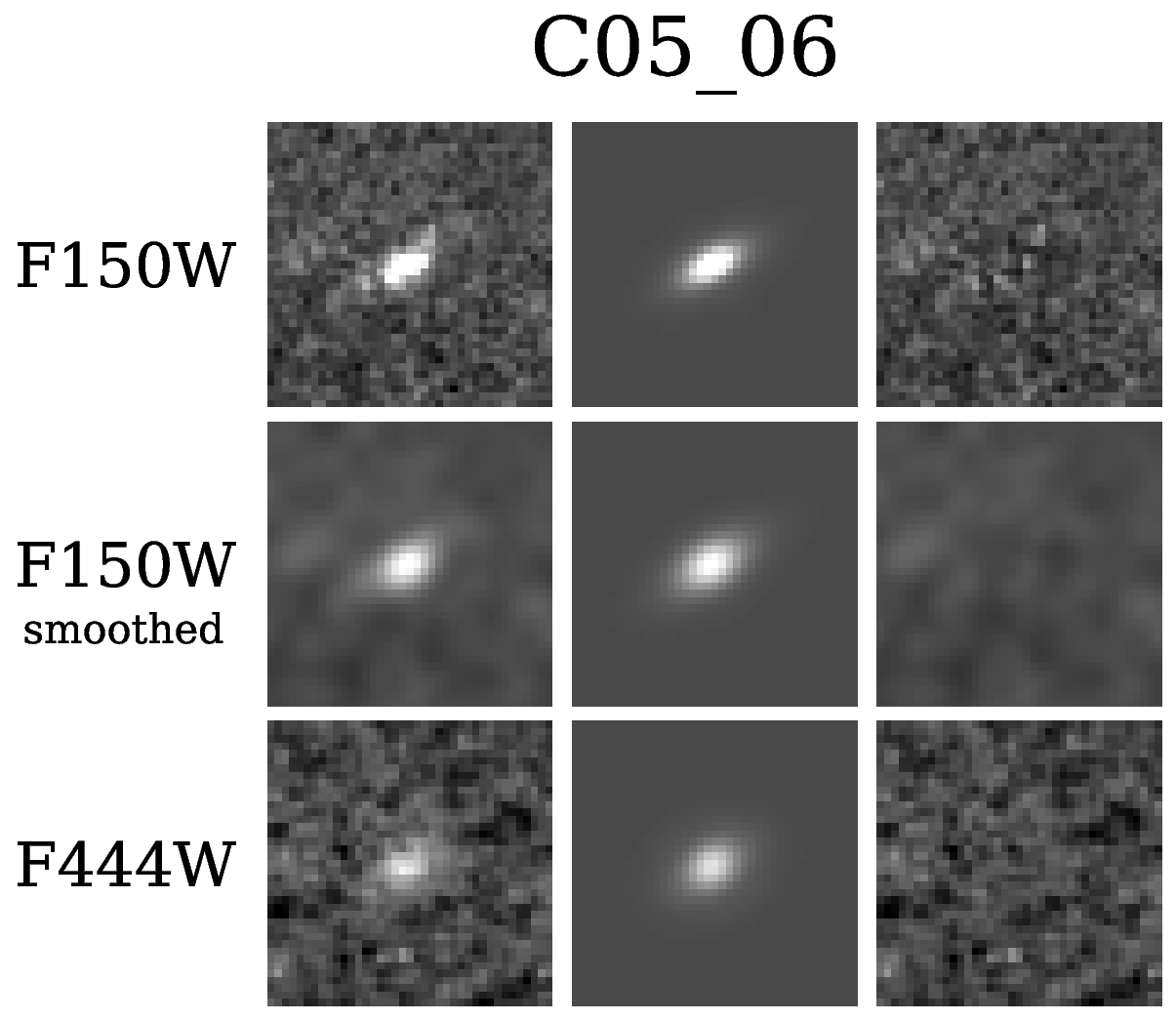}
   \includegraphics[height=0.14\textheight]{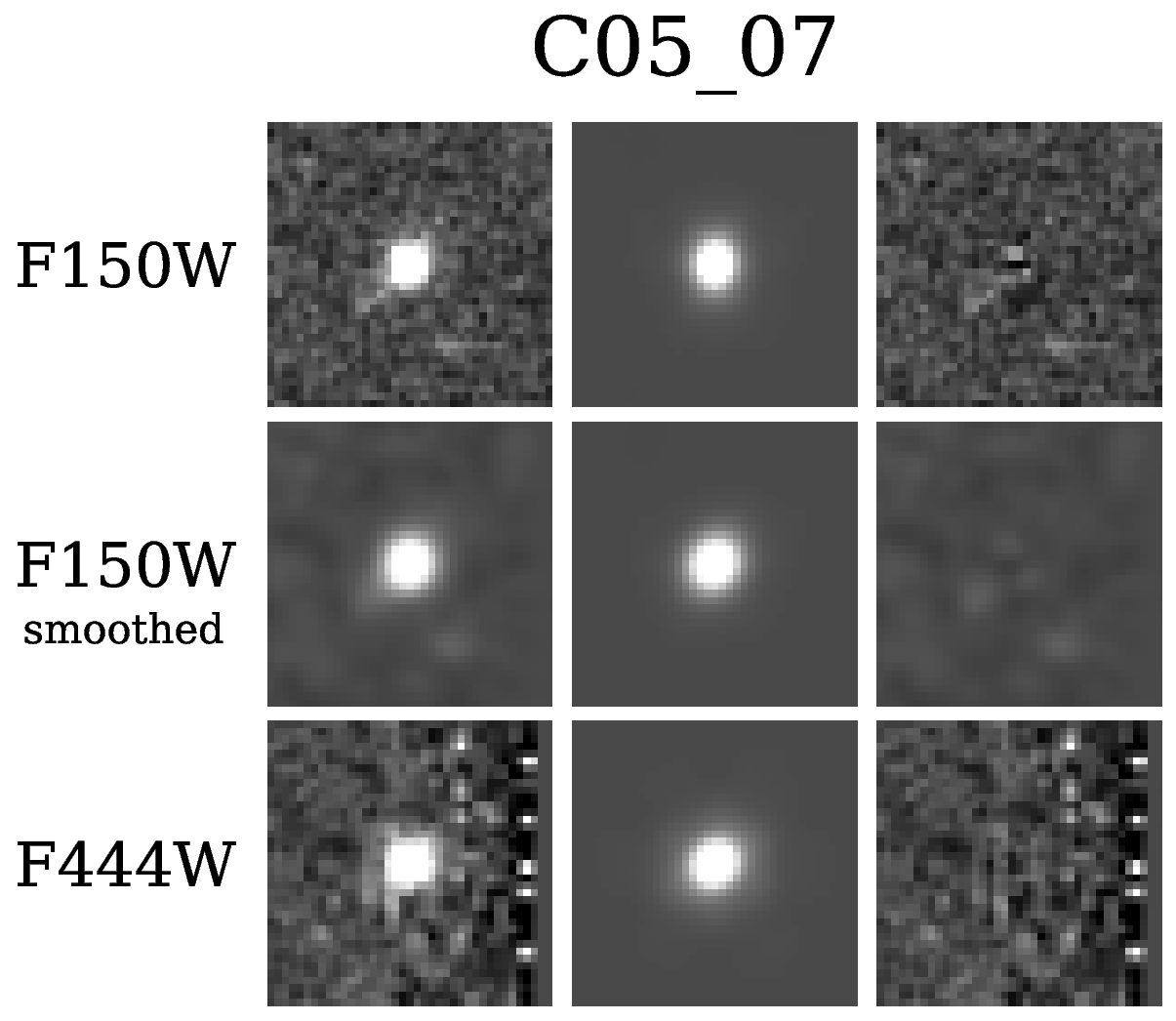}
   \includegraphics[height=0.14\textheight]{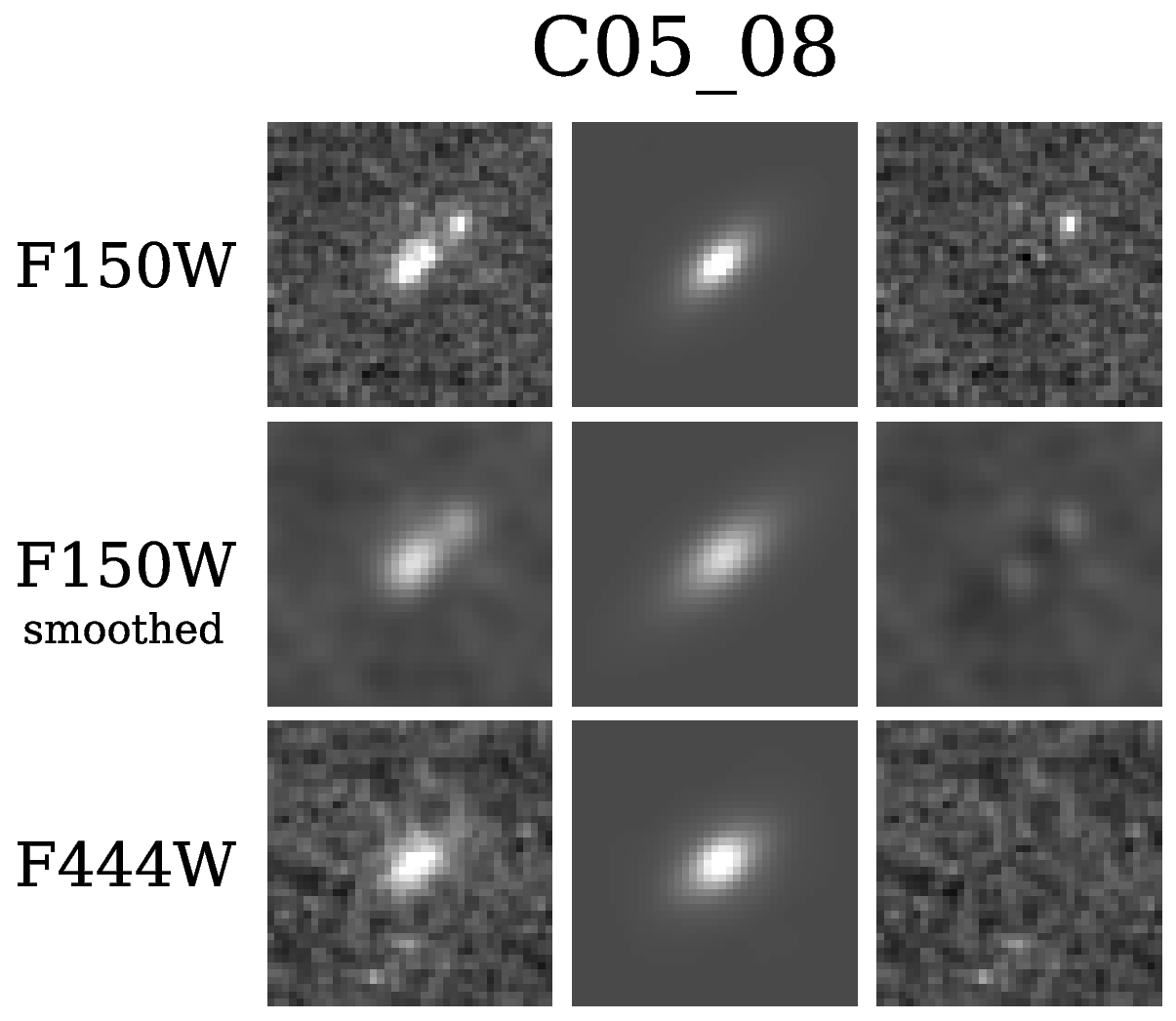}
   \includegraphics[height=0.14\textheight]{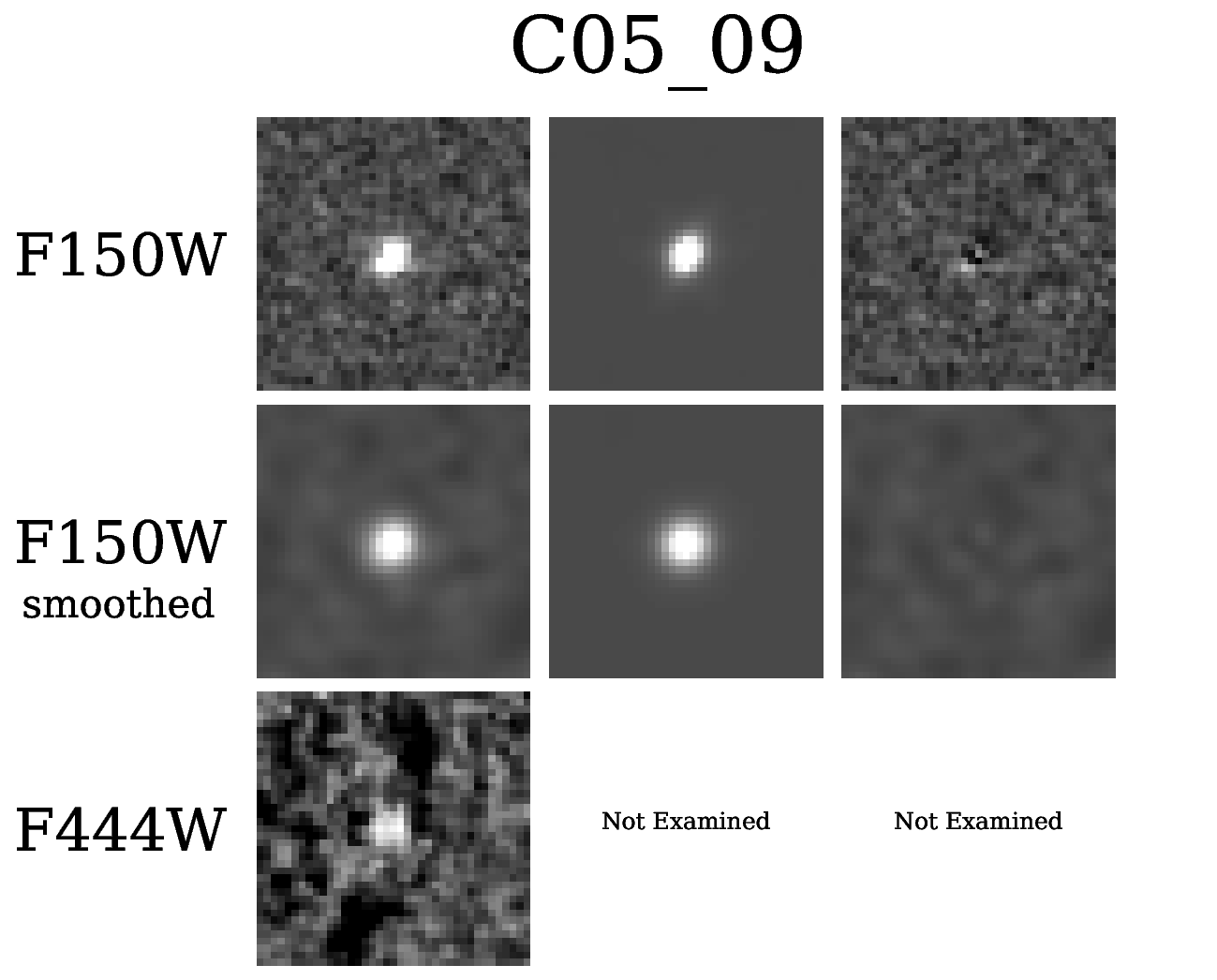}
   \includegraphics[height=0.14\textheight]{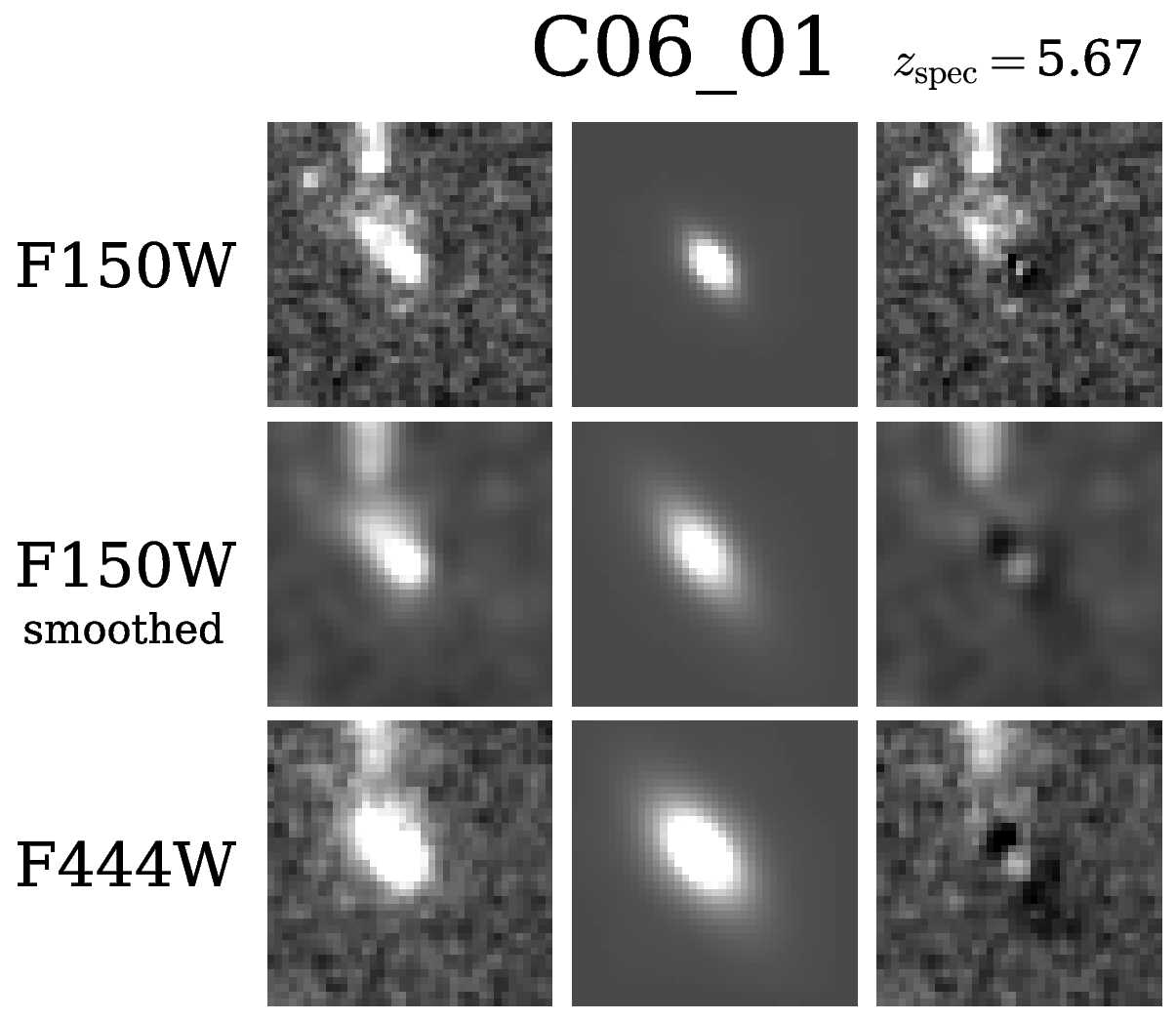}
   \includegraphics[height=0.14\textheight]{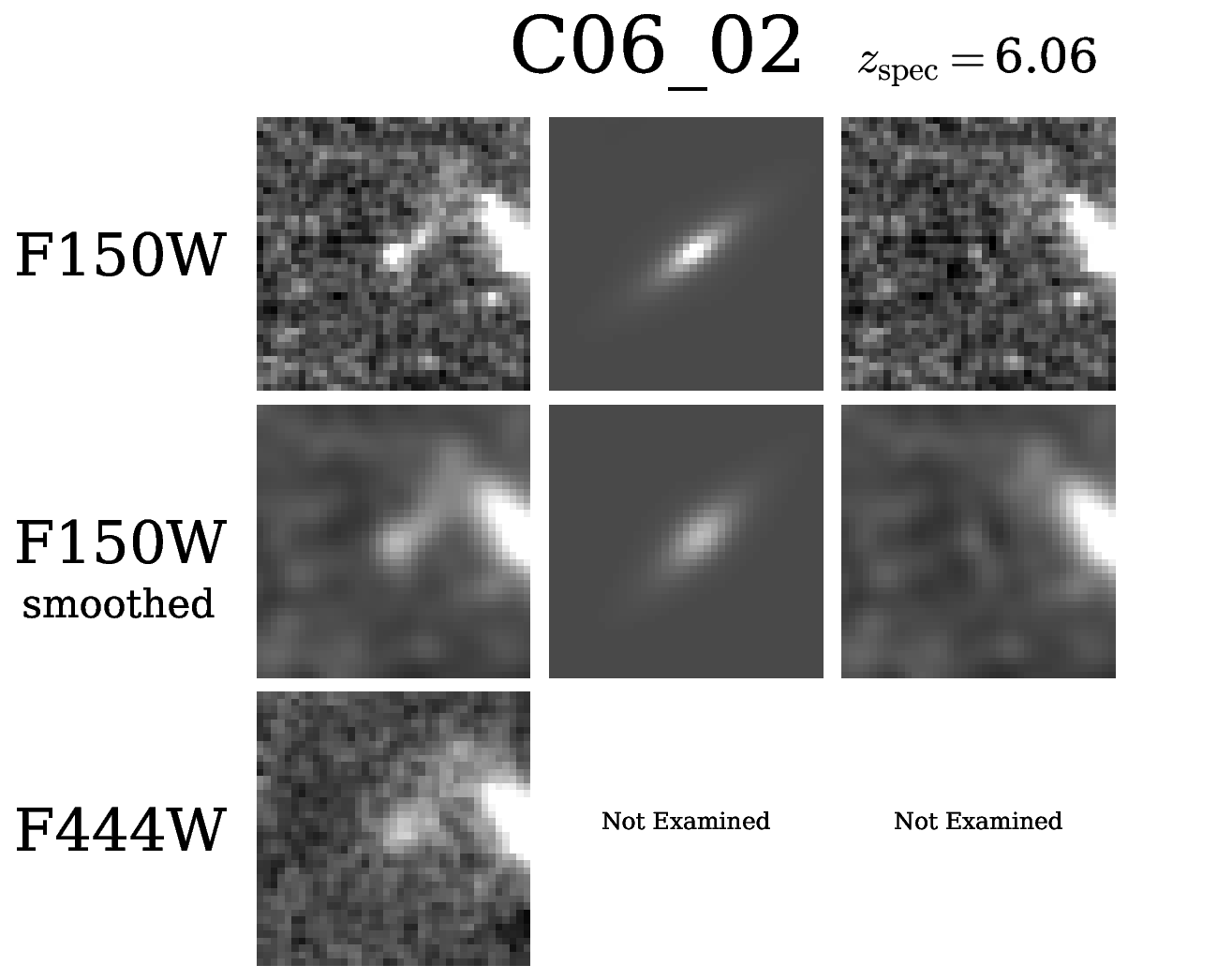}
   \includegraphics[height=0.14\textheight]{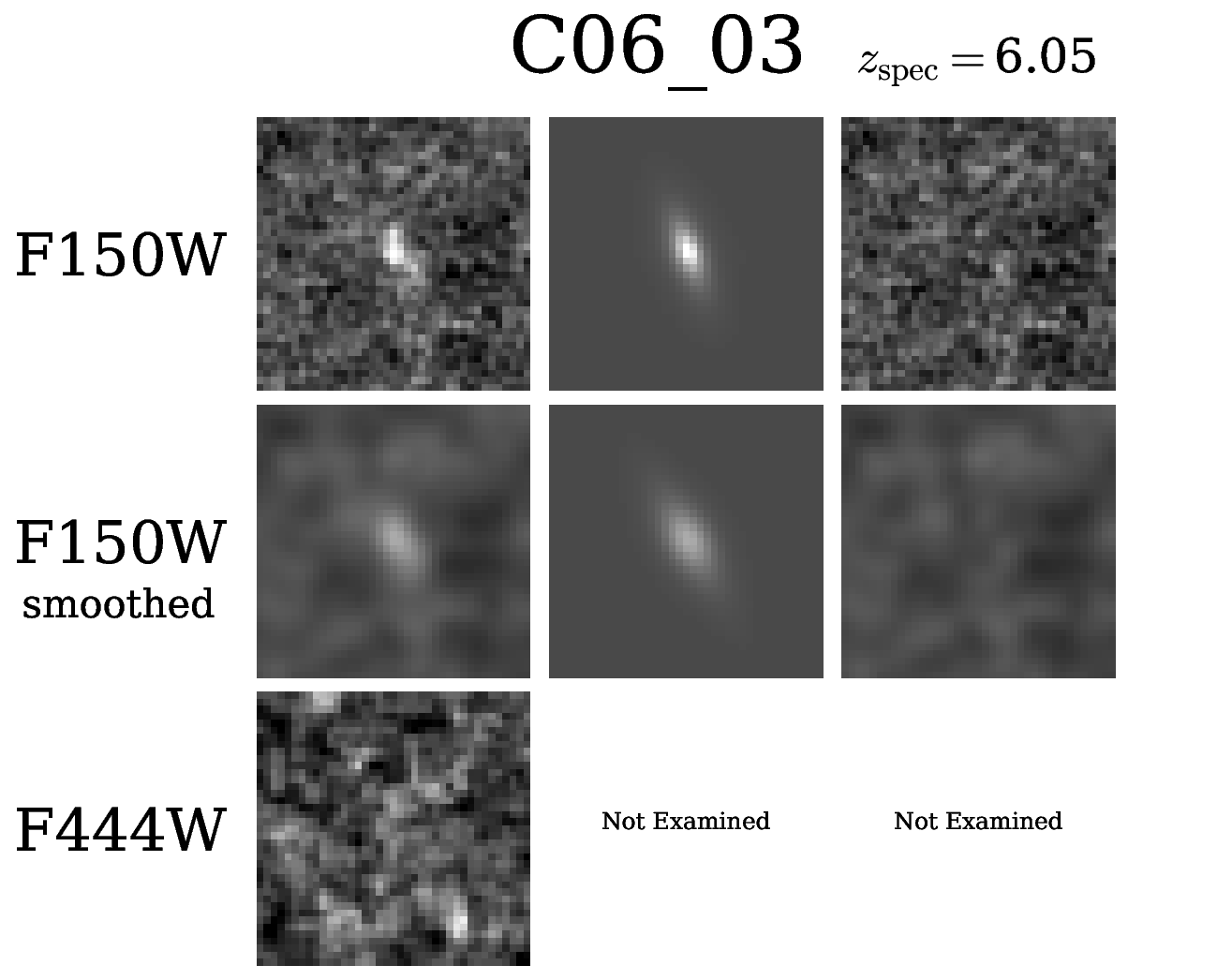}
   \includegraphics[height=0.14\textheight]{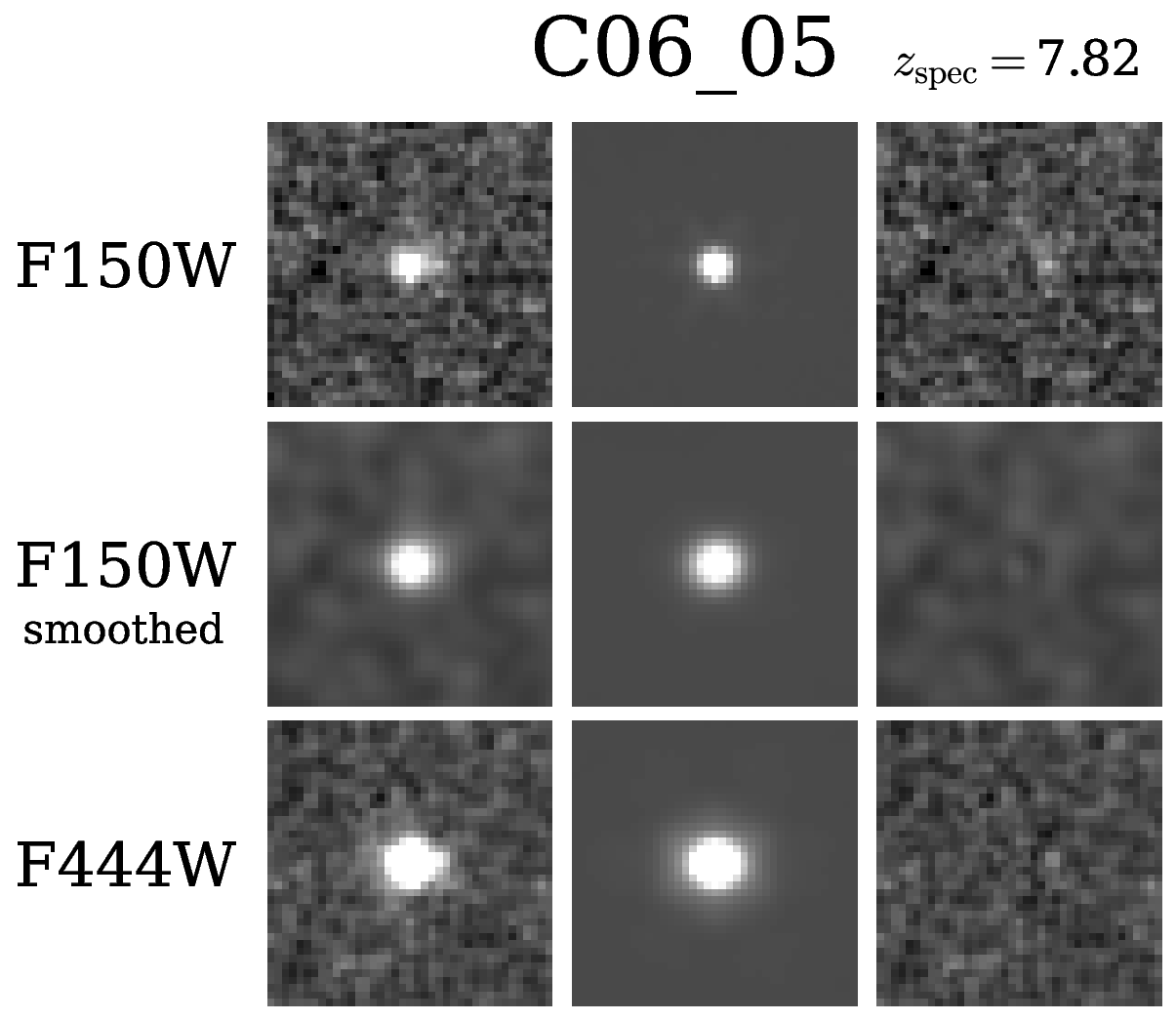}
   \includegraphics[height=0.14\textheight]{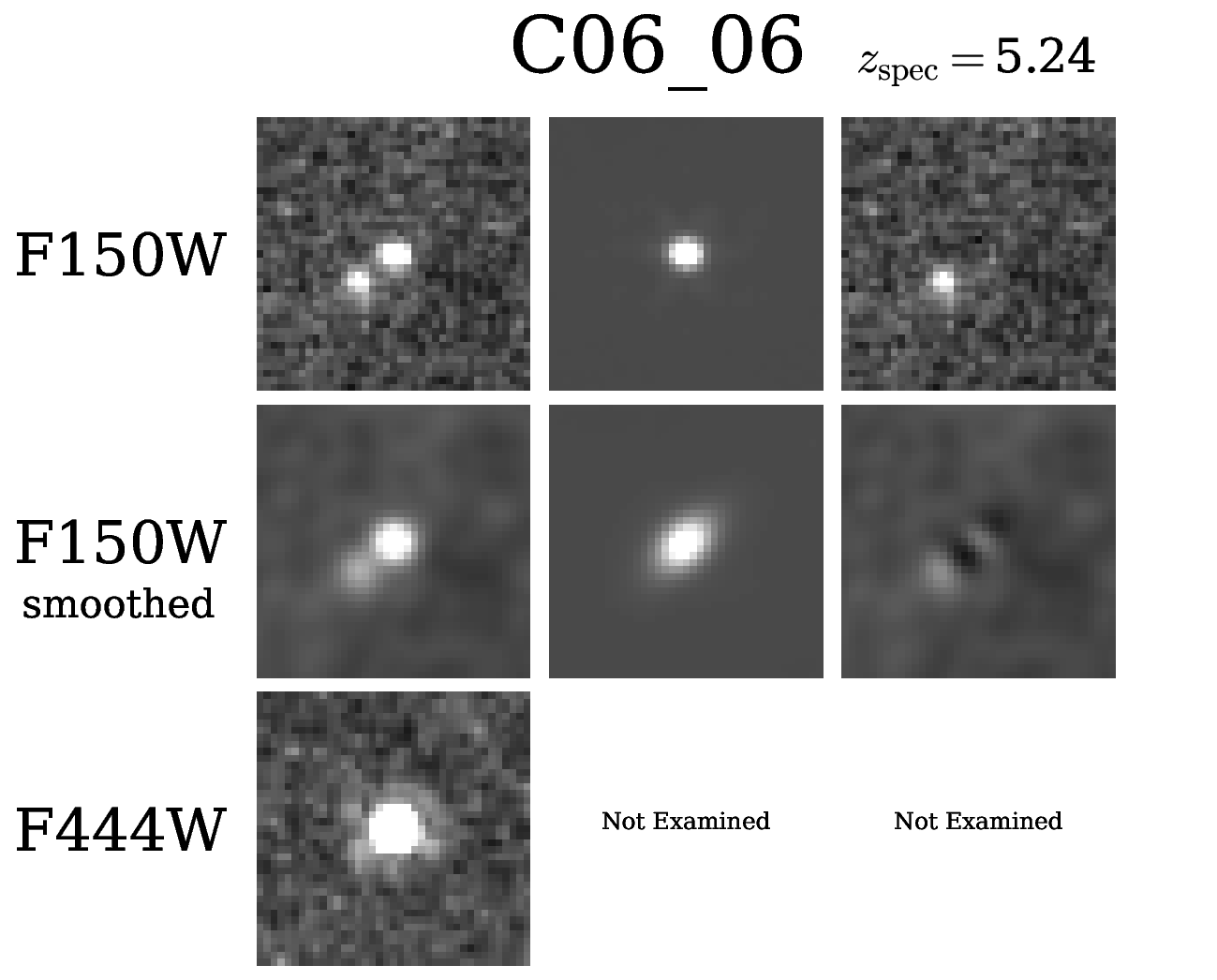}
   \includegraphics[height=0.14\textheight]{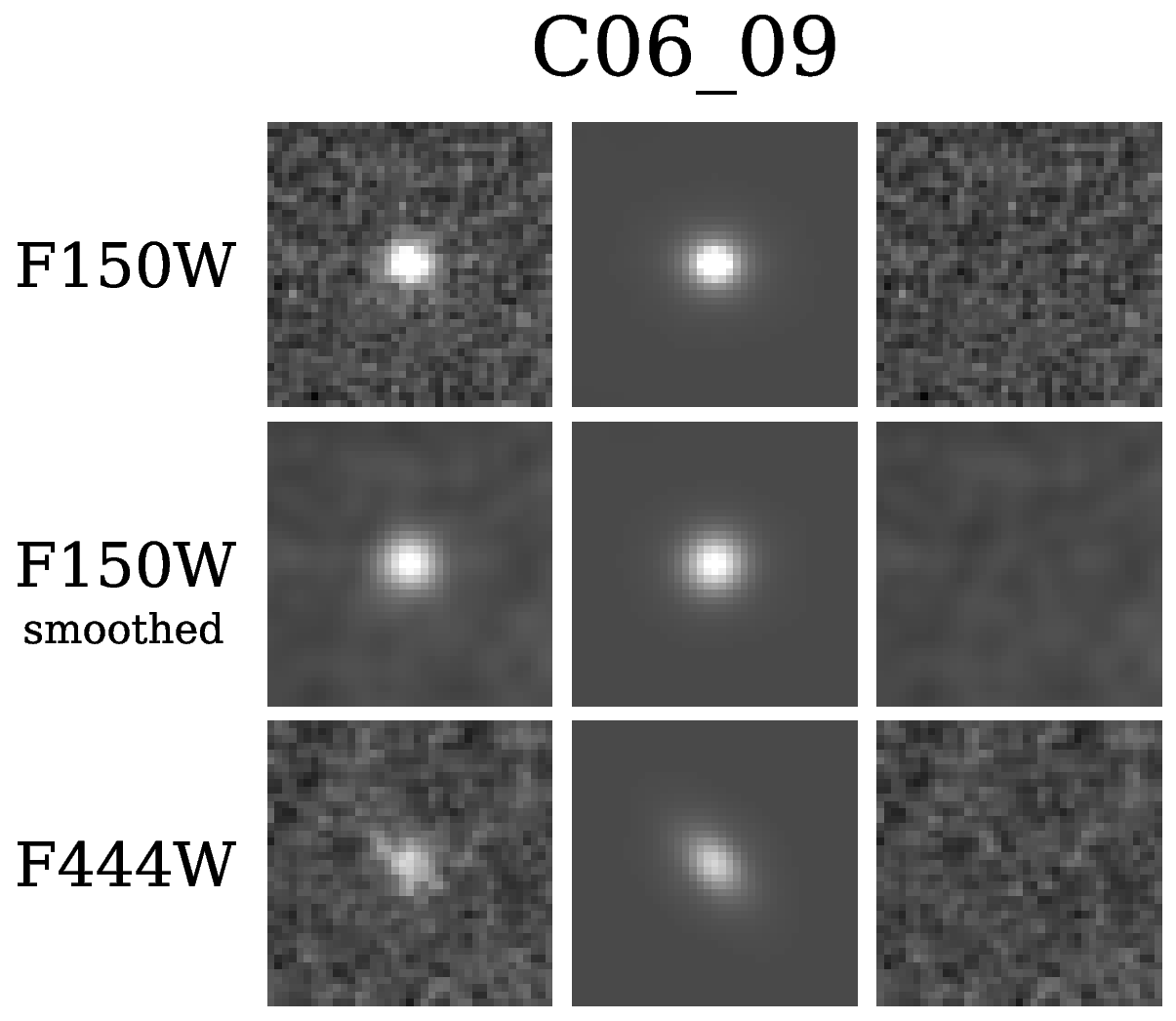}
   \includegraphics[height=0.14\textheight]{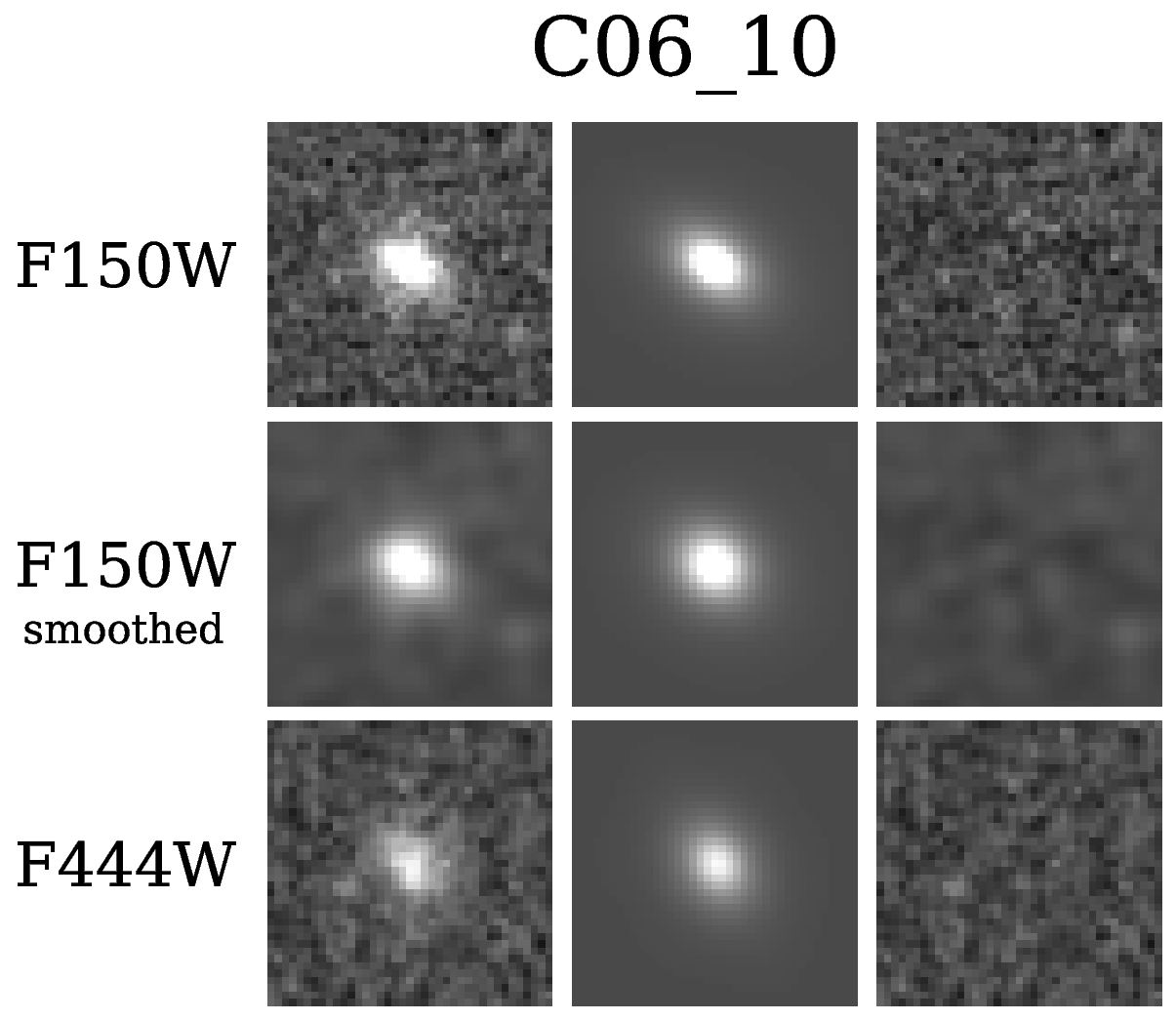}
   \includegraphics[height=0.14\textheight]{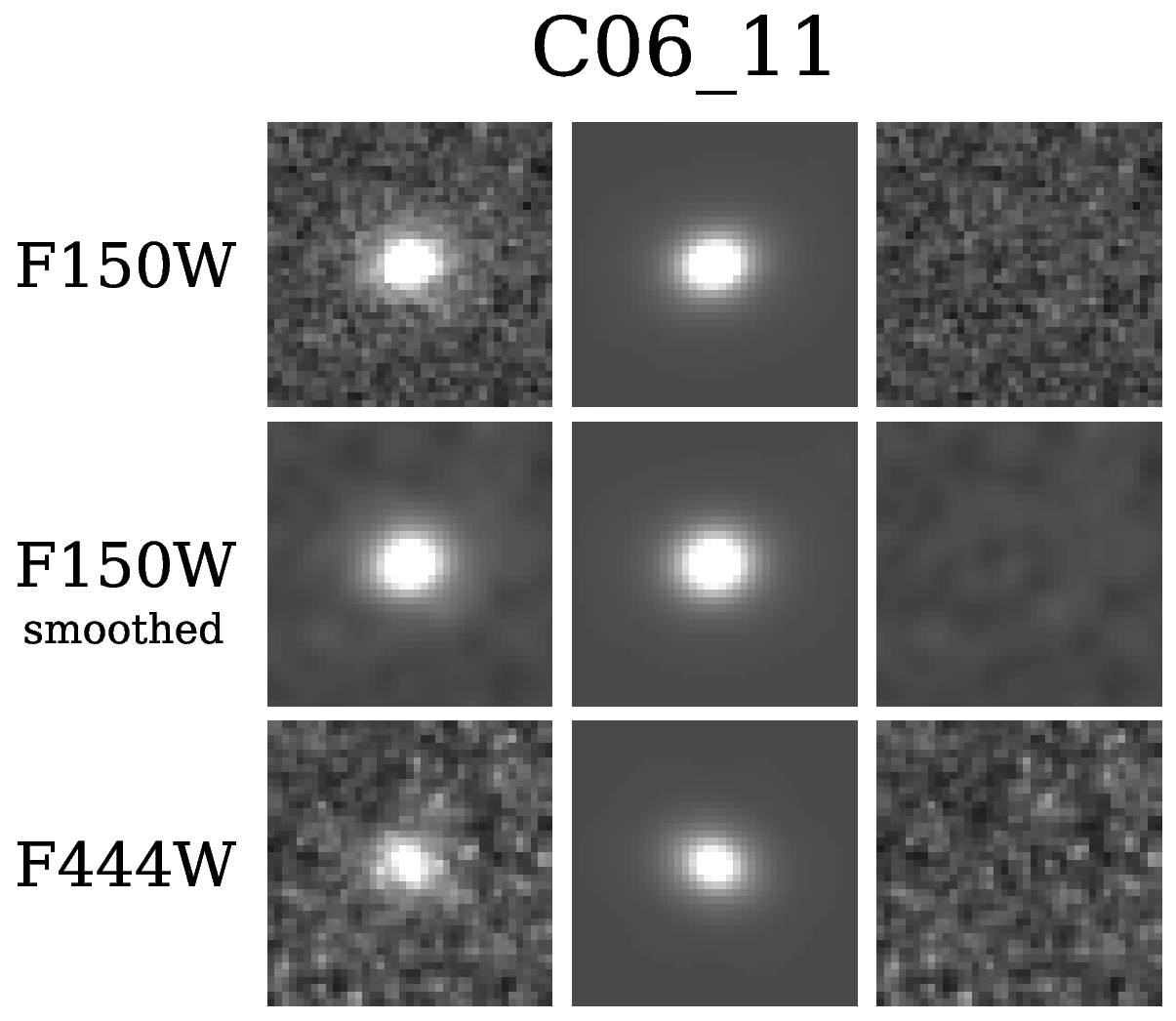}
   \includegraphics[height=0.14\textheight]{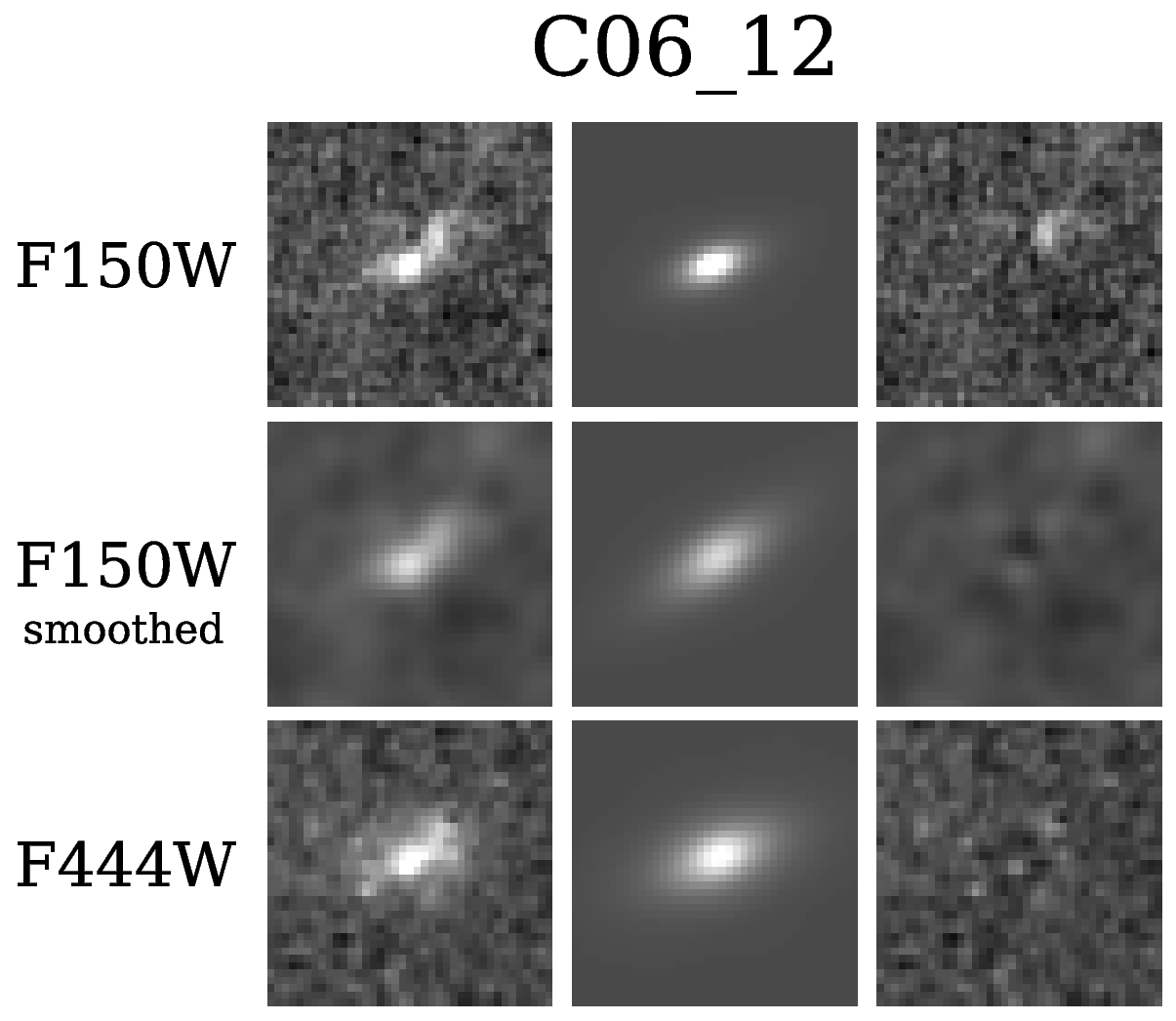}
   \includegraphics[height=0.14\textheight]{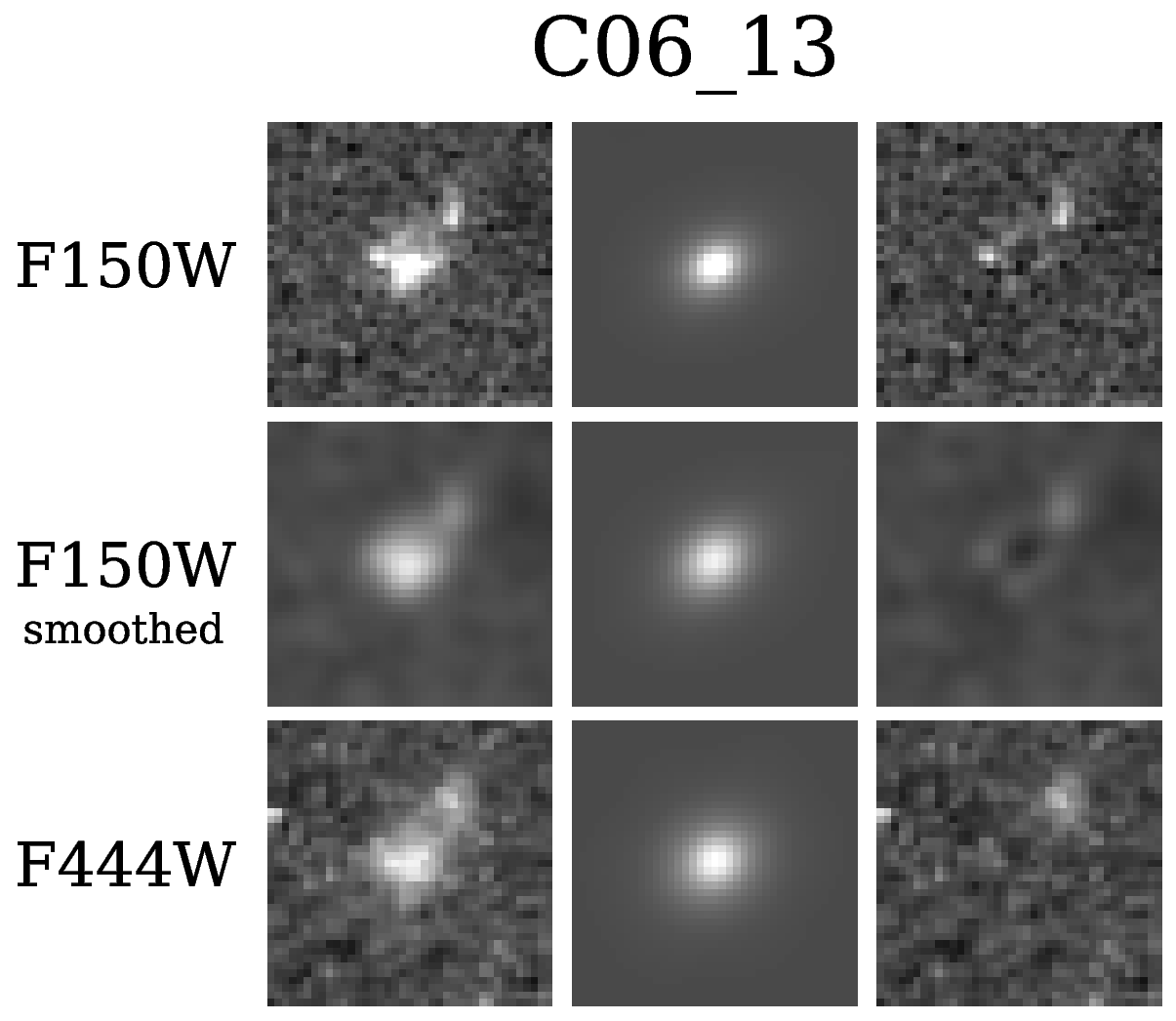}
   \includegraphics[height=0.14\textheight]{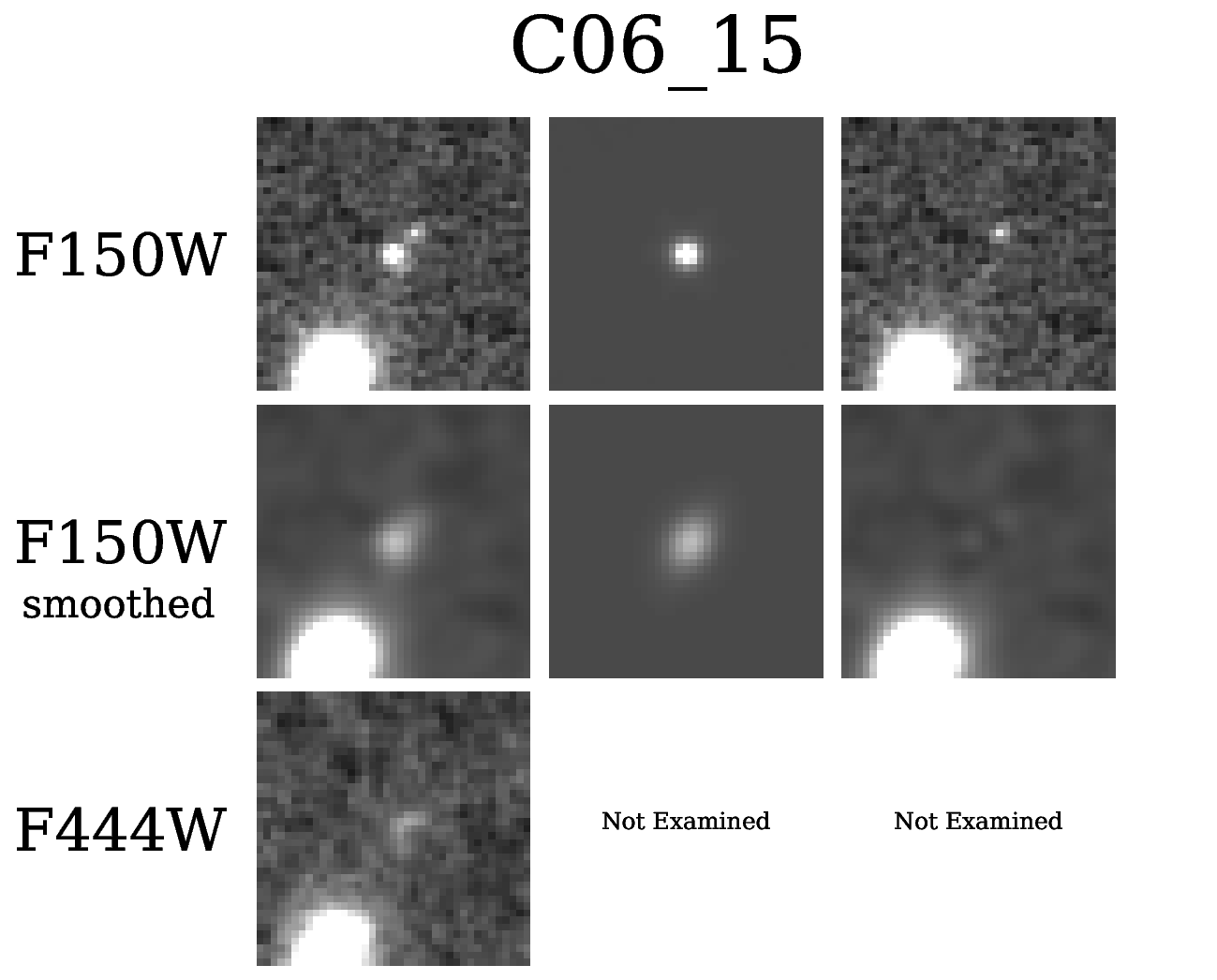}
   \includegraphics[height=0.14\textheight]{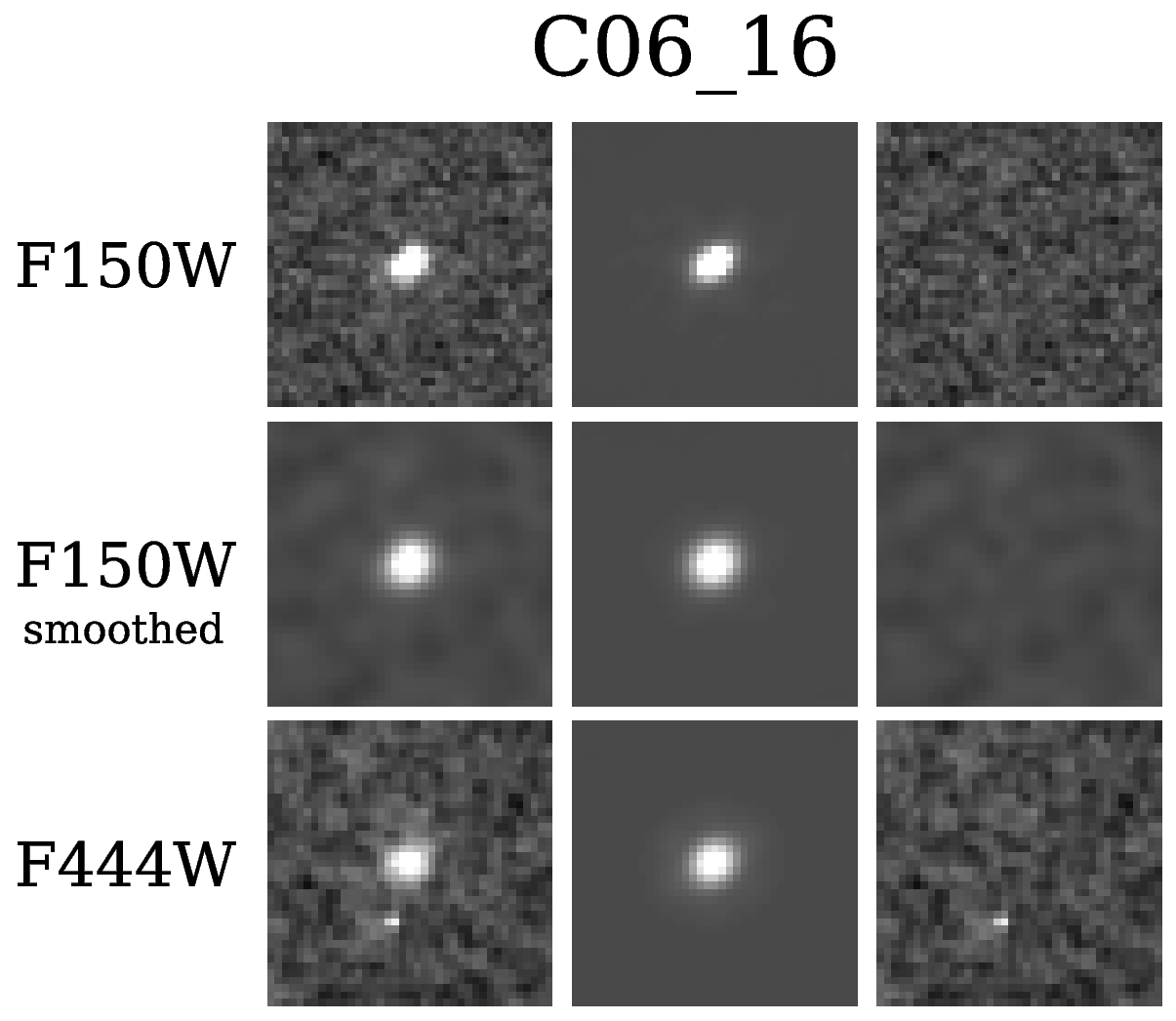}
   \includegraphics[height=0.14\textheight]{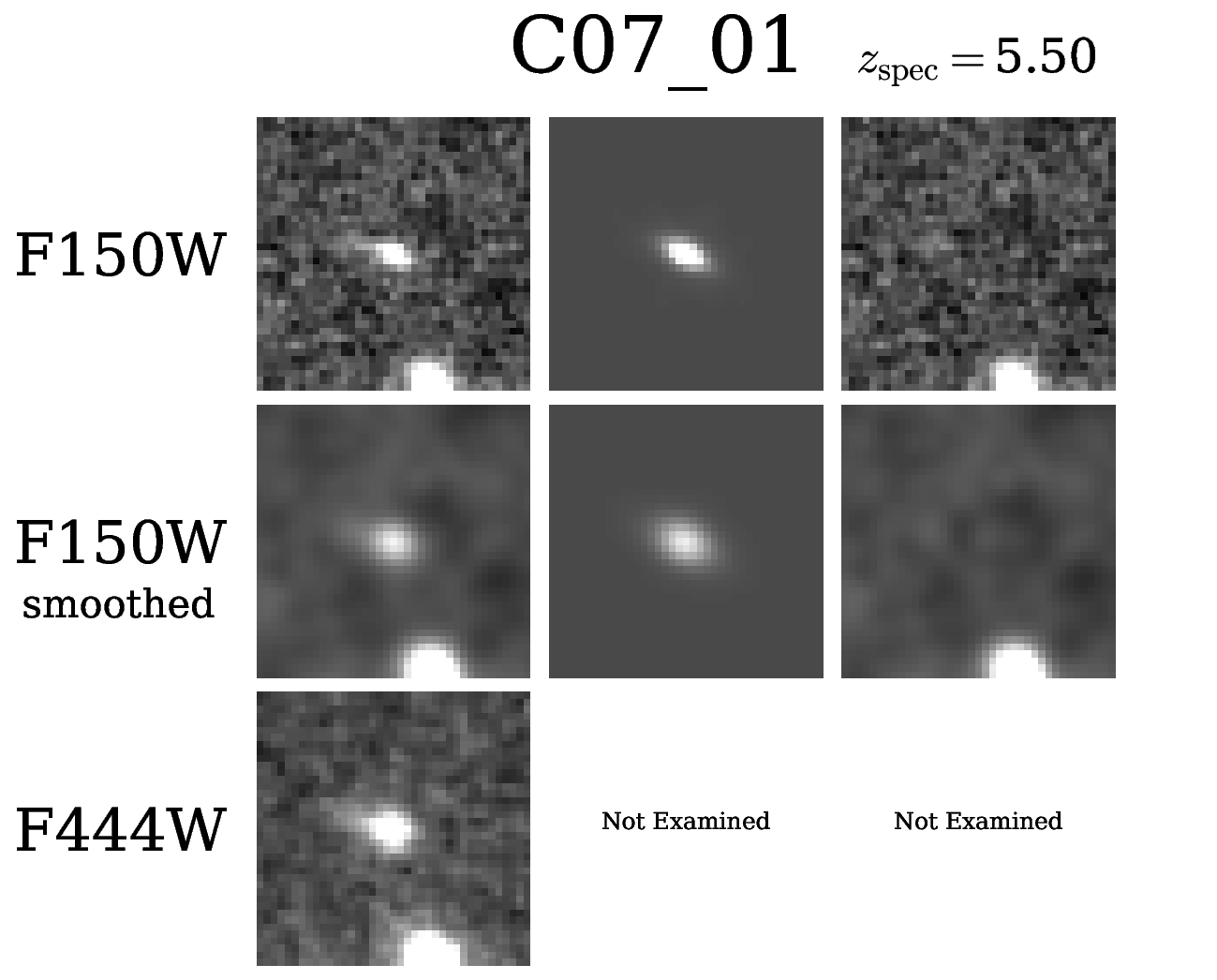}
   \includegraphics[height=0.14\textheight]{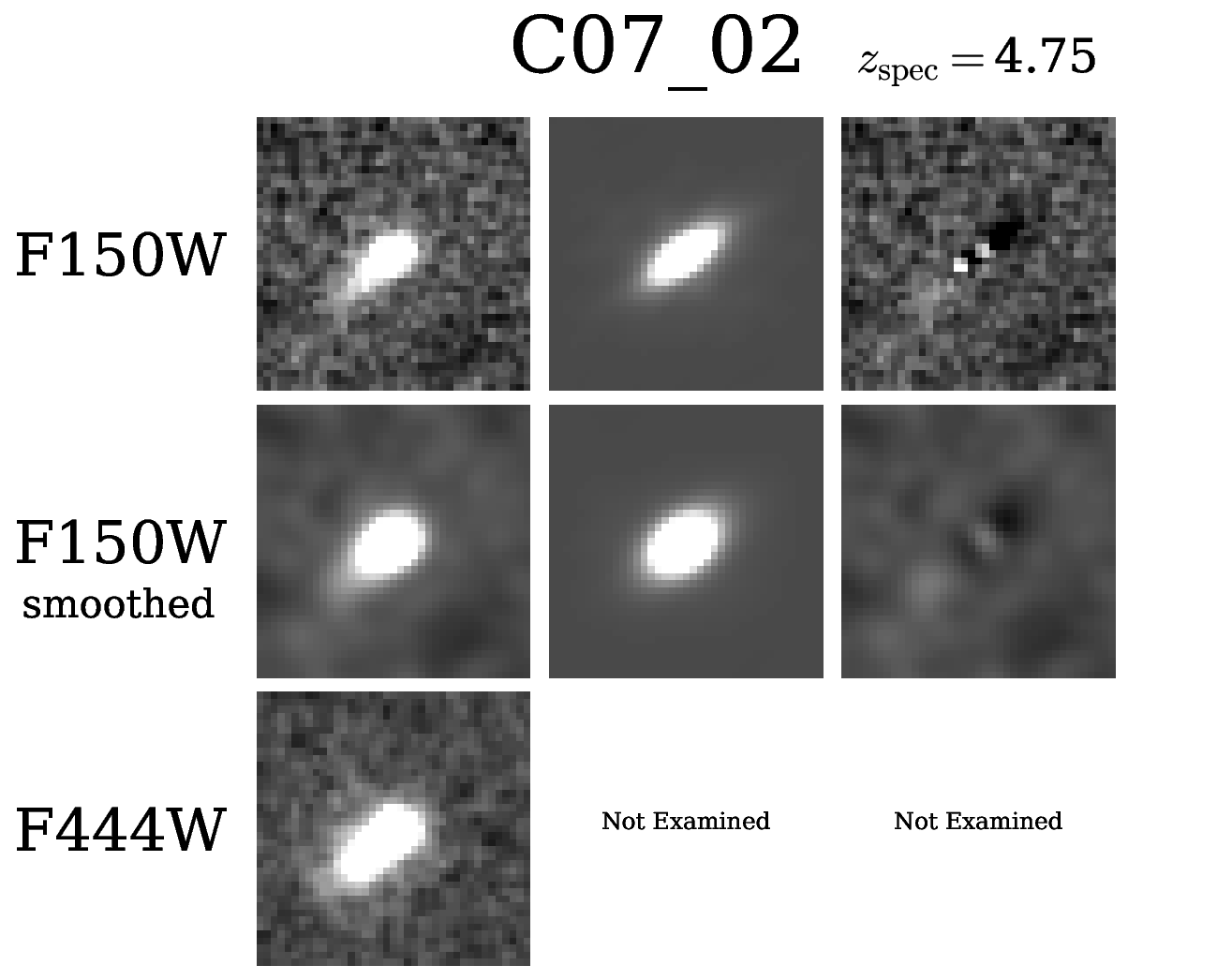}
   \includegraphics[height=0.14\textheight]{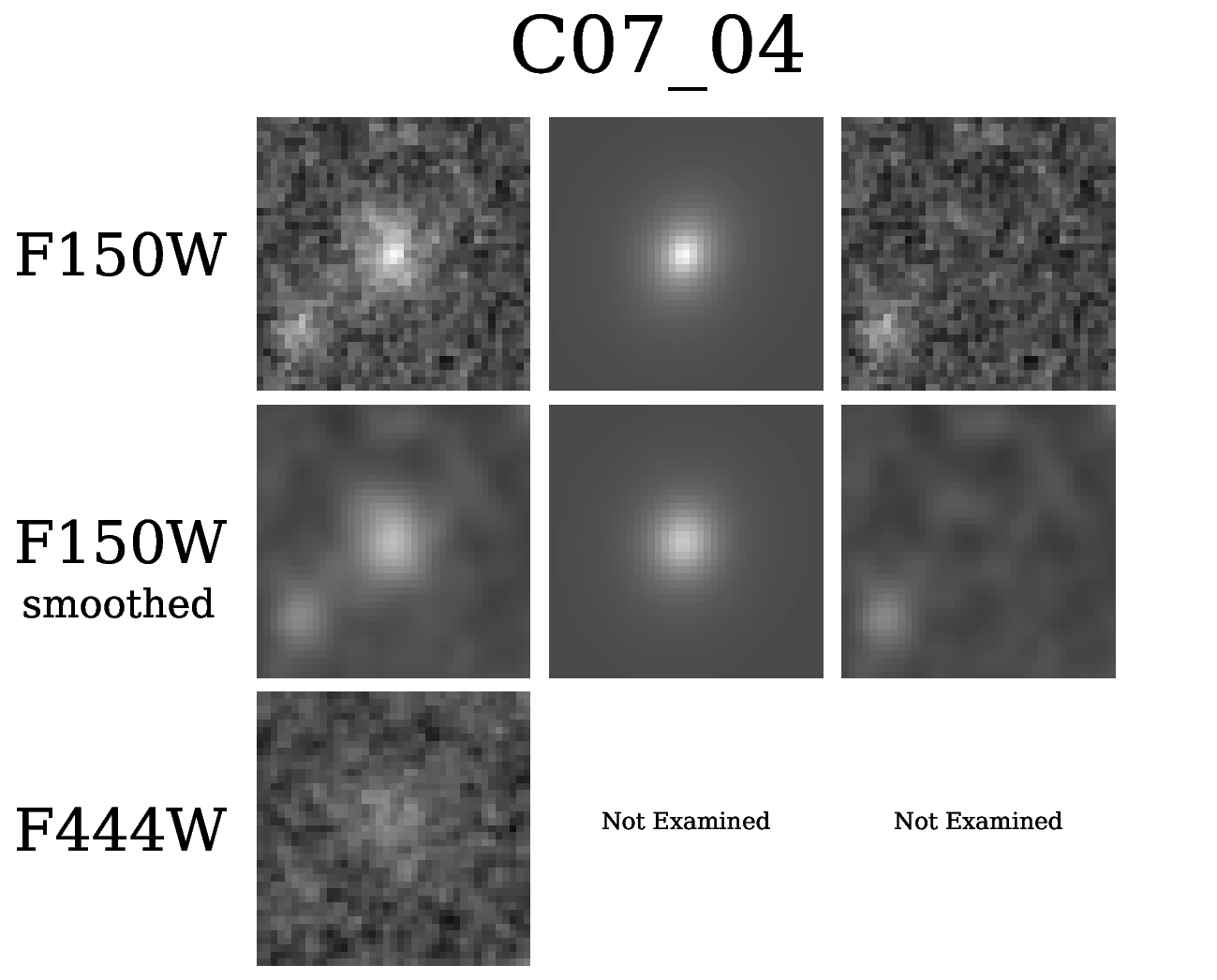}
   \includegraphics[height=0.14\textheight]{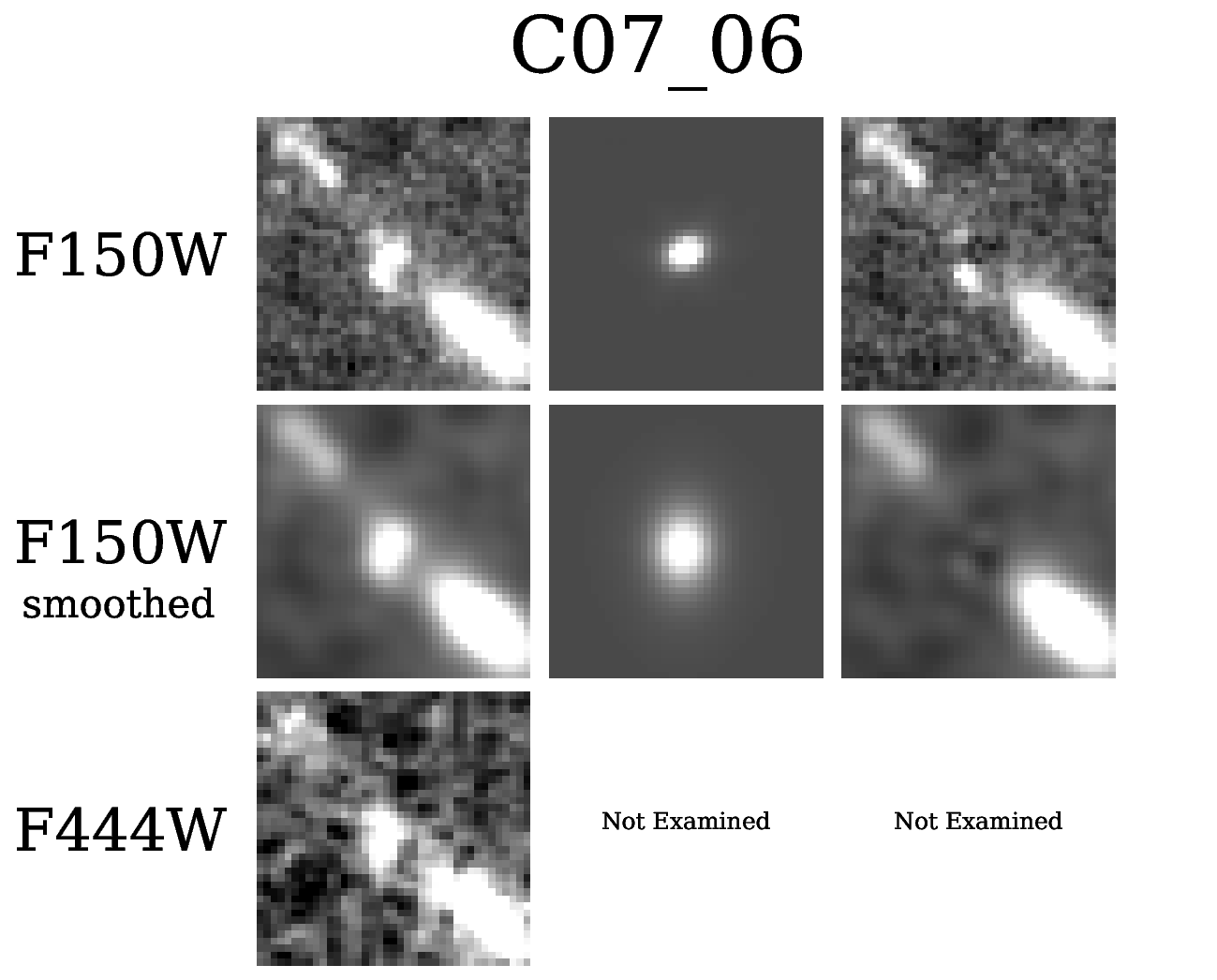}
   \includegraphics[height=0.14\textheight]{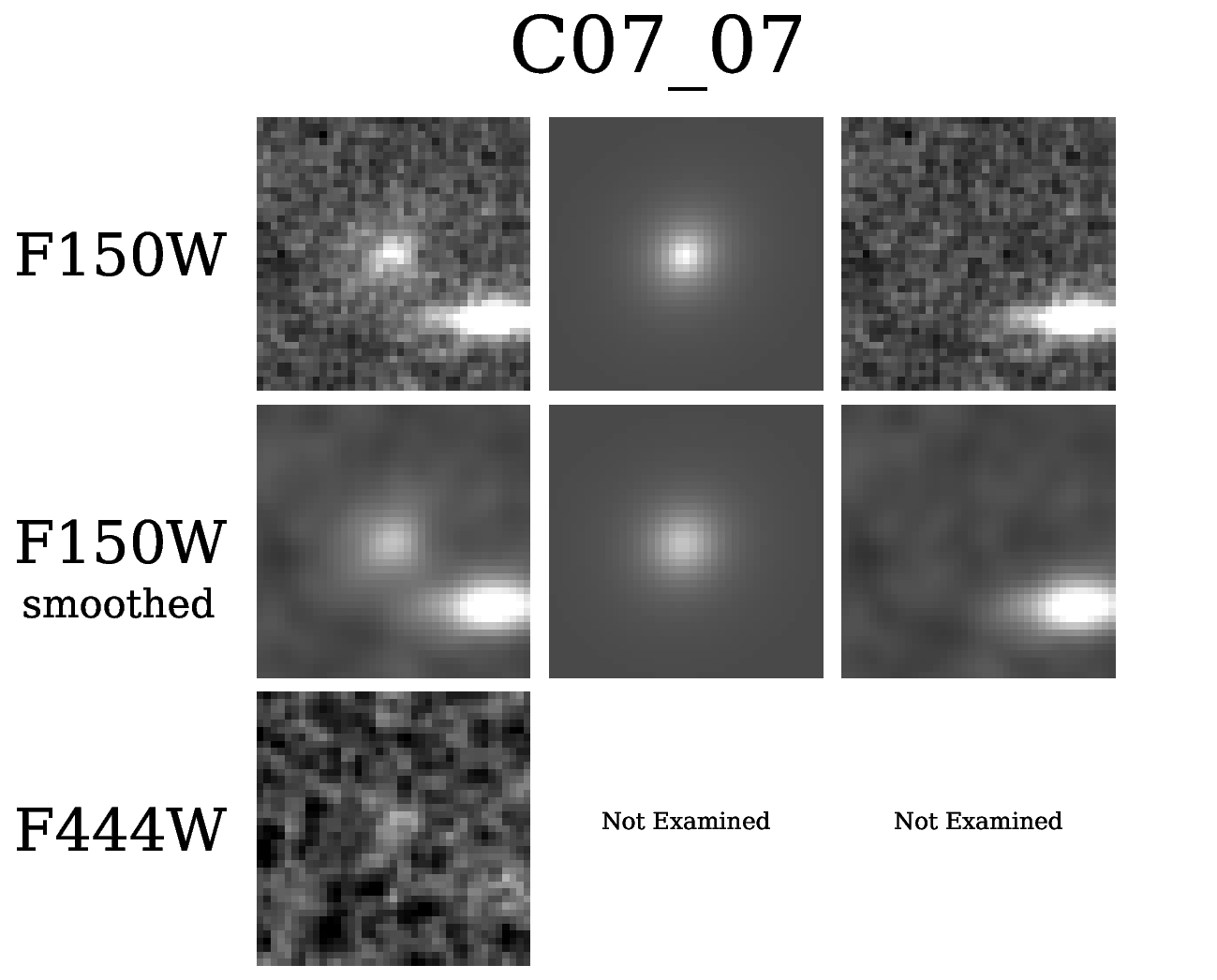}
   \includegraphics[height=0.14\textheight]{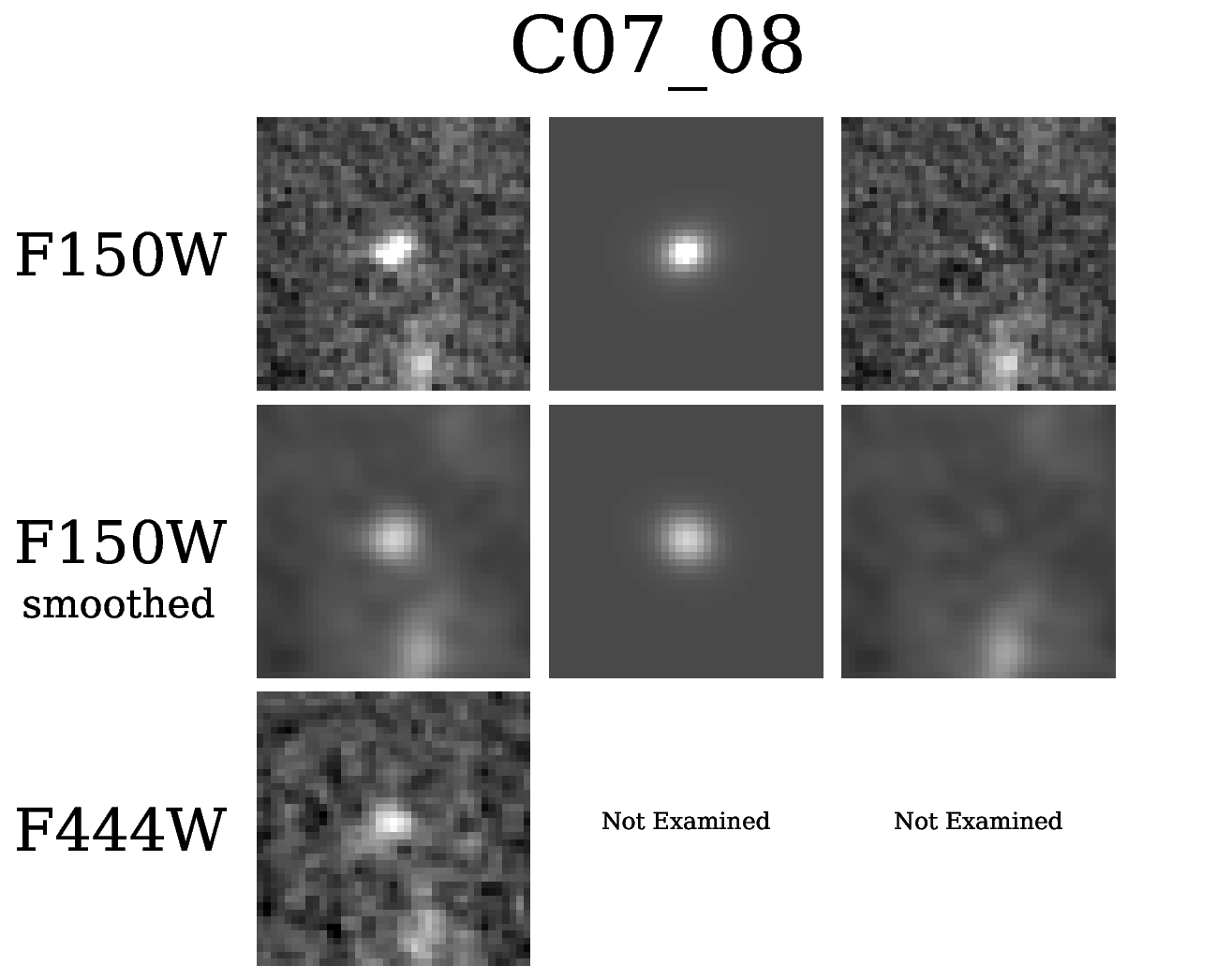}
   \includegraphics[height=0.14\textheight]{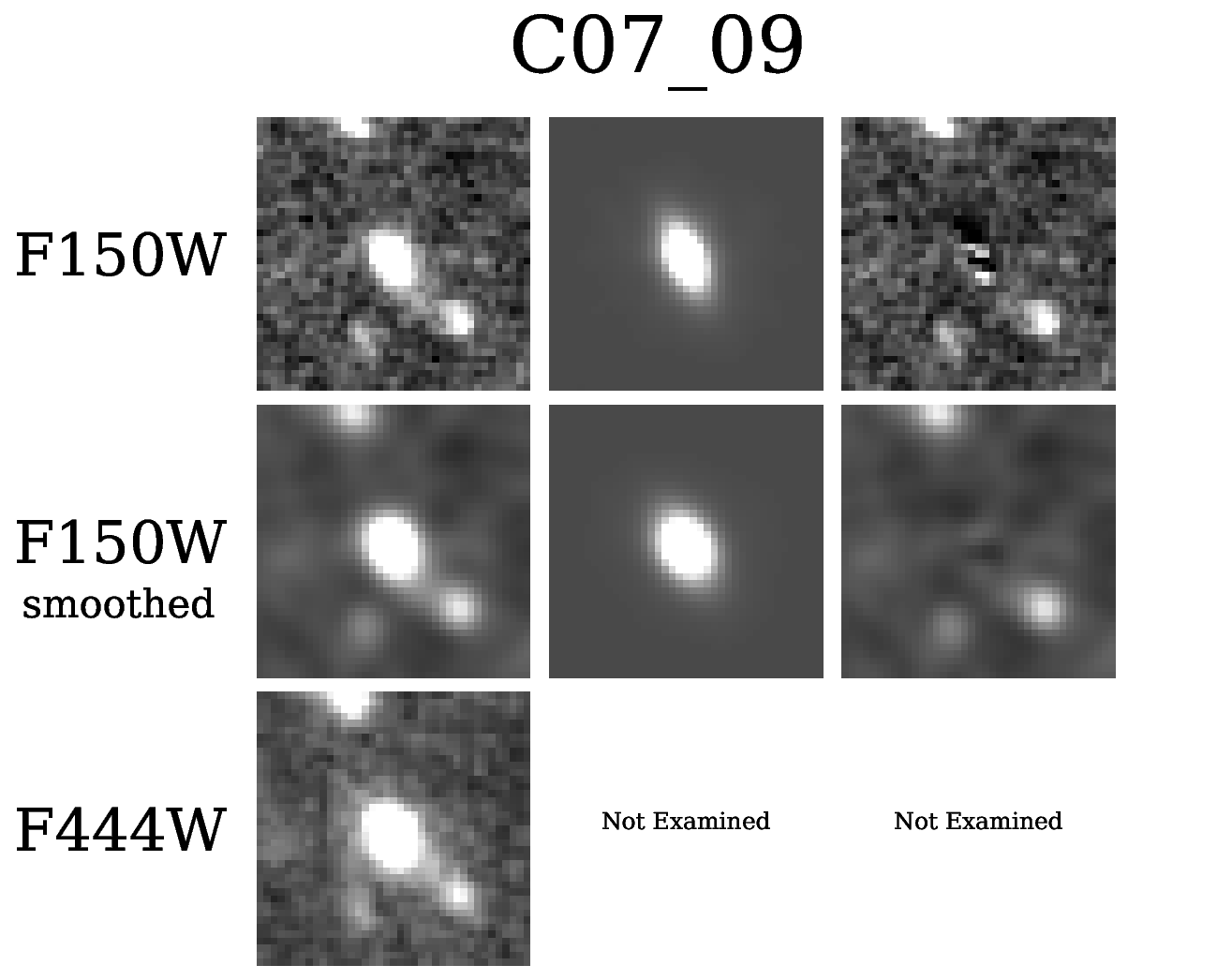}
   \includegraphics[height=0.14\textheight]{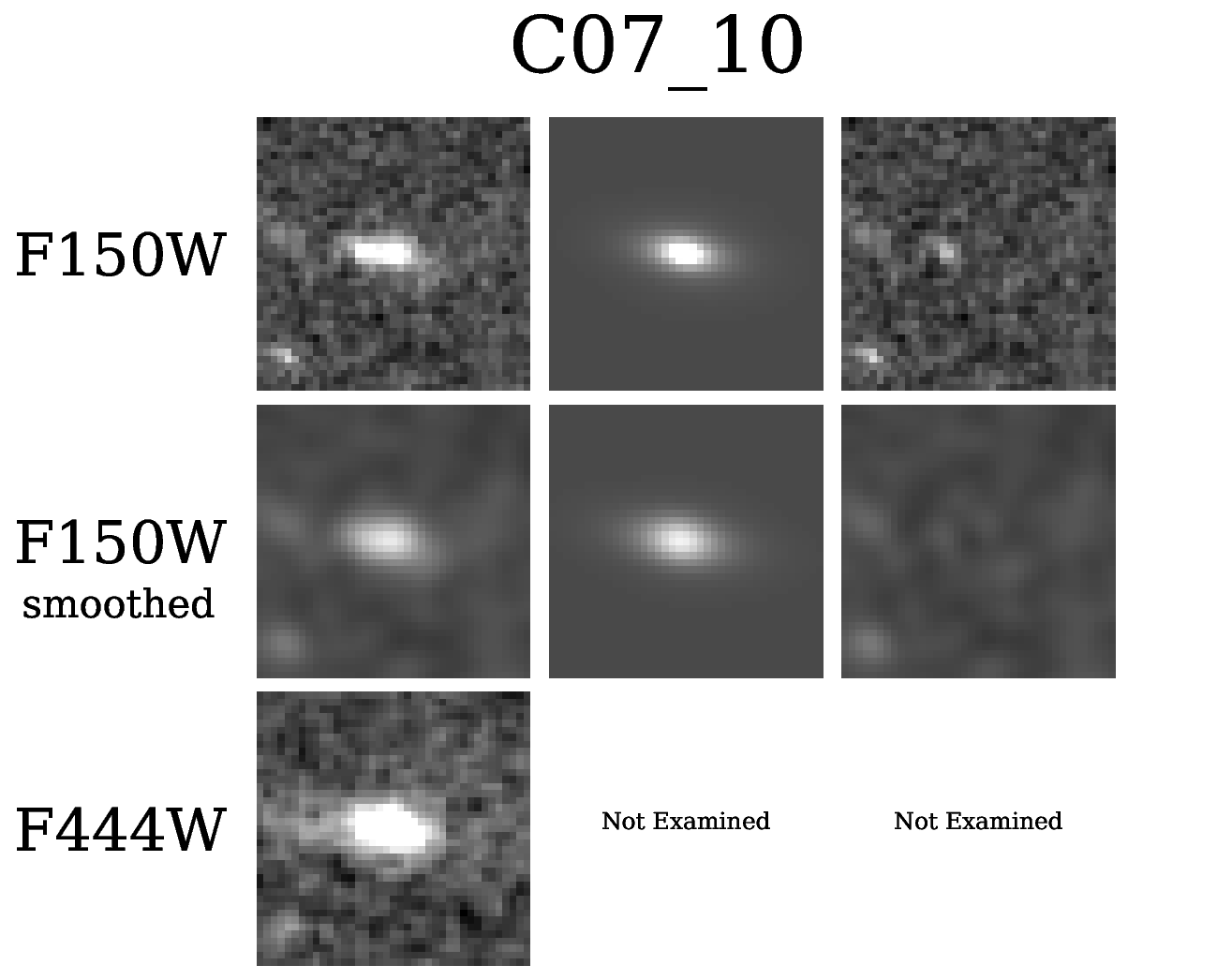}
\caption{
(Continued)
}
\end{center}
\end{figure*}

\addtocounter{figure}{-1}
\begin{figure*}
\begin{center}
   \includegraphics[height=0.14\textheight]{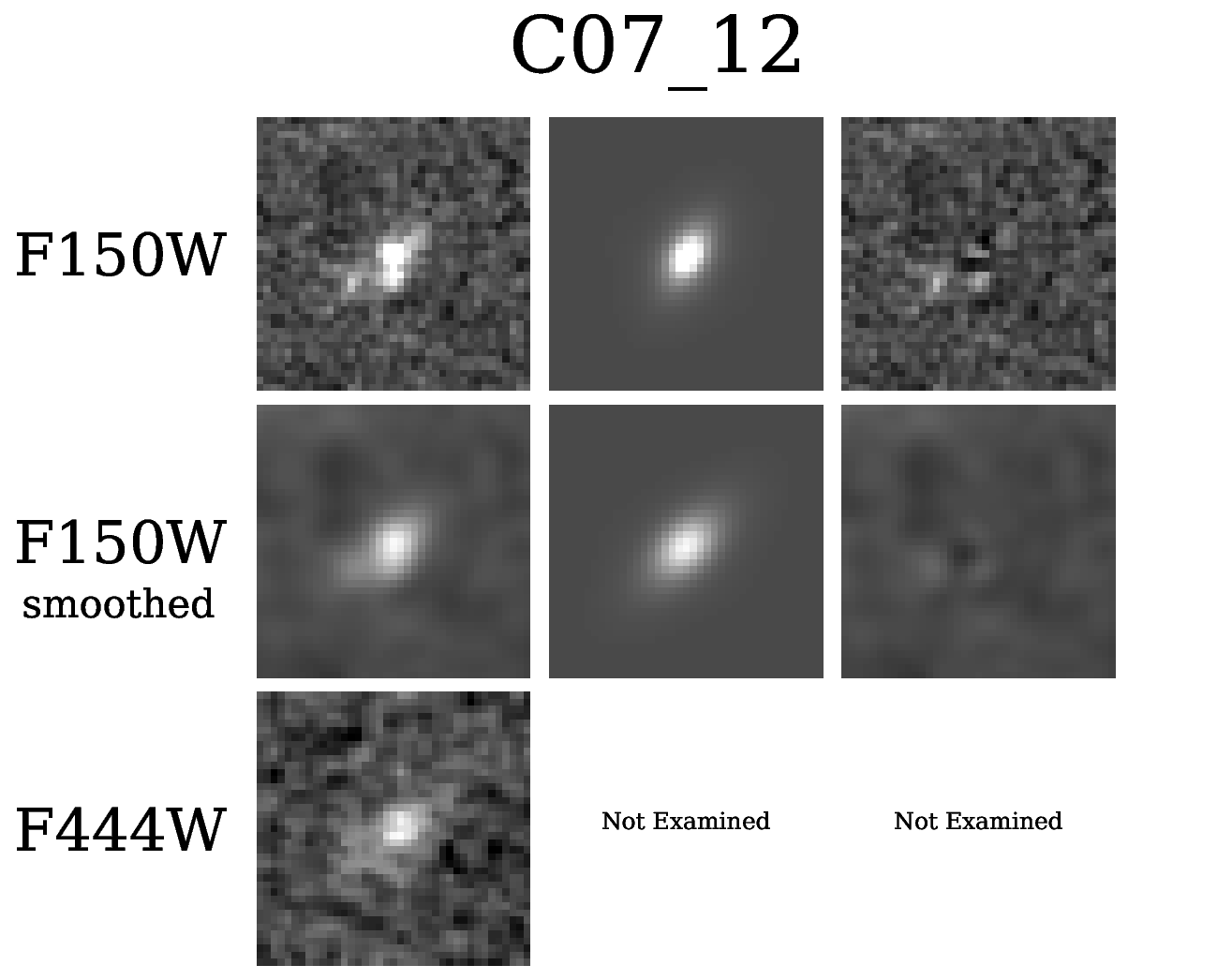}
   \includegraphics[height=0.14\textheight]{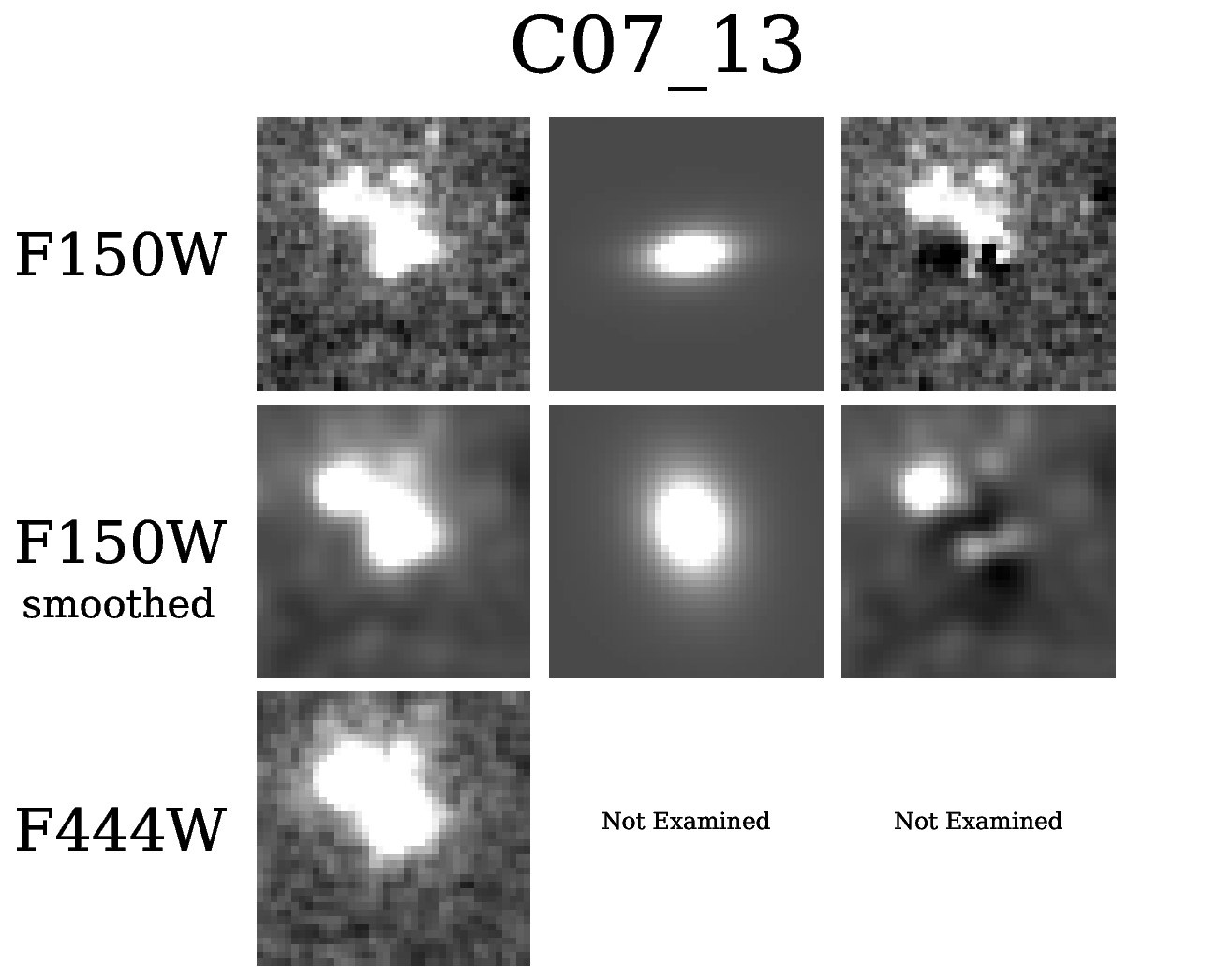}
   \includegraphics[height=0.14\textheight]{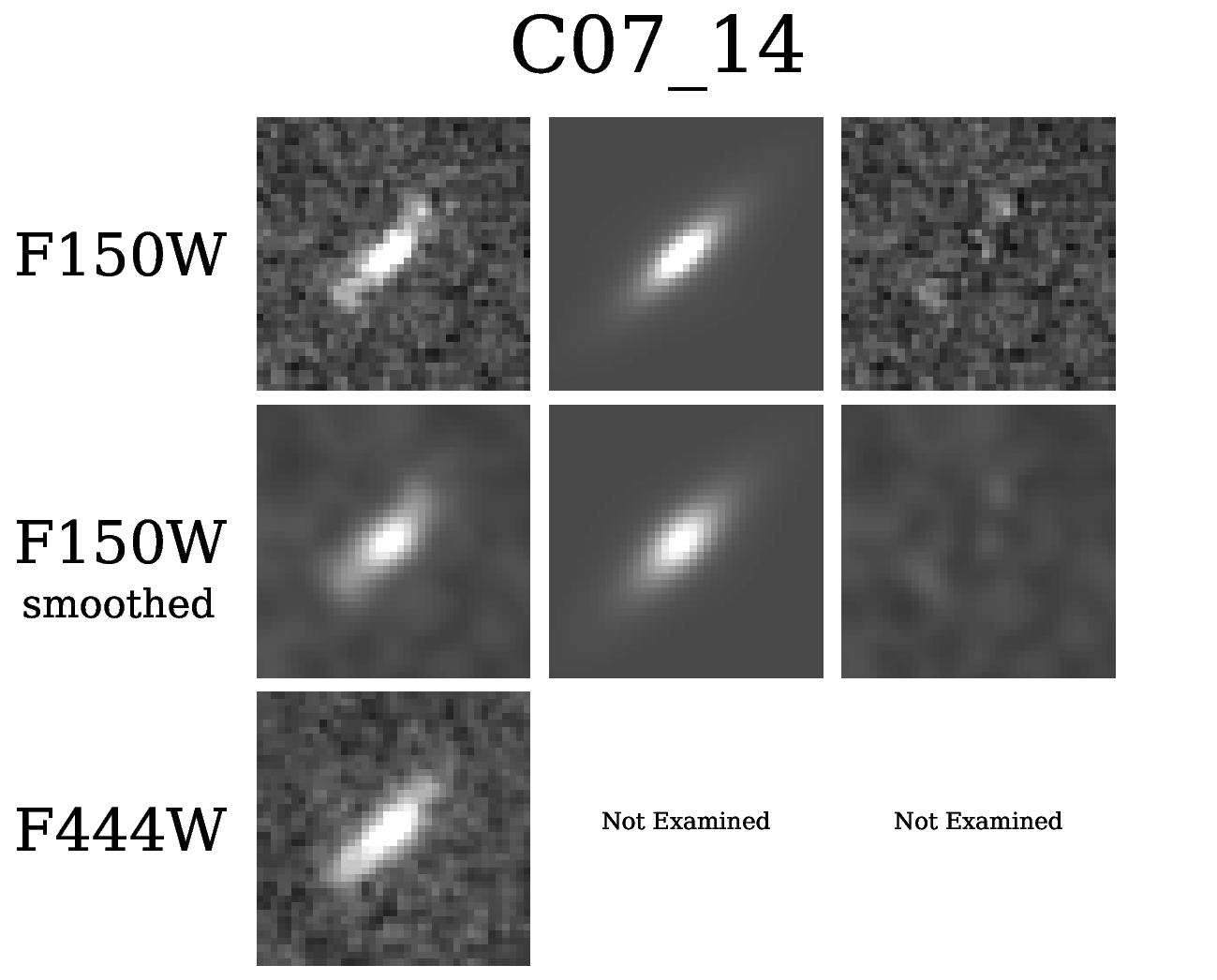}
   \includegraphics[height=0.14\textheight]{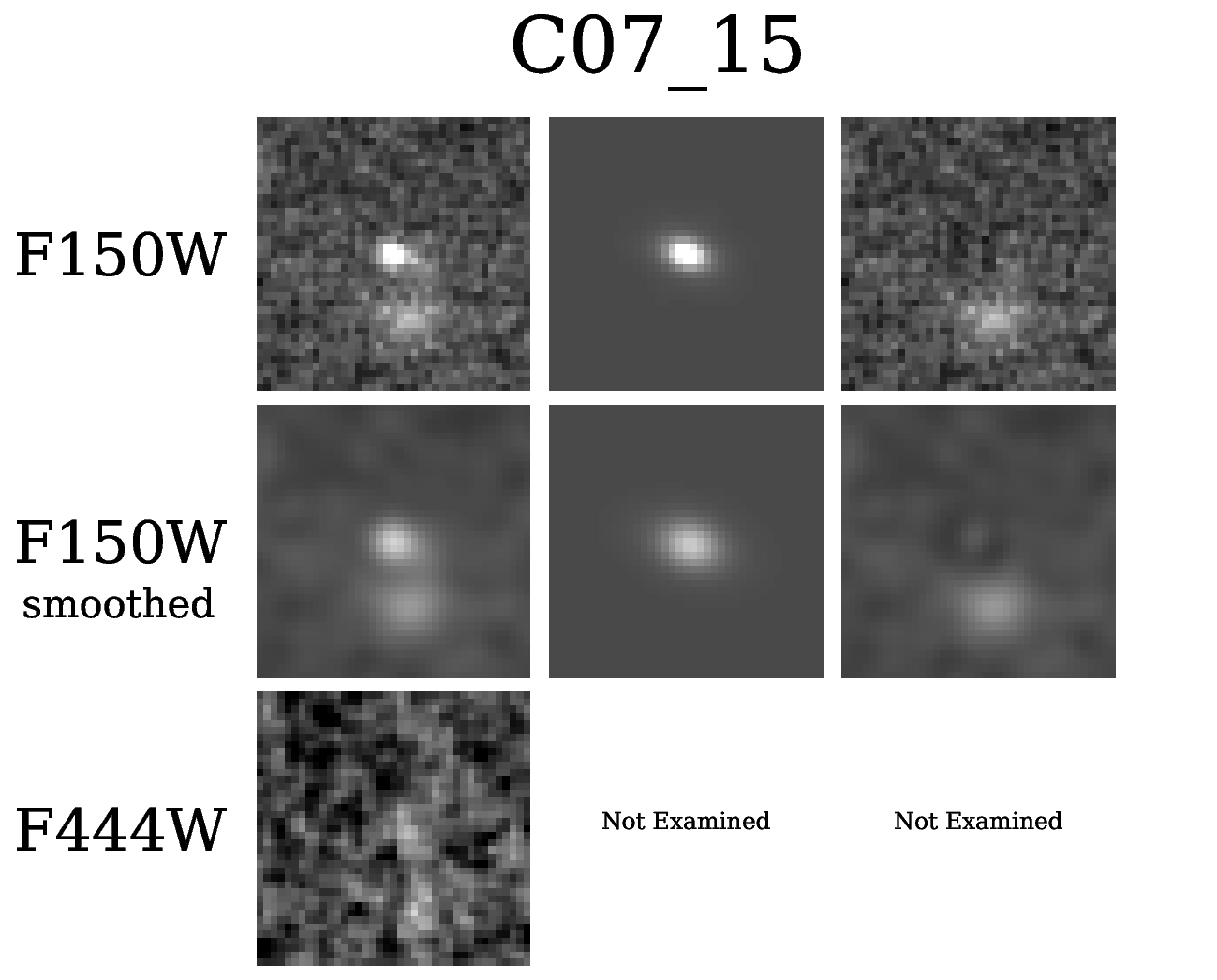}
   \includegraphics[height=0.14\textheight]{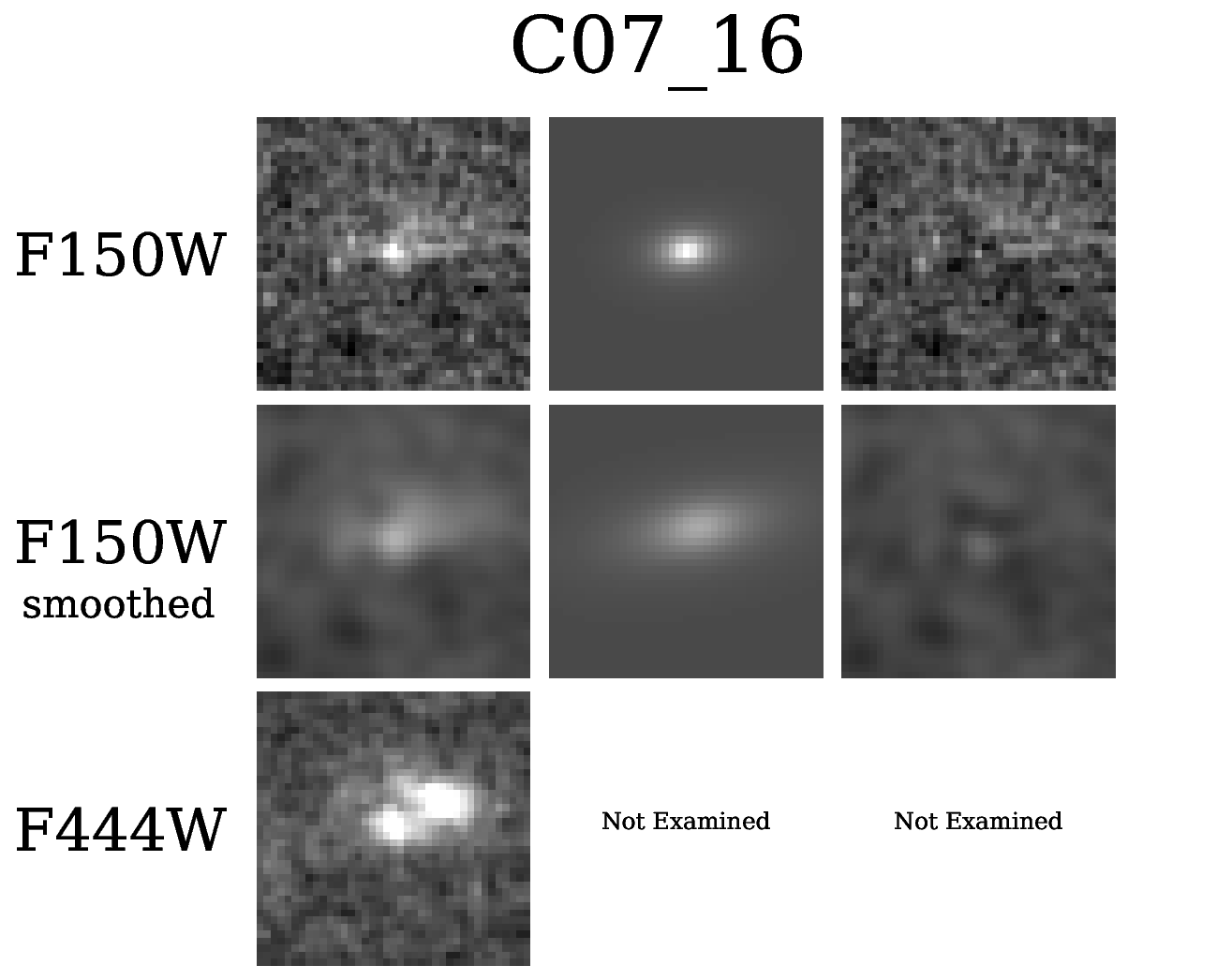}
   \includegraphics[height=0.14\textheight]{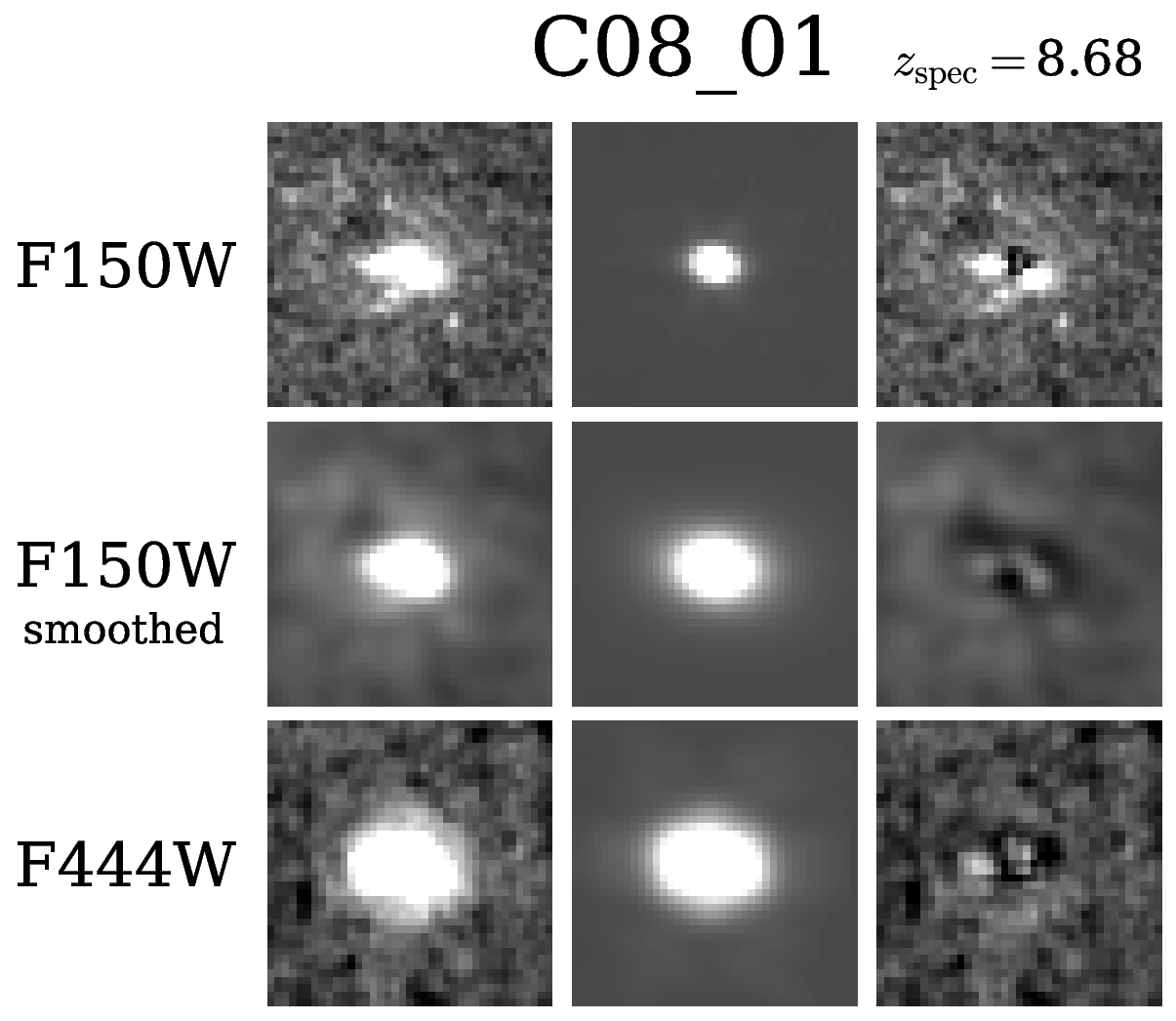}
   \includegraphics[height=0.14\textheight]{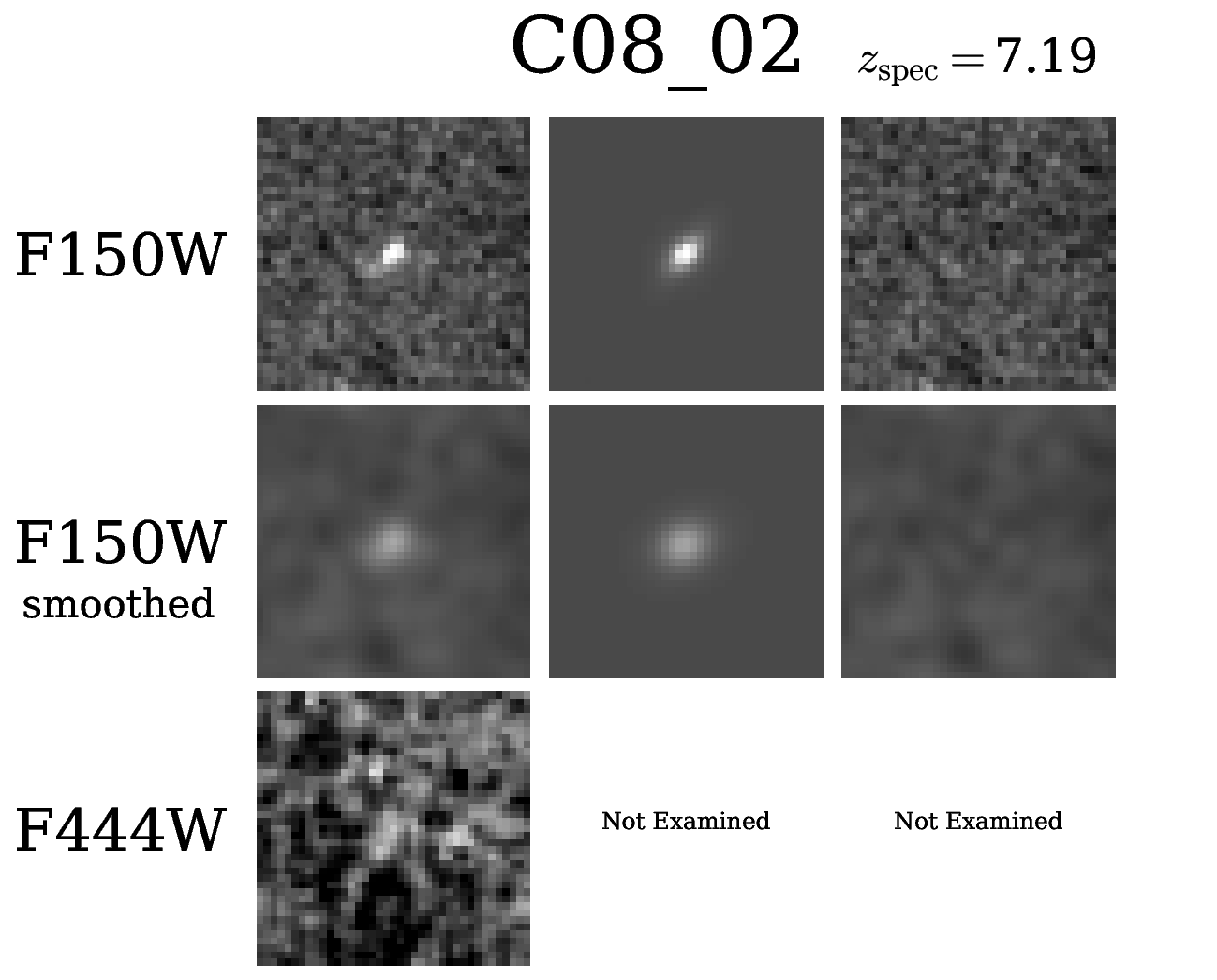}
   \includegraphics[height=0.14\textheight]{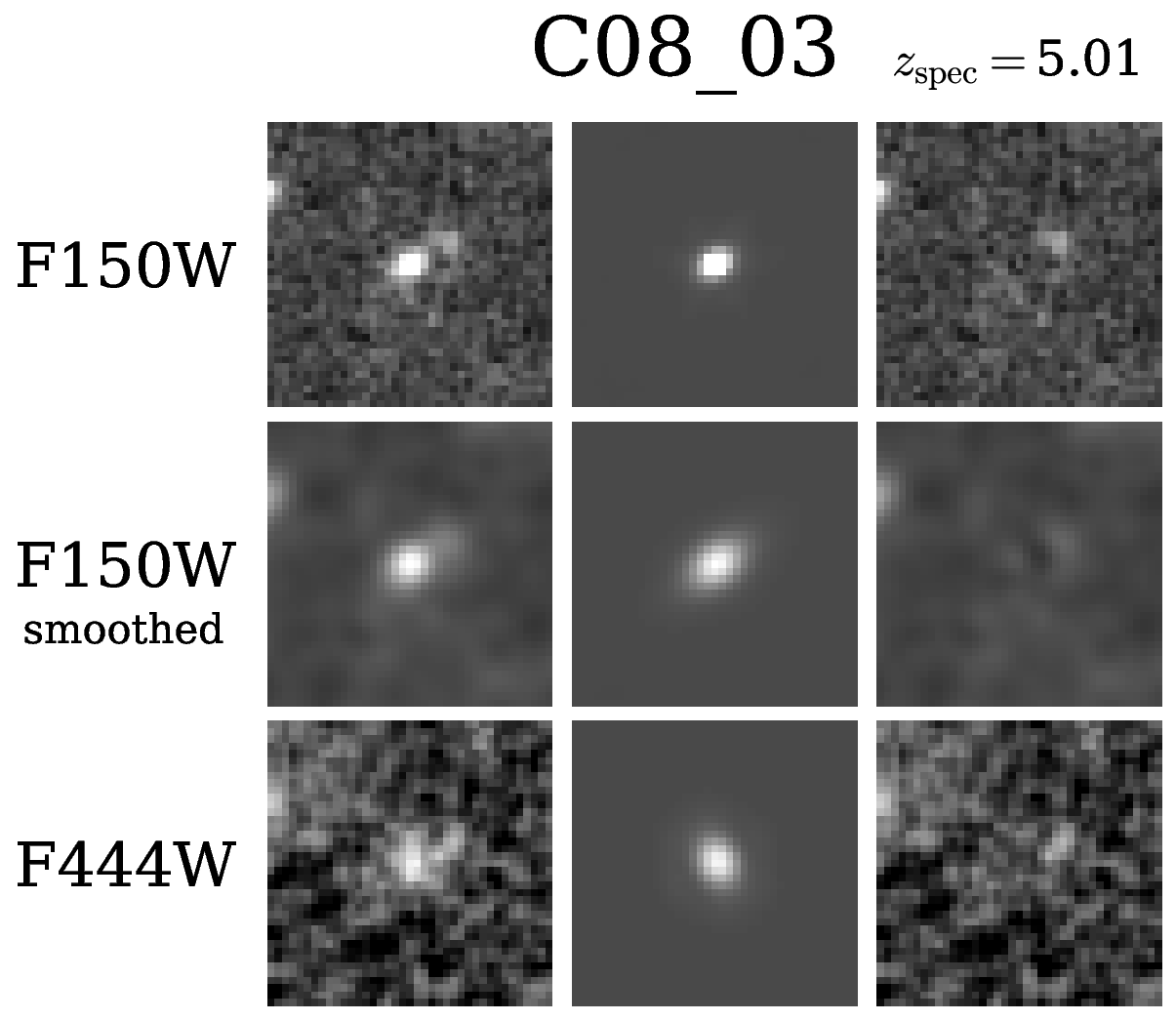}
   \includegraphics[height=0.14\textheight]{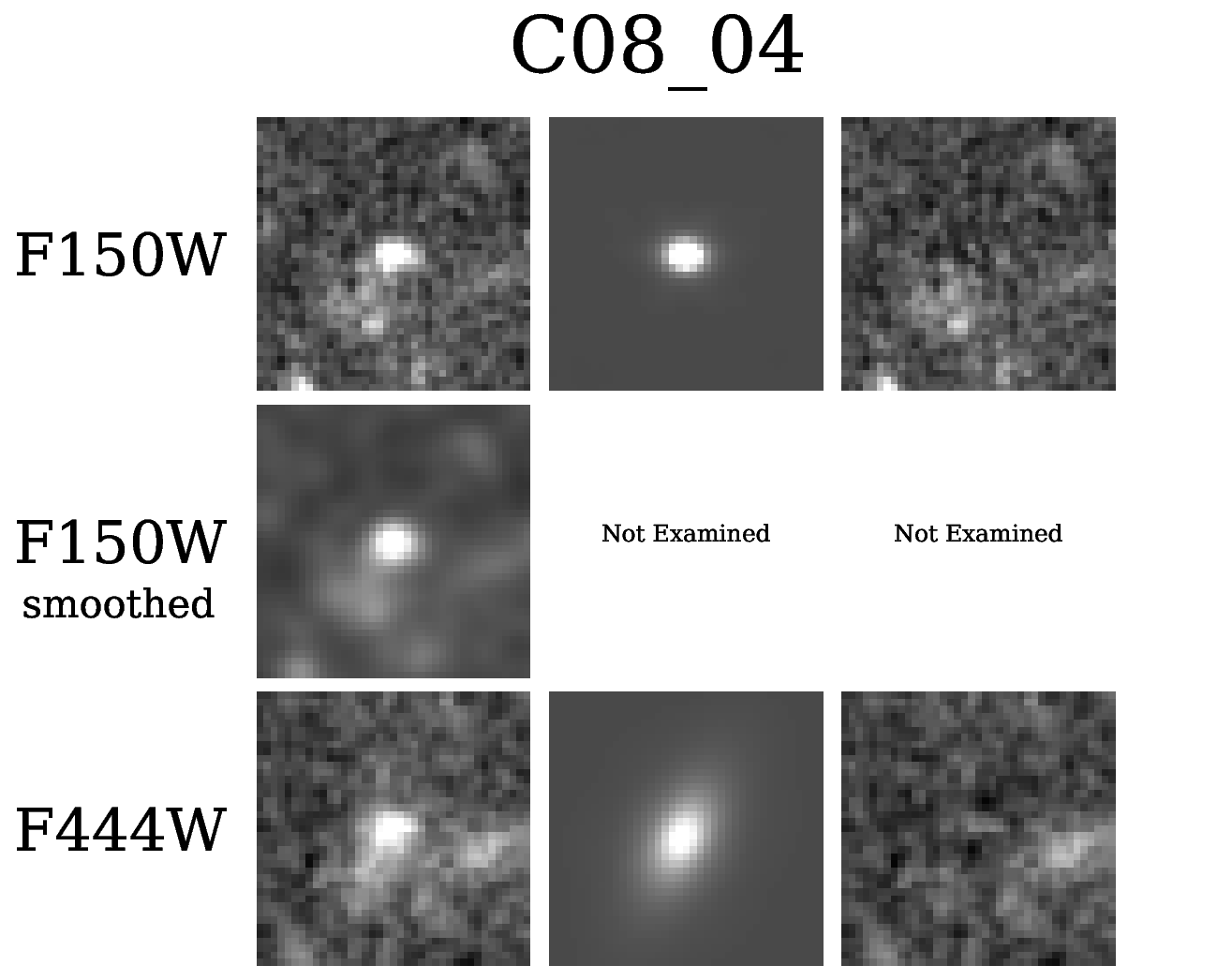}
   \includegraphics[height=0.14\textheight]{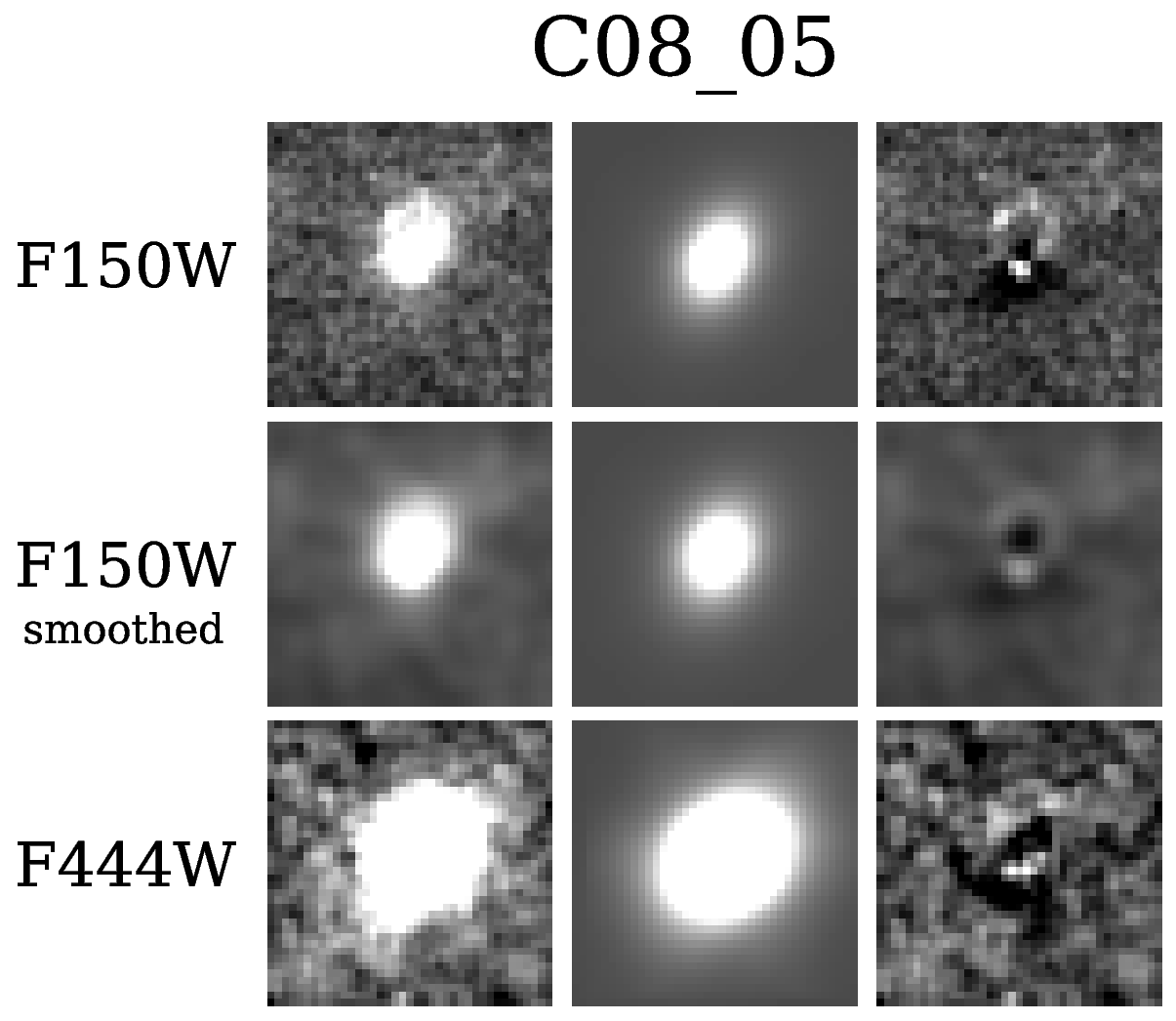}
   \includegraphics[height=0.14\textheight]{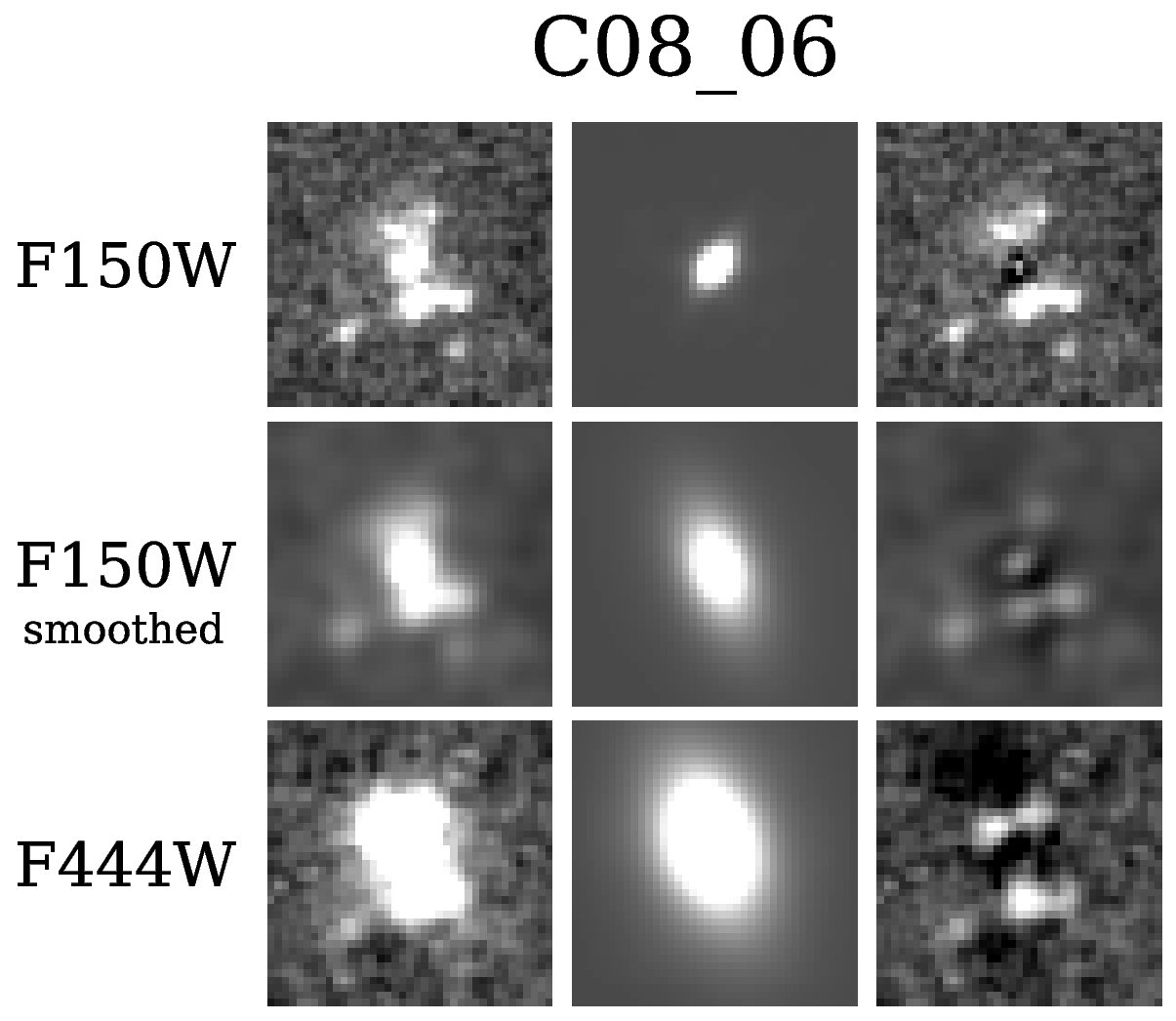}
   \includegraphics[height=0.14\textheight]{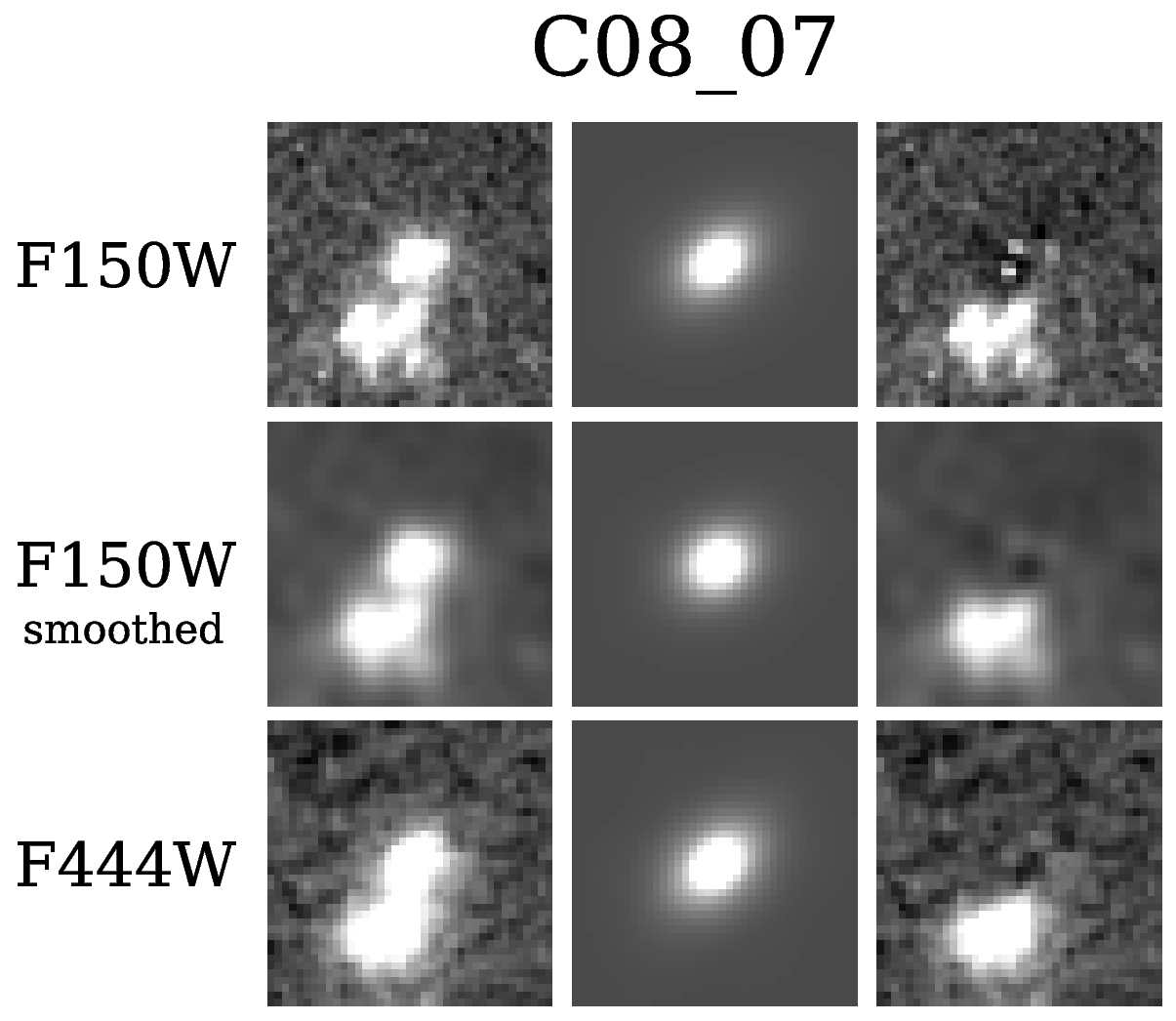}
   \includegraphics[height=0.14\textheight]{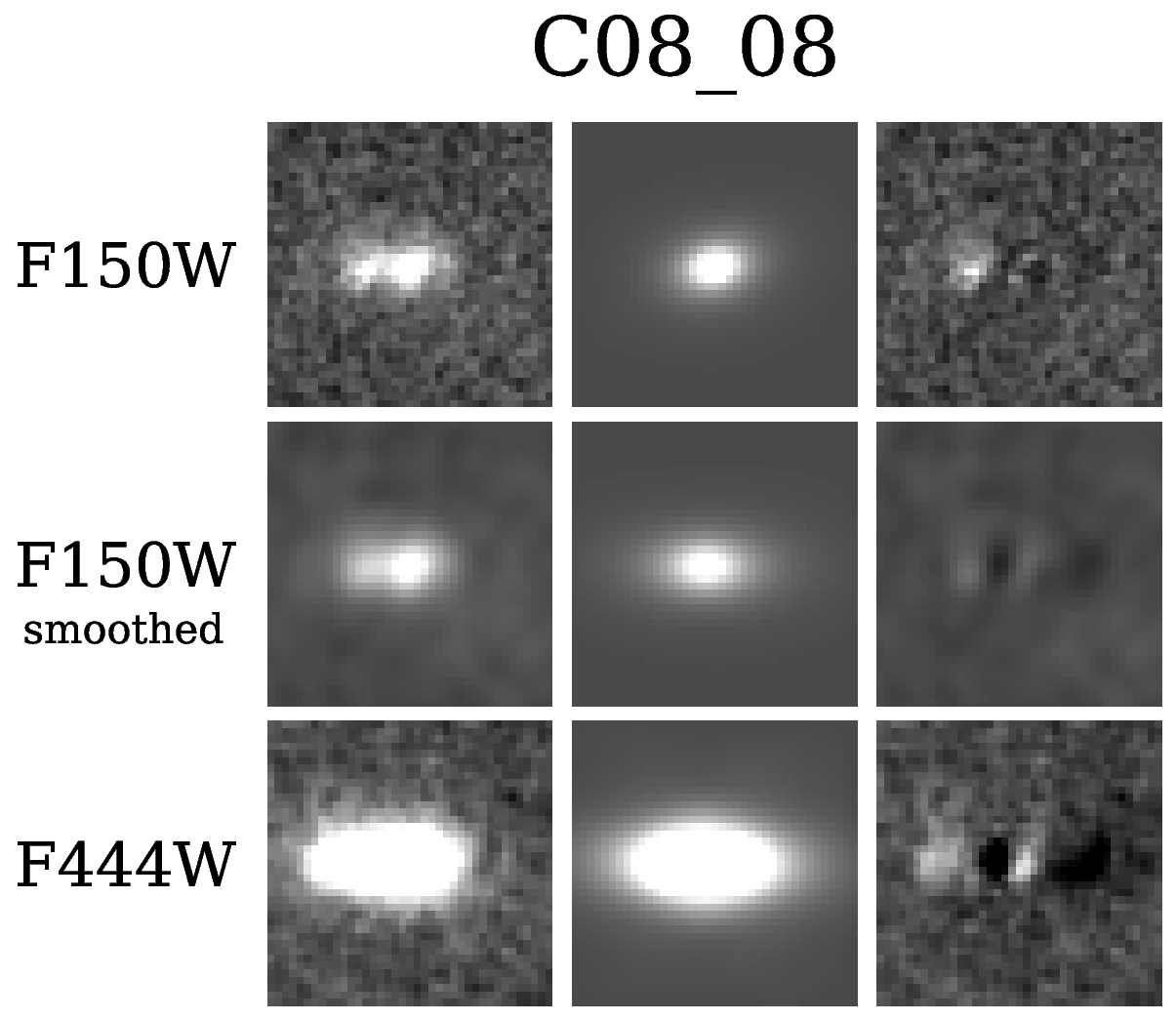}
   \includegraphics[height=0.14\textheight]{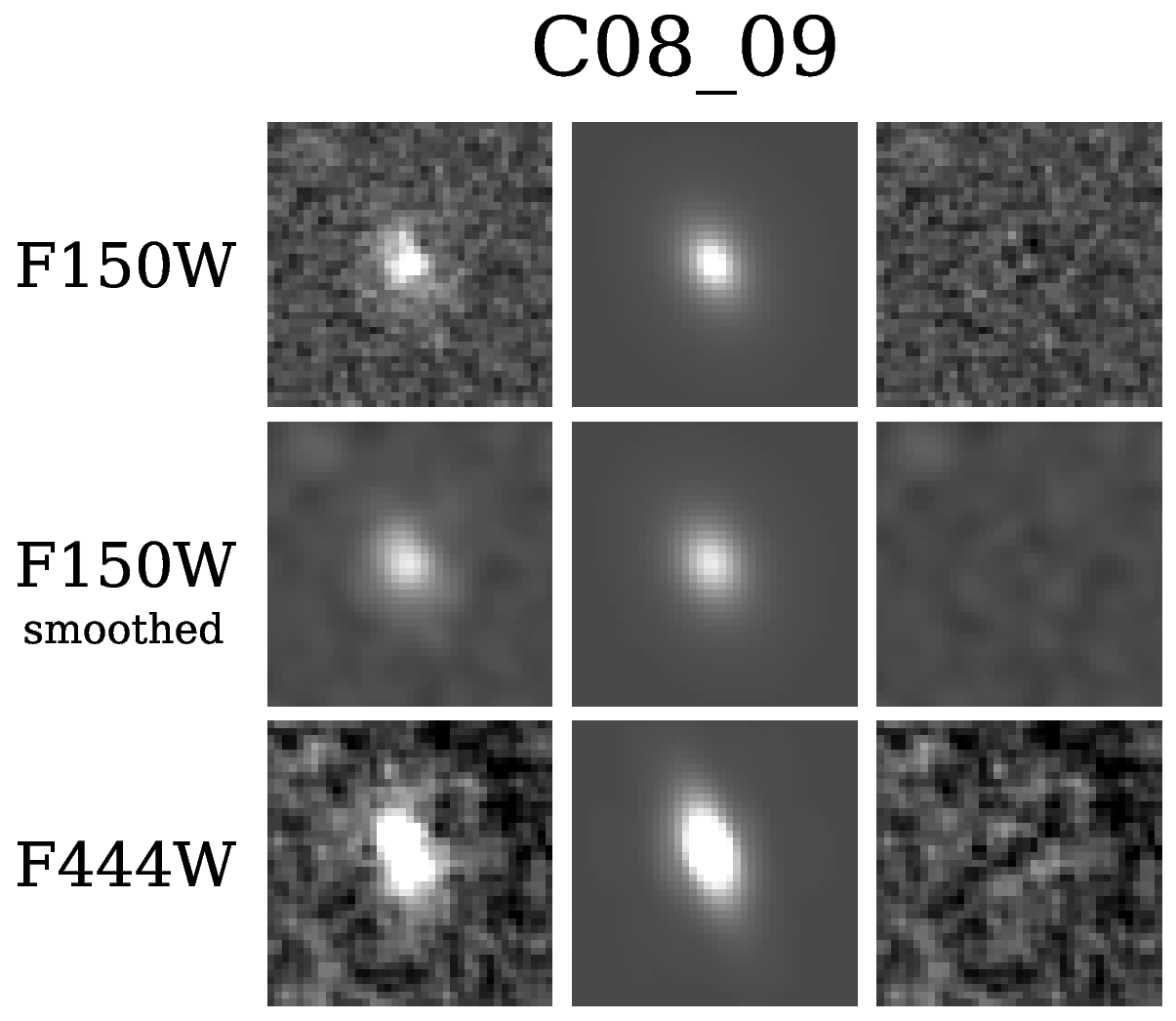}
   \includegraphics[height=0.14\textheight]{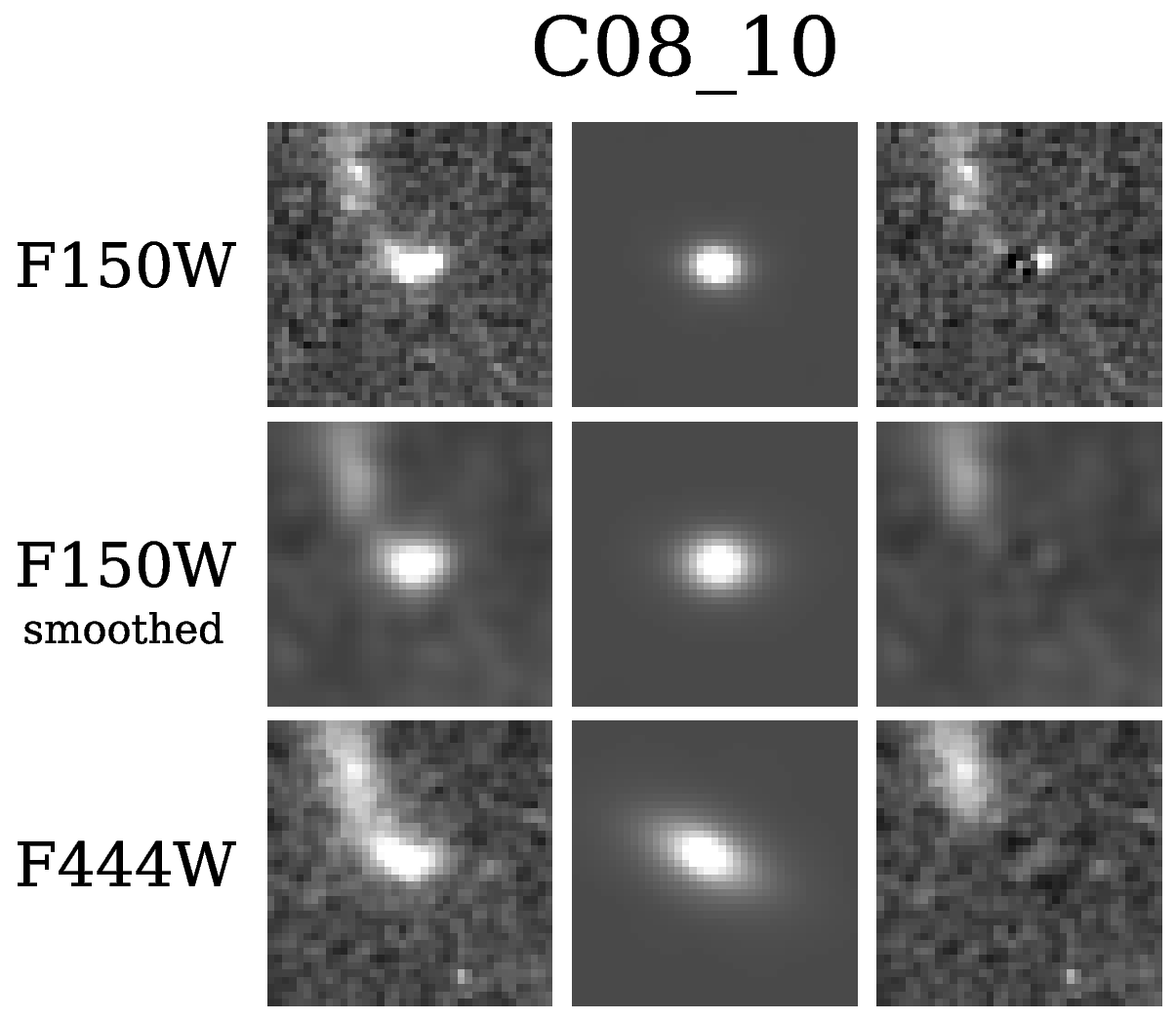}
   \includegraphics[height=0.14\textheight]{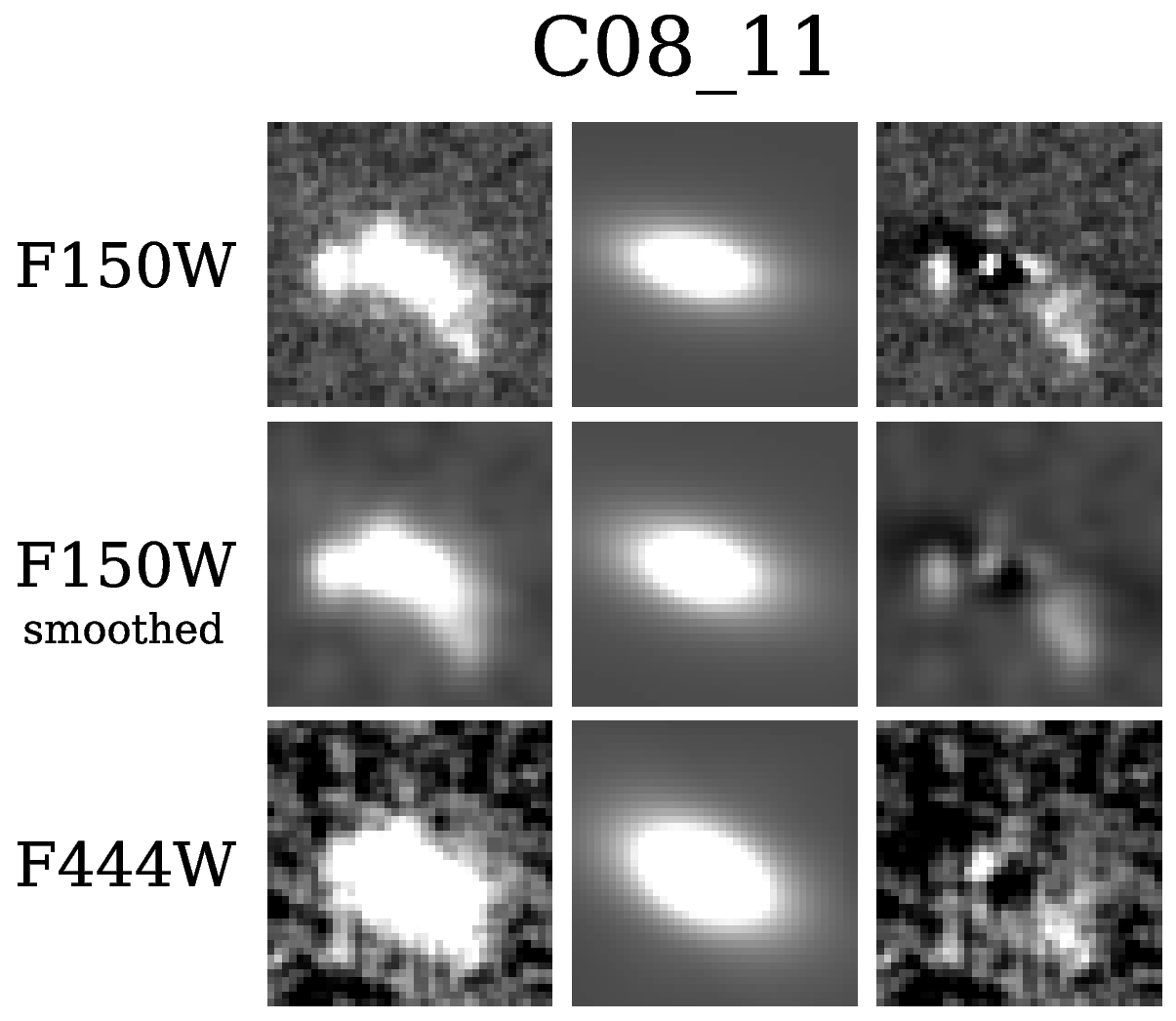}
   \includegraphics[height=0.14\textheight]{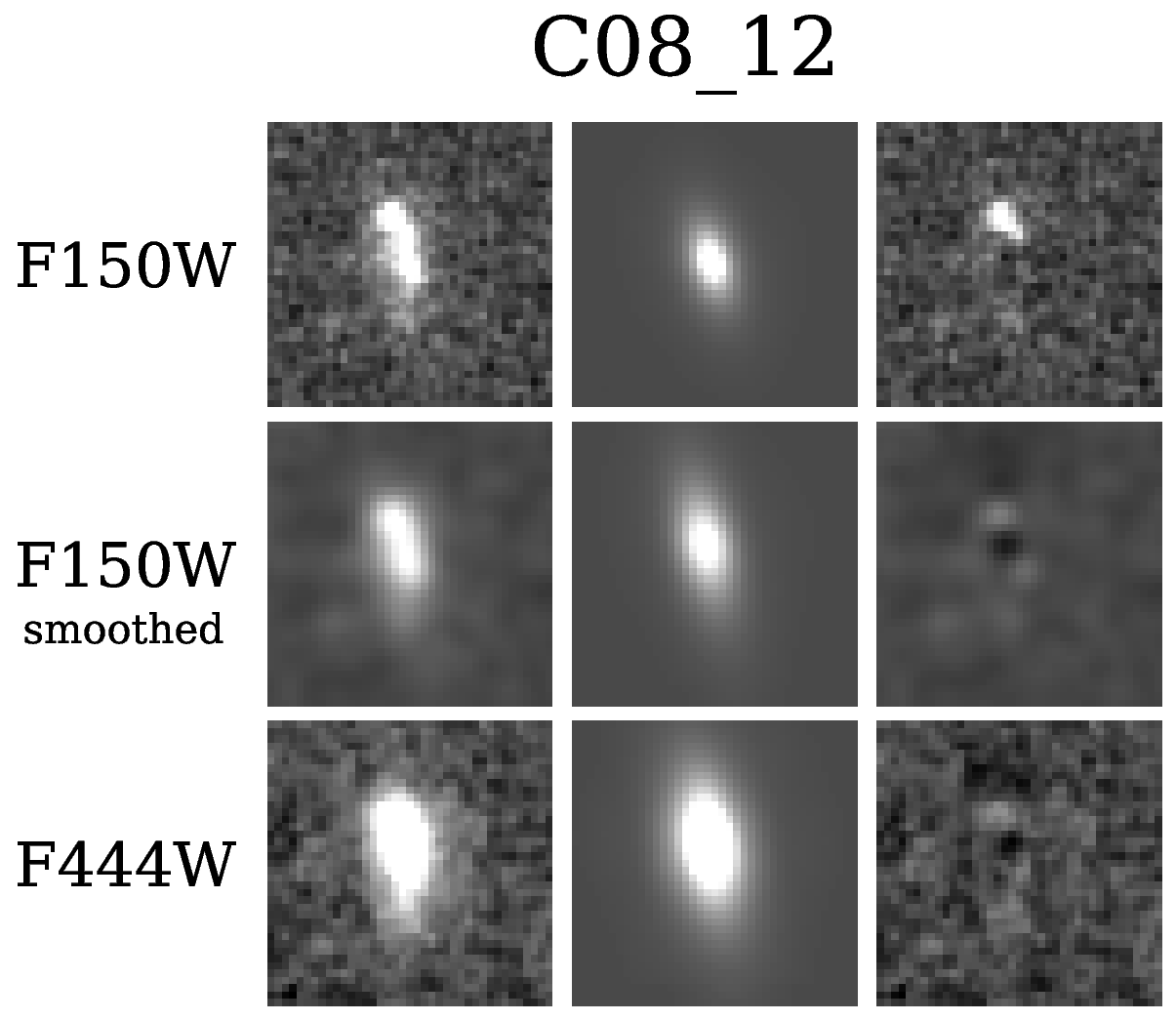}
   \includegraphics[height=0.14\textheight]{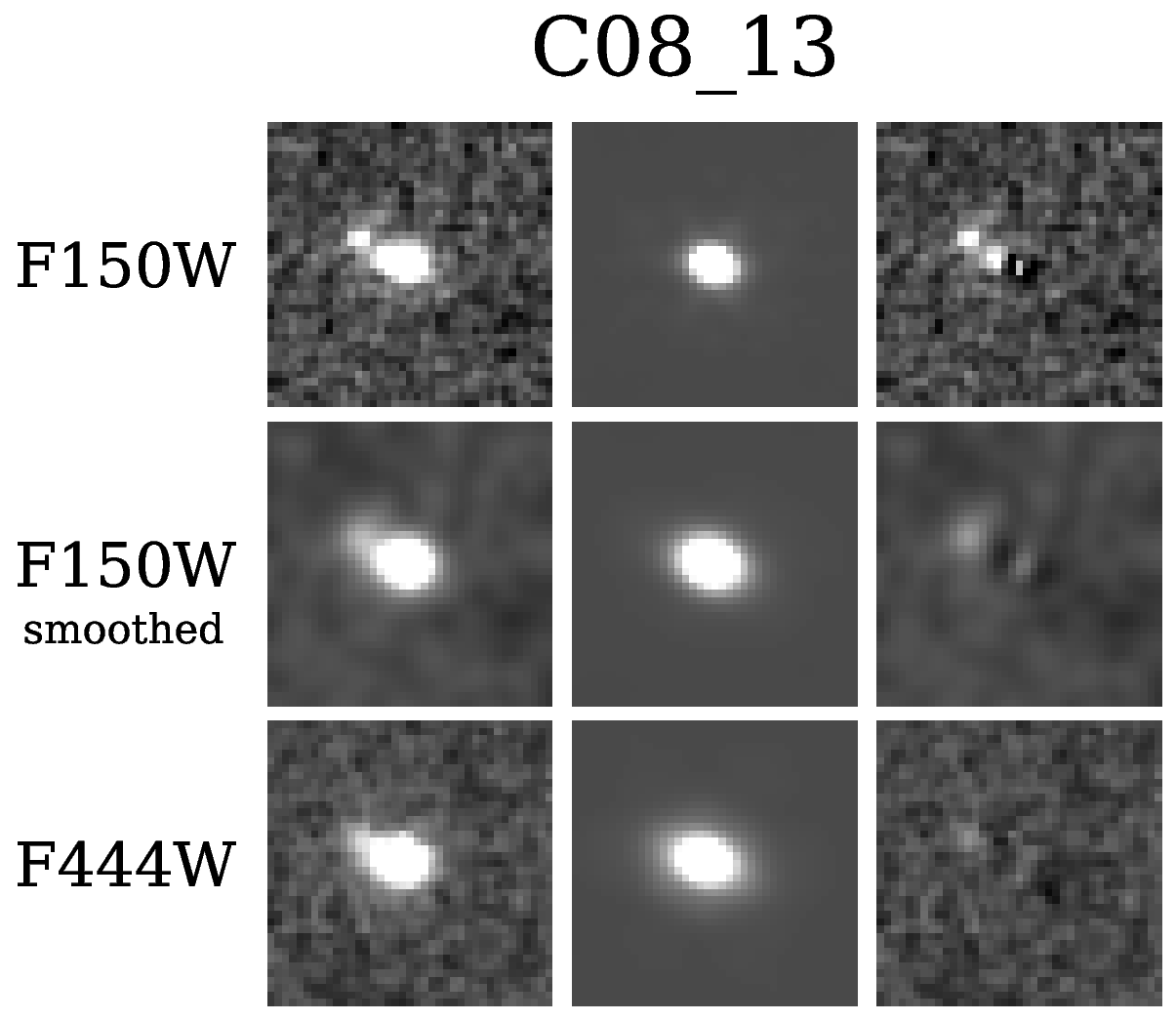}
   \includegraphics[height=0.14\textheight]{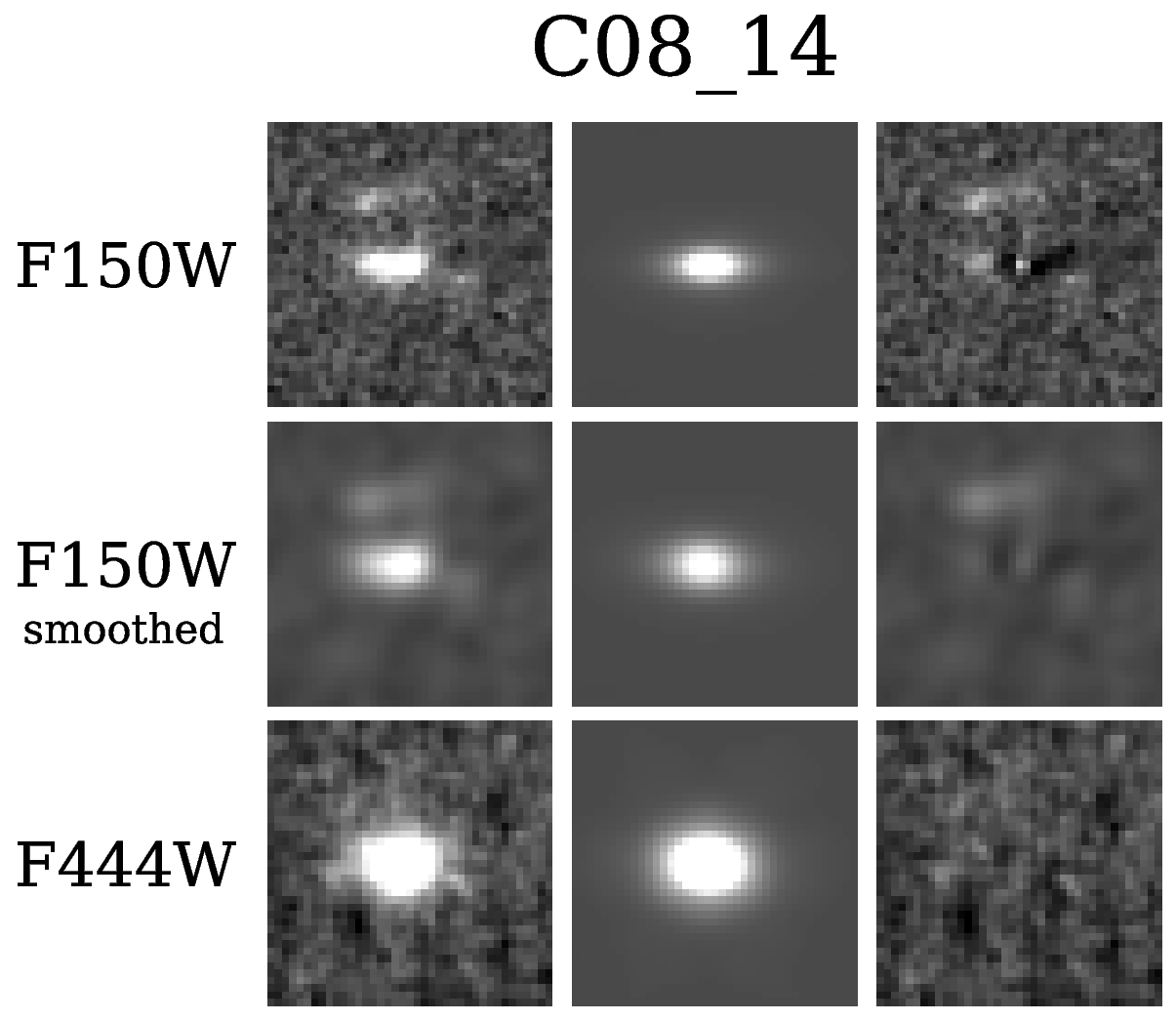}
   \includegraphics[height=0.14\textheight]{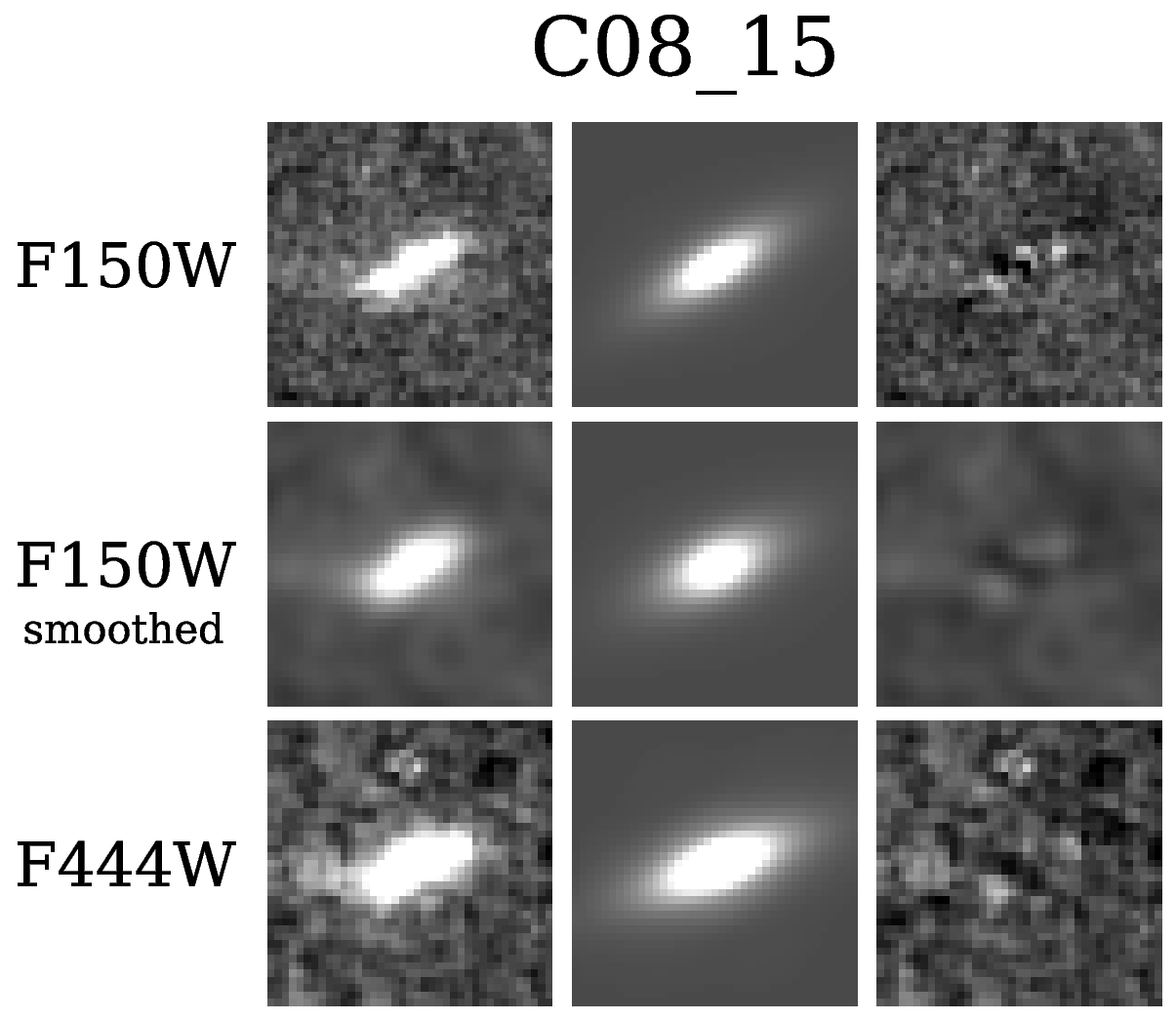}
   \includegraphics[height=0.14\textheight]{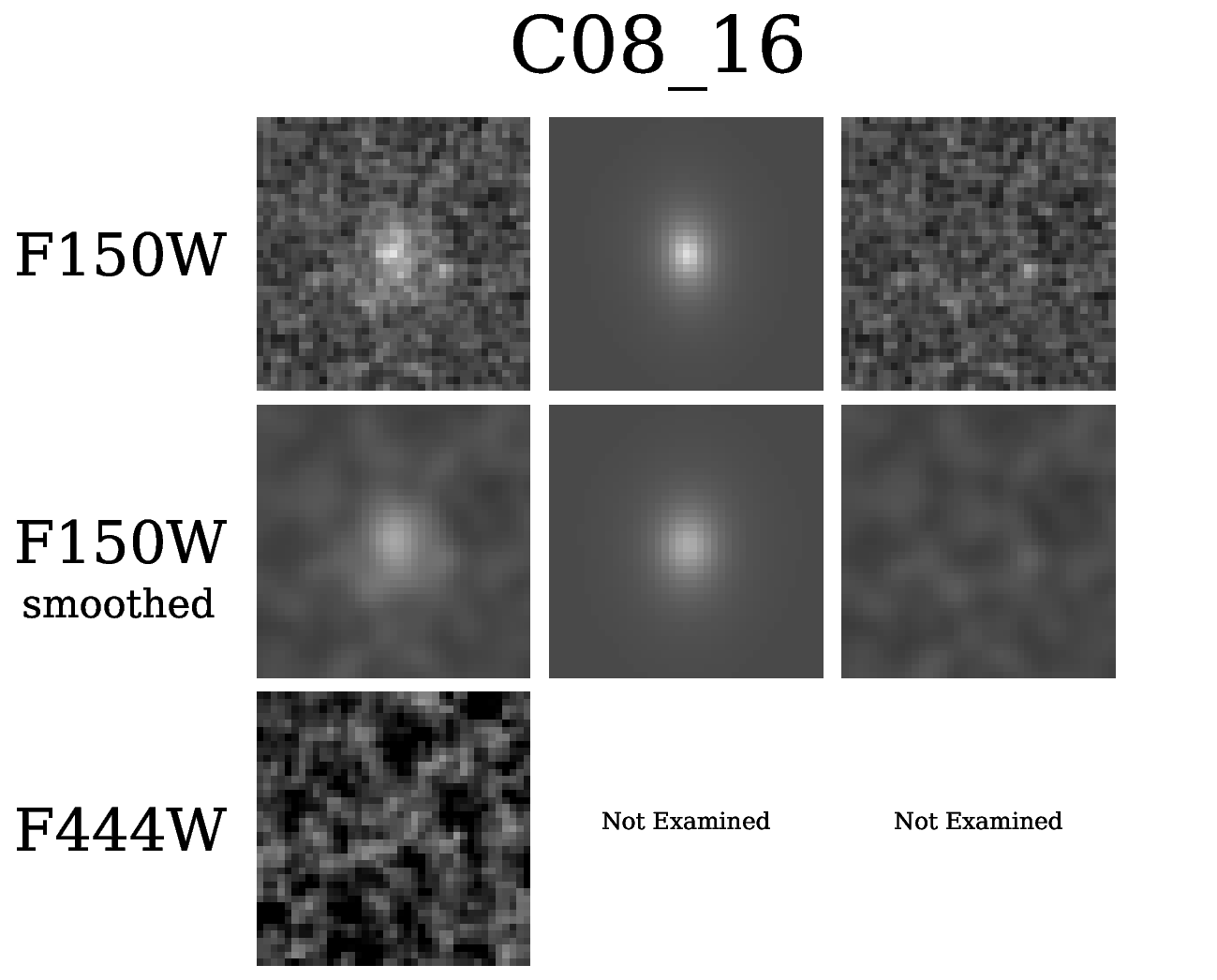}
   \includegraphics[height=0.14\textheight]{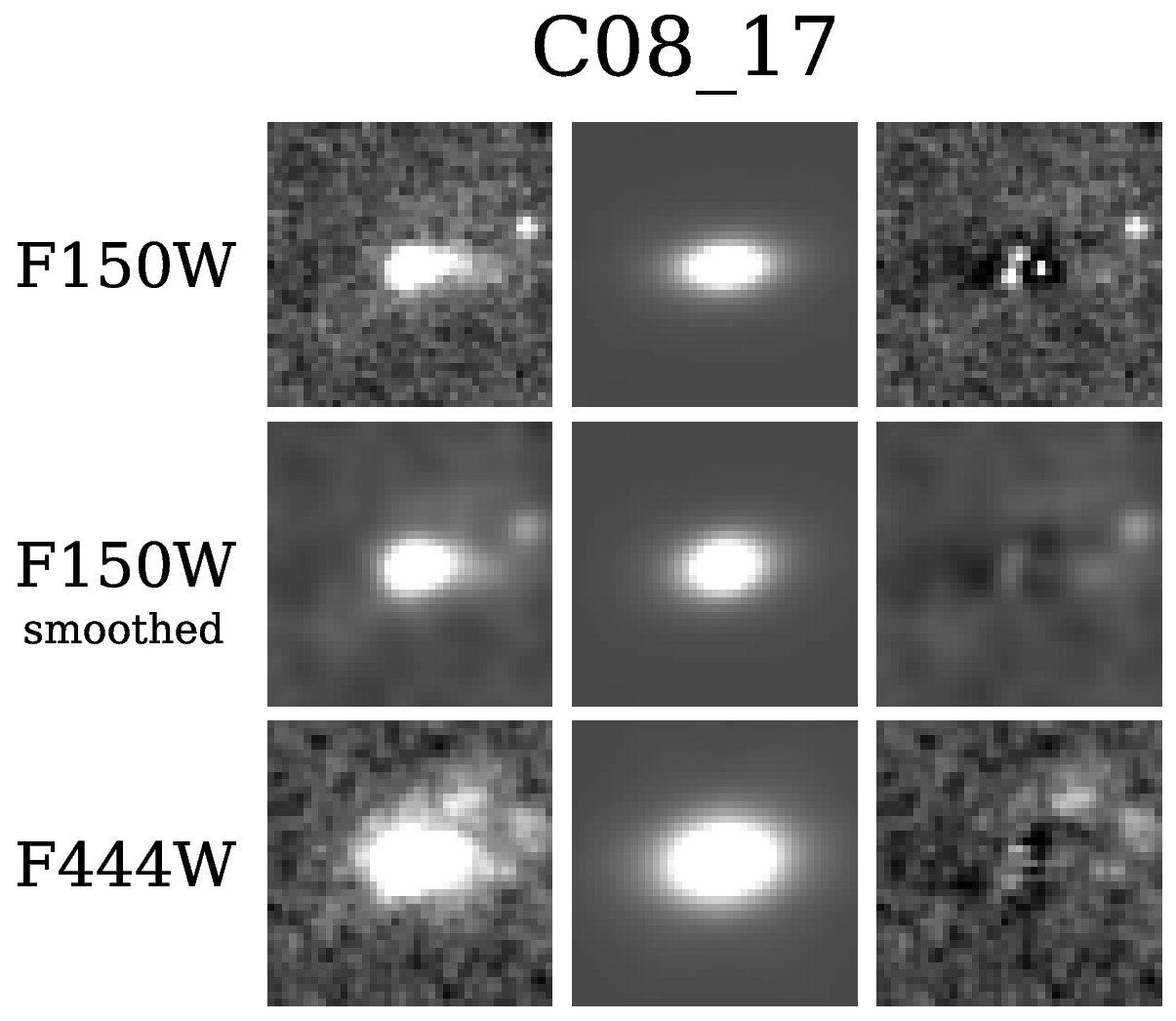}
   \includegraphics[height=0.14\textheight]{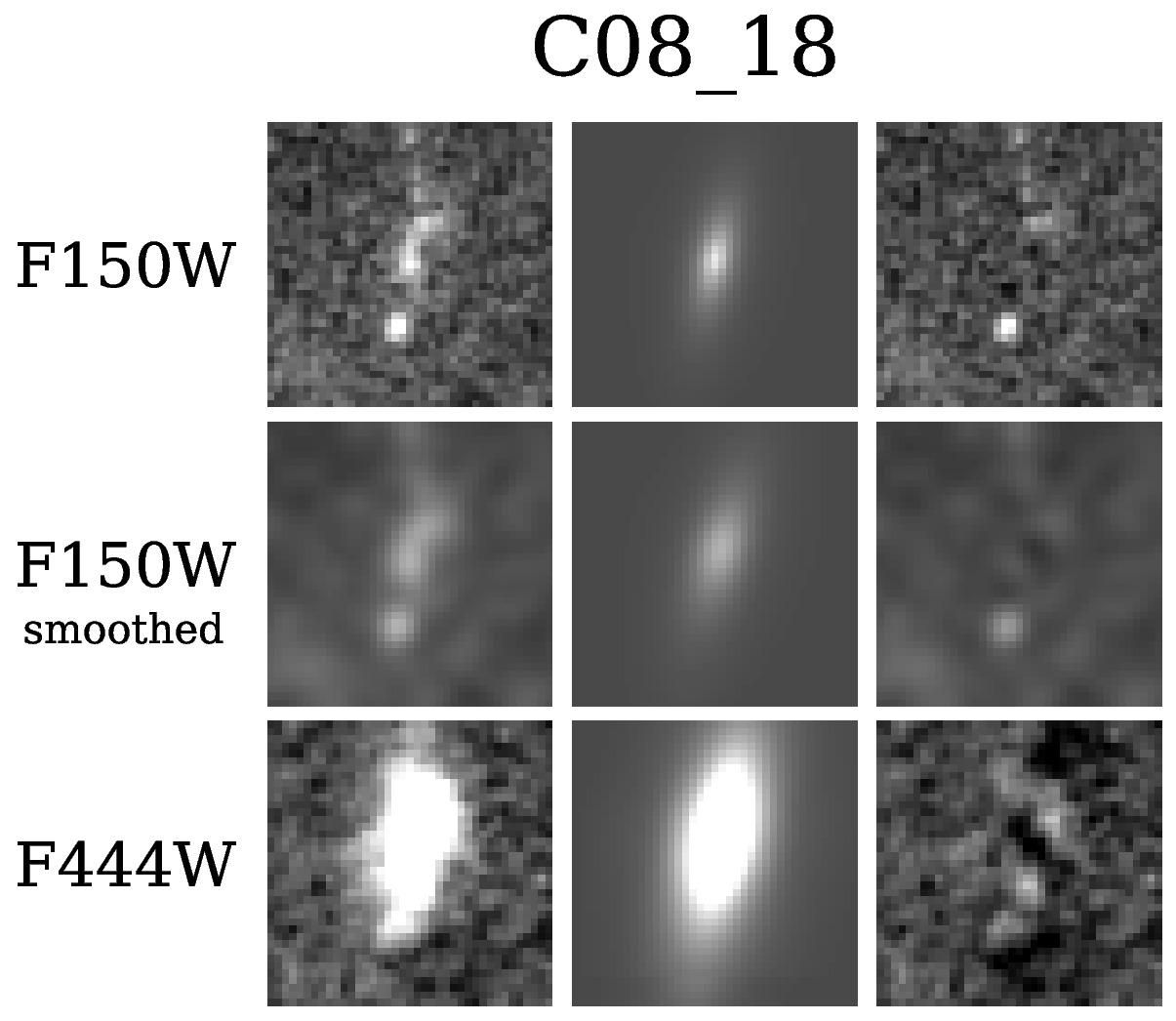}
   \includegraphics[height=0.14\textheight]{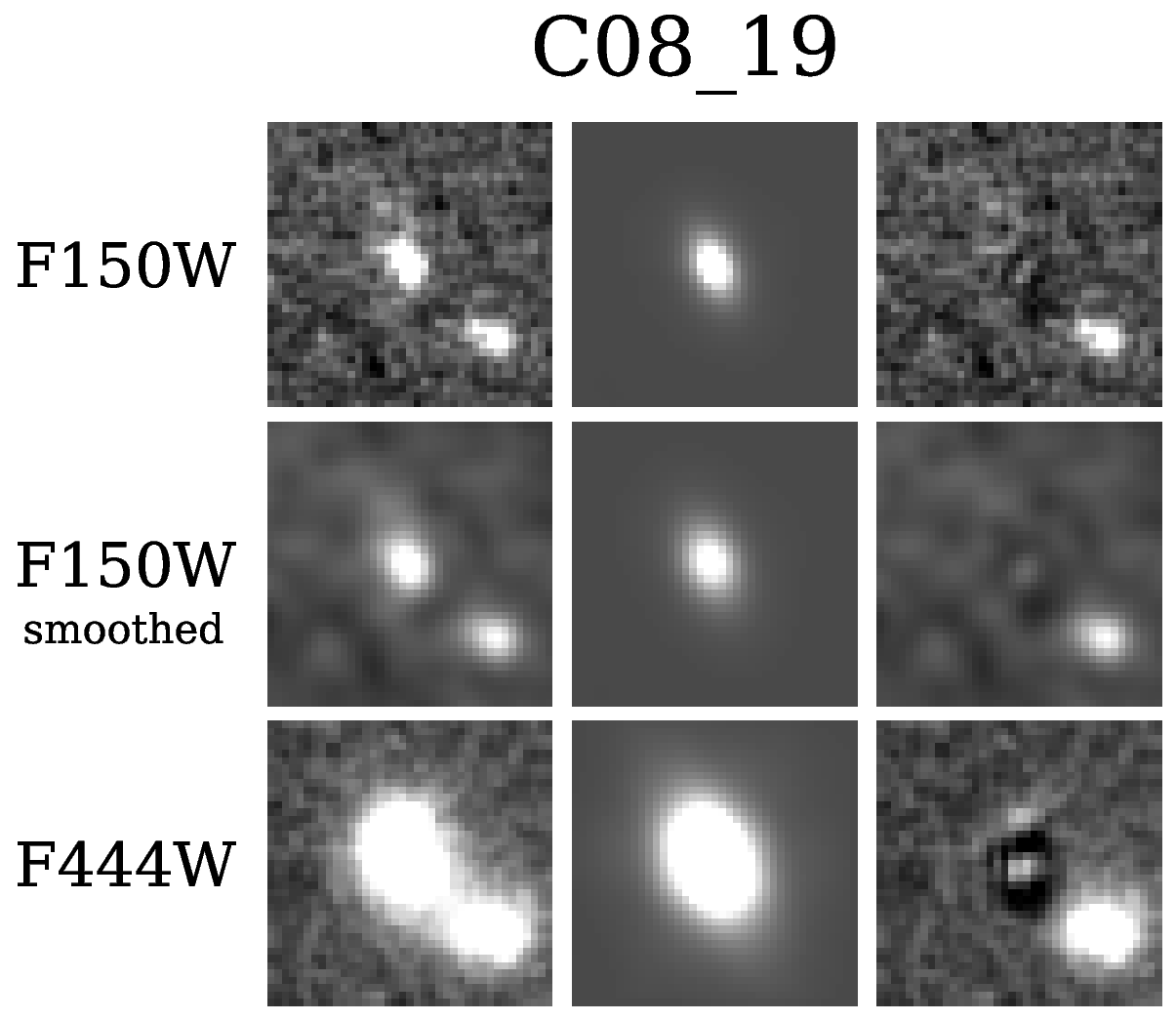}
\caption{
(Continued)
}
\end{center}
\end{figure*}

\addtocounter{figure}{-1}
\begin{figure*}
\begin{center}
   \includegraphics[height=0.14\textheight]{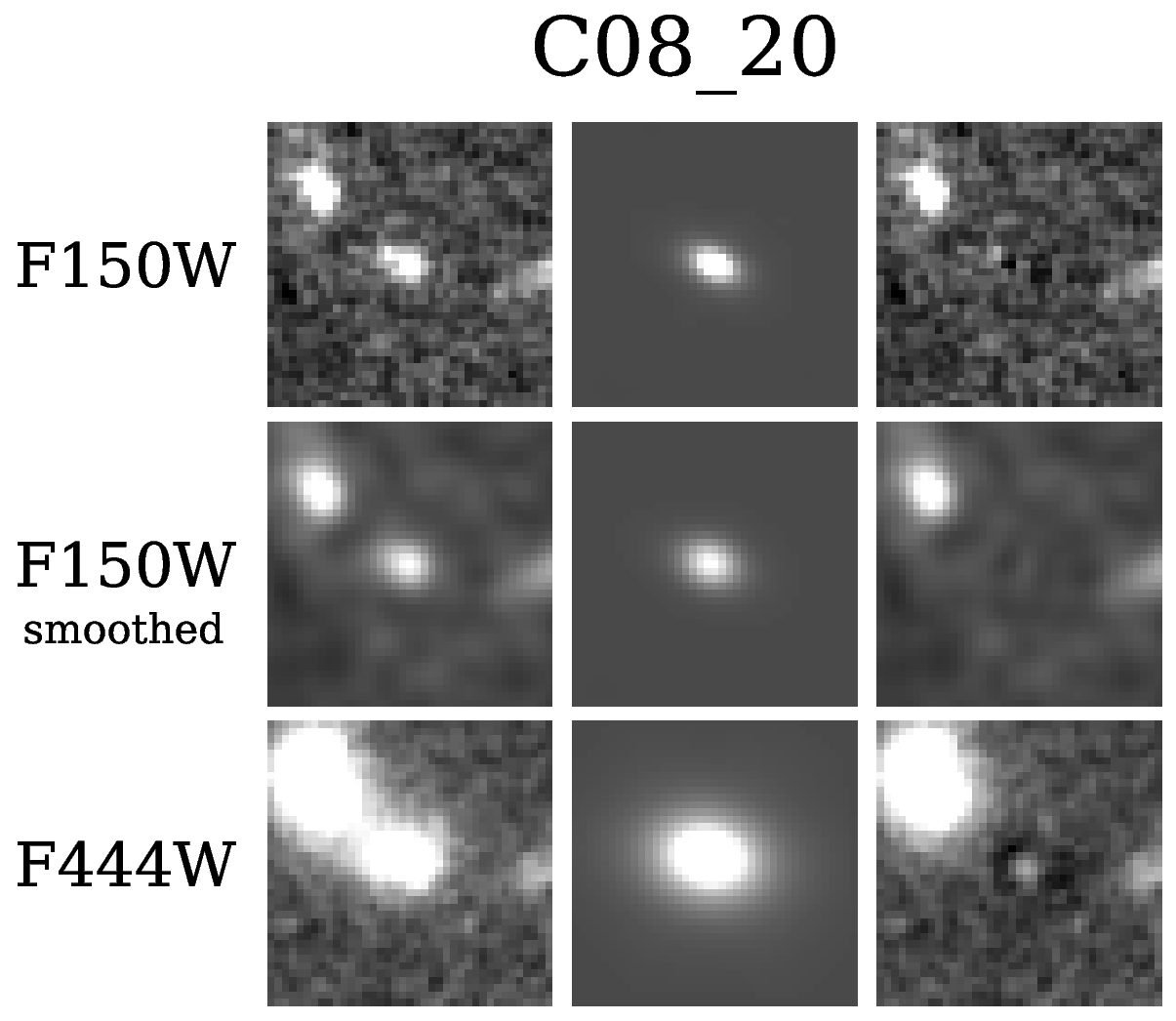}
   \includegraphics[height=0.14\textheight]{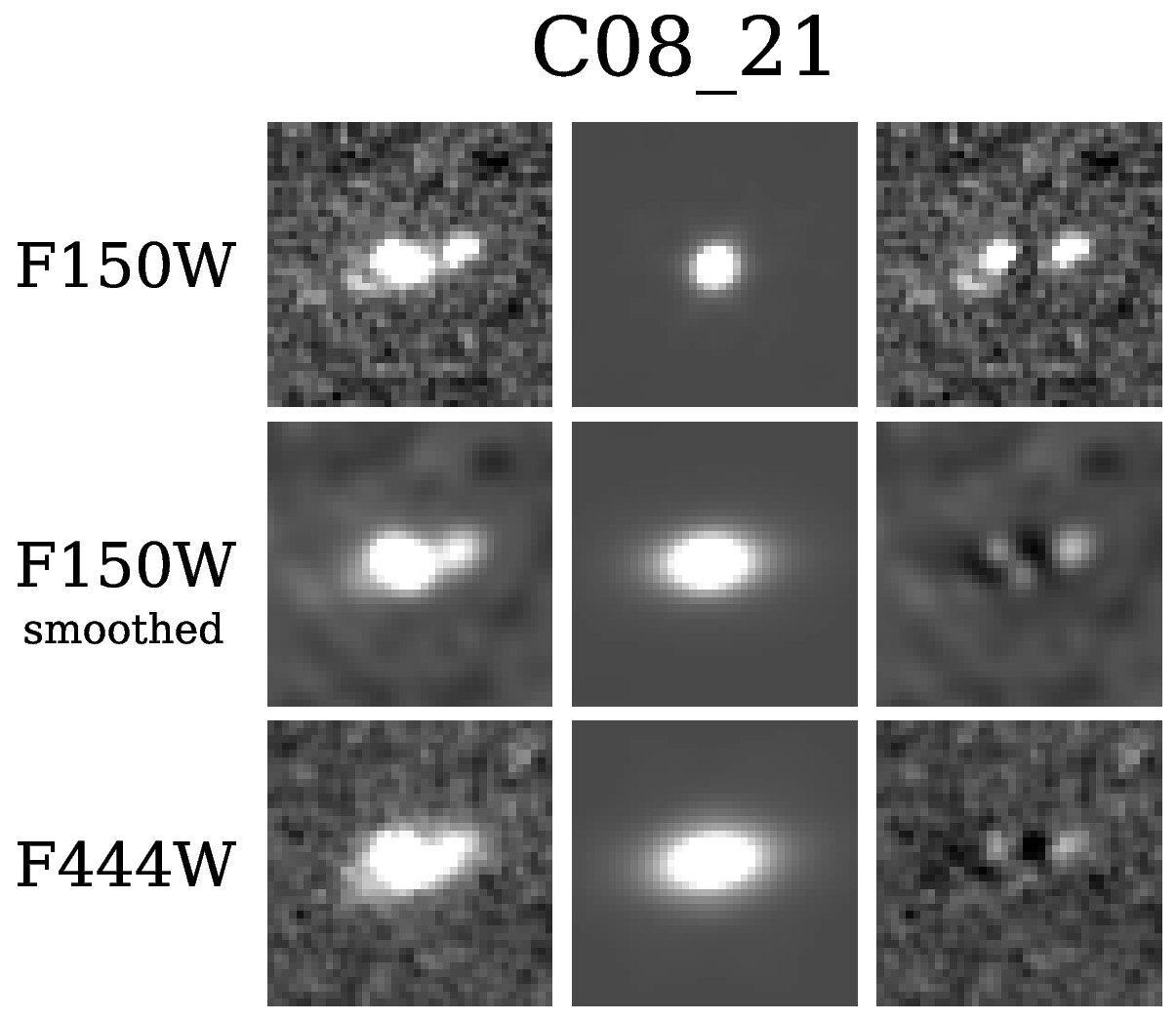}
   \includegraphics[height=0.14\textheight]{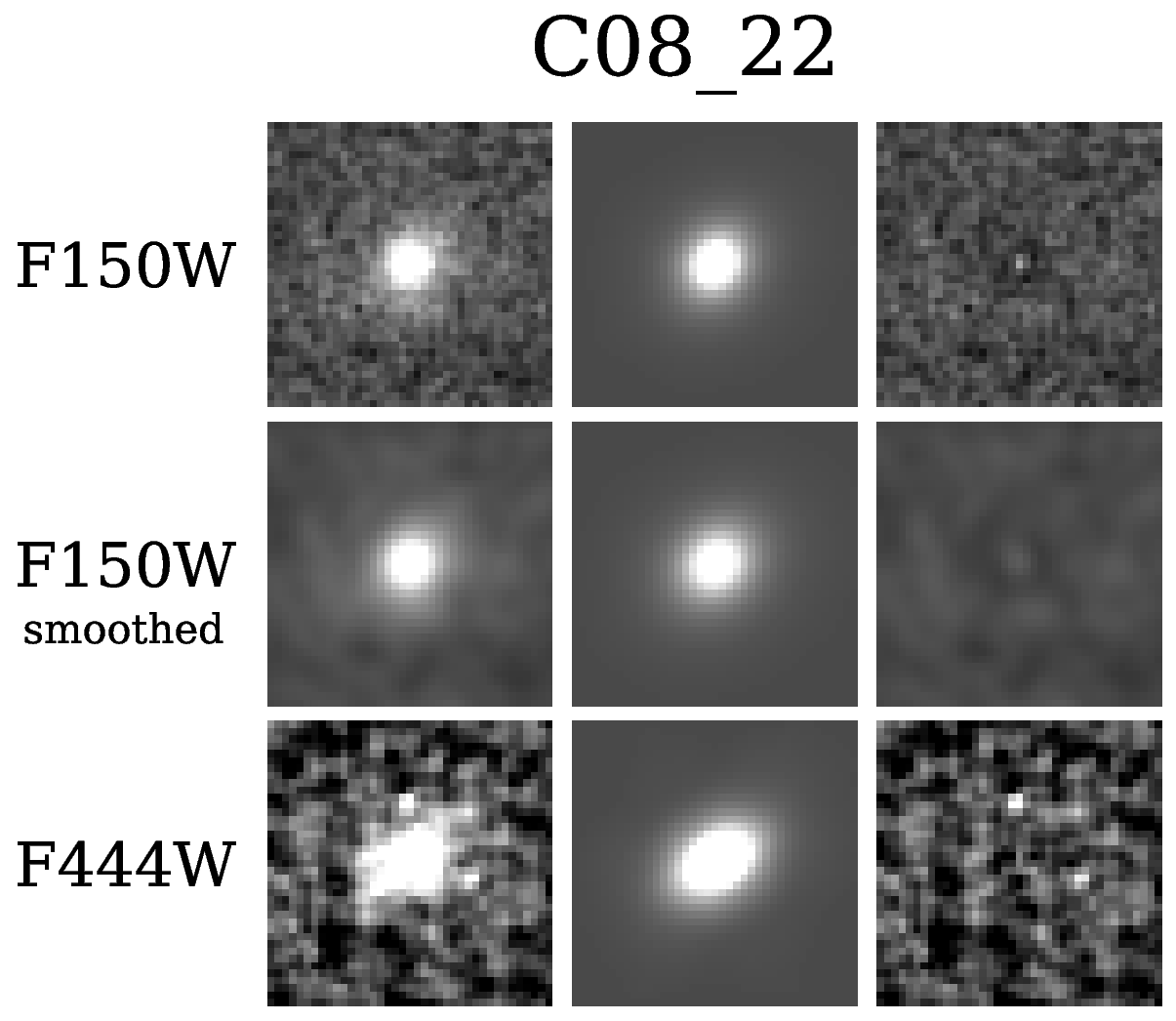}
   \includegraphics[height=0.14\textheight]{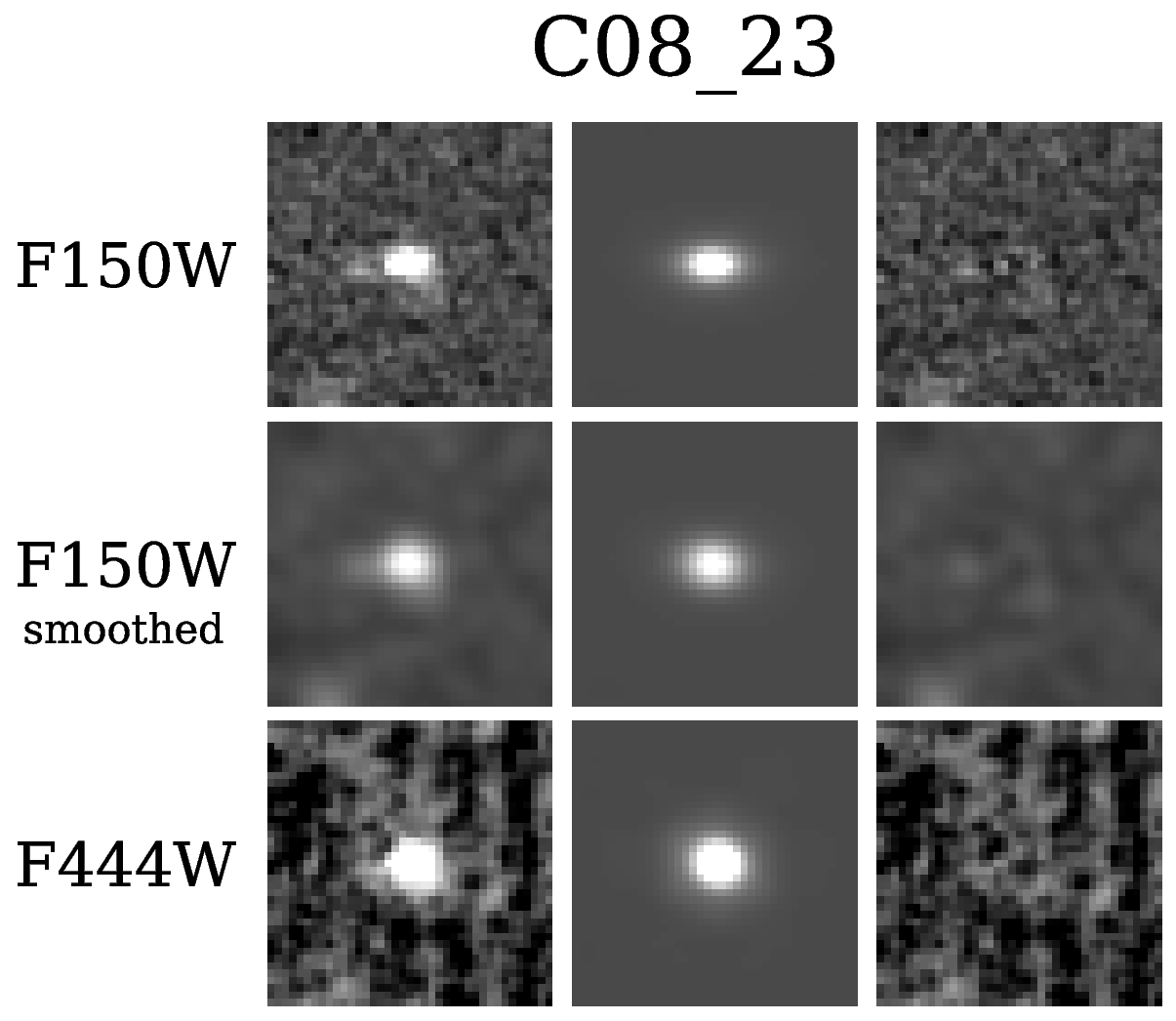}
   \includegraphics[height=0.14\textheight]{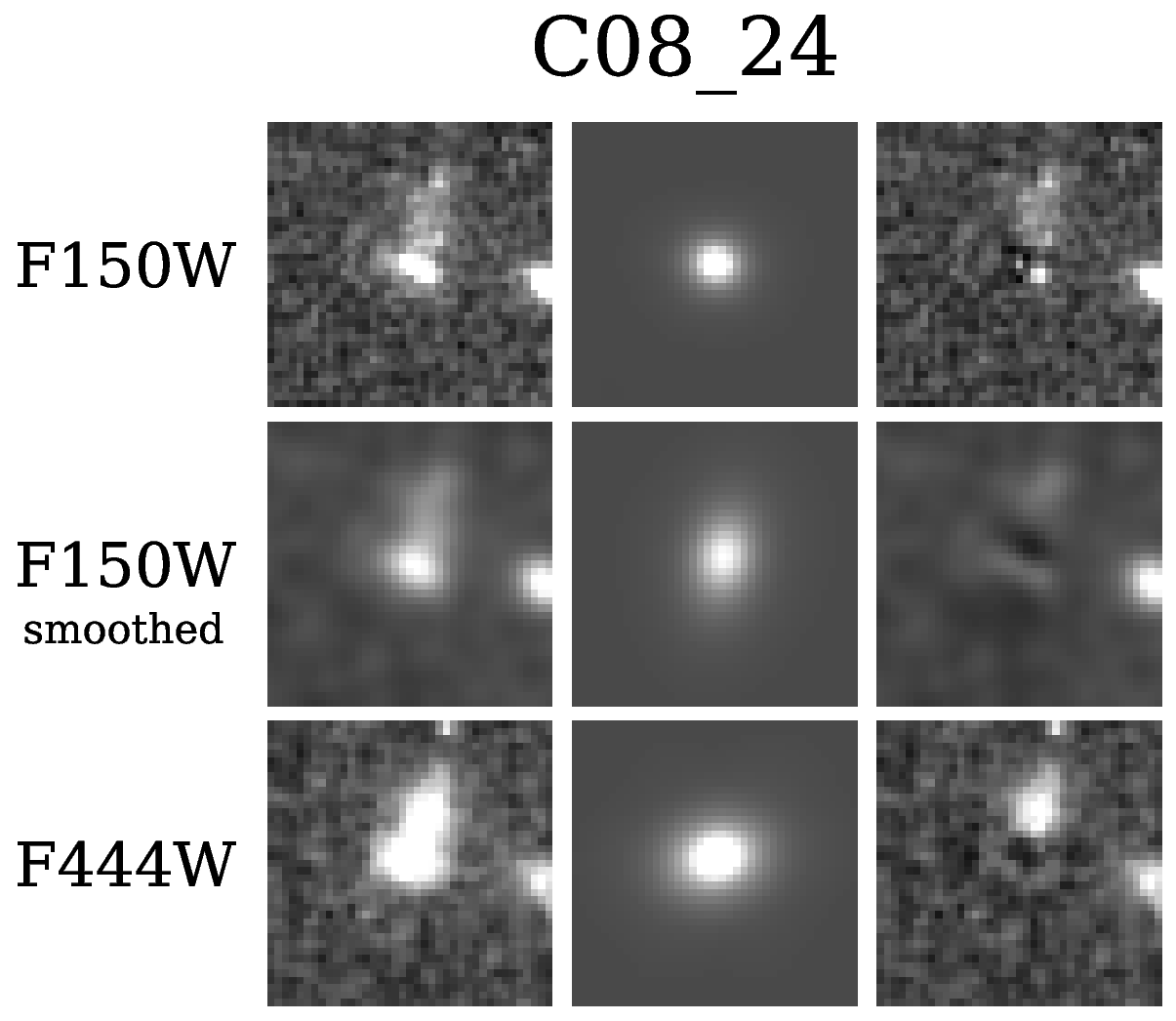}
   \includegraphics[height=0.14\textheight]{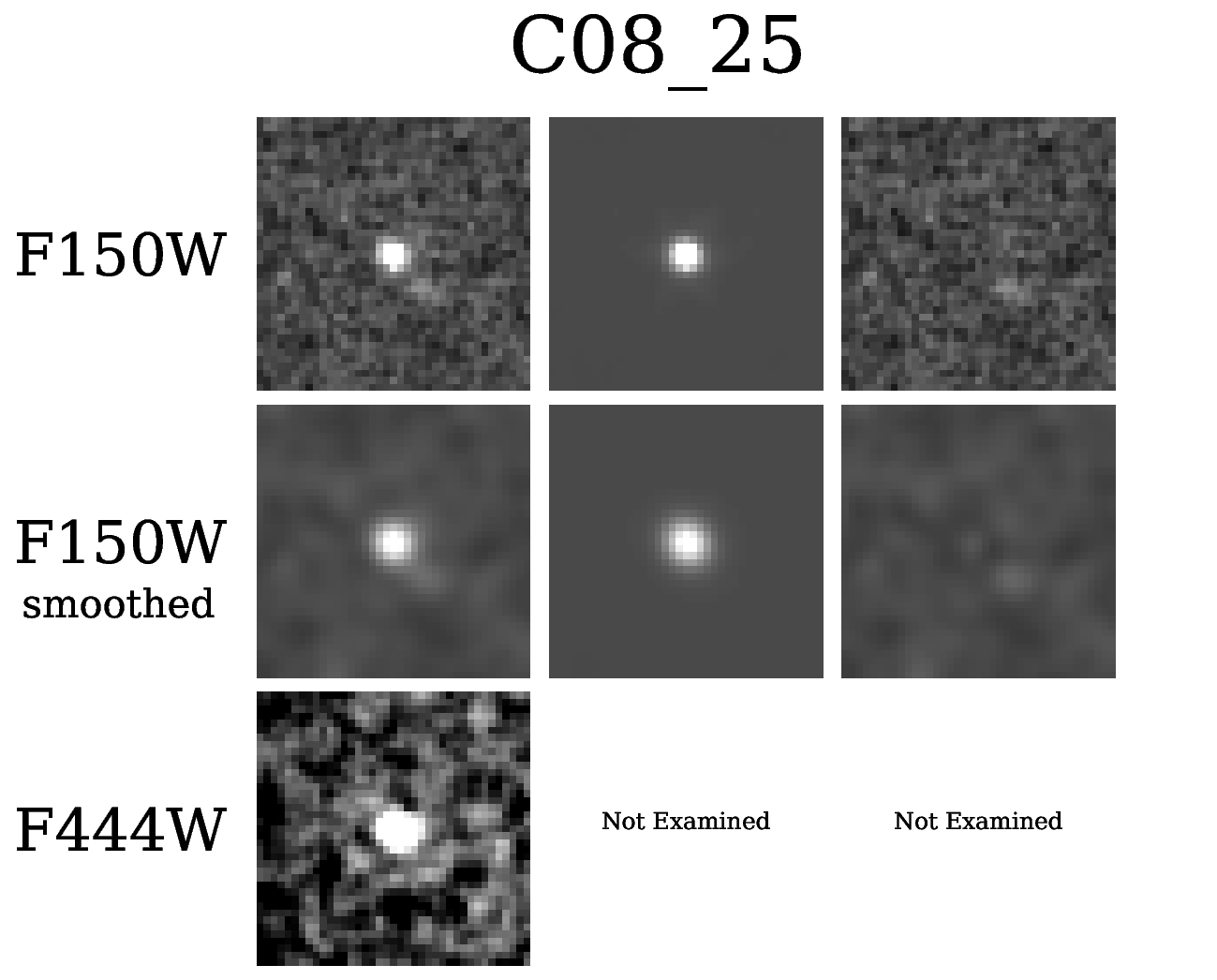}
   \includegraphics[height=0.14\textheight]{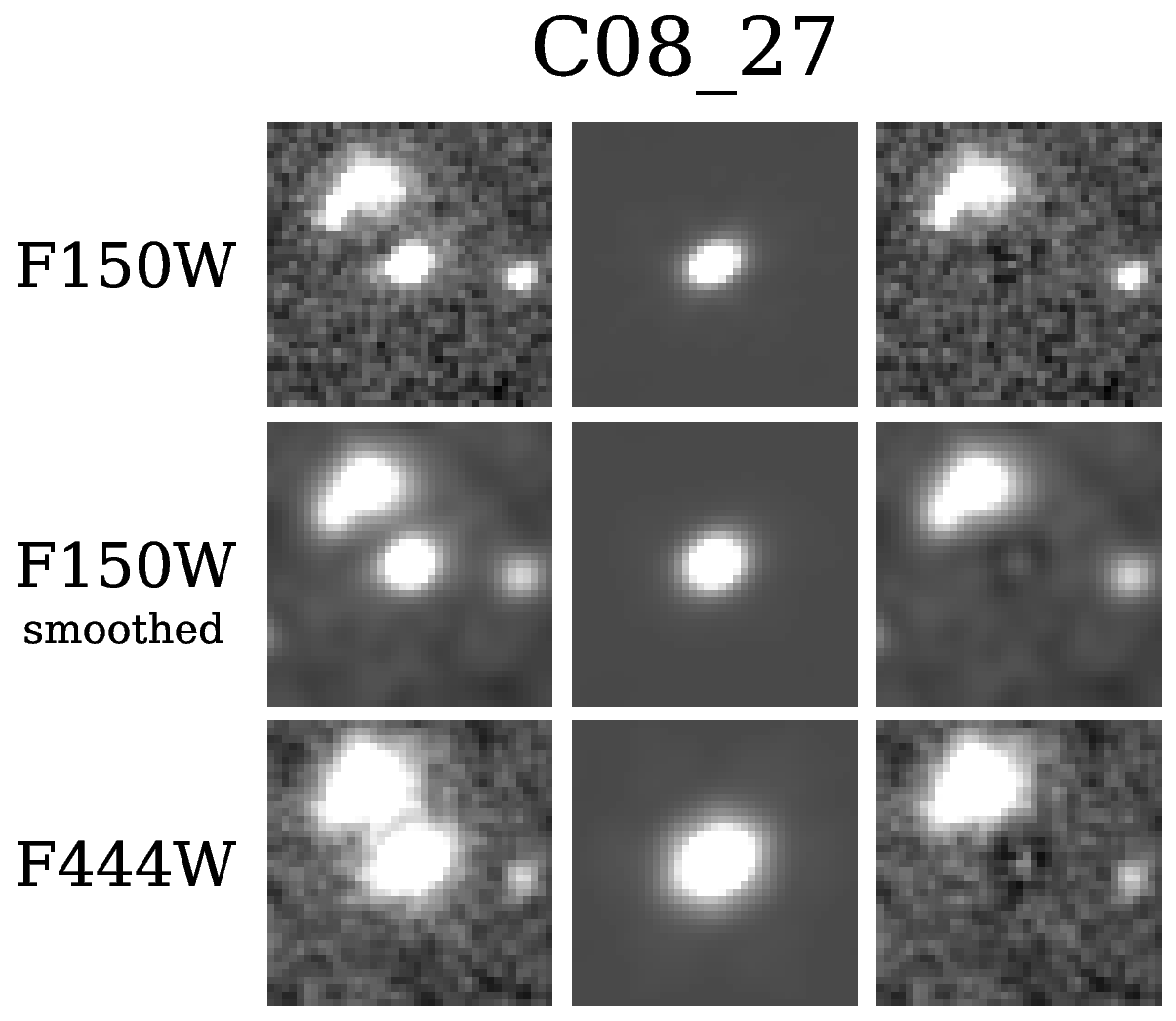}
   \includegraphics[height=0.14\textheight]{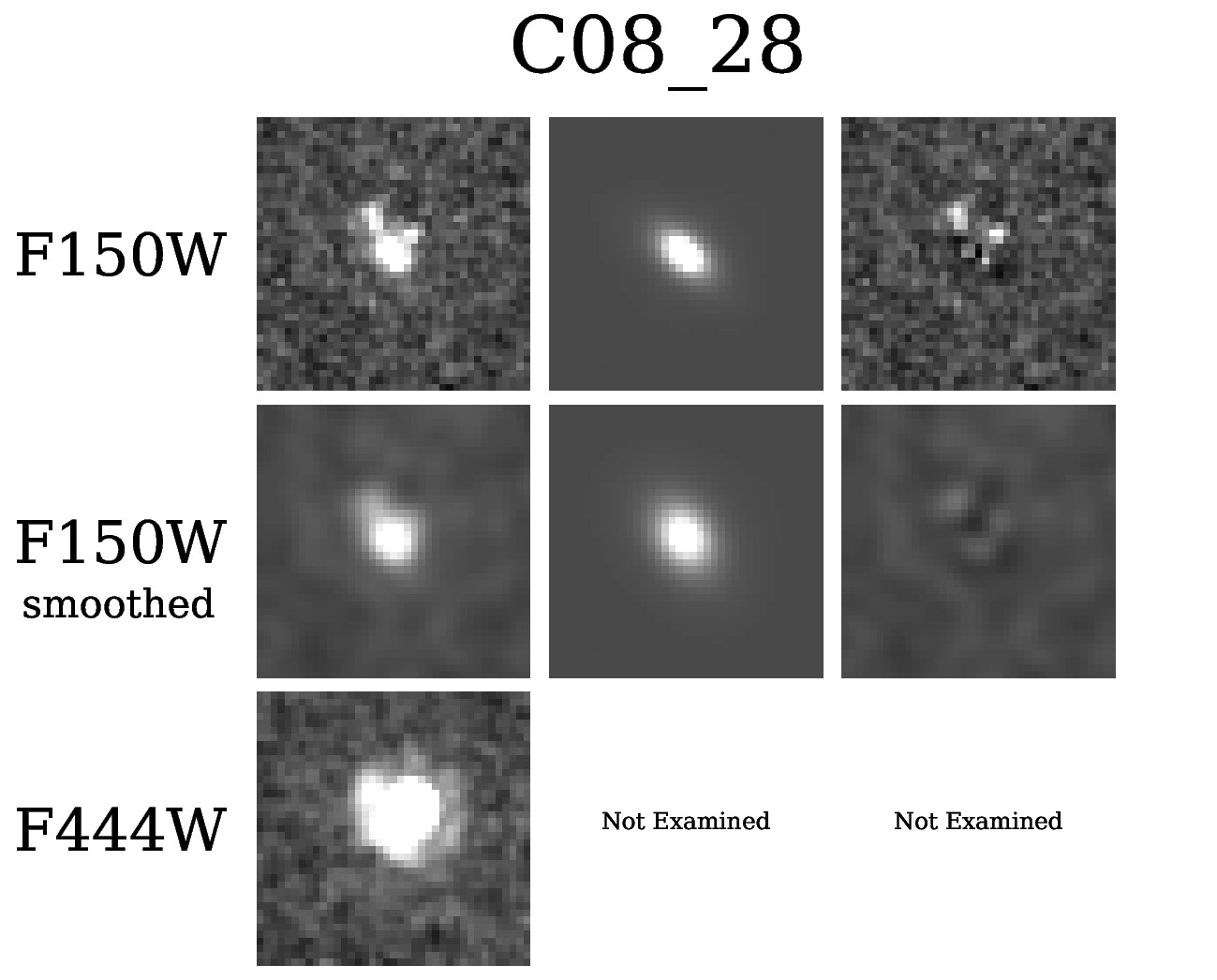}
   \includegraphics[height=0.14\textheight]{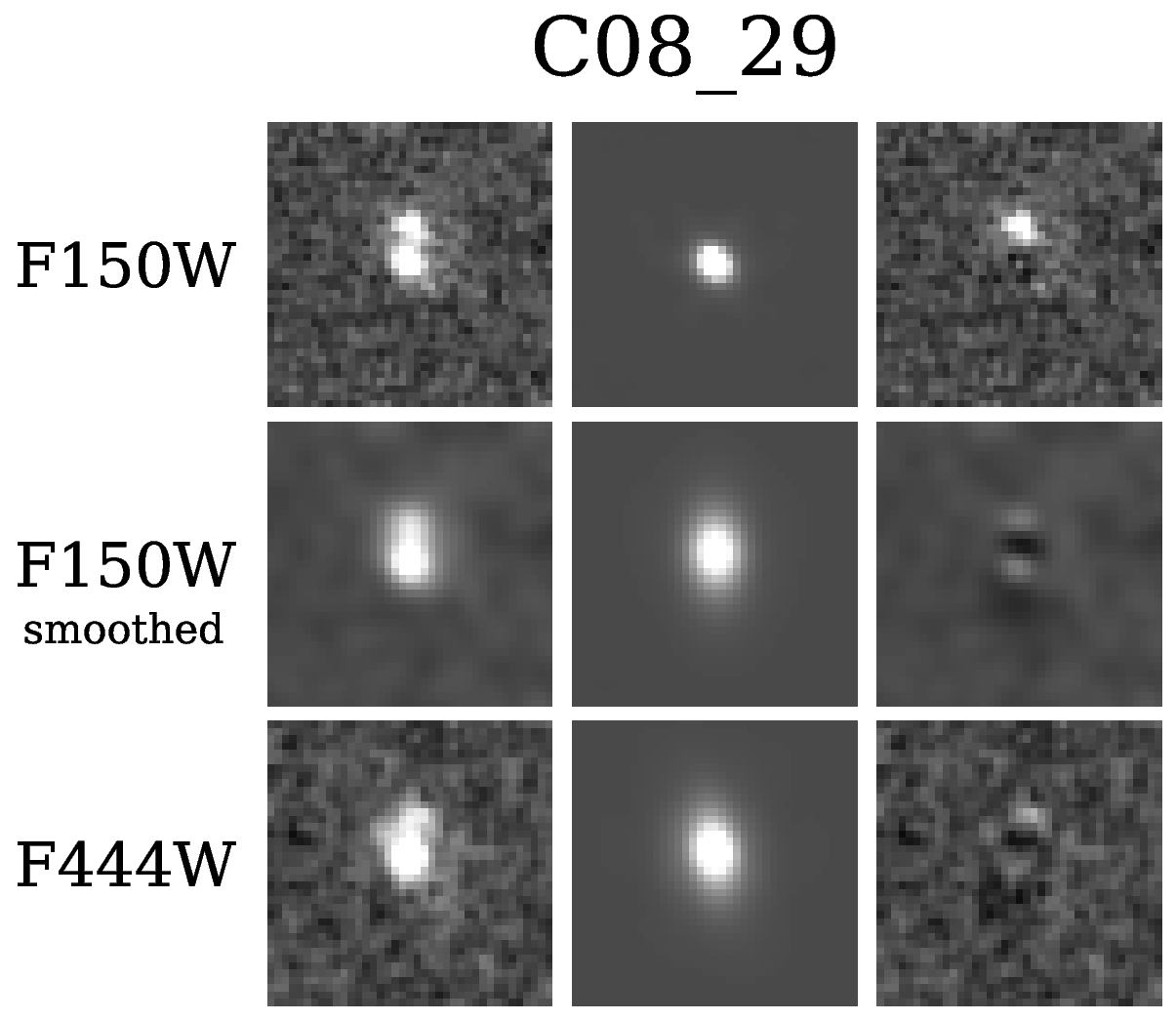}
   \includegraphics[height=0.14\textheight]{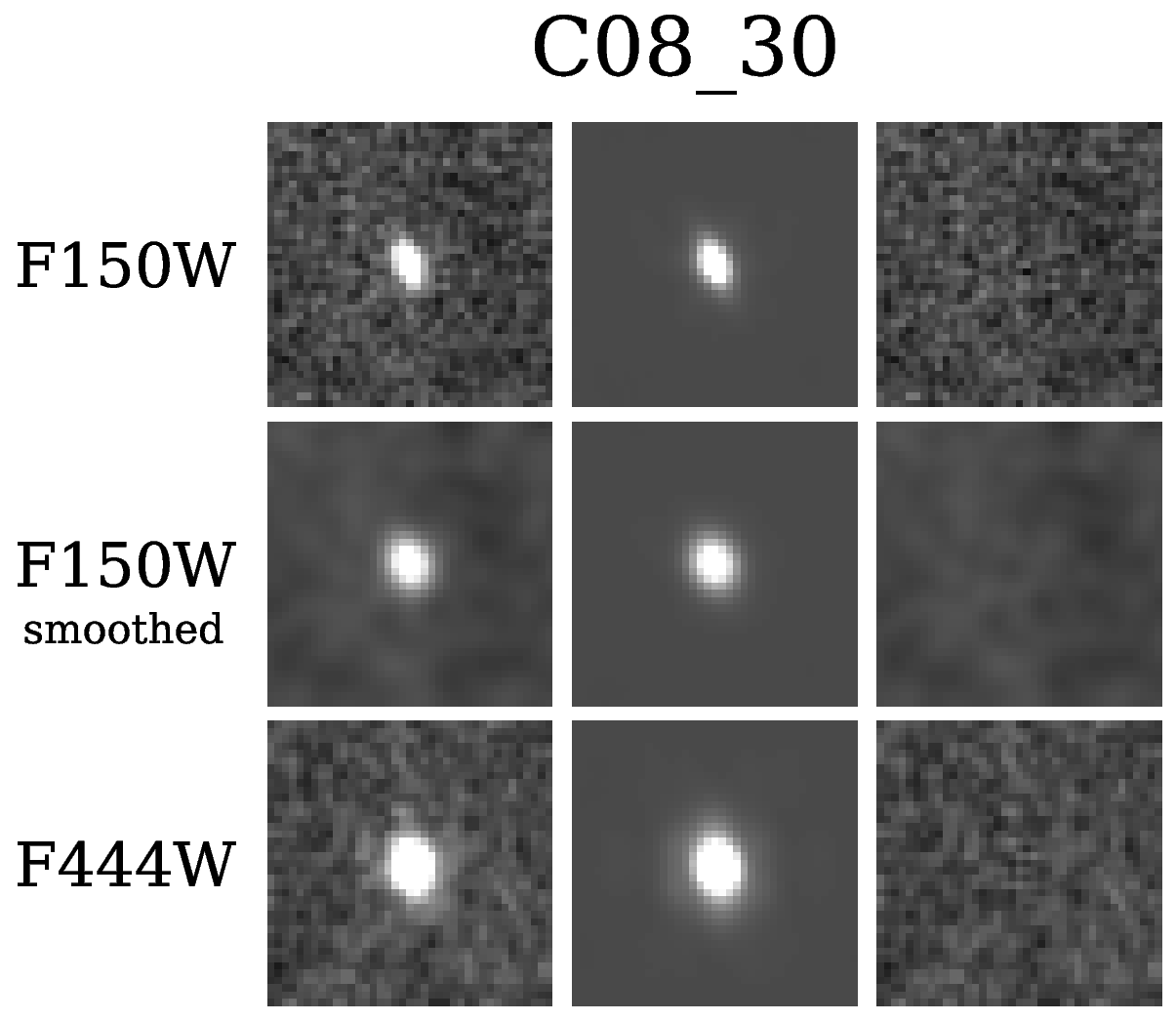}
   \includegraphics[height=0.14\textheight]{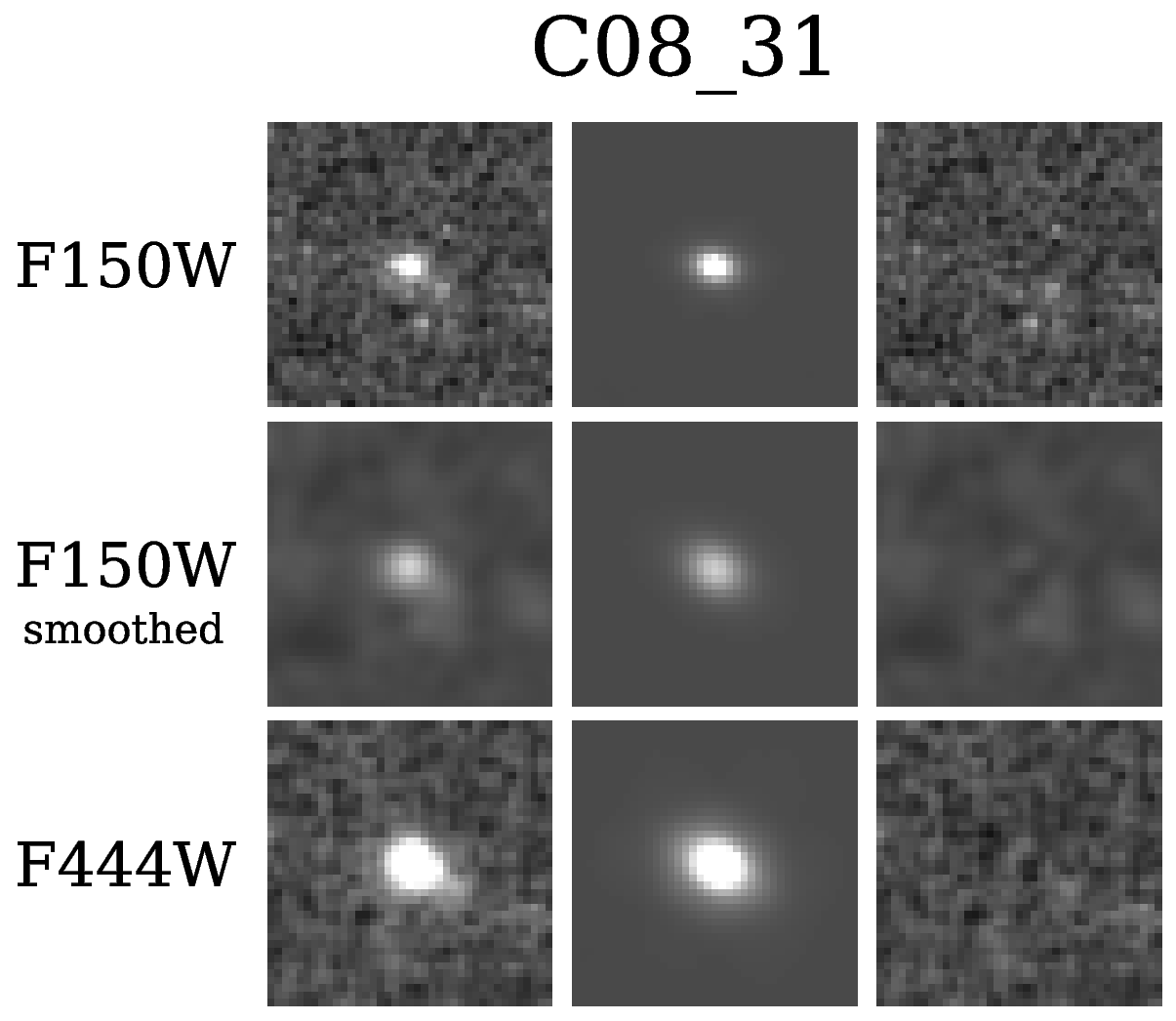}
   \includegraphics[height=0.14\textheight]{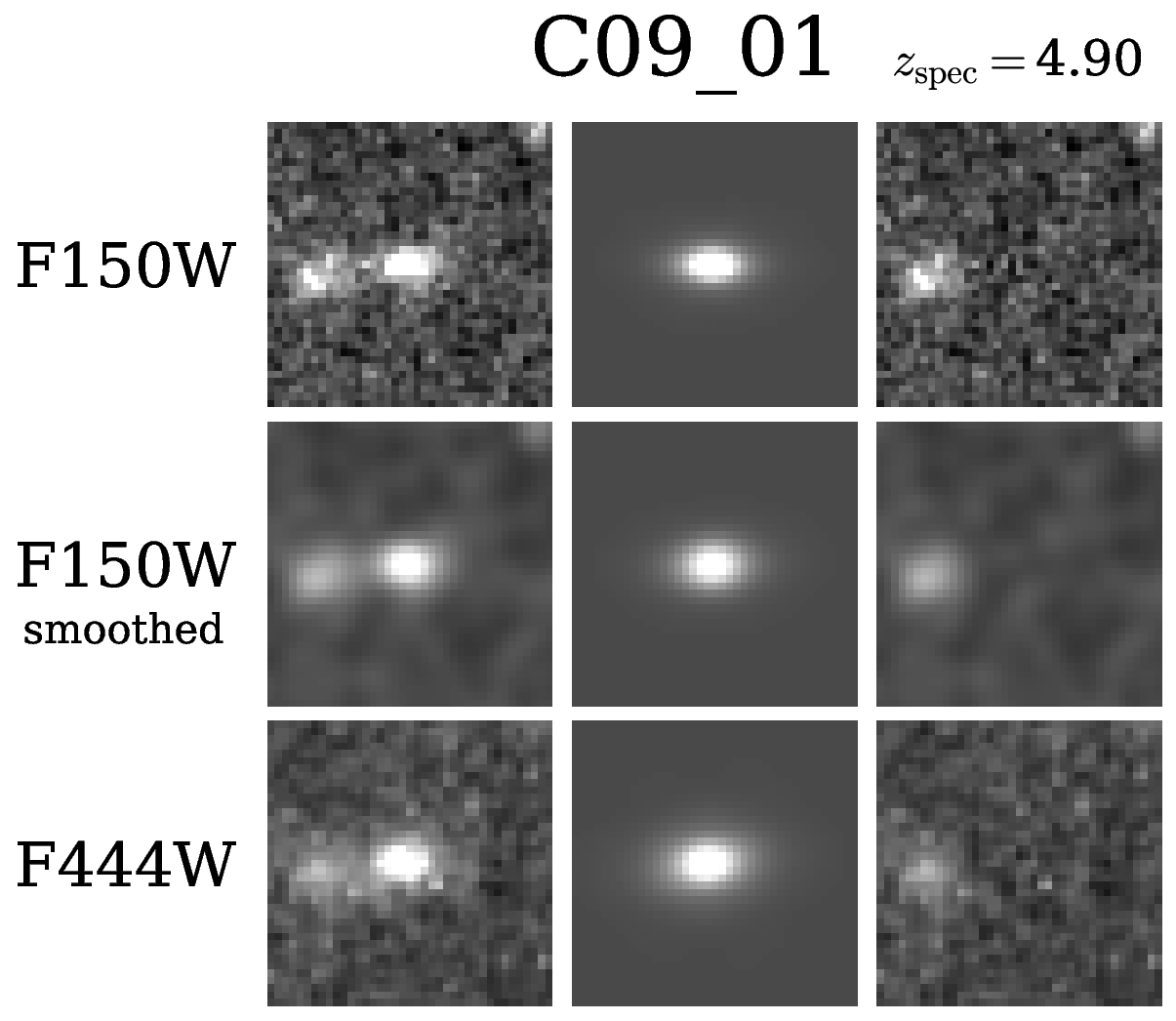}
   \includegraphics[height=0.14\textheight]{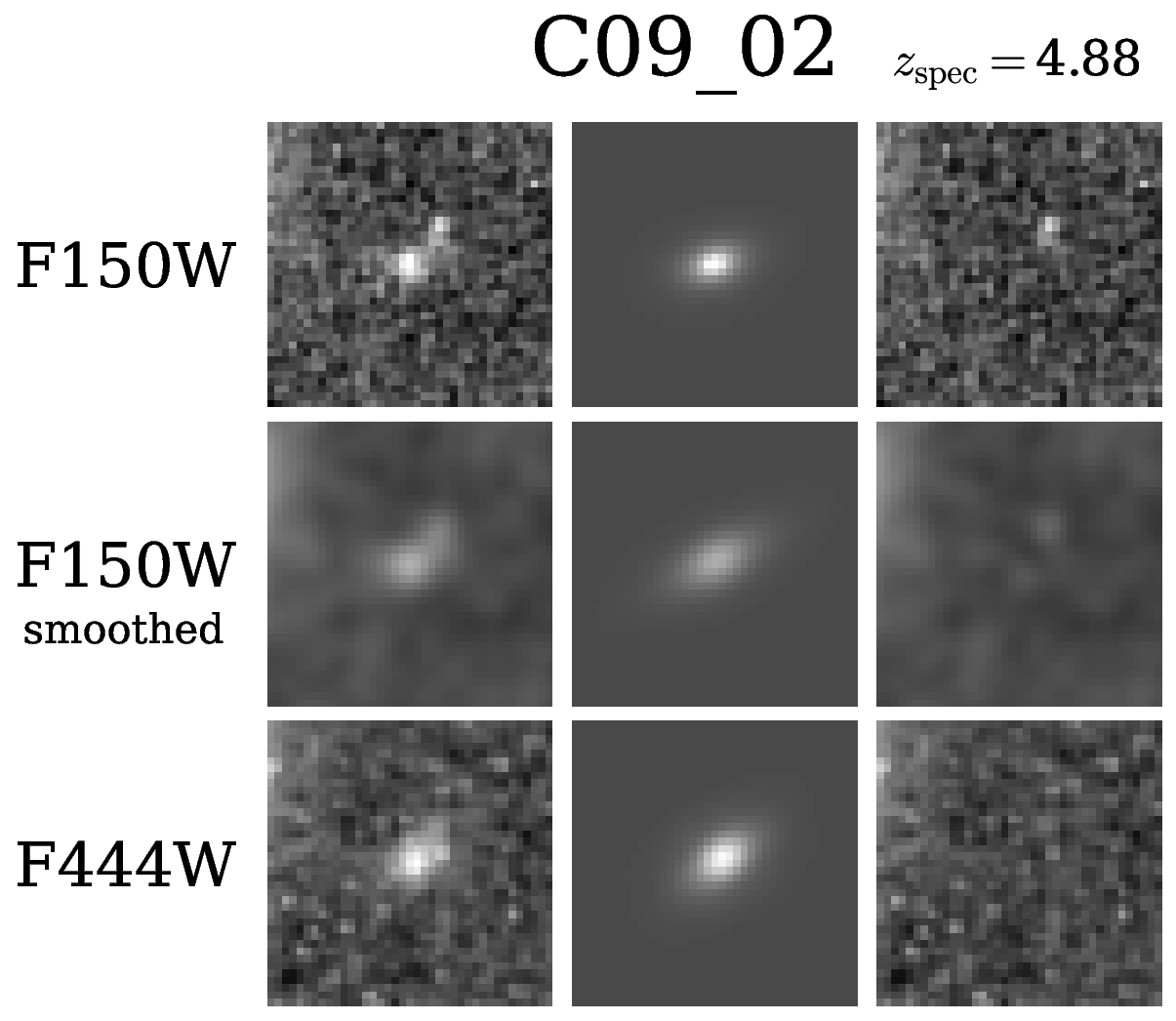}
   \includegraphics[height=0.14\textheight]{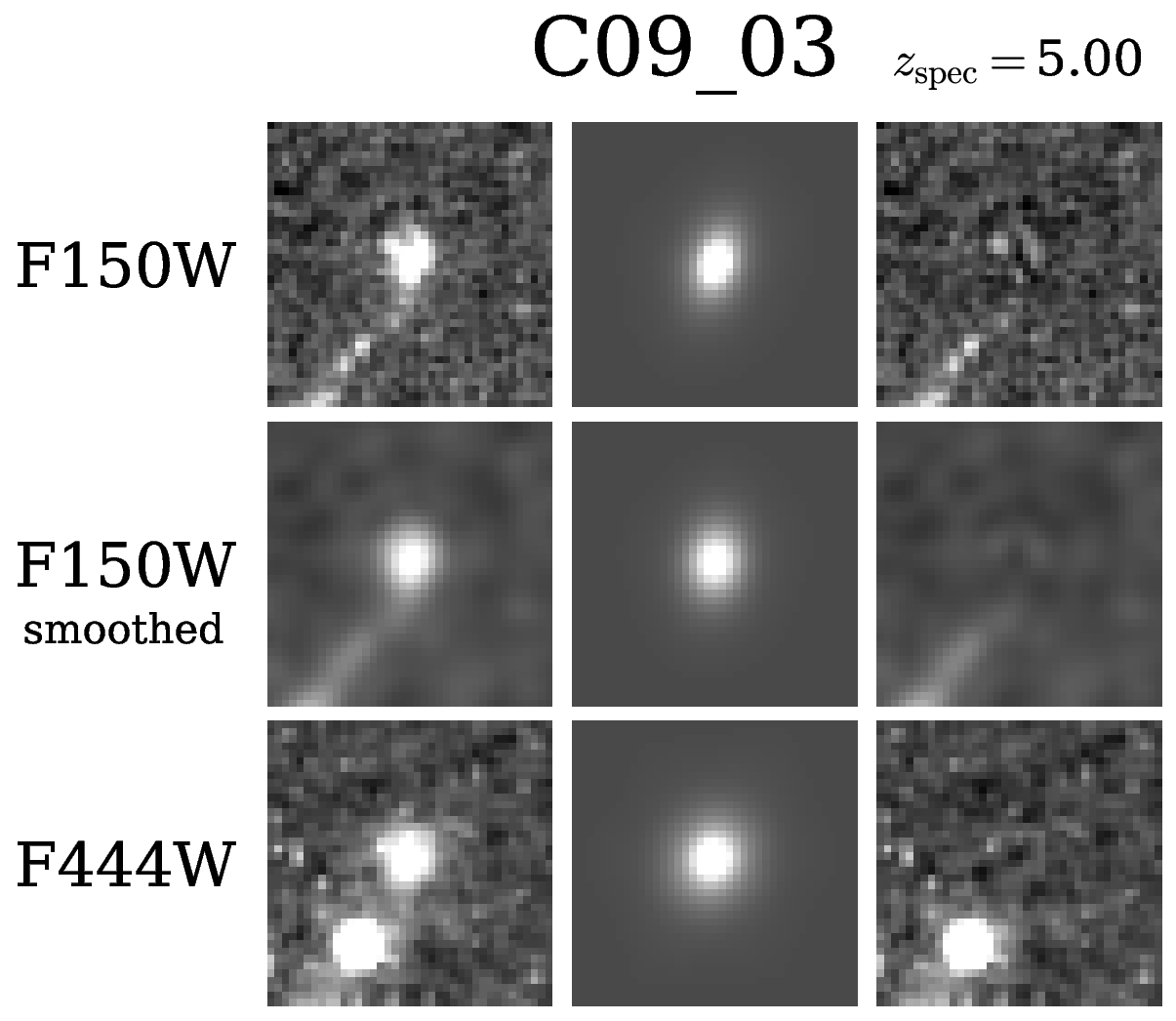}
   \includegraphics[height=0.14\textheight]{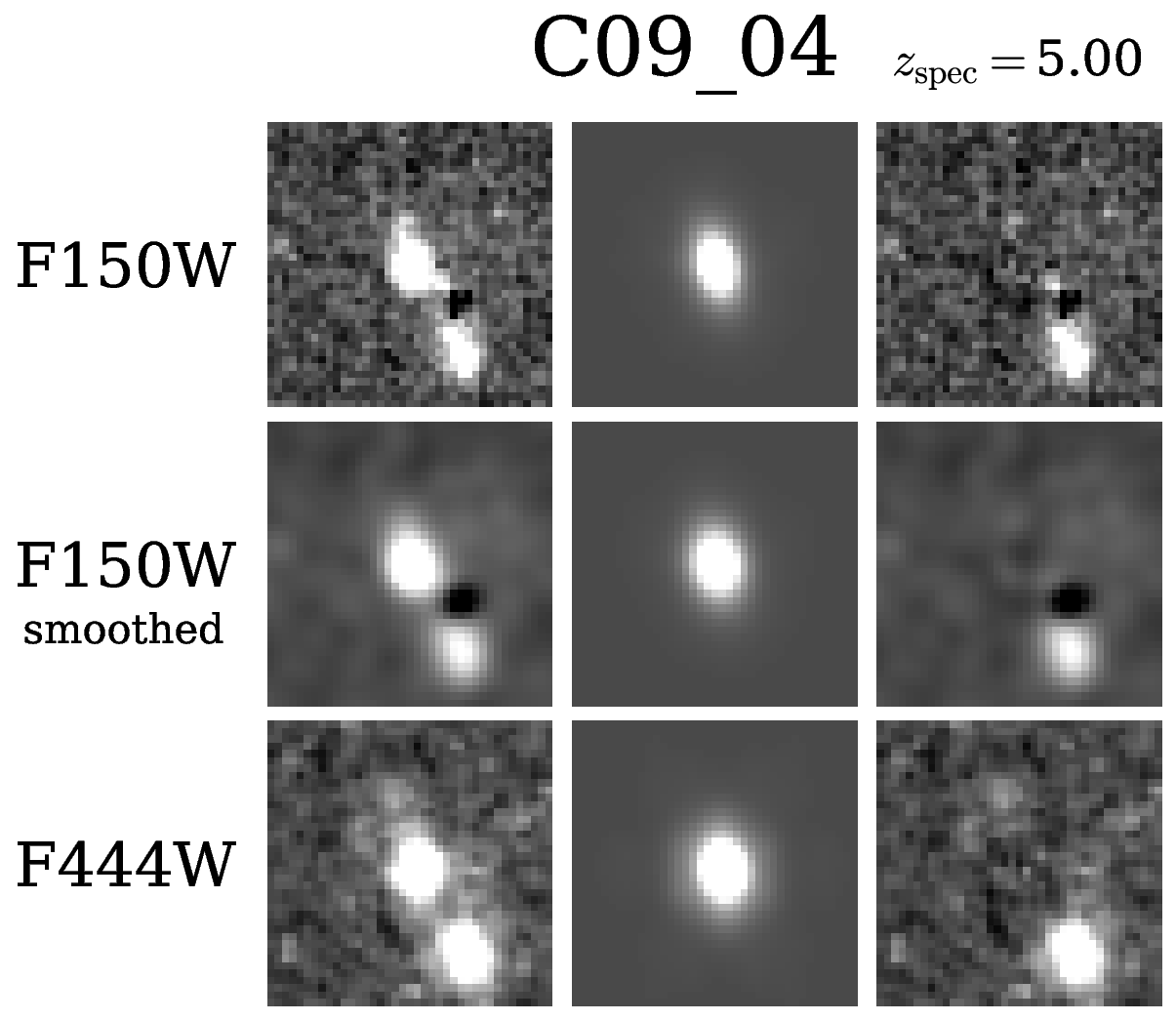}
   \includegraphics[height=0.14\textheight]{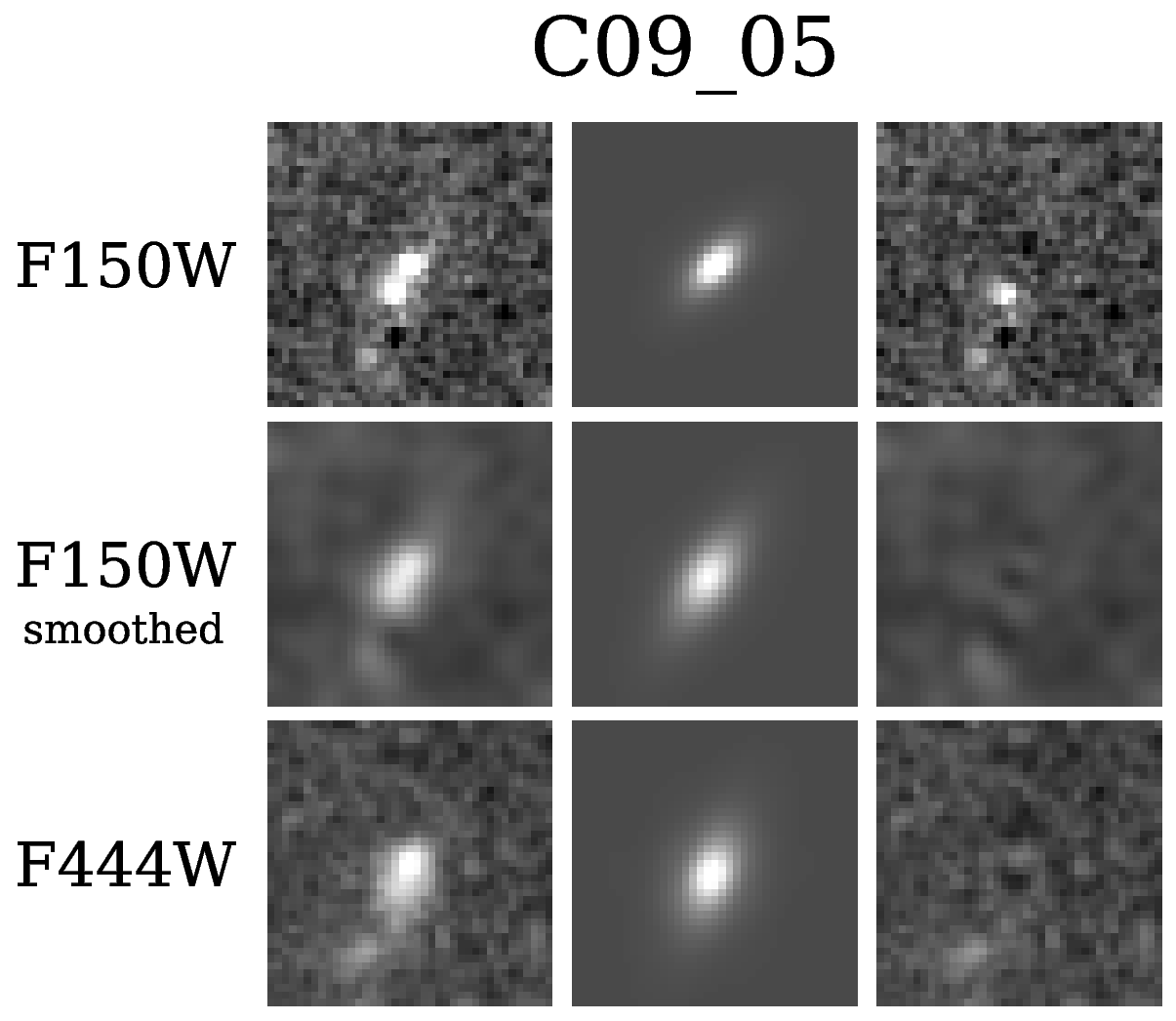}
   \includegraphics[height=0.14\textheight]{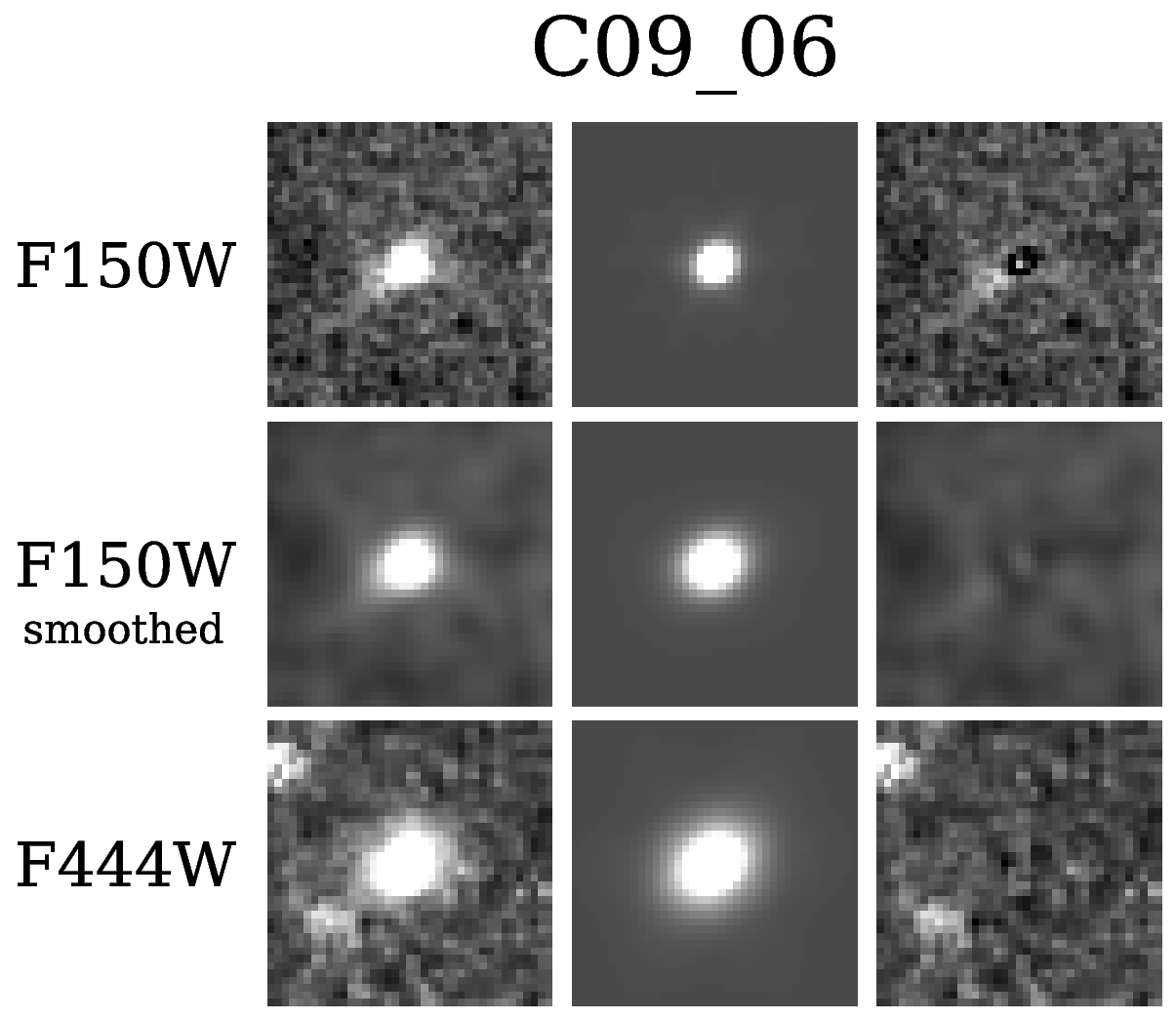}
   \includegraphics[height=0.14\textheight]{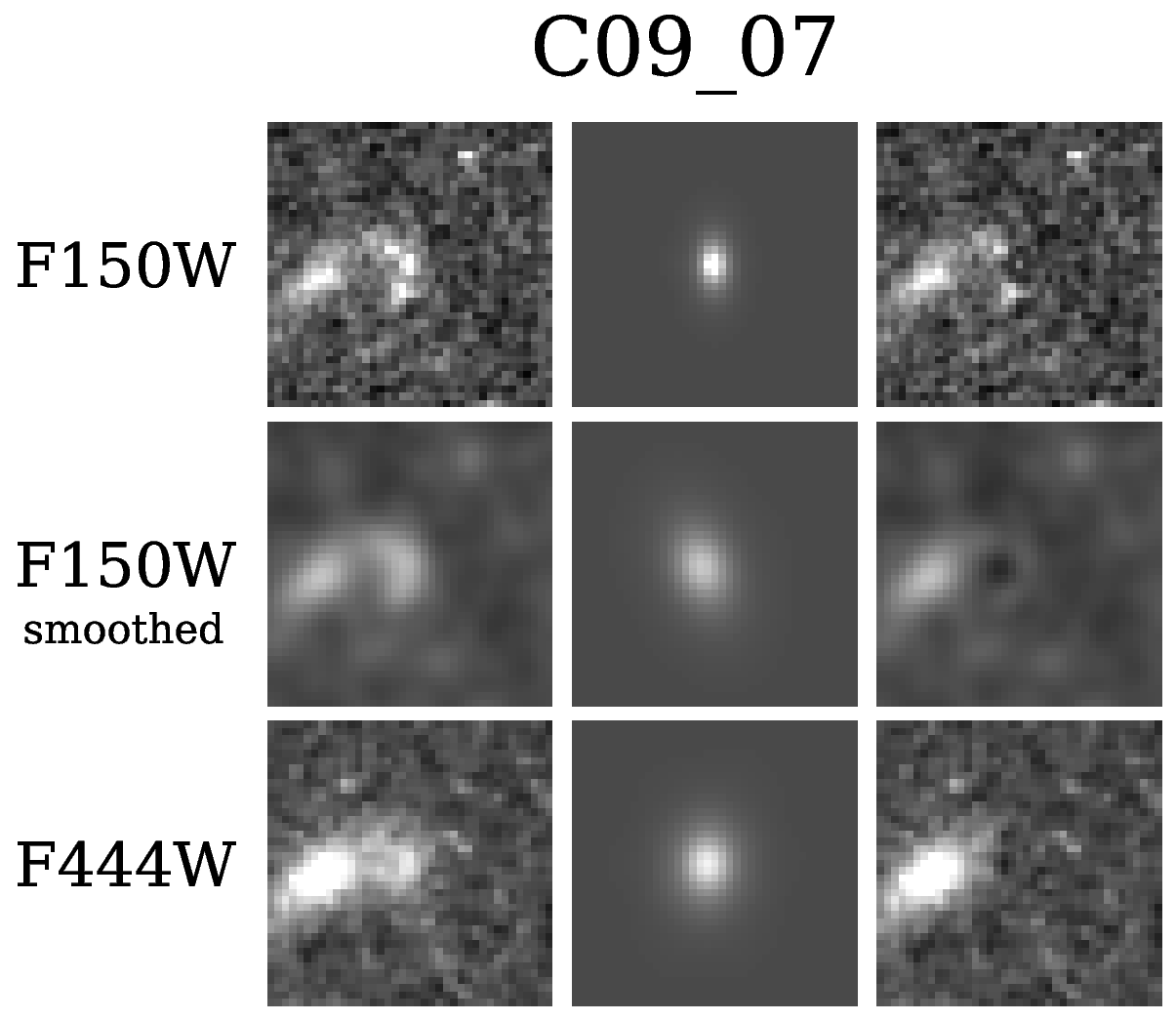}
   \includegraphics[height=0.14\textheight]{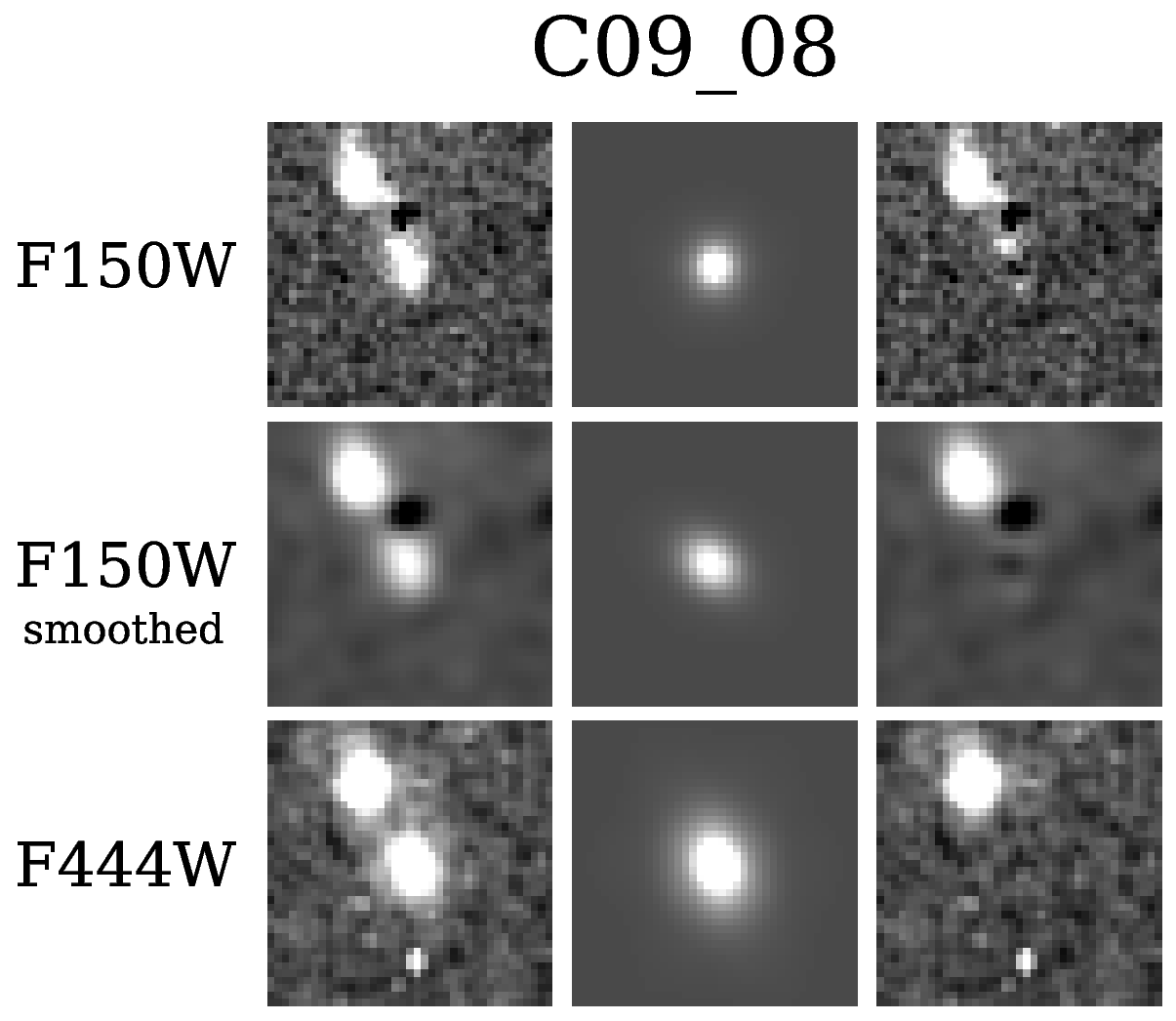}
   \includegraphics[height=0.14\textheight]{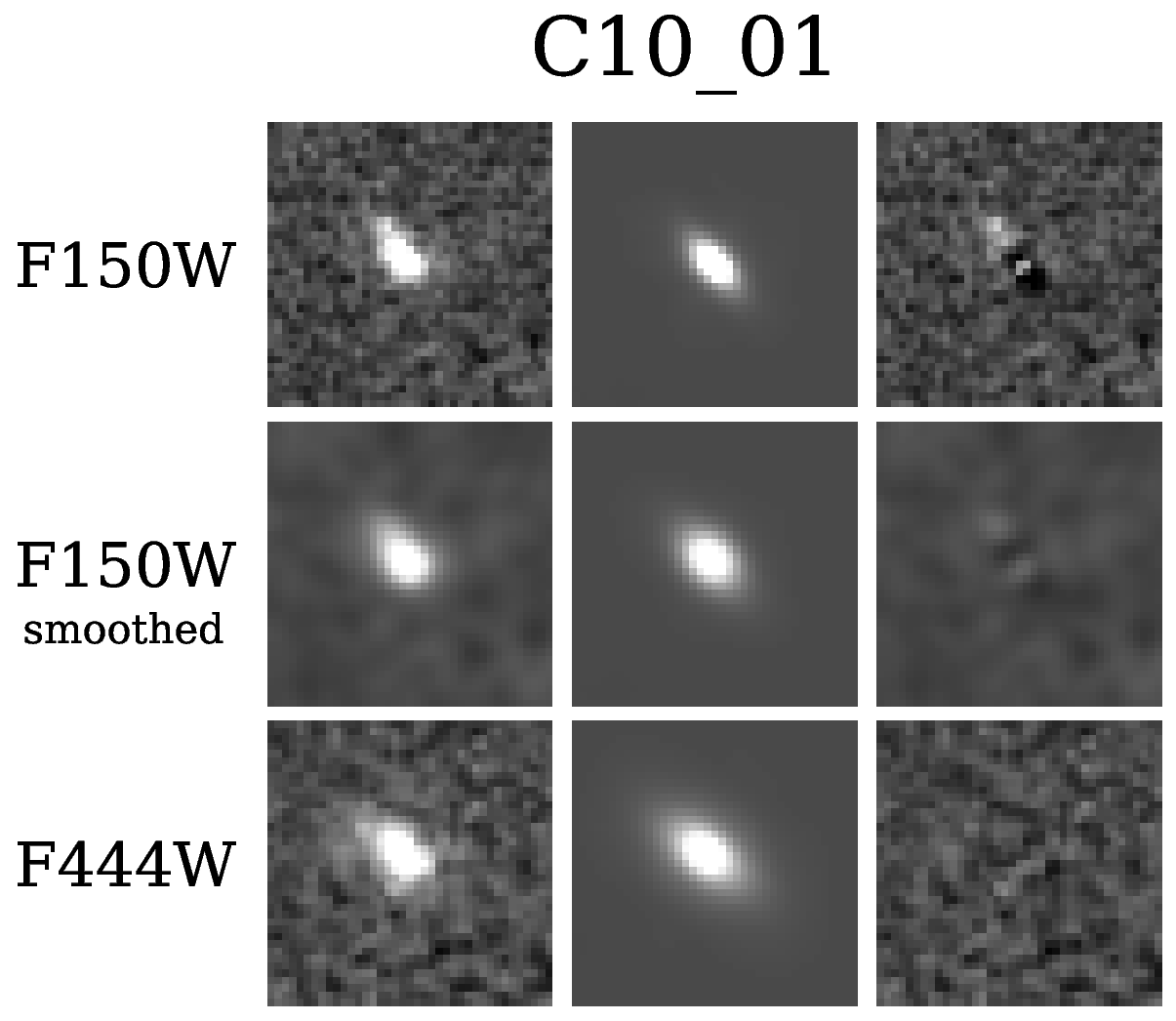}
   \includegraphics[height=0.14\textheight]{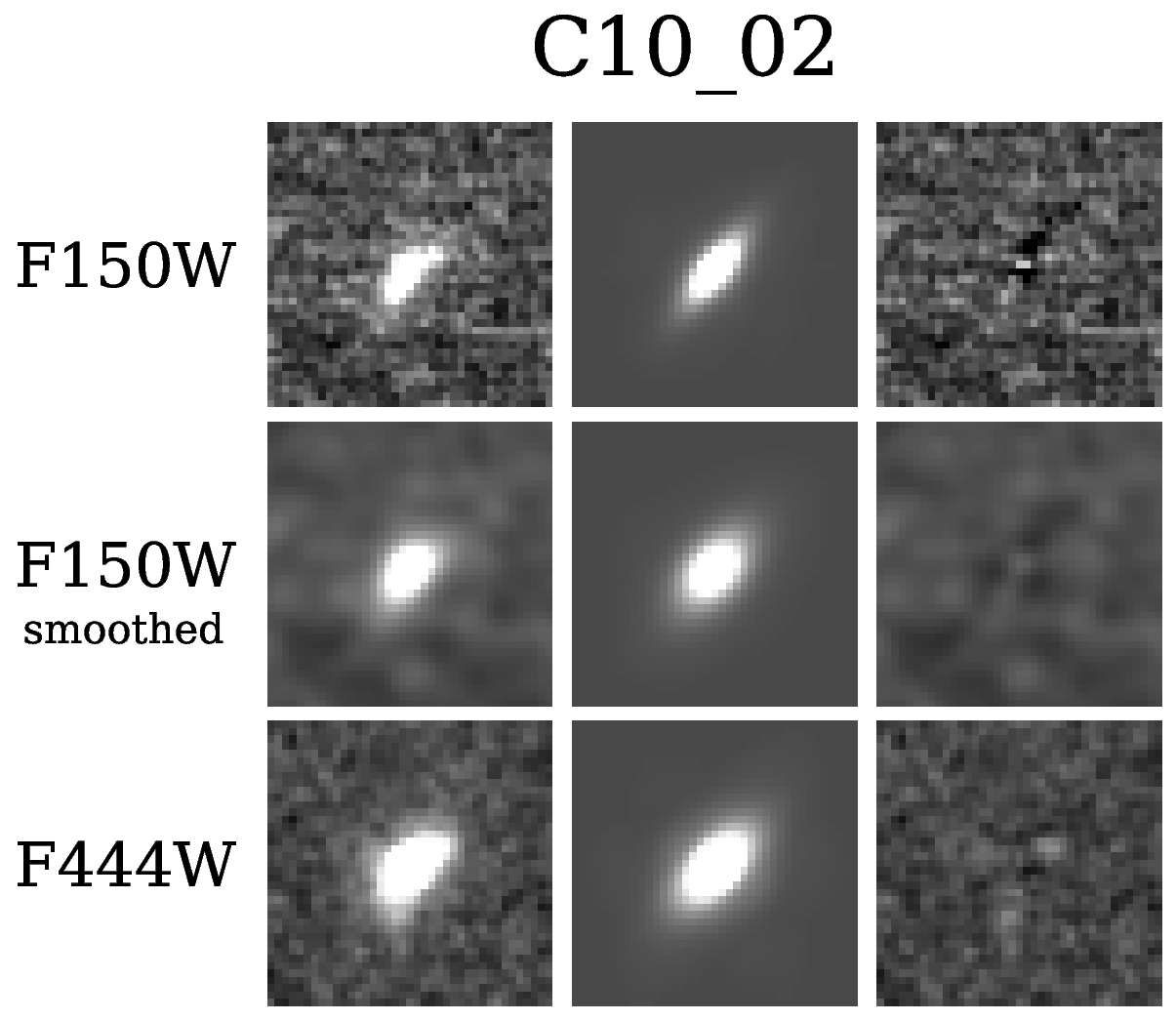}
   \includegraphics[height=0.14\textheight]{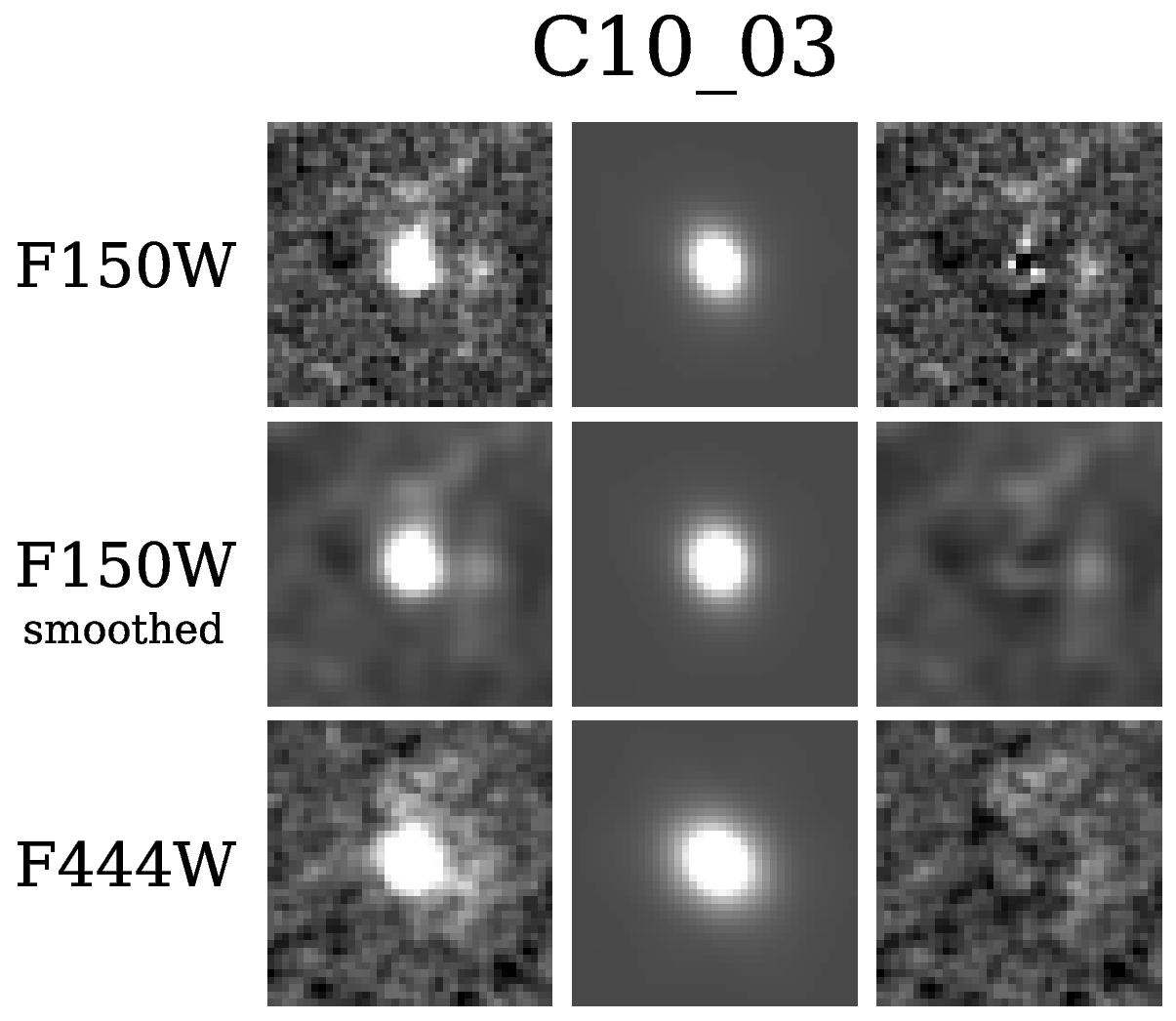}
   \includegraphics[height=0.14\textheight]{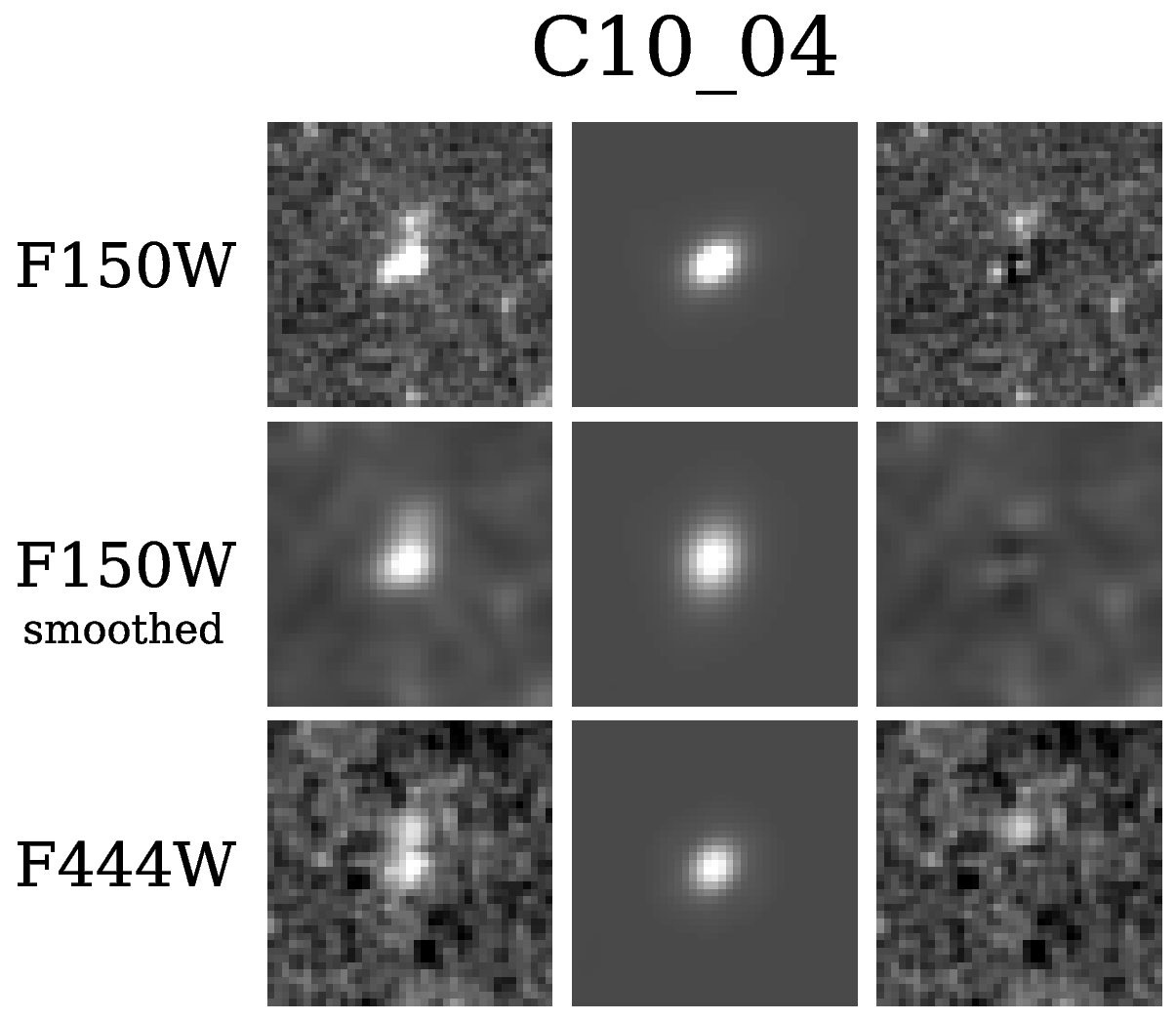}
   \includegraphics[height=0.14\textheight]{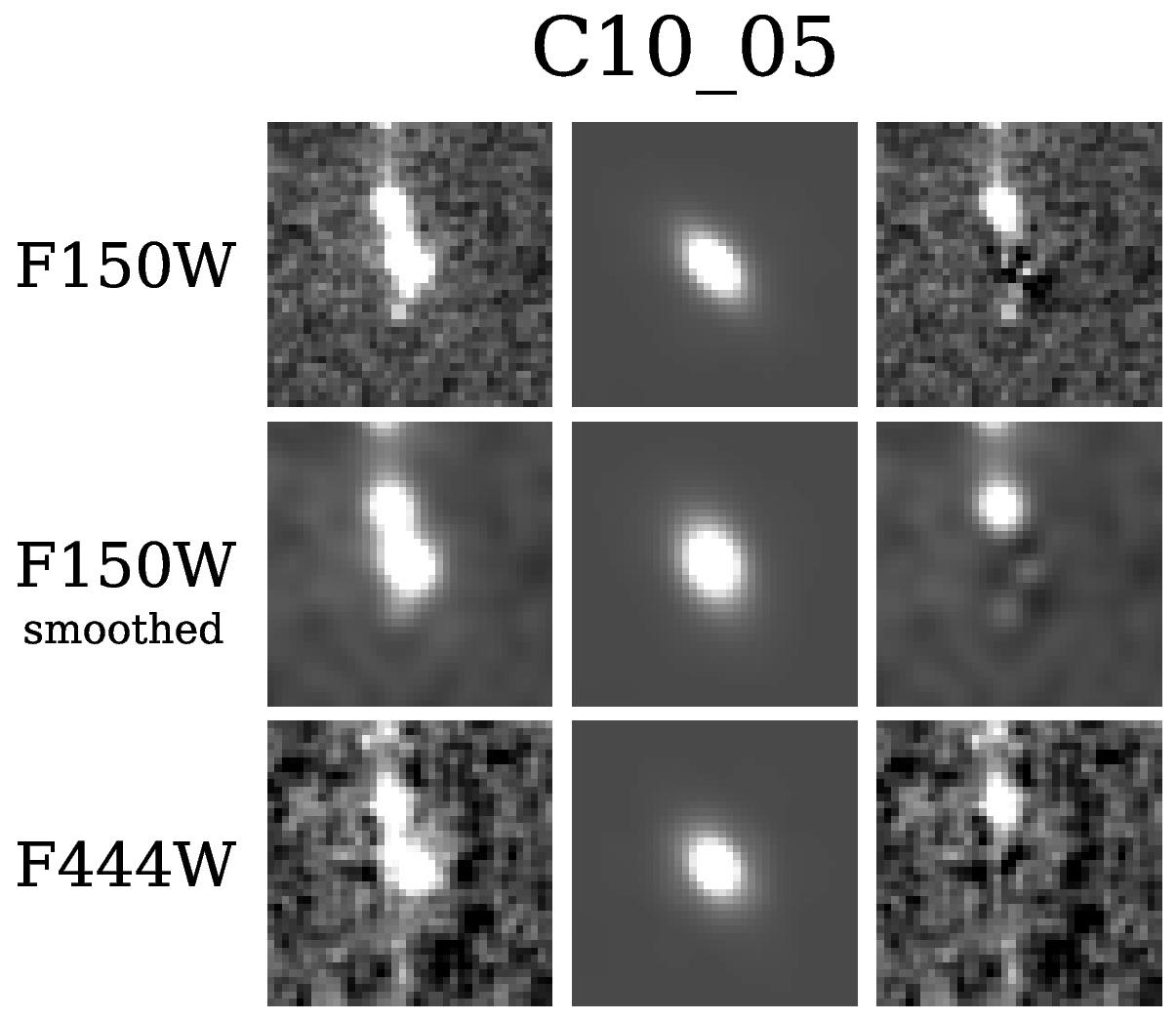}
\caption{
(Continued)
}
\end{center}
\end{figure*}

\addtocounter{figure}{-1}
\begin{figure*}
\begin{center}
   \includegraphics[height=0.14\textheight]{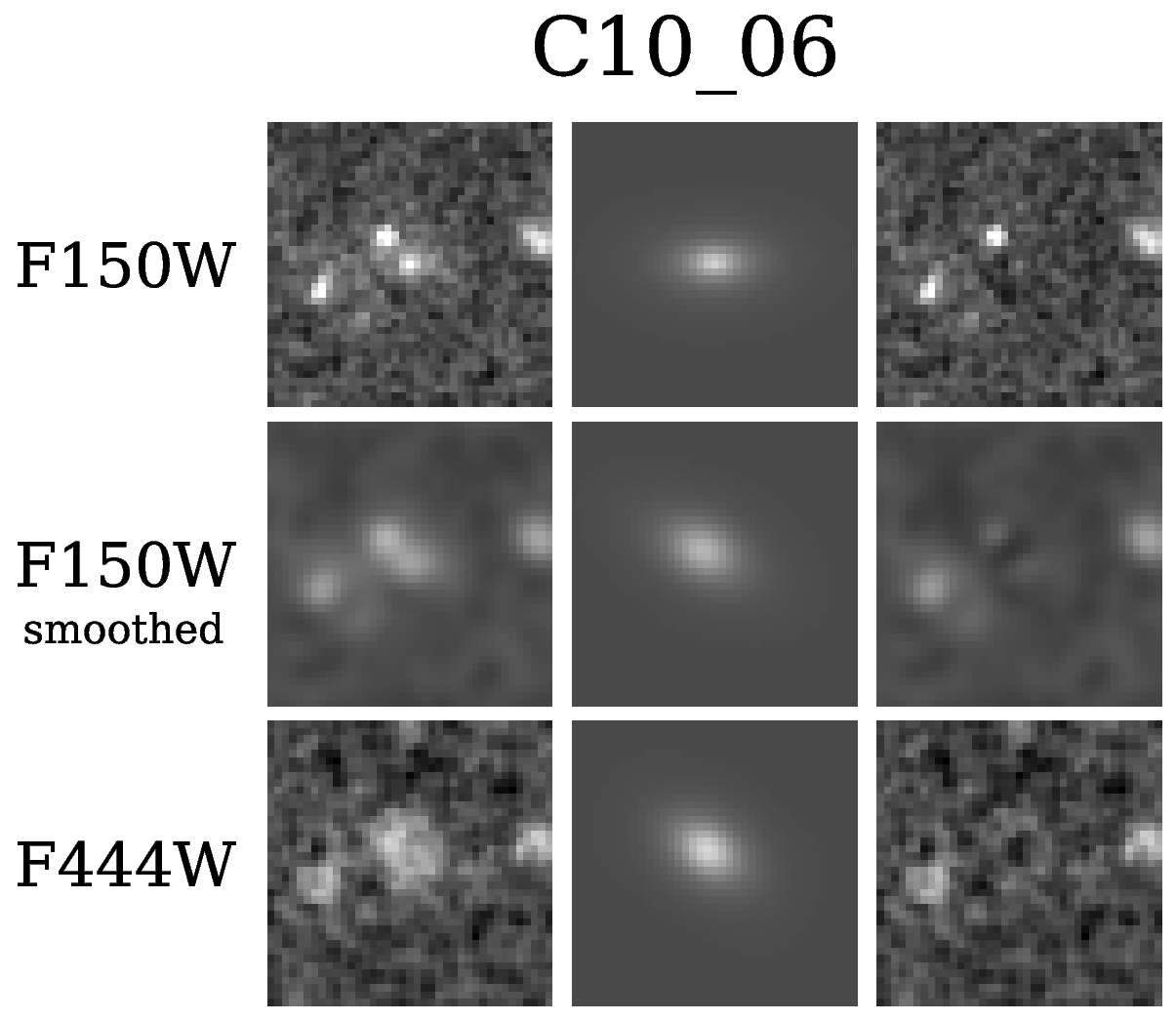}
   \includegraphics[height=0.14\textheight]{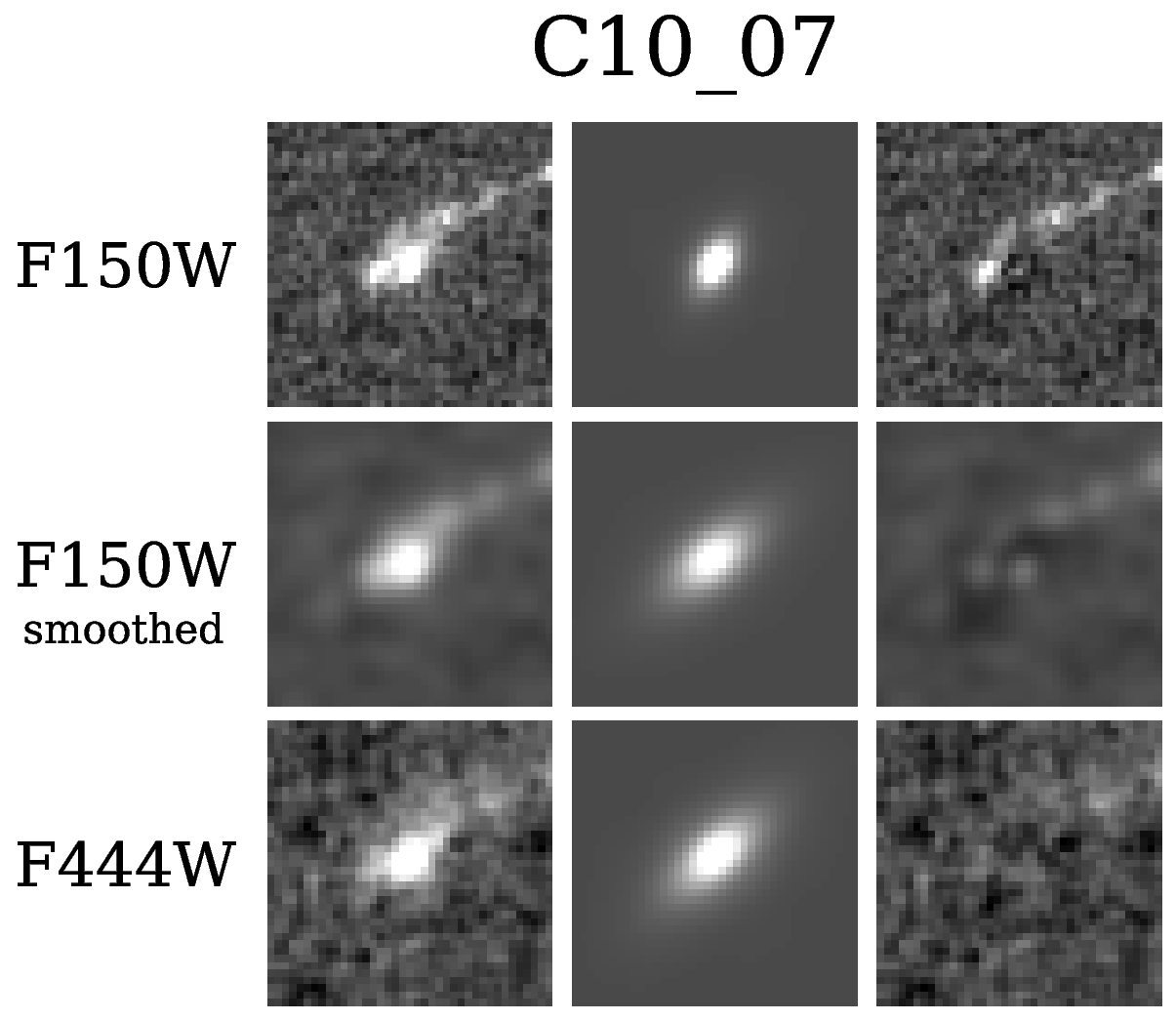}
   \includegraphics[height=0.14\textheight]{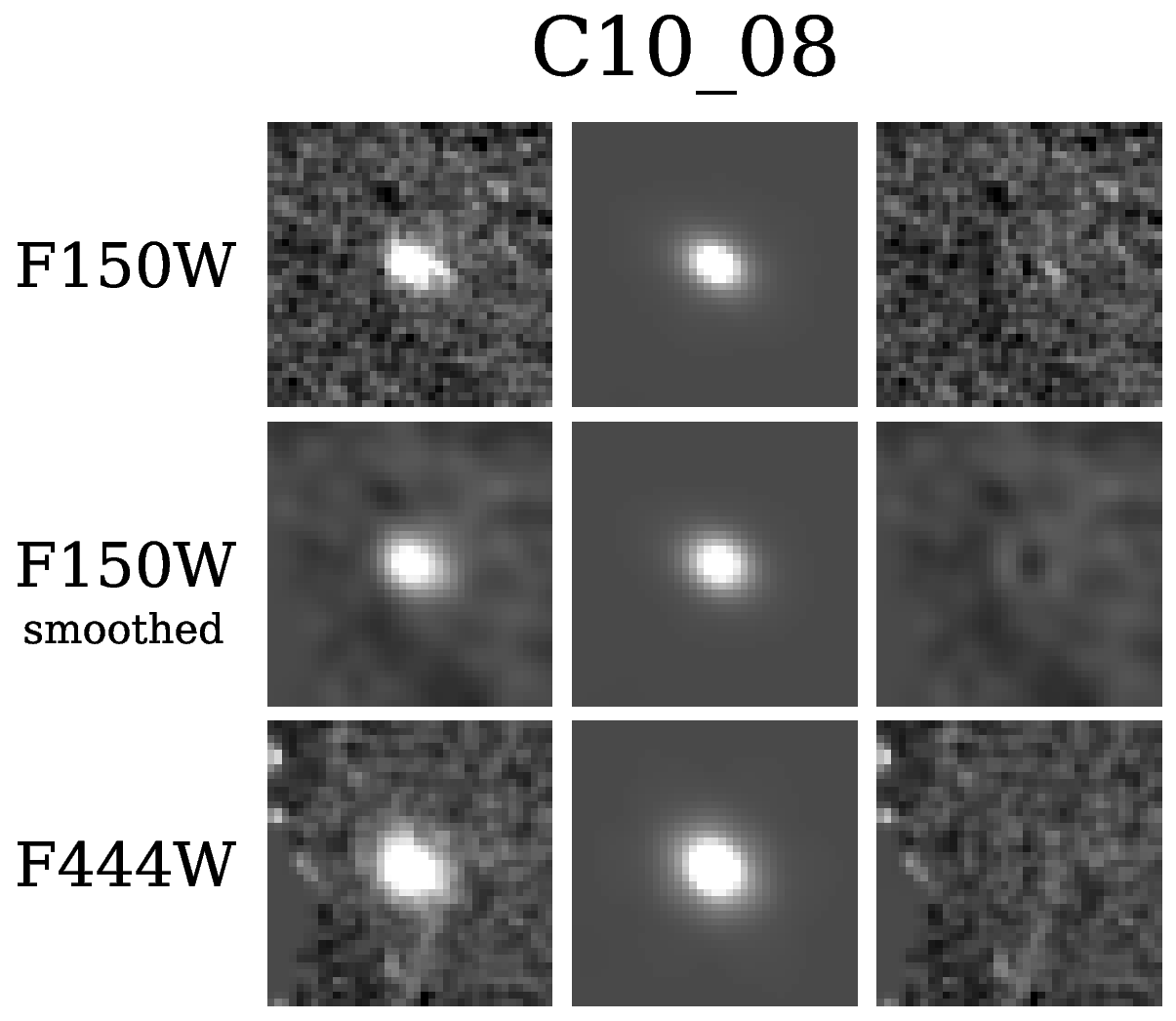}
   \includegraphics[height=0.14\textheight]{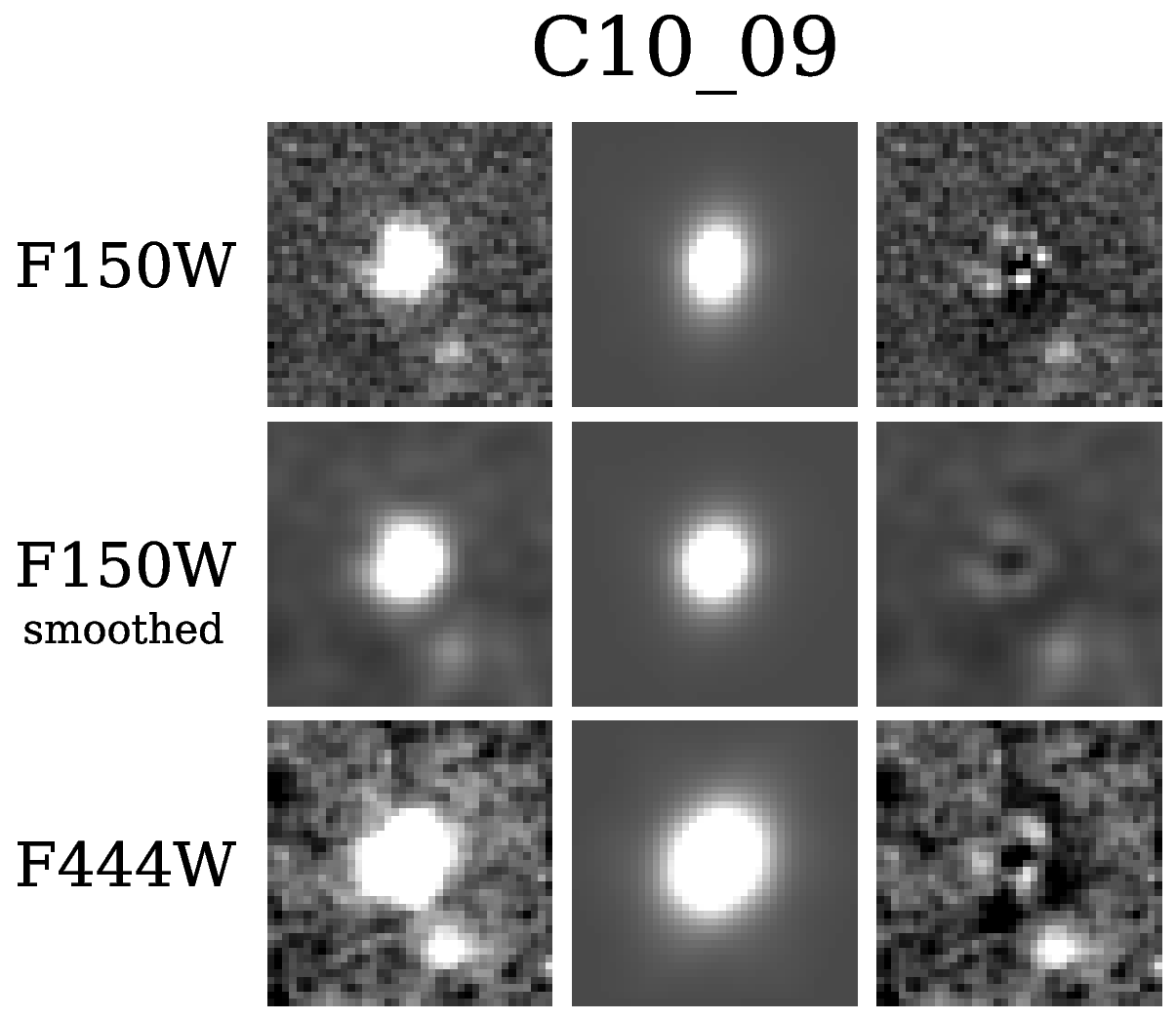}
   \includegraphics[height=0.14\textheight]{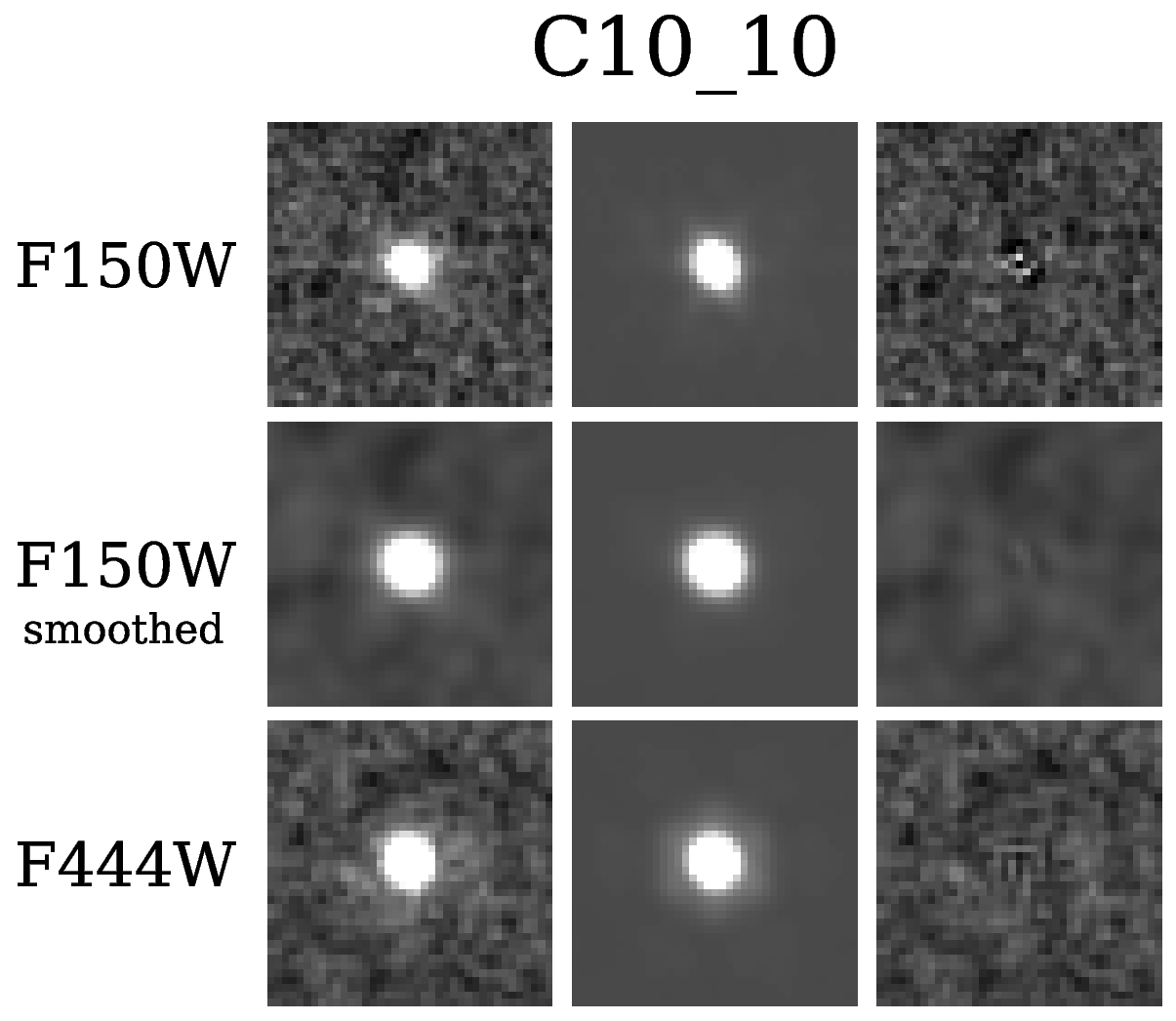}
\caption{
(Continued)
}
\end{center}
\end{figure*}


\begin{figure*}
\begin{center}
   \includegraphics[width=0.16\textwidth]{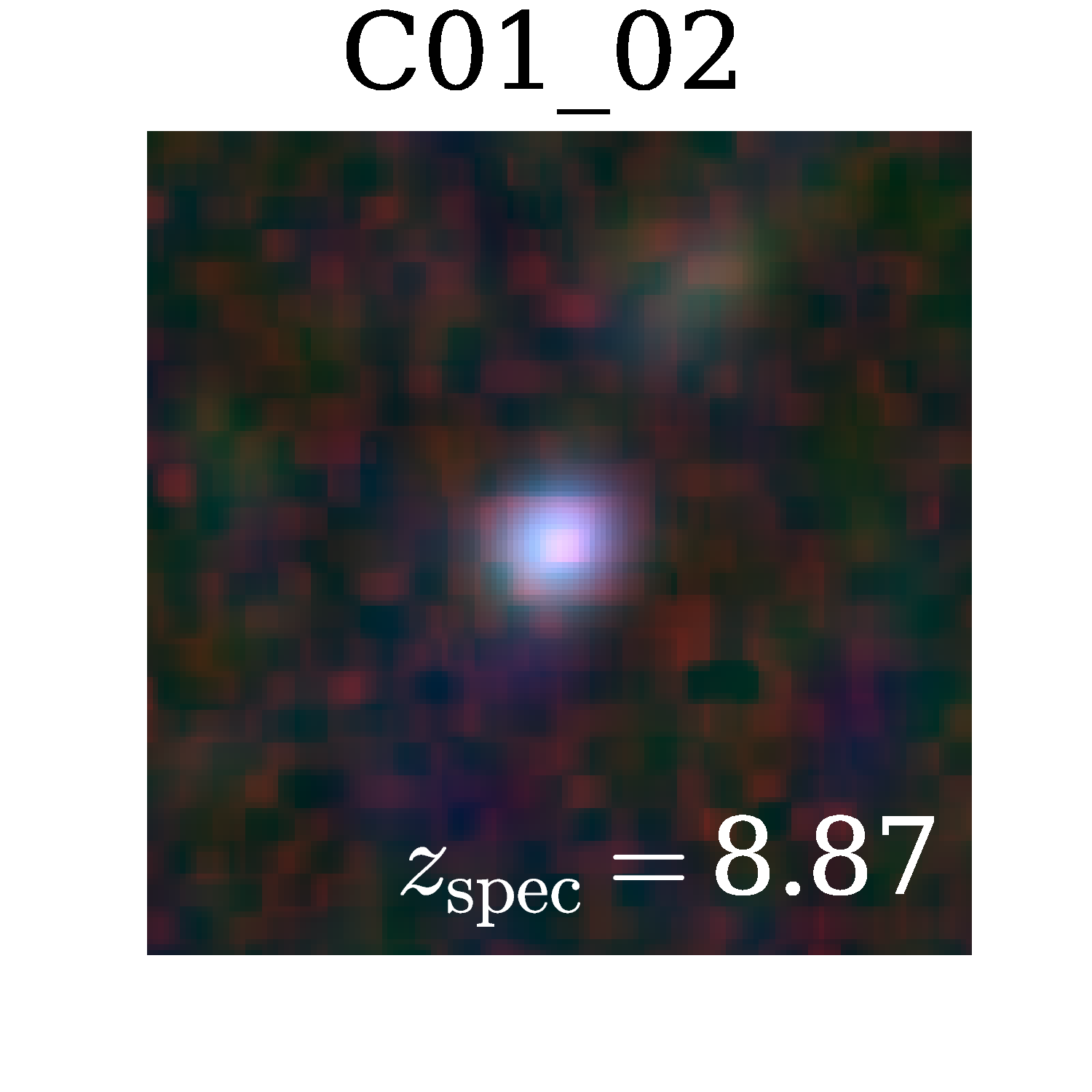}
   \includegraphics[width=0.16\textwidth]{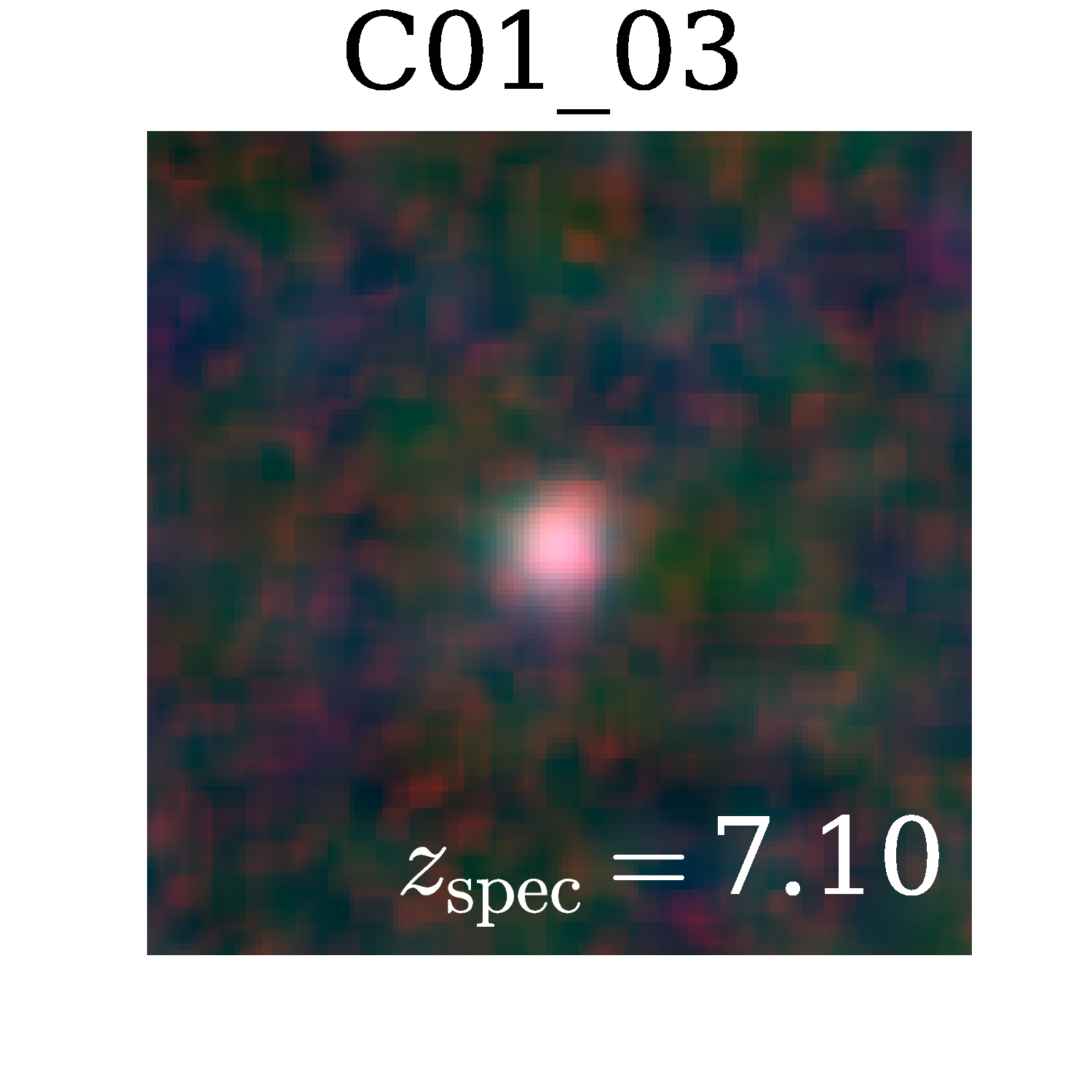}
   \includegraphics[width=0.16\textwidth]{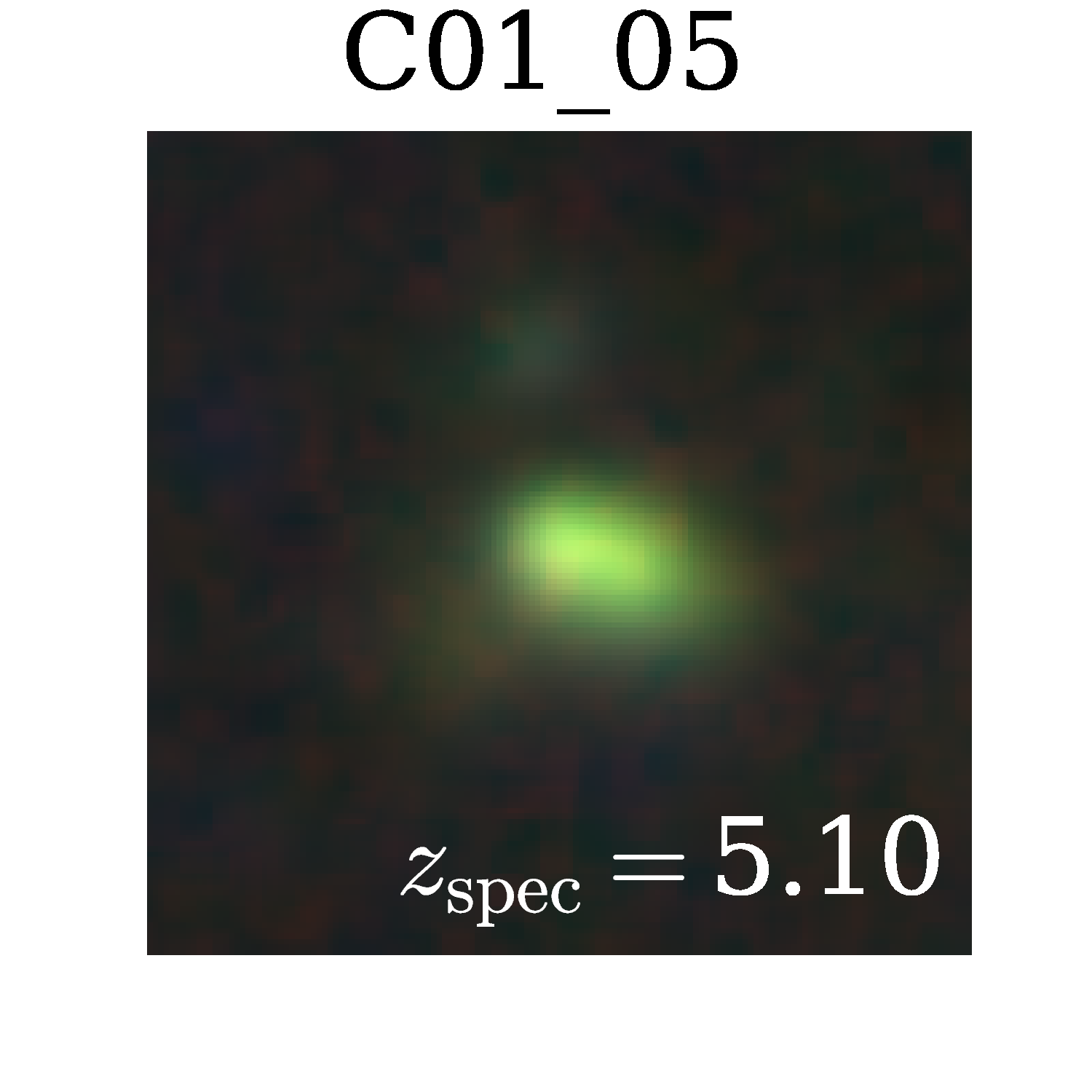}
   \includegraphics[width=0.16\textwidth]{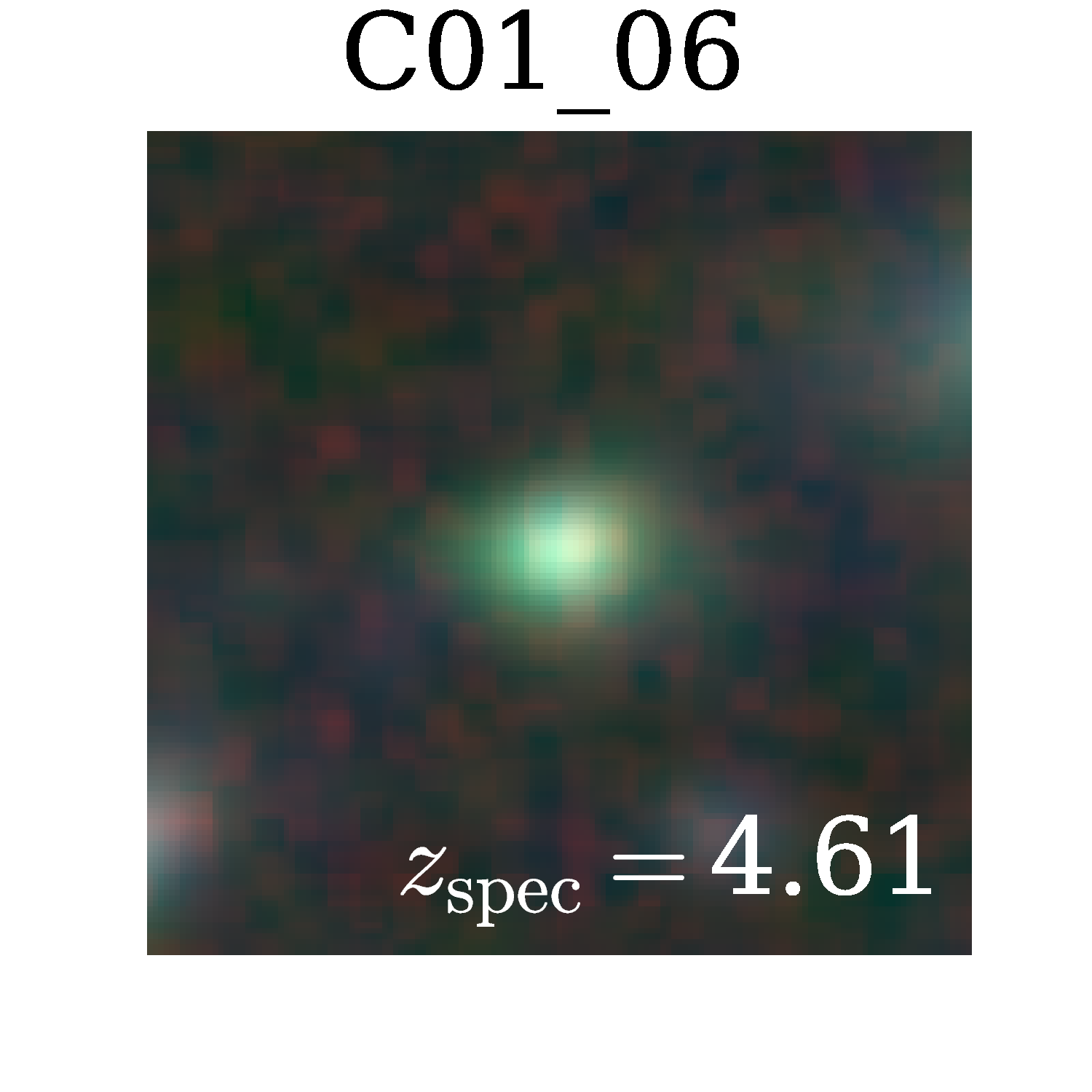}
   \includegraphics[width=0.16\textwidth]{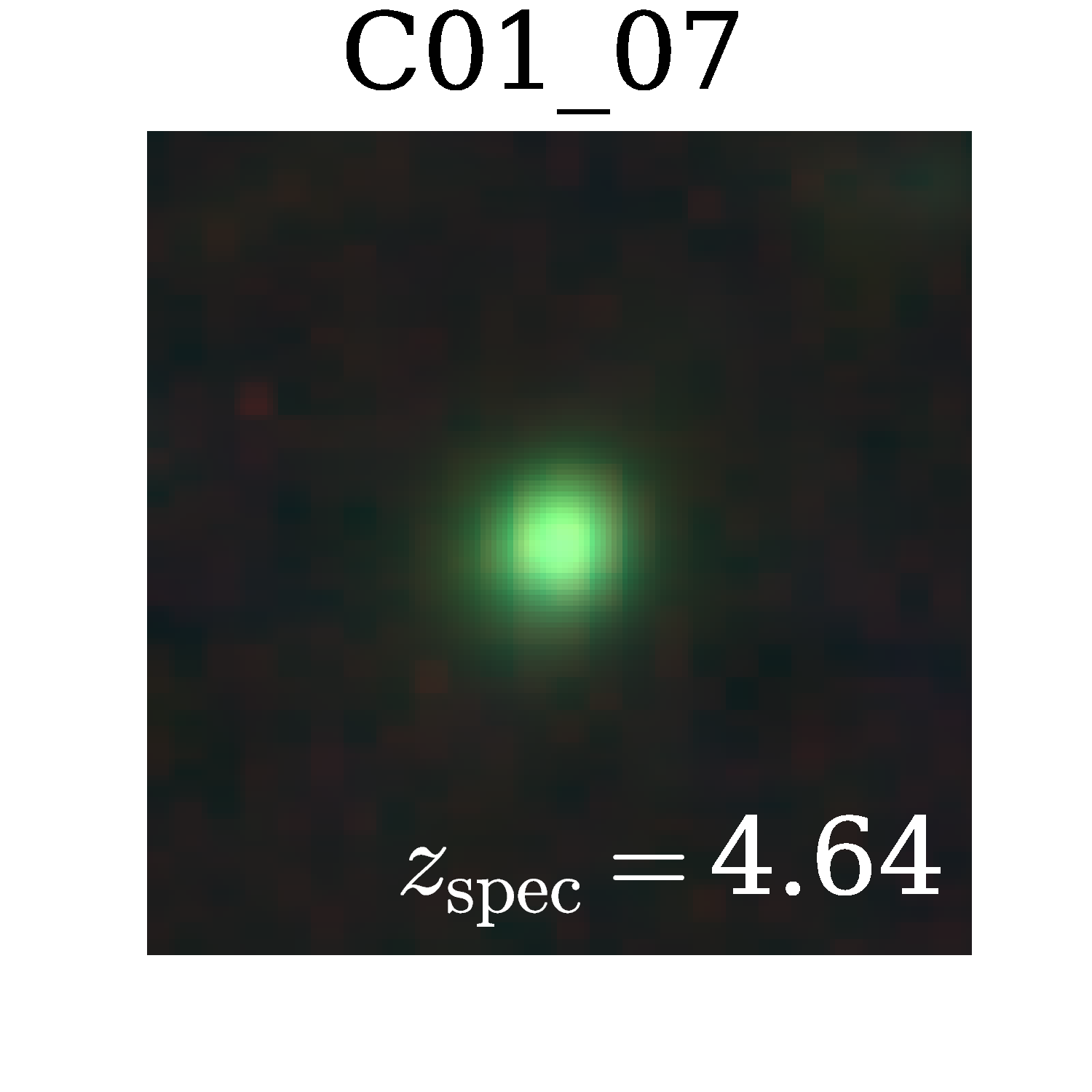}
   \includegraphics[width=0.16\textwidth]{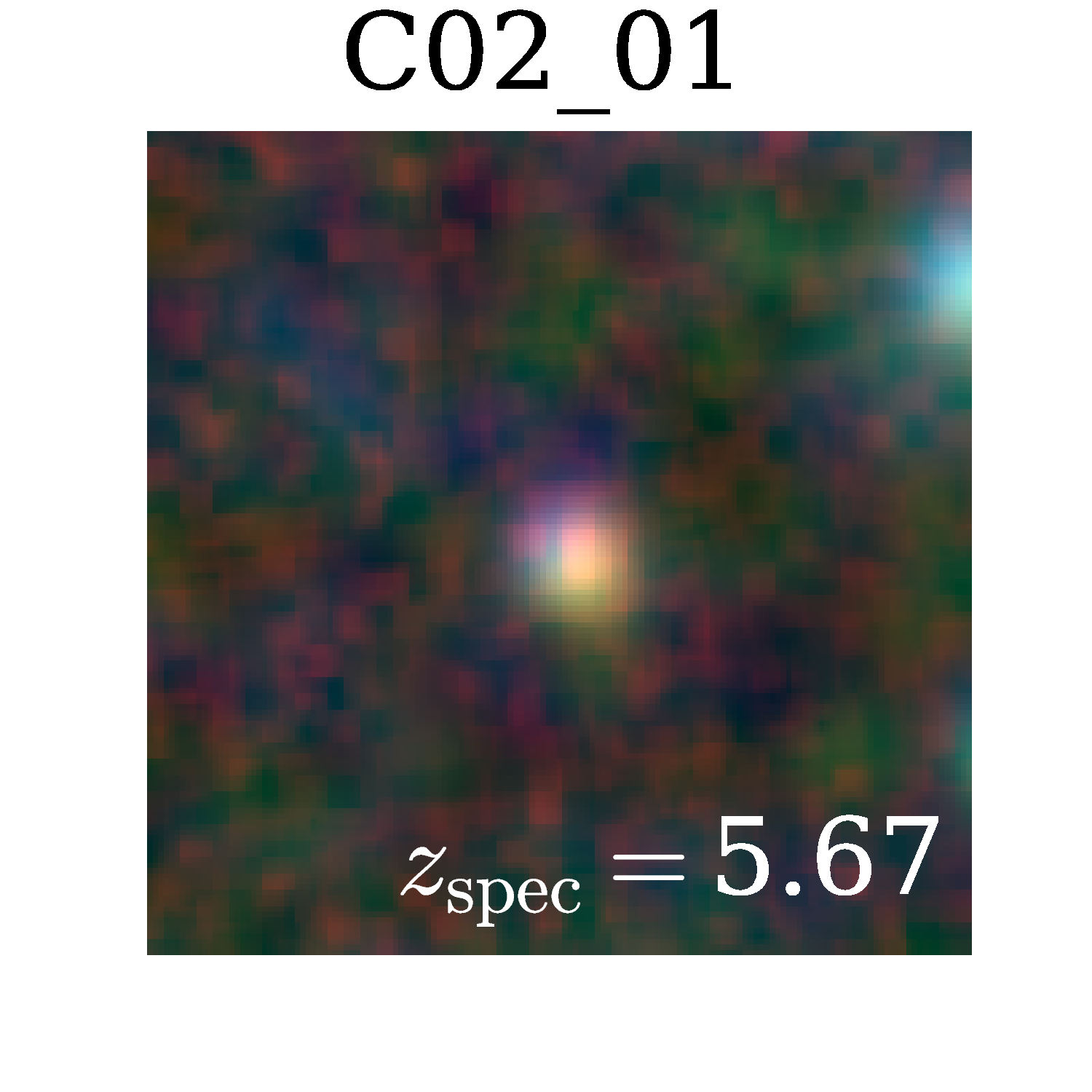}
   \includegraphics[width=0.16\textwidth]{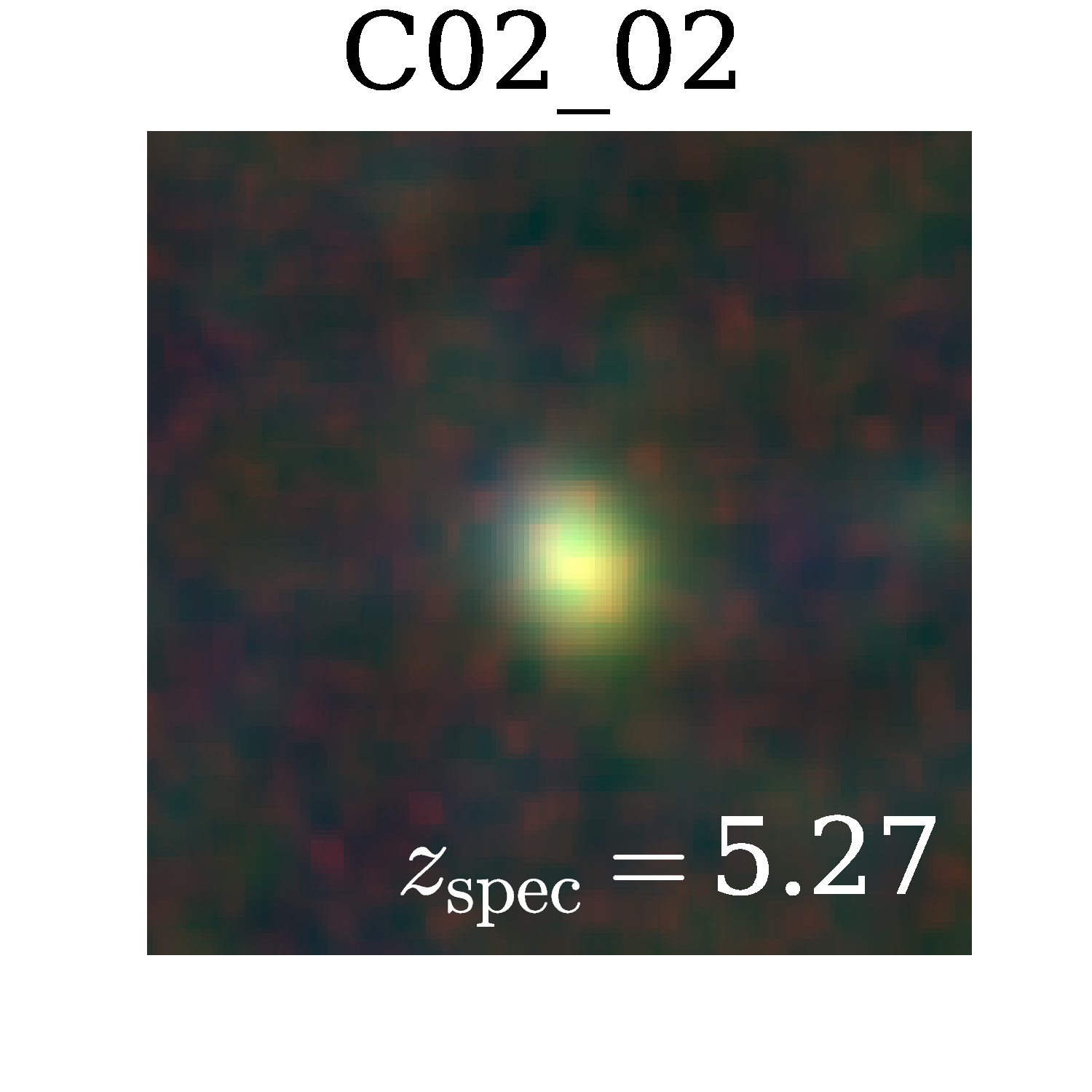}
   \includegraphics[width=0.16\textwidth]{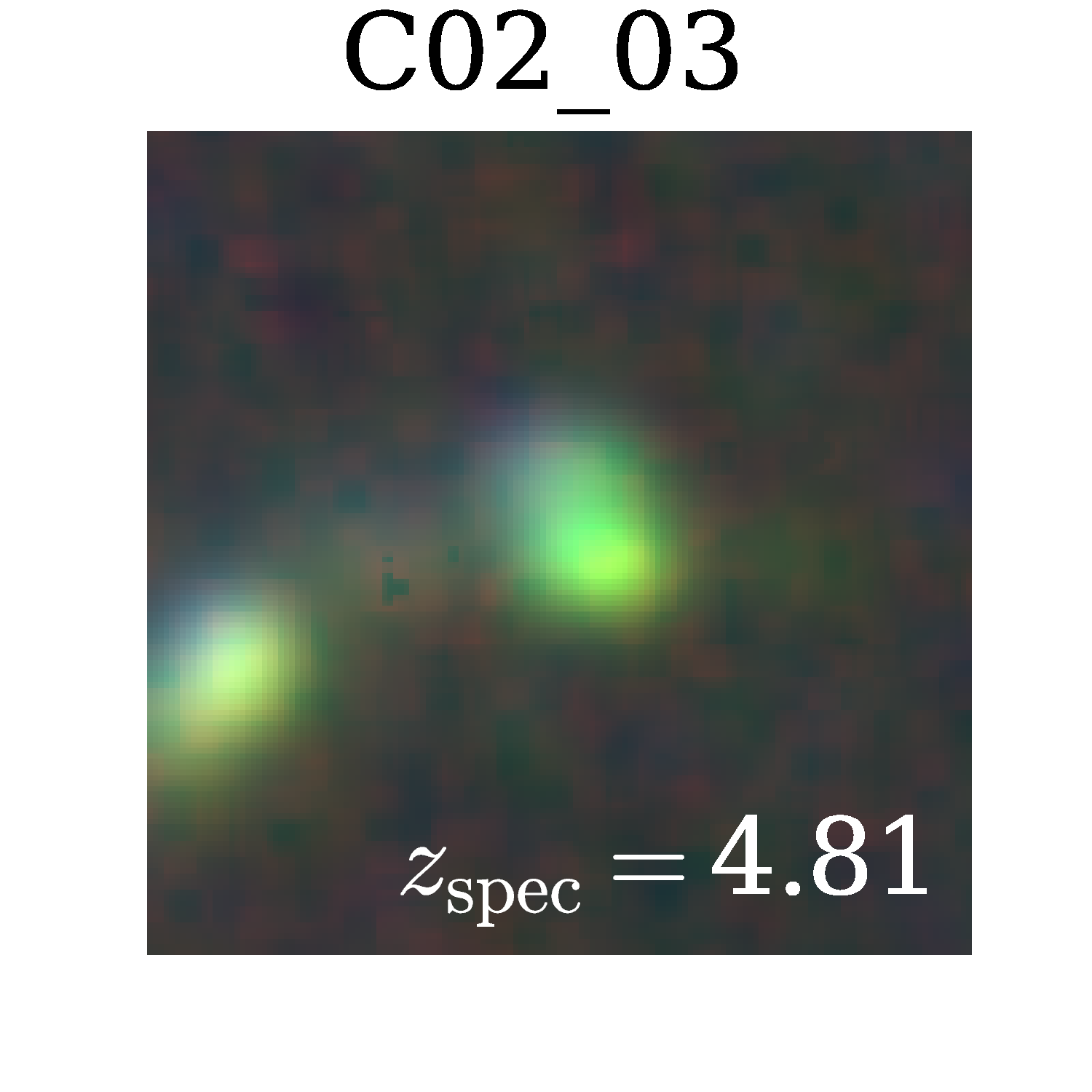}
   \includegraphics[width=0.16\textwidth]{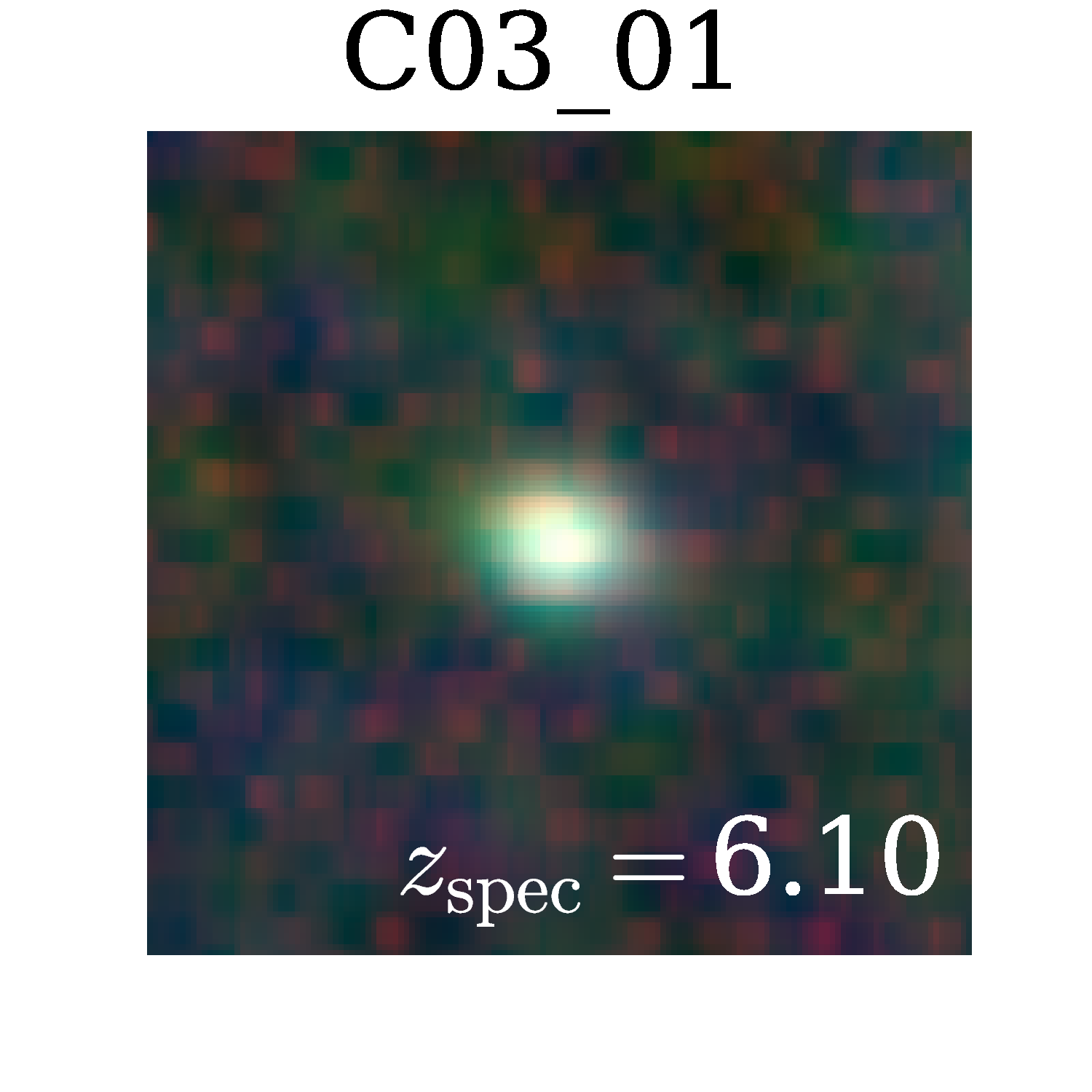}
   \includegraphics[width=0.16\textwidth]{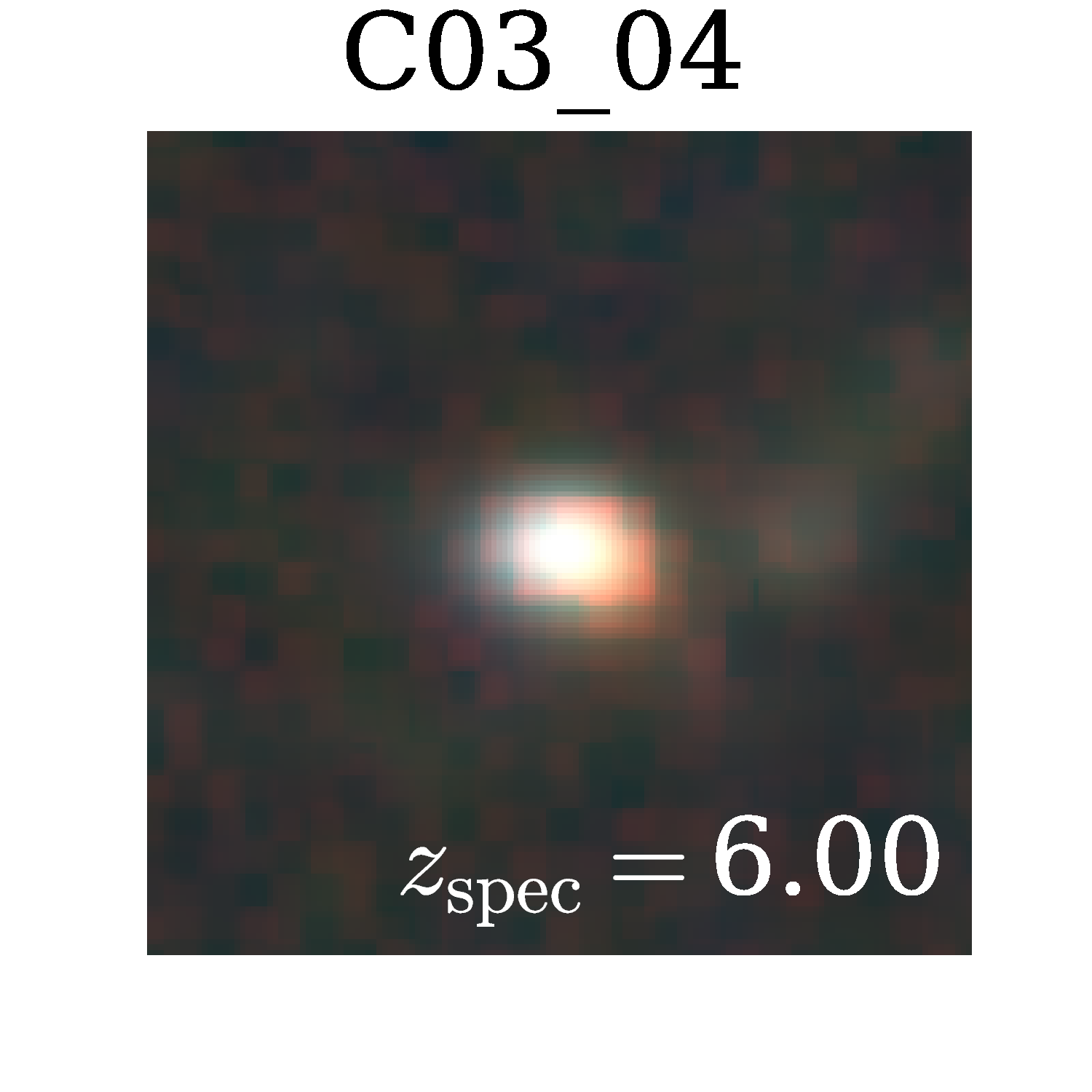}
   \includegraphics[width=0.16\textwidth]{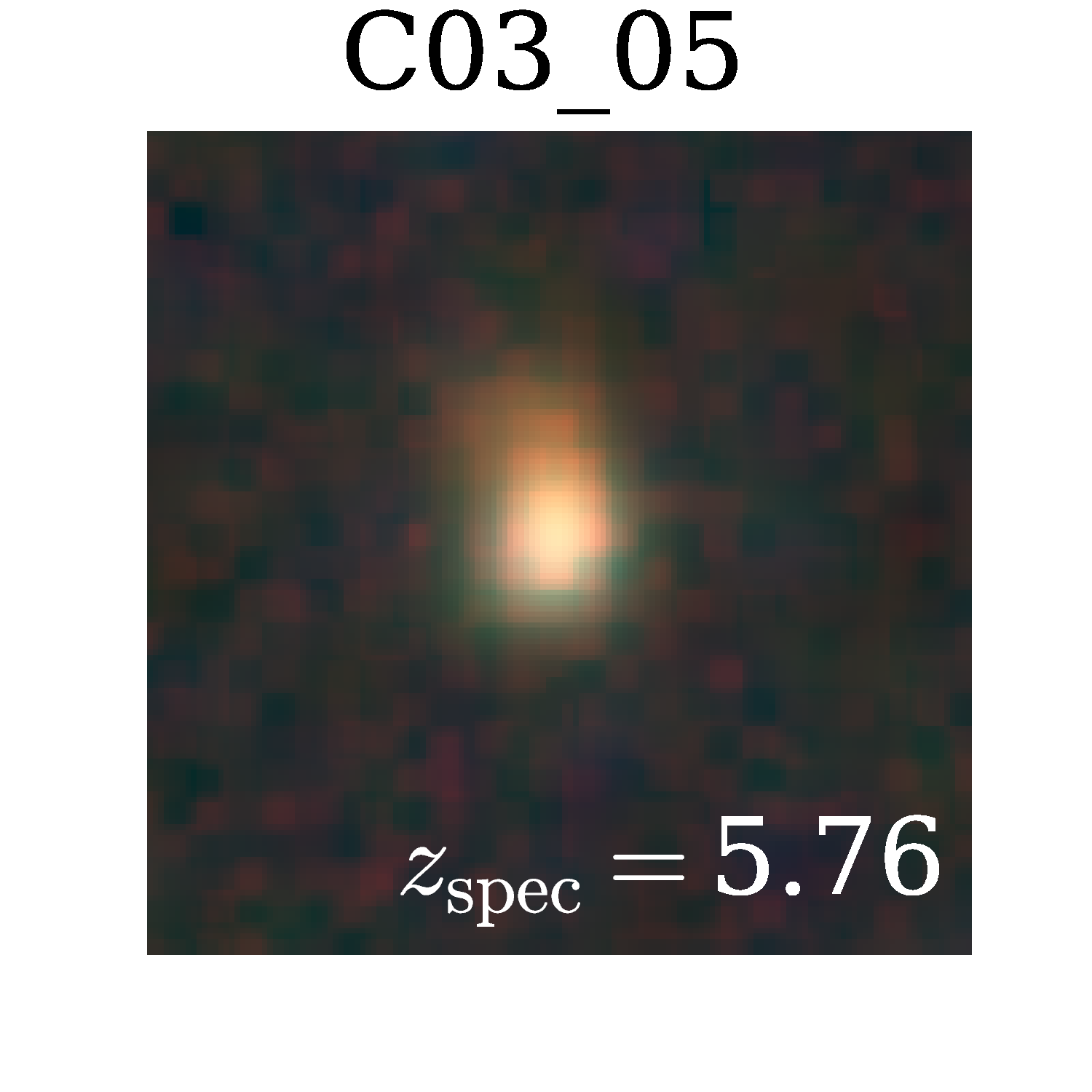}
   \includegraphics[width=0.16\textwidth]{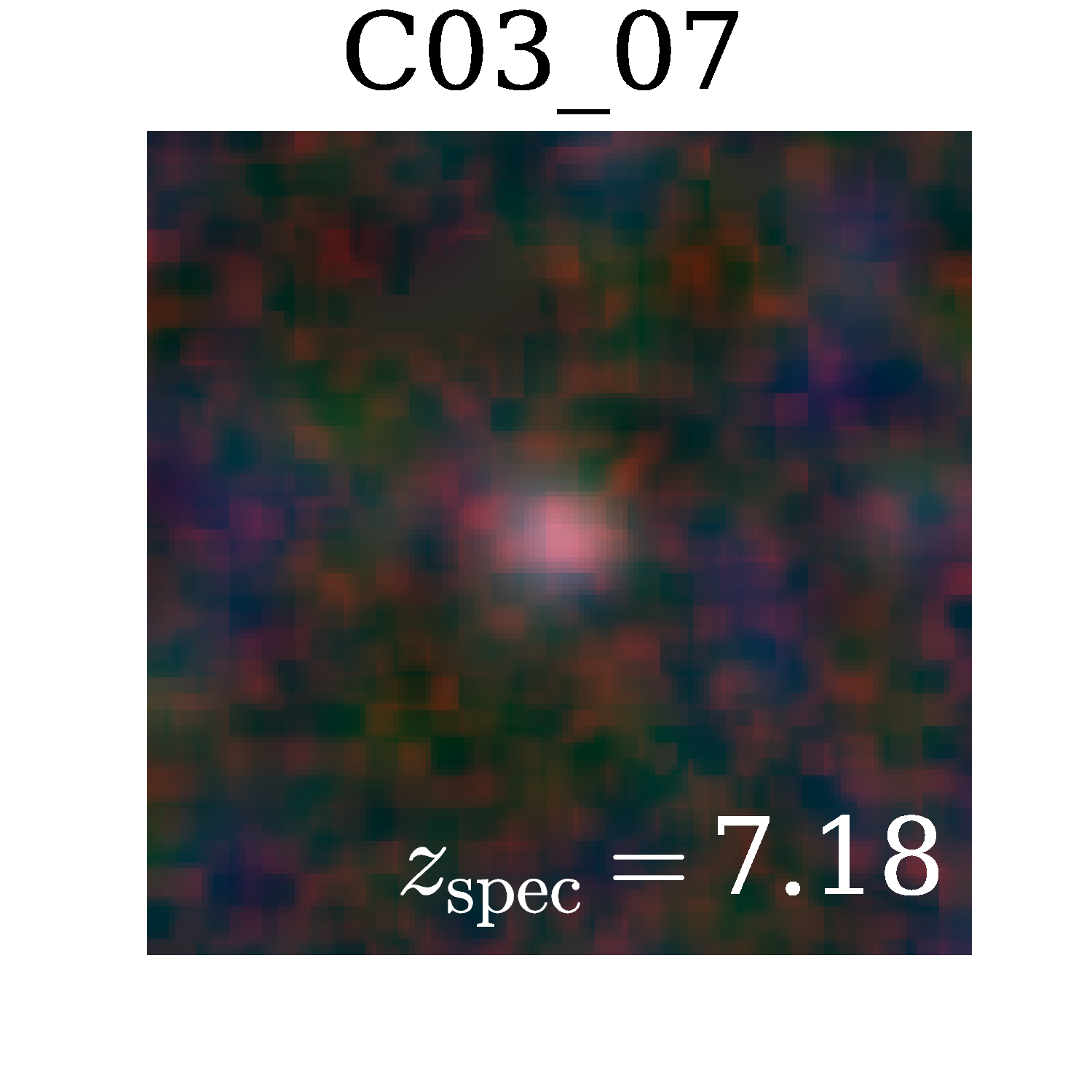}
   \includegraphics[width=0.16\textwidth]{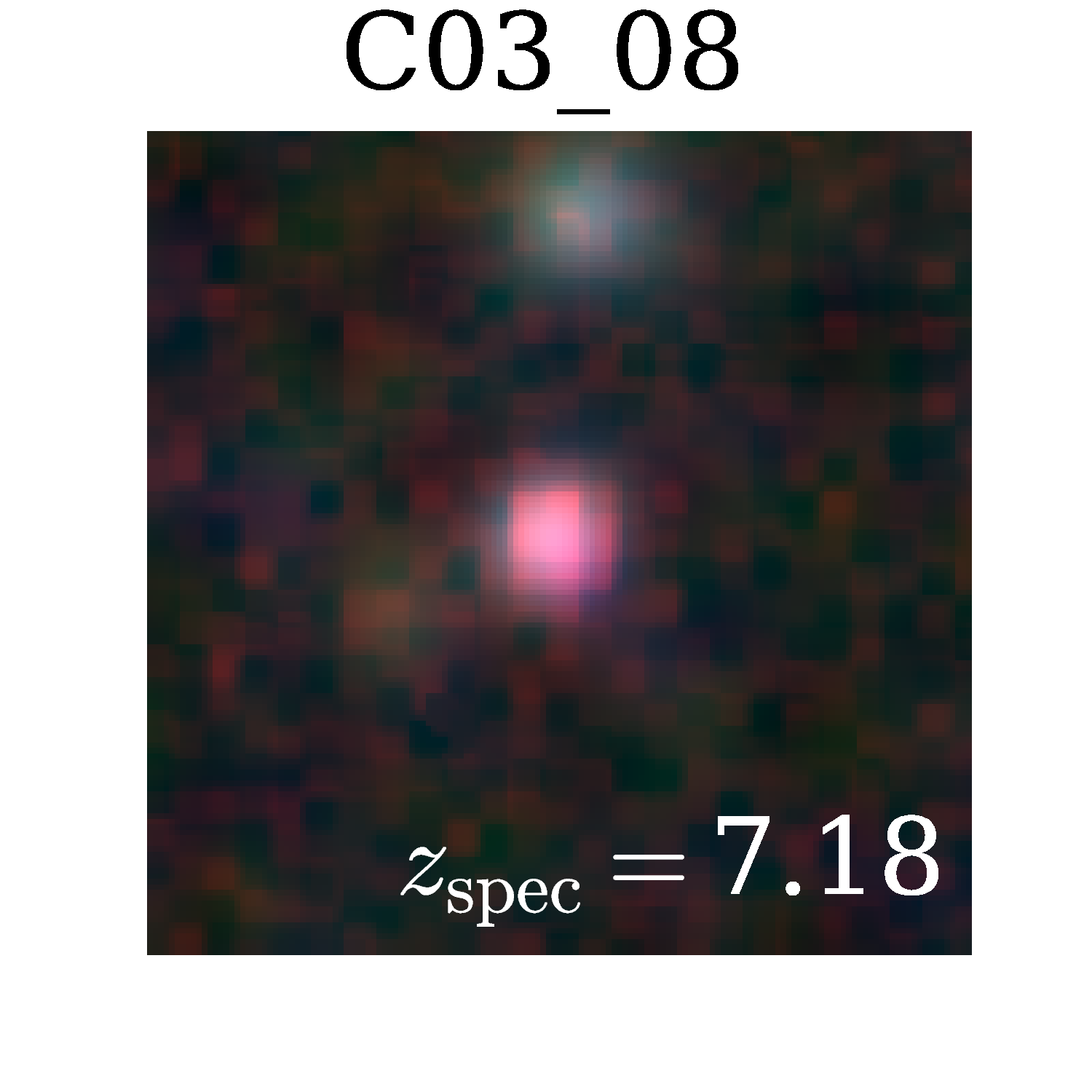}
   \includegraphics[width=0.16\textwidth]{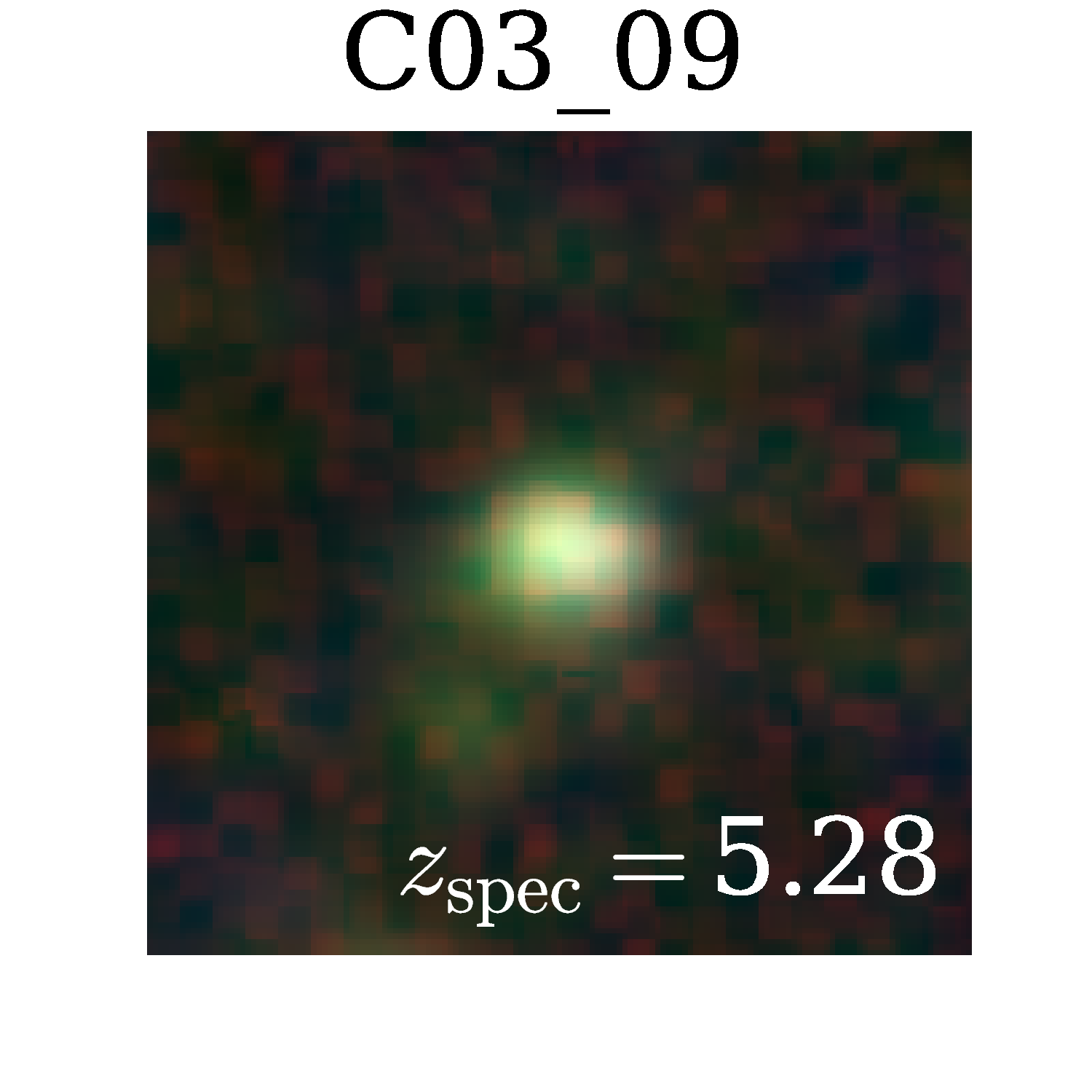}
   \includegraphics[width=0.16\textwidth]{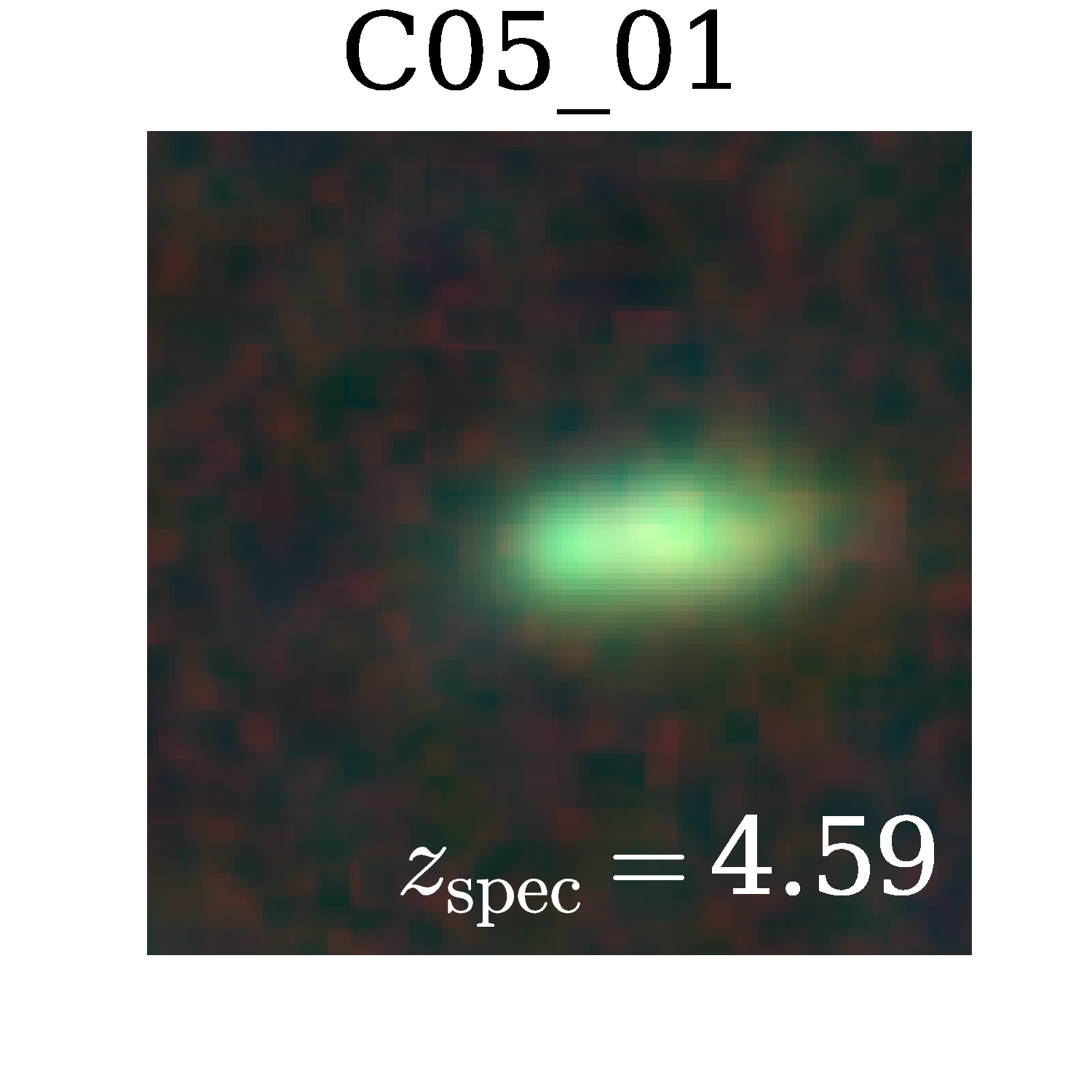}
   \includegraphics[width=0.16\textwidth]{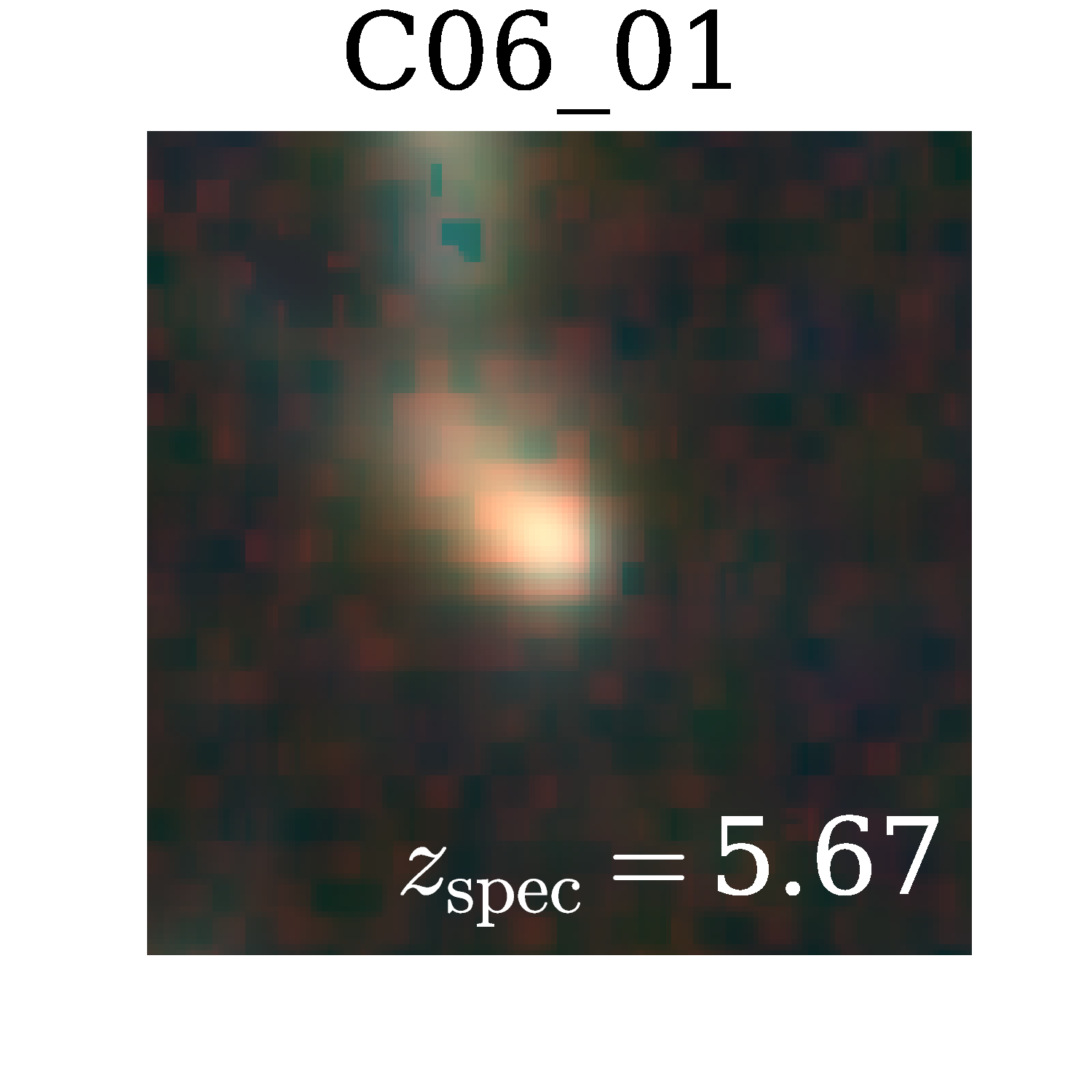}
   \includegraphics[width=0.16\textwidth]{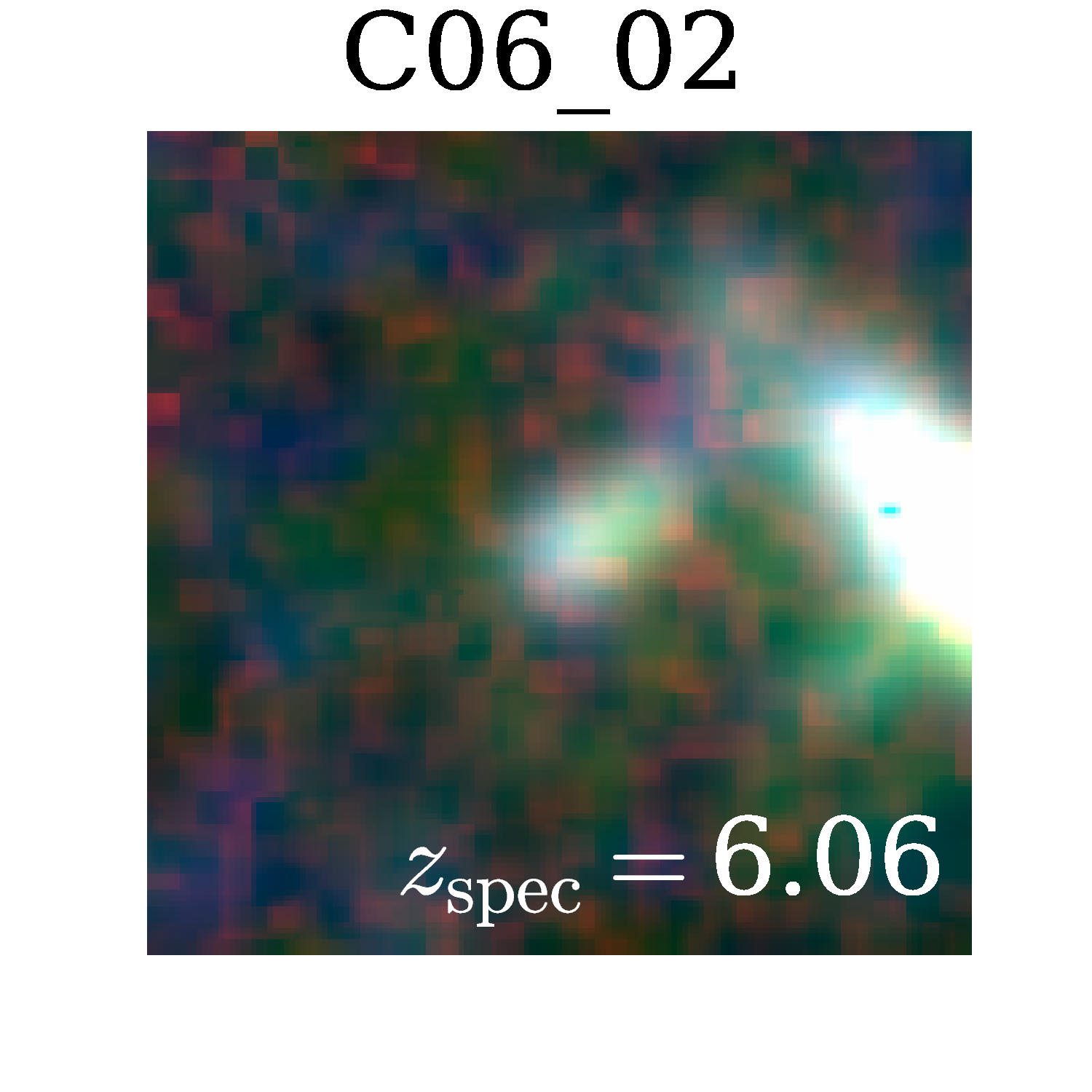}
   \includegraphics[width=0.16\textwidth]{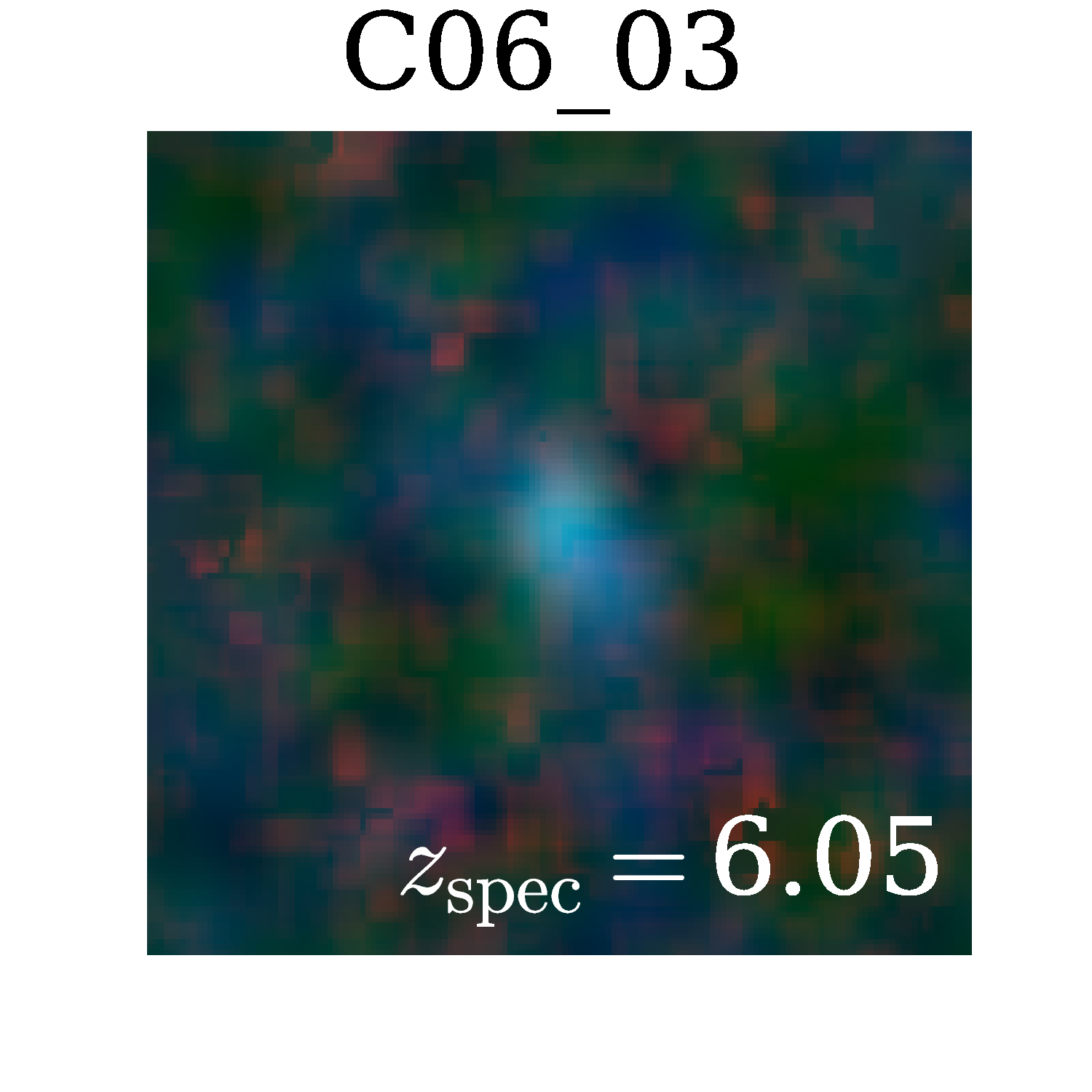}
   \includegraphics[width=0.16\textwidth]{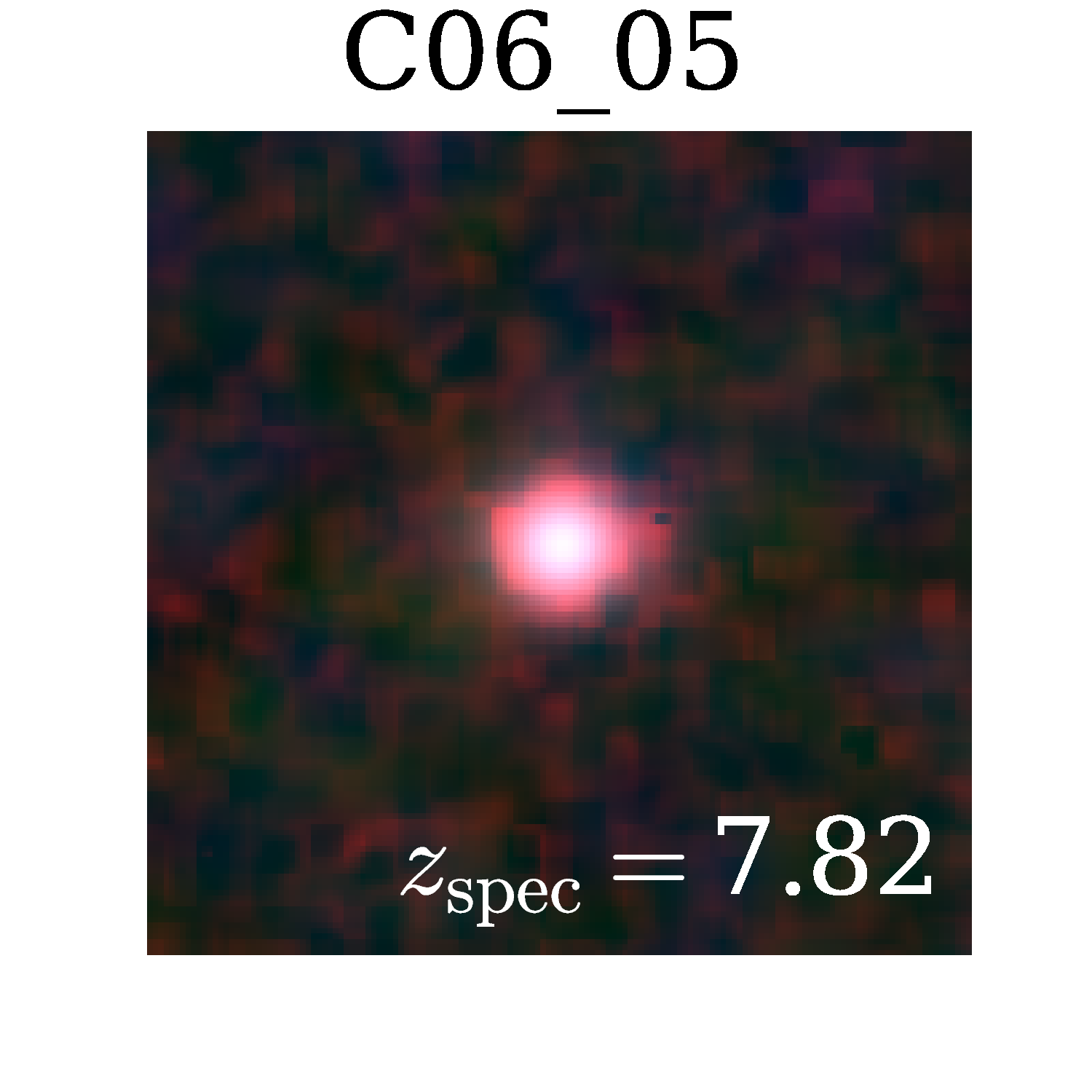}
   \includegraphics[width=0.16\textwidth]{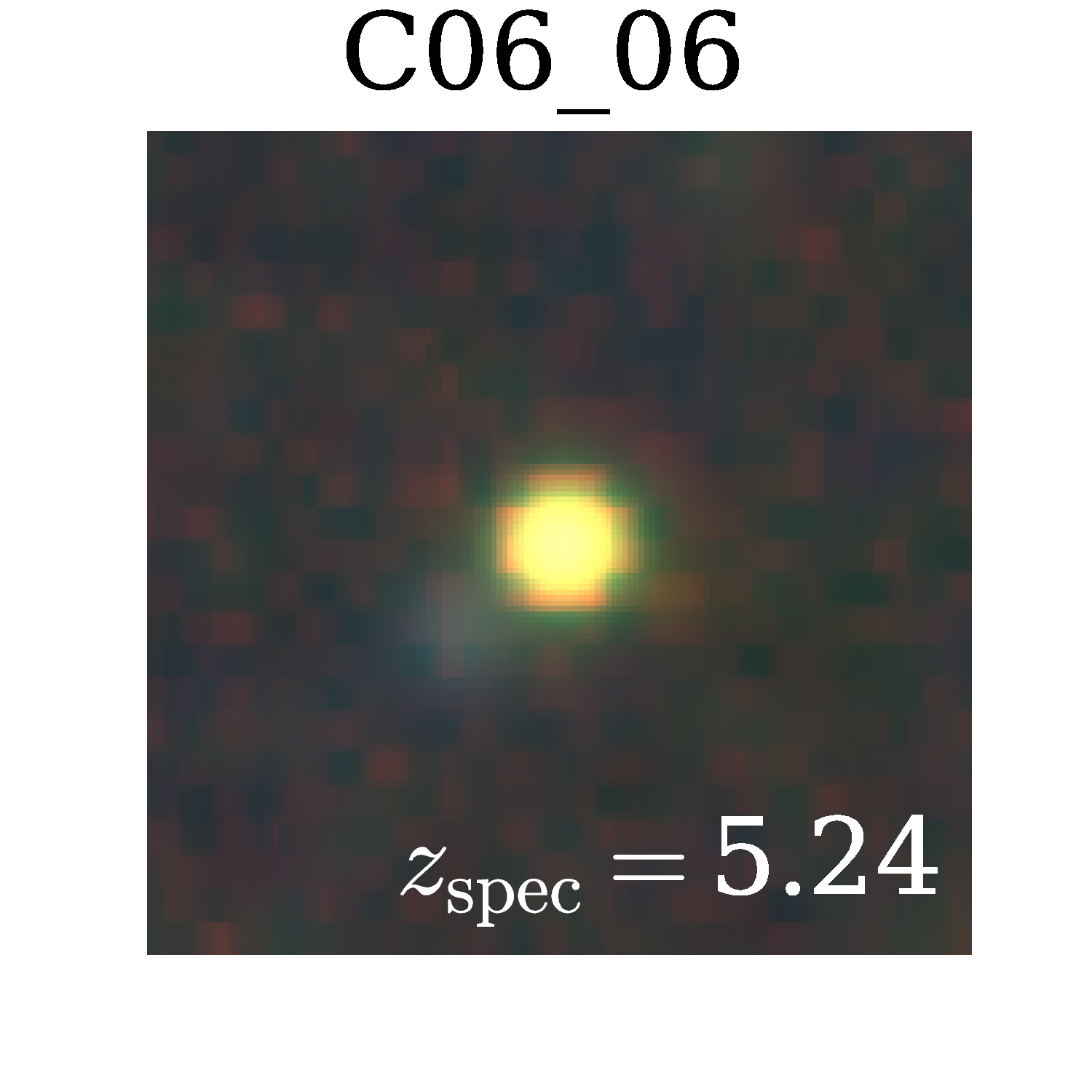}
   \includegraphics[width=0.16\textwidth]{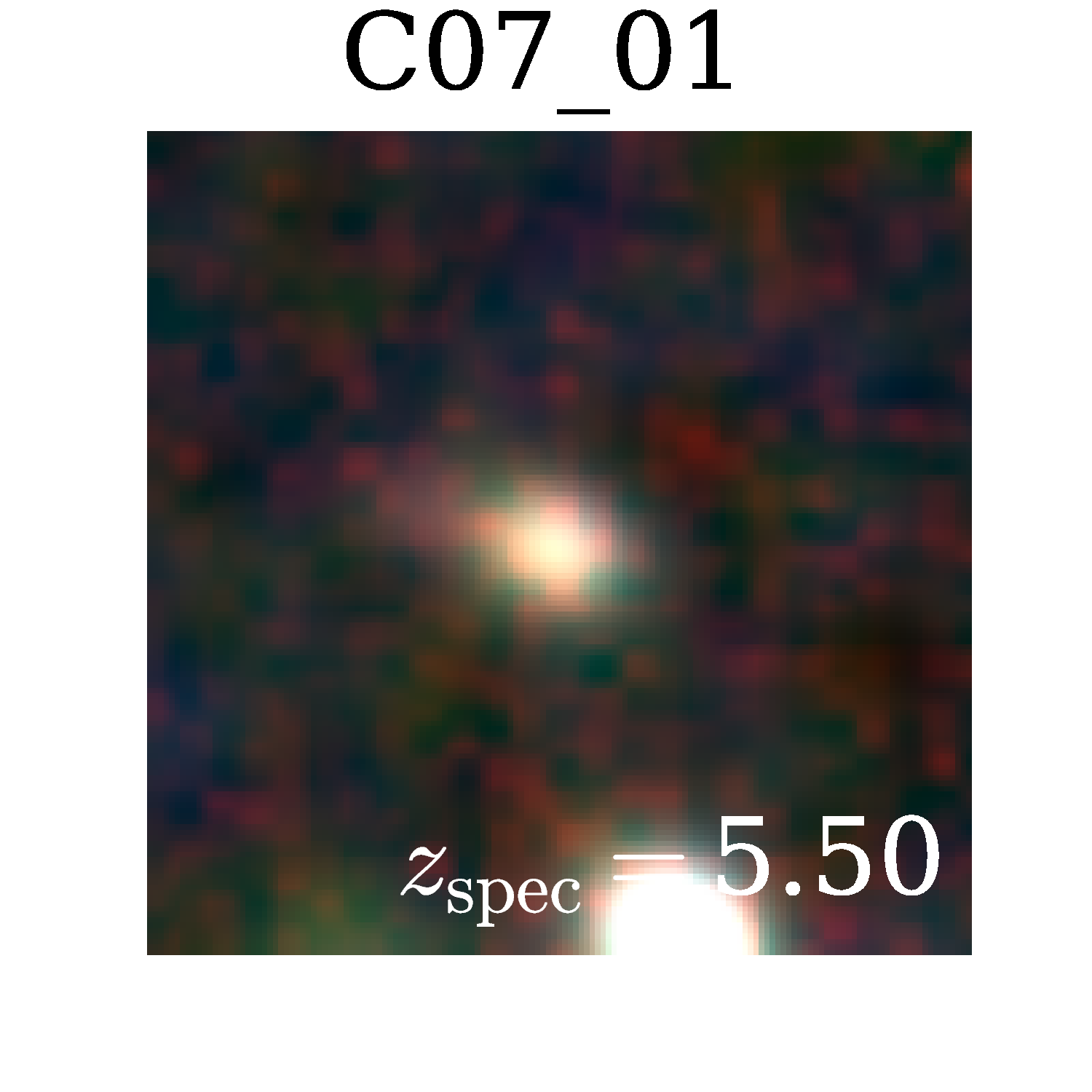}
   \includegraphics[width=0.16\textwidth]{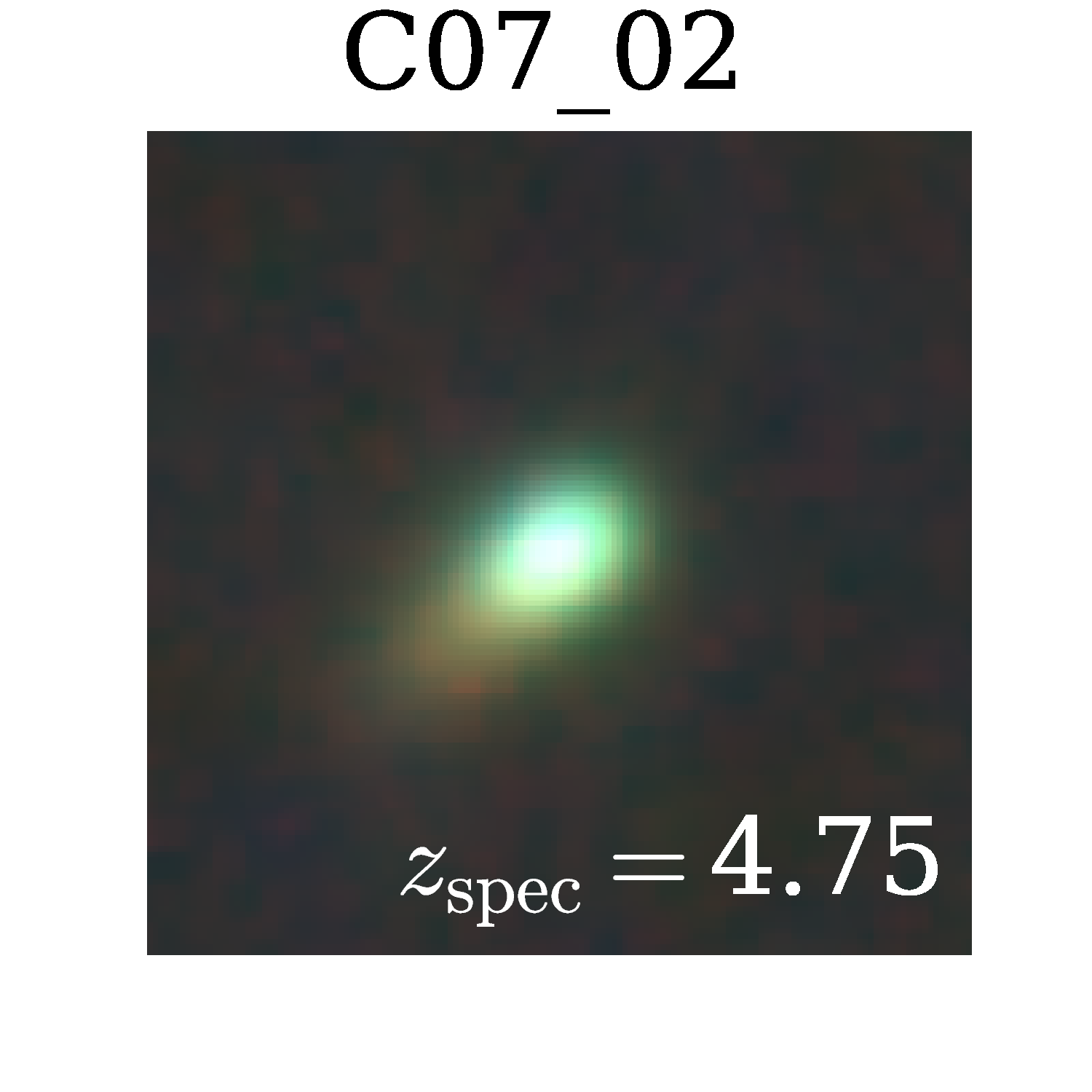}
   \includegraphics[width=0.16\textwidth]{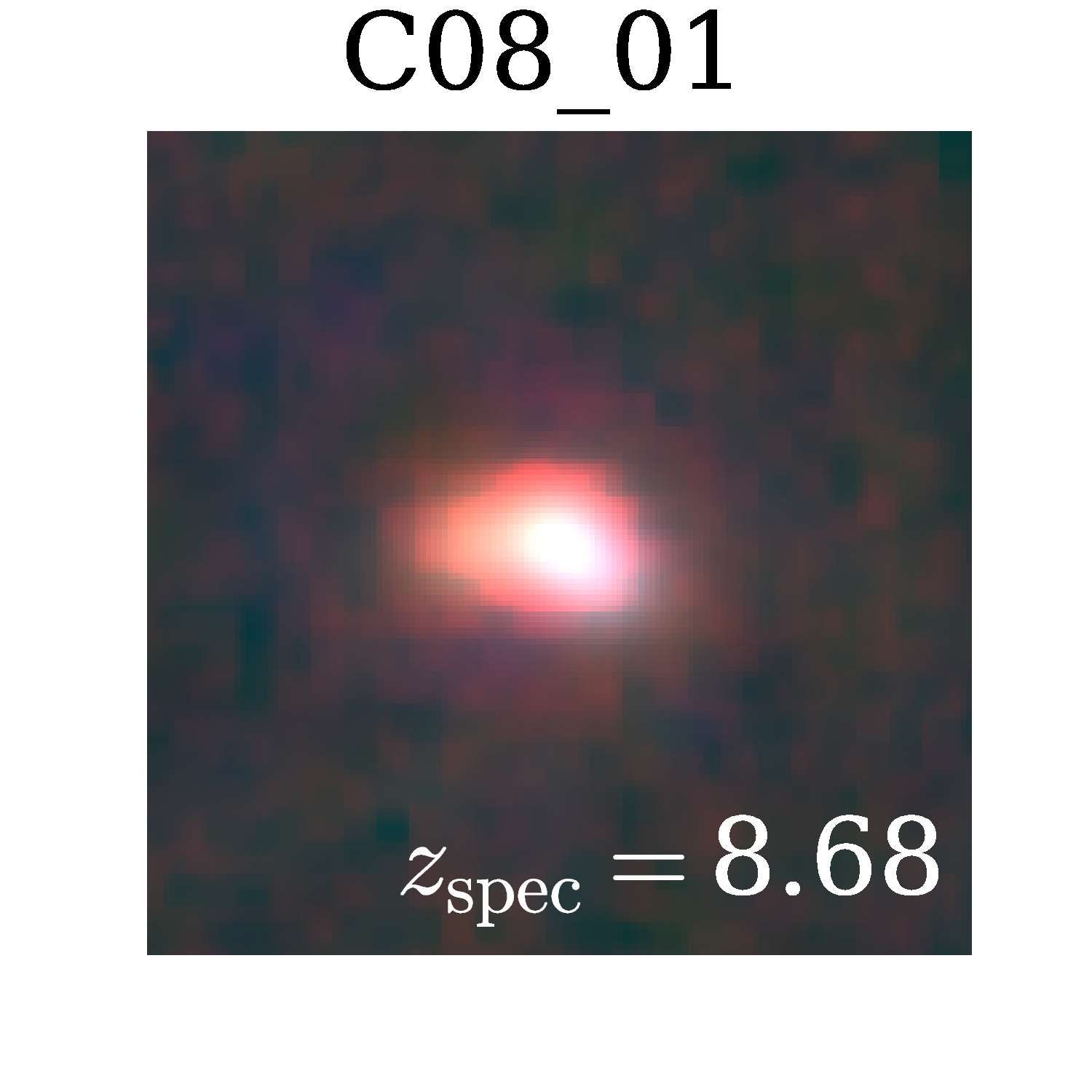}
   \includegraphics[width=0.16\textwidth]{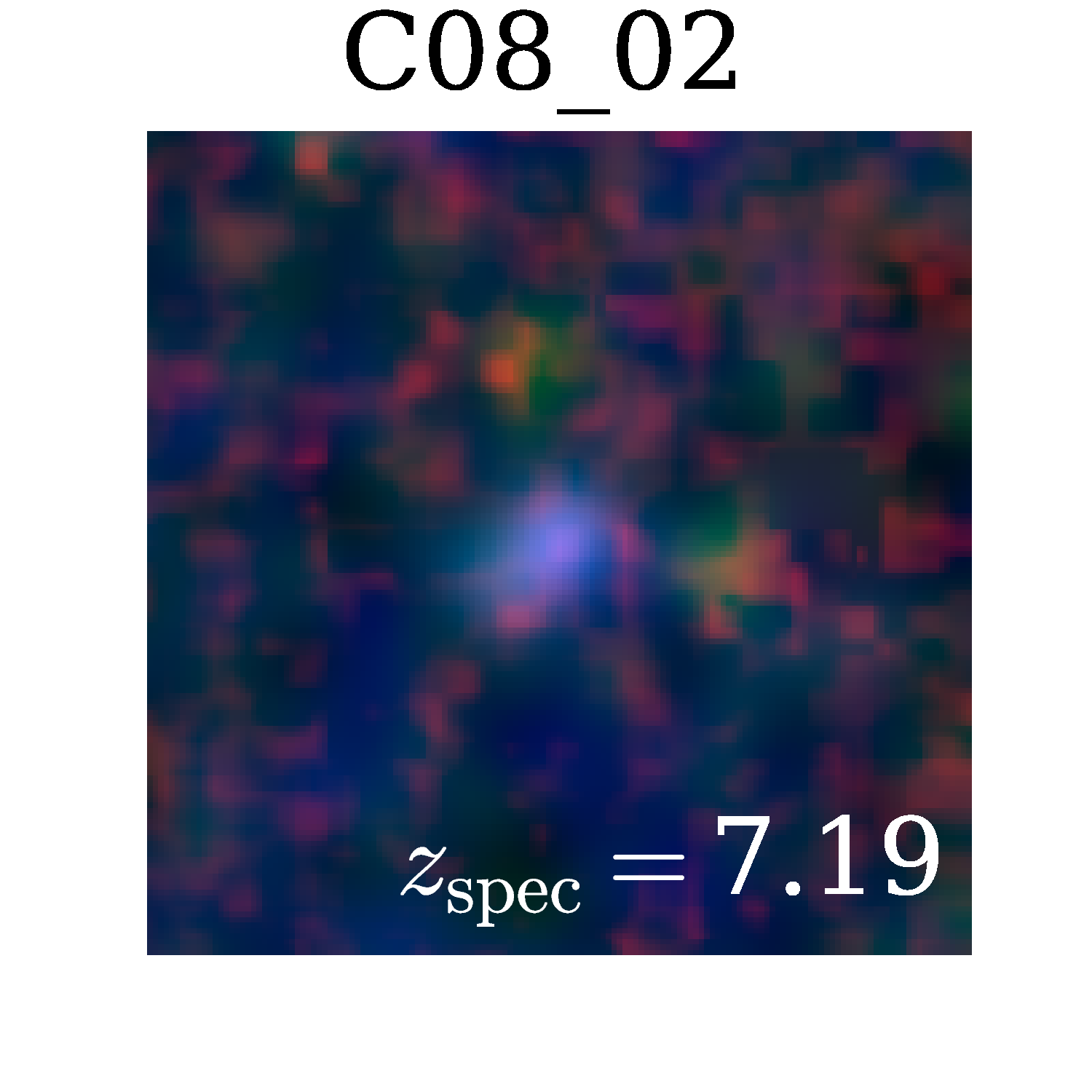}
   \includegraphics[width=0.16\textwidth]{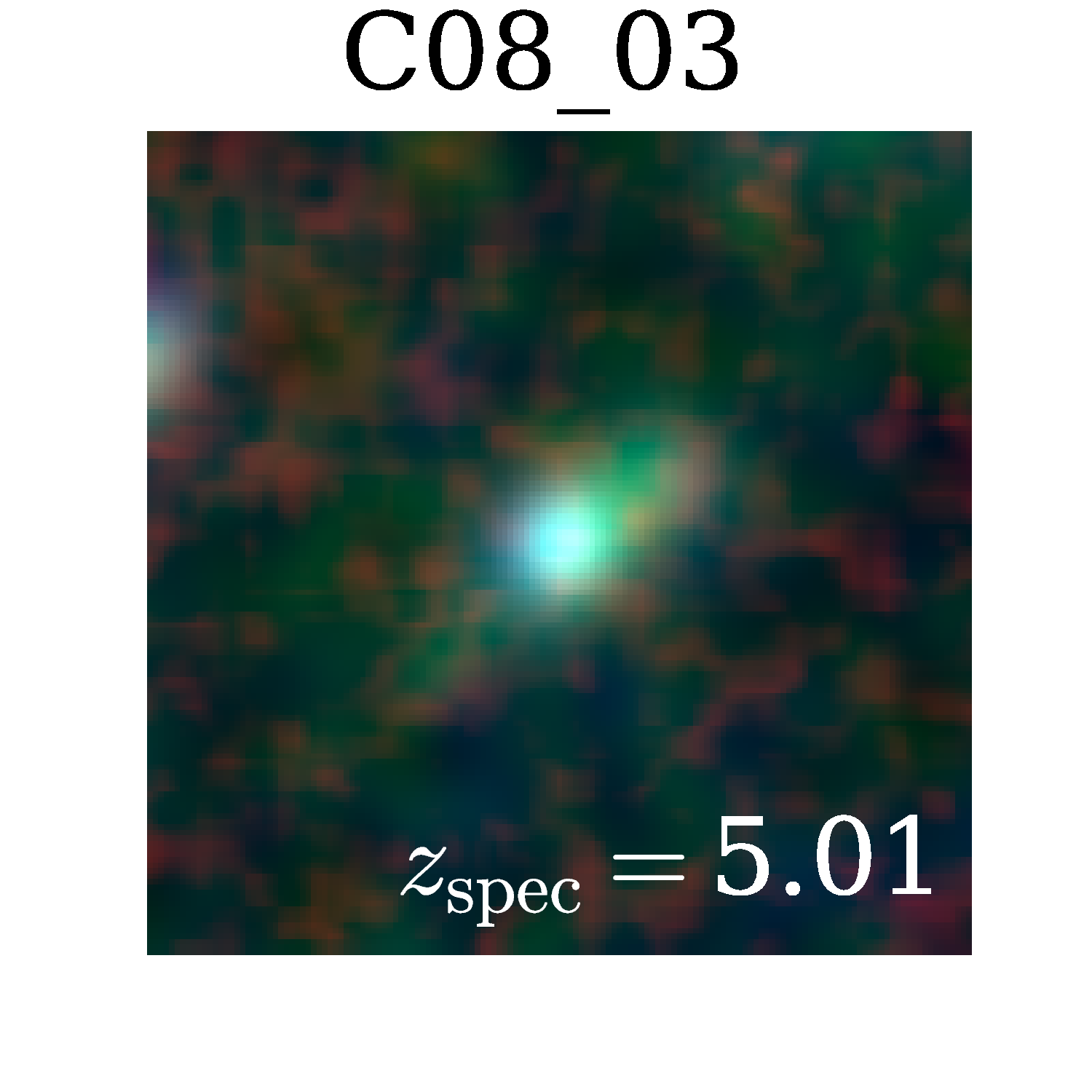}
   \includegraphics[width=0.16\textwidth]{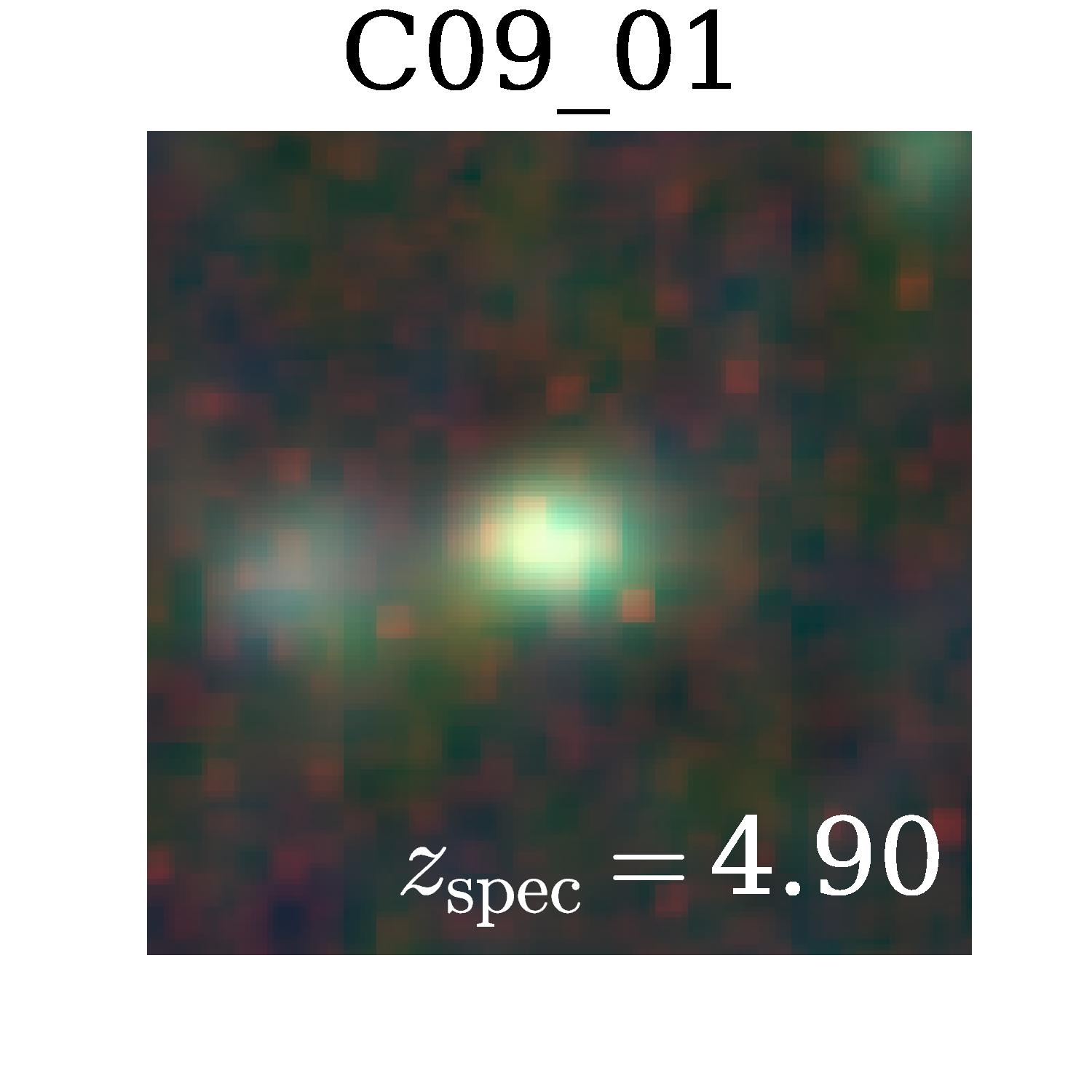}
   \includegraphics[width=0.16\textwidth]{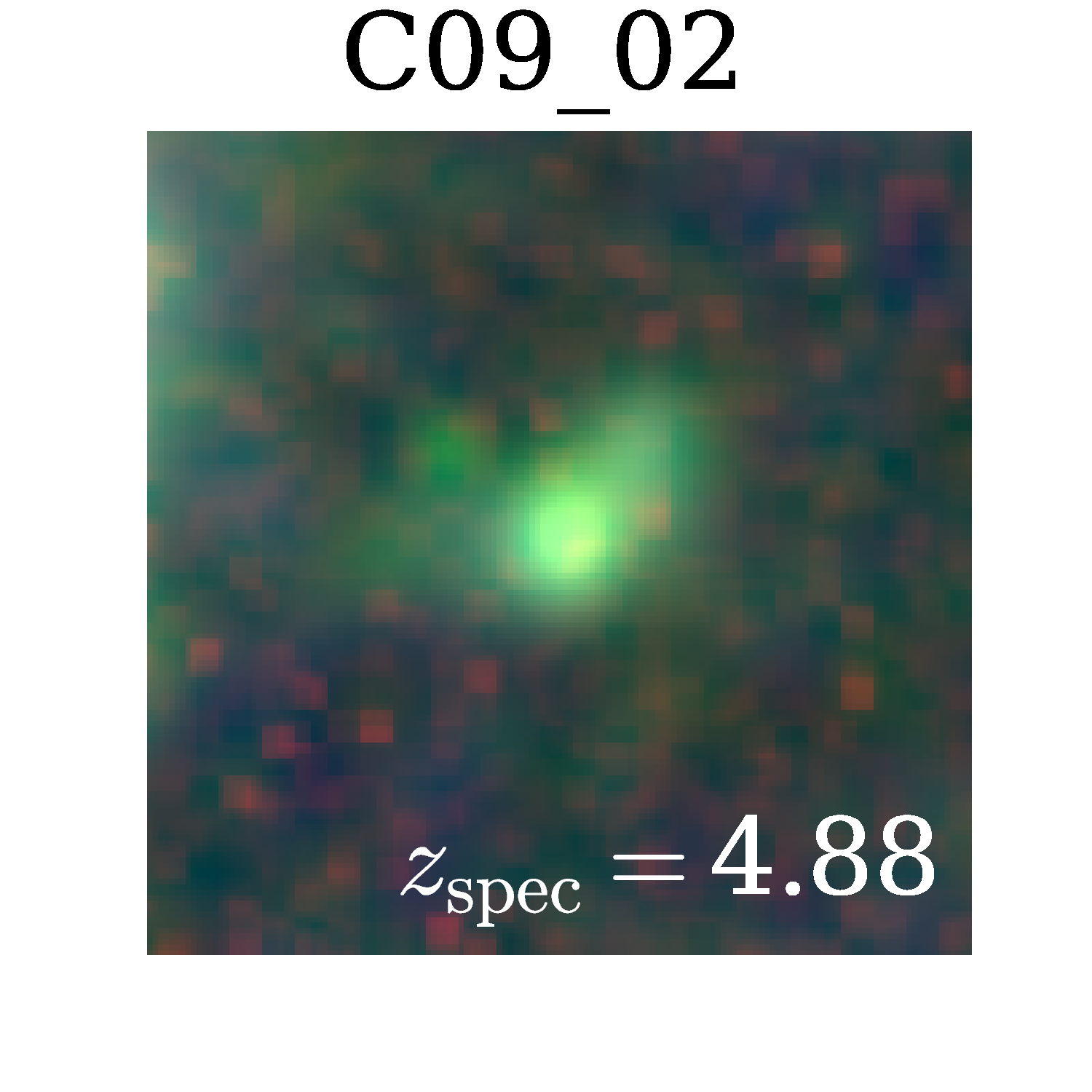}
   \includegraphics[width=0.16\textwidth]{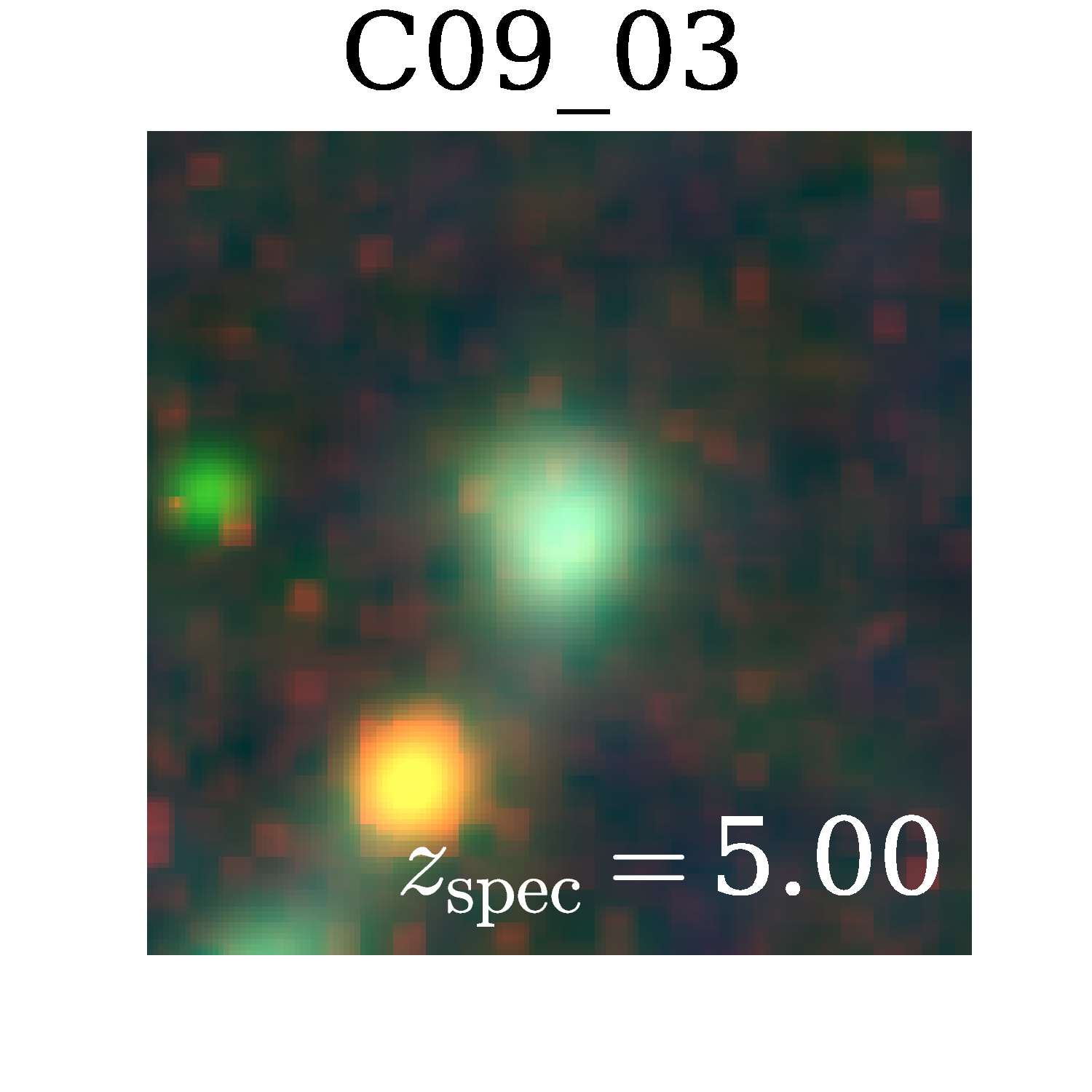}
   \includegraphics[width=0.16\textwidth]{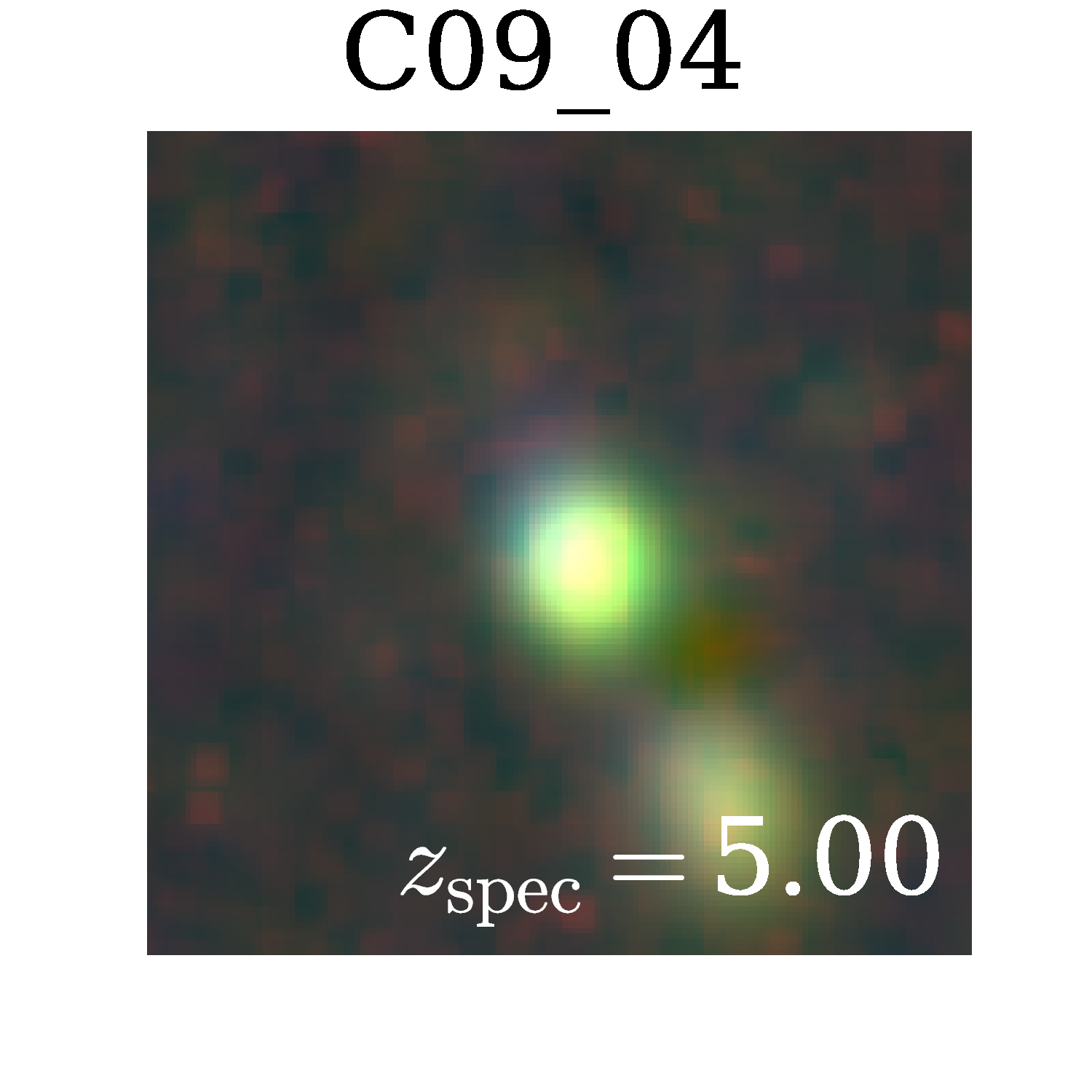}
\caption{
Pseudo-color images of spectroscopically confirmed galaxies in our sample 
whose sizes are measured in our SB profile fittings. 
The F150W image is assigned to blue, the F277W image to green, and the F444W image to red, 
after their PSFs are matched. 
The size of each image is $1\farcs5 \times 1\farcs5$.
}
\label{fig:PseudoColors}
\end{center}
\end{figure*}


\end{document}